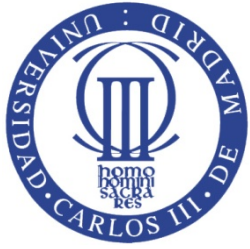

# UNIVERSIDAD CARLOS III DE MADRID

TESIS DOCTORAL

## Contribuciones a las Comunicaciones Ópticas en Espacio Libre: Utilización de Telescopios *Cherenkov* como Receptores y Corrección de *Beam Wander* en Comunicaciones Cuánticas

Autor:

**Alberto Carrasco Casado**

Director:

**José Manuel Sánchez Pena[1]**

Codirectores:

**Ricardo Vergaz Benito[1]**
**Verónica Fernández Mármol[2]**

1. Departamento de Tecnología Electrónica
   *Universidad Carlos III de Madrid*

2. Instituto de Tecnologías Físicas y de la Información
   *Consejo Superior de Investigaciones Científicas*

Leganés, 11 de junio de 2015

Universidad
Carlos III de Madrid
www.uc3m.es

TESIS DOCTORAL

# Contribuciones a las Comunicaciones Ópticas en Espacio Libre: Utilización de Telescopios *Cherenkov* como Receptores y Corrección de *Beam Wander* en Comunicaciones Cuánticas

Autor:
**Alberto Carrasco Casado**

Director:
**José Manuel Sánchez Pena**

Codirectores:
**Ricardo Vergaz Benito**
**Verónica Fernández Mármol**

Firma del Tribunal Calificador:

Firma

Presidente:   Santiago Mar Sardaña

Vocal:        Vicente Martín Ayuso

Secretario:   David Sánchez Montero

Calificación:

Leganés, 11 de junio de 2015

*The late invention of telescopes has so exercised*
*most of the geometers that they seem*
*to have left nothing unattempted in Optics,*
*no room for further improvements.*

Isaac Newton

*Optical Lectures* (1728)

# Resumen

Actualmente existe un consenso en que las comunicaciones ópticas en espacio libre constituirán la próxima generación de enlaces inalámbricos de alta velocidad en varios escenarios estratégicos. Esta tesis se centra en los dos donde esta tecnología puede producir un mayor impacto: comunicaciones interplanetarias y comunicaciones cuánticas. Consecuentemente, este trabajo está estructurado en dos bloques. En el primero, se ofrece una propuesta novedosa relacionada con la utilización de telescopios *Cherenkov* como estaciones receptoras. Las señales que se recibirán en la Tierra desde sondas en espacio profundo en futuras implementaciones de comunicaciones ópticas serán extremadamente débiles y se requerirán nuevas estaciones terrenas para dar soporte a estos enlaces. Esta tesis analiza la viabilidad de utilizar la tecnología desarrollada para los telescopios de rayos gamma que constituirán el observatorio CTA (*Cherenkov Telescope Array*) en la implementación de una nueva clase de estación terrena receptora. Entre las principales ventajas que brindan estos telescopios se encuentran las mayores aperturas necesarias para superar las limitaciones de potencia que comparte la astronomía terrestre de rayos gamma y las comunicaciones ópticas desde espacio profundo. Además, el elevado número de grandes telescopios que se construirán para CTA hará posible reducir costes unitarios gracias a una producción masiva. El segundo bloque de la tesis se enmarca en las comunicaciones cuánticas en espacio libre y en particular en la distribución de clave cuántica (QKD), que se ha convertido en un nuevo paradigma en el campo de la seguridad de la información. Esta técnica ofrece una forma teóricamente segura de comunicarse a través de un canal inseguro gracias a que es posible detectar la presencia de un intruso. QKD ha demostrado ser una forma fiable de transmitir datos sensibles mediante fibra óptica. Sin embargo, su alternativa no guiada también ofrece importantes ventajas, así como nuevos retos. El más importante es la necesidad de operar bajo una fuerte turbulencia atmosférica y altos niveles de ruido de fondo durante el día. Para mitigar estos efectos normalmente se lleva a cabo un compromiso al diseñar la óptica del receptor, ya que un campo de visión estrecho mejora el rechazo al ruido de fondo, pero aumenta las pérdidas relacionadas con las turbulencias y un campo de visión amplio produce el efecto opuesto. En este segundo bloque de la tesis se propone la utilización de un sistema de corrección de turbulencias para solucionar ambas limitaciones de forma simultánea, y se analizan y experimentan las distintas estrategias para llevar a cabo la implementación e integración con el sistema de QKD.

# Abstract

Free-space optical communication is widely regarded as the next-generation of high-speed wireless communication links in several scenarios. This thesis focuses on the two main applications where this technology can bring the most significant impact: interplanetary communications and quantum communications. Consequently, the dissertation is structured in two sections. In the first one, a novel proposal is suggested regarding to using Cherenkov telescopes as ground-station receivers. The signals that will be received on Earth from future lasercom transmitters in deep-space will be extremely weak and new ground stations will have to be developed in order to adequately support these links. This thesis addresses the feasibility of using the technology developed for the gamma-ray telescopes that will make up the Cherenkov Telescope Array (CTA) observatory in the implementation of a new kind of ground station. Among the main advantages that these telescopes provide are the much larger apertures needed to overcome the power limitation that ground-based gamma-ray astronomy and deep-space optical communication both have. Also, the large number of big telescopes that will be built for CTA will make it possible to reduce unitary costs by economy-scale production. The second section of the thesis is framed in the field of free-space quantum communications. In particular, in quantum key distribution (QKD), which has become a new paradigm in the discipline of information security. This technique offers a theoretically-secure way to communicate over an insecure channel since the presence of an eventual eavesdropper can be detected during the key transmission. QKD has proven to be a reliable way to transmit sensitive data through optical fiber. However, the free-space alternative also brings important advantages, as well as new challenges. The main challenge is the need to operate both under strong atmospheric turbulence and daylight background noise. In order to mitigate these effects, a trade-off is usually required when designing the receiver's optics, since a narrow field-of-view improves background noise rejection, but increases turbulence-related losses and a wide field-of-view produces the opposite effect. In this section of the thesis, a correction system for atmospheric turbulence is proposed to overcome both limitations at the same time, and different strategies are analyzed and experimented to carry out the implementation and integration within the QKD system.

# Índice

# Acrónimos

| Acrónimo | Inglés | Español |
|---|---|---|
| AES | Advanced Encryption Standard | Estándar de encriptación avanzada |
| AO | Adaptive Optics | Óptica adaptativa |
| APD | Avalanche PhotoDiode | Fotodiodo de avalancha |
| ASTRI | Astrofísica con Specchi a Tecnologia Replicante Italiana | Astrofísica con espejos con tecnología de réplica italiana |
| BER | Bit Error Rate | Tasa de error de bit |
| CCL | Compiled Command Language | Lenguaje de comandos compilados |
| COSTAR | Corrective Optics Space Telescope Axial Replacement | Reemplazo de óptica corrector axial para telescopio espacial |
| CSIC | - | Consejo Superior de Investigaciones Científicas |
| CTA | Cherenkov Array Telescope | Matriz de telescopios Cherenkov |
| DC | Davies-Cotton | Davies-Cotton |
| DES | Data Encryption Standard | Estándar de encriptación de datos |
| DLS | Damped Least Squares | Mínimos cuadrados amortiguados |
| DSN | Deel Space Network | Estación de espacio profundo |
| EAS | Extensive Air Shower | Cascada atmosférica extensa |
| EDRS | European Data Relay System | Sistema repetidor de datos europeo |
| EFL | Effective Focal Length/Equivalent Focal Length | Longitud focal efectiva/equivalente |
| ESA | European Space Agency | Agencia europea del espacio |
| ETS | Engineering Test Satellite | Satélite de test de ingeniería |
| EXPRESS | Expedite the PRocessing of Experiments to the Space Station | Acelerar el procesado de experimentos a la estación espacial |
| FSM | Fast/Fine Steering Mirror | Espejo de desvío rápido/fino |
| FSOC | Free-Space Optical Communication | Comunicaciones ópticas en espacio libre |
| FWHM | Full Width at Half Maximum | Anchura complete a la mitad del máximo |
| GATE | GAmma-ray Telescope Elements | Elementos de telescopio de rayos gamma |
| GM-APD | Geiger Mode Avalanche PhotoDiode | Fotodiodo de avalancha en modo Geiger |
| GOLD | Ground Orbiter Lasercomm Demonstration | Demostración de comunicaciones por láser entre tierra y orbitador |
| GOPEX | Galileo Optical Experiment | Experimento óptico Galileo |
| GRB | Gamma-Ray Burst | Estallido de rayos gamma |
| GTC | - | Gran Telescopio de Canarias |
| HESS | High Energy Stereoscopic System | Sistema estereoscópico de alta energía |
| HET | Hobby-Eberly Telescope | Telescopio Hobby-Eberly |
| IACT | Imaging Atmospheric Cherenkov Telescopes | Telescopio Cherenkov atmosférico de imágenes |
| INSA | - | Ingeniería y Servicios Aeroespaciales |
| ISS | International Space Station | Estación espacial internacional |
| ITEFI | - | Instituto de Tecnologías Físicas y de la Información |
| JAXA | Japan Aerospace Exploration Agency | Agencia japonesa de exploración aeroespacial |
| JCR | Journal Citation Report | Informe de citas en revistas |
| JWST | James Webb Space Telescope | Telescopio Espacial James Webb |
| KARI | Korea Aerospace Research Institute | Instituto de investigación aeroespacial coreano |
| **LADEE** | Lunar Atmosphere and Dust Environment Explorer | Explorador del ambiente de polvo y atmósfera lunar |
| **LBT** | Large Binocular Telescope | Gran telescopio binocular |
| **LCRD** | Laser Communication Relay Demonstration | Demostración de repetidor de comunicación por láser |
| **LEO** | Low Earth Orbit | Órbita baja terrestre |
| **LEP** | Lateral Effect Photodiode | Fotodiodo de efecto lateral |
| **LLCD** | Lunar Laser Communication Demonstration | Demostración de comunicaciones por láser en la Luna |
| **LLGT** | Lunar Laser Ground Terminal | Terminal de tierra para láser lunar |
| **LMA** | Levenberg–Marquardt Algorithm | Algoritmo Levenberg–Marquardt |
| **LOS** | Line Of Sight | Línea de visión |
| **LRO** | Lunar Reconnaissance Orbiter | Orbitador de reconocimiento lunar |
| **LST** | Large-Size Telescope | Telescopio de gran tamaño |
| **MAGIC** | Major Atmospheric Gamma-ray Imaging Cherenkov | Cherenkov atmosférico de imagen rayos gamma |
| **MATLAB** | MATrix LABoratory | Laboratorio de matrices |
| **MLCD** | Mars Laser Communication Demonstration | Demostración de comunicaciones por láser en Marte |
| **MODTRAN** | MODerate resolution atmospheric TRANsmission | Transmisión atmosférica con resolución moderada |
| **MOPA** | Master Oscillator Power Amplifier | Oscilador principal/amplificador de potencia |
| **MRO** | Mars Reconnaissance Orbiter | Orbitador de reconocimiento marciano |
| **MST** | Medium-Size Telescope | Telescopio de tamaño medio |
| **NASA** | National Aeronautics and Space Administration | Agencia nacional aeronáutica y espacial |
| **NASDA** | National Space Development Agency | Agencia nacional de desarrollo espacial |
| **NFIRE** | Near Field Infrared Experiment | Experimento de infrarrojo cercano |
| **OA** | - | Óptica adaptativa |
| **OCDHRLF** | Optical Communication Demonstration and High-Rate Link Facility | Demostración de comunicaciones ópticas e instalación para enlace de alta velocidad |
| **OICETS** | Optical Inter-orbit Communications Engineering Test Satellite | Satélite de test de ingeniería para comunicaciones ópticas entre órbitas |
| **OLSG** | Optical Link Study Group | Grupo de estudio para enlace óptico |
| **OLT** | Overwhelmingly Large Telescope | Telescopio abrumadoramente grande |
| **OOK** | On-Off Keying | Codificación de encendido y apagado |
| **OPALS** | Optical PAyload for Lasercomm Science | Carga útil óptica para ciencia con comunicaciones por láser |
| **OPD** | Optical Path Difference | Diferencia de caminos ópticos |
| **OSA** | Optical Society of America | Sociedad americana de óptica |
| **OSETI** | Optical Search for ExtraTerrestrial Intelligence | Búsqueda óptica de inteligencia extraterrestre |
| **OSLO** | Optics Software for Layout and Optimization | Software de óptica para diseño y optimización |
| **PID** | Proportional Integral Derivative | Proporcional, integral, derivativo |
| **PMT** | PhotoMultiplier Tube | Fotomultiplicador |
| **PNSA** | Photon Number Splitting Attack | Ataque por división del número de fotones |
| **PPM** | Pulse Position Modulation | Modulación por posición de pulsos |
| **PSD** | Position Sensitive Detector | Detector sensible a la posición |
| **PSF** | Point Spread Function | Función de dispersión puntual |
| **P-V** | Peak to Valley | Pico a valle |

| QBER | Quantum Bit Error Rate | Tasa de error de bit cuántico |
|---|---|---|
| QD | Quadrant Detector | Detector de cuadrantes |
| QKD | Quantum Key Distribution | Distribución de clave cuántica |
| RF | Radio Frequency | Radio frecuencia |
| RMS | Root Mean Square | Valor cuadrático medio |
| SALT | Southern African Large Telescope | Gran telescopio sudafricano |
| SC | Schwarzschild-Couder | Schwarzschild-Couder |
| SEP | Sun-Earth-Probe | Sol-Tierra-Sonda |
| SETI | Search for ExtraTerrestrial Intelligence | Búsqueda de inteligencia extraterrestre |
| SILEX | Semi-conductor Inter satellite Link EXperiment | Experimento de enlace intersatélite de semiconductor |
| SINTONIA | - | Sistemas No Tripulados Orientados al Nulo Impacto Ambiental |
| SKR | Secret Key Rate | Tasa de clave secreta |
| SNR | Signal to Noise Ratio | Relación señal a ruido |
| SOHO | SOlar and Heliospheric Observatory | Observatorio solar y heliosférico |
| SOL | Spanish Optical Link | Enlace óptico español |
| SPAD | Single Photon Avalanche Detector | Detector de avalancha de fotones individuales |
| SPE | Sun-Probe-Earth | Sol-Sonda-Tierra |
| SPICE | Simulation Program with Integrated Circuits Emphasis | Programa de simulación con énfasis en circuitos integrados |
| SST | Small-Size Telescope | Telescopio de pequeño tamaño |
| SZA | Solar Zenith Angle | Ángulo cenital solar |
| TDRSS | Tracking and Data Relay Satellite System | Sistema de satélites repetidores de seguimiento y datos |
| TMO | Table Mountain Observatory | Observatorio de *Table Mountain* |
| TMT | Thirty-Meter Telescope | Telescopio de treinta metros |
| UAV | Unmanned Aerial Vehicle | Vehículo aéreo no tripulado |
| UAH | - | Universidad de Alcalá |
| UC3M | - | Universidad Carlos III de Madrid |
| UMA | - | Universidad de Málaga |
| UPM | - | Universidad Politécnica de Madrid |
| UHE | Ultra High Energy | Ultra alta energía |
| VCSEL | Vertical-Cavity Surface-Emitting Laser | Láser de emisión superficial con cavidad vertical |
| VERITAS | Very Energetic Radiation Image Telescope Array System | Sistema en array de telescopios para imagen de radiación muy energética |
| VHE | Very High Energy | Muy alta energía |
| VLT | Very Large Telescope | Telescopio muy grande |
| WCP | Weak Coherent Pulses | Pulsos débiles coherentes |

# Índice de figuras

# Índice de tablas

# 1. Introducción

Las telecomunicaciones inalámbricas han conocido en los últimos años una explosión sin igual. Desde las redes globales de telefonía móvil hasta los enlaces vía satélite, las comunicaciones no guiadas han traído consigo una enorme cantidad de nuevos servicios y suponen uno de los principales pilares de la actual Sociedad de la Información. En particular las comunicaciones ópticas en espacio libre [1], pese a su reciente aparición, proporcionan una serie de ventajas que han permitido crear aplicaciones completamente nuevas como la distribución cuántica de claves [2] o protagonizar un cambio de paradigma en aplicaciones tradicionales como las comunicaciones vía satélite en general [3] o los enlaces en espacio profundo en particular [4].

Esta tesis se centra en las que podrían considerarse las dos aplicaciones más prometedoras de esta disciplina: las comunicaciones ópticas en espacio profundo y la distribución de clave cuántica en espacio libre. Ambas comparten una historia similar: propuestas originalmente hace décadas, no ha sido hasta los últimos años que han empezado a desarrollarse con fuerza. No obstante, por diferentes motivos se puede decir que en los dos casos aún no se ha superado completamente la fase de prueba de concepto. Aún no existen en la actualidad sistemas de comunicaciones que operen rutinariamente utilizando ninguna de las dos tecnologías, si bien ya han tenido lugar un buen número de demostraciones exitosas. Esto convierte a ambas aplicaciones en campos de investigación muy fértiles y activos, en los que continuamente se alcanzan nuevos hitos y donde la obsolescencia llega muy rápidamente.

El trabajo de esta tesis se ha dividido en dos bloques, cada uno dedicado a una de estas dos tecnologías mencionadas. Si bien ambas comparten un marco común muy específico como son las comunicaciones ópticas en espacio libre, existen importantes diferencias debido a sus distintos objetivos y ámbitos de aplicación. No obstante, quizás la mayor diferencia entre los dos bloques se basa en el enfoque metodológico seguido: más teórico en el primero y más experimental en el segundo (sin dejar de lado uno de los dos aspectos en ningún bloque). Este capítulo se dedica a hacer una breve introducción a las disciplinas en las que se enmarca la presente tesis, algunos fundamentos teóricos básicos sobre los que se apoyan ambos bloques, así como un resumen de las motivaciones que han dado lugar a esta tesis y las aportaciones en las que ha desembocado, y por último, un breve resumen de contenidos.

## 1.1. COMUNICACIONES NO GUIADAS

Se engloba dentro del concepto de comunicaciones en espacio libre, comunicaciones inalámbricas o comunicaciones no guiadas, a aquellas en las que se hace uso del espacio libre como medio de transmisión existente entre el emisor y el receptor, transfiriendo la información a distancia sin usar cables o cualquier otro tipo de medio de guiado. Las comunicaciones no guiadas han sido el segmento de la industria de las telecomunicaciones de más rápido crecimiento en los últimos años [5]. Las redes de telefonía móvil han experimentado un crecimiento exponencial y actualmente la penetración mundial asciende al 96 % [6]. Más recientemente, las redes de datos inalámbricas están sustituyendo a las tradicionales redes de cable tanto en hogares como en empresas y grandes organizaciones. La tecnología inalámbrica ha traído una innumerable cantidad de aplicaciones que van más allá de los sistemas de comunicaciones, tales como geoposicionamiento, telemedicina, pago electrónico, robótica, vigilancia, seguridad, etc.

### 1.1.1. Origen de las comunicaciones no guiadas

Los primeros enlaces de comunicaciones no guiadas datan de épocas prehistóricas, en las que el hombre usaba señales de humo, antorchas, etc. para transmitir alguna información a distancia. Posteriormente se sucedieron un gran número de mejoras basadas en el mismo principio pero hubo que esperar hasta la década de 1860 para que Maxwell postulara las leyes básicas del electromagnetismo y 20 años más tarde Heimrich Hertz demostrara su existencia. En 1893 Nikola Tesla demuestra por primera vez la transmisión y recepción de energía electromagnética sin cables y en 1901 Marconi realiza la primera comunicación transatlántica mediante ondas de radio. Marconi logró transmitir la letra S mediante código Morse a través de 3360 km de océano, si bien ignorando los principios de la propagación electromagnética y utilizando dispositivos inventados por otros [7]. Desde entonces la tecnología de radio evolucionó rápidamente, permitiendo transmisiones a mayores distancias, con mayor calidad, menos potencia y empleando sistemas más pequeños.

Al principio todas las transmisiones introducían información modulando la amplitud de una portadora. No es hasta 1933 que es patentada la radio FM en la que se modula la frecuencia de la señal y en esta década comienzan las transmisiones de TV en Europa y más tarde en Norteamérica. En la década de 1950 hacen su aparición los transistores que acabarán sustituyendo a las válvulas y una década más tarde comienzan a verse las primeras transmisiones digitales. En 1957 comienza la era espacial con el lanzamiento del primer satélite artificial por parte de la Unión Soviética y con ella las comunicaciones por satélite. Durante los años siguientes se suceden varios satélites repetidores y en 1962 el Telstar 1 se convierte en el primer satélite en transmitir señales de televisión. El primer satélite geoestacionario fue el Syncom 2 (1963), y en 1965 el Intelsat 1 (conocido como *early bird*) fue el primer satélite geoestacionario comercial de comunicaciones. Desde entonces, la sucesión de satélites de comunicaciones cada vez más sofisticados ha sido constante. Cabe señalar que sólo transcurrieron once años entre el lanzamiento del primer satélite artificial y la realización efectiva de un sistema global de comunicaciones por satélite plenamente operacional (Intelsat 3, en 1968) [8].

## 1.1.2. Ventajas e inconvenientes

Las comunicaciones no guiadas presentan una serie de importantes ventajas en relación a los sistemas guiados en un buen número de aplicaciones. Por una parte, su uso es imprescindible o muy necesario en determinados escenarios en que es imposible o poco práctico el establecimiento de un enlace mediante una conexión física, como un cable. Esto es válido sobre todo cuando los equipos están separados por grandes trechos de agua, escarpadas montañas o inhóspitos desiertos, o al comunicarse con transpondedores alojados en satélites. También se hace imprescindible el uso de sistemas inalámbricos cuando los terminales son móviles.

Las comunicaciones no guiadas se basan en la transmisión de información modulada en ondas electromagnéticas propagándose en el espacio. El frente de ondas en el que se concentra la potencia va expandiéndose siguiendo la ley del cuadrado de la distancia (Figura 1), según la cual la densidad de potencia existente a una determinada distancia desde la fuente es inversamente proporcional al cuadrado de la misma. Este fenómeno trae consigo una importante desventaja en el uso de comunicaciones no guiadas dado que siempre se perderá potencia si esta no incide en su totalidad sobre la antena receptora, lo que será tanto más difícil cuanto mayor sea la distancia de propagación.

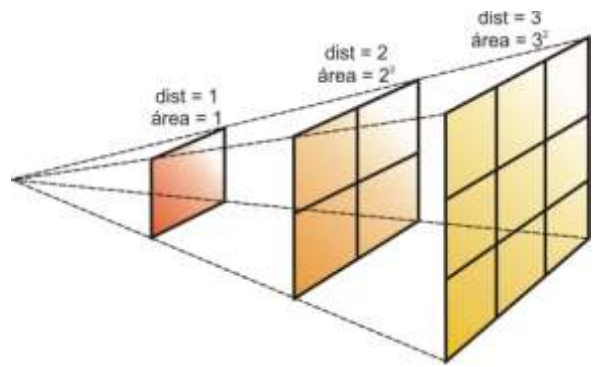

**Figura 1. Ley del cuadrado de la distancia.**

Como se explicará más adelante, el fenómeno de la divergencia de las ondas electromagnéticas en general depende, para una distancia dada, de dos factores: la longitud de onda y el tamaño de la antena transmisora. A menor longitud de onda y mayor tamaño de antena, la divergencia sufrida por la señal es menor, lo que significa que las transmisiones son más directivas. Por ello, en enlaces no difusivos interesa siempre elegir una frecuencia alta y/o una antena grande. La cantidad de información que se puede transmitir en un sistema de comunicación está directamente relacionada con el ancho de banda de la señal modulada y este ancho de banda viene limitado en último término por la frecuencia de la señal portadora. Al incrementarse esta, se puede incrementar el ancho de banda transmitido. Debido a la divergencia, en comunicaciones no guiadas se suelen emplear altas frecuencias, lo que a su vez proporciona grandes anchos de banda.

También conviene recordar que el medio utilizado es un medio compartido, lo que da lugar a una serie de características que se han de considerar. Por una parte, en un medio compartido puede haber muchos enlaces simultáneos, lo que puede dar lugar a interferencias. Este problema se suele solventar realizando una gestión apropiada del espectro electromagnético en las diferentes bandas de utilización. Además de otros sistemas de comunicaciones, existe ruido de muchos tipos que se añade a la señal recibida, haciéndola más difícilmente distinguible. Y por último, cuando el medio de transmisión es la atmósfera existen fenómenos tales como la absorción, el *scattering*, la lluvia, las nubes, etc. que afectan al enlace en una medida u otra según diversos factores como la longitud de onda, el emplazamiento, etc.

## 1.2. COMUNICACIONES ÓPTICAS NO GUIADAS

Los sistemas actuales de comunicación utilizan una onda electromagnética sobre la que se superpone la información que se desea transmitir. A esta señal que no contiene información útil se le conoce como portadora y la técnica habitual consiste en la modulación de la portadora mediante una señal de información. Una vez que la señal se ha propagado hasta su destino, se extrae la información de esta a través de un proceso de demodulación. Cuando la frecuencia de la señal portadora cae dentro del rango de frecuencias visibles por el ojo humano o cercano a ellas, al tipo de comunicaciones que resulta del empleo de estas frecuencias se le conoce como comunicaciones ópticas. La banda propia de las comunicaciones ópticas (Figura 2) se extiende desde el infrarrojo próximo ($\nu \approx 3 \cdot 10^{13}$ Hz) hasta el ultravioleta próximo ($\nu \approx 1,5 \cdot 10^{15}$ Hz).

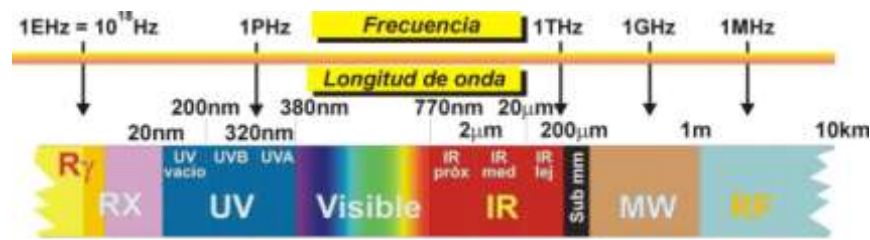

**Figura 2. Espectro electromagnético (no a escala).**

A diferencia de las comunicaciones ópticas guiadas, en las que las señales ópticas son conducidas a través de un camino establecido en el interior de fibras ópticas, las comunicaciones ópticas no guiadas constituyen el tipo de comunicación en el que se usan portadoras a frecuencias ópticas desplazándose en un medio abierto, como puede ser la atmósfera, el agua o el vacío.

## 1.2.1. Reseña histórica

Los orígenes de las comunicaciones ópticas no guiadas coinciden con el origen de las comunicaciones ópticas mismas. Desde épocas prehistóricas, el hombre ha usado señales ópticas para comunicarse, si bien esta comunicación consistía originalmente en muy poca información y cubría muy cortos alcances. Las comunicaciones ópticas no guiadas al menos cuentan con cerca de tres milenios de historia, ya que ha sido verificado el empleo sobre el 800 A. C. de señales ópticas creadas con fuego para transmitir información formada por un limitado número de mensajes conocidos previamente por ambas partes [9]. Sobre el 200 A.C. el historiador griego Polibio desarrolló un sistema que era capaz de transmitir letras en lugar de mensajes fijos. El funcionamiento se basaba en una tabla de códigos como la de la Figura 3. En función del número de antorchas que hubiera encendidas a la derecha y a la izquierda, se seleccionaba una letra determinada. Con operadores entrenados se lograba una comunicación de unas ocho letras por minuto [10].

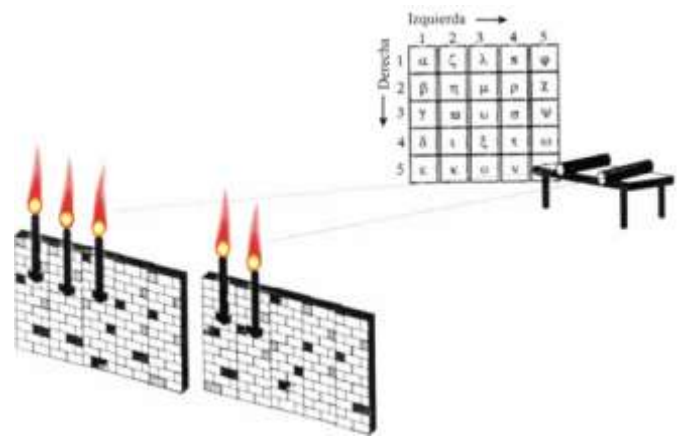

**Figura 3. Sistema de comunicaciones ópticas de Polibio [10].**

El avance de las comunicaciones ópticas prácticamente se detuvo hasta el siglo XVII, con el desarrollo del telescopio. En 1791 se dio un gran paso con la invención del primer telégrafo (que fue óptico, no eléctrico) por parte de Claude Chappe. Consistía en un mástil fijo que soportaba dos piezas móviles con las que se podían adaptar una gran cantidad de configuraciones diferentes. La disposición era observada a través de un telescopio por un operador situado en otra estación, actuando de repetidora a la siguiente estación. Este sistema propició la creación de una red de comunicaciones ópticas en Francia que en 1844 cubría unos 5000 km con más de 500 estaciones [11]. Aunque este tipo de comunicación tuvo un gran éxito en su época, el descubrimiento de la electricidad y el telégrafo eléctrico propició su final. El primer sistema de comunicaciones ópticas no guiadas en el sentido actual fue el fotófono, construido por Alexander Graham Bell en 1880 [12], cuatro años después del teléfono. Consistía en un espejo reflector de la luz del Sol que podía ser modulado acústicamente. La señal modulada, después de avanzar unos 200 metros, incidía sobre una placa de selenio cuya resistencia eléctrica dependía de la intensidad luminosa que incidiera sobre ella. De esta forma se obtenía una corriente eléctrica que podía ser convertida, mediante un altavoz, en una señal acústica.

En 1916 Albert Einstein presentó su trabajo teórico sobre la emisión estimulada de la radiación electromagnética. Hasta entonces se creía que un fotón sólo podía interaccionar con un átomo de dos formas: podía ser absorbido, elevando el átomo a un nivel de energía superior, o podía ser emitido, de forma que el átomo pasase a un nivel inferior. Einstein propuso una tercera posibilidad: que un fotón con energía correspondiente al valor de una transición entre niveles podía estimular que un átomo pasase a un nivel energético inferior, emitiendo otro fotón con idéntica energía que el primero. Este trabajo teórico supuso la base para el posterior desarrollo del láser. Sin embargo, tuvieron que pasar más de 40 años hasta que el primer láser de gas fuese construido. Sólo dos años más tarde, se construyó el primer láser de semiconductor.

      Un primer experimento de comunicaciones ópticas no guiadas (llevado a cabo por la NASA) se realizó en 1967 [10] usando un láser de $CO_2$ y técnicas coherentes, y un segundo (por parte de la Fuerza Aérea de los EEUU) empleando un láser Nd:YAG y técnicas de detección directa. Sin embargo, los efectos de la atmósfera sobre las señales hicieron decaer el interés quedando la técnica relegada a unas pocas aplicaciones marginales. En su lugar, en los años posteriores a 1970 se produjo un enorme desarrollo en las fibras ópticas. De esta manera quedó casi totalmente detenido el desarrollo de las comunicaciones ópticas no guiadas hasta el resurgimiento en la actualidad de una serie de proyectos que las sitúan como una ventajosa alternativa en una variedad de escenarios y aplicaciones, especialmente la transmisión de clave cuántica y las comunicaciones en espacio profundo, como alternativa a la fibra óptica y a las microondas, respectivamente.

## 1.2.2. Ventajas e inconvenientes

Las comunicaciones ópticas no guiadas presentan ventajas significativas en relación a los sistemas de microondas o de fibra óptica en aplicaciones tales como las comunicaciones por satélite o en enlaces urbanos punto a punto, respectivamente. La mejora conseguida con este tipo de enlaces es tanto mayor cuanto mayor sea la distancia a superar y la capacidad requerida del enlace. La principal ventaja de las comunicaciones ópticas en espacio libre proviene del hecho de que a frecuencias ópticas la divergencia del haz es mucho más reducida que a frecuencias menores, y por lo tanto también es menor el tamaño del spot detectado en el receptor (Figura 4). El pequeño tamaño de spot se traduce en un gran aumento de la densidad de potencia recibida, consiguiéndose así una importante mejora en las prestaciones del sistema. Sin embargo, a grandes distancias de propagación, esta gran ventaja se traduce en la necesidad de disponer de un sistema de apuntamiento extremadamente preciso.

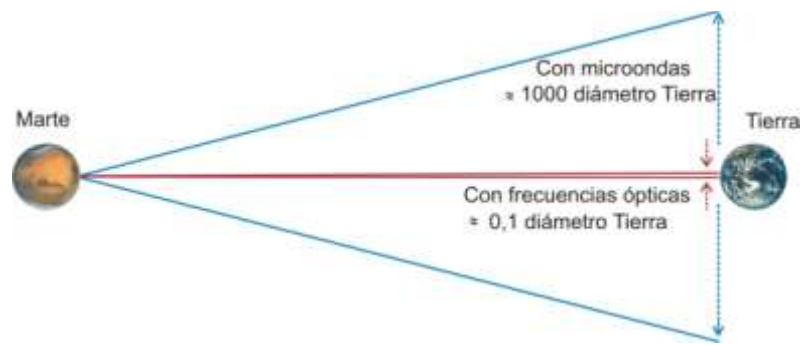

**Figura 4. Divergencia de un haz de frecuencias ópticas
y de microondas en un enlace Marte-Tierra.**

Otra consecuencia de la menor longitud de onda de la señal portadora es que son necesarias antenas (telescopios) de menor tamaño que a frecuencias de microondas para obtener la misma ganancia. A esto se le suma el hecho de que los equipos son más ligeros y pequeños que los empleados en microondas (alrededor de un 60 % más pequeños [13, p. 19]). Esta reducción de carga supone una gran ventaja económica en el caso de situarlos a bordo de satélites o naves espaciales, donde cada kilogramo de más eleva considerablemente el presupuesto necesario. En comunicaciones por fibra óptica, el empleo de las fuentes y detectores ópticos viene impuesto por las limitaciones de la propia fibra. En comunicaciones ópticas no guiadas se dispone de un rango más amplio de elección, aunque hay que señalar que si el medio de propagación incluye la atmósfera, la selección de la longitud de onda se verá condicionada. La fibra óptica impone limitaciones adicionales en la distribución cuántica de claves al limitar la longitud máxima de los enlaces (ya que de momento no existen los repetidores cuánticos) y modificar la polarización (aspecto frecuentemente crítico para este tipo de sistemas). Por el contrario, la atmósfera (o en su caso el vacío) es un medio con muy pocas pérdidas y no birrefringente, lo que se adapta mucho mejor a las comunicaciones cuánticas [14, p. 161].

Frente a las virtudes comentadas de los sistemas ópticos no guiados, hay que mencionar algunos inconvenientes. El principal lo constituye la atmósfera terrestre, que provoca determinados efectos indeseados en las señales ópticas, deteriorando las prestaciones del enlace, y en ciertas condiciones bloqueándolo por completo. Dejando aparte el efecto de la absorción (que se puede minimizar haciéndose despreciable con una correcta elección de la longitud de onda), las turbulencias atmosféricas consiguen que, al distorsionar el frente de ondas por variaciones aleatorias en el índice de refracción, la intensidad y fase de la señal recibida en el telescopio fluctúen. Por último, el empleo de longitudes de onda ópticas podría llegar a ser una amenaza para la visión de las personas si no se toman las medidas oportunas. No obstante, esto no suele suponer un problema debido a la densidad de potencia de los haces y al empleo de longitudes de onda seguras para la visión. En este sentido, las comunicaciones ópticas explotan una zona del espectro electromagnético no usada por otras aplicaciones, por lo que la gestión del mismo es mucho más simple que en el caso del congestionado espectro radioeléctrico.

## 1.3. ORIGEN Y APORTACIONES DEL TRABAJO REALIZADO

Como se explicó al inicio de este apartado, esta tesis se ha dividido en dos bloques, ambos compartiendo el marco común de las comunicaciones ópticas en espacio libre. En este apartado se pretende ofrecer un repaso del origen de las motivaciones que han dado lugar a cada parte de esta tesis y resumir las aportaciones a la comunidad científica a las que ha dado lugar el trabajo realizado.

El primero de los bloques se basa en el trabajo desarrollado a partir de la propuesta de aprovechar los esfuerzos llevados a cabo en el diseño del proyecto CTA para la construcción de uno o varios telescopios adaptados para funcionar como receptores en Tierra de comunicaciones ópticas desde espacio profundo. Se trata de una propuesta original de la que no se conoce ningún antecedente previo en la literatura científica. Para desarrollar este trabajo se partió de la experiencia obtenida durante la realización del Proyecto Fin de Carrera titulado "Diseño de un enlace de comunicaciones ópticas con Marte" para la obtención en la Universidad de Málaga (UMA) del título de Ingeniero Técnico en Telecomunicación. Este trabajo se basó en información obtenida de un proyecto de la NASA, cancelado más tarde por cuestiones presupuestarias, en el que se pretendía llevar a cabo en 2009 la primera demostración de un enlace de comunicaciones ópticas desde espacio profundo (Marte en este caso). En el momento de realizar dicho proyecto, las especificaciones del orbitador marciano estaban bien definidas, a falta por completar el diseño de las estaciones terrenas. Por ello, el trabajo se centró en el análisis de misión y el diseño de alto nivel de los requisitos necesarios para dar soporte en Tierra a este enlace. Posteriormente, mientras se cursaba el Máster Oficial en Ciencia y Tecnología desde el Espacio en la Universidad de Alcalá (UAH), se elaboró en la Universidad Carlos III de Madrid (UC3M) un estudio sobre adaptación de telescopios astronómicos para operar como estaciones de comunicaciones ópticas, correspondiente a un paquete de trabajo del proyecto SOL (*Spanish Optical Link*), contratado a la UC3M por la empresa INSA (Ingeniería y Servicios Aeroespaciales). La experiencia acumulada en estos dos proyectos culminó con la publicación del capítulo "*Development of optoelectronic sensors and transceivers for spacecraft applications*" del libro "*Advances in spacecraft technologies*" [15]. A raíz de la participación de INSA en el proyecto CTA, surgió más tarde la colaboración con la UC3M para realizar el estudio de viabilidad que fue el germen del primer bloque de esta tesis. Tras la finalización de este proyecto de colaboración se procedió a la publicación (como primer autor) de sus resultados en la revista *Applied Optics*, incluida en el JCR (*Journal Citation Report*), titulado "*Feasibility of utilizing the CTA gamma-ray telescopes as free-space optical communication ground stations*" [16]. Por último, en esta tesis se presenta además el resultado de la continuación de este trabajo.

El segundo bloque de esta tesis tiene su origen en el trabajo realizado durante el Proyecto Fin de Carrera titulado "Enlace de comunicaciones ópticas no guiadas aire-tierra basado en retromodulador" para la obtención en la Universidad de Alcalá (UAH) del título de Ingeniero Superior en Telecomunicación. Este proyecto se realizó en la Universidad Politécnica de Madrid (UPM) en el marco del mencionado proyecto SOL, financiado por INSA. Consistió en el desarrollo de un transmisor de comunicaciones ópticas basado en retromodulador: un modulador de cristal líquido que introduce información en la luz incidente en un retrorreflector con el objetivo de prescindir en el terminal remoto del sistema láser y de apuntamiento. Este trabajo dio lugar a una publicación JCR titulada "*V-shape liquid crystal-based retromodulator air to ground optical communications*" [17], a varias contribuciones a congresos nacionales e internacionales [18], [19], [20], [21] y al 1er premio de Ingeniería en el Certamen Universitario Arquímedes [22], de ámbito nacional. Posteriormente, como continuación a la colaboración entre INSA, UC3M y UPM en el proyecto SOL, y a raíz de la participación de INSA en el proyecto SINTONIA (Sistemas No Tripulados Orientados al Nulo Impacto Ambiental), liderado por Boeing Research & Technology Europe, SL, INSA subcontrató a la UC3M la tarea "T8.1.9 Enlaces ópticos". El trabajo desarrollado en esta tarea consistió en el diseño y desarrollo de un sistema de comunicaciones por láser entre una estación óptica terrena y un UAV (vehículo aéreo no tripulado, del inglés *Unmanned Aerial Vehicle*), con una mayor orientación al sistema de apuntamiento desde Tierra. Este trabajo dio lugar a una publicación JCR (como primer autor) titulada "*In-axis reception by polarization discrimination in a modulating-retroreflector-based free-space optical communication link*" [23] y varias contribuciones a congresos internacionales [24], [25]. Durante una de estas contribuciones [26], se entró en contacto con el grupo de Investigación en Criptología y Seguridad de la Información del Consejo Superior de Investigaciones Científicas (CSIC) y surgió la oportunidad de colaborar en el desarrollo de un sistema de corrección de turbulencia atmosférica para el sistema de distribución cuántica de claves desarrollado por este grupo debido a las grandes similitudes entre este trabajo y el desarrollado con INSA-UC3M durante el proyecto SINTONIA. De esta colaboración resultó el trabajo de la segunda parte de esta tesis, culminando en una contribución a un congreso nacional [27] y la publicación en la revista *SPIE Optical Engineering* de un artículo JCR (como primer autor) titulado *"Correction of beam wander for a free-space quantum key distribution system operating in urban environment"* [28], un capítulo en un libro editado por *Springer* de próxima aparición titulado *"Design and implementation of a high-speed free-space quantum communication system for daylight operation in urban environment under atmospheric turbulence"* [29] y otro artículo en proceso de escritura a fecha de finalización de este documento.

## 1.4. RESUMEN DEL CONTENIDO DE LA TESIS

Tal como se ha comentado, este trabajo está dividido en dos bloques diferentes, que representan el apartado 2 (Telescopios *Cherenkov* para comunicaciones en espacio profundo) y el apartado 3 (Óptica activa en distribución cuántica de claves) de la tesis. El primer bloque consiste en un estudio de viabilidad para reutilizar telescopios *Cherenkov* como estaciones terrenas receptoras en enlaces de comunicaciones ópticas desde espacio profundo, y en el segundo se presenta una solución para adaptar un sistema de comunicaciones cuánticas en espacio libre para operación diurna en presencia de ruido de fondo y turbulencia atmosférica. En general, se puede decir que el primer bloque es un trabajo teórico y el segundo un trabajo experimental, si bien ambos bloques incluyen los dos tipos de tratamientos en mayor o menor medida. A continuación se presenta un breve resumen pormenorizado del contenido de esta tesis.

El primer bloque se inicia con una introducción al problema en el apartado 2.1 donde se plantea la conveniencia de estudiar la viabilidad de utilizar de los telescopios *Cherenkov* del proyecto CTA para comunicaciones ópticas desde espacio profundo debido a que ambos comparten una necesidad común: compensar la reducida potencia óptica recibida mediante una gran apertura de recepción. En al apartado 2.2 se introducen una serie de conceptos relacionados con la disciplina de estudio que hace uso de los telescopios *Cherenkov*: la astronomía de altas energías. Estos conceptos son necesarios para comprender adecuadamente la operación de este tipo de telescopios, por lo que únicamente se tratan de forma sucinta los conceptos necesarios para llevar a cabo el trabajo de esta tesis. El apartado 2.3 continúa en la línea del anterior, pero en este caso centrándose en la parte técnica de la disciplina y en especial en el proyecto CTA, bajo el cual se planean construir los telescopios que se estudian en esta tesis. El bloque de introducción de conceptos se completa con el apartado 2.4, donde se expone brevemente la disciplina que hará uso de la propuesta de adaptación sugerida en este trabajo: las comunicaciones ópticas espaciales. Para ello, se realiza un recorrido histórico de la evolución de este incipiente campo, así como un repaso de sus fundamentos más importantes para llevar a cabo el estudio de adaptación. En el apartado 2.5 se enumeran las motivaciones que justifican realizar este estudio, destacando las numerosas coincidencias que en su implementación técnica muestran dos disciplinas tan alejadas como la astrofísica terrestre de rayos gamma y las comunicaciones ópticas desde espacio profundo. El apartado 2.6 incluye la única parte experimental de este bloque de la tesis, consistente en una serie de medidas de reflectividad sobre muestras de espejos de telescopios *Cherenkov*. Estas medidas se realizaron con el objetivo de evaluar la viabilidad de la utilización en comunicaciones de la tecnología de espejos desarrollada para CTA, con un interés especial en el análisis espectral y en los procesos de fabricación. En el apartado 2.7 se analiza una de las principales diferencias de un telescopio *Cherenkov* y un telescopio de comunicaciones: el enfoque. Esta diferencia se debe a la distinta distancia de la que proviene el objeto a enfocar y hace necesario un reenfoque para adaptar su uso a comunicaciones. El apartado 2.8 muestra una simulación óptica del telescopio MAGIC II que se realizó en la fase inicial de este estudio, cuando la información sobre los telescopios de CTA era aún escasa. El resultado de esta simulación fue muy importante porque sirvió para extraer conclusiones preliminares sobre cual sería la principal limitación de este tipo de telescopios: sus aberraciones ópticas. Basándose en las conclusiones extraídas en este apartado, en el 2.9 se estudia cómo influye el peor comportamiento óptico de estos telescopios en las prestaciones de un enlace de comunicación. Se explica que esta limitación impone la necesidad de emplear campos de visión más amplios a los ideales en comunicaciones bajo un entorno de ruido de fondo solar, lo que impacta de forma negativa en la relación señal a ruido del enlace. El apartado 2.10 enumera las limitaciones al mínimo campo de visión utilizable en orden de menor a mayor relevancia. La importancia de este análisis estriba en la necesidad de identificar qué efectos predominan en cada caso para centrar los esfuerzos en ellos, antes de pasar a los menos relevantes. Para estudiar específicamente los telescopios de CTA concentrándose en el comportamiento óptico, al haber identificado las aberraciones como su primera limitación, se realizaron una serie de modelos utilizando el software OSLO. Los resultados de estas simulaciones se muestran en el apartado 2.11 y serán utilizados en el apartado 2.12 para evaluar el comportamiento de un enlace de comunicaciones en términos de relación señal a ruido. Para ello, se consideraron una serie de diferentes escenarios, desde enlaces con satélites de órbita baja hasta sondas en Marte, y se analizaron las diferentes opciones de utilización de los telescopios de CTA. Por último, en el apartado 2.13 se proponen una serie de alternativas orientadas a mejorar la calidad óptica de los telescopios de CTA con el objetivo de reducir su campo de visión y así optimizar sus prestaciones en un enlace de comunicaciones desde espacio profundo en presencia de ruido solar.

En cuanto al segundo bloque, tras una breve introducción en el apartado 3.1 al problema a abordar, se continúa en el apartado 3.2 con una revisión de fundamentos de criptografía clásica para comprender la importancia de la criptografía cuántica, que se introduce en el apartado 3.3. Para dar una visión general de la disciplina se presentan los conceptos fundamentales, necesarios para describir el protocolo implementado por el sistema de comunicaciones cuánticas desarrollado en el CSIC. En el apartado 3.4 se particulariza la visión general de los apartados anteriores para describir los distintos elementos del sistema experimental de QKD en espacio libre (divididos en el sistema transmisor y receptor: Alice y Bob, respectivamente), cuyos principios de funcionamiento fue necesario dominar para analizar posteriormente sus limitaciones. En los dos siguientes apartados se presentan los fundamentos matemáticos necesarios para abordar la teoría utilizada en el diseño de la solución propuesta más adelante. Para ello, el apartado 3.5 introduce una serie de conceptos relacionados con la propagación de haces gaussianos que serán necesarios en el apartado 3.6, donde se analiza más a fondo la teoría de la turbulencia atmosférica relacionada con el problema a resolver: el *beam wander* (las fluctuaciones de posición en el frente de onda al llegar al receptor). En el apartado 3.7 se analiza la influencia del ruido de fondo en el sistema de QKD y se relaciona con la influencia del *beam wander* para pasar a proponer una solución que afronte ambos problemas de forma simultánea. Esta aproximación es necesaria porque las soluciones que precisa cada uno de estos problemas por separado influyen mutuamente con efectos opuestos, de forma que mitigar un problema empeora el otro y viceversa. Para solucionarlo se propone la implementación de un sistema automático de corrección de *beam wander* y se sugieren dos distintas configuraciones. Por último, se analiza la conveniencia de cada una de ellas, seleccionando la más apropiada para su implementación experimental y su integración en el sistema de QKD. El apartado 3.8 se dedica a analizar los diferentes aspectos de interés de los componentes fundamentales del sistema de corrección propuesto, específicamente el espejo modulable y el detector de posición. Para ello, se ofrece una caracterización de los componentes y una discusión de las consideraciones importantes a tener en cuenta al seleccionarlos y utilizarlos en el sistema de corrección. En el apartado 3.9 se estudia la influencia de la longitud de onda en el *beam wander*, para lo que se describe un experimento realizado con este propósito al no haber encontrado en la literatura científica ninguna referencia a esta dependencia más que de forma superficial y tratarse de un aspecto fundamental en la estrategia de corrección propuesta. Con el objetivo de realizar simulaciones ópticas del sistema de corrección adaptado al sistema de QKD, en el apartado 3.10 se realiza un modelado del telescopio receptor que precisó un análisis en profundidad debido a la escasa información de este telescopio por parte de su fabricante. En los tres últimos apartados de la tesis se tratan las diferentes alternativas experimentales que se exploraron para implementar el sistema de corrección: en bucle abierto (apartado 3.11), en bucle cerrado (apartado 3.12) y basada en doble corrección (apartado 3.13). La corrección en bucle abierto se trata de la estrategia más simple, si bien adolece de una serie de inconvenientes que no la hacen apropiada para las condiciones de operación del enlace. La corrección en bucle cerrado solventa los problemas de la estrategia en bucle abierto, si bien presenta una serie de dificultades técnicas descritas en el apartado que no la presentan como la más aconsejable. No obstante, a diferencia de la corrección en bucle abierto, no presenta problemas irresolubles y por ello se propone una serie de soluciones a los mismos. Por último, se presenta la que se identificó como la estrategia óptima para realizar la corrección de *beam wander* así como de cualquier otro tipo de desalineamiento del haz. Esta estrategia se basa en un sistema de corrección basado en doble espejo, que a cambio de una asumible mayor complejidad, permite integrar una corrección mucho más robusta en el sistema de QKD sin perjudicar sus prestaciones.

# 2. Telescopios *Cherenkov* para Comunicaciones en Espacio Profundo

## 2.1. INTRODUCCIÓN

Existe un consenso creciente acerca de la utilización de las comunicaciones ópticas en enlaces de espacio profundo, cuyas ventajas se hacen más patentes cuanto mayor es la distancia y los requisitos en cuanto a ancho de banda [30]. La divergencia de las señales ópticas, de varios órdenes de magnitud por debajo de las señales de radio frecuencia, posibilita una transmisión de la potencia mucho más eficiente desde el emisor hacia el receptor. Además, es posible el aprovechamiento de la madura tecnología de fibra óptica para conseguir tasas binarias mucho mayores que en radio frecuencia, entre otras ventajas, como la menor potencia consumida, peso y tamaño de los terminales [31]. Sin embargo, las señales de los futuros enlaces de comunicaciones desde espacio profundo serán extremadamente débiles, exigiendo el máximo en las prestaciones de cada elemento del enlace. Las medidas de optimización relacionadas con las estaciones terrenas tienen un gran potencial para mejorar la recepción de estas señales [32], y además son la opción más recomendable por trasladar la complejidad desde los terminales remotos a los terminales terrestres, donde cualquier dificultad o limitación es más fácilmente acometible. La medida más directa es el aumento de la apertura de los telescopios receptores, con el objetivo de capturar un mayor número de fotones. La misma medida ha sido llevada a cabo tradicionalmente en su equivalente en radio frecuencia, siendo las antenas de 70 metros los principales receptores de la estación de espacio profundo de la NASA. Y lo mismo cabe decir en el caso de los telescopios astronómicos, con los sucesivos aumentos del área recolectora para maximizar la sensibilidad de los mismos [33, p. 1].

El proyecto CTA (del inglés *Cherenkov Telescope Array*) [34] es una colaboración multinacional para construir durante los próximos años una nueva generación de telescopios *Cherenkov*, basados en los telescopios que han revolucionado la astronomía en rayos gamma desde la Tierra en los años recientes. El proyecto CTA conseguirá una mejora cualitativa en la sensibilidad de la detección de eventos mediante un *array* de varias decenas de telescopios de tres diferentes tamaños (~6 metros, ~12 metros y ~24 metros de diámetro) en dos localizaciones diferentes, una en cada hemisferio terrestre, cuyos emplazamientos están todavía por determinar. Actualmente se halla en fase de diseño y prototipado y se planea comenzar la construcción en 2016 para alcanzar la operatividad a partir de 2020 [35, p. 297]. La astronomía de rayos gamma comparte la misma limitación fundamental que las comunicaciones en espacio profundo: las señales extremadamente débiles que deben ser recibidas en la Tierra. La solución llevada a cabo en los telescopios *Cherenkov* ha sido tradicionalmente, y así será también en CTA, incrementar la apertura de recepción y replicar los telescopios usando topologías de *array*. Ambas soluciones pueden ser aprovechadas de forma directa por las comunicaciones ópticas.

Los telescopios *Cherenkov* no detectan la radiación de rayos gamma directamente. En su lugar, detectan sus efectos tras interaccionar con la alta atmósfera. Cuando las partículas que resultan de esta interacción viajan a través de un medio a una velocidad mayor a la de la luz en el medio, se produce el efecto *Cherenkov*, consistente en una lluvia de fotones con forma de cono que determina la dirección de las partículas originales [36]. Estos fotones se reciben en un intervalo espectral desde unos 300 nm hasta varios metros, limitado por la transmisión atmosférica [37, p. 59]. Sin embargo, la intensidad es proporcional a $\lambda^2$, y por ello las componentes azul y ultravioleta predominan. Por esta razón, aunque el objetivo de CTA es la astronomía de rayos gamma, sus telescopios son de hecho telescopios ópticos, y un número de circunstancias favorables justifican el estudio de la posibilidad de aprovechar esta tecnología en estaciones ópticas de comunicaciones. En esta tesis se desarrolla un estudio de viabilidad para llevar a cabo esta propuesta, de la que no se conoce ningún precedente en la bibliografía científica. Para ello se ha utilizado en lo posible la última información del proyecto CTA disponible de forma oficial, y en el caso de diseños no finalizados o no accesibles se ha empleado información relativa a los telescopios actualmente operativos más parecidos a los que se usarán en CTA, especialmente los telescopios MAGIC I y II en el observatorio Roque de los Muchachos de La Palma.

## 2.2. ASTRONOMÍA DE ALTAS ENERGÍAS

En astronomía se denomina altas energías a la correspondiente a la radiación X y gamma, es decir, el extremo más energético del espectro electromagnético. El estudio de los rayos cósmicos y los neutrinos también se incluye dentro de esta disciplina debido a su alta energía asociada. Para cuantificar estas energías se emplea el electrón-voltio (eV), que es la energía que adquiere un electrón cuando es acelerado por una diferencia de potencial de un voltio. La luz visible tiene energías en torno al eV, los rayos x están mil veces por encima (keV) y los rayos gamma más de un millón de veces (MeV), pudiendo alcanzar casi los 100 billones de eV (~100 TeV) [38]. Esta disciplina es relativamente moderna, con unos comienzos paralelos a los de la era espacial, alrededor de los años 60. Su desarrollo ha dado lugar al descubrimiento de un gran número de nuevos objetos astronómicos, en general relacionados con los sucesos más violentos y energéticos del universo.

En este apartado se introduce una serie de conceptos relacionados con la disciplina de estudio que hace uso de los telescopios *Cherenkov*, con el objetivo de presentar el ámbito de trabajo en el que se ha desarrollado esta parte de la tesis. Si bien se ofrece una panorámica del entorno científico en el que operan estos telescopios, la atención se centra específicamente en los conceptos más directamente involucrados en el trabajo desarrollado: la adaptación de los telescopios terrestres de rayos gamma como terminales receptores de comunicaciones ópticas.

## 2.2.1. Rayos cósmicos y rayos gamma

La Tierra es constantemente alcanzada por partículas extraterrestres de alta energía (en el amplio rango que cubre desde 1 GeV hasta unos $10^{12}$ GeV), llamadas de forma genérica rayos cósmicos. Su composición es bien conocida: principalmente protones de alta energía, y en menor medida otros núcleos atómicos y electrones. Sin embargo, su origen siempre fue una incógnita, aunque últimamente se han determinado posibles escenarios como las ondas de choque a que dan lugar las supernovas o los intensos campos magnéticos alrededor de los púlsares. Esta incógnita se debe a que al estar los rayos cósmicos cargados eléctricamente, son desviados por los campos magnéticos del medio interestelar e intergaláctico, destruyéndose así la información relativa a su procedencia. Por otra parte, los rayos gamma son fotones de alta energía sin carga, por lo que no sufren ninguna perturbación en su trayectoria debido a los campos magnéticos. Esto permite utilizarlos como trazadores sobre el origen de los rayos cósmicos, ya que el origen de ambos está relacionado. Además, los rayos gamma por sí mismos han generado gran interés al permitir estudiar objetos únicamente detectables en esa zona del espectro, fenómenos de especial interés a altas energías, así como modernas teorías como la materia oscura o las supercuerdas. Por estas razones, la astronomía de rayos gamma se ha convertido en uno de los campos de investigación científica más prometedores de los últimos años [39, p. 2].

Los rayos gamma forman la parte del espectro electromagnético de mayor energía y menor longitud de onda (considerados a partir de 0,001 nm) e históricamente ha sido la última región espectral descubierta. Su existencia está relacionada con emisiones no térmicas, ya que, desde el Big Bang, no existe nada tan caliente como para generar radiación electromagnética tan energética. En esta parte del espectro de alta y muy alta energía (respectivamente VHE y UHE, del inglés *Very High Energy* y *Ultra High Energy*), de energías mayores a 100 GeV, existe una conocida relación entre el flujo de partículas y su energía, que establece que las de mayor energía se presentan en menor proporción. Por ejemplo, a $10^{11}$ eV los rayos cósmicos tienen un flujo por metro cuadrado y estereorradián de uno por segundo, a $10^{16}$ eV de uno cada año y a $10^{21}$ eV de uno por siglo (Figura 5). Esto evidencia que una única técnica no puede servir para observar en todo el rango de energías. En el caso de los rayos gamma, en la práctica se establece una diferenciación de técnicas cada tres órdenes de magnitud de energía [40, p. 4]. Además de presentarse en un gran rango, las elevadas energías imponen otra dificultad añadida: si bien el resto del espectro puede ser captado con algún tipo de antena o telescopio, en general la radiación gamma no permite actualmente su detección de forma directa (a excepción de algunos satélites que captan rayos gamma en la parte energéticamente más baja del espectro) y han de usarse técnicas indirectas de reconstrucción como la explicada en el siguiente apartado.

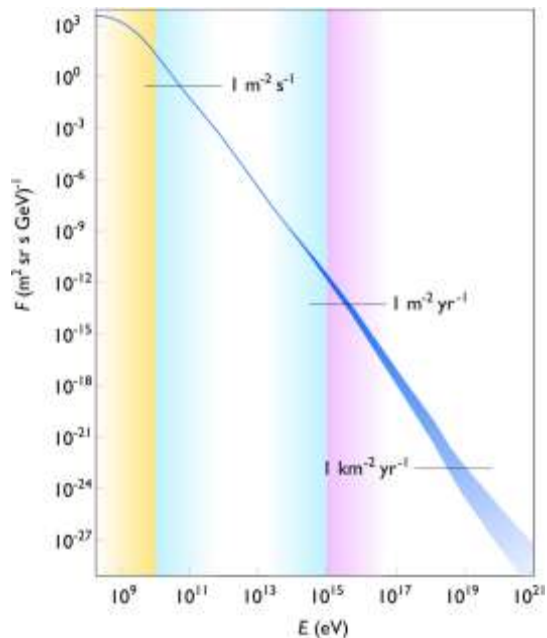

**Figura 5. Flujo de rayos cósmicos frente a energía.**

## 2.2.2. Efecto *Cherenkov* y detección en tierra

El efecto *Cherenkov* fue descubierto por Pavel Alekseyevich Cherenkov en 1934 [41], gracias a lo cual le fue concedido el premio nobel en 1958. Consiste en la generación de una luz característica llamada radiación *Cherenkov* producida cuando partículas cargadas interaccionan con un medio a una velocidad mayor a la de la luz en ese medio [42, p. 46]. Cuando los fotones de la radiación gamma interaccionan con las moléculas de la atmósfera, se producen pares electrón-positrón. Estas partículas vuelven a interaccionar con la atmósfera y a través de *Bremsstrahlung* pierden parte de su energía creando fotones gamma secundarios. Estas partículas secundarias a su vez producen otros pares electrón-positrón, y así sucesivamente, mientras los fotones permanezcan con una energía superior a 1,022 MeV. El resultado es una cascada de pares electrón-positrón y fotones que viajan por la atmósfera hasta llegar a la superficie (Figura 6, izquierda). Las partículas resultantes poseen una velocidad inicial mayor a la de la luz en la atmósfera, lo que produce una onda de choque de una forma cónica y compuesta por luz azulada, conocida como radiación *Cherenkov* [43, p. 539]. La luz *Cherenkov* también es producida por los rayos cósmicos con un efecto diferenciable por sus distintas características (Figura 6, derecha). Solo el 0,1 % de las lluvias se generan por rayos gamma; el resto tiene su origen en los rayos cósmicos y suponen la principal fuente de ruido de fondo en la astronomía gamma terrestre.

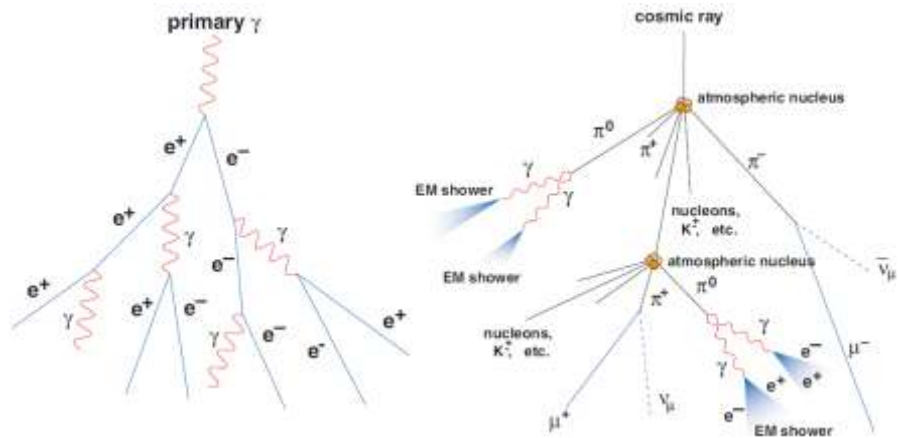

**Figura 6. Lluvia de partículas generada por rayos gamma (izquierda) y por rayos cósmicos (derecha).**

Todo el proceso desde que la partícula de alta energía impacta en la atmósfera hasta que los fotones resultantes llegan a la superficie terrestre se denomina cascada atmosférica extensa (Figura 7). Debido a las continuas variaciones en las condiciones atmosféricas y a las distintas energías de las partículas que interactúan con ella, el cono de luz *Cherenkov* puede tener multitud de orígenes, con alturas variando entre 6 y 20 km [44, p. 236]. Cuanto mayor es la altura a la que se genera, menor ángulo describirá el cono, aunque se suele considerar una altura típica de 10 km con un cono de 0,7°, lo que equivale a un área de unos 120 km [45, p. 5], conocida como piscina de luz *Cherenkov*, donde existe un máximo de partículas alrededor de 1 TeV. Registrando imágenes de la distribución de estas cascadas es posible deducir el desarrollo de las mismas y así determinar la dirección de llegada de los rayos gamma originales, así como su energía. Esta técnica es la más eficiente para el estudio de los rayos gamma en la Tierra y se basa en telescopios ópticos, diferenciados de los convencionales por una serie de características especiales, como por ejemplo que no enfocan al infinito, sino precisamente a una altura de 10 km.

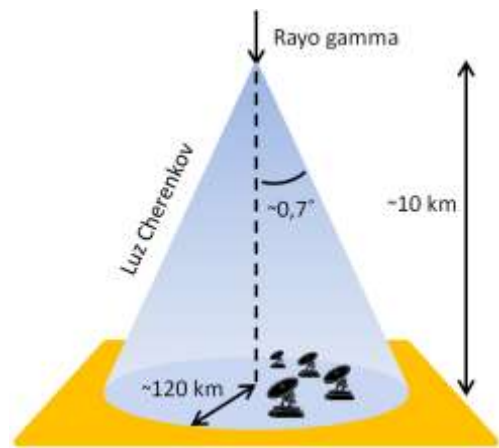

**Figura 7. Cascada atmosférica de luz *Cherenkov*.**

## 2.2.3. Telescopios terrestres de rayos gamma

Existen varias técnicas para estudiar los rayos cósmicos y gamma que llegan a la Tierra: la detección del efecto *Cherenkov* que se produce al interactuar los rayos gamma con el agua almacenada en grandes tanques con fotomultiplicadores como elementos detectores [46]; la detección de la emisión de fluorescencia que sucede junto a la emisión *Cherenkov* al interactuar las partículas cargadas con la atmósfera, de cuya intensidad y separación temporal se puede deducir la energía y dirección de las partículas originales [47]; la detección directa de la cascada de partículas secundarias mediante *arrays* de decenas de pequeños detectores de centelleo cubriendo extensas áreas de terreno [48]; y por último la detección de la luz *Cherenkov* con grandes telescopios mediante su focalización en grandes cámaras. Esta última es la técnica de interés en esta tesis por ser la empleada en el proyecto CTA, cuyos telescopios se pretenden adaptar para FSOC.

Los telescopios *Cherenkov* o IACT (del inglés *Imaging Atmospheric Cherenkov Telescope*) son telescopios ópticos terrestres, mucho más simples que los convencionales utilizados en astronomía visible e infrarroja debido a las menores exigencias respecto a la formación de imágenes por una serie de razones que se explicarán más adelante. Están diseñados para detectar la luz *Cherenkov* en noches oscuras debido a la baja intensidad de dicha radiación y en muy buenas condiciones atmosféricas, por lo que se sitúan en emplazamientos de similares características a los elegidos para los observatorios astronómicos convencionales o directamente junto a ellos. En una primera aproximación se puede decir que están compuestos por un reflector segmentado de gran apertura y una cámara consistente en varios cientos de fotodetectores en el plano imagen, cuyo plano objeto se encuentra a unos 10 km de altura en la atmósfera, todo ello expuesto a la intemperie y montado en una estructura más parecida a la de un radiotelescopio que a la de un telescopio óptico.

En términos generales, cualquier telescopio busca recolectar el mayor número de fotones posible, dada una extensión angular determinada por la aplicación, y redirigirlos hacia algún tipo de dispositivo fotodetector sensible a la longitud de onda de interés. La apertura vendrá determinada por la cantidad de fotones que se precise recolectar para obtener una detección válida. Por ello, debido a la débil y dispersa radiación *Cherenkov*, los IACT son telescopios con reflectores mucho mayores a los empleados en astronomía óptica, con tamaños típicos entre los 6 y los 30 metros. Respecto a la resolución angular, pese a tratarse de telescopios de imágenes, los requerimientos de calidad óptica son poco estrictos, a medias entre un simple recolector de luz (simple y barato) y un sistema formador de imagen (complejo y caro). Si un telescopio astronómico convencional puede alcanzar resoluciones en torno al arco segundo, los telescopios *Cherenkov* están alrededor de los 15 arco minutos, unos tres órdenes de magnitud por debajo.

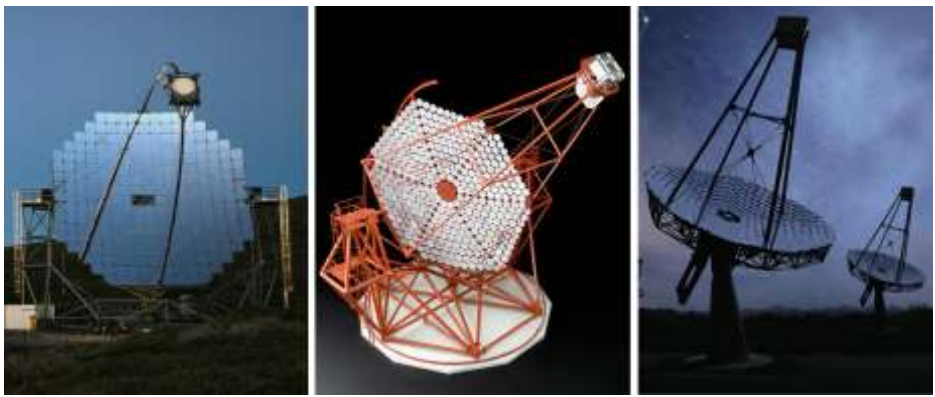

**Figura 8. Principales observatorios de telescopios *Cherenkov*:
MAGIC (izquierda), HESS (centro) y VERITAS (derecha).**

La astronomía basada en IACT ofrece una mejora en la sensibilidad y en la resolución angular a costa de un menor campo de visión. Esta técnica se demostró en 1988 con la primera observación astronómica en rayos gamma basada en un telescopio *Cherenkov* [49]. Se pudo registrar una fuerte señal proveniente de la nebulosa del Cangrejo en el rango de los TeV mediante un IACT de 10 metros de diámetro. Desde entonces la técnica ha mejorado manifiestamente y actualmente existen tres principales observatorios de rayos gamma: MAGIC (*Major Atmospheric Gamma Imaging Cherenkov*) en La Palma con dos telescopios de 17 metros de diámetro, VERITAS (*Very Energetic Radiation Image Telescope Array System*) en Arizona con cuatro telescopios de 12 metros y HESS (*High Energy Stereoscopic System*) [50] en Namibia con cuatro telescopios de 12 metros y uno de 28 metros [1] (Figura 8).

## 2.3. PROYECTO CTA

### 2.3.1. Descripción del proyecto

La última generación de telescopios *Cherenkov* (principalmente MAGIC, VERITAS y HESS) han convertido a la astronomía de rayos gamma en una disciplina consolidada, con más de 100 fuentes identificadas en el espacio, como supernovas, púlsares, sistemas binarios, galaxias activas, etc. Sin embargo, con la tecnología actual haría falta más de una década para realizar un sondeo completo del cielo, de lo que ha surgido la necesidad de llevar a cabo un esfuerzo conjunto internacional para potenciar el desarrollo de esta disciplina. De las técnicas existentes para observación de rayos gamma en tierra, la detección de la radiación *Cherenkov* mediante *arrays* de detectores ofrece la mayor cobertura angular de cielo y el mayor ciclo de trabajo extendiendo la operación durante el día. Sin embargo, la técnica basada en IACT es muy superior en cuanto a sensibilidad, fundamental para la realización de un mapeo detallado del cielo, por lo que ha sido la elegida para el futuro CTA (del inglés *Cherenkov Telescope Array*).

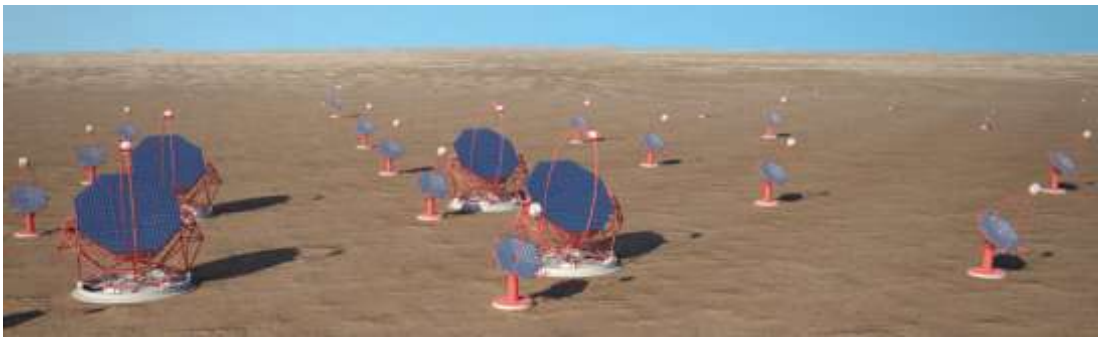

**Figura 9. Ilustración del futuro *Array* de Telescopios *Cherenkov* (CTA) [51].**

Concebido en 2005 por miembros de los observatorios MAGIC y HESS, y con una puesta en funcionamiento inicialmente prevista para 2020, el desarrollo del proyecto CTA (Figura

---

[1] El diseño de los telescopios Cherenkov en general se basa en un reflector segmentado, con diferentes topologías en cuanto a tamaño y forma de los espejos individuales. El diámetro de apertura se refiere al equivalente de un reflector circular único y se emplea por tradición y simplicidad al compararlos con telescopios más tradicionales, si bien sería más correcto hablar de área recolectora.

9) será el resultado de una colaboración multinacional de todos los grupos del mundo que trabajan actualmente con telescopios *Cherenkov* (más de 600 científicos de 25 países del mundo), y por primera vez en la disciplina ofrecerá un modo de operación abierto a toda la comunidad científica. Pretende ofrecer una cobertura total del cielo, para lo cual contará con dos ubicaciones, una por hemisferio, siendo la del Sur la principal ubicación debido a la abundancia de fuentes de interés en la zona central de la galaxia. Se ha previsto un presupuesto total en torno a los 150 millones de euros, recientemente actualizado a 200, que se considera un coste modesto para una instalación de la magnitud y potencial científico de CTA. Una vez operativo, el proyecto CTA promete un salto cualitativo en cuanto a [39, p. 23]:

- Sensibilidad: la mejora será de más de un factor 10 sobre la de cualquier instrumento actual, lo que permitirá incrementar el número de fuentes de rayos gamma detectadas en un orden de magnitud, hasta unas mil previstas.

- Rango espectral: con el mismo observatorio podrá cubrirse un rango de energías de hasta cuatro órdenes de magnitud, aproximadamente el doble que cualquiera de los principales observatorios actuales.

- Resolución angular: unas 5 veces mejor gracias no solo al aumento de la resolución de cada telescopio, sino a su operación conjunta, que permite la detección de la misma cascada desde diferentes telescopios.

- Resolución temporal: podrá resolver temporalmente eventos de rayos gamma en rangos inferiores al minuto cuando actualmente los mejores observatorios ofrecen resoluciones de varios minutos.

- Flexibilidad de operación: el gran número de telescopios y sus distintas características permitirá una variedad de configuraciones para maximizar la sensibilidad, el campo de visión o combinaciones de ambos.

## 2.3.2. Tipos de telescopios y configuración del *array*

Debido al amplio rango de energías que se precisa detectar, no es posible utilizar un diseño único para todos los telescopios. Por ello, en CTA se han distinguido tres rangos energéticos, diseñando un tipo distinto de telescopio para cada uno de ellos y un criterio de despliegue en el *array*:

- Telescopios de pequeño tamaño SST (del inglés *Small Size Telescope*). Destinados a cubrir el rango de mayores energías, por encima de varios TeV. Se prevé un *array* de unos 100 telescopios con aperturas alrededor de los 6 metros de diámetro (ver Figura 10, derecha) y unos 10° de campo de visión, cubriendo un área de hasta 10 km². La luz *Cherenkov* originada en este rango de energía proporciona un flujo de fotones suficientemente alto como para producir una detección fiable de un evento utilizando una superficie colectora relativamente pequeña y a distancias mayores que el radio típico de la piscina de luz *Cherenkov*. Por ello, es preferible utilizar un mayor número de telescopios más pequeños y así cubrir un área mayor. Así se asegura que siempre habrá telescopios situados en la distancia óptima de detección (algo más de 70 metros del centro de la piscina), donde la intensidad de luz es mayor, hay menores fluctuaciones y el ángulo facilita una mejor reconstrucción de la dirección original de los rayos. La filosofía de desarrollo de estos telescopios difiere de los de mayor tamaño, ya que debido a su gran número se requerirá llevar a cabo una producción masiva, para lo cual el proceso deberá optimizarse. Por ello, se realizarán varios prototipos y se elegirá el más eficiente en la fabricación final. Se barajan dos opciones para el diseño óptico: reflector simple o doble.

- Telescopios de tamaño medio MST (del inglés *Medium Size Telescope*). Destinados a cubrir el rango de energías entre 100 GeV y 1 TeV. Los telescopios tendrán unas aperturas de alrededor de 12 metros de diámetro (ver Figura 10, centro) y entre 6° y 8° de campo de visión y formarán un *array* de varias decenas de unidades, con espaciados de alrededor de 100 metros y cubriendo un área aproximada de 1 km². El objetivo de agruparse en una superficie de detección mayor que la piscina de luz *Cherenkov* es compartido con los telescopios SST, si bien en este caso es necesaria una mayor área colectora en cada IACT para producir una detección fiable, debido al menor flujo de fotones en este rango de energía. Estos telescopios formarán el núcleo de CTA, por ser la tecnología más conocida gracias a la experiencia de los actuales observatorios como MAGIC. Sin embargo, se pretende ofrecer un salto cualitativo respecto a los actuales, por lo que se construirá un prototipo y, tras su evaluación, se seguirá con la cadena de producción del grueso de unidades. Si bien este tipo de telescopios se ha venido utilizando desde hace años, e incluso en formaciones de varios elementos, esta será la primera vez que el tamaño del *array* será mayor que la piscina de luz *Cherenkov*, lo que permitirá obtener una gran mejora en la resolución. Esto ya se ha comprobado en los infrecuentes casos en los que un mismo evento ha logrado disparar simultáneamente la detección de los cuatro telescopios HESS.

- Telescopios de gran tamaño LST (del inglés *Large Size Telescope*). Destinados a cubrir el rango de baja energía, por debajo de 100 GeV. Se prevé un *array* de 4 telescopios de unos 24 metros de diámetro (ver Figura 10, izquierda), lo que supone alrededor de un 10 % del área colectora total de CTA, entre 4° y 5° de campo de visión y un espaciado similar al de los MST, de unos 100 metros. El gran tamaño se debe a que en este rango energético no existe suficiente flujo de fotones *Cherenkov* para producir una señal de disparo fiable utilizando una superficie colectora pequeña. Una alternativa a un gran reflector sería emplear conjuntamente varios telescopios de menor tamaño, pero esto exigiría superponer de forma simultánea las imágenes de todos los telescopios para producir el disparo, lo que supondría un gran desafío tecnológico debido al gran ancho de banda requerido. A diferencia de los SST y MST, dado el gran tamaño y coste de estos telescopios así como su menor número, inicialmente no se prevé desarrollar ningún prototipo. En su lugar, se construirán los telescopios finales directamente en la ubicación definitiva, para lo cual se llevará a cabo un detallado estudio con el objetivo de minimizar el coste y maximizar sus prestaciones.

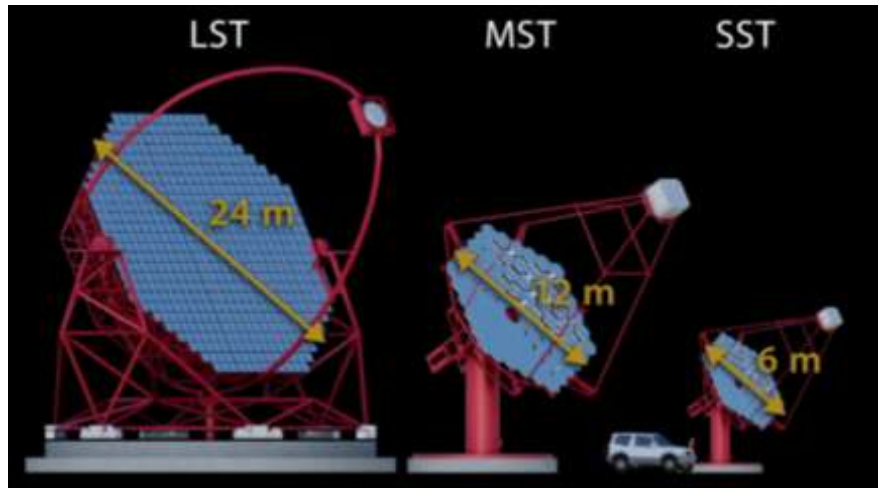

**Figura 10. Ilustración de prototipos para los telescopios LST, MST y SST [52].**

El número exacto de telescopios para cada uno de los dos *arrays* de CTA aún no está establecido, ya que depende de las configuraciones de los mismos en los *arrays* definitivos. La determinación de estas configuraciones precisa detalladas y costosas simulaciones, con cientos de telescopios de distintas características, dependientes de las cascadas previstas y de la tecnología de los telescopios, aún no completamente definida. Las simulaciones son fundamentales para optimizar la sensibilidad del *array*, así como la eficiencia en la discriminación de cascadas hadrónicas (de rayos cósmicos) y electromagnéticas (de rayos gamma). En este sentido, en todas las configuraciones se prevé un gran número de SST al facilitar la distinción entre cascadas incluso en energías solo accesibles a LST [53, p. 876]. Se barajan 14 configuraciones distintas, aunque solo 4, los llamados *subarray* E, I, J y K, son las candidatas por ofrecer un compromiso entre prestaciones y coste. El *subarray* E (el correspondiente al CTA Sur en la Figura 11) lo compondrían 59 telescopios: 4 LST, 23 MST y 32 SST; el I estaría formado por 77 telescopios: 3 LST, 18 MST y 56 SST; el J por 49 telescopios: 3 LST y 46 MST con distintos campos de visión; y el K por 76 telescopios: 5 LST y 71 SST [54, p. 57]. Cada hemisferio precisará una configuración específica, según sus objetivos científicos: CTA Sur estará orientado a observar fuentes de la zona central de la galaxia, por lo que será sensible al rango completo de energías; en cambio, CTA Norte se dedicará a observaciones extragalácticas y no requerirá detectar el espectro superior de energías, lo que se traduce en un menor número de telescopios SST.

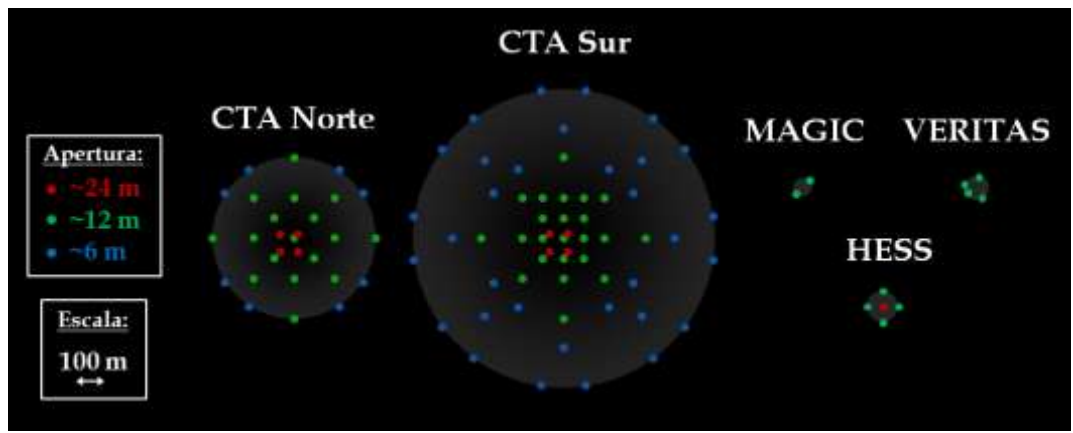

**Figura 11. Posibles alternativas para la configuración de CTA Norte y Sur [52] y comparación con los principales *arrays* de IACT operativos en la actualidad [55].**

### 2.3.3. Resumen de fundamentos tecnológicos

La tecnología necesaria para lograr los objetivos científicos establecidos para el proyecto CTA existe y es bien conocida desde hace tiempo, especialmente debido a la experiencia obtenida en observatorios como MAGIC, VERITAS y HESS. En la medida de lo posible, se aprovecharán los últimos avances tecnológicos realizados en este ámbito y además se han realizado importantes progresos orientados a mejorar ciertas especificaciones críticas. Sin embargo, el principal reto de CTA no es mejorar el estado del arte mediante la innovación con nuevas técnicas, sino que está relacionado con aplicar la tecnología conocida de forma eficiente para hacer uso de la economía de escala durante la fabricación y despliegue de los observatorios y aprovechar las ventajas que proporciona la configuración en *array* [39, p. iii]. La mayor innovación de CTA será precisamente la aplicación de las tecnologías existentes a este tipo de operación para cumplir con los propósitos últimos de la astronomía de rayos gamma. Estos propósitos en general dependen de la capacidad para recolectar luz y la eficiencia en la detección de los fotones recolectados y su conversión en fotoelectrones. Además, el sistema óptico debería ser capaz de proporcionar una resolución suficiente dentro del campo de visión requerido. El campo de visión limitará el tamaño máximo de las imágenes obtenidas y la calidad óptica de los telescopios limitará la resolución con que pueden observarse estas imágenes. Por último, la electrónica y las comunicaciones involucradas en el procesamiento de las señales generadas deben proporcionar un ancho de banda suficiente para resolver las cortas duraciones de los pulsos de luz *Cherenkov*.

El diseño óptico de los telescopios varía según el tamaño de su apertura con el objetivo de lograr el mejor compromiso entre varios parámetros. Por una parte, una relación focal f/D (siendo f la longitud focal y D el diámetro de la apertura) elevada proporciona una mejor resolución en los telescopios al disminuir sus aberraciones. Sin embargo, la simplicidad buscada para los IACT exige, en la medida de lo posible, el diseño de telescopios de reflector único con foco primario, lo que es incompatible con una elevada relación focal, la cual precisaría unas estructuras complejas para proporcionar la suficiente estabilidad en un foco muy alejado del reflector principal. Por ello, en este tipo de telescopios se utilizan relaciones focales entre f/1 y f/2, mucho menores que las de telescopios astronómicos convencionales [33, p. 135], lo que redunda en una peor resolución en las imágenes producidas, debido al mayor efecto de las aberraciones geométricas.

La forma del reflector es otro parámetro importante: en grandes telescopios segmentados como los IACT, un perfil muy adecuado es el *Davies-Cotton* [56], ya que todos los espejos tienen el mismo radio de curvatura esférico R = 2f y están dispuestos según un perfil esférico con radio f (Figura 12, izquierda). Esta simple estrategia proporciona un buen comportamiento ante aberraciones con los amplios campos de visión que precisa esta disciplina. El inconveniente que presentan es una alta dispersión temporal: los fotones que impactan en el extremo del telescopio llegan antes que los centrales debido a la diferencia de longitud de los caminos. Para reflectores relativamente pequeños, esta dispersión es menor a los ~3 ns de duración de los pulsos *Cherenkov*, pero al aumentar el tamaño de la apertura el retardo es mayor y el diseño *Davies-Cotton* no es válido. En este tipo de grandes telescopios se hace más apropiado un perfil parabólico, que al ser isócrono no introduce dispersión temporal, a cambio de presentar frente a los anteriores un comportamiento algo peor ante aberraciones para los mismos campos de visión. Así, para los diseños de reflector único SST y MST es preferible el perfil *Davies-Cotton* para proporcionar una buena resolución sobre un campo de visión amplio, siendo el parabólico el único posible para LST debido a su menor dispersión temporal.

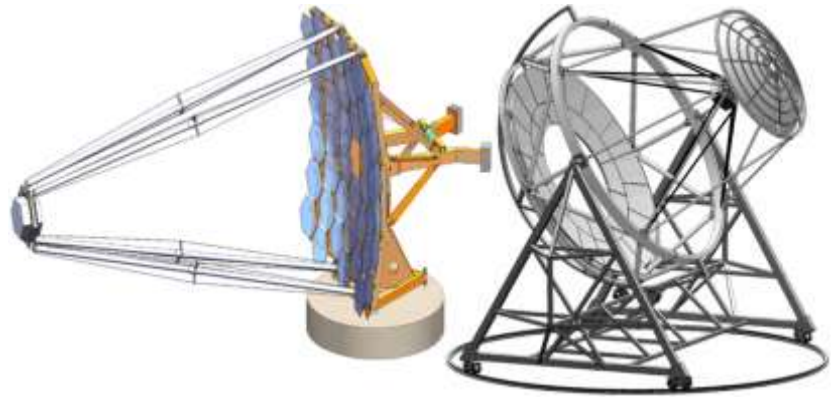

**Figura 12. Diseño *Davies-Cotton* [57] (izquierda)
y *Schwarzschild-Couder* [58] (derecha).**

Los diseños de doble espejo han sido investigados específicamente para CTA, ya que nunca antes se ha construido un IACT de este tipo. Estos diseños permiten una importante mejora de la calidad óptica de las imágenes observadas en un mayor campo de visión. El diseño de doble espejo contemplado para CTA es el *Schwarzschild-Couder* [59], que consiste en dos reflectores asféricos con un perfil basado en las ecuaciones de *Schwarzschild* para eliminar la aberración esférica y de coma. El reflector principal tiene topología segmentada y el secundario es un espejo monolítico, con un plano focal esférico para minimizar el astigmatismo [60]. A diferencia de los telescopios parabólicos y *Davies-Cotton*, en los que la cámara debe ser soportada en una estructura muy alejada del reflector principal, los *Schwarzschild-Couder* son diseños más compactos (Figura 12, derecha) gracias a que el esquema de doble espejo permite reducir la longitud focal de forma importante (resultando en una focal de menos de la mitad de su equivalente *Davies-Cotton*). El despliegue final de CTA probablemente incluirá para el caso de SST y MST un número similar de diseños de foco primario (*Davies-Cotton*) y de doble espejo (*Schwarzschild-Couder*) [61].

En los diseños con perfil parabólico y *Davies-Cotton* cada espejo estará construido con un perfil esférico con el objetivo de minimizar el coste de fabricación. En el caso de LST, la forma parabólica se obtiene situando a los espejos siguiendo esta curva, y construyendo los espejos esféricos con varios radios de curvatura distintos con valores cercanos a los que tendría la parábola en cada punto (esta es por ejemplo la estrategia seguida en MAGIC [62, p. 7]). La construcción de los espejos para los diseños de doble reflector, así como los SST tipo *Davies-Cotton*, presentan mayores exigencias debido a la asfericidad en el primer caso y a los cortos radios de curvatura (de menos de 10 metros) [63]. La predominancia de reflectores esféricos es una de las principales diferencias con los telescopios astronómicos convencionales, cuyos espejos describen perfiles derivados de secciones cónicas (parábolas, elipsoides e hipérbolas).

El gran tamaño de los telescopios de CTA hará que la estructura del reflector principal sufra deformaciones de origen gravitatorio (a lo que se sumarán las debidas al viento y los gradientes térmicos). Estas deformaciones impactan negativamente en la calidad óptica de las imágenes obtenidas al introducir una importante fuente de error en el frente de onda. Para evitarlo tradicionalmente los IACT de gran tamaño han contando con sistemas de alineamiento activo de los espejos para compensar la deformación global de la forma del reflector. Este es el caso de MAGIC [64] o HESS [65] que incorporan sistemas de dos actuadores longitudinales por espejo, con uno de los tres puntos de sujeción fijo. Estos sistemas se utilizan para el alineamiento inicial de los espejos utilizando láser o imágenes de estrellas brillantes, y durante la operación normal se pueden realizar ajustes automáticos basados de forma determinista en el ángulo de elevación del telescopio (estrategia prevista para SST), o activamente, monitorizando la deformación de la estructura periódicamente (para MST y especialmente LST).

La calidad óptica de los telescopios se expresa utilizando la función de dispersión puntual PSF (del inglés *Point Spread Function*), que cuantifica cómo de bien se concentra la luz de una fuente puntual recibida por el sistema óptico. En general, la PSF proporcionada por los telescopios debe ser equivalente al tamaño de cada píxel[2]. Si fuera mayor se estaría desaprovechando la máxima resolución de la cámara y si fuera menor, sería la calidad óptica del telescopio la desaprovechada por una cámara que no puede registrar esa resolución. Por ello, el diseño óptico del telescopio y el diseño de la cámara están íntimamente relacionados. El tamaño de píxel depende de la tecnología de detección utilizada, que en IACT se concentra principalmente en los fotomultiplicadores o PMT (del inglés *PhotoMultiplier Tube*). Estos detectores ofrecen una buena eficiencia cuántica (hasta >40 % en la última generación de PMT [66]) con unas ganancias elevadas (hasta ~$10^6$), con bajo ruido y una buena respuesta temporal para detectar los rápidos pulsos *Cherenkov* (desde varios nanosegundos) a partir de hasta un solo fotón. El material empleado ha sido tradicionalmente un compuesto bialcalino (Sb-Rb-Cs, Sb-K-Cs) por proporcionar una curva de eficiencia cuántica en la banda de 300-600 nm bien adaptada al espectro *Cherenkov*. El tamaño de estos píxeles será de entre algo menos de 1 cm (para los diseños de doble reflector) y unos 5 cm (en el caso de LST). Ubicada ante la matriz de píxeles formada por los PMT, se utiliza una matriz de conos *Winston*, que sirven para concentrar la luz en cada detector, y permiten reducir su tamaño en un factor 3-4, además de minimizar el espacio perdido entre detectores hasta un máximo de 1 mm [67]. La abundante experiencia en PMT hacen a estos detectores los principales candidatos para CTA. Sin embargo, existe una línea de investigación sobre otro tipo de detectores basados en los tradicionales fotodiodos de avalancha APD (del inglés *Avalanche PhotoDiode*) operados en modo *geiger* como contadores de fotones. Con el doble de eficiencia cuántica que los PMT y un menor tamaño, estos detectores de Silicio ofrecen una prometedora alternativa, especialmente en los SST de doble espejo, cuya mejor resolución precisa píxeles de menor tamaño a los que se pueden fabricar con PMT.

Las cámaras consisten en una estructura cilíndrica construida con materiales ligeros que contiene en uno de los extremos la matriz de PMT dispuestos para optimizar la recolección de la luz recibida. Con matrices de cientos de detectores y una sección transversal de más de dos metros, serán las mayores cámaras que se han construido, y por ello se tratará de instrumentos muy pesados, de hasta 2,5 toneladas [68]. Ensamblada al soporte del reflector principal mediante un entramado de brazos metálicos, la estructura de la cámara está en una posición fija, si bien interiormente es posible desplazarla a lo largo de un carril longitudinal de hasta 40 cm [67, p. 4] por delante y detrás del plano focal para corregir errores de alineamiento. En previsión de posibles actualizaciones futuras de fotodetectores, electrónica de preprocesado de señal, electrónica de disparo, etc. el diseño de cada cámara es modular (Figura 13).

---

[2] Si bien esta es una buena regla, en los diseños ópticos se definen unas relaciones más precisas como que la parte de la PSF que contiene el 40 % de la potencia debe abarcar la mitad del tamaño de píxel [39, p. 74], o un tercio de su tamaño para un 80 % de confinamiento [63, p. 16].

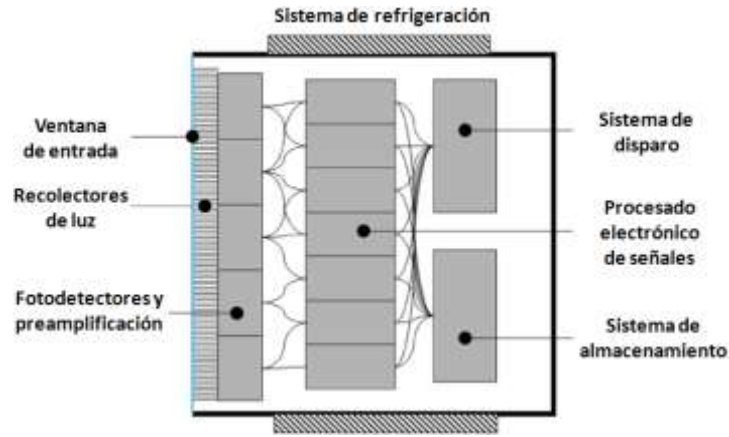

**Figura 13. Diseño modular de las cámaras [39, p. 99].**

Para distinguir un evento astronómico captado por un IACT por encima del ruido de fondo es necesario diseñar un sistema de disparo. En una configuración tipo *array* como la de CTA se requiere un disparo basado en coincidencias entre telescopios. El criterio clave es cuánta información se intercambia entre telescopios para llegar a un compromiso óptimo entre fiabilidad en el disparo y simplicidad en la sincronización. La estrategia más simple consiste en dejar que cada telescopio proporcione un disparo de forma independiente y comparar los disparos entre telescopios cercanos (esta es la estrategia de HESS [69]). En este caso el mínimo umbral para producir un disparo depende del umbral de cada telescopio por separado. En el otro extremo de complejidad estaría la estrategia consistente en comparar píxel a píxel las imágenes de cada uno de los telescopios, sea de forma directa con las señales analógicas (menos retardo pero más complejo) o con las señales digitales una vez procesadas (más retardo a costa de simplificar la sincronización). En este caso el umbral del disparo podría estar muy por debajo del umbral de cada telescopio de forma separada, lo que aumentaría el número de eventos detectados a costa de aumentar la complejidad en la sincronización. Entre estos dos extremos existe una variedad de estrategias intermedias, como comparar a nivel de píxel únicamente las zonas candidatas a producir un disparo. En cualquier caso, un sistema de disparo flexible que permita múltiples configuraciones diferentes precisará un ancho de banda excepcional: por ejemplo, un único telescopio de 1000 píxeles muestreado solo a 200 MHz con 8 bits por píxel genera un flujo binario de más de 1 Tbit/s. La infraestructura de comunicaciones que se requiere para dar soporte a este flujo de datos entronca perfectamente con su utilización para sustentar un sistema de FSOC de alta velocidad como el que se pretende integrar.

## 2.4. COMUNICACIONES ÓPTICAS EN EL ESPACIO

Los vehículos y asentamientos espaciales del futuro van a exigir un cambio de paradigma en la forma en que se transmite y recibe la información. Esto vendrá determinado por un incremento exponencial de la cantidad de datos de diferente naturaleza que se precisará intercambiar entre emplazamientos cada vez más remotos, sumado al hecho de que las características de esta comunicación en general se degradan con la distancia. Las telecomunicaciones por radiofrecuencia en espacio profundo (entendiendo espacio profundo por distancias más allá de la Luna) suponen ya en la actualidad un cuello de botella en el volumen de datos que pueden transmitirse hacia la Tierra. Por ejemplo, el orbitador MRO (del inglés *Mars Reconnaissance Orbiter*) de la NASA es la sonda con mayor capacidad de comunicación en espacio profundo y su máxima tasa de transmisión utilizando su transmisor de microondas en condiciones ideales está por debajo de los 6 Mbit/s [70, p. 46]. Para satisfacer la creciente demanda de ancho de banda, tradicionalmente la estrategia ha sido la migración a frecuencias de portadora cada vez mayores [71, p. 33]. Así, las bandas L y S (1-4 GHz) usadas durante la década de los 60 dieron paso a la banda X (8-12 GHz) a finales de los 70 y a las primeras experiencias en banda Ka (26,5-40 GHz) en la década de 2000. El traslado a frecuencias ópticas supone un salto cualitativo, porque significa pasar de decenas de GHz (banda Ka) a cientos de miles de GHz (equivalente a una longitud de onda de 1 μm), lo que son unos cuatro órdenes de magnitud. Traducido a capacidad de transmisión, transmitir un mapeo completo de Marte de ~30 cm de resolución precisaría 9 años utilizando la mejor tecnología de microondas y 9 semanas con comunicaciones ópticas.

El desarrollo de las comunicaciones ópticas en espacio profundo o FSOC (del inglés *Free-Space Optical Communication*) se ha visto inicialmente ralentizado por la conjunción de una serie de factores. Entre ellos destacan la barrera de entrada que en la práctica supone competir con una tecnología que está totalmente consolidada en el ámbito aeroespacial, como es la basada en radiofrecuencias. Otras dificultades para superar esta fase inicial tienen que ver con las propias del sector espacial, muy reticente a integrar nuevas tecnologías si no es con pequeños saltos cuantitativos incrementales a partir de lo que se ha probado con anterioridad. No obstante, el despliegue de las comunicaciones ópticas en el espacio será inevitable a medida que las necesidades de ancho de banda se vayan haciendo más evidentes. Las principales agencias aeroespaciales coinciden en que esta tecnología será estratégica en las telecomunicaciones del futuro [72]. En este apartado se repasa el desarrollo histórico de este tipo de comunicaciones y se presenta una serie de principios básicos que serán de interés para entender la adaptación que precisa un telescopio *Cherenkov* para desempeñar la función de terminal de comunicaciones ópticas en enlaces desde espacio profundo.

## 2.4.1. Principales proyectos

En 1945, más de 10 años antes de la creación de la NASA, el escritor de ciencia ficción Arthur C. Clarke ya imaginó naves espaciales que transmitían mensajes hacia la Tierra utilizando haces de luz [73, p. 266]. Sin embargo, no será hasta la década de los 90 [74, p. 134] cuando se comiencen a realizar los primeros estudios sobre la utilización de frecuencias ópticas para transmitir información en el espacio, especialmente motivados por la necesidad de ahorrar masa y energía en los terminales de comunicaciones, siguiendo la evolución del aumento de frecuencia portadora que había tenido lugar en las décadas anteriores [13, p. 2]. La primera demostración de que las frecuencias ópticas podrían utilizarse en el futuro para comunicar terminales remotos en espacio profundo la llevó a cabo la NASA en diciembre de 1992. Bajo el denominado experimento GOPEX (*Galileo Optical Experiment*), desde el observatorio de *Table Mountain* en California se emitieron pulsos a 532 nm utilizando un láser Nd:YAG pulsado a 15-30 Hz con varias decenas de MW de potencia de pico y un telescopio reflector de 60 cm (para transmitir un haz de 15 cm sin truncamiento). Este haz se dirigió hacia la cámara de la sonda Galileo, que pudo detectar la transmisión como la imagen de destellos luminosos al observar la Tierra hasta una distancia de 6 millones de km [75]. Entre 1995 y 1996 la NASA en colaboración con la NASDA (la entonces agencia espacial japonesa, predecesora de la actual JAXA) llevó a cabo una segunda demostración con el proyecto GOLD (*Ground Orbiter Lasercom Demonstration*). En este proyecto se estableció por primera vez una comunicación bidireccional (enlace de subida a 514,5 nm y de bajada a 830 nm) de hasta 1 Mbit/s entre el observatorio de *Table Mountain* y el satélite geoestacionario japonés ETS-VI [76]. En 2001 se llevó a cabo el primer enlace entre dos satélites (el japonés de órbita baja OICETS y el europeo geoestacionario Artemis) como parte del proyecto SILEX (*Semi-conductor Inter satellite Link EXperiment*) de la ESA [77]. Este proyecto (Figura 14) involucró el uso de terminales con telescopios de 25 cm y diodos láser de 800-850 nm con una potencia pico de 120 mW, alcanzando tasas binarias de hasta 50 Mbit/s a distancias de más de 45000 km, lo que fue un gran hito histórico. En 2006, el terminal óptico del satélite Artemis también protagonizó el primer enlace entre un satélite y un avión cuando éste se desplazaba a 500 km/h a una altura de entre 6 y 10 km [78]. En 2008 se consiguió un nuevo récord de velocidad de transmisión entre el satélite americano de baja órbita NFIRE y el geoestacionario alemán TerraSAR-X alcanzando 5,6 Gbit/s a una distancia de 5000 km, lo que supuso una mejora de un factor 20× en relación a las máximas tasas binarias entre satélites con tecnología de microondas [79].

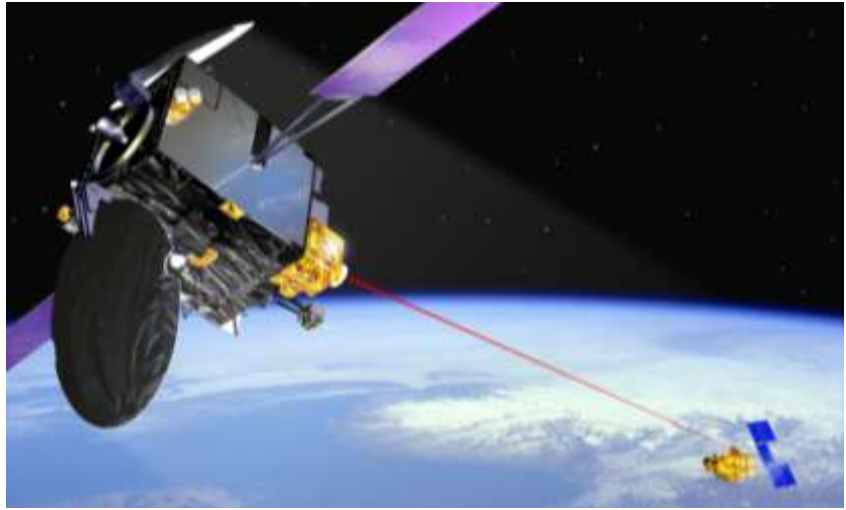

**Figura 14. Ilustración artística del satélite Artemis transmitiendo un enlace de comunicación por láser al satélite OICETS (proyecto SILEX de la ESA) [80].**

Otros proyectos nunca fueron más allá de la fase de diseño, tales como el proyecto OCDHRLF (*Optical Communication Demonstration and High-Rate Link Facility*) [81], que en 2002 pretendía instalar a bordo de la ISS (*International Space Station*) un terminal de comunicaciones ópticas basado en componentes comerciales con una capacidad máxima de 2,5 Gbit/s. Su diseño consistía en un láser en configuración MOPA de 1550 nm y 200 mW de potencia de salida y otro de 980 nm y 10 mW para apuntamiento. De forma paralela, el proyecto EXPRESS (*Expedite the PRocessing of Experiments to the Space Station*) pretendía utilizar el terminal de la ISS a modo de repetidor para incrementar la capacidad de transmisión del trasbordador espacial, limitada a 50 Mbit/s en radiofrecuencia, instalando un terminal óptico con capacidad de 10 Gbit/s [82]. Otro ambicioso proyecto que fue cancelado por cuestiones presupuestarias fue el MLCD (*Mars Laser Communication Demonstration*), con el que la NASA pretendía llevar a cabo en 2009 la primera demostración funcional de un enlace de comunicaciones ópticas desde espacio profundo, Marte en este caso. El proyecto consistía en un terminal óptico de bajo consumo y masa y capacidad de hasta 100 Mbit/s instalado en el orbitador marciano MTO (*Mars Telecom Orbiter*), que nunca llegó a lanzarse [83].

Desde la demostración del proyecto GOPEX en 1992, no se había vuelto a intentar ningún experimento en espacio profundo hasta 2013, cuando se estableció un enlace de subida entre una estación terrestre y el satélite LRO (*Lunar Reconnaissance Orbiter*) [84]. El experimento, liderado por la NASA, consistió en la transmisión de la imagen de la Mona Lisa, convirtiéndose en la primera comunicación óptica en espacio profundo de la historia, si bien la tasa de transmisión fue muy baja (300 bit/s) porque se realizó haciendo uso del sistema de *ranging* a 532 nm de la LRO, no diseñado para comunicaciones. A finales de 2013, el proyecto LLCD (*Lunar Laser Communication Demonstration*) de la NASA se convertiría en la primera demostración exitosa de un sistema de comunicaciones ópticas en espacio profundo (Figura 15). Este proyecto consiguió establecer un enlace bidireccional (622 Mbit/s de bajada y 20 Mbit/s de subida) entre la Luna y la Tierra, basado en los diseños del cancelado proyecto MLCD. El terminal óptico se alojó en el satélite LADEE (*Lunar Atmosphere and Dust Environment Explorer*) y consistió en un láser de 0,5 W de potencia media a 1550 nm transmitido a través de un telescopio de 10 cm [85]. El proyecto LLCD logró multiplicar por más de seis la máxima velocidad de transmisión alcanzada desde la Luna reduciendo en la mitad la masa destinada a las comunicaciones.

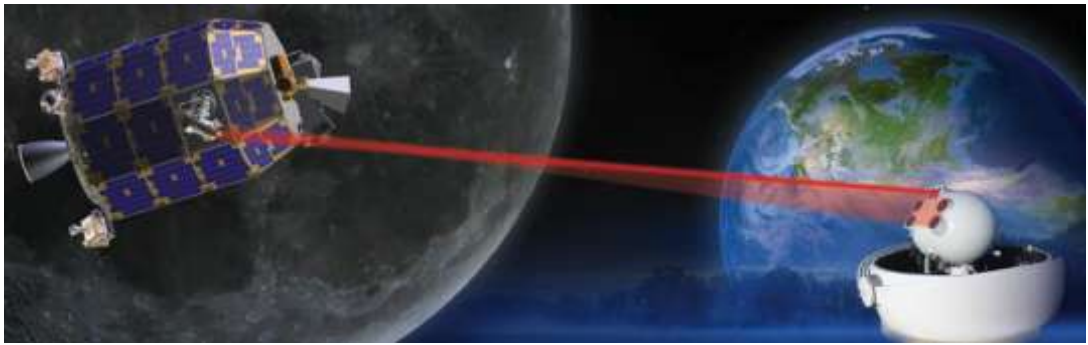

**Figura 15. Ilustración artística del satélite LADEE comunicándose por láser con la estación terrena LLGT (proyecto LLCD de la NASA) [86].**

En Europa, el satélite geoestacionario Alphasat fue lanzado en 2013 con un terminal óptico embarcado para actuar de repetidor de alta velocidad para satélites de baja órbita. El objetivo general es dar un primer paso en la integración de las comunicaciones ópticas en las redes de comunicación globales, por ejemplo para descongestionar la capacidad de comunicación de las órbitas bajas, cuyos satélites solo son visibles por las estaciones terrestres unos 10 minutos en cada una de las 4-5 órbitas diarias. El terminal óptico de Alphasat tiene una capacidad bidireccional de 1,8 Gbit/s utilizando una longitud de onda de 1064 nm con un telescopio de 13,5 cm [87]. Sendos terminales ópticos serán instalados a bordo de los sucesivos satélites Sentinel de la ESA para comenzar a integrar las comunicaciones ópticas en la red europea EDRS (*European Data Relay System*) a partir de 2015. En 2014 fue instalado en la ISS el módulo OPALS (*Optical PAyload for Lasercomm Science*) de la NASA, que es un sistema simple de bajo coste que consiguió con éxito demostrar su funcionalidad con un enlace hacia la Tierra de 50 Mbit/s [88]. El proyecto LCRD (*Laser Communication Relay Demonstration*) de la NASA dará soporte a partir de 2018 a otros satélites como repetidor de alta velocidad a bordo de un satélite geoestacionario parecido a Alphasat, pero con dos módulos basados en el de LLCD, con hasta 1,24 Gbit/s [89]. Este proyecto será el primer paso para integrar las comunicaciones ópticas en la red americana TDRSS (*Tracking and Data Relay Satellite System*). La NASA contempla para 2018 una demostración de hasta 250 Mbit/s desde Marte [90] y la ESA estima unos 100 Mbit/s desde una distancia de 1 Unidad Astronómica para dar soporte a las próximas misiones de espacio profundo [91]. El gran éxito de todos estos experimentos ha hecho posible la aparición de una serie de nuevos proyectos para el futuro, con una proyección cada vez más global. Por ejemplo, tanto Facebook como Google están trabajando en sendos proyectos para ofrecer acceso generalizado a Internet desde cualquier parte del mundo mediante una red que incluye un segmento óptico espacial para dar soporte al enorme ancho de banda que se prevé será necesario [92].

## 2.4.2. Revisión de fundamentos

La transición desde las frecuencias de radio al espectro óptico involucra un cambio de paradigma en la forma en que se transmite la información en espacio libre porque en el nivel físico todas las partes del sistema de comunicación son diferentes a las usadas en los enlaces basados en microondas. A continuación se hace un breve repaso de los fundamentos de esta tecnología, con un especial énfasis en los conceptos relacionados con el ámbito de interés de esta parte de la tesis. Dado que el nivel físico fue identificado como el más diferencial con las comunicaciones ópticas, también en la operación como telescopio *Cherenkov*, es este al que se dedica una mayor atención, evitando entrar en gran detalle en los niveles superiores como la modulación, codificación, etc., más similares a los enlaces tradicionales.

Naturalmente, la característica más distintiva de las comunicaciones ópticas es la longitud de onda. Este tipo de comunicaciones espaciales se planea desplegarlas haciendo uso del infrarrojo cercano, si bien también pueden tener lugar en la zona visible [75] [84] e incluso en la ultravioleta [93]. Más específicamente, dos longitudes de onda se barajan como las principales candidatas: 1064 nm y 1550 nm [72, p. 103]. Actualmente la diferencia entre una y otra es más territorial que de otro tipo: Estados Unidos ha apostado por 1064 nm y Europa por 1550 nm, aunque últimamente se está tendiendo a generalizar esta última. Si bien la segunda presenta una divergencia algo mayor en un transmisor limitado por difracción, tiene una serie de importantes ventajas que en general la hacen preferible sobre la primera. Entre ellas destaca la gran cantidad de tecnología desarrollada para la tercera ventana de mínima atenuación en fibra óptica: desde fuentes hasta detectores, pasando por amplificadores y demás dispositivos electroópticos; también es una ventaja su mayor seguridad para la visión incluso a potencias muy altas, debido a que en el rango de 1150-1450 nm la mayor parte de la radiación es absorbida por la córnea y el resto se propaga absorbiéndose principalmente por el humor acuoso del ojo, quedando protegida la retina [94, p. 52][3]; también presenta unas mejores características para la propagación atmosférica [4, p. 152] al no existir absorción debida al vapor de agua u otros gases, presentar menores fluctuaciones de intensidad por la turbulencia atmosférica, que disminuye con la longitud de onda según $\lambda^{-7/6}$ [95, p. 7] y por ser el *scattering* Rayleigh también inversamente proporcional a la longitud de onda según $\lambda^{-4}$ [96, pp. 22-17] reduciéndose el ruido de fondo debido a la radiancia del cielo. Por último, gracias a que la eficiencia cuántica de detección es actualmente tan buena en 1550 nm como en longitudes de onda inferiores, en la práctica presenta una mayor sensibilidad. Esto se debe a que longitudes de onda más largas implican una menor energía electromagnética y por lo tanto a igualdad de potencia recibida, la producción de fotones será mayor. Por ejemplo, a 1550 nm se producen aproximadamente el doble de fotones que a 800 nm para la misma potencia óptica y eficiencia cuántica. Por todas estas razones, la longitud de 1550 nm se presenta como la principal candidata a ser adoptada de forma general y por ello en esta tesis, en caso de particularizar la longitud de onda, se optará por 1550 nm.

La consecuencia más importante de las pequeñas longitudes de onda utilizadas en FSOC en comparación con la radiofrecuencia es la menor divergencia de los haces, como ya se introdujo en el apartado 1.2.2. En un telescopio limitado por difracción, la divergencia del haz transmitido es directamente proporcional a la longitud de onda (ver apartado 2.10.1). En el caso de grandes distancias este es un factor fundamental porque a mayor divergencia, mayor es el área sobre la que se reparte la potencia transmitida en el receptor. Con longitudes de onda ópticas es posible aumentar la densidad de potencia recibida en varios órdenes de magnitud: por ejemplo, transmitiendo al límite de difracción, una apertura de 40 cm desde Neptuno produciría un haz equivalente al diámetro de la Tierra con una longitud de onda de ~1 μm y de 10000 veces el diámetro de la Tierra con una frecuencia de 30 GHz (banda Ka), lo que supone una reducción del área, y por lo tanto un aumento de la densidad de potencia recibida, de ocho órdenes de magnitud. Esta gran directividad exige una gran precisión de apuntamiento para mantener el enlace operativo, tras la fase inicial de adquisición en la que se establece una línea de visión común entre ambos terminales. Si en radiofrecuencia la precisión de apuntamiento está en el orden de los miliradianes en la banda Ka (lo que puede conseguirse sin ningún sistema específico, solo utilizando el control de actitud de las propias naves), un enlace FSOC desde espacio profundo exigiría una precisión por debajo del microradián [97].

Si se compara la radiofrecuencia con las comunicaciones ópticas a nivel de tasas de transmisión, es necesario hablar de límites potenciales ya que la tecnología optoelectrónica actual aún está lejos de alcanzarlos. La velocidad binaria está limitada por una fracción de la frecuencia portadora, por lo que a frecuencias tan altas como las de la luz, las tasas potenciales están muy por encima de los Tbit/s, resultando en otra mejora de varios órdenes de magnitud en relación a las microondas. Actualmente, en un solo canal (una única longitud de onda sin multiplexación) se pueden alcanzar velocidades de varios Tbit/s [98], y combinando diferentes técnicas de multiplexación para la transmisión de muchos canales de forma simultánea se ha llegado a superar los 100 Tbit/s en un enlace de FSOC [99] y los 200 Tbit/s en fibra óptica [98], con aspiraciones de superar próximamente la barrera del Pbit/s.

---

[3] Esta ventaja de la longitud de onda de 1550 nm es aplicable especialmente a enlaces de subida y a enlaces entre satélites en los que haya astronautas involucrados, ya que la potencia de un enlace de bajada será en general despreciable a efectos de visión, debido a su elevada dispersión espacial.

En general los esquemas de codificación de información para detección y corrección de errores causados por el canal no son cualitativamente diferentes a los de radiofrecuencia (códigos convolucionales tipo *Reed-Solomon* y codificación en bloque como los turbocódigos). Sin embargo, las técnicas de modulación de la información en la señal electromagnética sí pueden diferir bastante. El formato más simple es la modulación OOK (del inglés *On-Off Keying*) (Figura 16, arriba), que consiste en encender y apagar el transmisor a alta velocidad. Esta técnica presenta algunos inconvenientes en enlaces de espacio profundo: por una parte se precisa una alta potencia de pico para compensar las pérdidas por espacio libre y por otra la potencia media debe ser reducida para minimizar el consumo. Las familia de técnicas de modulación por posición de pulsos PPM (del inglés *Pulse Position Modulation*) (Figura 16, abajo) solucionan estos inconvenientes mediante la codificación de más de un bit en cada pulso [100]. Estas modulaciones consisten en dividir la duración de cada secuencia de n bits en m = $2^n$ slots, correspondiente a los m símbolos que se pueden codificar. Cada vez que se transmite un pulso, se sitúa en uno de estos slots de forma que su valor se define por la posición en el intervalo temporal, reduciendo la relación entre la potencia pico $P_{pico}$ y la potencia media $P_{med}$ según la ecuación (2-1).

$$\frac{P_{pico}}{P_{med}} = \frac{m}{n} \qquad (2\text{-}1)$$

De esta forma, se consigue reducir el ciclo de trabajo del láser y mejorar la relación señal a ruido, a cambio de exigir una mayor velocidad de modulación para conseguir la misma tasa binaria (la duración mínima de estos pulsos dependerá de una serie de parámetros del enlace, pero puede llegar a los 200 ps [101, p. 78]).

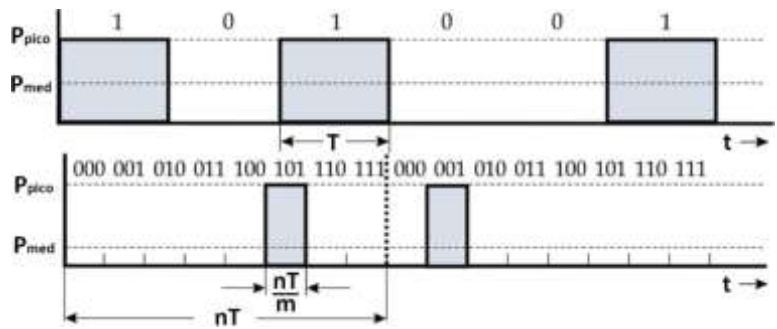

**Figura 16. Modulación de la secuencia 101001
en OOK (arriba) y 8-PPM (abajo).**

Alternativamente a estos esquemas, conocidos como de detección directa (porque se basan en transformar la potencia óptica directamente en una corriente eléctrica), existen otros esquemas basados en la detección coherente (consistentes en la superposición entre la señal recibida y la de un oscilador local). Estos últimos presentan unas características superiores en términos de relación señal a ruido [102, pp. 107-116], sin embargo su implementación práctica impone una serie de inconvenientes, siendo el principal de ellos que la turbulencia atmosférica hace que el frente de onda pierda coherencia espacial, lo que supone un importante hándicap en este tipo de modulaciones. Existen soluciones, como el uso de óptica adaptativa en los telescopios, si bien son complejas y costosas, por lo que estas modulaciones no son las óptimas para el caso en que las señales deban atravesar la atmósfera [103, p. 39]. Por ello, el formato adoptado por las principales agencias espaciales para los distintos escenarios de FSOC es PPM [72, p. 104].

En cuanto a la tecnología de detección de la luz, los mayores requisitos de sensibilidad respecto a las comunicaciones ópticas convencionales han implicado el desarrollo de una tecnología específica. Esta tecnología está en general basada en el uso de fotodetectores con amplificación interna en el dominio óptico como los fotodiodos de avalancha APD (del inglés *Avalanche Photodiode*) o los tubos fotomultiplicadores PMT (del inglés *PhotoMultiplier Tube*). La contribución de ruido se minimiza por diferentes técnicas, como la refrigeración hasta temperaturas criogénicas (especialmente importante en infrarrojo) y/o con elevadas tensiones de polarización para producir altas ganancias de amplificación. Con este tipo de detectores, conocidos como contadores de fotones, es posible distinguir la llegada de fotones individuales. Los más habituales han sido los APD modo *Geiger* o GM-APD (también conocidos como SPAD, del inglés *Single Photon Avalanche Detector*), que trabajan de una forma similar a los contadores *Geiger* y se implementan llevando la tensión de polarización del APD a valores cercanos a su saturación. De esta forma, la llegada de un solo fotón dispara una avalancha de portadores que proporcionan un pulso muy intenso. El ancho de banda de estos dispositivos está limitado por el tiempo de recuperación (en el orden de los ns) que debe transcurrir tras cada avalancha para volver a estar preparado para el siguiente evento, estando en el orden de las decenas de Mbit/s. Este tipo de detectores han demostrado mejoras de la eficiencia de hasta 40× en términos de fotones por bit comparados con los detectores tradicionales [104]. La limitación temporal de esta tecnología puede solucionarse con *arrays* de detectores [4] si bien esta estrategia tiene un recorrido limitado y en los últimos años se ha desarrollado una nueva tecnología de detección que promete sustituir por completo a las anteriores en este tipo de enlaces. Se trata de detectores basados en un nanohilo refrigerado hasta temperaturas criogénicas de algunos Kelvin para convertirlo en superconductor y alimentado con una corriente eléctrica justo por debajo de la necesaria para romper la superconductividad. En este estado, la llegada de un solo fotón que incide en el nanohilo (dispuesto en forma de laberinto con sucesivos bucles para maximizar el área), hace que la temperatura se eleve lo suficiente para que la superconductividad desaparezca temporalmente [105] y se pueda registrar el evento. Estos detectores, que fueron utilizados por primera vez en el proyecto LLCD en 2013 [106], han demostrado eficiencias de hasta el 90 % en 1550 nm con tiempos de respuesta que permiten alcanzar tasas binarias por encima del Gbit/s.

La atmósfera puede presentar efectos perjudiciales en los enlaces de FSOC, como se introdujo en el apartado 1.2.2. Uno de estos efectos es el debido a la turbulencia atmosférica, que se origina por variaciones en la temperatura del aire. Al trasladarse las masas de viento de diferentes temperaturas, por tanto, el índice de refracción fluctúa de forma aleatoria en el tiempo y el espacio, alterando los frentes de onda. Pese a que en general el enlace de bajada es el que presenta las mayores dificultades técnicas en espacio profundo, en el caso de la turbulencia atmosférica es el de subida el más afectado ya que el efecto se da en los primeros kilómetros y se amplifica en el resto del camino de propagación. Por ello, no se incidirá aquí en este fenómeno, si bien cabe señalar que en sistemas coherentes será imprescindible el empleo de alguna técnica de corrección de turbulencia y en sistemas de detección directa su empleo dependerá de si el telescopio tiene una calidad óptica suficiente como para no enmascarar la corrección de la turbulencia (esto se repasará en el apartado 2.10).

---

[4] El esquema de *array* de GM-APD presenta ventajas adicionales al aumento del ancho de banda, como la posibilidad de extracción de información relativa al seguimiento o a las condiciones atmosféricas y la adaptación dinámica del campo de visión, en función del número de elementos empleados.

Otro efecto de la atmósfera es el de la absorción de la radiación electromagnética debida a los gases (Nitrógeno y Oxígeno principalmente), aunque la presencia de ventanas de mínima absorción permite evitar casi completamente este efecto mediante una correcta selección de la longitud de onda, como es el caso para 1064 nm y 1550 nm.

La presencia de nubes es otro de los efectos que más impacto tiene sobre los enlaces de FSOC, llegando a bloquear temporalmente la comunicación si la línea de visión se ve obstruida. Su variabilidad y aleatoriedad hacen que solo existan dos métodos para mitigar sus efectos: una correcta selección del emplazamiento y su replicación con técnicas de diversidad. Los emplazamientos más adecuados coinciden con los observatorios astronómicos, normalmente situados por encima de 2 km de altura, precisamente para evitar el impacto de la primera capa de la atmósfera. Se puede llegar a conseguir una disponibilidad superior al 90 % en el enlace de comunicación si se utilizan al menos tres lugares distintos [107].

Por último, el *scattering* es otro efecto que impacta negativamente a las comunicaciones ópticas desde el espacio. Se debe a la presencia de partículas en suspensión, o aerosoles, que causan que la luz se disperse de una forma que depende de la relación entre el tamaño de la partícula y la longitud de onda, de su forma y de la relación entre el índice de refracción de la partícula y el del medio. El efecto más perjudicial del *scattering* para las FSOC no es el que afecta a la señal de comunicaciones, sino el que afecta a la luz del Sol durante el día: los fotones solares son dispersados por los aerosoles atmosféricos de forma que su dirección puede pasar a coincidir con la línea de visión directa del enlace de comunicación y acoplarse en forma de ruido de fondo, aunque originalmente tuvieran una dirección diferente. Para mitigar este efecto es necesario bloquear la mayor cantidad de luz para rechazar todos los ángulos fuera del campo de visión y reducir este al mínimo posible, así como emplear técnicas de filtrado espacial y espectral.

## 2.5. MOTIVACIONES PARA ESTE ESTUDIO

Como se introdujo en el apartado 2.1, la astronomía terrestre de rayos gamma comparte con las comunicaciones ópticas en espacio profundo un número importante de características tecnológicas originadas por la coincidencia de una serie de objetivos comunes muy parecidos. A raíz de estas significativas similitudes, en este apartado se presenta una enumeración de las motivaciones que justifican la propuesta de emplear telescopios *Cherenkov* para dar soporte a enlaces de comunicaciones ópticas en espacio libre. Si bien muchas de ellas son extensibles a telescopios *Cherenkov* en general, aquí se ha particularizado para el caso de los previstos para el proyecto CTA.

- Las comunicaciones ópticas en espacio profundo están limitadas por la reducida potencia óptica recibida en Tierra, limitación que se hace más presente con la distancia. Una mejora significativa en el rendimiento de estos enlaces viene del aumento de la apertura de recepción, y los telescopios de CTA proporcionan aperturas mucho mayores de las que se han considerado hasta la fecha para estaciones terrenas (que aún en la teoría nunca han superado los 10 metros).

- Los telescopios de CTA operan en una topología de *array* de manera nativa, lo que podría ser aprovechado para aumentar de forma arbitraria la apertura de recepción por medio de la replicación de elementos en caso de ser necesario. La capacidad de disponer de un mayor número de terminales receptores hace que la principal ventaja de esta propuesta (las grandes aperturas de los telescopios) sea además escalable.

- El aumento en Tierra de la apertura receptora equivalente va en la dirección de la tendencia natural de las comunicaciones por satélite, que en lo posible tiende a trasladar la complejidad desde el terminal remoto (en el espacio) hacia el terminal local (en la Tierra). Esto se debe a que cualquier esfuerzo tecnológico en cuanto a masa, volumen o consumo eléctrico es más fácilmente asumible en una estación terrena que en una sonda espacial.

- La fase de desarrollo de CTA será de tal magnitud que podría ser equiparada con una cadena de montaje de grandes telescopios, en la que los costes de producción de unidades adicionales serán mucho menores a los que se incurrirían en el caso de un único desarrollo ad hoc. Los costes de las infraestructuras de soporte también serán menores, por ser compartidos, a los de una dedicada.

- En general la calidad óptica de los IACT es mucho menos estricta que en telescopios astronómicos, lo que implica una disminución de costes respecto a estos. En principio esta menor exigencia es compartida por los telescopios de comunicaciones y de rayos gamma, si bien los primeros tendrán unos requisitos algo mayores que los IACT, por lo que estos podrían precisar ser modificados para adaptar determinadas características.

- Los IACT realizan sus observaciones astronómicas empleando campos de visión muy amplios, al contrario que los reducidos campos de visión de los telescopios de comunicaciones. Dado que la resolución angular de los IACT se establece considerando estos amplios campos de visión, el mismo telescopio empleado para comunicaciones tendrá automáticamente una mejor resolución debido a las menores aberraciones asociadas al menor campo de visión.

- La región espectral de operación es diferente en comunicaciones y en astronomía *Cherenkov*, por lo que la reflectividad de los espejos, que depende de la longitud de onda, podría ser un parámetro limitante. Como se verá, de hecho la reflectividad es incluso mejor en las longitudes de onda de comunicaciones, lo que permite reutilizar la tecnología de fabricación de espejos sin modificación ni pérdidas de potencia.

- La rapidez de las señales recibidas en astronomía de rayos gamma exige una infraestructura electrónica y de comunicaciones que satisface sobradamente los requisitos de las señales de comunicaciones. Muestreos a GHz de cámaras con cientos de píxeles cada una y transmisión por fibra óptica a Tbit/s desde los telescopios ofrecen una capacidad más que suficiente para soportar las señales de comunicaciones actuales y futuras.

- La astronomía *Cherenkov* comparte con las comunicaciones ópticas los mismos requisitos en cuanto a las condiciones de la atmósfera en las estaciones terrenas: elevada altitud de emplazamientos para mitigar los efectos de la turbulencia atmosférica, ausencia de fuentes luminosas con el objetivo de reducir el ruido acoplado al sistema receptor y máxima inmunidad ante la nubosidad orientada a reducir los tiempos muertos de operación de los telescopios.

- Se prevén dos emplazamientos para CTA, uno en cada hemisferio, separados de forma aproximadamente equidistante en longitud. Esto es un valor añadido en misiones de espacio profundo en relación al efecto de bloqueo de las señales de comunicaciones debido a la rotación de la Tierra. La presencia de terminales de comunicaciones en los dos emplazamientos proporcionaría una alta disponibilidad de seguimiento en misiones de espacio profundo.

- Uno de los principales objetivos de CTA es la detección de brotes de rayos gamma, cuyo origen es imprevisible y su duración puede ser tan corta como algunos segundos. Detectar estos eventos precisa de un reposicionamiento muy rápido de los telescopios, lo que puede ser una ventaja al utilizar los telescopios como terminales de comunicaciones, ya que podrían realizar seguimiento de satélites cercanos a la tierra con órbitas muy rápidas.

- La forma de los reflectores de los IACT se elige de acuerdo con dos criterios: minimizar la dispersión temporal de los fotones reflejados en dos puntos muy distantes del telescopio y minimizar la aberración de coma en el plano focal. El primer criterio es compartido en comunicaciones, ya que la resolución temporal de las señales es similar, y el segundo es aún más exigente en los telescopios *Cherenkov* debido a la incertidumbre en la dirección de origen de las señales.

## 2.6. VIABILIDAD DE ESPEJOS DE IACT PARA FSOC

Adaptar un IACT para su uso en FSOC exige hacer un repaso de los diferentes componentes del telescopio, evaluando la viabilidad de su aplicación directamente, o en su caso la adaptación necesaria con el objetivo de convertirlo en apto para la nueva aplicación. Los espejos de un telescopio son el primer elemento que encuentra la radiación electromagnética incidente y por ello tienen una gran importancia. En este apartado se hace un repaso de las técnicas de fabricación de espejos para telescopios *Cherenkov* y se estudia cómo se podrían reutilizar o adaptar para la fabricación de los espejos necesarios para comunicaciones. También se presentan medidas experimentales llevadas a cabo en esta tesis sobre muestras de espejos fabricadas con las técnicas más importantes utilizadas para el desarrollo de IACT.

### 2.6.1. Estrategias de fabricación en CTA

Es un hecho bien establecido que el coste de fabricación del espejo primario de un gran telescopio astronómico aumenta exponencialmente con el área reflectora [108]. Por ello, las grandes aperturas precisadas en los IACT en general y en CTA en particular exigen la segmentación del reflector principal en múltiples espejos de menor tamaño. En CTA se precisará una superficie colectora de aproximadamente 10000 m² [109, p. 1], por lo que es crítico llegar a un compromiso entre la calidad óptica y el coste de producción. En astronomía de rayos gamma con IACT no se precisa una gran calidad óptica en los instrumentos ya que la luz *Cherenkov* producida por los eventos de interés se origina a una distancia desconocida por su aleatoriedad, variando en un rango entre 6 y 20 km de altura [44]. Esta incertidumbre obliga a establecer un enfoque a una distancia intermedia de unos 10 km, lo que hace que las observaciones estén siempre desenfocadas. Esto es diferencia del resto de telescopios astronómicos, enfocados en el infinito de forma permanente. El resultado es que la precisión de enfoque en IACT está unos dos órdenes de magnitud por debajo de la habitual y la resolución de la distancia desde cada espejo al plano focal se expresa en centímetros en lugar de cercana a la longitud de onda o incluso por debajo de ella. Por ello, los espejos fabricados para IACTs son siempre de peor calidad óptica que los fabricados para otro tipo de telescopios astronómicos. Se sacrifican las prestaciones ópticas a costa de otro tipo de cualidades necesarias como su mayor resistencia a las condiciones de intemperie, su ligereza o su menor coste de fabricación, que queda aproximadamente unos dos órdenes de magnitud por debajo de los de telescopios astronómicos [33, p. 155].

Aunque los grandes espejos monolíticos que precisarían los IACT son prohibitivos, una gran segmentación también influye negativamente en el coste del telescopio, ya que se precisarían demasiados puntos de soporte y elementos de alineamiento. Por otra parte, espejos muy grandes afectan a la calidad óptica al usar configuraciones como el *Davies-Cotton*, por lo que es necesario alcanzar un compromiso. Los segmentos de CTA serán probablemente hexagonales y de entre 1 y 2 m² [110, p. 2]. No obstante, los que se utilicen en ópticas dobles pueden presentar formas distintas (por ejemplo, algunos tendrán forma de pétalo alrededor de un espejo central). En general tendrán perfiles esféricos para reducir costes, ya que proporcionarán una calidad óptica suficiente en la mayoría de los casos. Los requisitos básicos de estos espejos serán una reflectividad mayor que el 80 % en el rango 300-600 nm y una durabilidad de varios años preservando sus características sin grandes deterioros. Este requisito es muy importante ya que los espejos están permanentemente expuestos a la intemperie. Otros requisitos incluyen un peso reducido para evitar cambios en la forma de las aperturas, alta rigidez para evitar deformaciones y estabilidad en sus características en un rango amplio de temperaturas (de -20°C a 40°C).

Existen varias técnicas de fabricación de espejos barajándose actualmente para CTA. Algunas parten de experiencias anteriores de otros IACT y otras están en fase de estudio y validación. Estas técnicas no son exactamente como las usadas en telescopios astronómicos convencionales, ya que los objetivos difieren. En CTA se planea fabricar una enorme cantidad de espejos, por lo que la velocidad y coste de fabricación va generalmente en detrimento de la calidad óptica. Hay dos grupos de técnicas candidatas para fabricar los espejos de CTA, las basadas en las técnicas clásicas de pulido, bien conocidas por su aplicación en IACT anteriores, y las basadas en métodos de replicación, muy eficiente especialmente para la fabricación masiva necesaria en CTA. El primer grupo lo constituyen dos técnicas:

- Espejos de vidrio con recubrimiento de aluminio (técnica ya empleada en HESS y VERITAS): es la técnica más clásica y mejor conocida, y se basa en un sustrato de vidrio pulido con la forma deseada al que se añade una capa frontal de aluminio y un recubrimiento transparente protector como óxido de aluminio (VERITAS) o cuarzo (HESS). Proporciona una buena PSF y alta reflectividad a cambio de fragilidad y peso considerables (especialmente en los tamaños considerados en CTA). Además, los espejos presentan una rápida degradación de los recubrimientos frontales (pueden llegar a perder un 5 % de reflectividad anual [111, p. 134]) y el proceso de fabricación es lento y laborioso. El coste actual de producción se encuentra en torno a los 1650 €/m².

- Espejos de aluminio pulido con diamante (empleada en MAGIC I y en el 75 % de los espejos de MAGIC II): se basa en un sándwich (Figura 17, izquierda) de dos capas de aluminio separadas por una estructura de aluminio de panal de abeja para dar rigidez al conjunto y permitir la circulación térmica. Tras un pulido grueso para dar el radio de curvatura aproximado a la capa frontal de aluminio, el pulido final se realiza mediante el giro de un cabezal de diamante. Finalmente, se añade un recubrimiento protector basado en cuarzo. Proporciona una menor degradación (pérdida de 1-2 % de reflectividad anual) a cambio de una reflectividad algo menor. El coste actual de producción se encuentra en torno a los 2450 €/m².

El otro grupo de técnicas se basan en la optimización de métodos de replicación para la producción en masa de espejos. El objetivo es minimizar el coste y el tiempo de fabricación en la producción de espejos ligeros y con una calidad óptica suficiente y fácilmente reproducible. Por ello, estas técnicas de réplica se consideran las principales candidatas en CTA, si bien aún se encuentran en desarrollo. La favorita de estas técnicas es el conformado de vidrio, y está basada en la utilización de un molde que se usa para dar la curvatura requerida a una lámina delgada de vidrio que se deposita sobre el molde en vacío para forzarla a copiar su curvatura, y sobre la que se adhiere un material de soporte (por ejemplo en panal de abeja) para dar rigidez al conjunto y otra capa plana de vidrio para cerrar el sándwich (Figura 17, derecha). Tras el curado del adhesivo que une todas las capas, el panel se libera del molde y la parte frontal se aluminiza y se aplica un recubrimiento similar a un espejo de vidrio. Un 25 % de los espejos de MAGIC II (los centrales) se fabricaron con esta técnica. Su coste actual de producción se encuentra en torno a los 2000 €/m².

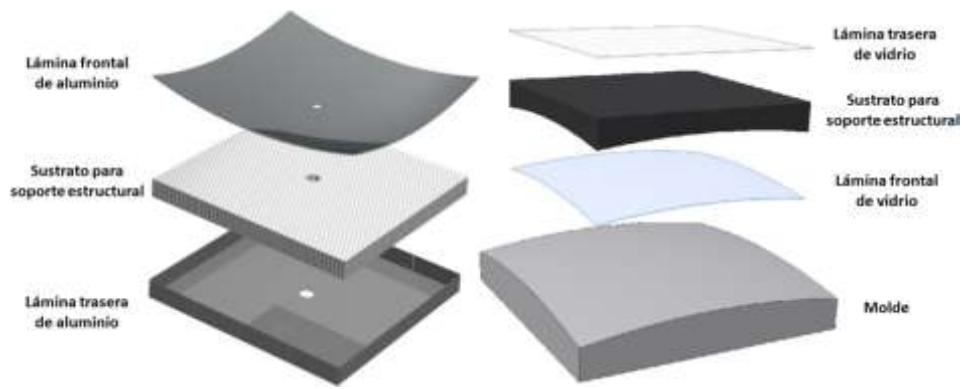

**Figura 17. Espejo de aluminio pulido con diamante (izquierda) y espejo de deposición de vidrio (derecha) [112].**

El grupo de técnicas de réplica está siendo intensamente investigado para su aplicación en CTA y será el preferido por defecto debido a la disminución de costes y la rapidez de producción de espejos que ofrece. No obstante, presenta algunos inconvenientes que merece la pena mencionar. En estos espejos la capa reflexiva no forma parte de la estructura principal, como sí es el caso en los espejos de aluminio pulido, sino que es un recubrimiento superficial, y por ello la reflectividad se degrada más rápidamente. Para solucionarlo, se prevé la aplicación de recubrimientos cada 4 años aproximadamente, lo que supone un gasto añadido [5]. Sin embargo, el principal problema de esta técnica es que no permite radios de curvatura reducidos, debido a que la tensión mecánica que soporta el vidrio sobre el molde hace que el espejo final tienda a volver a su forma original. Por ello, se prevé utilizar la técnica en la versión *Davies-Cotton* de MST y quizás de SST, pero en ningún caso para *Schwarzschild-Couder* que precisa grandes curvaturas, además de que con las formas asimétricas de estas configuraciones se obtienen peores resultados. Posiblemente tampoco se pueda utilizar para LST debido a que los resultados no son tan buenos en espejos grandes, y reducir el tamaño de los espejos de LST supone un aumento excesivo de puntos de fijación y complejidad en la estructura. No obstante, se sigue trabajando en esta línea con variaciones de esta técnica para intentar solventar estas dificultades, aprovechándose de que la instalación de espejos es un paso posterior en el despliegue de los telescopios, y contando con las técnicas clásicas en los casos en que no pueda aplicarse.

## 2.6.2. Reflectividad en espejos de aluminio

Una dificultad para estudiar la viabilidad de la reutilización de los espejos de telescopios *Cherenkov* para comunicaciones ópticas se basa en los distintos rangos espectrales en que opera cada aplicación. El espectro de la luz *Cherenkov* se encuentra en el rango de 300 a 600 nm, y contando con este parámetro se optimizan gran parte de los aspectos técnicos de los IACT, incluyendo la reflectividad de los espejos. En cambio, las comunicaciones ópticas en espacio profundo se desplegarán en el infrarrojo cercano, específicamente en 1064 nm y preferiblemente en 1550 nm (ver apartado 2.4.2). Como ambas longitudes de onda se encuentran lejos del espectro *Cherenkov*, las medidas de reflectividad de los espejos de IACT nunca llegan a esta región, por lo que se hace necesario un estudio ad-hoc sobre estas longitudes de onda para analizar su conveniencia en FSOC.

A través de la empresa INSA (Ingeniería y Servicios Aeroespaciales, S.A.), que financió la fase inicial del estudio realizado en la primera parte de esta tesis, se pudo acceder a las instalaciones de los telescopios MAGIC en el observatorio astronómico de Roque de los Muchachos de La Palma (Canarias) para tomar medidas in situ de reflectividad sobre uno de los espejos de 1 m² de superficie del telescopio MAGIC I. Como se explicó en el apartado anterior, los espejos de este telescopio están fabricados con la tecnología de pulido con diamante de un sustrato de aluminio, cuyo espectro en reflexión es en general conocido (ver apartado 2.6.3). Sin embargo, se desconoce el comportamiento del recubrimiento protector (especialmente importante en este tipo de telescopios permanentemente expuestos a la intemperie) en el infrarrojo cercano, que en todo caso estaría optimizado únicamente para el espectro de luz *Cherenkov*.

Las medidas de reflectividad especular sobre este espejo se realizaron utilizando dos espectrómetros portátiles del fabricante *Ocean Optics*, cada uno diseñado para un rango espectral específico: 200-1100 nm y 900-1700 nm. El primero es el USB2000+ (Figura 18, izquierda), destinado a cubrir principalmente el espectro visible y utilizado con el objetivo de validar las medidas comparándolas con otras anteriores. El segundo es el NIRQuest (Figura 18, segunda desde la izquierda), destinado a cubrir el espectro infrarrojo cercano y utilizado para realizar las medidas sobre las longitudes de onda de interés para FSOC. Para completar las medidas se utilizó una lámpara halógena HL2000-HP (Figura 18, derecha), una sonda R400-7-UV/VIS para medidas de reflectancia (Figura 18, tercera por la izquierda), un soporte RPH-1 para calibración de la distancia de medida (Figura 18, segunda por la derecha) y un patrón estándar de calibración STAN-SSH (Figura 18, tercera por la derecha), todo también del fabricante *Ocean Optics*.

---

[5] Este inconveniente es compartido por las técnicas clásicas de espejos fabricados en vidrio con cualquier recubrimiento reflectivo.

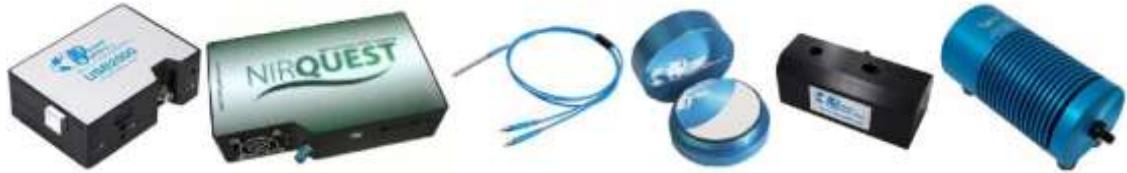

**Figura 18. De izquierda a derecha: espectrómetro USB2000, espectrómetro NIRQuest, sonda R400-7-UV/VIS, patrón STAN-SSH, soporte de medida RPH-1 y fuente halógena HL2000-HP.**

Para comparar la validez de la técnica y la instrumentación utilizada para tomar las medidas de reflectividad, se caracterizó el espejo de MAGIC I en el espectro *Cherenkov* utilizando el espectrómetro USB2000 para la zona del visible. Se puede comprobar que existe un buen acuerdo entre las medidas realizadas (Figura 19, derecha) y las de referencia tomadas de la bibliografía [62, p. 9] (Figura 19, izquierda), donde la reflectividad queda siempre por encima de 80 % en ambos casos, y también concuerda con el resultado numérico de otras medidas de referencia de los mismos espejos, con reflectividades en un rango de $0{,}77 \pm 0{,}04$ [113, p. 360]. Las pequeñas desviaciones observadas se pueden atribuir a variaciones en el recubrimiento, que pueden llegar a alcanzar un 8 % [37, p. 57] dependiendo de la región del espejo donde se realice la medida (la medida mostrada es un promedio de la reflectividad de la zona central, de un borde y de una esquina; se desconoce sobre qué parte fue tomada la medida de referencia). También se sabe que en la fase más temprana de producción de los espejos de MAGIC I existía una mayor diferencia en el grosor del recubrimiento entre las diferentes zonas de su superficie (pudiendo llegar hasta varias decenas de nanómetros), que se consiguió homogeneizar en las fases más tardías de producción [37, p. 59].

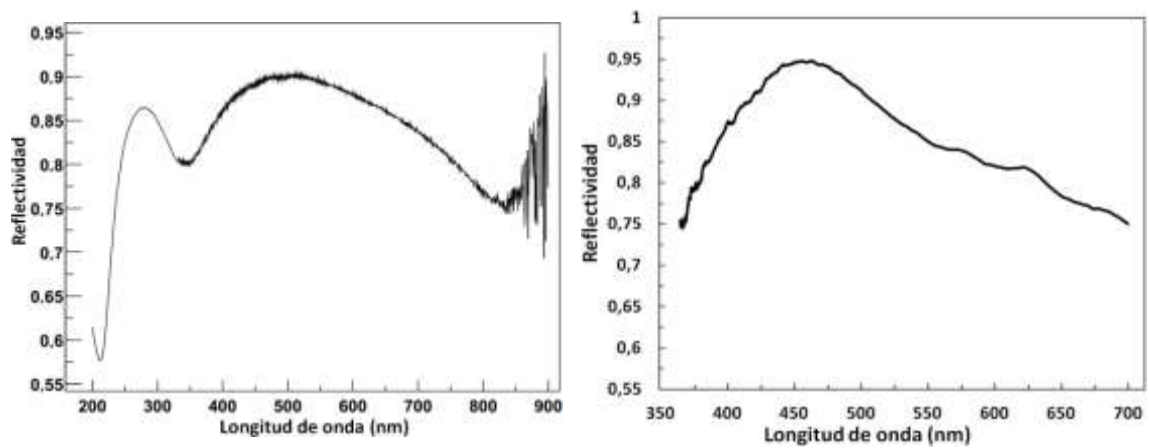

**Figura 19. Medida de la reflectividad de un espejo de MAGIC I en el espectro visible de acuerdo con [62] (izquierda) y las realizadas en esta tesis (derecha).**

En la Figura 20 se muestra la medida de la reflectividad especular tomada en las diferentes regiones del mismo espejo de MAGIC I en la zona espectral del infrarrojo cercano. Se aprecian diferencias de hasta un 4 % en la reflectividad de las diferentes regiones del espejo. Pese a no haber sido optimizados para estas longitudes de onda, la reflectividad es superior a la correspondiente al espectro *Cherenkov*, quedando por encima del 90 % en las regiones de interés para FSOC.

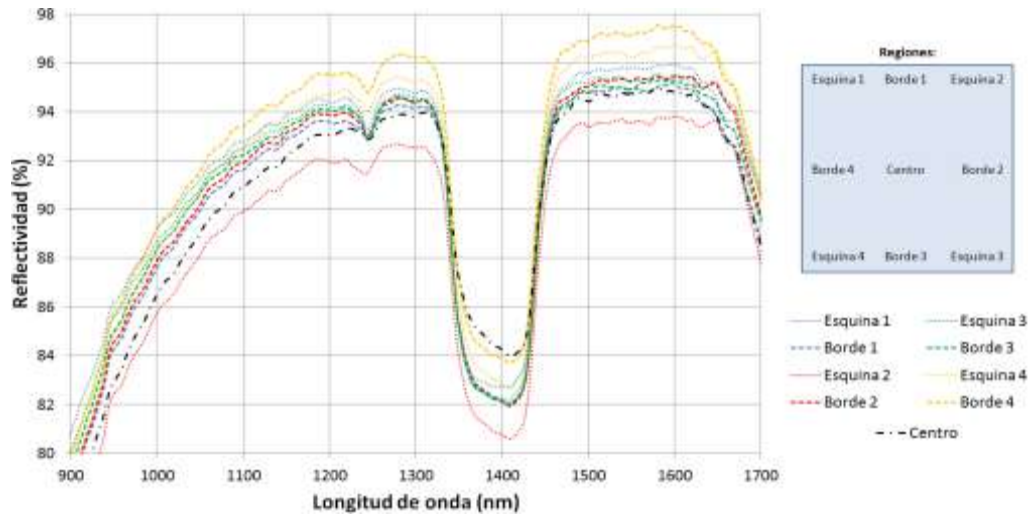

**Figura 20. Medida de la reflectividad de un espejo de MAGIC I tomada sobre diferentes regiones en el infrarrojo cercano.**

En la Figura 21 se muestra la reflectividad promediada en varias regiones de la superficie del espejo en todo el espectro, desde 400 nm hasta 1700 nm, como superposición de las medidas realizadas con los dos espectrómetros (esta superposición es posible gracias a que la reflectancia de cada aparato se calibró con el mismo patrón). Se pueden comprobar los buenos resultados en la región de interés en FSOC, especialmente en la longitud de onda de mayor interés, 1550 nm, donde la reflectividad alcanza un 94,8 ± 0,1 %, siendo de 91 ± 1,7 % para 1064 nm. Estos resultados confirman que las técnicas de fabricación de espejos de aluminio pulido utilizadas para IACT son directamente aplicables a los telescopios adaptados para FSOC. Por ello, independientemente de si se utiliza esta técnica u otra diferente en CTA, no hay necesidad de recurrir a técnicas más sofisticadas como las utilizadas en telescopios astronómicos, y se obtendrán resultados próximos a estos.

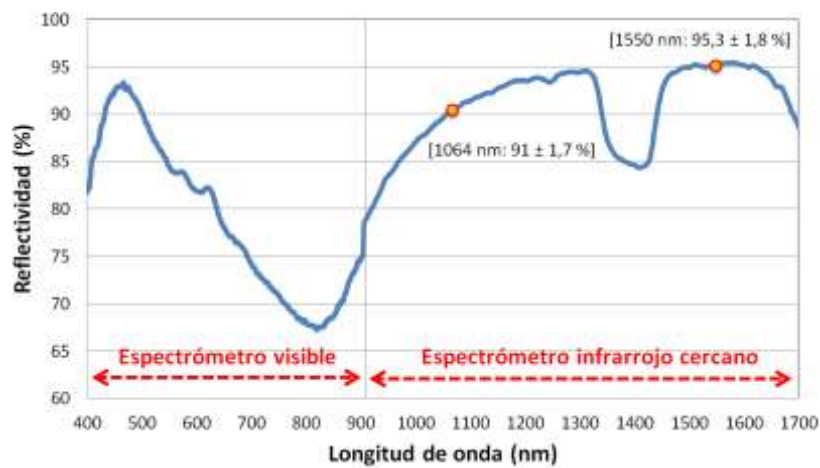

**Figura 21. Medida de la reflectividad promedio de un espejo de MAGIC I en el espectro visible y el infrarrojo cercano.**

### 2.6.3. Reflectividad en espejos de vidrio

En el caso de los espejos de aluminio del apartado 2.6.2, se desconocía el comportamiento en reflexión en el infrarrojo cercano del recubrimiento protector, ya que la propiedad reflexiva la aportaba el material del propio espejo. La técnica de réplica es otra técnica candidata para la fabricación de los espejos de CTA. En este caso, la propiedad reflexiva la aporta el recubrimiento aplicado sobre el vidrio [6] (y sobre el cual a su vez se aplica otro recubrimiento protector), por lo que existe una mayor variedad de posibilidades. El objetivo en IACT es maximizar la reflectividad en el espectro *Cherenkov* de 300 a 600 nm, lo que en principio convierte al aluminio en la elección natural de entre los materiales más habituales (Figura 22). En este caso la solución más común es una capa reflexiva en aluminio y otra protectora más espesa, normalmente de cuarzo. El problema que presenta esta alternativa está relacionado con su degradación a la intemperie, perdiendo alrededor de un 5 % al año, lo que exige aplicar nuevos recubrimientos cada 5 años aproximadamente [111, p. 134]. Esto se debe a que el aluminio no se adhiere perfectamente al vidrio y el agua puede penetrar entre las dos capas [39, p. 80], provocando además la oxidación de la capa metálica.

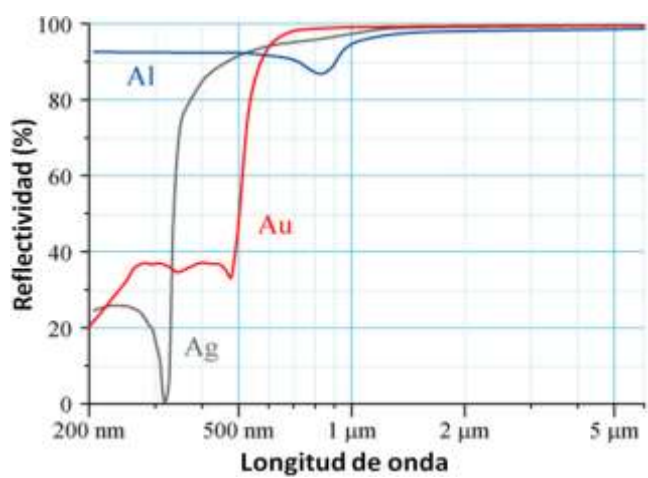

**Figura 22. Reflectividad en función de la longitud de onda para el aluminio (Al), la plata (Ag) y el oro (Au) [114].**

Para evitar la degradación de los recubrimientos de aluminio, actualmente existe una variedad de líneas de investigación sobre distintas tecnologías para fabricar los recubrimientos de espejos de vidrio. Todas ellas están orientadas a maximizar la reflectividad con la mínima degradación posible al exponer los espejos a la intemperie, y conseguir una fabricación simple para adaptarse bien a las técnicas de réplica de producción masiva. Una de estas tecnologías consiste en recubrimientos a base de una capa metálica de aluminio y sobre ella una multicapa dieléctrica. Esta multicapa, al tiempo que protege la capa de aluminio, aumenta la reflectividad según la ley de Fresnel, alternando capas de diferentes espesores e índices de refracción [115, p. 2]. Esta multicapa además se comporta como un filtro interferencial, siendo capaz de aumentar selectivamente la reflectividad de forma que se maximice en el espectro *Cherenkov* (hasta en un 5 % [116]) y se minimice fuera de él para rechazar el ruido de fondo ajeno a la banda de interés. La otra alternativa consiste en recubrimientos puramente dieléctricos sin capa de aluminio, por lo que la adhesión al vidrio es mucho mejor, disminuyendo su degradación. Esta alternativa será en general preferible al contar con las ventajas de la primera y sin los inconvenientes que presenta la capa metálica. No obstante, su coste de fabricación es superior al de los espejos metálicos, aunque podría ser compensado con su mayor tiempo de vida. Un problema detectado en este tipo de espejos puramente dieléctricos es que presentan una condensación significativamente mayor sobre su superficie que los espejos metálicos. Esto se debe a que la emisividad de estos espejos es unas diez veces mayor en el infrarrojo medio (entre 8 y 14 μm), por lo que tienden a enfriarse muy rápidamente [117], favoreciendo la formación de rocío sobre su superficie.

---

[6] Existe una alternativa orientada a alargar la vida útil del espejo que consiste en aplicar el recubrimiento reflexivo entre el sustrato de soporte y la capa frontal de vidrio, por detrás de ésta. En este caso no existe degradación de la capa reflexiva al estar totalmente protegida, pero al atravesar el vidrio en dos ocasiones, se producen pérdidas. Para minimizarlas sería necesario un vidrio demasiado delgado, por lo que esta técnica se ha descartado para CTA.

Al igual que en el caso de los espejos metálicos del apartado 2.6.2, los espejos basados en sustrato de vidrio investigados para CTA nunca se caracterizan en la zona del infrarrojo cercano, llegando solo hasta el límite superior del espectro *Cherenkov*. Con el objetivo de analizar la viabilidad de esta tecnología para la zona de interés en FSOC, a través de Andreas Förster, responsable de testeo de espejos en CTA, se pudieron obtener muestras de los tres tipos de espejos construidos con técnicas de réplica: el AR100 con recubrimiento de aluminio y de cuarzo (Figura 23, centro), el DH100 con recubrimiento de aluminio y multicapa de $SiO_2$+$HfO_2$+$SiO_2$ (Figura 23, derecha) y el DD040 de dieléctrico puro (Figura 23, izquierda).

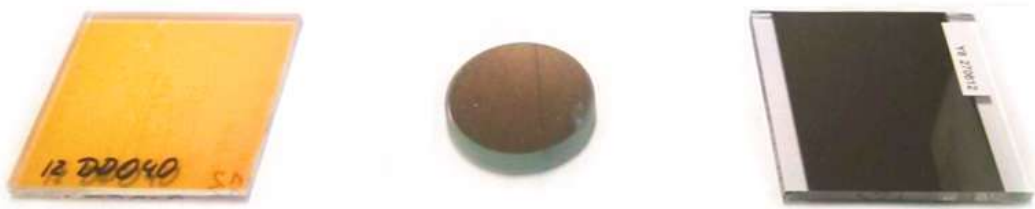

**Figura 23. Muestras caracterizadas de espejos basados en sustrato de vidrio. DD040: dieléctrico puro (izquierda); AR100: aluminio + cuarzo (centro); DH100: aluminio + multicapa dieléctrica (derecha).**

Las medidas de reflectividad de los espejos de la Figura 23 se han realizado utilizando el espectrómetro Lambda 900 de *Perkin Elmer* del Departamento de Metrología del Consejo Superior de Investigaciones Científicas (CSIC), en el Instituto de Tecnologías Físicas y de la Información (ITEFI) "Leonardo Torres-Quevedo", Madrid. En todos los casos, la reflectividad se mide para el ángulo estándar de incidencia de 8°. En la Figura 24 se muestra la reflectividad de los tres espejos cubriendo desde el espectro *Cherenkov* hasta la zona de interés en FSOC en un barrido con pasos de 10 nm. Se puede comprobar que los dos recubrimientos basados en aluminio como material reflector proporcionan una buena reflectividad en el infrarrojo cercano, especialmente en la longitud de onda de mayor interés, 1550 nm. El espejo DH100 proporciona una reflectividad algo mayor en la zona del espectro *Cherenkov* gracias al recubrimiento dieléctrico. El DD040 es el que mayor reflectividad muestra en esta región, entre el 90 % y el 99 %, cayendo abruptamente para longitudes de onda superiores a 580 nm, donde pasa a reflejar solamente entre el 4 % y el 20 % de la luz incidente.

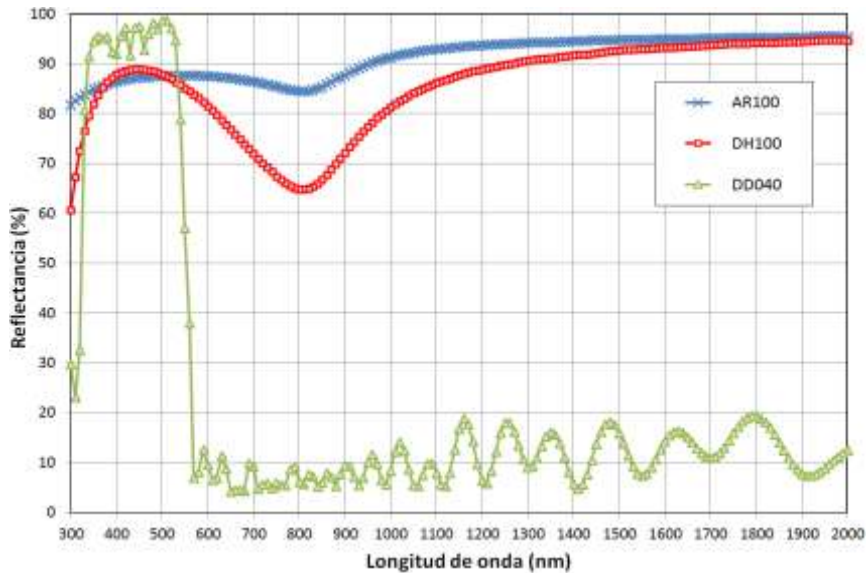

**Figura 24. Reflectividad de los espejos AR100, DH100 y DD040.**

En la Tabla 1 se muestran los rangos de reflectividades de los tres espejos correspondientes al rango genérico del espectro *Cherenkov* (300-600 nm), las correspondientes al rango para el cual la reflectividad queda por encima del 90 % en el DD040 (339-533 nm) y las reflectividades específicas para las longitudes de onda de interés en FSOC (1064 nm y 1550 nm). Para tomar estas medidas se realizó otra tanda de medidas con pasos más pequeños de 1 nm.

**Tabla 1. Reflectividad de los espejos AR100, DH100 y DD040 en distintas regiones espectrales.**

| Región<br>Espejo | 300-600 nm | 339-533 nm | 1064 nm | 1550 nm |
|---|---|---|---|---|
| **AR100** | 81,2-87,5 % | 83,9-87,4 % | 92,5 % | 94,9 % |
| **DH100** | 60,4-88,8 % | 79,1-88,8 % | 84,5 % | 93,0 % |
| **DD040** | 6,7-98,8 % | 90,1-98,8 % | 5,3 % | 7,3 % |

Analizando los resultados de la Tabla 1 se puede concluir que el mejor espejo para FSOC es el clásico de aluminio con recubrimiento de cuarzo (AR100), que alcanza una reflectividad cercana al 95 % en 1550 nm y el peor es el DD040, con una reflectividad prohibitivamente reducida. Sin embargo, este espejo es claramente superior a los otros en la región *Cherenkov*, si bien en un rango algo inferior al considerado habitualmente (aunque esto se puede modular ajustando los índices de refracción y espesores de las capas de dieléctro). Su buen comportamiento reflexivo en 339-533 nm, unido por una parte a su menor degradación en intemperie debido a la buena adherencia sobre vidrio y por otra parte a su rechazo a bandas ajenas a la radiación *Cherenkov*, le hacen el candidato más probable para CTA si se lograra solucionar el problema de condensación que presentan actualmente.

La tecnología de aluminizado de vidrio y recubrimiento protector es sobradamente conocida y ofrece el menor coste y una gran facilidad de fabricación, por lo que podría utilizarse directamente para fabricar los espejos de un terminal de comunicaciones basado en IACT. Cabe mencionar que la Universidad de Sao Paulo está trabajando en una línea de investigación con el objetivo de mejorar esta técnica para su aplicación en CTA [118, p. 4]. En el caso de la técnica de espejos dieléctricos, estos no podrían ser reutilizados directamente para FSOC, aunque es ciertamente posible modificar la banda reflexiva ajustando los índices de refracción y espesores de dieléctrico. De esta forma puede conseguirse trasladar el rango de máxima reflectividad al entorno de 1550 nm, con el aliciente que supone el rechazo del resto de longitudes de onda para reducir el ruido de fondo. En 2012 se comenzó una campaña de reemplazo de espejos en los cuatro telescopios originales de HESS: de los 1520 nuevos espejos que se instalaron, 100 fueron puramente dieléctricos [116], mostrando la reflectividad de la Figura 25. Se puede comprobar cómo precisamente la región alrededor de 1550 nm está optimizada para una reflectividad cercana al 100 %. Se desconoce si este comportamiento fue diseñado con este objetivo, pero sin duda se trata de una demostración de que incluso los espejos dieléctricos de IACT pueden ser reutilizados para FSOC con ganancias excelentes. Además, los propios fabricantes de estos espejos para HESS confirmaron que el proceso de fabricación admite este tipo de ajustes sin ningún coste adicional [119].

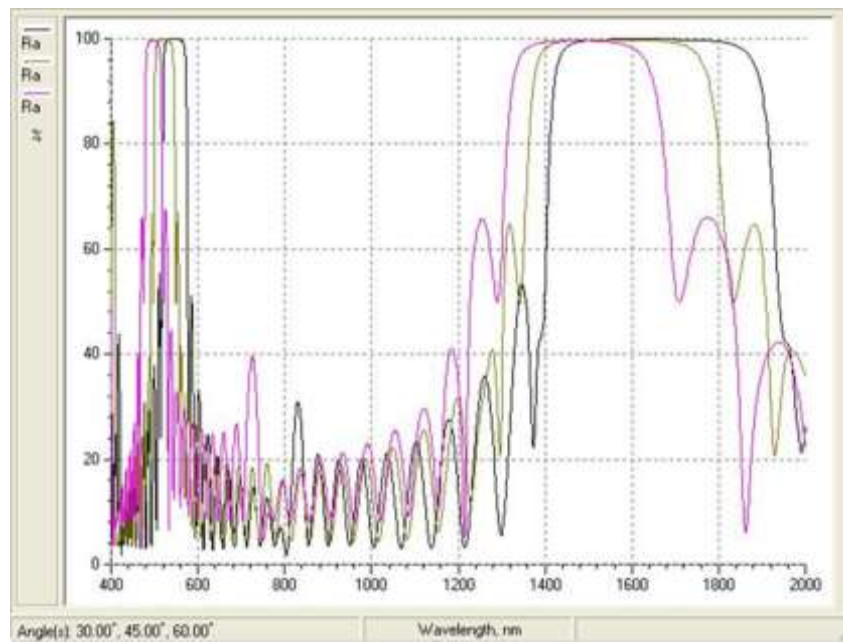

**Figura 25. Reflectividad de un espejo dieléctrico de HESS para diferentes ángulos de incidencia desde el espectro visible hasta el infrarrojo cercano [119].**

# 2.7. ENFOQUE EN UN IACT

En este apartado se estudian las consecuencias que tiene desde un punto de vista óptico sobre la adaptación para FSOC el hecho de que los telescopios *Cherenkov* no enfocan al infinito, como el resto de telescopios astronómicos y de comunicaciones, sino a una altura determinada en la atmósfera, donde tiene lugar el efecto *Cherenkov*.

## 2.7.1. Fórmula de Gauss para una lente delgada

Cualquier sistema óptico se considera enfocado cuando la luz de los puntos de un objeto observado converge en una imagen lo mejor posible, y en otro caso el sistema estará desenfocado. La imagen de un sistema óptico estará entonces en el plano donde se cortan los rayos de un haz que se propaga en la misma dirección que el eje óptico tras refractarse sobre las superficies del sistema óptico, que de forma general se puede modelar como una lente delgada. La longitud focal se refiere a la distancia entre el centro de la lente y el punto de convergencia de los rayos cuando el sistema enfoca al infinito.

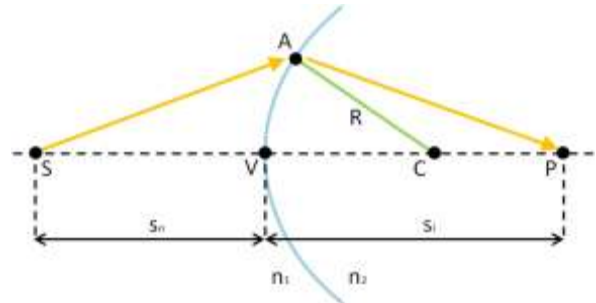

**Figura 26. Refracción en un interfaz esférico entre dos medios.**

Partiendo de las leyes de Snell, es posible [120, p. 154] llegar a la expresión de la ecuación (2-2), que describe la refracción de un interfaz esférico entre dos medios, como el de la Figura 26, donde R es el radio de curvatura del interfaz, $n_1$ y $n_2$ son los índices de refracción de cada medio (siendo $n_2 > n_1$), $s_o$ es la distancia SV, conocida como distancia objeto, y $s_i$ la distancia VP, conocida como distancia imagen. Esta expresión es válida independientemente de la posición de A en la curva, asumiendo óptica gaussiana o paraxial, por lo tanto se cumplirá para todos los rayos.

$$\frac{n_1}{s_o} + \frac{n_2}{s_i} = \frac{n_2 - n_1}{R} \qquad (2\text{-}2)$$

Para extender su aplicación a una lente delgada genérica donde habría dos interfaces, la ecuación (2-2) puede transformarse [120, p. 158] en la famosa fórmula del fabricante de lentes o fórmula de lentes delgadas de la ecuación (2-3), donde $R_1$ y $R_2$ son los radios de curvatura de cada interfaz y $n_l$ es el índice de refracción de la lente, mayor que el del medio, que se asume que es aire, $n_a = 1$.

$$\frac{1}{s_o} + \frac{1}{s_i} = (n_l - 1)\left(\frac{1}{R_1} - \frac{1}{R_2}\right) \qquad (2\text{-}3)$$

Si $s_o \to \infty$, entonces $s_i = f$, siendo f la longitud focal, y viceversa, si $s_i \to \infty$, entonces $s_o = f$, por lo que la ecuación (2-3) puede convertirse en la ecuación (2-4), conocida como la fórmula de Gauss para una lente delgada.

$$\frac{1}{f} = \frac{1}{2R} = \frac{1}{s_o} + \frac{1}{s_i} \qquad (2\text{-}4)$$

## 2.7.2. Enfocar un IACT al infinito

Una diferencia fundamental, explicada en el apartado 2.6.1, entre telescopios *Cherenkov* y telescopios astronómicos es la distancia de los objetos a observar. Si en los telescopios astronómicos esta distancia puede considerarse infinita por la lejanía de los objetos, la luz

*Cherenkov* que deben captar los IACT proviene de una distancia indeterminada. Esto implica que por una parte las imágenes de un telescopio *Cherenkov* nunca estarán enfocadas por la incertidumbre del origen de la luz, y por otra que el telescopio debe enfocarse de una forma distinta a la que lo está un telescopio astronómico convencional. Normalmente los IACT se enfocan considerando una distancia típica de origen del efecto *Cherenkov* igual a 10 km, lo que significa que una fuente de luz situada a otra distancia, incluido un transmisor de comunicaciones, que equivale a una fuente en el infinito, quedaría desenfocada, es decir, lo que idealmente debería ser un punto en el plano focal (aunque en la práctica esto nunca es así), será en su lugar una extensión borrosa. Esto tiene consecuencias muy negativas en un receptor de comunicaciones (explicadas en el apartado 2.9), que es necesario evitar reenfocando el telescopio para observar objetos en el infinito. Este nuevo enfoque es posible realizarlo desplazando longitudinalmente el detector una determinada distancia desde donde se sitúa la cámara de un IACT enfocada a 10 km.

La utilidad en telescopios reflectores de la fórmula de Gauss para una lente delgada vista en el apartado 2.7.1 es que es aplicable no solo a lentes delgadas, sino también a espejos [121, p. 40], aunque en este caso el criterio de signos cambia: un espejo cóncavo se comporta como una lente convergente y un espejo convexo como una lente divergente: en un espejo cóncavo como el reflector de los IACT las distancias medidas en la dirección de la propagación de la luz incidente hacia el espejo serán positivas y las medidas en la dirección de la luz reflejada serán negativas.

$$\frac{1}{f} = \frac{1}{s_o} + \frac{1}{s_i} = \frac{1}{10\,\text{km}} + \frac{1}{f+\varepsilon} \qquad (2\text{-}5)$$

Con la ecuación (2-5) es posible calcular cual debería ser el desplazamiento $\varepsilon$ que habría que trasladar el detector de comunicaciones respecto a la posición de la cámara *Cherenkov*. Si se asume una longitud focal f de 17 metros, como el telescopio MAGIC II, este desplazamiento es de 2,9 cm en dirección al reflector (Figura 27), lo que se encuentra dentro del rango permitido de desplazamiento mecánico para la cámara en estos telescopios (en el caso de MAGIC II este rango es de 30 cm [122]), por lo que aunque en principio los IACT no están diseñados para enfocar al infinito, el reenfoque es posible mediante un pequeño movimiento del detector, cuya viabilidad ya está reflejada en el diseño mismo de los IACT.

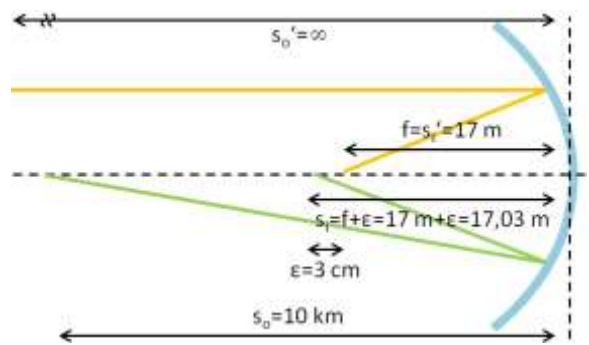

**Figura 27. Enfoque al infinito en un IACT de f= 17 m.**

En la Tabla 2 se muestra el desplazamiento necesario $\varepsilon$ para enfocar un IACT en el infinito, considerando los tres tipos de telescopios de CTA y el MAGIC. Se puede comprobar que todos son desplazamientos reducidos para las dimensiones de los telescopios: el mayor de ellos, de 9 cm, relativo a CTA-LST, se refiere a una cámara de 3,14 × 3,14 m [123, p. 3]).

Además, aunque en el caso de un IACT este desplazamiento implica trasladar completamente la voluminosa estructura de la cámara, en un telescopio de comunicaciones no existiría tal cámara. En su lugar, únicamente habría que ubicar el plano focal en su posición correspondiente, lo que es una maniobra perfectamente viable.

**Tabla 2. Desplazamiento necesario ε para enfocar un IACT al infinito.**

| Parámetro Telescopio | Diámetro D | Relación focal f/D | Longitud focal f | Desplazamiento ε |
|---|---|---|---|---|
| MAGIC I/II | 17 m | 1 | 17 m | 2,9 cm |
| CTA-SST | 6 m | 0,5 | 12 m | 1,4 cm |
| CTA-MST | 12 m | 1/0,75 | 16 m | 2,5 cm |
| CTA-LST | 24 m | 1,25 | 30 m | 9 cm |

## 2.8. SIMULACIÓN ÓPTICA DE MAGIC II

Tras verificar la viabilidad del empleo directo de las técnicas de fabricación de espejos en IACT para un telescopio dedicado a FSOC, el siguiente paso lógico es analizar la calidad óptica de este tipo de telescopios, sabiendo de antemano que es muy inferior a la de los telescopios astronómicos convencionales, para evaluar si estos diseños pueden aún así ser directamente reutilizables o precisan alguna adaptación. Ante la escasez de información sobre los telescopios de CTA durante la fase inicial de este estudio, se comenzó realizando una simulación simplificada del telescopio MAGIC II, utilizando óptica geométrica en dos dimensiones. Se eligió este telescopio por tratarse del más avanzado en el campo de la astronomía terrestre de rayos gamma y por lo tanto el más parecido a los futuros telescopios de CTA. A continuación se describen los resultados extraídos de dicho trabajo por ser muy ilustrativos del comportamiento óptico de este tipo de telescopios, pese a tratarse de una simulación en primera aproximación.

### 2.8.1. Telescopio MAGIC II

Los telescopios MAGIC (del inglés *Major Atmospheric Gamma-ray Imaging Cherenkov*) son telescopios *Cherenkov* de última generación situados en el observatorio de Roque de los Muchachos en La Palma, Canarias. MAGIC I (Figura 28, izquierda) lleva en operación desde 2003 y MAGIC II (Figura 28, derecha) desde 2009. Ambos son telescopios segmentados con una apertura reflectora de 17 metros de diámetro, si bien el primero lo forman 964 espejos cuadrados de 0,495 metros de lado con un área total de 236 m$^2$ y el segundo lo forman 247 espejos cuadrados de 1 metro de lado con un área total de 247 m$^2$ [124]. Eran los más grandes del mundo hasta finales de 2012, cuando se inauguró el HESS II de 28 metros en Namibia, con 875 espejos y un área total de 614 m$^2$ [125].

La relación focal de los telescopios MAGIC es de f/D = 1, por lo que al ser un telescopio de foco primario, la cámara (con un peso de una media tonelada y compuesta por 576 fotomultiplicadores) enfocada en el infinito estaría situada a 17 metros del reflector, definiendo su longitud focal f. Los espejos de ambos telescopios están montados sobre paneles de 1 metro de lado (4 espejos por paneles en el caso de MAGIC I y un único espejo en MAGIC II), que permiten alinear cada segmento en lo que se conoce como cofaseo en telescopios segmentados, además de poder corregir en tiempo real las deformaciones del disco inducidas por la gravedad al cambiar su orientación.

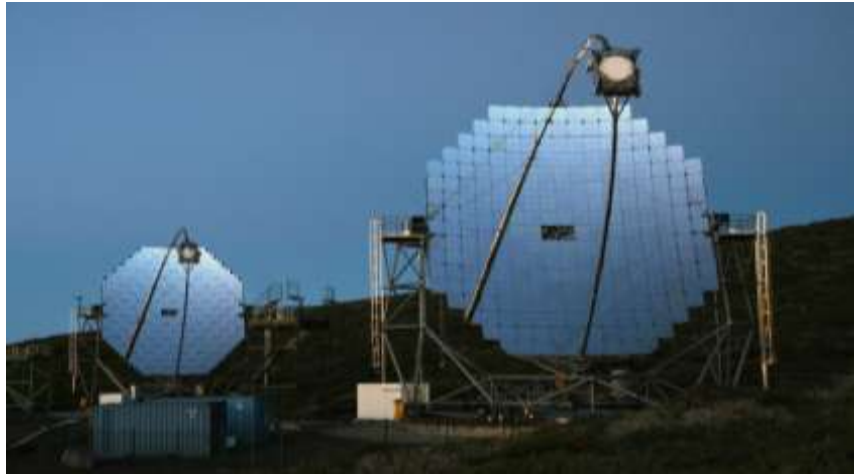

**Figura 28. Telescopios MAGIC I (izquierda) y MAGIC II (derecha) en el observatorio de Roque de los Muchachos en La Palma, Canarias [126].**

El perfil del reflector describe una curva parabólica, elegida para mantener la coherencia temporal de los fotones que llegan al plano focal [127, p. 4]. En el diseño de IACT de diámetros mayores a unos 10 metros en los que no es factible aumentar la relación focal, la forma parabólica es fundamental para conseguir las resoluciones temporales que se requieren, en el orden de los nanosegundos. Por ejemplo, si MAGIC fuera esférico en lugar de parabólico, introduciría un inaceptable retardo de decenas de nanosegundos entre los fotones axiales y marginales [40, p. 48]. Si bien el perfil global es parabólico, cada espejo individual tiene una superficie con forma esférica, con radios de curvatura constantes por segmentos pero variables de uno a otro. La curvatura de cada espejo es dependiente de la posición que ocupe en la curva parabólica, con el objetivo de aproximar el perfil al de la parábola que se pretende emular. De esta forma se mantienen los requisitos temporales de un telescopio parabólico, pero manteniendo los reducidos costes asociados a la óptica esférica.

## 2.8.2. Modelado del telescopio

Los telescopios parabólicos describen superficies que se derivan de las conocidas como secciones cónicas: curvas formadas al intersectar un cono y un plano que no pasa por el vértice (Figura 29, izquierda). La propiedad fundamental de las cónicas es que la normal en cualquier punto coincide con la bisectriz del ángulo formado por los radios que unen ese punto con los dos focos (Figura 29, derecha). Esto significa, en términos de propagación y reflexión de la luz, que todos los rayos cuyo origen sea uno de los focos, tras reflejarse, convergerán en el otro foco, formando una imagen perfecta del primero (propiedad conocida como stigmatismo). El caso más simple es el de la parábola, que es una elipse degenerada en la que uno de los focos está en el infinito. Todos los rayos que provengan de este foco en el infinito, o lo que es lo mismo, paralelos al eje de la parábola, después de la reflexión convergerán en el foco de la misma.

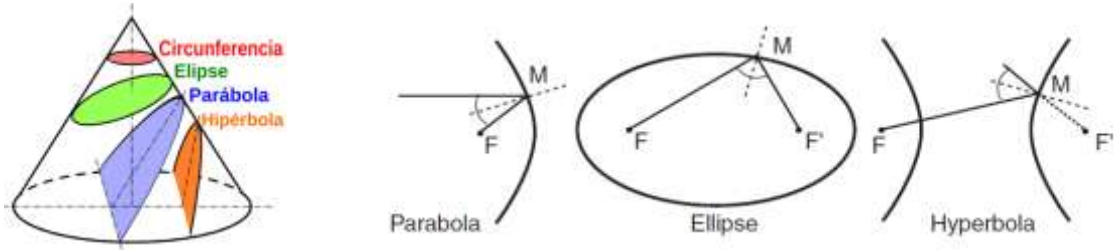

**Figura 29. Secciones cónicas (izquierda) y su propiedad de stigmatismo (derecha).**

Si bien matemáticamente se puede demostrar que un reflector parabólico consigue una imagen perfecta de una fuente en el infinito al coincidir el eje de la fuente y el del telescopio, cuando existe una desviación entre ambos, los telescopios basados en parábolas presentan aberración de coma.

$$y^2 - 2Rx + \left(1 - e^2\right)x^2 = 0 \qquad (2\text{-}6)$$

La ecuación (2-6) define todas las secciones cónicas en dos dimensiones {x, y}, siendo R el radio de curvatura y e la excentricidad [128]. De esta ecuación de segundo grado se puede despejar x en la ecuación (2-7) y desarrollarla en una serie de Taylor según se muestra en la ecuación (2-8).

$$x = \frac{R - \sqrt{R^2 - (1+\kappa)y^2}}{1 + \kappa} \qquad (2\text{-}7)$$

$$x = \frac{y^2}{2R} + (1+\kappa)\frac{y^4}{8R^3} + (1+\kappa)^2\frac{y^6}{16R^5} + \frac{5}{128}(1+\kappa)^3\frac{y^8}{16R^7} + \ldots \qquad (2\text{-}8)$$

donde $\kappa = -e^2$ y define la familia de cónicas. Para una parábola, $\kappa = -1$, y así se llega a la conocida expresión de la ecuación (2-9), donde f es la longitud focal, igual a 2R.

$$x = \frac{y^2}{2R} \Rightarrow y^2 = 2Rx = 4fx \qquad (2\text{-}9)$$

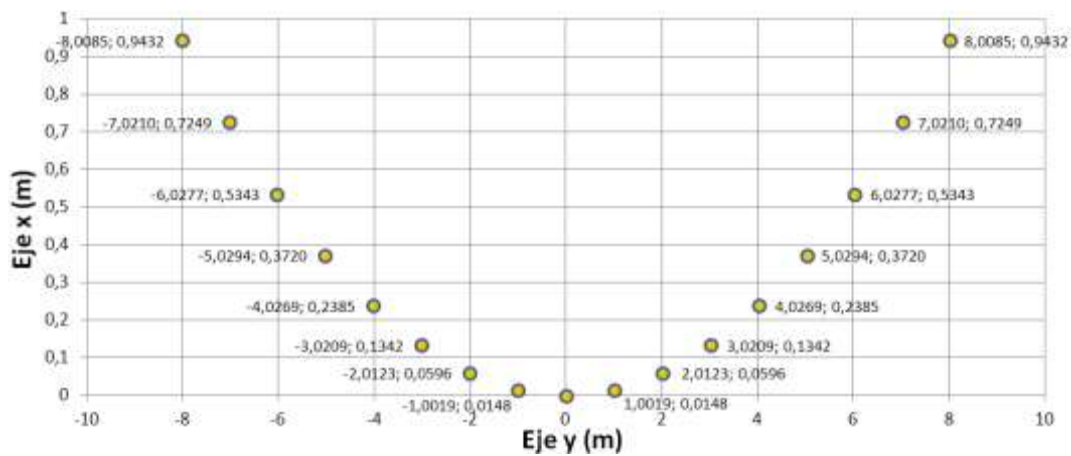

**Figura 30. Parábola centrada en {0, 0}, con foco en {17, 0} y 17 espejos de 1 metro distribuidos de forma equidistante desde el centro.**

La ecuación (2-9) representa una parábola bidimensional centrada en el punto {0, 0} y con foco en el punto {f, 0}. Para MAGIC II, f = 17 m, por lo que $y^2 = 68\,x$. Discretizando el eje x para conseguir una representación de dicha parábola, se ha calculado la posición de cada uno de los segmentos del telescopio comenzando desde el central, situado en {0, 0}, y desplazándose por la curva con distancias de 1 metro entre posiciones consecutivas. En la Figura 30 puede apreciarse la distribución de los 17 espejos según esta parábola con separaciones equidistantes entre ellos. Para el cálculo de estas posiciones 2D se ha empleado una precisión de 0,2 mm respecto a los 17 metros de apertura de la parábola, es decir, el error es menor que 0,001 %.

Los espejos de MAGIC son esféricos, construidos con un radio de curvatura idealmente constante a lo largo de toda la superficie de cada segmento. Sin embargo, una parábola ideal tiene infinitos radios de curvatura que cambian de un punto a otro de su superficie. Por ello, es necesario determinar un radio de curvatura único para cada espejo del telescopio. Además, el radio de curvatura no solo varía con la posición, sino que en cada punto de una parábola tridimensional también existen infinitos radios de curvatura en función de qué plano corte la curva. Un criterio razonable es elegir el promedio $r_{c\ med}$ entre el radio de curvatura máximo $r_{c\ máx}$ y el mínimo $r_{c\ mín}$ de cada segmento.

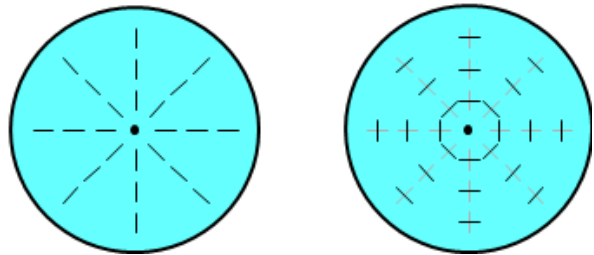

**Figura 31. Planos tangenciales (izquierda) y planos sagitales (derecha).**

En cada punto, el radio de curvatura máximo $r_{c\ máx}$ lo determina el corte de la parábola con los planos tangenciales (Figura 31, izquierda), definidos por los meridianos que se alejan desde el eje óptico si se observa la parábola de frente. El radio de curvatura mínimo $r_{c\ mín}$ lo determina el corte de la parábola con los planos sagitales (Figura 31, derecha), definidos por los meridianos que circunscriben el eje óptico y son perpendiculares a cada plano tangencial en ese punto. En las ecuaciones (2-10), (2-11) y (2-12) se definen estos radios de curvatura $r_{c\ máx}$, $r_{c\ mín}$ y $r_{c\ med}$ [40, p. 50] calculados en función de la longitud focal f y la distancia al centro r.

$$r_{c\ máx} = 2f\left(1 + \frac{r^2}{4f}\right)^{\frac{3}{2}} \qquad (2\text{-}10)$$

$$r_{c\ mín} = 2f\left(1 + \frac{r^2}{4f}\right)^{\frac{1}{2}} \qquad (2\text{-}11)$$

$$r_{c\ med} = \frac{r_{c\ máx} + r_{c\ mín}}{2} \qquad (2\text{-}12)$$

En la Figura 32 se muestran los radios de curvatura máximos $r_{c\ máx}$, mínimos $r_{c\ mín}$, y medio $r_{c\ med}$ calculados para los nueve tipos de espejos distintos que hay en una sección longitudinal del telescopio MAGIC II. Se representa el valor de cada segmento en función de la distancia del centro del espejo al centro de la parábola.

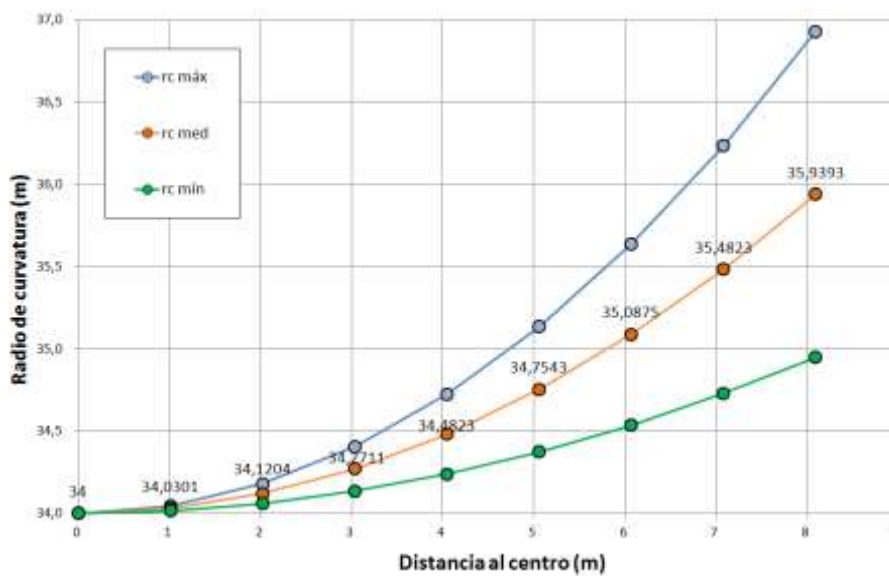

**Figura 32. Radios de curvatura máximo, medio y mínimo para los 9 tipos de distintos espejos del corte longitudinal del telescopio MAGIC II en función de la distancia al centro de la parábola.**

### 2.8.3. Simulación con óptica geométrica

Partiendo del modelo anterior de MAGIC II, se ha simulado su comportamiento utilizando el software de óptica geométrica *OpticsLab* de *Science Lab Software*. Este software permite realizar modelados rápidos de sistemas ópticos relativamente simples a cambio de algunas limitaciones. Aunque el entorno de trabajo es 3D, el número máximo de elementos y la incapacidad para automatizar su procesado impide realizar una simulación del telescopio completo, que exigiría manejar 249 espejos en el caso de MAGIC II (Figura 33, izquierda) o 198 en el caso de CTA-LST (Figura 33, derecha). Por ello, se ha modelado únicamente un corte bidimensional de la diagonal más larga, que permite que el número de elementos disminuya considerablemente. Se escoge esta diagonal por contener los espejos más alejados del centro, que son los que han de introducir mayores aberraciones (se trata de un caso peor en un telescopio no limitado por difracción). Esta aproximación es válida en tanto el telescopio sea simétrico respecto al eje óptico, como es el caso de MAGIC II[7].

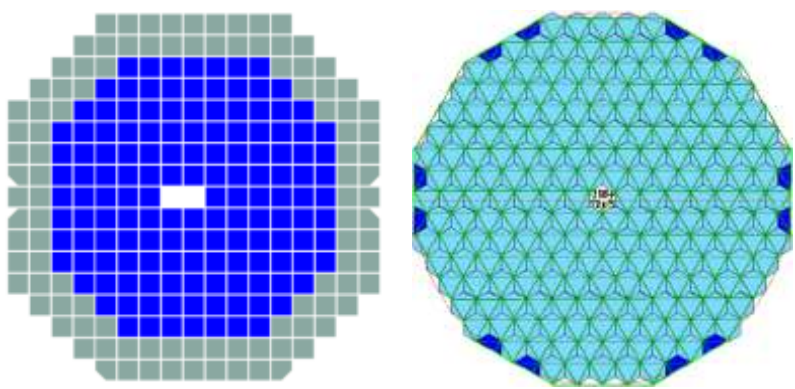

**Figura 33. Vista frontal (no a escala) de la superficie reflectora de los telescopios MAGIC II (izquierda) y CTA-LST (derecha).**

---

[7] En la práctica, al tratarse de telescopios del tamaño de MAGIC II pueden existir deformaciones en función de la orientación, principalmente por efecto de la gravedad, aunque son parcialmente compensadas con los sistemas de control automático de espejos.

En los IACT, el alineamiento de los espejos es más simple que en un telescopio segmentado astronómico convencional, pero igualmente fundamental para optimizar la PSF. Para implementarlo en la simulación de *OpticsLab* se ha programado un algoritmo de control de bucle en lazo cerrado tipo PID que actúa sobre la orientación de cada segmento hasta conseguir tras 30 iteraciones que el el haz colimado incidente acabe en el plano focal del telescopio. La función de error a minimizar en el bucle es la diferencia entre la posición instantánea del centroide del *spot diagram* de cada segmento en un detector sensible a la posición colocado en el foco y el centro geométrico del mismo. El modelo de MAGIC II (Figura 34) queda completado tras realizar el alineamiento de cada espejo, obteniendo un error de posición del *spot diagram* en el orden de los picómetros.

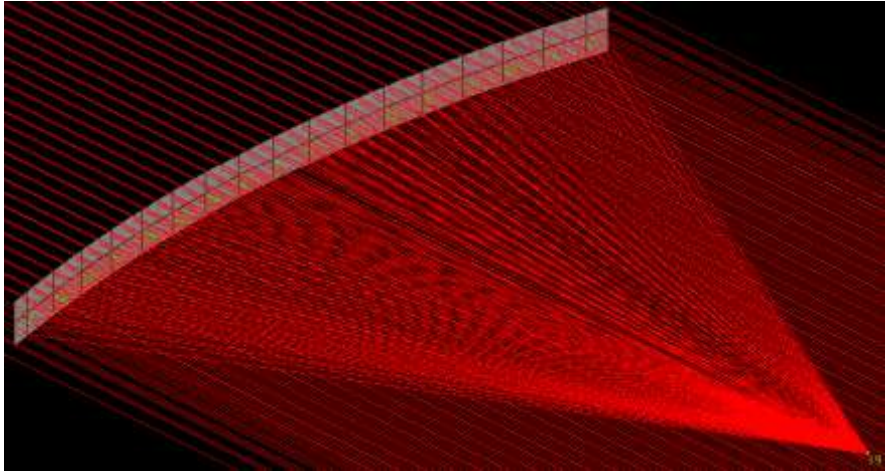

**Figura 34. Modelo en *OpticsLab* de un corte longitudinal de MAGIC II.**

En la Figura 35 se muestra a la izquierda una vista lateral del modelo de MAGIC II y a la derecha una ampliación en el plano focal a 17 metros del centro de la parábola. En la Figura 36 se muestra el *spot diagram* simulado con 61517 rayos.

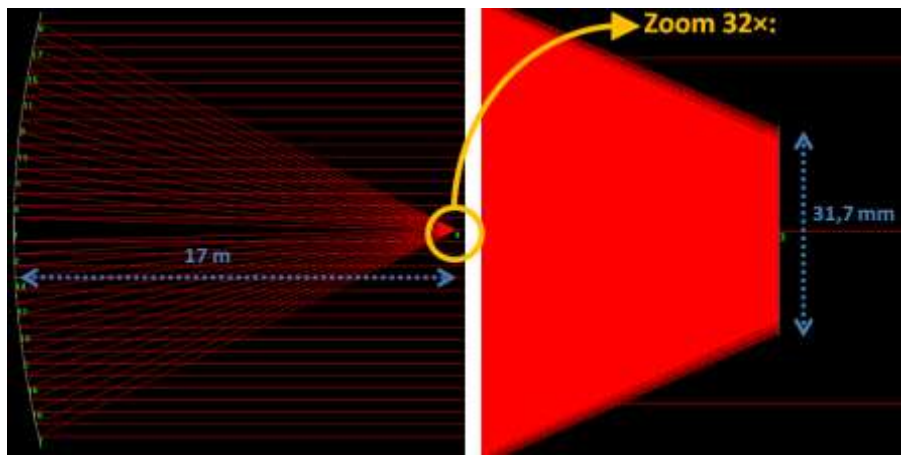

**Figura 35. Vista lateral del modelo de MAGIC II en *OpticsLab*
(izquierda) y ampliación ×32 del plano focal (derecha).**

Se puede comprobar que el *spot diagram* tiene una extensión de 31,7 mm en el plano focal. Para comparar el resultado de esta simulación con el valor real de la PSF del MAGIC II se ha empleado una medida experimental realizada en el propio telescopio y reportada en [113, p. 360], donde se modeló según una gaussiana con σ = 10,5 mm. Según la definición de la distribución gaussiana de la ecuación (2-13), para el valor de σ medido experimentalmente, la anchura del *spot diagrm* de 31,7 mm se correspondería con un área, o equivalentemente con una energía contenida, del 86,9 %, lo que puede considerarse como una buena aproximación entre la medida experimental y la simulada. Tal parecido entre una simulación con simple óptica geométrica y una medida experimental (que incluye más efectos que los contemplados en el modelo de *OpticsLab*), sugieren que la calidad óptica de MAGIC II está limitada por las fuertes aberraciones que introducen los espejos esféricos.

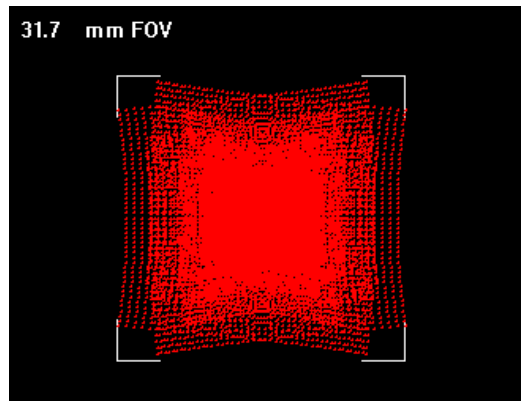

**Figura 36. Simulación en *OpticsLab*
del *spot diagram* de MAGIC II.**

$$f(x) = \frac{1}{\sqrt{2\pi\sigma^2}} e^{\frac{(x-\mu)^2}{2\sigma^2}} \qquad (2\text{-}13)$$

# 2.9. PSF, CAMPO DE VISIÓN Y RUIDO DE FONDO

La simulación del telescopio MAGIC del apartado anterior en términos de PSF revela que la calidad óptica de este tipo de telescopios está muy por debajo de la habitual en el tipo de telescopios utilizados para astronomía. La resolución óptica de estos telescopios tienda a aproximarse al límite de difracción, especialmente en los telescopios de última generación donde se supera la limitación de la turbulencia atmosférica mediante sistemas de óptica adaptativa. Por ejemplo, el límite de difracción de un telescopio de 10 metros (del tipo Keck o GTC) se correspondería con una PSF nominal en el visible de ~0,1 µm de diámetro, unos cinco órdenes de magnitud mejor que la resolución óptica de MAGIC II calculada en el apartado 2.8. En este apartado se analiza qué influencia tienen estas peores prestaciones en el enlace de comunicación y se estudian las estrategias para adaptar un IACT minimizando el impacto negativo de su reducida calidad óptica.

## 2.9.1. Scattering y ruido de fondo

En el apartado 2.4.2 se introdujo que en los enlaces de FSOC que incluyen estaciones receptoras en Tierra el *scattering* atmosférico suponía un inconveniente, con la capacidad para impactar muy negativamente en la comunicación. El *scattering* es un efecto de dispersión de la luz originado por partículas en suspensión en la atmósfera. Según sea el tamaño de estas partículas, el *scaterring* se divide en Rayleigh (partículas mucho menores que la longitud de onda) y Mie (partículas similares o mayores que la longitud de onda). El último se debe a los diferentes tipos de aerosoles en suspensión en la atmósfera y para cuantificarlo se utilizan diferentes modelos según la ubicación del receptor. La concentración de este tipo de aerosoles tiene una dependencia exponencial con la altitud [129, p. 444], quedando la mayor concentración y variabilidad por debajo de una altura de 1-2 km. Por ello, para un receptor situado en un observatorio astronómico, generalmente por encima de estas alturas, como los contemplados para CTA, la contribución de *scattering* Mie se puede considerar despreciable (por ejemplo, considerando una altura de 2 km, en una latitud media con un ángulo cenital igual a cero, las pérdidas a 1550 nm estarían por debajo de los 0,2 dB [4, p. 148]). Por otra parte, el *scattering* Rayleigh se debe principalmente a las moléculas de los gases atmosféricos, especialmente Nitrógeno y Oxígeno, y su capacidad de dispersión tiene una dependencia con $\lambda^{-4}$, lo que hace que el efecto sobre una señal de 1550 nm sean unas pérdidas de potencia prácticamente despreciables, por debajo de 0,1 dB [4, p. 144].

El efecto que tiene el *scattering* sobre el enlace de FSOC no es directo, sino indirecto: la dispersión de la luz de diferentes fuentes a la del láser de comunicaciones, como el Sol, la Luna, los planetas, etc. acaba entrando en el campo de visión del receptor terrestre incluso aunque su orientación originalmente esté alejada de las fuentes de luz. La dispersión de la luz acoplada al receptor tiene los dos orígenes comentados en el párrafo anterior: en este caso la dispersión Rayleigh será la principal contribución cuando el receptor esté orientado con ángulos lejanos a la fuente, y la dispersión Mie dominará cuando el receptor apunte cerca del origen de la fuente. Esta radiación se denomina radiancia del cielo y al entrar en el telescopio receptor, se suma a la señal de comunicaciones en forma de ruido de fondo, que reduce la relación señal a ruido degradando las prestaciones del enlace. Esta contribución de potencia de ruido $N_S$ (W) está definida por la ecuación (2-14) [4, p. 154], donde $L$ ($\lambda$, $\theta$, $\varphi$) representa la radiancia espectral del cielo por unidad de área (que depende de la longitud de onda $\lambda$, del ángulo cenital $\theta$ del receptor y del ángulo $\varphi$ entre el receptor y el Sol) y se expresa en W/(cm² · srad · μm). Para una radiancia espectral dada, la potencia de ruido depende del área $A_r$ (cm²) presentada por la apertura receptora, del campo de visión $\Omega_{FOV}$ (srad) y de la anchura $\Delta\lambda$ (μm) del filtrado espectral de la señal.

$$N_S = L(\lambda, \theta, \varphi) A_r \Omega_{FOV} \Delta\lambda \qquad (2\text{-}14)$$

De la ecuación (2-14) se pueden deducir las estrategias para reducir el ruido de fondo debido al *scattering* de la luz solar. Por una parte, la radiancia del cielo es un término variable sobre el que no se puede actuar, ya que los enlaces de FSOC en espacio profundo deben operar durante el día, y su valor puede llegar a ser muy elevado cuando el Sol está cerca del campo de visión del telescopio. La apertura no se debe reducir con el objetivo de disminuir el ruido de fondo, ya que se estaría también limitando la potencia de señal, que es el parámetro más importante de un enlace de este tipo (que siempre está limitado por potencia). Con el filtrado espectral es posible eliminar el ruido fuera de la banda de comunicaciones, pero no el de esta. Por lo tanto, el factor sobre el que deberían centrarse los esfuerzos es el campo de visión, haciendo que este sea lo más reducido posible y así minimizar la contribución de esta fuente de ruido de fondo.

No obstante, cabe mencionar que la radiancia del cielo que queda fuera del campo de visión del telescopio pero que entra a través de su apertura también puede acoplarse en forma de ruido cuando se refleje en la estructura o el espejo primario y se difunda en muchas direcciones, alguna de las cuales coincida con el campo de visión. La fuente más importante de esta contribución será la reflexión difusa en el espejo primario. Estas reflexiones pueden deberse a las microrrugosidades y al polvo: en el caso del infrarrojo cercano, el efecto de las microrrugosidades es despreciable y es el polvo el predominante [33, p. 193]. Dado que un telescopio de comunicaciones debe operar durante el día, la aportación de este ruido se considera despreciable frente a la radiancia del cielo que se acopla directamente al campo de visión.

### 2.9.2. Campo de visión

El campo de visión describe la extensión angular que muestra el plano objeto en el plano imagen de un determinado sistema óptico. No solo depende de las características del sistema óptico sino también del detector destinado a registrar la luz capturada. En la Figura 37 se muestra el campo de visión $\theta_{FOV}$ de un sistema óptico genérico de primer orden caracterizado por su longitud focal f y un tamaño del elemento detector igual a d. A efectos de campo de visión, se trabaja con una única longitud focal equivalente EFL (del inglés *Effective Focal Length*) aunque el sistema óptico esté compuesto por más de un elemento. Esta EFL es el equivalente a la longitud focal de una lente delgada con la misma apertura que el primer elemento del sistema y que produzca un haz refractado convergente en el plano focal con el mismo ángulo que el producido por el sistema de varios elementos.

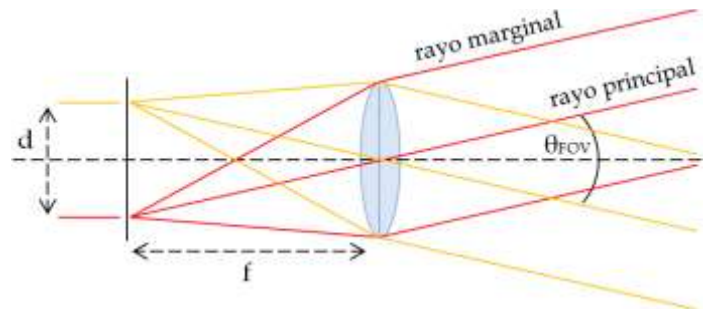

**Figura 37. Campo de visión de un sistema óptico genérico definido por su longitud focal f y con un tamaño de detección d.**

La Figura 37 representa la convergencia en el plano focal de los dos haces colimados que describen el mayor ángulo posible dada la extensión d del elemento detector, situado en el foco equivalente del sistema óptico. Para ello se ha dibujado en cada extremo del campo de visión el rayo principal (el que atraviesa el sistema óptico originado en el extremo del objeto, que al tratarse de un haz colimado equivale al infinito y por lo tanto al centro del haz, y termina en el extremo de la imagen) y los dos rayos marginales (los que atraviesan los extremos de la pupila de entrada originados en el centro del objeto) que describen completamente un haz colimado atravesando un sistema óptico. Según este esquema, es fácil deducir el ángulo completo que describiría el campo de visión $\theta_{FOV}$ dado por la ecuación (2-15).

$$\theta_{FOV} = 2\arctan\left(\frac{d}{2f}\right) \qquad (2\text{-}15)$$

Para el cálculo del campo de visión, en esta tesis se asume un detector de forma circular con diámetro d. De lo contrario, el campo de visión sería distinto siendo necesario especificar si se trata del campo de visión horizontal (calculado usando el lado horizontal del detector $d_H$), el vertical (usando el lado vertical $d_V$) o el diagonal (usando la diagonal $d_D = \sqrt{(d_H^2 + d_V^2)}$)). Para evitar una confusión muy común al referirse al campo de visión de forma genérica, en adelante se estará hablando siempre del ángulo completo $\theta_{FOV}$, a no ser que se indique lo contrario, en cuyo caso se hablará de semiángulo, expresado como $\pm\theta_{FOV}/2$.

En la ecuación (2-15) se puede comprobar cómo el campo de visión es proporcional al tamaño del detector y queda determinado por los rayos principales correspondientes a los extremos del mismo (Figura 38). Por ello, dado un sistema óptico con una EFL fija, la única [8] estrategia para reducir el campo de visión es reducir el tamaño del área activa dedicada a la detección en el plano focal. Adicionalmente, el tamaño del fotodetector está inversamente relacionado con la velocidad de detección a través de la capacitancia equivalente del área activa. Por ello, altas tasas de transmisión también requieren un tamaño reducido del fotodetector (si bien este requisito es menos exigente al poder recurrir a una agrupación de detectores pequeños).

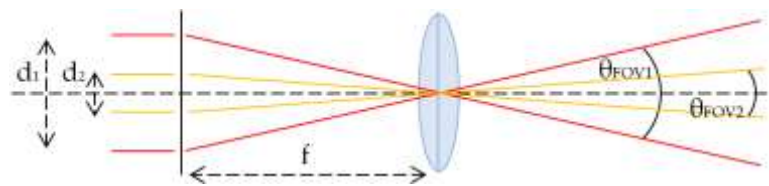

**Figura 38. Campo de visión $\theta_{FOV}$ en función del tamaño detector d.**

## 2.9.3. Adaptación de un IACT

Los campos de visión de los telescopios *Cherenkov* son en general relativamente grandes. El campo de visión de los telescopios astronómicos es típicamente de una fracción de grado [72, p. 25]; en comparación, los IACT suelen diseñarse con campos de visión que pueden llegar hasta los 10°. En el caso de CTA, los SST deben tener unos amplios campos de visión para soportar el modo de operación conjunta de observación de los mismos eventos, al estar muy espaciados unos de otros. En el caso de los MST y LST este requisito no es tan importante, pero también requieren campos de visión muy extensos para detectar lluvias electromagnéticas sin truncamiento y optimizar el tiempo de observación cubriendo mayores extensiones de cielo. Las cascadas *Cherenkov* suceden a una distancia media de 10 km, lo que equivale a un cono que cae sobre la superficie con un ángulo de unos 0,7° (Figura 7). Por ello, estadísticamente la mayor parte de la luz *Cherenkov* llegará a los IACT con un determinado ángulo respecto al eje óptico, lo que exige a estos telescopios realizar sus observaciones con un campo de visión amplio. Como se vio en la ecuación (2-15), para disponer de un gran campo de visión se puede aumentar el tamaño del área de detección y/o reducir la longitud focal. En los IACT se dan ambas características a la vez. Las cámaras de estos telescopios tradicionalmente han estado construidas por agrupaciones de detectores en los que cada píxel se corresponde con un fotomultiplicador, por lo que se trata de cámaras de gran tamaño. Por ejemplo, la cámara de MAGIC tiene 3,5° de campo de visión [130], que con una longitud focal de 17 metros equivale a un área de detección circular de más de 1 metro de diámetro en el plano focal. Por otra parte, las relaciones focales f/D son muy reducidas, algo mayores que f/1, lo que implica una mayor presencia de aberraciones. Con estas relaciones focales se puede conseguir una resolución suficiente para los limitados requisitos de imagen *Cherenkov* con campos de visión hasta unos 4° [131]. Por encima de esto, se hace necesario aumentar la relación focal por encima de f/2 para llegar a campos de visión de hasta ~10°. Sin embargo, esta estrategia está muy limitada en telescopios de espejo único y gran apertura, ya que un plano focal muy alejado del reflector principal requiere un complejo sistema de soporte para no producir deformaciones en el resto de la estructura. Para evitar estos problemas es necesario recurrir a configuraciones de doble espejo con perfiles asféricos, lo que encarece excesivamente los telescopios, especialmente los MST y LST.

---

[8] Estrictamente también se podría reducir el campo de visión creando un viñeteo artificial, es decir, haciendo que la pupila de entrada tenga un tamaño menor a la apertura del sistema óptico. Sin embargo, esta estrategia que puede tener sentido en otras aplicaciones, no debe utilizarse cuando el objetivo es recolectar la mayor cantidad de fotones posible.

En el apartado 2.9.1 se concluyó que una estrategia fundamental para optimizar la relación señal a ruido en telescopios receptores terrestres dedicados a FSOC con operación diurna es minimizar el campo de visión. En este sentido, los requisitos de telescopios para FSOC y los IACT son opuestos. No obstante, no hay que confundir los requisitos de detección en el campo de visión de un IACT con el campo de visión de un telescopio de FSOC. En el primer caso se refieren a una cámara compuesta por cientos de detectores individuales o píxeles, y en el segundo a un único píxel, el fotodetector. Como se ha explicado, el campo de visión depende tanto de la longitud focal efectiva como del tamaño del detector. Por ello, el campo de visión de un IACT empleado en comunicaciones con un único detector habrá de heredar la focal debida a su construcción como IACT, pero al tener un área de detección mucho más pequeña, tendrá un campo de visión menor al nominal del telescopio incluyendo la cámara.

En cualquier caso, la reducida focal de estos telescopios sí presenta un inconveniente. Las configuraciones de foco primario como las usadas en IACT ofrecen relaciones focales en el rango de f/1 y f/3, en comparación con la configuración *Cassegrain*, en principio más apropiada para FSOC, con relaciones focales entre f/8 y f/15, o configuraciones Coudé que pueden llegar hasta f/100 [33, p. 135]. Esto hace que para una misma apertura, los IACT proporcionen longitudes focales muy reducidas, por lo que solo queda actuar sobre el tamaño del área detectora. Sin embargo, esta área no puede reducirse arbitrariamente, sino que en general dependerá de la extensión de la PSF en el plano focal. También en IACT, la PSF está relacionada con el tamaño de cada píxel de la cámara. Desde un punto de vista estrictamente científico, el tamaño de píxel debería ser el menor posible para aumentar la resolución de la imagen. Si la PSF es suficientemente pequeña, el criterio último para decidir el tamaño del píxel depende del coste, ya que para un mismo campo de visión, un menor tamaño de píxel involucra una cámara con un elevado número de píxeles, con sus canales electrónicos asociados. Sin embargo, en IACT generalmente el tamaño del píxel depende directamente de la PSF, por ser esta de tamaño considerable, usándose algún criterio que contemple incluir la mayor parte de la energía de una fuente puntual en el área activa del fotodetector. Por ejemplo, que la PSF que contiene el 40 % de la potencia tenga la mitad del tamaño de píxel [39, p. 74], o un tercio de su tamaño para un 80 % de confinamiento [63, p. 16].

La consecuencia de un detector mayor que la PSF del telescopio es el desaprovechamiento de la máxima resolución óptica en un IACT, y en FSOC la degradación de la relación señal a ruido, debido a la captación de un mayor nivel de ruido de fondo. Un tamaño de píxel menor que el de la PSF supondría en un IACT un desperdicio en el número de fotodetectores, al producir una imagen con la misma resolución que utilizando menos píxeles de mayor tamaño. En el caso de FSOC, se estaría perdiendo potencia óptica de la señal de comunicación, lo que supone un desaprovechamiento de la apertura receptora. No obstante, en presencia de altos niveles de ruido de fondo podría tener sentido un truncamiento de la señal para reducir la radiancia del cielo, aunque en general y de forma ideal la solución óptima es reducir el tamaño de la PSF para adaptarla a una menor área detectora.

## 2.10. LIMITACIONES AL CAMPO DE VISIÓN

En el apartado anterior se concluyó que para reducir el ruido de fondo en FSOC, especialmente debido a la dispersión de la luz del Sol durante el día, es necesario minimizar el campo de visión del telescopio, para lo cual si no se puede actuar sobre la longitud focal, debe reducirse el área detectora en el plano focal, que viene limitada en general por la resolución óptica del telescopio. Esta puede venir determinada por una serie de causas de diferente naturaleza. En este apartado se analizan los factores que se han identificado como limitaciones al campo de visión, ordenados de menor a mayor relevancia en un telescopio de comunicaciones y en uno dedicado a astronomía *Cherenkov* (Figura 39). Se trata de un análisis fundamental para canalizar los esfuerzos de la adaptación de los IACT hacia las soluciones más eficaces, ignorando efectos que puedan no tener relevancia en su funcionamiento como receptor de FSOC. Dado que la resolución final del telescopio será la convolución de las resoluciones debidas a cada factor limitante [132, p. 14], los esfuerzos deberán concentrarse, al menos en principio, en la mayor de las contribuciones.

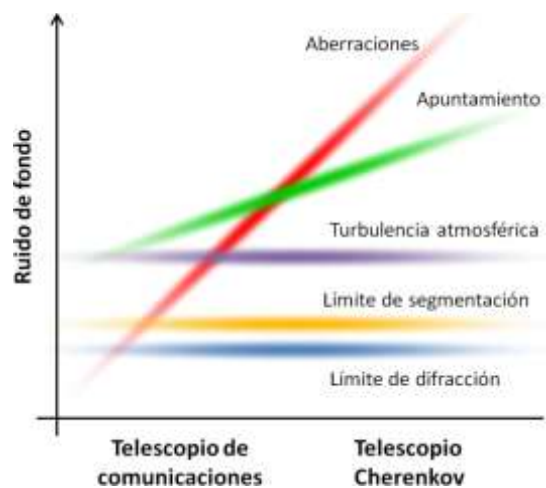

**Figura 39. Limitaciones al campo de visión en IACT y en FSOC.**

Así, la primera limitación que todo telescopio enfrenta es el límite de difracción, que es la causa última de que la PSF no sea un punto en el plano focal. Por encima de esta, se encontrarían los efectos producidos en la PSF por la segmentación del reflector primario del telescopio y, según se utilicen o no sistemas de óptica adaptativa, la limitación debida a la turbulencia atmosférica (tras la corrección del frente de onda se puede llegar a alcanzar el límite de difracción, incluso en grandes telescopios segmentados). Por último, enmascarando a los efectos anteriores, se encontrarían la precisión de apuntamiento y la limitación debida a las diferentes aberraciones ópticas, en ambos casos muy variables según se trate de un telescopio para FSOC o un IACT. En el último caso, como se ha comprobado en la simulación del telescopio MAGIC, las aberraciones inherentes al diseño óptico del telescopio imponen el límite a la mínima resolución óptica alcanzable.

## 2.10.1. Límite de difracción

En una primera aproximación, la luz puede ser considerada como un conjunto de rayos propagándose de forma geométrica. Si bien este análisis es válido en muchos casos, es fundamental delimitar cuándo lo es y cuándo no, ya que frecuentemente esta simplificación no es suficientemente precisa para la descripción de un sistema óptico. En esos casos es necesario acudir a la teoría de la difracción, que es la que describe la luz de una forma más cercana a la realidad, al considerar su naturaleza ondulatoria. Existen diferentes teorías que explican la naturaleza ondulatoria de la luz con aproximaciones de diferente complejidad y aplicabilidad. Sin embargo, la más simple y sorprendentemente exacta en la mayoría de los casos es la de Huygens. Según esta teoría, cada punto de un frente de onda puede ser considerado a su vez una fuente secundaria, resultando en un nuevo frente de onda que es un patrón de difracción a causa de la interferencia de las fuentes secundarias [133, pp. 191-202].

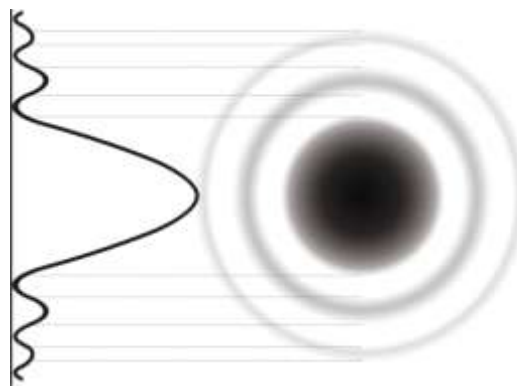

**Figura 40. Patrón de difracción producido
por una apertura circular (disco de Airy).**

En general, los procesos de difracción de la luz se pueden agrupar en dos tipos: difracción de Fresnel y difracción de Fraunhofer. La primera se refiere al campo cercano, es decir, a sistemas ópticos en los que la fuente y la apertura de difracción están a una distancia finita. A longitudes de onda tan reducidas como las del espectro óptico, el caso más común es el de Fraunhofer o de campo lejano y se suele considerar que la fuente de luz ilumina a la apertura con una onda plana. La difracción de Fraunhofer de una apertura circular se utiliza para determinar el límite fundamental de un sistema óptico con simetría circular, como puede ser considerado cualquier telescopio genérico. El patrón de difracción de menor dispersión angular se produce cuando dicha apertura se ilumina homogéneamente con una onda plana y consiste en un conjunto de anillos concéntricos conocido como disco de Airy (Figura 40).

La máxima resolución de un telescopio viene determinada por el límite de difracción, debido a la naturaleza ondulatoria de la luz y al carácter finito de las aperturas que muestra el sistema óptico. Si el diámetro de la apertura es D y λ la longitud de onda, la variación angular de la intensidad de radiación viene dada por la ecuación (2-16) [134, p. 124].

$$\frac{I(\theta)}{I(0)} = \left[ 2 \frac{J_1\left(\frac{\pi D}{\lambda} \operatorname{sen}(\theta)\right)}{\frac{\pi D}{\lambda} \operatorname{sen}(\theta)} \right]^2 \qquad (2\text{-}16)$$

En la ecuación (2-16), $J_1(x)$ es la función de Bessel de primer orden de x. Su primer mínimo corresponde a x = 3,83. Usando este valor, y utilizando la aproximación sen(θ) ≈ θ, se obtiene el límite de difracción del telescopio, que viene dado por la ecuación (2-17)[9].

$$\theta = 1,22 \left(\frac{\lambda}{D}\right) \qquad (2\text{-}17)$$

En la representación gráfica de la ecuación (2-16) que se muestra en la Figura 41 se puede observar cómo la anchura del lóbulo principal de la función de Bessel, donde está contenida la mayor parte de la energía de la señal, aumentará al disminuir la apertura y/o aumentar la longitud de onda. Es importante tener en cuenta que el límite de difracción representa el radio del disco de Airy. El diámetro del mismo, con un factor 2,44 en lugar de 1,22, determinaría la mínima PSF que cualquier sistema óptico puede tener, lo que quiere decir que el sistema está limitado por difracción. La PSF se usa para cuantificar la calidad de un sistema óptico y se define como la respuesta espacial del sistema en el plano focal a una fuente puntual en el infinito, equivalente a un frente de ondas planas.

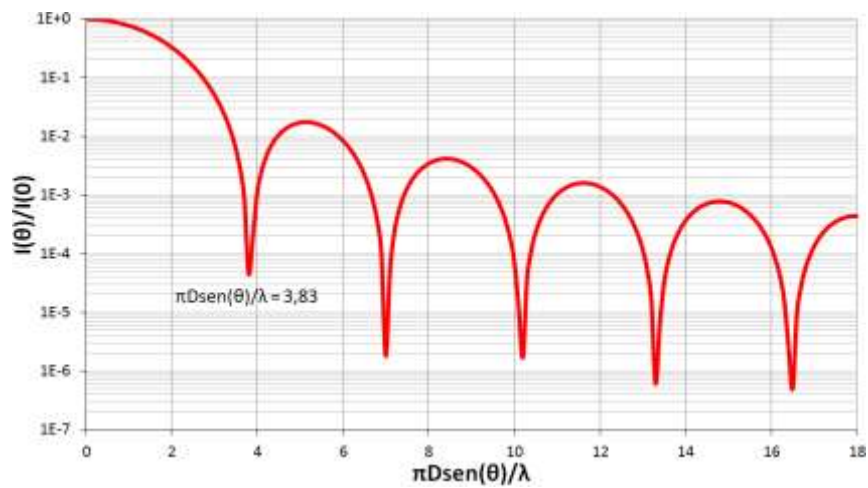

**Figura 41. Representación gráfica de la intensidad de radiación en función de la longitud de onda λ, el diámetro de apertura D y la variación angular θ.**

En la Figura 42 se muestra la dependencia de la PSF limitada por difracción en función de la longitud de onda para los tres tamaños de telescopio de CTA (SST, MST y LST). Se puede comprobar que a mayor apertura, menor es la PSF, que en todos los casos se mantiene por debajo de 1 μrad y que equivale a 5 μm en el plano focal si se asume una relación focal idéntica para todos los telescopios de f/1,3. Este será el límite último de resolución del telescopio, si bien en IACT este límite nunca se alcanza ni aspira a alcanzarse debido a la existencia de otros límites por encima de este, ya que los requisitos de formación de imágenes de la radiación *Cherenkov* son mucho menores.

---

[9] Esta fórmula se corresponde con el criterio del primer mínimo de la función de Bessel, igual a πDsen(θ)/λ = 3,83. Si se usara otro criterio, el factor multiplicador de λ/D en la ecuación (2-17) sería diferente. Por ejemplo [10, p. 666], si en lugar de tomar el punto en el que se encuentra el primer cero, se tomase el punto en el que la potencia cae a la mitad, el factor multiplicador sería 1,03 en lugar de 1,22.

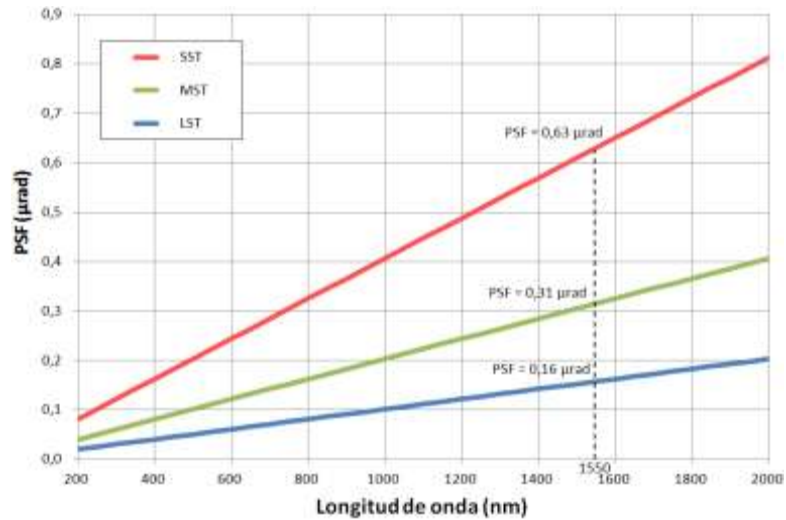

**Figura 42. PSF limitada por difracción para los tres tamaños de telescopio de CTA.**

## 2.10.2. Límite por segmentación

El tamaño y la calidad de la óptica de los telescopios astronómicos apenas mejoraron durante las décadas previas a los años 80. Entonces se produjo uno de los avances más importantes de esta disciplina al surgir la técnica de segmentación de los espejos primarios. Con esta nueva estrategia fue posible construir áreas reflectoras mucho mayores que las conocidas hasta entonces de forma eficiente en coste. Así, se sucedieron una serie de telescopios que fueron marcando la pauta, entre los que destacan el *Hobby-Eberly* (HBT) de 9,2 metros en Texas [135] y su gemelo SALT (*Southern African Large Telescope*) en Sudáfrica [136], los Keck I y II de 10 metros en Hawai [137] y el Gran Telescopio de Canarias (GTC) de 10,4 metros [138], todos ellos operativos actualmente, y el TMT (Thirty Meter Telescope) de 30 metros en Hawai [139] y el EELT (European Extremely Large Telescope) de 39,3 metros en Chile [140], actualmente en fase de diseño. En estos telescopios es muy importante realizar un buen alineamiento y cofaseo de los espejos individuales para obtener una superficie equivalente lo más parecida posible a la ideal sin segmentar. Así, cada espejo tendrá generalmente tres grados de libertad: la traslación longitudinal (pistón) y la rotación alrededor de los dos ejes normales al longitudinal (*tip/tilt*).

La evaluación de la calidad óptica de un telescopio limitado por difracción se suele establecer en función del *seeing* del emplazamiento del observatorio (ya que como se verá este efecto enmascara al límite de difracción). Este parámetro, explicado más profusamente en el apartado 2.10.3, se cuantifica con el parámetro de Fried $r_0$ y evalúa la influencia de la turbulencia atmosférica en el frente de onda. Sin embargo, en telescopios segmentados, por debajo de esta limitación puede encontrarse la del efecto de la propia segmentación del espejo primario. El problema que presentan los telescopios segmentados es que introducen discontinuidades en el frente de onda. Su principal efecto se manifiesta en la forma y la distribución de energía del patrón de difracción producido por la apertura reflectora.

Estrictamente, este efecto debería clasificarse dentro de las limitaciones por aberración, ya que es una aberración del frente de onda. Sin embargo, la clasificación que se hace aquí sobre las limitaciones al campo de visión es más cuantitativa que cualitativa, por lo que su lugar debe situarse entre el límite de difracción y la turbulencia atmosférica. En telescopios monolíticos la separación de estos dos límites se realiza mediante el parámetro de Fried $r_0$, de forma que si el diámetro D del primario es mayor que $r_0$ (como es normalmente el caso en grandes telescopios), se puede decir que el telescopio está limitado por la turbulencia. Sin embargo, en telescopios segmentados cada espejo individual sí puede ser menor que $r_0$, por lo que la influencia de la turbulencia depende del cofaseo entre los segmentos. Así, un telescopio correctamente cofaseado estará limitado por turbulencia, y en caso contrario podría estarlo por el límite de difracción de los espejos individuales, dependiendo de varios factores.

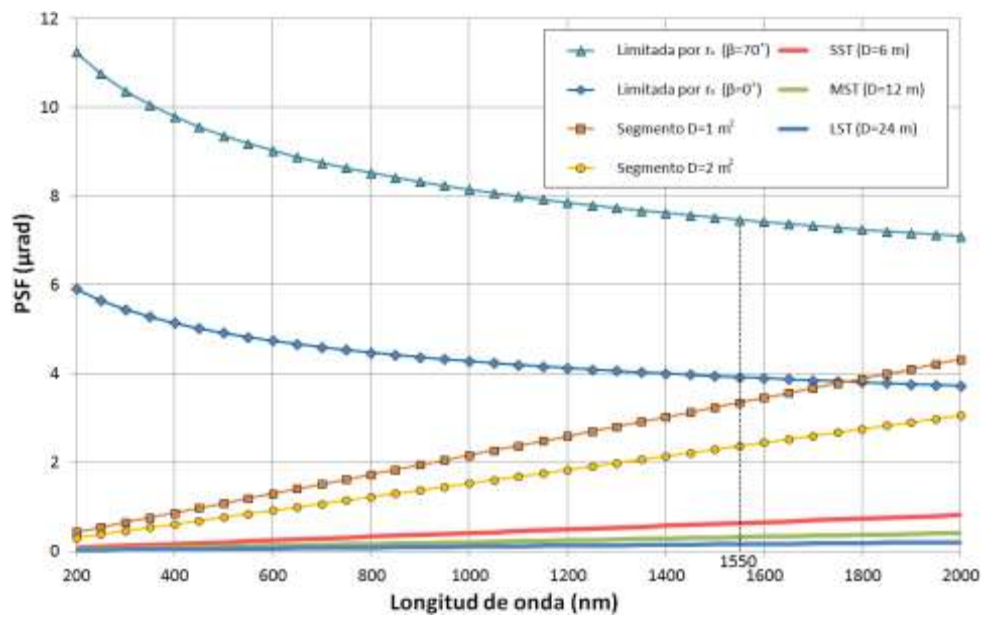

**Figura 43. PSF limitada por turbulencia para dos ángulos cenitales (máximo y mínimo) y limitada por difracción para los tres tamaños de telescopio de CTA y para los tamaños de segmento máximo y mínimo.**

En la Figura 43 se presenta el resultado de la simulación de la PSF limitada por difracción para los tres telescopios de CTA (de 6, 12 y 24 metros de diámetro, tal como aparecía en la Figura 42), de la PSF limitada por difracción para los segmentos individuales de estos telescopios (utilizando un tamaño mínimo de 1 m² y uno máximo de 2 m² [110, p. 2], y considerados circulares en lugar de hexagonales por simplicidad) y la PSF limitada por turbulencia para un ángulo cenital[10] mínimo de 0° y uno máximo de 70° (este es un margen habitual considerado para FSOC en espacio profundo [4, p. 155]). El cálculo de la PSF limitada por turbulencia se ha realizado utilizando la ecuación (2-25), explicada en el siguiente apartado. Se puede comprobar que incluso en un telescopio con un cofaseo incorrecto en el que prevalezca el límite de difracción de sus segmentos individuales, para una longitud de onda de 1550 nm (y también por debajo de esta) el peor caso en el que el segmento tiene el mínimo tamaño contemplado para CTA resulta en un límite enmascarado por la turbulencia (incluso por su caso más favorable en el que la señal de comunicaciones realiza el mínimo recorrido en la atmósfera).

---

[10] Definido el ángulo cenital como el ángulo subtendido por la orientación del telescopio respecto al cénit. Es decir, un ángulo cenital mayor implica que las señales realizan un recorrido más largo a través de la atmósfera, y por tanto se verán afectadas en mayor medida.

Se ha demostrado que en telescopios astronómicos con un correcto cofaseo la contribución a la PSF debida a la segmentación no es mucho mayor que la producida por los brazos de la propia estructura en telescopios de óptica simple [132, p. 16]. En el caso de IACT, como se comprueba en la Figura 43, el efecto de la segmentación es completamente despreciable en presencia de turbulencia atmosférica, hasta el punto que la calidad de imagen admite graves errores en el cofaseo de los segmentos. Por ejemplo, el telescopio MAGIC I presenta un importante y sistemático error de pistón debido a un sobredimensionamiento de los paneles que impidió instalarlos de forma adyacente. En su lugar, se tuvo que forzar un error de pistón (Figura 44) para poder acomodar todos los espejos en el espacio disponible, y aún así la PSF resultante es similar a la de MAGIC II, que carece de este error. Esto puede explicarse por la menor influencia del error de pistón si se realiza un buen cofaseo de *tip/tilt* [132, p. 58] y por la existencia de otros factores más limitantes de la resolución óptica, que enmascaran el efecto de la segmentación.

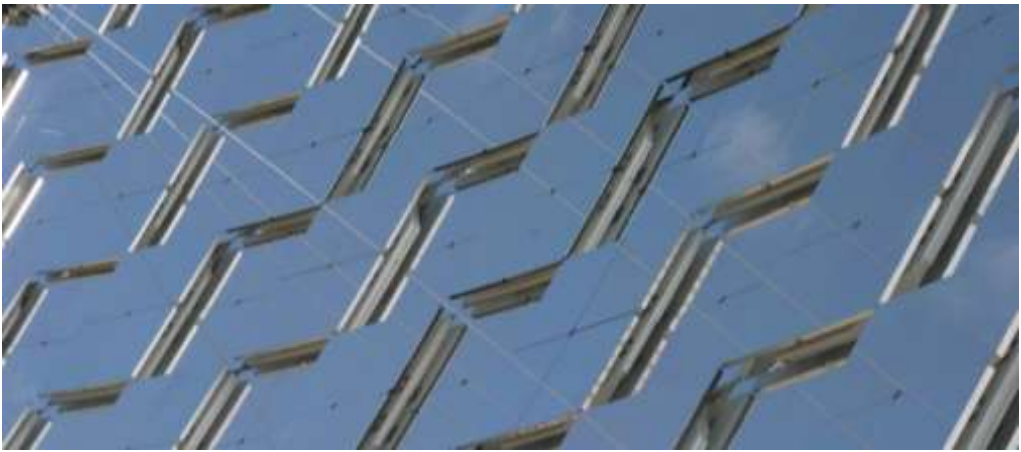

**Figura 44. Detalle del error de pistón en la superficie reflectora del telescopio MAGIC I [126].**

## 2.10.3. Límite por turbulencia

En este apartado se introducen algunos conceptos fundamentales relacionados con la turbulencia atmosférica. Aunque en la primera parte de esta tesis este efecto tiene una importancia menor, se considera oportuno presentar aquí una serie de nociones al ser comunes a ambas partes. La turbulencia tiene distintos efectos en cada aplicación, debido tanto a las diferencias de los haces empleados en cada tipo de enlace como a las diferencias en el propio canal atmosférico. En el apartado 3.6 se tratarán los efectos de la turbulencia relacionados con el canal horizontal usado en la distribución cuántica de claves, y en este apartado se tratarán los efectos relacionados con el canal vertical usado en las comunicaciones en espacio profundo.

El origen de la turbulencia atmosférica se debe a variaciones en la temperatura del aire de la atmósfera. Dichas variaciones pueden tener una multiplicidad de causas, pero el efecto resultante es que se forman masas de aire de distintos tamaños a distintas temperaturas. El aire que se calienta adquiere menor densidad y asciende, y el aire frío, de mayor densidad, cae por efecto de la gravedad, desplazando al caliente. De esta forma, se producen turbulentos movimientos de masas de aire. Al variar la densidad del medio, estos movimientos conllevan a una variación del índice de refracción, creando campos de fluctuaciones aleatorias tanto en tiempo como en espacio. Según la teoría de propagación de ondas electromagnéticas, las señales que atraviesan dicho medio sufren refracciones, que se traducen en un efecto acumulativo a lo largo del trayecto y produce variaciones en la fase y la amplitud de las señales recibidas (Figura 45).

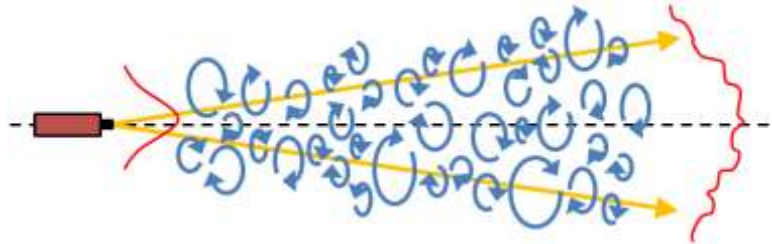

**Figura 45. Deformación de un haz al propagarse por la atmósfera.**

El efecto predominante de la turbulencia atmosférica en enlaces verticales es el aumento de la divergencia del haz, que origina un ensanchamiento del spot recibido en el plano focal del telescopio [4, p. 53]. El efecto es idéntico al sufrido por los telescopios astronómicos [141, pp. 410-415], donde la máxima resolución angular pasa de estar limitada por difracción (o segmentación) a estar limitada por turbulencia, si bien este efecto tiene distintas consecuencias: en astronomía el empeoramiento de la resolución angular impacta en la capacidad para discriminar detalles pequeños en el plano imagen, ya que la distancia mínima resoluble entre dos fuentes puntuales será mayor, al aumentar la PSF relativa a cada una de dichas fuentes; en FSOC el impacto recae sobre el ruido de fondo recibido, ya que es necesario aumentar el campo de visión para que pueda recolectarse toda la luz procedente de la única fuente de interés (el transmisor remoto).

Como se ha visto, el índice de refracción n varía de forma turbulenta en la atmósfera. Esta variación, que depende del medio, fue recogida por A. L. Cauchy en una expresión verificada experimentalmente cuando el medio es el aire, y que E. Lorenz extendió para tener en cuenta los efectos de la humedad, resultando en la fórmula de Cauchy-Lorenz [142, p. 552] de la ecuación (2-18).

$$n - 1 = \left( \frac{77{,}6 \cdot 10^{-6}}{T} \right) \left( 1 + \frac{7{,}52 \cdot 10^{-3}}{\lambda^2} \right) \left( p + 4810 \frac{v}{T} \right) \qquad (2\text{-}18)$$

En la ecuación (2-18), $\lambda$ es la longitud de onda en $\mu$m, p y v son respectivamente la presión del aire seco y del vapor de agua en mb y T es la temperatura en K. En observatorios astronómicos se suele considerar que el índice de refracción solo depende de las variaciones de temperatura, no de la presión. Estas variaciones de temperatura presentan un espectro proporcional a $k^{-5/3}$ (siendo k el número de onda $k = 2\pi/\lambda$) según la teoría clásica de la turbulencia de Kolmogorov [143]. Esta dependencia se expresa mediante una función, llamada "de estructura", que promedia cuadráticamente la fluctuación sobre un intervalo r, como la de la ecuación (2-19), donde T(x) es la temperatura en un punto determinado y T(x+r) la temperatura en otro punto alejado una distancia r.

$$D_T(r) = \left\langle \left( T(x+r) - T(x) \right)^2 \right\rangle \qquad (2\text{-}19)$$

Si r representa una separación pequeña (menor a la denominada escala interior $l_0$, correspondiente al tamaño de los remolinos más pequeños) respecto a x, la función de estructura toma la forma de la ecuación (2-20) [144, p. 28], donde $C_T^2$ es el denominado parámetro de estructura de la temperatura, que tiene su homólogo para el índice de refracción en la ecuación (2-21) [145, p. 392].

$$D_T(r) = C_T^2 r^{2/3} \tag{2-20}$$

$$D_n(r) = C_n^2 r^{2/3} \tag{2-21}$$

El parámetro de estructura del índice de refracción $C_n^2$ de la ecuación (2-21) se definiría según la ecuación (2-22) [146, p. 65].

$$C_n^2 = C_T^2 \left[ \left( \frac{77,6 \cdot 10^{-6}}{T} \right) \left( 1 + \frac{7,52 \cdot 10^{-3}}{\lambda^2} \right) \left( \frac{P}{T} \right) \right]^2 \approx C_T^2 \left[ 79 \cdot 10^{-6} \frac{P}{T^2} \right]^2 \tag{2-22}$$

El parámetro de estructura del índice de refracción describe la intensidad de las fluctuaciones del índice de refracción y por ello se utiliza para diferenciar entre distintos regímenes turbulentos: régimen débil ($C_n^2 \leq 10^{-17}$ m$^{-2/3}$), moderado ($10^{-17}$ m$^{-2/3} \leq C_n^2 \leq 10^{-13}$ m$^{-2/3}$) y fuerte ($C_n^2 \geq 10^{-13}$ m$^{-2/3}$) [146, p. 65]. Si bien este parámetro se suele considerar constante para caracterizar la atmósfera en un instante determinado, en realidad cambia continuamente con las condiciones atmosféricas y además tiene una dependencia con la altura, como se verá a continuación.

El parámetro $C_n^2$ representa una medida local de la amplitud de las inhomogeneidades del índice de refracción, o lo que es lo mismo, de la intensidad de la turbulencia, que va variando continuamente. Para conocer exactamente su valor en un instante determinado sería necesario integrarlo a lo largo del camino de propagación. Normalmente se utilizan modelos empíricos, como el modelo clásico de Hufnagel [147] de la ecuación (2-23), donde A es una constante arbitraria para escalar la dependencia con el régimen de turbulencia en superficie, $V/V_m$ es la relación de la velocidad del viento de capas altas y su media y z la altura en km.

$$C_n^2(z) = A \left[ 2,2 \cdot 10^{-23} z^{-10} e^{-z} \left( \frac{V}{V_m} \right) + 10^{-16} e^{-z/1,5} \right] \tag{2-23}$$

En la Figura 46 se muestra la dependencia de $C_n^2$ con la altura, escalada para tres regímenes de turbulencia en superficie alrededor del valor considerado como típico de régimen moderado ($C_n^2(0) = 10^{-15}$ m$^{-2/3}$ en rojo, y alrededor $C_n^2(0) = 10^{-14}$ m$^{-2/3}$ en verde y $C_n^2(0) = 10^{-16}$ m$^{-2/3}$ en azul). También se muestran distintas velocidades típicas de vientos de capas altas. Al aumentar la altura, la temperatura desciende y con ella la densidad del aire. Con menor densidad y mayor uniformidad, hay menos movimientos de masas de aire y menos turbulencias, aunque generalmente hay que ascender hasta varios kilómetros, donde se sitúan los observatorios astronómicos, para lograr un régimen de turbulencia débil [148].

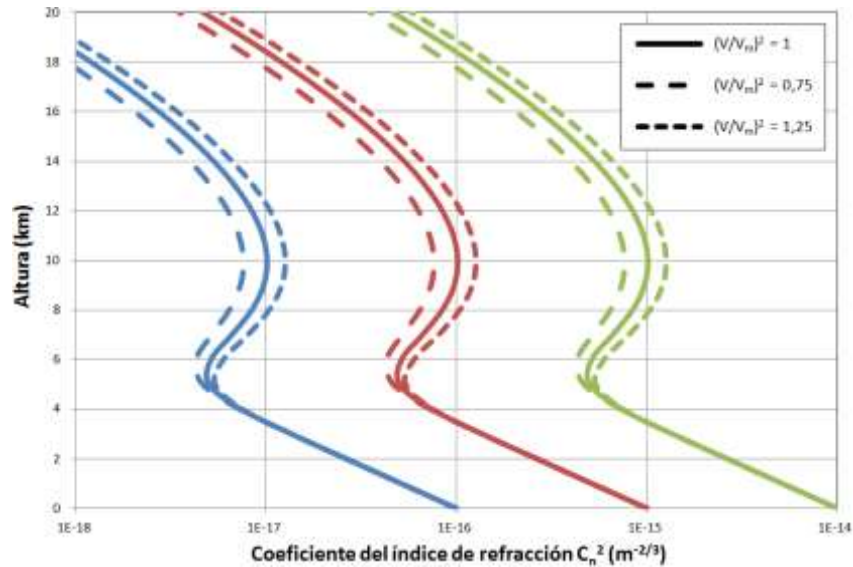

**Figura 46. Perfil del parámetro de estructura del índice de refracción según la altura para distintos regímenes de turbulencia en superficie (distintos colores) y distintas velocidades de vientos de capas altas.**

Además de con la altura, $C_n^2$ cambia con las estaciones; cambia en periodos de meses, días horas y minutos; y también cambia con el emplazamiento. Pese a que su valor puede variar hasta en más de un orden de magnitud en unos pocos metros, se suele utilizar un valor promedio. Para estimarlo, lo habitual es llevar a cabo mediciones sobre un haz que se propaga a través de la atmósfera en un momento y un lugar determinado. Para ello, se pueden utilizar las variaciones en el ángulo de llegada del haz o las fluctuaciones de la intensidad de la onda electromagnética recibida [149, p. 12].

El parámetro más útil para evaluar el efecto de la turbulencia en enlaces de espacio profundo es el parámetro de Fried o longitud de coherencia $r_0$, también conocido en astronomía como *seeing*. Se define como el diámetro expresado en cm de un área circular sobre la cual el error de frente de onda debido a la turbulencia deja de ser apreciable (el criterio puede variar, pero se suele considerar que esto sucede cuando el error cuadrático medio del frente de onda es de 1 radián). Este criterio se utiliza para determinar que los telescopios con aperturas menores a $r_0$ pueden llegar a operar cercanos al límite de difracción incluso en presencia de turbulencia atmosférica (aunque el efecto del *beam wander*, analizado en el apartado 3.6.2, persistiría). Por el contrario, los que cuenten con aperturas mayores se verán siempre limitados por la turbulencia, lejos del límite de difracción que de otro modo proporcionarían sus grandes aperturas.

Si bien el parámetro de Fried es comúnmente utilizado en la evaluación de la turbulencia para observación astronómica, fue desarrollado por David Fried en un estudio de un receptor heterodino de comunicaciones [150]. Encontró que el máximo diámetro que podía tener una apertura antes de que su rendimiento pudiera verse seriamente limitado debido a la turbulencia venía dado por la expresión (2-24) [151, pp. 29-31], donde k es el número de onda, β es el ángulo cenital, L la distancia total de propagación y $C_n^2$ es el parámetro de estructura del índice de refracción. Se puede comprobar que regímenes de turbulencia fuerte supondrán una disminución del parámetro de Fried, con lo que el área de mínimo error de frente de onda disminuirá y la turbulencia afectará también a telescopios de aperturas menores.

$$r_0 = \left( \frac{0{,}423k^2}{\cos\beta} \int_0^L C_n^2(z)\,dz \right)^{-3/5} \qquad (2\text{-}24)$$

Para cuantificar el parámetro de Fried es necesario integrar la expresión de $C_n^2$ dependiente de la altura de la ecuación (2-23) según la expresión de la ecuación (2-24). Este cálculo es complejo e inexacto, por lo que normalmente se recurre a una simplificación. Si se considera la luz recibida como una onda plana (suposición válida para fuentes muy lejanas, tanto en astronomía como en comunicaciones en espacio profundo), la expresión (2-24) se puede simplificar [152] en la expresión (2-25), de la mediana de $r_0$, que asume condiciones de *seeing* típicas de observatorios astronómicos basadas en medidas experimentales.

$$[r_0]_{mediana} = 0{,}114\left(\cos\beta\right)^{3/5}\left(\frac{\lambda}{5{,}5\cdot10^{-7}}\right)^{6/5} \qquad (2\text{-}25)$$

Así, para $\lambda = 1{,}55$ μm, el parámetro de Fried que cabría encontrar mirando al cénit en un emplazamiento similar al de CTA sería de unos 40 cm, e iría reduciendo su valor con el aumento del ángulo cenital. La validez de esta estimación se ha podido verificar comparando el resultado teórico con las medidas experimentales realizadas en unas circunstancias lo más parecidas posible a las de CTA. Para ello, se ha elegido una campaña de medidas [153] que se llevó a cabo durante el día en el observatorio solar de Sacramento Peak en Nuevo México. El resultado estimado con la ecuación (2-25) para esta campaña resulta en una mediana de $r_0$ de 10,41 cm en el cénit y las medidas experimentales proporcionaron una mediana de 8,7 cm, que coincide con el resultado teórico utilizando la ecuación (2-25) relativo a un ángulo cenital de unos 40°. Resulta evidente pues que para telescopios del tamaño de cualquier IACT, la turbulencia supondrá siempre un límite mucho mayor que la difracción. Cuando esto sucede, la PSF del telescopio ya no será $2{,}44\,\lambda/D$, como se vio en el apartado 2.10.1, sino $0{,}98\,\lambda/r_0$ [154][11]. En la Figura 47 se muestra la dependencia del parámetro de Fried con el ángulo cenital para una longitud de onda de 1550 nm, y la PSF limitada por turbulencia que resulta del mismo, así como la PSF limitada por difracción para los distintos tipos de telescopios de CTA. Se comprueba cómo esta última queda en todos los casos muy por debajo de la PSF limitada por turbulencia.

Como se verá, la limitación por turbulencia no afecta a la resolución óptica en astronomía *Cherenkov*, quedando esta resolución limitada por otros efectos de mayor relevancia que cualquier efecto atmosférico. Sin embargo, en comunicaciones ópticas en espacio profundo sí es posible realizar realizar un diseño tal que la limitación última acabe siendo la turbulencia. Es decir, los límites estudiados en los apartados 2.10.4 y 2.10.5 se podrían evitar partiendo de un diseño inicial diferente, y la máxima resolución, que determina el mínimo campo de visión utilizable, vendría impuesta por la turbulencia atmosférica [155]. Para esta situación se sugiere el empleo de sistemas de óptica adaptativa con el objetivo de minimizar el efecto de la turbulencia atmosférica, reduciendo así la PSF del receptor por debajo de las limitaciones impuestas por ella [156].

---

[11] Hay que señalar que esta comparación no es exacta sino aproximada, ya que se basa en dos criterios diferentes: el límite de $2{,}44/D$ se refiere a la anchura total entre los primeros nulos y el límite de $0{,}98/r_0$ se refiere a la anchura total entre los puntos en que la intensidad cae a la mitad. La comparación correcta sería con $2{,}06/D$ (ver nota 9 del apartado 2.10.1), que es el límite de difracción considerado con el mismo criterio que en $r_0$

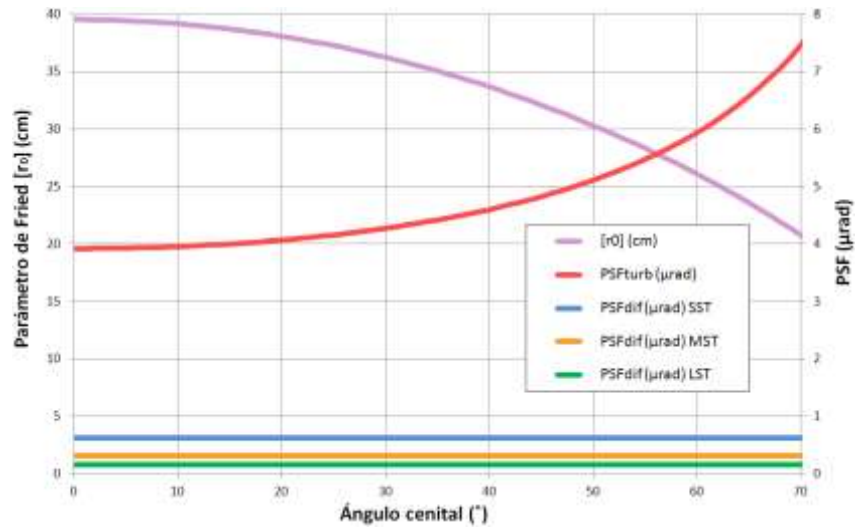

**Figura 47. Representación del parámetro de Fried $r_0$ y de la PSF limitada por turbulencia y por difracción para los tres tipos de telescopios de CTA en función del ángulo cenital para 1550 nm.**

## 2.10.4. Límite por apuntamiento

En general, el objetivo del sistema de apuntamiento de un telescopio es dirigir el eje óptico hacia la dirección deseada de observación y seguir al objetivo durante la duración de la misma (esta última tarea puede diferenciarse con la denominación de seguimiento, pero aquí se la incluirá como una parte del apuntamiento). Para llevarlo a cabo, los primeros telescopios astronómicos se basaban en una única rotación ecuatorial en bucle abierto alrededor de la estrella polar a un ritmo constante de 15 segundos de arco por segundo de tiempo (la velocidad de rotación de la Tierra). Actualmente, las mayores precisiones requeridas exigen un bucle cerrado de retroalimentación donde se controla constantemente la posición real de cada motor encargado del apuntamiento y la posición deseada para reducir el error y maximizar la precisión. La limitación impuesta por el apuntamiento de la estructura del telescopio no es una limitación de la respuesta óptica del telescopio (la resolución óptica será idéntica independientemente de la precisión de apuntamiento). Sin embargo, sí es correcto hablar del apuntamiento como una limitación al campo de visión mínimo que se puede utilizar en la práctica. Una precisión de apuntamiento peor que el campo de visión hará que el objeto a observar (el terminal transmisor en el caso de un enlace de comunicaciones) desaparezca de la línea de visión.

El apuntamiento de un telescopio depende en último término de la estructura que soporta la óptica del mismo. Aunque los primeros IACT se implementaron con monturas ecuatoriales, su mayor complejidad dio paso a las monturas altacimutales que son la base de todos los IACT de la actualidad. Para construir este tipo de monturas básicamente hay dos diseños posibles, aunque existen otros como propuestas teóricas [39, p. 66]. El primero, utilizado en HESS y MAGIC (Figura 48, izquierda), está basado en un carril circular para el movimiento acimutal soportando el reflector principal mediante dos torres que facilitan el movimiento en altitud sin necesidad de emplear contrapesos para equilibrar la estructura. Tiene la ventaja de permitir amplios movimientos a costa de ocupar una mayor extensión y dejar a los mecanismos más expuestos. El segundo, utilizado en VERITAS (Figura 48, derecha), está basado en un posicionador central, cerca del cual se sitúa el reflector principal, que necesita de un contrapeso trasero para equilibrar el conjunto. En este caso los mecanismos están contenidos dentro del posicionador, protegidos del exterior. La segunda opción es ampliamente utilizada en antenas, radiotelescopios y concentradores solares con cortas relaciones focales. Sin embargo, en telescopios de mayores relaciones focales y pesadas cámaras en el foco como los IACT, se precisa de grandes contrapesos que limitan su aplicabilidad a telescopios pequeños y medianos. Por ello, para los telescopios LST de CTA solo es aplicable el diseño circular, siendo posibles ambas estrategias para MST y SST.

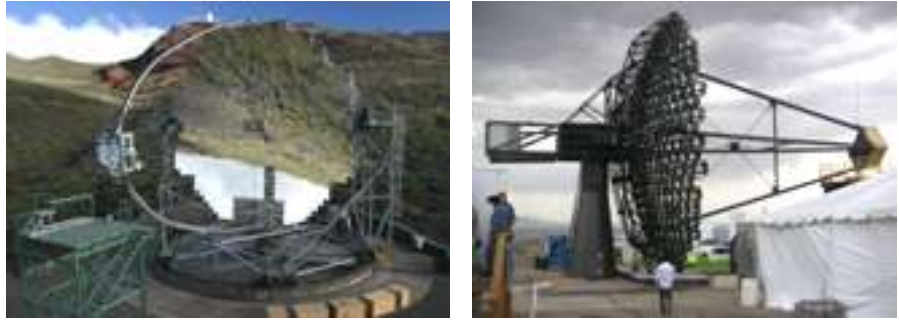

**Figura 48. Telescopio MAGIC (izquierda) y VERITAS (derecha).**

Como se ha indicado, el sistema de apuntamiento de los telescopios actuales se basa en un bucle cerrado de retroalimentación (Figura 49). El objetivo es minimizar el error entre la posición u orientación deseada para el apuntamiento del telescopio y la medida a través de los sensores de posición. Los elementos principales de este sistema son los sensores de posición, el controlador PID y el sistema actuador (encargado de convertir la señal digital de control en una señal analógica que alimenta los motores, consiguiendo el movimiento deseado en la estructura del telescopio). El objetivo del controlador es conseguir la señal de actuación necesaria para llevar el telescopio hacia la posición deseada, basándose para ello en la información medida de la posición real del telescopio. Se diseña de forma que consiga este objetivo de la forma más rápida y estable posible y por ello el sistema de control está basado en un bucle PID (Proporcional, Integral y Derivativo). La forma más simple de conseguir reducir el error entre la señal medida y la deseada es generar una señal de control proporcional a este error. El problema de basar el control solo en un factor proporcional es que llegado a un punto cercano al objetivo el error residual no se elimina por completo, permaneciendo un error residual estacionario. Para ello, se incluye un factor integral que hace que la señal de control sea proporcional al tiempo durante el cual existe un error. Esto logra eliminar el error residual, pero puede dar lugar a inestabilidades que se solucionan reduciendo su peso en relación al factor proporcional. Así se consigue suavizar la corrección en el tiempo, eliminándose el error de baja frecuencia del factor proporcional. Por último, al reducir el peso del factor integral, los errores de alta frecuencia no se corrigen y para solucionarlo se suele incluir un factor derivativo. Este último factor elimina los cambios rápidos de la señal de error, con lo que ajustando los tres factores se consigue una minimización del error de apuntamiento de forma muy rápida y precisa.

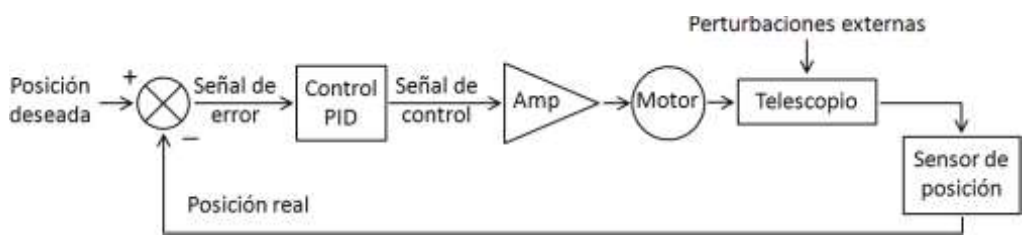

**Figura 49. Diagrama de bloques del sistema de control de apuntamiento de un telescopio.**

Conviene distinguir la diferencia entre la precisión de apuntamiento y la resolución de apuntamiento (Figura 50). La primera es la diferencia entre la dirección de apuntamiento deseada para orientar el telescopio hacia un determinado objetivo y la dirección de apuntamiento real hacia la que está orientado el telescopio. La segunda define el mínimo movimiento del telescopio en los ejes que definen su apuntamiento (generalmente altitud y acimut). En la práctica, lo normal es que ambas tiendan a converger o bien se centren los esfuerzos en el factor más limitante. Por ello, es muy común hablar indistintamente de precisión o resolución para referirse a la capacidad de un telescopio de apuntar en una dirección deseada, refiriéndose en este caso al parámetro que actúe como cuello de botella. Atendiendo a la propia definición de la precisión de apuntamiento, se puede deducir que el elemento del sistema de apuntamiento encargado de garantizarla es el sistema de control PID. Al tratarse de una técnica muy conocida sin mayores exigencias, este parámetro no es una limitación en el apuntamiento de un IACT.

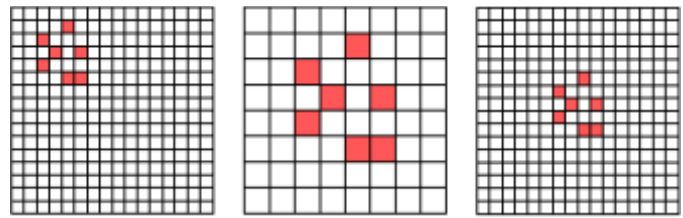

**Figura 50. Ilustración del apuntamiento de un sistema con buena resolución y mala precisión (izquierda), mala resolución y buena precisión (centro) y buena resolución y buena precisión (derecha).**

En cuanto a la resolución de apuntamiento, son dos los elementos sobre los que se apoya su implementación: los propios motores de la estructura y los sensores de posición. Los primeros no imponen en principio un factor limitante en la resolución. En telescopios SST y MST basados en posicionador central se contempla emplear directamente sistemas comerciales o bien desarrollos basados en sistemas comerciales [39, p. 67]. Para los telescopios MST y LST basados en carril circular, probablemente se implementarán motores con transmisión de piñón y cremallera (MAGIC), motores de fricción (HESS I) o bien motores de acción directa (HESS II). Las dos últimas son las tecnologías empleadas en los grandes telescopios astronómicos convencionales y ofrecen unas buenas prestaciones, especialmente la última de ellas, por carecer de sistemas mecánicos de transmisión. Incluso la tecnología de MAGIC, actualmente en desuso en grandes telescopios astronómicos, es capaz de proporcionar resoluciones cercanas al µrad [157, p. 4]. Todas ellas son tecnologías muy probadas, por lo que se suelen utilizar directamente sistemas comerciales o basados en ellos para adaptarlos a cada telescopio particular [33, p. 275].

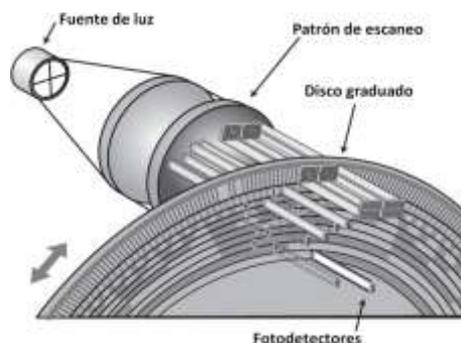

**Figura 51. Codificador de posición [33, p. 281].**

El otro elemento fundamental en el sistema de apuntamiento lo constituye el sensor de posición. La resolución de apuntamiento del telescopio estará limitada en último término por estos dispositivos. Integrados en los motores que controlan el movimiento del telescopio, estos sensores se denominan codificadores de posición y su objetivo es convertir la posicion angular del eje del motor en un código digital único para determinar su posición de una forma muy exacta. Consisten en una serie de marcas de referencia alrededor del eje a medir y una cabeza lectora óptica para registrarlas (Figura 51). Como la resolución de apuntamiento de los IACT no presenta ningún requisito especial en comparación con los telescopios convencionales, se suelen utilizar sistemas comerciales [157, p. 95]. La resolución de estos sensores determina el número máximo de posiciones que pueden codificarse. Así, es posible comparar varios telescopios (Figura 52) asumiendo una resolución nominal de apuntamiento calculada como $360°/2^N$, siendo N la resolución del codificador en bits.

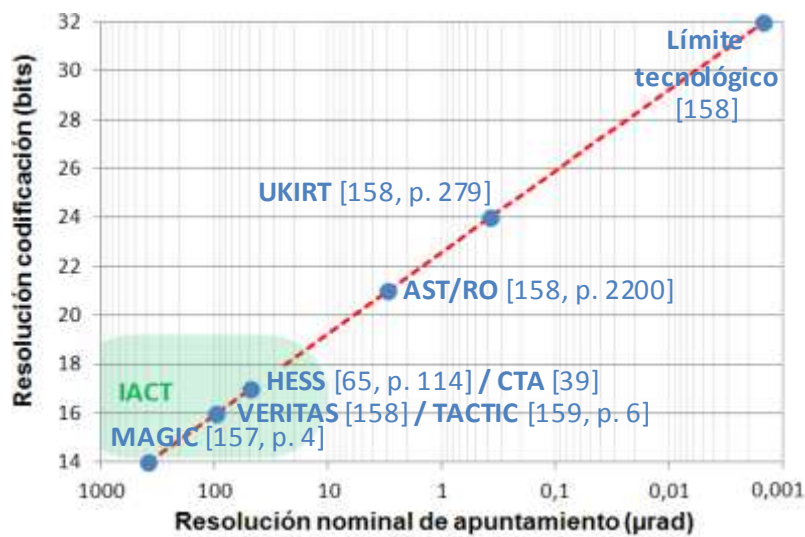

**Figura 52. Resolución de codificación y resolución de apuntamiento para varios telescopios *Cherenkov* y telescopios convencionales.**

Resulta evidente solo atendiendo a la resolución nominal, cómo el apuntamiento en IACT es muy inferior al de los telescopios astronómicos convencionales. Esto se explica, como se verá más adelante, porque su peor calidad óptica hace inútil una mejor capacidad de apuntamiento, que tiende a relajarse hasta cumplir con los requisitos mínimos de la astronomía de rayos gamma. Esta resolución determina la capacidad de apuntamiento absoluto del telescopio, conocido como apuntamiento ciego. En telescopios astronómicos convencionales se sitúa alrededor del µrad y se considera suficiente para buena parte de las observaciones, aunque es posible mejorarla, como también se hace en IACT partiendo de una resolución nominal menor, en más de un orden de magnitud, recurriendo a diferentes técnicas como la monitorización de estrellas conocidas de referencias para rectificar la orientación real del telescopio [33, pp. 35-36] o la aplicación de correcciones basadas en el conocimiento experimental que se tiene de la deformación de la estructura según las diferentes orientaciones del telescopio [39, p. 68].

La tecnología tanto de motores como de codificadores de posición es suficientemente madura como para poder utilizar directamente la de los propios IACT, o bien realizar alguna adaptación basada en sistemas comerciales para mejorar la capacidad de apuntamiento en caso de ser necesario. Esta tesis se limita a constatar que existe margen de mejora, si bien no analiza qué adaptaciones serían necesarias en su caso. El campo de visión debería ser al menos igual que la resolución de apuntamiento para asegurar que el objetivo siempre permanece visible por el telescopio. Sin embargo, normalmente se trabaja utilizando un margen de seguridad para contemplar escenarios más realistas. De hecho, frecuentemente el diseño del mínimo campo de visión utilizable viene limitado por el apuntamiento más que por los otros factores vistos en los anteriores apartados. En la Tabla 3 se muestra que tanto para telescopios de rayos gamma como de comunicaciones se suele establecer una relación similar entre el campo de visión y la resolución de apuntamiento. Sin embargo, se puede comprobar cómo el campo de visión en telescopios FSOC es mucho más estrecho. Esto se explica en parte por los mejores sistemas de apuntamiento de los telescopios convencionales utilizados hasta ahora para soportar enlaces de comunicaciones en comparación con los IACT. Estos sistemas facilitan realizar seguimientos de campos de visión más estrechos, aunque para ello la resolución óptica debe ser suficiente como para permitir reducir tanto el campo de visión, lo que sucede en los telescopios usados para FSOC, pero no en los IACT sin ninguna adaptación.

Tabla 3. Margen de seguridad en función del apuntamiento y del campo de visión de varios telescopios.

| Telescopio | Tipo de telescopio | Resolución de apuntamiento | Campo de visión | Margen de seguridad |
|------------|-------------------|---------------------------|-----------------|---------------------|
| MAGIC | IACT | 34,9 µrad [157, p. 5] | 174,5 µrad [126] | 5 |
| VERITAS | IACT | 17,5 µrad [158, p. 34] | 192 µrad [158, p. 34] | 11 |
| CTA-LST | IACT | 5,2 µrad [39, p. 68] | 122,2-209,4 µrad [159] | 23,3-40 |
| OCTL | FSOC | 1,4 µrad [160, p. 12] | 48,9 µrad [160, p. 3] | 33,7 |
| OLSG | FSOC | 0,5 µrad [72, p. 51] | 2,4 µrad [72, p. 51] | 5,1 |
| 10mOGS | FSOC | 9,9 µrad [161, p. 2] | 9,9 µrad [161, p. 2] | 1 |

En cuanto a la capacidad de seguimiento, los requisitos de la estructura de los telescopios están relacionados con la velocidad a la que se puede seguir a un objetivo móvil, que en el caso de un telescopio convencional puede ser cualquier objeto celeste, y en el caso de un telescopio de comunicaciones es el terminal remoto. Aunque en general los telescopios FSOC basados en IACT tendrían como principal aplicación enlaces de muy larga distancia (y por lo tanto muy lentos movimientos), una estación de comunicaciones capaz de dar soporte a distintos tipos de terminales remotos proporcionaría un valor añadido a la red en la que trabajara. Por ello, este tipo de terminales podría ser desde satélites LEO de órbita baja de muy rápidos desplazamientos hasta sondas en espacio profundo moviéndose lentamente. Afortunadamente, uno de los principales objetivos de CTA es el estudio de estallidos de rayos gamma [162, pp. 211-212] y para poder detectarlos los IACT deben ser reorientados muy rápidamente tras producirse una señal de advertencia desde un satélite. MAGIC puede orientarse hacia cualquier punto del cielo en un máximo de 30 segundos [163, p. 359], HESS lo hace a una velocidad de 100º por minuto [65, p. 114] y los telescopios de CTA se moverán hasta 180º en 20 segundos [162, p. 66], que es una velocidad de 9º/s, o según cifras más conservadoras 30º en 90 segundos [164, p.74], equivalente a 0,3º/s.

$$T = 2\pi\sqrt{\frac{r^3}{G \cdot M}} = 2\pi\sqrt{\frac{(R_{LEO} + R_T)^3}{G \cdot M}} \qquad (2\text{-}26)$$

Considerando 160 km como la mínima altura $R_{LEO}$ de una órbita LEO, según la ecuación (2-26) [165], donde $R_T$ es el radio de la Tierra, G es la constante gravitacional y M la masa terrestre, se puede obtener un periodo orbital T de 5261,29 s, que determinaría la máxima velocidad de rotación de un satélite de comunicaciones. Este periodo resulta en una velocidad angular de 0,068°/s, que es más de cuatro veces más lenta que la peor capacidad de seguimiento prevista para CTA, por lo que se puede deducir que esta funcionalidad estará directamente disponible sin necesidad de adaptación. Esto demuestra que la capacidad de los motores de los telescopios de CTA es suficiente para realizar el seguimiento necesario en FSOC. De forma similar al apuntamiento, en el seguimiento además de la capacidad de los motores, es necesario contar con una resolución suficiente de lectura de la posición. Si en el caso del apuntamiento se realizaba mediante codificadores de posición, en el caso del seguimiento se utilizan tacómetros para medir la velocidad de rotación, que a su vez se basan en las lecturas de los codificadores de posición. Esta tecnología es suficientemente madura como para detectar velocidades tan lentas como 1 giro/día [33, p. 282]. En general se asume que estos sistemas no impondrían mayores limitaciones ya que la estabilidad del seguimiento se suele considerar entre uno y dos órdenes de magnitud mejor que la capacidad de apuntamiento absoluta [33, p. 253].

## 2.10.5. Límite por aberraciones

Si bien la función principal de un telescopio es recolectar la luz de un objeto y focalizarla en un punto del plano imagen, en la realidad no es posible cumplir con este objetivo debido a una serie de errores en la forma en que el sistema óptico focaliza la luz. En todo sistema de formación de imágenes se habla de dos limitaciones fundamentales que convierten la PSF en una mancha en lugar de un punto: la difracción, estudiada en el apartado 2.10.1, y las aberraciones geométricas. Si bien la difracción es un efecto inevitable causado por la naturaleza de la formación de imágenes, la aberración es la consecuencia del diseño del sistema óptico y se puede minimizar recurriendo a una variedad de estrategias. Estas estrategias dependerán de la aplicación, pero en general se puede decir que con un correcto diseño es posible conseguir que el efecto de las aberraciones sea tan pequeño que quede por debajo del límite de difracción.

En 1856 Philip Ludwig von Seidel publicó un trabajo [166] en el que recopiló por primera vez con un tratamiento matemático las cinco aberraciones ópticas primarias o de tercer orden. Por ello, estas aberraciones son comúnmente conocidas por aberraciones de Seidel. Estas aberraciones son aplicables a los sistemas ópticos que operan con luz monocromática, como lo son los de comunicaciones ópticas. Además, los telescopios empleados en comunicaciones utilizan óptica reflexiva y las leyes de reflexión, a diferencia de las de refracción, son independientes de la longitud de onda. Por estos motivos la aberración cromática es obviada aquí.

La teoría de óptica geométrica de primer orden (óptica Gaussiana) se basa en la suposición de que el sistema óptico se limita a operar en una región muy próxima al eje óptico, lo que es conocido por aproximación paraxial. Matemáticamente, esta suposición proporciona las equivalencias $\text{sen}(x) \approx x$ y $\cos(x) \approx 1$, lo que supone una gran simplificación del cálculo matemático. En óptica de primer orden los órdenes mayores a uno en los desarrollos en series de Taylor de las ecuaciones (2-27) y (2-28) se desprecian, asumiendo que los elementos ópticos son perfectos. Por ello, la óptica Gaussiana carece del efecto de la aberración geométrica y solo debe usarse como una aproximación a la realidad que puede distar mucho de esta cuando los ángulos no son muy reducidos.

$$\text{sen}(x) \approx x - \frac{x^3}{3!} + \frac{x^5}{5!} - \frac{x^7}{7!} + ... \tag{2-27}$$

$$\cos(x) \approx 1 - \frac{x^2}{2!} + \frac{x^4}{4!} - \frac{x^6}{6!} + ... \tag{2-28}$$

Si se consideran el segundo y tercer orden [167], aparecen una serie de efectos más próximos al comportamiento real que dan lugar a las cinco aberraciones de tercer orden o de Seidel (si se obvia la aberración cromática, también de tercer orden). Dichas aberraciones son las siguientes (Figura 53): esférica, coma, astigmatismo, curvatura de campo y distorsión.

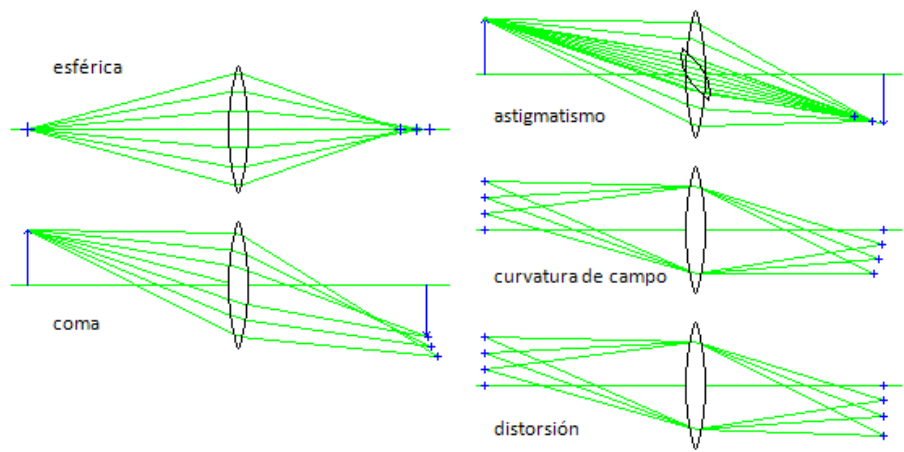

**Figura 53. Aberraciones ópticas primarias, de Seidel o de tercer orden.**

Existen otras aberraciones de orden superior: las 9 de quinto orden o de *Schwarzschild*, las 14 de séptimo orden, etc. sin embargo, en este estudio resultarán despreciables dada la magnitud de la resolución requerida y la prevalencia de las aberraciones presentes de orden inferior. Todas las aberraciones se pueden corregir hasta cierto punto y el objetivo de cualquier diseño óptico será conseguir un compromiso entre esta desviación de su comportamiento ideal y los propósitos de la aplicación, en términos de resolución óptica.

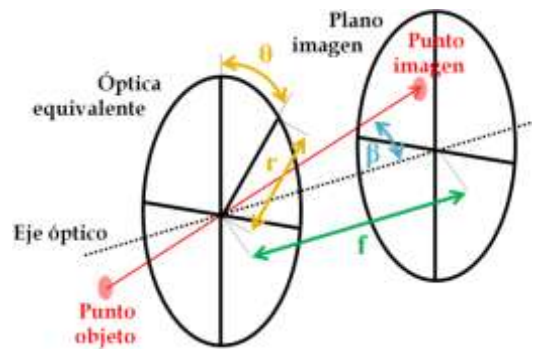

**Figura 54. Correspondencia entre plano imagen y objeto a través de un sistema óptico simétrico.**

Para sistemas ópticos simétricos respecto al eje óptico, las aberraciones pueden describirse a través de la función $W(r, \theta, h')$ de la ecuación (2-29) [40, p. xxii], que depende de las coordenadas polares $r$ y $\theta$ donde el reflector es iluminado y de la distancia $h' = f\beta$ de la luz reflejada en el plano focal, donde $f$ es la longitud focal y $\beta$ el ángulo de desviación respecto al eje óptico con que incide la luz en el reflector (Figura 54).

$$W(r,\theta,h') = \overbrace{a_e r^4}^{\text{esférica}} + \overbrace{a_c h' r^3 \cos\theta}^{\text{coma}} + \overbrace{a_a h'^2 r^2 \cos^2\theta}^{\text{astigmatismo}} + \overbrace{a_c h'^2 r^2}^{\text{c. de campo}} + \overbrace{a_d h'^3 r \cos\theta}^{\text{distorsión}} \qquad (2\text{-}29)$$

siendo los parámetros de cada aberración los definidos en la ecuación (2-30), para un sistema óptico con la apertura coincidente con un reflector cóncavo de radio de curvatura $R$, del tipo de cualquier telescopio convencional definido por su longitud focal equivalente $f$ [168].

$$a_e = \frac{1}{4R}\left(\frac{1}{R} + \frac{1}{f}\right)^2 ; \quad a_c = \frac{R+f}{R^2 f^2} ; \quad a_a = \frac{1}{Rf^2} ; \quad a_c = a_d = 0 \qquad (2\text{-}30)$$

La aberración esférica se puede minimizar aumentando la focal. Depende de la distancia radial según $r^4$, de forma que los rayos marginales sufren de mayor aberración esférica, focalizándose ante el plano focal, a diferencia de los paraxiales (cercanos al eje) que se focalizan tras el plano focal. Esta dependencia, en distintos grados, se repite en todas las aberraciones de Seidel, pero las otras cuatro son genéricamente aberraciones fuera de eje. Esto quiere decir que también dependen del ángulo de incidencia $\beta$ (que viene determinado por el campo de visión del telescopio), por lo que si este es cero (luz alineada con el eje óptico), las cuatro se anulan. Sin embargo, la esférica puede estar presente incluso en sistemas perfectamente alineados.

$$W(R,\theta,\beta,f) = \overbrace{\left[\frac{9}{32f^3}\right]}^{\text{esférica}} r^4 + \overbrace{\left[\frac{3}{4f^2}\right]}^{\text{coma}} \beta r^3 \cos\theta + \overbrace{\left[\frac{1}{2f}\right]}^{\text{astigmatismo}} \beta^2 r^2 \cos^2\theta \qquad (2\text{-}31)$$

Usando para telescopios reflectores la equivalencia $R = 2f$, y combinando y simplificando las ecuaciones (2-29) y (2-30) en la ecuación (2-31), se comprueba que todas las aberraciones aumentan con el tamaño de la apertura al aumentar $r$; que solo la coma y el astigmatismo tienen dependencia con el ángulo incidente; y que mayores focales hacen disminuir todas las aberraciones. Así, los telescopios con mayores aberraciones serán los de una focal más corta y una apertura más grande. Es decir, los que cuenten con una menor relación focal $f/D$, que es precisamente el caso de los IACT, como se explicó en el apartado 2.9.3. Como además los requisitos de resolución no son muy exigentes en astronomía *Cherenkov*, en este tipo de grandes telescopios segmentados con un número tan elevado de espejos, se recurre habitualmente a la óptica esférica para reducir los costes, lo que es una gran fuente de aberraciones. Por todo ello, la mayor limitación al campo de visión de un IACT viene impuesta por la presencia de aberraciones geométricas, y sobre su reducción deberían centrarse los mayores esfuerzos de adaptación de estos telescopios como receptores de comunicaciones ópticas.

# 2.11. MODELOS EN OSLO DE LOS TELESCOPIOS CTA

Dentro del objetivo de reducir el campo de visión de los IACT para minimizar el ruido de fondo que se acopla al sistema de comunicaciones, se han analizado todos los factores que están presentes en este tipo de telescopios limitando su resolución óptica. Se ha concluido que la primera y mayor limitación la constituyen las aberraciones geométricas, por lo que la adaptación de estos telescopios debería centrarse en minimizar estas antes que cualquier otra limitación. En este apartado se pretende estudiar el comportamiento óptico de los telescopios de CTA, para obtener una caracterización lo más realista posible de su resolución óptica, a fin de poder diseñar el campo de visión más adecuado y calcular su comportamiento en un enlace de comunicaciones ópticas. Además, disponer de un modelo óptico detallado es el primer paso para, mediante el estudio de las aberraciones presentes en cada telescopio, proponer soluciones para su reducción.

Por la razón explicada en el párrafo anterior, es importante conocer la resolución óptica de los telescopios y esta caracterización es diferente para un telescopio *Cherenkov* que para uno de comunicaciones. La principal diferencia reside en el hecho de que los primeros deben considerar grandes campos de visión y los segundos no. Por ejemplo, en el telescopio HESS se puede comprobar que es posible reducir la PSF en casi un factor 5 al reducir el campo de visión desde 5° hasta 0,1° [169, p. 16], un valor aún mucho mayor que los utilizados en comunicaciones. Debido a los reducidos campos de visión de interés en comunicaciones, especialmente en comparación con los empleados en astronomía *Cherenkov*, interesará estudiar el comportamiento de los telescopios con rayos en eje. El aumento de las aberraciones para ángulos fuera de eje en estos rangos de campos de visión será despreciable en comparación con el resto de aberraciones, justamente el caso contrario al de cualquier IACT empleado en astronomía de rayos gamma.

Para todas las simulaciones se ha utilizado el entorno OSLO (del inglés *Optics Software for Layout and Optimization*), que es un software de simulación óptica profesional con más de 40 años de historia. Su utilización se dirige tanto al diseño como al análisis de todo tipo de sistemas ópticos usando para ello óptica geométrica y óptica física. Es una herramienta de referencia en el sector y ha sido usado en el diseño de numerosos grandes telescopios segmentados del tipo de los IACT, como por ejemplo el TMT (*Thirty-Meter Telescope*) [170] o el JWST (*James Webb Space Telescope*) [171]. OSLO es un entorno de simulación de sistemas ópticos secuenciales [12]. Cada sistema secuencial se define como una serie de superficies que uno o más rayos interceptan una tras otra al atravesarlo. La luz entra al sistema desde la izquierda a la derecha y se encuentra secuencialmente con superficies numeradas desde cero en adelante, siendo la última de ellas el plano imagen (Figura 55). Por ello, las distancias, definidas por un grosor de una superficie neutra, serán positivas hacia la derecha y negativas hacia la izquierda, como en el caso de producirse reflexiones, y los radios de curvatura serán positivos si su centro está hacia el lado derecho y negativos hacia el izquierdo. De esta forma, se puede modelar cualquier sistema dividiéndolo en sucesivas superficies secuenciales definidas principalmente por un grosor, una apertura y un radio de curvatura, así como por un gran número de características opcionales (formas distintas a la esfera, perfiles de curvatura asféricos, material de la superficie, etc.).

---

[12] Opcionalmente es posible crear grupos compuestos por elementos no secuenciales. Como se verá, esto es necesario para definir una superficie reflectora segmentada como la de los IACT a la que todos los rayos deben llegar simultáneamente pese a tratarse de diferentes superficies.

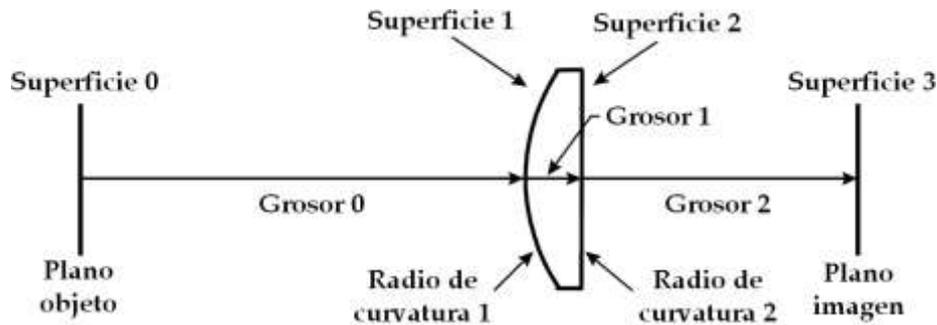

**Figura 55. Esquema secuencial de un sistema óptico básico en OSLO.**

## 2.11.1. Modelado de espejos realistas

La mayor parte de los espejos que previsiblemente constituirán los principales telescopios de CTA, es decir, SST y MST de tipo *Davies-Cotton* y LST (Tabla 4), están basados en un diseño de topología hexagonal con apotema de entre unos 0,5 y 1 metros (o lo que es lo mismo, entre unos 1 y 2 metros de distancia entre lados opuestos). En el caso de LST, la apotema prevista es de 75,5 cm; en MST de 60 cm; y en SST de 39 cm [63, p. 5]. En todos los casos el perfil de los espejos es esférico con un radio de curvatura fijo a lo largo de toda su superficie (aunque variable de unos a otros espejos, según la zona de la parábola en el caso de LST y de distinto valor para MST y SST).

**Tabla 4. Características de los espejos de los principales telescopios de CTA.**

| Telescopio | Topología | Apotema | Distancia lado-lado | Área | Perfil | Radio de curvatura |
|---|---|---|---|---|---|---|
| CTA-LST | Hexagonal | 75,5 cm | 151 cm | 1,96 m$^2$ | Esférico | ~62 m |
| CTA-MST | Hexagonal | 60 cm | 120 cm | 1,25 m$^2$ | Esférico | 32 m |
| CTA-SST | Hexagonal | 39 cm | 78 cm | 0,53 m$^2$ | Esférico | 11,2 m |

El primer paso para el modelado de los espejos es replicar su topología hexagonal. En OSLO no existen superficies hexagonales, pero es posible crear cualquier forma como la intersección de otro tipo de superficies más simples. Para recrear un hexagono se utilizaron tres rectángulos con centros coincidentes y una rotación de 60° entre ellos. Las formas compuestas modeladas en OSLO se derivan del área resultante de la superposición de todas las formas básicas empleadas, apareciendo el resto del área como una superficie inexistente en el modelo. Así, los tres rectángulos de la izquierda de la Figura 56 no producen una superficie con la forma del hexágono exterior sino como la del interior, mostrado a la derecha. Es importante tener esto en cuenta, ya que en OSLO no es posible visualizar la forma final de la superficie.

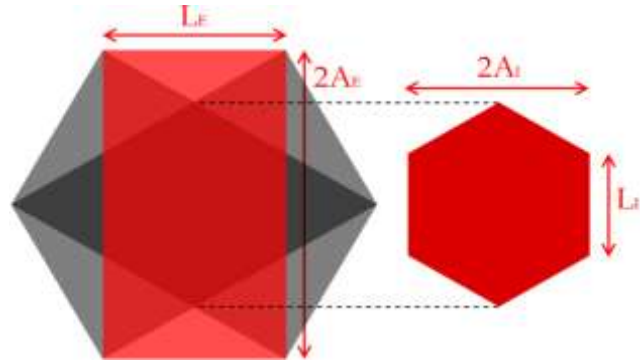

**Figura 56. Rectángulos necesarios (izquierda) para modelar un hexágono en OSLO (derecha).**

Cada rectángulo se modela en OSLO definiendo la longitud de cada uno de sus lados. Para producir un hexágono con apotema $A_I$ y lado $L_I$ como el de la Figura 56 se precisarán tres rectángulos de lados $L_E$ y $2A_E$. Así, partiendo del valor de la distancia lado-lado $2A_I$ correspondiente a cada tipo de espejo según la Tabla 4, se pueden obtener los valores $L_E$ y $2A_E$ de los rectángulos mediante las ecuaciones (2-32) y (2-33), obtenidas por trigonometría básica.

$$L_E = 2A_I \tag{2-32}$$

$$2A_E = \frac{L_E}{\tan(30°)} \tag{2-33}$$

En la Figura 57 se muestra el resultado de simular en OSLO con una longitud de onda de 1550 nm el frente de onda (arriba), el diagrama de spot (centro) y la PSF (abajo) correspondientes a cada uno de los tres espejos básicos de CTA cuando se ilumina su superficie homogéneamente, utilizando para ello un haz circular de radio igual a $2A_I/\cos(30°)$. El resultado de estas simulaciones es interesante para mostrar los indicadores que se usan en la evaluación óptica de los telescopios, si bien aquí por simplicidad se utilizará solo uno de ellos (el diagrama de spot) para evaluar los telescopios modelados. Para ello, en estas simulaciones es posible determinar si cada espejo está limitado por difracción según los tres criterios clásicos: según el primero, el límite de difracción lo define un OPD[13] P-V (*Peak to Valley*) igual a λ/4; el segundo criterio se establece en función de si el diagrama de spot queda dentro o fuera del disco de Airy (ver apartado 2.10.1); y el tercero, o criterio de Maréchal, especifica un valor igual a 0,8 del cociente de *Strehl*[14].

---

[13] El OPD (*Optical Path Difference*), o diferencia de caminos ópticos, es el parámetro habitual para evaluar la aberración presente en un frente de onda, comparando la diferencia de caminos ópticos en relación a un frente de onda perfectamente esférico [172, p. 49].

[14] El cociente de *Strehl* se define como la relación entre la intensidad de pico en el patrón de difracción mostrado por el sistema de estudio comparada con la intensidad de pico del sistema equivalente funcionando en el límite de difracción [172, p. 783].

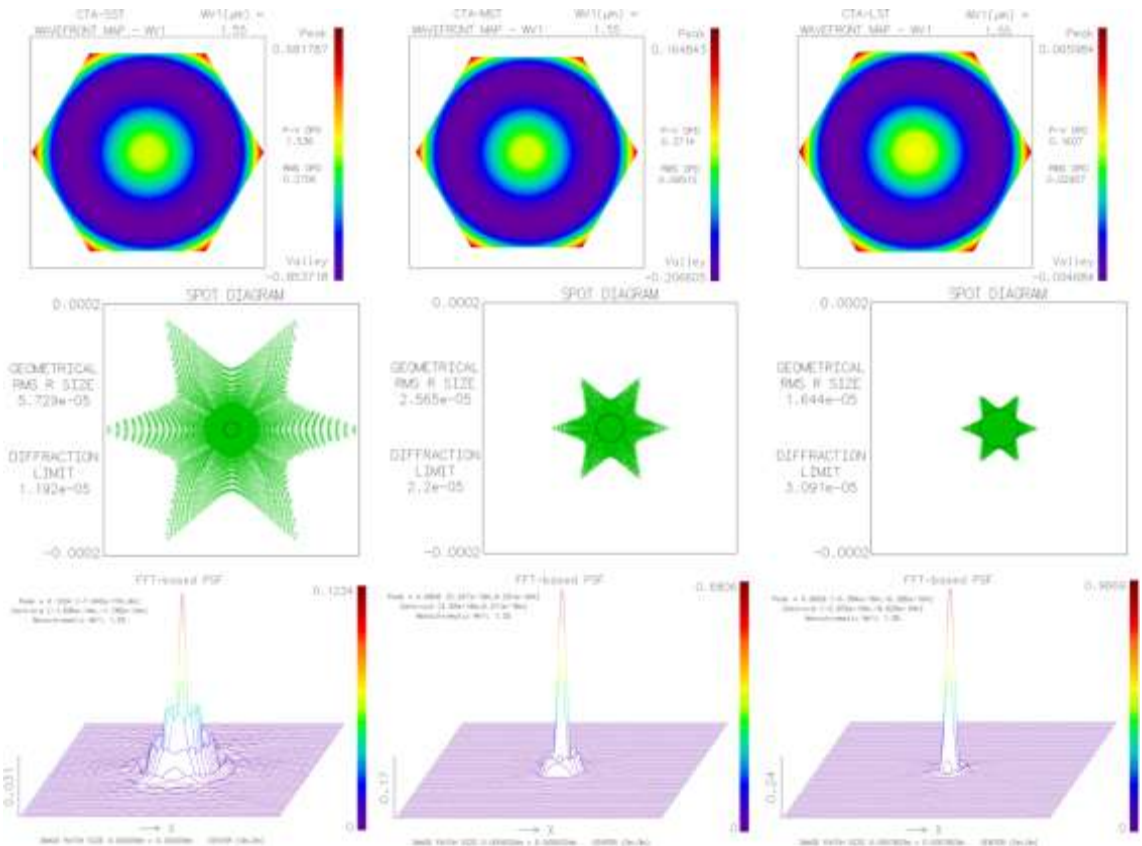

**Figura 57. Simulación del frente de onda (arriba), diagrama de spot (centro) y PSF (abajo) de los espejos de SST (izquierda), MST (centro) y LST (derecha) para λ = 1,55 µm.**

En la Tabla 5 se muestran los resultados numéricos generados en la simulación (obtenidos de los resultados mostrados de la Figura 57). En cada parámetro se ha indicado su valor simulado y a la derecha, entre paréntesis, la referencia del límite de difracción según cada uno de los tres criterios mencionados en el párrafo anterior. Se puede comprobar que la resolución óptica de los espejos de SST está limitada por sus aberraciones, haciendo que el OPD P-V quede por encima de λ/4, el diagrama de spot sea mayor que el disco de Airy y el coeficiente de *Strehl* quede por debajo de 0,8. El caso de MST se podría considerar aproximadamente en el límite de difracción debido a su proximidad a los valores de referencia (y se visualiza bien en el aspecto de la PSF de la Figura 57). Y por último, los valores de LST sí lo sitúan claramente por debajo del límite de difracción. Estos resultados se explican porque de LST a SST se reduce el radio de curvatura (mayores curvaturas) debido a la menor focal (recuérdese que la focal es igual a la mitad del radio de curvatura), y también se disminuye el tamaño de la apertura (ver la dependencia de las aberraciones con la focal y la apertura en el apartado 2.10.5).

**Tabla 5. Determinación del límite de difracción para los espejos de los principales telescopios de CTA. Entre paréntesis se indica la referencia del límite de difracción según cada criterio.**

| Telescopio | OPD *Peak-to-Valley* | Radio del diagrama de spot | Coeficiente de *Strehl* |
|------------|----------------------|----------------------------|-------------------------|
| CTA-SST | 1,536λ (>0,25λ) | 57,29 µm (>11,92 µm) | 0,12 (<0,8) |
| CTA-MST | 0,371λ (~0,25λ) | 25,65 µm (~22 µm) | 0,68 (~0,8) |
| CTA-LST | 0,161λ (<0,25λ) | 16,44 µm (<30,91 µm) | 0,97 (>0,8) |

En los siguientes apartados se tendrá en cuenta cómo se distorsiona el frente de onda según la topología de los telescopios con el objetivo de evaluar su calidad óptica real. Sin embargo, antes de modelar los telescopios, es necesario disponer de un modelo realista de los espejos. El simulado anteriormente es un modelo ideal, que únicamente considera sus especificaciones nominales. El proceso de fabricación de los espejos no proporciona espejos ideales, sino que se producen imperfecciones que podrían contribuir a la distorsión del frente de onda cuando se simule el telescopio completo. Será interesante incluir modelos no ideales en las simulaciones de los telescopios para determinar si la resolución de estos viene determinada por la provocada por los espejos o bien queda enmascarada por el diseño del propio telescopio. Para tratar de obtener un modelo más realista de los espejos, se ha partido de las medidas del frente de onda realizadas tras la fabricación de algunos prototipos de espejos para CTA que aparecen en la literatura científica (Tabla 6). Comparando los resultados experimentales de prototipos reales con los modelos ideales, se observa que el límite de difracción ya queda muy lejano en todos los casos. En el caso menos favorable de los modelos ideales, se obtenía un OPD P-V de $1{,}536\lambda = 2{,}38$ µm, algo más de un orden de magnitud por debajo de las medidas experimentales. Esto da lugar a unas PSF mucho mayores. Para comparar modelos se puede usar la medida RMS del diagrama de spot de las simulaciones y la medida del $D_{80}$ (que define el diámetro del spot que contiene el 80 % de la luz en el foco) de los prototipos. En este caso, de nuevo la medida menos favorable de las simulaciones proporcionaba un diagrama de spot de 115 µm de diámetro, en comparación con las medidas experimentales del $D_{80}$ de varios mm, también por encima de un orden de magnitud peor.

**Tabla 6. Medidas experimentales de OPV P-V y D80 sobre prototipos de espejos de CTA.**

| LST Sanko (Japón) [172] | LST Sanko (Japón) [173] | SST ASTRI (Italia) [174] |
|---|---|---|
| 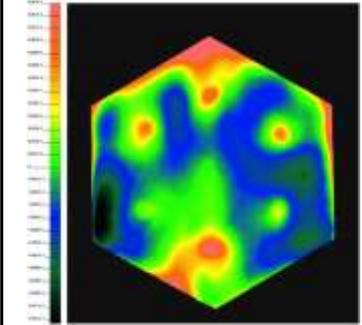 | 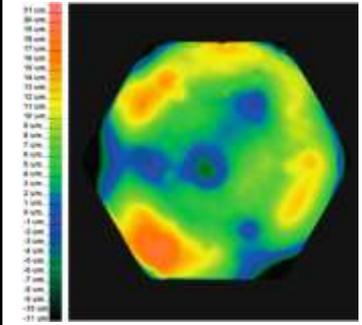 | 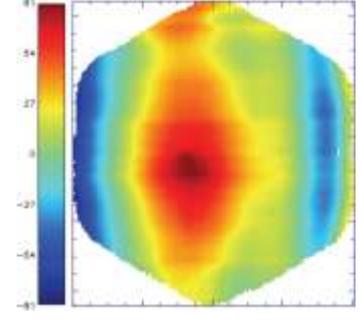 |
| OPD P-V = 25,65 µm<br>$D_{80}$ = 8 mm | OPD P-V = 32 µm<br>$D_{80}$ = 8,15 mm | OPD P-V = 29,752 µm<br>$D_{80}$ = 5,5 mm |

Se propuso simular en OSLO estos efectos reales de espejos individuales para poder propagar su efecto en las simulaciones de los telescopios de CTA completos. En OSLO no es posible introducir el error de frente de onda de una superficie directamente caracterizado por el OPD P-V [15], pero sí es posible asociar una deformación determinada a una de las superficies, en este caso a los espejos. Para ello, es necesario definir el perfil de la deformación mediante un archivo de texto asociado a la superficie. Esta definición puede llevarse a cabo mediante los polinomios de Zernike o bien mediante una matriz bidimensional de datos. Dado que se ignoran los polinomios de Zernike que caracterizan la superficie experimental, se optó por generar una matriz a partir de una de las medidas de los prototipos para obtener el perfil definido por las microrrugosidades de estos espejos. Se seleccionó la medida del espejo desarrollado por el fabricante Sanko en Japón, por suponer un caso peor (ver Tabla 6, centro). Sobre la medida del frente de onda se superpuso una cuadrícula de 32×32 (Figura 58), y se asignó un valor a cada celda según el código de colores definido por la escala de medida de Sanko, obteniendo un total de 1024 valores con los que definir la superficie.

---

[15] Esto no debe verse necesariamente como una limitación del *software*, ya que OSLO trabaja con modelos ópticos completos y la medida de OPD P-V solo aporta una parte limitada de la información del frente de onda. Solo con este dato se ignora por completo la forma del frente de onda, que es lo que lo define completamente.

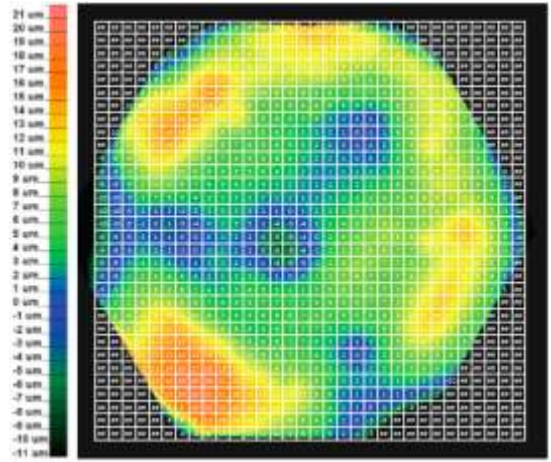

**Figura 58. Asignación de una matriz de 32×32 valores sobre la deformación del frente de onda para modelar de forma realista la superficie reflectora ideal.**

Una vez generada la matriz de valores que definen la deformación de la superficie reflectora e introducida cierta información para el correcto procesamiento de esta deformación en OSLO, es posible asociarla a la superficie del espejo. En la Figura 59 se muestra el resultado de la simulación del modelo realista de LST a la derecha. Observando la medida experimental original a la izquierda puede comprobarse su gran similitud (nótese que la escala de colores es diferente). El OPD P-V obtenido en la simulación es de 18,4λ, que es igual a 28,52 μm, muy cercano a las 32 μm del prototipo medido, por lo que puede considerarse un modelo válido en el que basarse para simular el comportamiento real de los espejos.

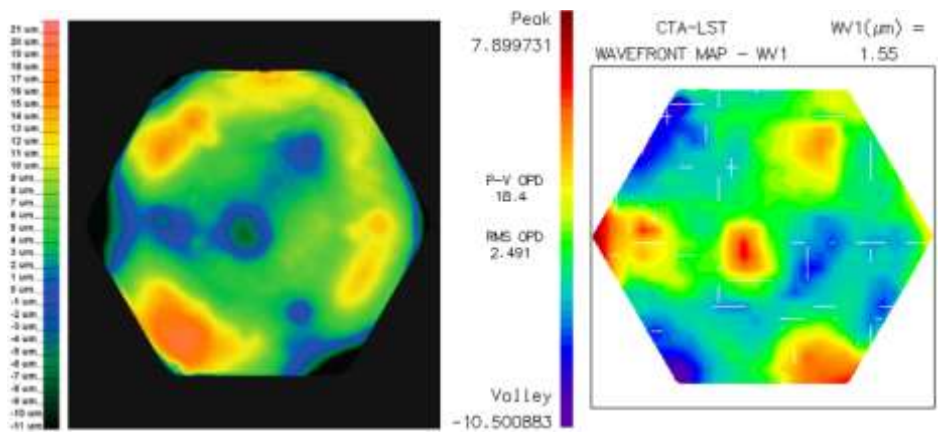

**Figura 59. Medida experimental del prototipo desarrollado por Sanko (izquierda) y simulación del modelo realista de espejo LST en OSLO (derecha).**

En las simulaciones de telescopios de los siguientes apartados se va a simular un elevado número de espejos y emplear para todos ellos el mismo modelo, basado en una medida experimental sobre un espejo específico, podría dar lugar a ciertas regularidades en los resultados que conviene evitar. Por ello, se decidió generar patrones aleatorios de la deformación de la superficie de los espejos basándose en los datos del modelo caracterizado. Inicialmente se programó en *Matlab* una rutina simple para generar una matriz de 64×64 con el objetivo de obtener una superficie con las mismas características que el modelo de Sanko. Sin embargo, si se pretende obtener un OPD P-V equivalente a la medida experimental, la generación aleatoria simple da lugar a grandes variaciones entre puntos adyacentes (Figura 60) que no se corresponden con una superficie realista.

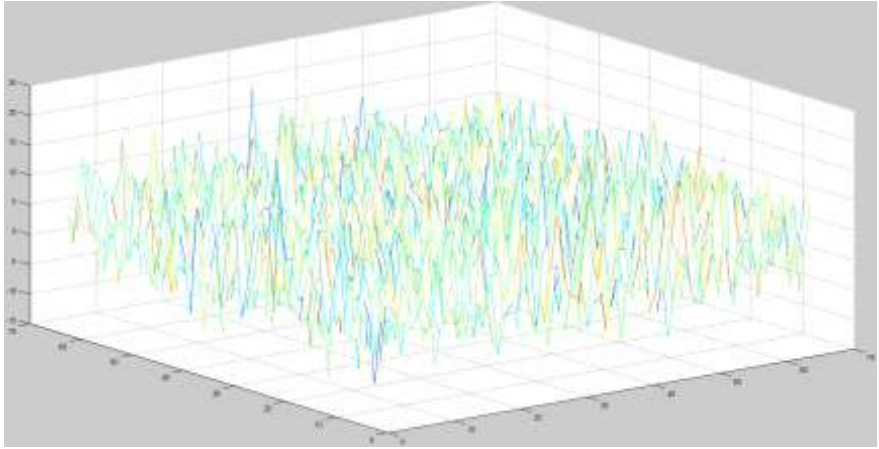

**Figura 60. Matriz de 64×64 generada inicialmente en *Matlab* para modelar el perfil Sanko.**

Para evitar el problema del modelo de la Figura 60, se incluyó en la rutina de *Matlab* un filtro para promediar zonas de puntos adyacentes. Se sabe que la media del perfil de Sanko es de 5,8 µm y su desviación típica 6,6 µm [173]. Por ensayo y error se fue ajustando la ventana del filtro hasta obtener unos valores de media y desviación típica equiparables a la medida experimental. En la Figura 61 se muestra un ejemplo del perfil generado aplicando esta rutina con una ventana de 13×13, resultando en una media de 5 µm y una desviación típica de 1,75 µm. De esta forma se puede generar de una vez una gran cantidad de perfiles aleatorios, cada uno almacenado en un archivo distinto, que debe ser asociado en OSLO a cada espejo del telescopio a simular.

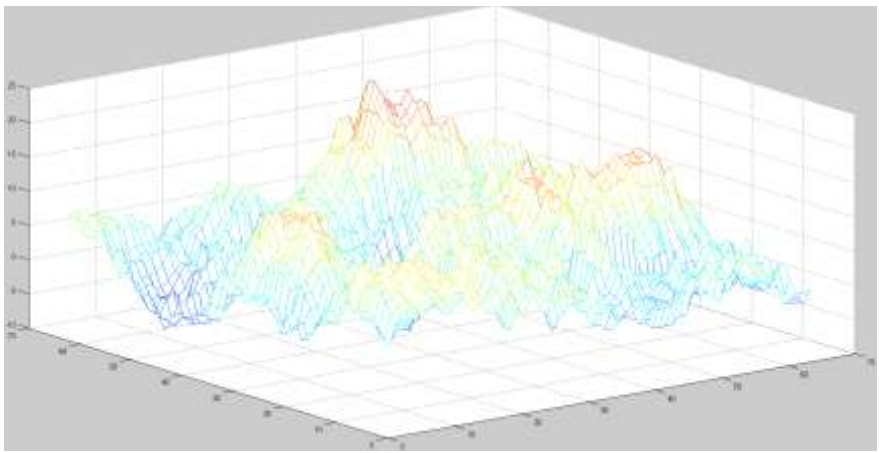

**Figura 61. Matriz de 64×64 valores aleatorios generada en *Matlab* para modelar el perfil Sanko tras aplicar un filtro promediado con ventana de 13×13.**

A continuación se muestra el código utilizado para generar estos perfiles en *Matlab*. Este código genera 128 perfiles aleatorios con un filtro tipo promedio con ventana 13×13 y los almacena en diferentes archivos de texto, donde cada línea es una fila de la matriz de 64×64 correspondiente a cada perfil. Las dos primeras líneas del archivo son las cabeceras necesarias para la correcta interpretación del archivo de tipo deformación interferométrica de OSLO. Además, el código permite obtener para cada superficie la media y desviación típica, de forma que se pueda monitorizar que los resultados se mantienen dentro de un rango razonablemente parecido al modelo experimental de referencia.

```matlab
media=[1:128];
desviacion=[1:128];
peaktovalley=[1:128];

for i=1:128
nombrefich=char(strcat('ESPEJO_',
   int2str(i),'.int'));
a=floor(10+10*randn(64));

h2=fspecial('average',[13,13]);
a=floor(imfilter(a,h2));

 fid=fopen(nombrefich,'w');
 fprintf(fid,'ESPEJO NÚMERO %i \r\n', i);
 fprintf(fid,'GRD 64 64 SUR WVL 1 SSZ 1 NDA 333
  \r\n');
 fprintf(fid, '%i %i %i %i %i %i %i %i %i %i %i %i %i
  %i %i %i %i %i %i %i %i %i %i %i %i %i %i %i %i
  %i %i %i %i %i %i %i %i %i %i %i %i %i %i %i %i
  %i %i %i %i %i %i %i %i %i %i %i %i %i %i %i %i
  \r\n', a);
 fclose(fid);
 mesh(a);

 med=mean(a);
 media(i)=mean(med);
 desv=std(a);
 desviacion(i)=std(desv);
 maximo=max(max(a));
 minimo=min(min(a));
 peaktovalley(i)=maximo-minimo;
end

media_de_medias=mean(media)
desviacion_de_medias=std(media)
maximo_de_medias=max(media)
minimo_de_medias=min(media)

media_de_PeakToValley=mean(peaktovalley)
desviacion_de_PeakToValley=std(peaktovalley)
maximo_de_PeakToValley=max(peaktovalley)
minimo_de_PeakToValley=min(peaktovalley)
```

En la Tabla 7 se muestran, a modo de ejemplo, las simulaciones de un espejo de SST, otro de MST y otro de LST, empleando algunos de los perfiles generados utilizando el código anterior y con los valores descritos.

**Tabla 7. Simulaciones de frente de onda sobre prototipos de espejos de CTA aplicando deformaciones aleatorias en una escala basada en medidas experimentales.**

| LST con perfil realista | MST con perfil realista | SST con perfil realista |
|---|---|---|
| 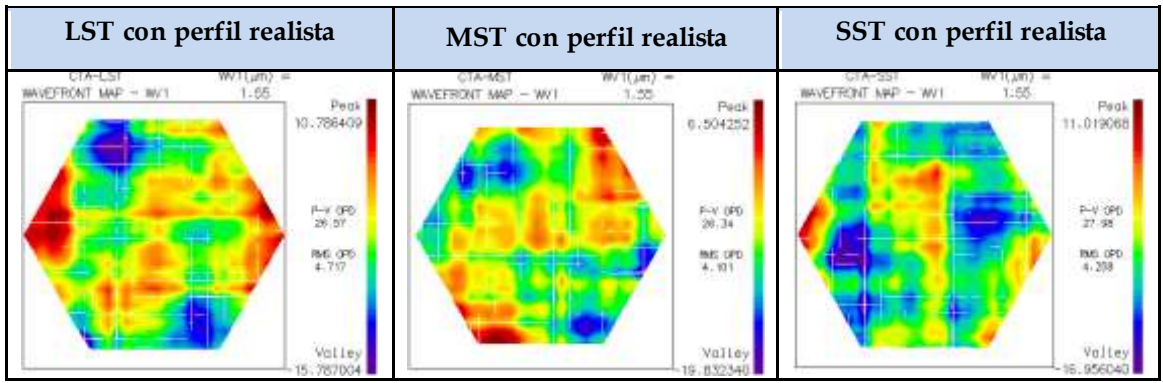 | | |

## 2.11.2. CTA-LST parabólico

El telescopio LST es el mayor de los telescopios de CTA, con una masa de unas 70 toneladas y un área reflectora prevista de 387 m² formada por 198 espejos hexagonales de 151 cm de distancia lado-lado y un área de 1,96 m² (Figura 62). La relación focal de este telescopio es de f/1,2, que con un diámetro equivalente de 23 metros, da lugar a una focal de 27,6 metros [172]. El perfil del reflector es parabólico, lo que se consigue disponiendo los espejos según este tipo de curva, aunque para reducir el coste de fabricación se utilizan espejos esféricos con radio de curvatura constante a lo largo de su superficie.

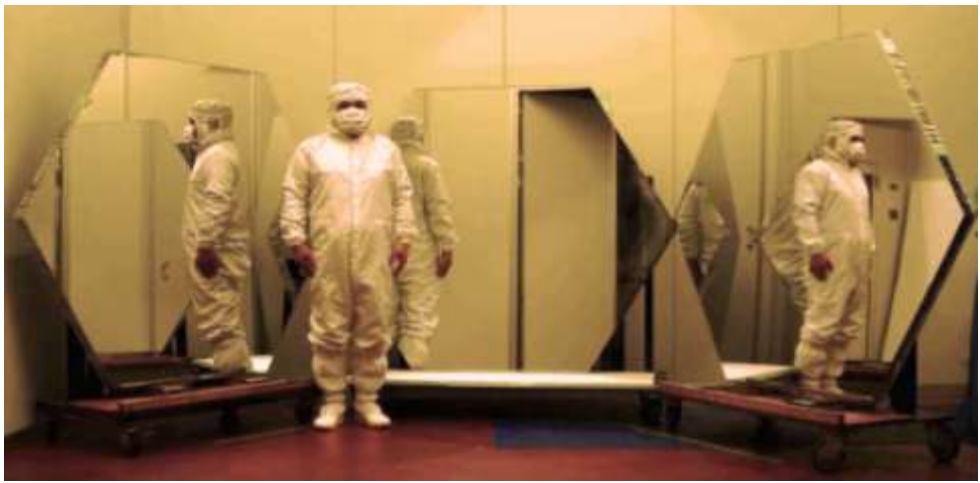

**Figura 62. Muestra de tres espejos CTA-LST fabricados en Sanko, Japón [172].**

Para introducir el modelo del telescopio en OSLO es necesario definir, para cada uno de los espejos, la posición {x, y, z} de su centro, el radio de curvatura $R_c$ y la inclinación referida a los dos planos de altitud y acimut. El primer paso es el cálculo de las posiciones centrales, para el cual se siguió un método basado en el descrito en el apartado 2.8.2 para el telescopio MAGIC II, basado en aplicar la ecuación paramétrica de la parábola utilizando los valores propios del telescopio como la longitud focal y el tamaño de cada espejo. En este caso, para cada espejo no solo se precisa su posición en dos dimensiones, sino en tres, ya que ahora se va a simular el telescopio completo y no solo una fila, por lo que el método de cálculo tiene algunas diferencias.

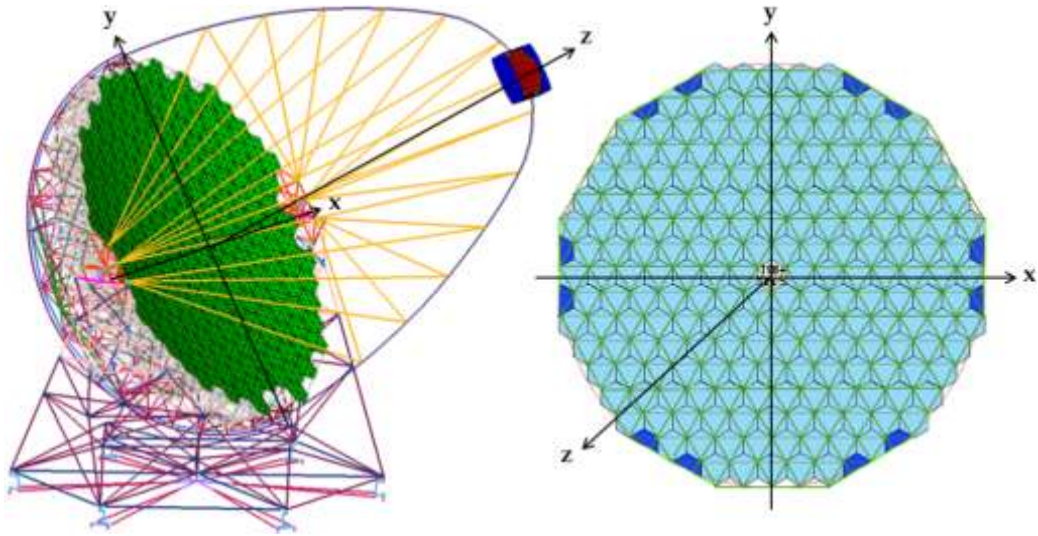

**Figura 63. Esquema de la disposición de los espejos de un telescopio CTA-LST.**

En la Figura 63 se muestra un esquema de la disposición de los espejos del telescopio. Se puede apreciar que sigue una colocación en red hexagonal, a diferencia de la red cuadrada de MAGIC II. Esto hace que los segmentos se vayan agrupando en anillos alrededor del centro. El resultado es que las posiciones {x, z} no serán las simétricas a las {y, z}, como ocurría con MAGIC II. Por ello, se han calculado por separado las posiciones {x, z} y {y, z} de las filas centrales del eje x y del eje y. Además, de cada uno de estos ejes existen dos filas centrales diferentes también debido a la disposición hexagonal de los espejos.

En la Figura 64 se muestran las posiciones centrales de los espejos de la fila central del eje x, que se corresponde con la fila en la que la coordenada y es igual a cero para todos los espejos. En la Figura 65 se muestran las posiciones de los espejos de la segunda fila del eje x, cuya coordenada x se ha calculado como la media de las coordenadas x de los espejos anterior y posterior de la fila central.

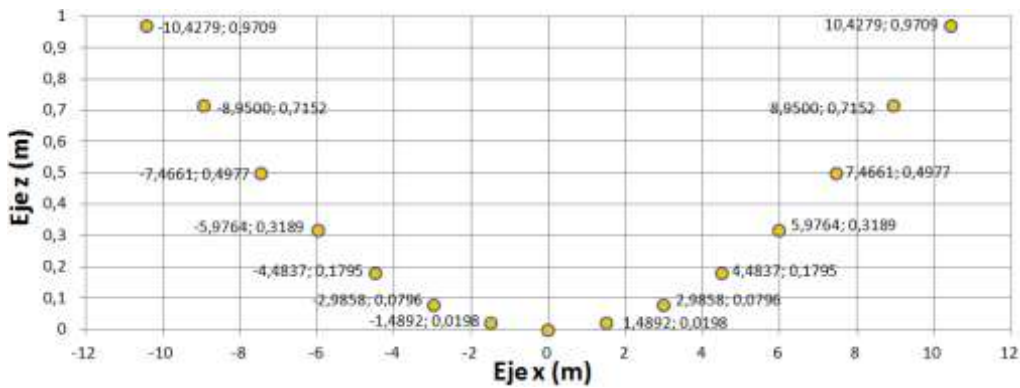

**Figura 64. Parábola centrada en el origen con foco en {x, z} = {0, 28} describiendo las posiciones centrales (en m) de los 15 espejos de CTA-LST en la fila central del eje x.**

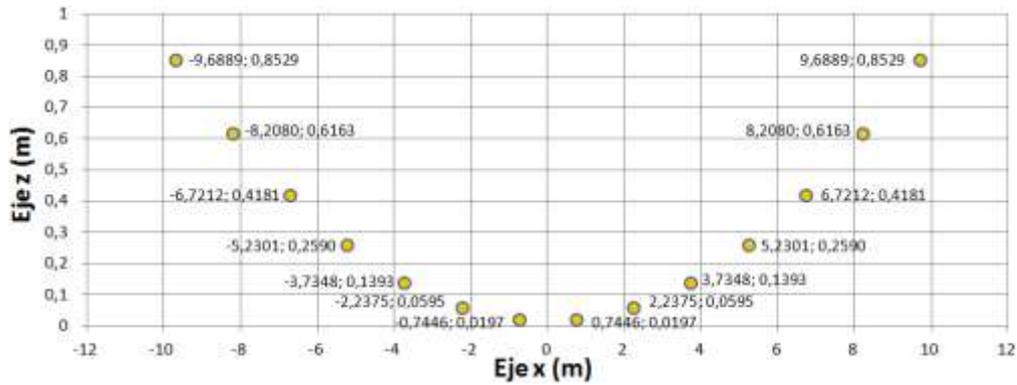

**Figura 65. Parábola centrada en el origen con foco en {x, z} = {0, 28} describiendo las posiciones centrales (en m) de los 14 espejos de CTA-LST en la segunda fila del eje x.**

En la Figura 66 y en la Figura 67 se presentan las posiciones equivalentes en el eje y, en el primer caso para la fila central y en el segundo caso para la segunda fila.

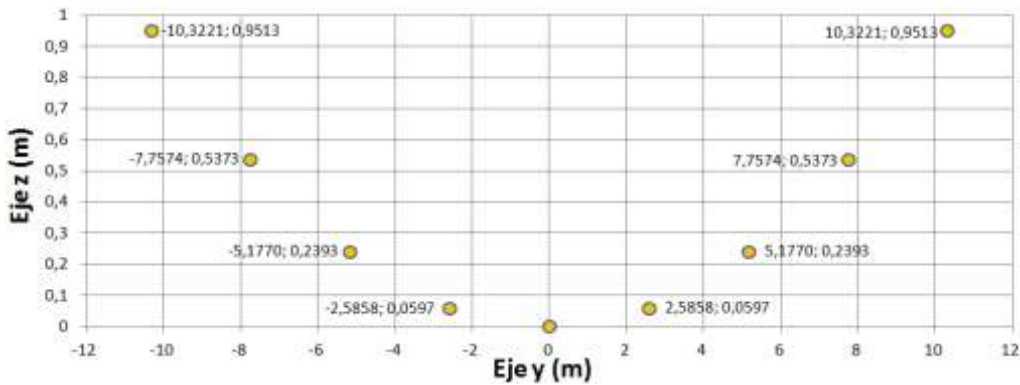

**Figura 66. Parábola centrada en el origen con foco en {y, z} = {0, 28} describiendo las posiciones centrales (en m) de los 9 espejos de CTA-LST en la fila central del eje y.**

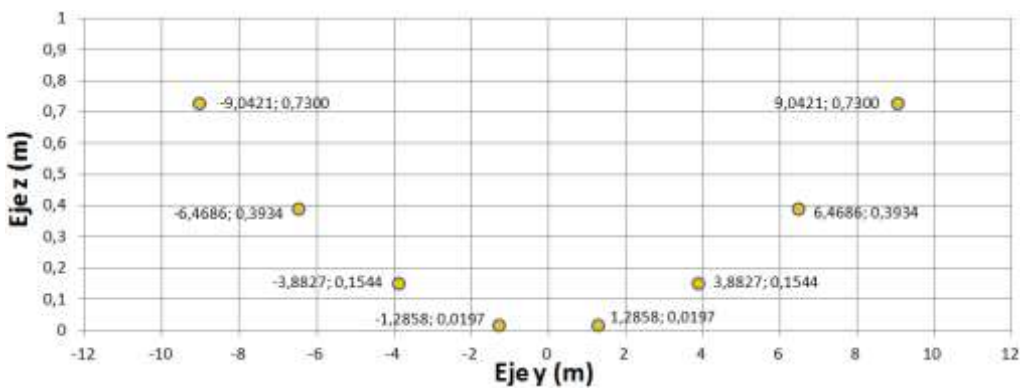

**Figura 67. Parábola centrada en el origen con foco en {y, z} = {0, 28} describiendo las posiciones centrales (en m) de los 8 espejos de CTA-LST en la segunda fila del eje y.**

A partir de las posiciones {x, y} de los espejos de las dos primeras filas de cada eje, ya es posible calcular las posiciones del resto de espejos del telescopio. Las posiciones {x, y} se completan por simple simetría y para calcular la coordenada z de todos los espejos se emplea la ecuación (2-34) que define el perfil de un paraboloide de revolución con foco igual a f.

$$x^2 + y^2 = 4fz \qquad (2\text{-}34)$$

Nótese que la coordenada y de los espejos de la segunda fila del eje x y la coordenada x de los espejos de la segunda fila del eje y no es igual a cero. Por ello, para el cálculo de la coordenada z de estas posiciones en las Figura 65 y Figura 67 se utilizó también la ecuación (2-34). Así es posible calcular las coordenadas {x, y, z} de todos los espejos del telescopio, que se muestran en la Figura 68, la Figura 69 y la Figura 70.

```
                    -2986  -1489    0    1489   2986
              -5230  -3735  -2237  -745   745   2237   3735   5230
        -7466  -5976  -4484  -2986  -1489    0    1489   2986   4484   5976   7466
  -8208  -6721  -5230  -3735  -2237  -745   745   2237   3735   5230   6721   8208
  -8950  -7466  -5976  -4484  -2986  -1489    0    1489   2986   4484   5976   7466   8950
-9689  -8208  -6721  -5230  -3735  -2237  -745   745   2237   3735   5230   6721   8208   9689
-10428 -8950  -7466  -5976  -4484  -2986  -1489    0    1489   2986   4484   5976   7466   8950  10428
-9689  -8208  -6721  -5230  -3735  -2237  -745   745   2237   3735   5230   6721   8208   9689
-10428 -8950  -7466  -5976  -4484  -2986  -1489    0    1489   2986   4484   5976   7466   8950  10428
-9689  -8208  -6721  -5230  -3735  -2237  -745   745   2237   3735   5230   6721   8208   9689
-10428 -8950  -7466  -5976  -4484  -2986  -1489    0    1489   2986   4484   5976   7466   8950  10428
-9689  -8208  -6721  -5230  -3735  -2237  -745   745   2237   3735   5230   6721   8208   9689
  -8950  -7466  -5976  -4484  -2986  -1489    0    1489   2986   4484   5976   7466   8950
  -8208  -6721  -5230  -3735  -2237  -745   745   2237   3735   5230   6721   8208
        -7466  -5976  -4484  -2986  -1489    0    1489   2986   4484   5976   7466
              -5230  -3735  -2237  -745   745   2237   3735   5230
                    -2986  -1489    0    1489   2986
```

**Figura 68. Coordenada x (en mm) de las posiciones centrales de los espejos de CTA-LST.**

```
                    10322  10322  10322  10322  10322
                 9042  9042  9042  9042  9042  9042  9042
              7757  7757  7757  7757  7757  7757  7757  7757  7757
           6469  6469  6469  6469  6469  6469  6469  6469  6469  6469  6469
         5177  5177  5177  5177  5177  5177  5177  5177  5177  5177  5177  5177
        3883  3883  3883  3883  3883  3883  3883  3883  3883  3883  3883  3883  3883
   2586  2586  2586  2586  2586  2586  2586  2586  2586  2586  2586  2586  2586  2586  2586
     1286  1286  1286  1286  1286  1286  1286  1286  1286  1286  1286  1286  1286  1286
   0     0     0     0     0     0     0     0     0     0     0     0     0     0     0
    -1286 -1286 -1286 -1286 -1286 -1286 -1286 -1286 -1286 -1286 -1286 -1286 -1286 -1286
-2586 -2586 -2586 -2586 -2586 -2586 -2586 -2586 -2586 -2586 -2586 -2586 -2586 -2586 -2586
  -3883 -3883 -3883 -3883 -3883 -3883 -3883 -3883 -3883 -3883 -3883 -3883 -3883
    -5177 -5177 -5177 -5177 -5177 -5177 -5177 -5177 -5177 -5177 -5177 -5177
      -6469 -6469 -6469 -6469 -6469 -6469 -6469 -6469 -6469 -6469 -6469
        -7757 -7757 -7757 -7757 -7757 -7757 -7757 -7757 -7757
           -9042 -9042 -9042 -9042 -9042 -9042 -9042
            -10322 -10322 -10322 -10322 -10322
```

**Figura 69. Coordenada y (en mm) de las posiciones centrales de los espejos de CTA-LST.**

```
                    -1031  -971   -951   -971  -1031
                 -910  -810  -750  -730  -750  -810  -910  -1049
           -1035  -856  -717  -617  -557  -537  -557  -617  -717  -856  -1035
        -871  -693  -553  -553  -453  -393  -393  -453  -553  -693  -871  -1089
      -955  -737  -558  -419  -319  -259  -239  -259  -319  -419  -558  -737  -955
    -850  -632  -454  -454  -314  -214  -154  -154  -214  -314  -454  -632  -850  -1106
-1031  -775  -557  -379  -239  -139   -80   -60   -80  -139  -239  -379  -557  -775  -1031
   -853  -616  -418  -259  -139   -59   -20   -20   -59  -139  -259  -418  -616  -853
-971  -715  -498  -319  -180   -80   -20    0    -20   -80  -180  -319  -498  -715  -971
   -853  -616  -418  -259  -139   -80   -20   -20   -59  -139  -259  -418  -616  -853
-1031  -775  -557  -379  -239  -139   -80   -60   -80  -139  -239  -379  -557  -775  -1031
    -850  -632  -454  -454  -314  -214  -154  -154  -214  -314  -454  -632  -850  -1106
      -955  -737  -558  -419  -319  -259  -239  -259  -319  -419  -558  -737  -955
        -871  -693  -553  -453  -393  -393  -453  -553  -693  -871  -1089
           -1035  -856  -717  -617  -557  -537  -557  -617  -717  -856  -1035
                 -910  -810  -750  -730  -750  -810  -910  -1049
                    -1031  -971   -951   -971  -1031
```

**Figura 70. Coordenada z (en mm) de las posiciones centrales de los espejos de CTA-LST.**

Como se explicó previamente, los radios de curvatura son fijos en toda la superficie del espejo, aunque de valor variable de unos espejos a otros. Para el cálculo de los radios de curvatura r$_{cmed}$ (Figura 71) se han empleado las ecuaciones (2-10), (2-11) y (2-12), explicadas en el apartado 2.8.2, tomando como distancia r la distancia desde el centro de cada espejo hasta el centro de la parábola.

**Figura 71. Radio de curvatura medio (en mm) de cada espejo de CTA-LST.**

La inclinación φ de cada espejo para conformar la superficie parabólica del reflector principal se ha calculado mediante optimizaciones en OSLO, para llegar a la mejor alineación posible de todos los segmentos del telescopio. Sin embargo, para llevar a cabo dichas optimizaciones antes es necesario partir de un conjunto de valores iniciales suficientemente aproximados. Para calcular estos valores iniciales de φ$_x$ y φ$_y$ (Figura 72 y Figura 73), se utilizó la aproximación de las ecuaciones (2-35) y (2-38), que se puede deducir por simple trigonometría. De nuevo, solo es necesario calcular para cada eje la fila central y la segunda fila y el resto de espejos del telescopio se deducen por simetrías a partir de estos. Todos los IACT permiten llevar a cabo estas inclinaciones mediante un juego de actuadores sobre cada espejo, controlados remotamente, que permiten realinear en tiempo real para compensar las deformaciones del telescopio por la propia presión gravitatoria según la orientación [39, p. 82].

$$\varphi_x = \frac{\arctan(x/f)}{2} \tag{2-35}$$

$$\varphi_y = \frac{\arctan(y/f)}{2} \tag{2-36}$$

**Figura 72. Valor inicial para la inclinación φ$_x$ (en grados) de cada espejo de CTA-LST.**

| | | | | 10,4437 | 10,4437 | 10,4437 | 10,4437 | 10,4437 | | | | |
|---|---|---|---|---|---|---|---|---|---|---|---|---|
| | | 9,1722 | 9,1722 | 9,1722 | 9,1722 | 9,1722 | 9,1722 | 9,1722 | 9,1722 | | | |
| | 7,8867 | 7,8867 | 7,8867 | 7,8867 | 7,8867 | 7,8867 | 7,8867 | 7,8867 | 7,8867 | | | |
| | 6,5891 | 6,5891 | 6,5891 | 6,5891 | 6,5891 | 6,5891 | 6,5891 | 6,5891 | 6,5891 | 6,5891 | 6,5891 | |
| 5,2818 | 5,2818 | 5,2818 | 5,2818 | 5,2818 | 5,2818 | 5,2818 | 5,2818 | 5,2818 | 5,2818 | 5,2818 | 5,2818 | |
| 3,9662 | 3,9662 | 3,9662 | 3,9662 | 3,9662 | 3,9662 | 3,9662 | 3,9662 | 3,9662 | 3,9662 | 3,9662 | 3,9662 | 3,9662 |
| 2,6438 | 2,6438 | 2,6438 | 2,6438 | 2,6438 | 2,6438 | 2,6438 | 2,6438 | 2,6438 | 2,6438 | 2,6438 | 2,6438 | 2,6438 |
| 1,3153 | 1,3153 | 1,3153 | 1,3153 | 1,3153 | 1,3153 | 1,3153 | 1,3153 | 1,3153 | 1,3153 | 1,3153 | 1,3153 | 1,3153 |
| 0 | 0 | 0 | 0 | 0 | 0 | 0 | 0 | 0 | | | | |
| -1,3153 | -1,3153 | -1,3153 | -1,3153 | -1,3153 | -1,3153 | -1,3153 | -1,3153 | -1,3153 | -1,3153 | -1,3153 | -1,3153 | -1,3153 |
| -2,6438 | -2,6438 | -2,6438 | -2,6438 | -2,6438 | -2,6438 | -2,6438 | -2,6438 | -2,6438 | -2,6438 | -2,6438 | -2,6438 | -2,6438 |
| -3,9662 | -3,9662 | -3,9662 | -3,9662 | -3,9662 | -3,9662 | -3,9662 | -3,9662 | -3,9662 | -3,9662 | -3,9662 | -3,9662 | |
| -5,2818 | -5,2818 | -5,2818 | -5,2818 | -5,2818 | -5,2818 | -5,2818 | -5,2818 | -5,2818 | -5,2818 | -5,2818 | | |
| -6,5891 | -6,5891 | -6,5891 | -6,5891 | -6,5891 | -6,5891 | -6,5891 | -6,5891 | -6,5891 | | | | |
| -7,8867 | -7,8867 | -7,8867 | -7,8867 | -7,8867 | -7,8867 | -7,8867 | -7,8867 | | | | | |
| -9,1722 | -9,1722 | -9,1722 | -9,1722 | -9,1722 | -9,1722 | -9,1722 | | | | | | |
| -10,4437 | -10,4437 | -10,4437 | -10,4437 | -10,4437 | | | | | | | | |

**Figura 73. Valor inicial para la inclinación $\varphi_y$ (en grados) de cada espejo de CTA-LST.**

Con los datos de posiciones, curvaturas e inclinaciones para cada espejo ya es posible modelar el telescopio completamente. En la Figura 74 se muestra el aspecto de la ventana de OSLO con el modelo de CTA-LST introducido y su perfil simulado. Cada superficie se define con una fila en la ventana de *Surface Data*. Como se describió en el apartado 2.11, la primera de ellas (superficie 0) es el plano objeto (OBJ). En el plano objeto, situado en el infinito (a una distancia o *Thickness* de $10^{20}$ mm) está la fuente que produce en el telescopio una serie de rayos paralelos que se propagan hacia la derecha hasta llegar a la superficie 1 (AST o superficie de apertura).

**Figura 74. Captura de pantalla de la ventana de OSLO con el modelo de CTA-LST.**

En OSLO el modelado de un telescopio segmentado requiere la creación de un grupo no secuencial ya que el modo de funcionamiento es por defecto secuencial (ver introducción al apartado 2.11). Este grupo no secuencial lo constituirían todos los espejos que forman parte del reflector principal, ya que el trazado de rayos en diferentes segmentos debe seguir el mismo orden de secuencia. La primera y última superficie del grupo no secuencial son superficies sin ningún efecto, necesarias para indicar el tránsito de entrada y salida, y todas las demás van referenciadas a esa primera (en la Figura 74 la primera superficie del grupo no secuencial sería la superficie 2). Antes del grupo se requiere otra superficie, que es idéntica a la primera del grupo y que funciona como *stop* o diafragma de la apertura (AST), y después del grupo otra igual a la última con la misma función. Desde la superficie 3 en adelante se describen todos los espejos; por ejemplo, la superficie 3 es la correspondiente al espejo central [16], lo que puede comprobarse observando el radio de curvatura de -56 metros (nótese que todas las unidades están en mm). No visible en la captura de pantalla estarían definidas las superficies del resto de espejos, la superficie de salida del grupo no secuencial y por último el plano imagen (IMS).

Una vez modelado el telescopio, es posible comenzar a caracterizar su comportamiento óptico. En la Figura 75 se muestra el diagrama de spot en el plano focal del telescopio, con un tamaño de spot RMS cercano a 5 cm (las medidas en OSLO de tamaño de spot se refieren siempre al radio).

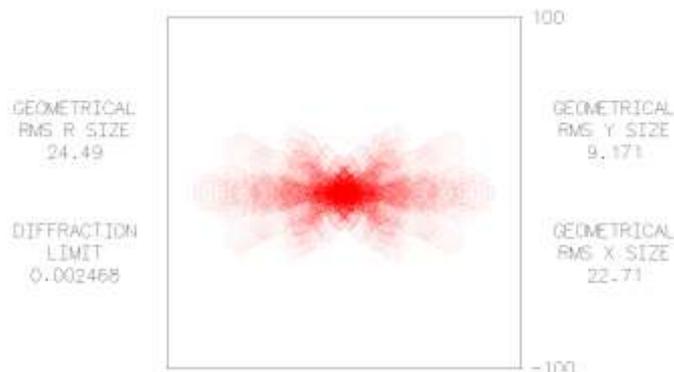

**Figura 75. Diagrama de spot (en mm) del modelo de CTA-LST en OSLO.**

Observando la forma del diagrama de spot de la Figura 75, se puede intuir que en la alineación de los espejos existe margen de mejora para concentrar la energía en un spot más pequeño. Para ello, se aprovechó la gran potencia de OSLO en cuanto a la capacidad de optimización de diseños ópticos. OSLO utiliza el algoritmo DLS (del inglés *Damped Least Squares*), también conocido como LMA (del inglés *Levenberg–Marquardt Algorithm*), que es el algoritmo de optimización más extendido en diseño óptico, empleado en alguna de sus variantes por todos los programas de óptica. En DLS la función de error $\varphi(x)$ se expresa como la suma ponderada de cuadrados según la ecuación (2-37) [175, p.194].

$$\varphi(x) = \sum_{i=1}^{m} w_i f_i^2(x) \qquad (2\text{-}37)$$

En la ecuación (2-37), el vector x representa un conjunto de variables de optimización $x = <x_1, x_2, x_3, \ldots, x_n>$ y $f_i$ son los llamados operandos. Cada operando $f_i$ contiene dos componentes que están relacionados por un operador matemático. Normalmente el primer componente es un desplazamiento de un rayo, el segundo componente es el valor objetivo y el operando es la resta. Los $w_i$ representan a los pesos que pueden usarse para ponderar la importancia relativa de cada operando.

---

[16] Nótese que el radio de apertura es de 866 mm, si bien los espejos tienen una distancia lado-lado de 1510 mm. Estos 866 mm se refieren al radio del círculo que circunscribe al hexágono que forma el espejo. El radio se calcula automáticamente en OSLO una vez se ha definido la forma de la superficie y solo describe el límite exterior, no la superficie reflectiva, que sería menor. Estos círculos se pueden observar en el diagrama del telescopio de la Figura 74. OSLO no muestra la forma real de las superficies especiales, sino únicamente este círculo exterior, lo que supone una dificultad adicional al diseñar este tipo de superficies.

El objetivo de DLS es minimizar la función de error $\varphi(x)$, que se consigue cuando todos los operandos se igualan a cero. La minimización de la función de error se lleva a cabo mediante un proceso iterativo basado en un modelo lineal ideal (en la realidad se utilizan derivadas de mayor orden para resolver las no linealidades) como el de la ecuación (2-38), que relaciona la dependencia entre variables y operandos.

$$f_i\left(x_j + \Delta x_j\right) = f_i\left(x_j\right) + \frac{\partial f_i}{\partial x_j} \Delta x_j \qquad (2\text{-}38)$$

Hay que señalar que DLS puede dar como resultado un mínimo local, que no necesariamente coincida con el mínimo global, por lo que es importante realizar una buena aproximación en la elección de las condiciones de partida (valores iniciales de las variables). En este sentido, el error más común en optimización, que lleva con frecuencia a la obtención de mínimos locales, es pretender eliminar alguna aberración en especial en lugar de balancear las existentes. Para ello, OSLO se encarga de definir unas relaciones entre operandos por defecto que consiguen solucionar este problema, y que dependen del tipo de optimización que se pretende obtener. En este caso, el tipo de optimización utilizada es la reducción del tamaño de spot, que está definida en una de las funciones de error ofrecidas por OSLO al usuario. Como variables se utilizaron las inclinaciones de los espejos en ambos ejes.

En el proceso de optimización no se obtuvieron buenos resultados intentando optimizar el telescopio completo de una vez. De hecho, los resultados empeoraron debido a que el número de variables es excesivo y la optimización de unas interfiere con las de otras sin lograr una optimización adecuada del conjunto. Para solucionarlo, se fue optimizando el telescopio por secciones u anillos salientes. Por ejemplo, una primera optimización se centró en el espejo central y los 6 contiguos del primer anillo, consiguiendo pasar del diagrama de spot de la izquierda de la Figura 76 al de la derecha. Así se redujo el radio del diagrama de spot desde 0,204 mm hasta 0,174 mm. Posteriormente se siguió con los 12 del segundo anillo, luego los 18 del tercero, los 24 del cuarto y así sucesivamente (ver la división del telescopio por anillos en la Figura 77).

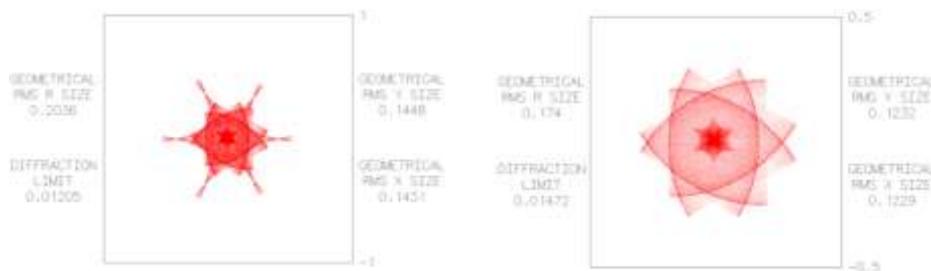

**Figura 76. Diagrama de spot (en mm) del primer anillo de espejos de CTA-LST antes ( izquierda) y después (derecha) de la optimización.**

En la optimización anterior se utilizaron como variables todas las inclinaciones de los espejos involucrados. Esto también puede influir negativamente en la optimización ya que la inclinación de espejos simétricos entre sí se compensa mutuamente de forma que no se obtiene un resultado óptimo. Para evitarlo, se establecieron relaciones entre espejos aprovechando la simetría del telescopio. Si se observa la Figura 72 (inclinaciones $\varphi_x$) y la Figura 73 (inclinaciones $\varphi_y$), se puede comprobar la existencia de estas simetrías. Por ejemplo, para el primer anillo de 6 espejos alrededor del central, se verifica que solo existen dos inclinaciones primarias y las demás, secundarias, están todas relacionadas. De esta forma, es posible minimizar el número de variables a modificar (en aproximadamente un factor cuatro) y se evitan las interferencias entre ellas en la optimización. En la Figura 77 se muestran los anillos en diferentes colores y en cada espejo el número de su superficie del modelo de OSLO, desde la 3 (espejo central) hasta la 201 (exterior derecha), lo que constituye los 198 espejos de LST. Los espejos en negro son los que contienen una inclinación primaria $\{\varphi_x, \varphi_y\}$, excepto los de la columna y la fila central, cuyas inclinaciones $\varphi_x$ y $\varphi_y$ son respectivamente iguales a cero.

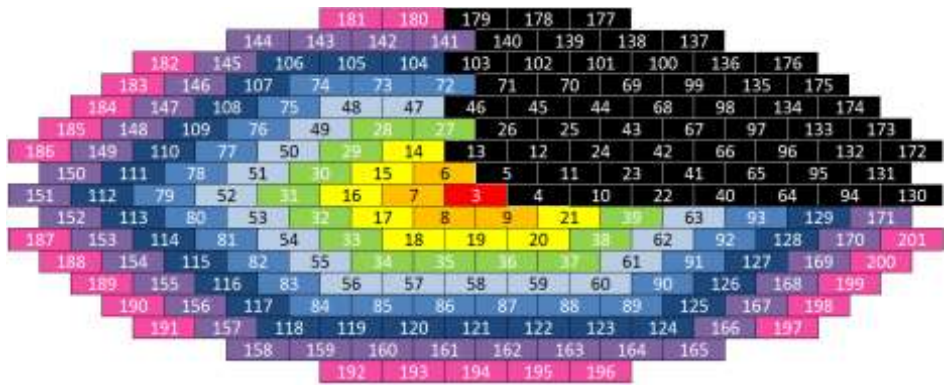

**Figura 77. Número de superficie en el modelo de OSLO por cada espejo y división del telescopio en anillos con las superficies en negro correspondientes a inclinaciones primarias.**

En cada anillo se establecen únicamente como variables las inclinaciones primarias, dejando fijas el resto de inclinaciones secundarias para que no interfieran con la optimización, y se va optimizando una variable cada vez, lo que resulta ser un proceso relativamente rápido al tratarse de un número relativamente reducido de variables. Una vez optimizada cada inclinación primaria, es necesario relacionar las secundarias con esta. Este proceso sería extremadamente tedioso realizarlo a mano dado el elevado número de espejos, cada uno con dos inclinaciones $\{\varphi_x, \varphi_y\}$, una por eje. Para evitarlo se recurrió a CCL, que es un lenguaje de programación derivado de C (se trata de un subconjunto de C) y que emplea OSLO para automatizar los diseños ópticos. El propio OSLO está programado en este lenguaje, por lo que las capacidades de automatización son considerables. Utilizando CCL, se programó una rutina que permitía relacionar todas las inclinaciones secundarias con las primarias. A continuación se muestra un ejemplo para el primer anillo:

```
cmd
LST_tilts_anillo_1 ()
        {
        Double V_TLA4;
        Double V_TLB4;
        Double V_TLA5;
        Double V_TLB5;

        V_TLA4 = tla[4];
        V_TLB4 = tlb[4];
        V_TLA5 = tla[5];
        V_TLB5 = tlb[5];

        tla 6  V_TLA5;
        tlb 6 -V_TLB5;
        tla 7  V_TLA4;
        tlb 7 -V_TLB4;
        tla 8 -V_TLA5;
        tlb 8 -V_TLB5;
        tla 9 -V_TLA5;
        tlb 9  V_TLB5;
        }
```

En esta rutina inicialmente se declaran las variables para inclinación de cada eje para los espejos correspondientes a las superficies 4 y 5 y se les asigna dicha propiedad mediante una lectura de los valores introducidos por el usuario en la interfaz del programa, ya descrita en la Figura 74. La primera capa la formarían los espejos de las superficies 4, 5, 6, 7, 8 y 9 (ver el esquema de la Figura 77 y la captura de pantalla de la Figura 74), siendo los primarios en cuanto a inclinación los espejos de las superficies 4 y 5 y los secundarios los de las superficies 6, 7, 8 y 9. Por ello, al final de la rutina se definen las inclinaciones para el eje x (tla = $\varphi_x$) y las del eje y (tlb = $\varphi_y$) de cada una de estas superficies secundarias en relación a las primarias definidas al principio. Este es el ejemplo del primer anillo, siendo homólogas las rutinas del resto de anillos en cuanto a las relaciones de los demás espejos del telescopio. Una vez seguido este método para optimizar el primer anillo, se pasó del diagrama de spot de la izquierda de la Figura 78 al de la derecha.

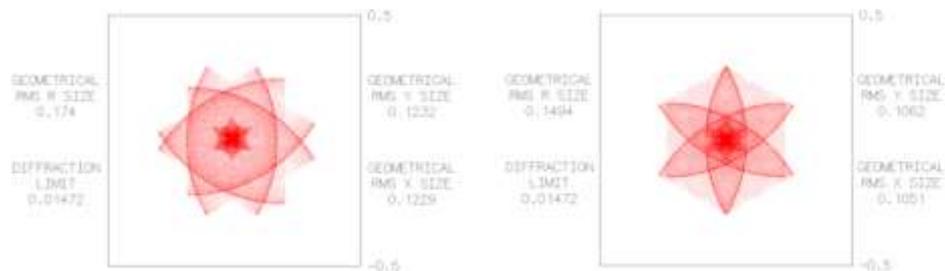

**Figura 78. Diagrama de spot (en mm) del primer anillo de espejos de CTA-LST con una optimización de todas las inclinaciones de forma independiente (izquierda) y de forma relacional (derecha).**

En la Figura 79 se muestra la mejora del segundo anillo antes y después de optimizar, ajustando en este caso una por una las variables primarias correspondientes a los espejos definidos en las superficies 10-13 y referenciando las inclinaciones de las superficies 14-21.

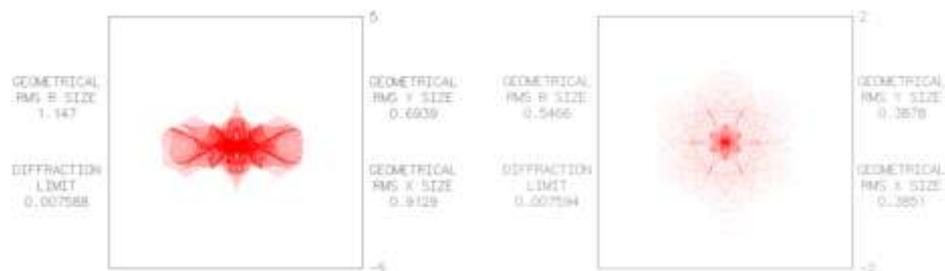

**Figura 79. Diagrama de spot (en mm) del segundo anillo de espejos de CTA-LST antes (izquierda) y después (derecha) de la optimización.**

En el resto de anillos el proceso es idéntico, obteniendo sucesivas mejoras del diagrama de spot. En la Tabla 8 se muestra el diagrama de spot conseguido tras realizar estas optimizaciones en cada anillo. Cada anillo se va circunscribiendo en el interior del siguiente, y se puede observar cómo el diagrama de spot va aumentando al ir añadiendo anillos cada vez más externos (porque equivale a ir disminuyendo progresivamente la relación focal del telescopio, al aumentar su apertura para una misma longitud focal). Precisamente por esta razón, es importante establecer la posición del plano focal utilizando únicamente el anillo periférico más externo: la mayor inclinación de los espejos situados en la periferia hace que su contribución al diagrama de spot sea mayor, por producir mayores aberraciones. Tras el proceso de optimización por anillos, se logra una reducción del diagrama de spot del telescopio completo desde 4,9 cm de diámetro hasta 1,2 cm.

**Tabla 8. Diagrama de spot (en mm) tras la optimización de cada anillo de CTA-LST.**

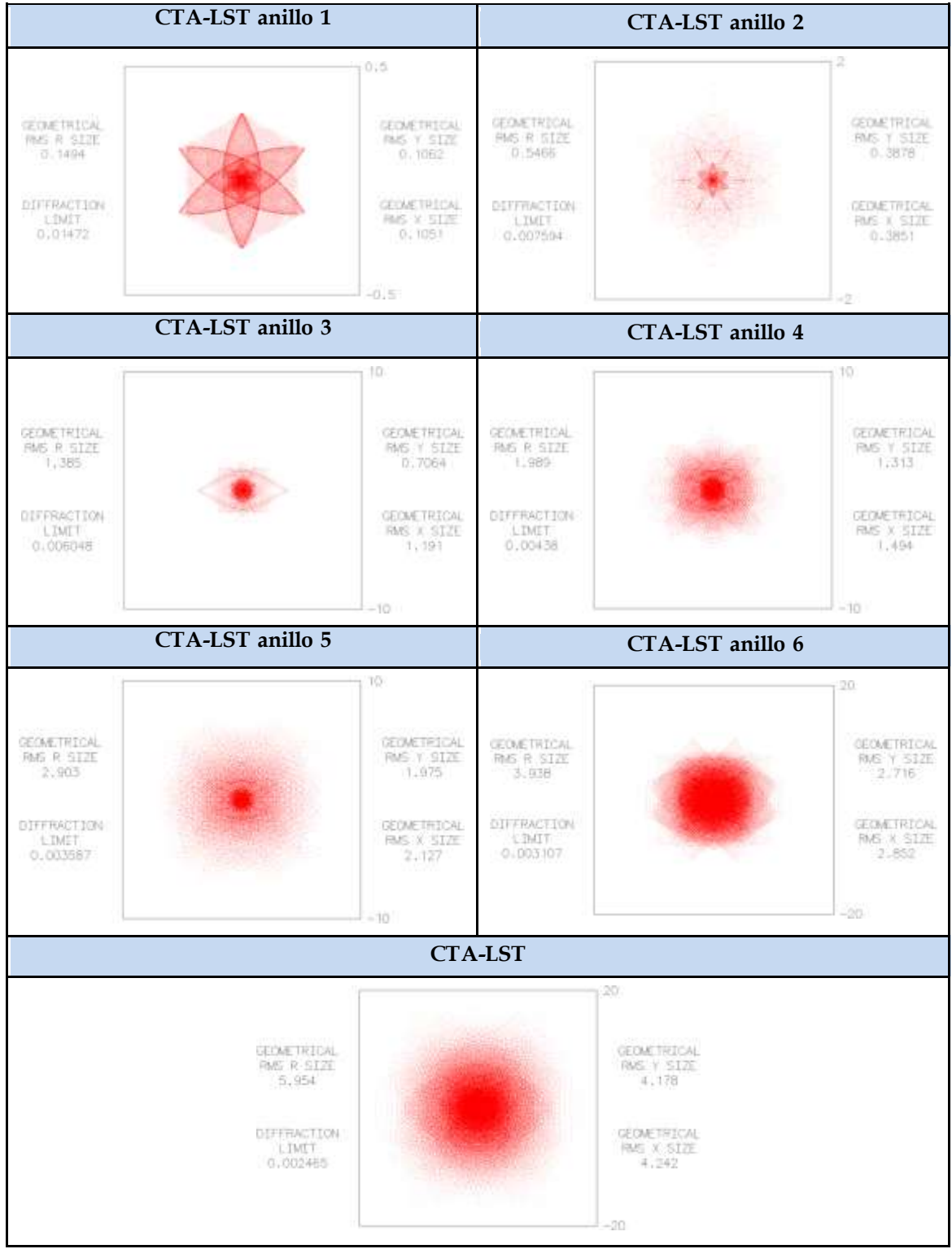

En algunos artículos [172, p. 2] [123, p. 3] se ha leído la propuesta de utilizar solo tres radios de curvatura diferentes en CTA-LST, de forma que los espejos de la parte central tendrían una longitud focal más corta y los de la zona externa más larga. Así, dividiendo a los espejos en tres grupos distintos, el radio de curvatura sería el mismo dentro de cada una de las tres zonas del telescopio, consiguiendo un compromiso entre coste y prestaciones ópticas. Dado que, aunque parece ser una propuesta firme, lo único que se he podido encontrar al respecto se refiere a este supuesto genérico sin especificar ningún radio de curvatura concreto, como una aproximación razonable se ha optado por dividir el telescopio en tres anillos con similar número de espejos y utilizar en cada anillo un promedio del radio de curvatura de todos los espejos del anillo al que pertenecen (Figura 80).

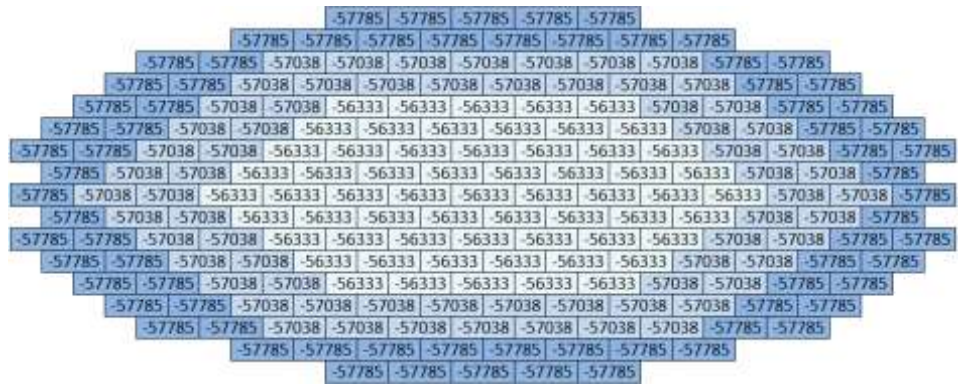

**Figura 80. Radio de curvatura (en mm) de cada espejo de CTA-LST según una división del telescopio en tres anillos.**

Tras llevar a cabo esta modificación, se puede apreciar (Figura 81) cierto aumento en el tamaño del diagrama de spot, como era de esperar. Sin embargo, este aumento queda limitado a poco más de un 4 %, demostrando ser una interesante estrategia para simplificar el proceso de fabricación de espejos sin degradar en exceso la calidad óptica.

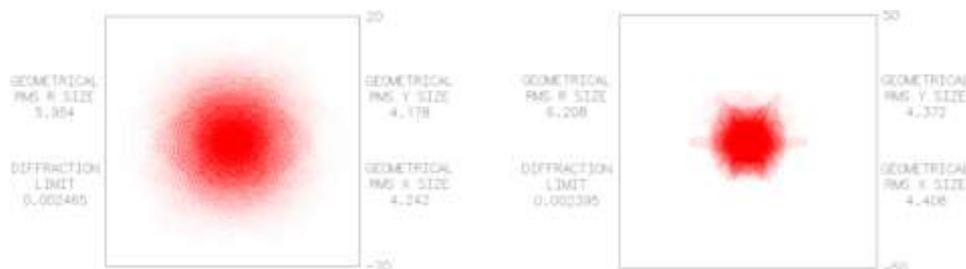

**Figura 81. Diagrama de spot del telescopio CTA-LST con múltiples radios de curvatura (izquierda) y con tres radios de curvatura (derecha).**

Para introducir los modelos de espejos realistas es necesario asociar a cada uno de los espejos del telescopio cada uno de los ficheros de deformación interferométrica generados según el algoritmo descrito en el apartado 2.11.1. Este sería un tedioso procedimiento a través de la interfaz de OSLO, por lo que se llevó a cabo editando directamente el archivo de definición del modelo del telescopio. Cada modelo en OSLO se almacena en un archivo de texto .len donde están descritas todas sus características: coordenadas, inclinaciones, curvaturas y demás propiedades de cada una de las superficies, así como los parámetros globales de configuración. Una de las líneas de código relativas a cada superficie se refiere a las deformaciones interferométricas, donde es posible indicar la ruta del archivo que le corresponde a cada superficie de los generados anteriormente con la rutina de Matlab.

Tras asociar una deformación diferente a cada espejo y simular su diagrama de spot se obtienen los resultados de la Figura 82, donde se puede apreciar un aumento, respecto al modelo con espejos ideales, de entre el 42 % y el 44 %, según se referencie el modelo con múltiples radios de curvatura o tres radios de curvatura. El tamaño final del diagrama de spot del telescopio LST resulta ser de 1,72 cm de diámetro en el primer caso y de 1,77 cm en el segundo, dos valores muy similares, por lo que se comprueba que al incluir el efecto de los espejos realistas en el modelo, el efecto de pasar de múltiples radios de curvatura a solo tres queda casi completamente enmascarado.

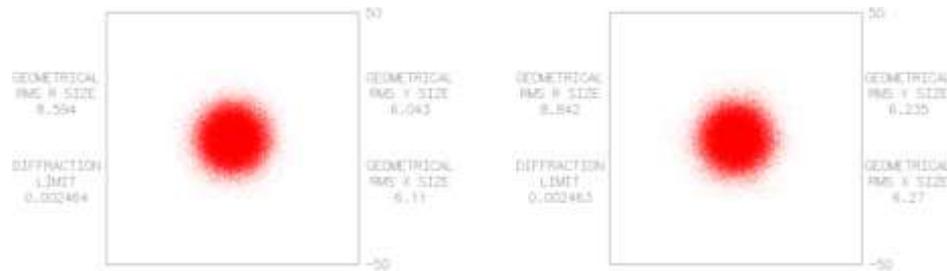

**Figura 82. Diagrama de spot del telescopio CTA-LST basado en el modelo de espejo realista con múltiples radios de curvatura (izquierda) y con tres radios de curvatura (derecha).**

Por último, hay que señalar que OSLO ofrece como medida del diagrama de spot únicamente un valor RMS, menor al equivalente a medir su tamaño geométrico en el plano imagen. Para utilizar un caso más desfavorable, y más parecido a otras simulaciones y medidas experimentales, se ha realizado en paralelo en *OpticsLab* y OSLO una serie de simulaciones de diferentes tipos de espejos en diferentes condiciones para comparar los resultados del diagrama de spot. De estas simulaciones se ha concluido que existe un factor de ~3,5 ± 10 % entre la medida de ambos diagramas de spot, por lo que los resultados de OSLO se han escalado con un factor de 3,5. En el caso de CTA-LST, el diámetro del diagrama de spot queda en 6,195 cm. Utilizando la ecuación (2-15), se puede calcular a qué campo de visión mínimo equivale esta resolución al considerar la longitud focal del telescopio, llegando a un campo de visión de 0,13° ó 2,24 mrad (ángulo completo).

### 2.11.3. CTA-MST *Davies-Cotton*

El tamaño de la apertura de los telescopios MST es el intermedio de los contemplados para CTA (Figura 83, izquierda), con un área reflectora efectiva de 104 m² formada por 86 espejos hexagonales de 120 cm de distancia lado-lado y un área de 1,25 m² cada uno. La relación focal de este telescopio es de f/1,3, con un diámetro equivalente de 12 metros, lo que da lugar a una focal de 15,6 metros [176, p. 2]. La forma del reflector sigue un perfil tipo Davies-Cotton [56] con todos los espejos esféricos y con igual radio de curvatura R = 2f = 31,2 m, dispuestos siguiendo una esfera de radio f = 15,6 m (Figura 83, derecha). Este diseño proporciona un buen comportamiento ante aberraciones geométricas cuando se utilizan los amplios campos de visión requeridos por la astronomía *Cherenkov*, pero a costa de presentar una alta dispersión temporal, por lo que no es aplicable en aperturas muy grandes como la de CTA-LST. En adelante se referirá a los telescopios MST tipo *Davies-Cotton* como MST-DC.

Como se hizo con el telescopio CTA-LST, para introducir el modelo en OSLO es necesario definir, para cada uno de los espejos, su posición {x, y, z}, su radio de curvatura R$_c$ y su inclinación referida a los dos planos de altitud y acimut. De nuevo, el primer paso es el cálculo de las posiciones, para lo cual se siguió el mismo método que en el caso del telescopio MAGIC II y CTA-LST. En este caso la curvatura de la estructura del telescopio donde se sitúan los espejos sigue un perfil esférico en lugar de parabólico, por lo que se empleó la ecuación (2-39), que define las coordenadas de una esfera centrada en el origen y con radio igual a R.

**Figura 83. Prototipo de un telescopio MST-DC (izquierda)
y esquema del perfil *Davies-Cotton* (derecha)**

$$x^2 + y^2 + z^2 = R^2 \qquad (2-39)$$

La topología de la superficie reflectora segmentada del telescopio MST-DC es la misma que la de CTA-LST (hexagonal), por lo que se repiten las simetrías de posición explicadas para aquel caso, y por ello basta con calcular las coordenadas de las dos primeras filas de cada eje, referenciando el resto de posiciones a estas. Así se obtuvieron las posiciones de los 86 espejos del telescopio MST-DC mostradas en la Figura 84 (eje x), Figura 85 (eje y) y Figura 86 (eje z).

**Figura 84. Coordenada x (en mm) de las posiciones de los espejos de MST-DC.**

**Figura 85. Coordenada y (en mm) de las posiciones de los espejos de MST-DC.**

**Figura 86. Coordenada z (en mm) de las posiciones de los espejos de MST-DC.**

Este telescopio CTA-MST es del tipo *Davies-Cotton*, por lo que sus espejos tienen un perfil esférico con idéntico radio de curvatura (31,2 metros) y no hay que calcularlo como en el caso de CTA-LST. Por lo tanto, solo falta el cálculo de los valores iniciales de las inclinaciones $\varphi_x$ (Figura 87) y $\varphi_y$ (Figura 88) para cada espejo. Para ello se siguió el mismo método descrito en el apartado anterior del telescopio LST.

**Figura 87. Valor inicial para la inclinación $\varphi_x$ (en grados) de cada espejo de MST-DC.**

**Figura 88. Valor inicial para la inclinación $\varphi_y$ (en grados) de cada espejo de MST-DC.**

Una vez obtenidos todos los parámetros característicos del telescopio MST-DC, se introdujeron en OSLO para modelar su comportamiento y se siguió el mismo método de optimización descrito en el apartado de CTA-LST. Tras este proceso, se obtuvieron los diagramas de spot de la Tabla 9, resultando para el caso del telescopio completo en un diámetro de spot igual a 4,3 mm.

**Tabla 9. Diagrama de spot (en m) tras la optimización de cada anillo de MST-DC.**

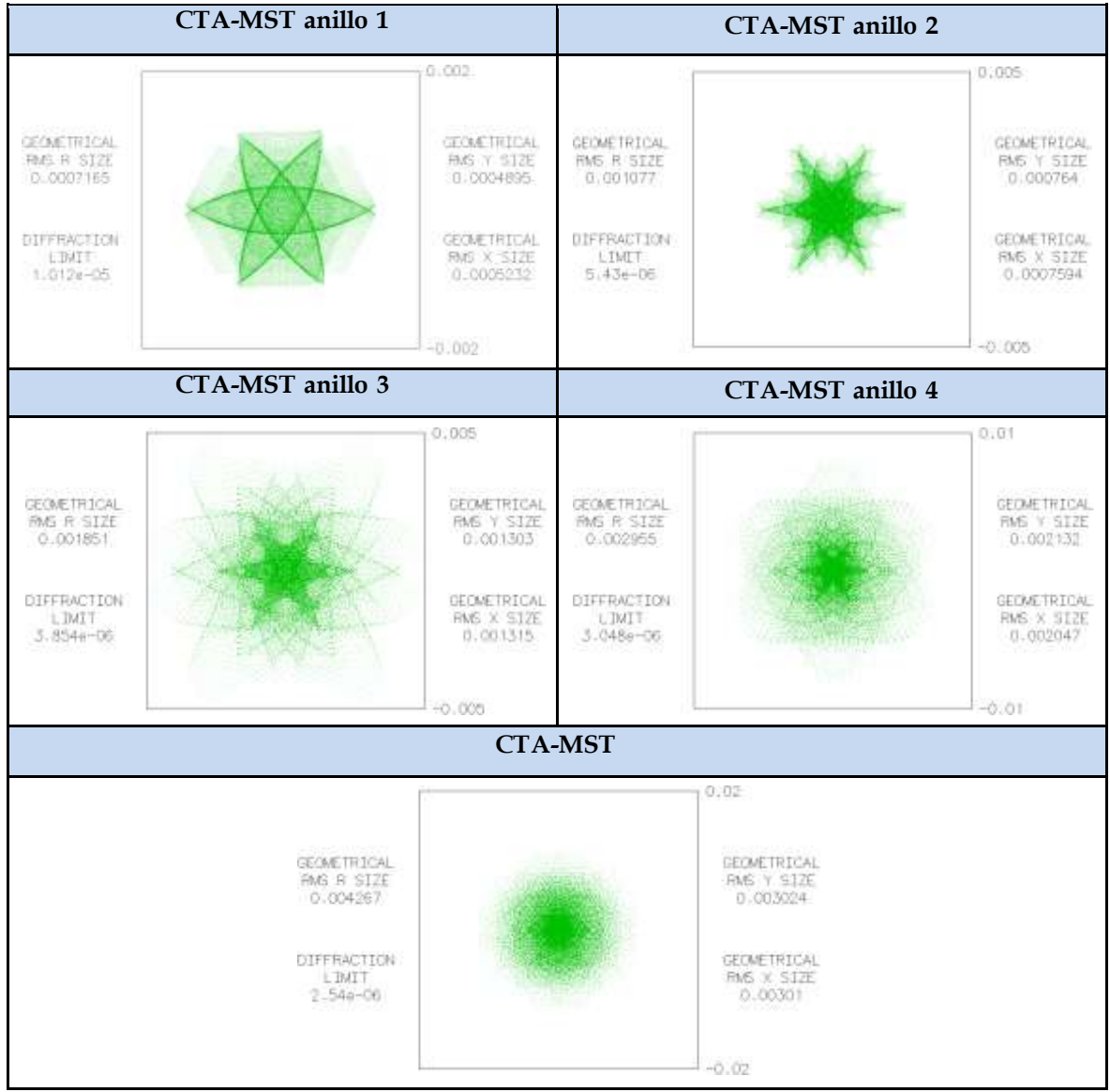

En la Figura 89 se muestra el diagrama de spot final del telescopio MST-DC al utilizar modelos realistas en cada uno de sus espejos. De nuevo se observa un empeoramiento de la resolución óptica, esta vez pasando de un diagrama de spot de 8,53 mm de diámetro (modelo de espejos ideales) a 9,77 mm. Aplicando la corrección de 3,5 para convertir su valor RMS, queda en 34,195 mm. Utilizando la ecuación (2-15), el campo de visión mínimo al que equivale esta resolución óptica es de 0,12° ó 2,19 mrad (ángulo completo).

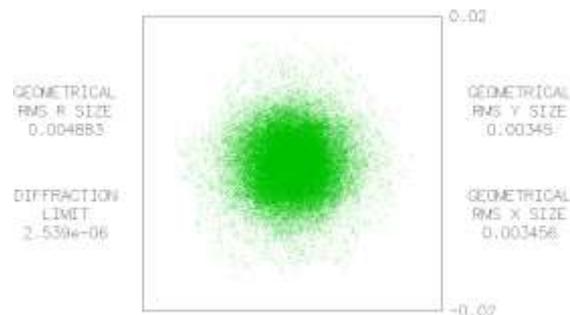

**Figura 89. Diagrama de spot del telescopio CTA-MST basado en el modelo de espejo realista.**

Por último, conviene señalar que como alternativa al diseño *Davies-Cotton*, el consorcio CTA propuso otro tipo de telescopio basado en una óptica de doble espejo en lugar de un reflector único. Este diseño es el *Schwarzschild–Couder*, explicado en el apartado 2.3.3. Pese a que se trata de un diseño innovador que nunca se ha empleado previamente en astronomía *Cherenkov*, sus buenos resultados preliminares lo han convertido en uno de los telescopios posibles para desplegar en CTA. En relación a su aplicación en FSOC, suponen una prometedora alternativa: por una parte, su diseño compacto los hace más apropiados para su adaptación a la operación diurna; por otra parte, gracias a su óptica asférica, este diseño brinda una menor PSF (si bien su objetivo de su diseño es reducir las aberraciones a lo largo de un gran campo de visión, más que reducirlas para los rayos en eje).

La mayor complejidad para caracterizar la óptica asférica, no ha hecho posible localizar la suficiente información como para llevar a cabo su modelado en OSLO. Por ello, se han utilizado directamente los datos finales de resolución óptica encontrados en la literatura científica disponible. La parte estadounidense del consorcio CTA es la encargada del diseño y desarrollo de los telescopio MST tipo *Schwarzschild–Couder* (en adelante MST-SC), para los que se prevé una apertura de 9,66 metros de diámetro con un bloqueo central debido al segundo espejo de 5,4 metros, una longitud focal de 5,6 metros y una PSF en eje de 0,017° ó 0,29 mrad [177]. Comparando estos datos con el MST-DC, se comprueba que el MST-SC proporciona una apertura con un área casi un 50 % menor (debido a su menor diámetro primario y al bloqueo central del segundo espejo), pero un campo de visión casi un orden de magnitud más reducido.

## 2.11.4. CTA-SST *Davies-Cotton*

El telescopio SST por defecto para CTA comparte con el MST el perfil *Davies-Cotton* y presenta la menor apertura de los tipos de telescopios descritos. El área reflectora efectiva es de 6,47 m² y está formada por 18 espejos hexagonales (Figura 90, izquierda) de 78 cm de distancia lado-lado y un área de 0,53 m². La relación focal de este telescopio es de f/1,4, lo que con un diámetro equivalente de 4 metros, da lugar a una focal de 5600 mm [178, p. 2]. Sus espejos son esféricos y de igual radio de curvatura R = 2f = 11,2 m, dispuestos siguiendo una esfera de radio f = 5,6 m (Figura 90, derecha).

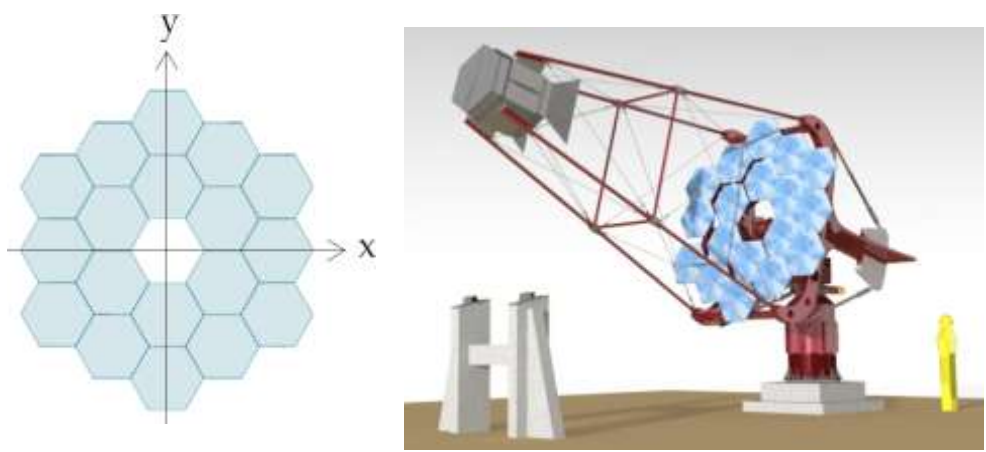

**Figura 90. Esquema de la apertura de un SST-DC (izquierda)
e ilustración de un prototipo del telescopio (derecha) [179].**

El cálculo de las posiciones {x, y, z}, radios de curvatura $R_c$ e inclinaciones referidas a los dos planos de altitud y acimut para cada espejo se realiza de forma idéntica a como se hizo en MST-DC, utilizando los valores característicos de SST-DC. Así se obtuvieron las posiciones de los 18 espejos del telescopio mostradas en la Figura 91 (eje x), Figura 92 (eje y) y Figura 93 (eje z), así como las inclinaciones de los espejos en el eje x (Figura 94) y en el eje y (Figura 95).

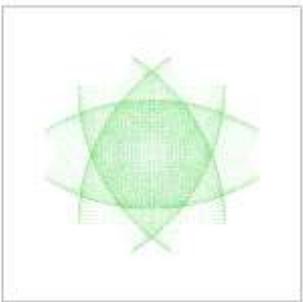

**Figura 91. Coordenada x (en mm) de las posiciones de los espejos de SST-DC.**

**Figura 92. Coordenada y (en mm) de las posiciones de los espejos de SST-DC.**

**Figura 93. Coordenada z (en mm) de las posiciones de los espejos de SST-DC.**

**Figura 94. Valor inicial para la inclinación $\varphi_x$ (en grados) de cada espejo de SST-DC.**

**Figura 95. Valor inicial para la inclinación $\varphi_y$ (en grados) de cada espejo de SST-DC.**

En la Tabla 10 se muestra el resultado de las simulaciones del diagrama de spot tras optimizar el primer anillo (izquierda) y el segundo (derecha) y en la Figura 89 tras incluir el modelo de espejo realista. Se comprueba que el diagrama de spot pasa de un diámetro de 6,59 mm a 9,81 mm, que tras aplicar la conversión de su valor RMS queda en 34,335 mm. Utilizando la ecuación (2-15), el campo de visión mínimo al que equivale esta resolución óptica es de 0,35° ó 6,13 mrad (ángulo completo).

**Tabla 10. Diagrama de spot (en m) tras la optimización de cada anillo de SST-DC.**

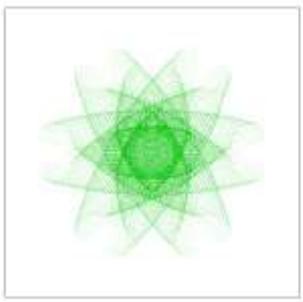

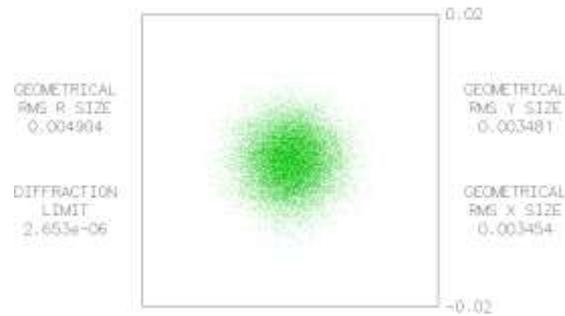

**Figura 96. Diagrama de spot del telescopio SST-DC basado en el modelo de espejo realista.**

Por último, al igual que en el caso del telescopio MST-SC, existe una propuesta para implementar una alternativa tipo *Schwarzschild–Couder* en los telescopios SST (en adelante referenciados como SST-SC). Por la misma razón que en MST-SC, no ha sido posible realizar el modelo en OSLO de estos telescopios, por lo que para las simulaciones posteriores se han utilizado directamente datos extraídos de la literatura científica relacionada. En el caso de los SST-SC, su diseño y desarrollo se asignó a dos grupos distintos, dando lugar a dos versiones alternativas: una de ellas es una colaboración entre Reino Unido y Francia, llamada GATE (del inglés *GAmma-ray Telescope Elements*) y la otra es una colaboración italiana, llamada ASTRI (del italiano *Astrofisica con Specchi a Tecnologia Replicante Italiana*).

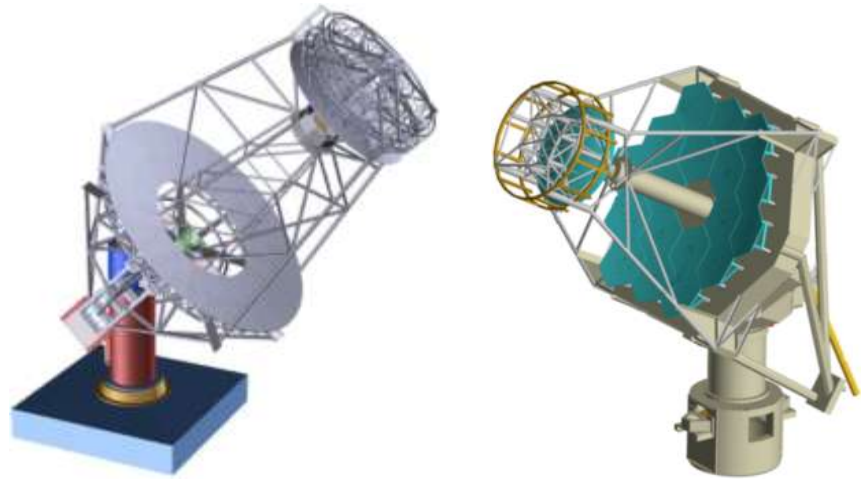

**Figura 97. Ilustración de un SST-SC tipo GATE (izquierda) y tipo ASTRI (derecha).**

El telescopio SST-SC de GATE (Figura 97, izquierda) está basado en un espejo primario segmentado de 4 metros de diámetro compuesto por seis espejos en forma de pétalo y un espejo secundario segmentado de 2 metros de diámetro con el mismo teselado de seis pétalos, una longitud focal de 2,283 metros y una PSF en eje de 0,04° ó 0,7 mrad [180] [181]. El telescopio SST-SC de ASTRI (Figura 97, derecha) está basado en un espejo primario segmentado de 4,306 metros de diámetro compuesto por 18 espejos hexagonales y un espejo secundario monolítico de 1,8 metros, una longitud focal de 2,15 metros y una PSF en eje de 0,06° ó 1,05 mrad [182] [183]. Comparado con el SST-DC, los telescopios SST-SC ofrecen una apertura con un área entre un 31 % (SST-SC GATE) y un 46 % (SST-SC ASTRI) menor, pero con una PSF entre 9 (SST-SC GATE) y 6 (SST-SC ASTRI) veces más reducida.

# 2.12. EVALUACIÓN DE UN ENLACE BASADO EN IACT

En los apartados anteriores se ha analizado en detalle cuáles serían las limitaciones de un telescopio *Cherenkov* al ser empleado como un receptor de comunicaciones ópticas con las pertinentes adaptaciones. Tras identificar la resolución óptica como el factor limitante en este tipo de telescopios, se han modelado los principales telescopios de CTA para caracterizar su resolución óptica. Aún siendo inferiores en este sentido a los telescopios convencionales, y en parte debido a esta razón, los IACT son mucho más baratos de producir. Esto será especialmente así en el marco de un proyecto como CTA en el que se va a llevar a cabo el equivalente a una producción en cadena de grandes telescopios. Por ello, una propuesta que sugiera la utilización de IACT para comunicaciones debería ser capaz de estimar cómo se comportarán estos telescopios en diferentes contextos de trabajo y si serán capaces de, pese a sus limitaciones, cumplir con su función de forma adecuada en un enlace de comunicaciones con condiciones realistas y desfavorables.

El objetivo de este apartado es precisamente ese: utilizar los telescopios de CTA analizados en una serie de escenarios como los planeados para las futuras estaciones receptoras terrenas de enlaces de FSOC de espacio cercano y espacio profundo y evaluar su desempeño. Así es posible predecir si estos telescopios son válidos tal cual, y en su caso comparar el resultado de los diferentes tipos para determinar la alternativa que mejor se adapte a esta aplicación. Para ello, se ha llevado a cabo el análisis de enlace habitual en FSOC, junto a una estimación de la calidad del enlace, asumiendo en cada caso unas condiciones desfavorables en cada uno de los diferentes escenarios. Estos escenarios se han basado en el informe del OLSG (del inglés *Optical Link Study Group*) publicado en diciembre de 2011 [72]. Este informe es el resultado de una colaboración multinacional entre los principales grupos de investigación en comunicaciones ópticas en espacio profundo. El OLSG lo componen los grupos de FSOC de las agencias espaciales estadounidense (NASA), europea (ESA), alemana (DLR), francesa (CNES), japonesa (JAXA) y coreana (KARI). En este importante estudio se proponía iniciar una colaboración entre estos grupos con el objetivo de armonizar tecnologías y compartir sus costes, así como la utilización de una infraestructura de comunicaciones ópticas común que pudiera dar soporte a las futuras misiones de espacio profundo. La tecnología de los terminales remotos se identificaba como suficientemente madura, y se señalaba la necesidad de investigar las diferentes soluciones para implementar las estaciones terrestres. Conviene resaltar que en la presentación final del informe del OLSG en junio de 2012 [184], se marcó como una de las líneas prioritarias a seguir la investigación relativa a estaciones terrenas, específicamente el estudio de adaptación de observatorios astronómicos existentes o fuera de uso para su uso como receptores de FSOC.

## 2.12.1. Cálculo de balance de potencia y relación señal-ruido

El balance de potencia es la herramienta más importante en el diseño de cualquier enlace de comunicaciones ópticas [17] ya que evalúa la capacidad del enlace para aprovechar en el receptor la potencia que se emplea en el transmisor para transmitir la información. Para calcularlo se utiliza la ecuación (2-40), conocida como ecuación de enlace o de balance de potencia. Esta ecuación [4, p. 89], particularizada en este caso para el enlace de bajada, pero válida en general para bajada y subida, toma en consideración todas las aportaciones de cada elemento del enlace.

---

[17] Si bien en los enlaces de comunicaciones ópticas guiadas es necesario realizar dos balances de enlace (de potencia y de dispersión), en FSOC siempre están limitados por potencia y por ello esta es la herramienta fundamental en este tipo de enlaces.

$$P_r(dBm) = P_t(dBm) + G_t(dB) - L_t(dB) - L_{tap}(dB) - L_{el}(dB)$$
$$- L_{atm}(dB) + G_r(dB) - L_r(dB) - L_{rap}(dB) \qquad (2\text{-}40)$$

En la ecuación (2-40),

- $P_r(dBm)$ es la potencia total a la entrada del receptor.

- $P_t(dBm)$ es la potencia media transmitida desde el terminal remoto.

- $G_t(dB)$ es la ganancia de la apertura transmisora.

- $L_t(dB)$ es la pérdida interna del transmisor.

- $L_{tap}(dB)$ es la pérdida por apuntamiento del transmisor.

- $L_{el}(dB)$ es la pérdida por espacio libre.

- $L_{atm}(dB)$ es la pérdida debida a la propagación atmosférica.

- $G_r(dB)$ es la ganancia de la apertura receptora.

- $L_r(dB)$ es la pérdida interna del transmisor.

- $L_{rap}(dB)$ es la pérdida por apuntamiento del receptor.

De estos términos, las ganancias $G_t$ y $G_r$ se calculan a partir de los diámetros $D_t$ y $D_r$ de sus respectivas aperturas y de la longitud de onda $\lambda$ mediante las ecuaciones (2-41) y (2-42), respectivamente. Las pérdidas de espacio libre $L_{el}$ se calculan a partir de la distancia del enlace d y la longitud de onda $\lambda$ mediante la ecuación (2-43).

$$G_t(dB) = 10\log\left(\frac{4\pi D_t}{\lambda^2}\right) \qquad (2\text{-}41)$$

$$G_r(dB) = 10\log\left(\frac{4\pi D_r}{\lambda^2}\right) \qquad (2\text{-}42)$$

$$L_{el}(dB) = 10\log\left(\frac{\lambda}{4\pi d}\right)^2 \qquad (2\text{-}43)$$

Cuando la comunicación tiene lugar en presencia de un elevado nivel de ruido de fondo debido a la necesidad de operación diurna, como es el caso estudiado, el parámetro fundamental para evaluar la calidad del enlace es la relación señal a ruido SNR (del inglés *Signal to Noise Ratio*). Para su cálculo no solo es necesario obtener la potencia de señal recibida $P_r$ a partir del balance de potencia y la potencia de ruido (a lo que se dedica el siguiente apartado), sino que hay que considerar el tipo de modulación que se utiliza para transmitir la información. Como se explicó en el apartado 2.4.2, existe un consenso entre las principales agencias espaciales en la utilización de la técnica PPM (del inglés *Pulse Position Modulation*) en enlaces de espacio profundo como los tratados en esta tesis, por lo que se asume esta modulación para el cálculo de la relación señal a ruido del enlace.

Es necesario calcular el número de fotones recibidos por pulso $n_r$ utilizando la ecuación (2-44) [185, p. 147], donde $P_r$ es la potencia de la señal recibida en Watios durante un tiempo T en segundos, $\eta_c$ es la eficiencia cuántica de la detección fotoeléctrica y $h\nu$ es la energía de un fotón (siendo h la constante de Plank y $\nu$ la frecuencia, ambas en unidades del Sistema Internacional), que se puede expresar como $\lambda/hc$ (siendo c la velocidad de la luz en el vacío).

$$n_r = \eta_c \, P_r \, T \, \frac{1}{h\nu} = \eta_c \, P_r \, T \, \frac{\lambda}{hc} \qquad\qquad (2\text{-}44)$$

Para distinguir entre el cálculo del número de fotones por pulso de señal $n_{rs}$ y de ruido $n_{rr}$, es necesario tener en cuenta el formato de la modulación. Según la ecuación (2-44), para conocer el número de fotones, hay que considerar la potencia recibida $P_r$ en un intervalo de tiempo T. En el caso de la potencia de señal, al usar una modulación PPM, la potencia se recibe únicamente durante el tiempo de slot $T_s$ y esa potencia recibida es la potencia pico $P_{pico}$. Así, utilizando la ecuación (2-1) en la ecuación (2-44), sabiendo que un intervalo de trama T se divide en m slots de duración $T_s$ (por lo tanto, $T_s = T/m$), y que en cada símbolo o pulso se codifican n bits, por lo que el régimen binario $R_b$ se calculará como $R_b = n/T$, se llega a la ecuación (2-45).

$$n_{rs} = \eta_c \, P_{pico} \, T_s \, \frac{\lambda}{hc} = \eta_c \, \frac{m P_{media}}{n} \, \frac{T}{m} \, \frac{\lambda}{hc} = \eta_c \, \frac{P_{media}}{n} \, \frac{n}{R_b} \, \frac{\lambda}{hc} = \eta_c \, \frac{P_{media}}{R_b} \, \frac{\lambda}{hc} \quad (2\text{-}45)$$

En cuanto al número de fotones de ruido por pulso $n_{rr}$, se parte de nuevo de la ecuación (2-44). En este caso, la potencia de ruido del cielo $N_s$ es una potencia media que se integra únicamente durante la duración del slot $T_s$, según la ecuación (2-46). En estos cálculos se comprueba la idoneidad de este tipo de modulaciones para un canal ruidoso y limitado por potencia. Al reducir el tiempo en que se transmite la información, se aprovecha mucho mejor la potencia óptica empleada y se minimiza el número de fotones de ruido.

$$n_{rr} = \eta_c N_s T_s \frac{\lambda}{hc} = \eta_c N_s \frac{T}{m} \frac{\lambda}{hc} = \eta_c \frac{n}{m} \frac{N_s}{R_b} \frac{\lambda}{hc} \qquad\qquad (2\text{-}46)$$

La tasa de fotones por pulso de señal en relación a los de ruido sirve para estimar la relación señal a ruido del enlace[18]. Además, es un parámetro cuyo límite es más fácil de visualizar: no se han considerado viables enlaces con menos de un fotón por pulso debido al gran número de errores o la reducción de la tasa binaria que implicaría en la práctica. La detección de un número de fotones por pulso igual o mayor a uno se ha demostrado experimentalmente factible utilizando detectores tipo SPAD (*Single Photon Avalanche Detector*), similares a los que se usan rutinariamente en enlaces de comunicaciones cuánticas como el descrito en el segundo bloque de esta tesis. El límite de un fotón por pulso es una frontera muy útil ya que depende de la tecnología y de un criterio de diseño fijo, no así como la potencia, cuyo límite dependerá de parámetros de diseño como la tasa binaria y la longitud de onda. Por ejemplo, usando la ecuación (2-45) se comprueba que asumiendo una eficiencia cuántica del 50 % y una longitud de onda de 1550 nm, para operar al mismo límite de un fotón por pulso, es necesaria una potencia $P_r$ de -125 dBm para transmitir a 1 kbps, de -95 dBm para 1 Mbps o de -65 dBm para 1 Gbps.

---

[18] En este cálculo se está asumiendo que la única aportación de ruido al enlace proviene de la radiancia del cielo. Pese a que en realidad existirán otras fuentes de ruido, en la estimación de esta tesis se está asumiendo que esta contribución será la predominante, muy por encima de otras como el ruido cuántico, ruido térmico, etc., cuya minimización se da por supuesta durante la construcción.

En la práctica, se exigirá un número de fotones por pulso mayor que el límite mencionado. Esto es especialmente cierto en presencia de ruido de fondo, ya que no todos los fotones recibidos son de señal. Por ello, el parámetro de calidad que caracteriza al enlace es la relación entre fotones de señal y ruido. Si se modela la llegada de fotones de ruido de fondo como el resultado de eventos independientes con una tasa de llegada constante en promedio, se suele utilizar una estadística de *Poisson* para describir el proceso. En la ecuación (2-47) se muestra la formulación de esta estadística, donde N es la tasa media de fotones de ruido por segundo y p (n, t) es la probabilidad de detectar n fotones de ruido en el área del detector en un tiempo t [33, p. 24].

$$p(n,t) = \frac{(Nt)^n\, e^{-Nt}}{n} \qquad\qquad (2\text{-}47)$$

Para discriminar una señal en presencia de ruido de fondo lo importante es la variación de ese ruido. Por ello, para evaluarlo se utiliza la desviación estándar, ya que el valor medio del ruido puede ser eliminado de la señal. Como se muestra en la ecuación (2-48), en una distribución de *Poisson* la desviación estándar es igual a la raíz del número medio de fotones.

$$\sigma = \sqrt{Nt} \qquad\qquad (2\text{-}48)$$

Asumiendo, como es habitual, que los errores en la detección se distribuyen siguiendo una estadística gaussiana, entonces al comparar el número de fotones de señal con la raíz del número de fotones de ruido, una SNR = 1 implica una probabilidad de error del 32 %; una SNR = 3, una tasa de errores de $10^{-3}$; una SNR = 6, una tasa de errores de $10^{-6}$; etc. En un enlace de comunicaciones se suele elegir una probabilidad de error nunca mayor que $10^{-3}$, por lo que se utilizará este límite para la estimación de la calidad del enlace.

## 2.12.2. Evaluación de la potencia de ruido de fondo

El objetivo de este apartado es evaluar la potencia de ruido de fondo, con la que relacionar la potencia de señal obtenida según el cálculo descrito en el apartado anterior relativo al balance de potencia, para obtener así una relación señal a ruido que caracterice el enlace de comunicación. En el apartado 2.9.1 se explicó cómo en la atmósfera tienen lugar dos tipos de dispersión de la luz o *scattering* (*Mie* y *Rayleigh*) según sea el tamaño de la partícula en suspensión en relación a la longitud de onda. En observatorios astronómicos, situados por encima de la zona de mayor concentración de aerosoles, el efecto de esta dispersión sobre la potencia de la señal se puede considerar despreciable.

El efecto indirecto sí es muy importante, especialmente durante el día [19]. El *scattering* que sufre la radiación de cualquier fuente lumínica, muy en especial el Sol, acaba haciendo que la luz dispersada se acople al campo de visión del receptor en forma de ruido de fondo, cuyo efecto será el de reducir la relación señal a ruido del enlace de comunicación. A esta aportación se la conoce como radiancia del cielo L ($\lambda$, $\theta$, $\varphi$) y se expresa en W/(cm² ·srad ·µm). A veces se la denomina radiancia espectral para distinguirla de la radiancia expresada en W/(cm² · srad). La radiancia espectral del cielo será el parámetro de interés, ya que conocido su valor, es posible calcular la potencia de ruido de fondo utilizando la ecuación (2-14). En esta ecuación será necesario además especificar el área de la apertura $A_r$, calculada con la ecuación (2-49) a partir de su diámetro D (cm), así como la anchura espectral $\Delta\lambda$ (µm) del filtro óptico y el campo de visión $\Omega_{FOV}$ (srad) de la apertura receptora. Este último se expresa en estereorradianes, por lo que se utilizará la ecuación (2-50) [186, p. 3] para convertir el campo de visión $\theta_{FOV}$ obtenido en las simulaciones de OSLO en radianes de geometría plana, resultando el cálculo de la potencia de ruido $N_S$ en la ecuación (2-51).

---

[19] Durante la noche también se produce cierta dispersión de la luz debida en este caso a cuerpos celestes. Es posible estudiar su influencia con cálculos derivados del albedo de las superficies de estos cuerpos y de su geometría respecto a la Tierra. Sin embargo, incluso en escenarios desfavorables, esta aportación queda varios órdenes de magnitud por debajo de la contribución solar durante el día [4, p. 168].

$$A_r = \frac{\pi D^2}{4} \qquad\qquad (2\text{-}49)$$

$$\Omega_{FOV}\left(\text{srad}\right) = \frac{\pi}{4}\,\theta_{FOV}^2 \qquad\qquad (2\text{-}50)$$

$$N_S = L(\lambda, \theta, \varphi)\left(\frac{\pi D \cdot \theta_{FOV}}{4}\right)^2 \Delta\lambda \qquad\qquad (2\text{-}51)$$

En la literatura científica relacionada con las comunicaciones ópticas desde espacio profundo existe abundante documentación de los niveles de radiancia del cielo para una variedad de escenarios y circunstancias [187] [188]. Sin embargo, el grueso de las publicaciones de esta disciplina proviene de los grupos de referencia de EEUU, que como se explicó en el apartado 2.4.2, tradicionalmente optaron por una longitud de onda de 1064 nm. Por ello, en esta tesis se ha simulado directamente la radiancia del cielo para obtener unos valores correspondientes a la longitud de onda de trabajo de 1550 nm. Para ello, se ha utilizado el *software* MODTRAN (del inglés *MODerate resolution atmospheric TRANsmission*), que se trata de un simulador desarrollado por *Spectral Sciences Inc.* [189] para modelar la propagación atmosférica de la radiación electromagnética, en un rango espectral que abarca desde el ultravioleta medio hasta el infrarrojo lejano. La base de datos atmosféricos de MODTRAN es la referencia habitual para los cálculos de ruido de fondo en enlaces de FSOC [4, pp. 151-168].

Con el objetivo de validar los resultados generados en MODTRAN, en la Figura 98 se muestra a la izquierda un resultado obtenido directamente del texto de referencia en FSOC [4, p. 156] y a la derecha el resultado de realizar la simulación en MODTRAN utilizando los mismos parámetros. Esta simulación se corresponde con la radiancia del cielo observada en un emplazamiento a nivel del mar caracterizado por el modelo rural de aerosoles (habitual en observatorios astronómicos) con una visibilidad de 23 km, un ángulo cenital solar de 45˚ (el ángulo que subtiende la posición relativa del Sol en el cielo respecto del cénit) y para ángulos cenitales de observación de 10˚, 40˚ y 70˚. Como se puede apreciar, el caso intermedio de 40˚ recibe la mayor radiancia al estar más próximo a la posición del Sol de 45˚. También sucede, como se comprobará más adelante, que el *scattering* induce mayores radiancias recibidas desde direcciones más próximas al horizonte que al cénit, si bien la mayor parte de la radiancia observada desde 70˚ en comparación con la observada desde 10˚ se debe a la menor distancia al Sol.

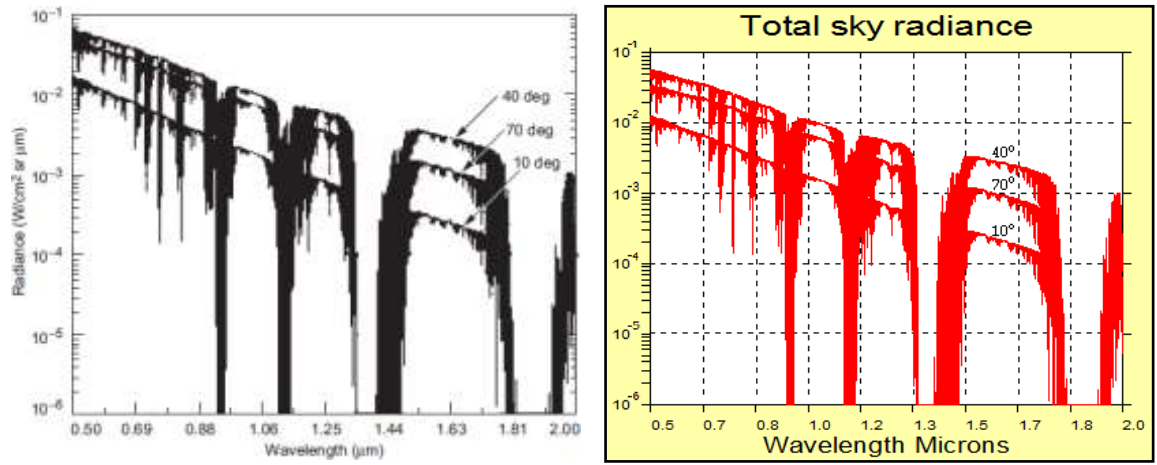

**Figura 98. Radiancia del cielo al nivel del mar al utilizar un modelo rural de aerosoles con visibilidad de 23 km, ángulo cenital solar de 45 ° a diferentes ángulos cenitales de observación, obtenido de [4, p. 156] (izquierda) y de MODTRAN (derecha).**

Para realizar las simulaciones es preciso establecer un conjunto de condiciones para la masa de aire modelada, la geometría de la propagación, etc. Siguiendo la metodología del OLSG, se establecerá un ángulo mínimo entre el Sol y el punto de observación (ángulo SEP, del inglés *Sun-Earth-Probe*) igual a 5° [20] y un ángulo cenital de observación máximo igual a 70°. Respecto al modelo de aerosol, se elegirá uno marítimo como el de los observatorios de Izaña o Roque de los Muchachos en las Islas Canarias, dado que estos son emplazamientos candidatos para CTA norte y buenos representantes del tipo de ubicación de estos telescopios. Para este tipo de modelo de aerosoles, Modtran establece una visibilidad de 23 km. Se sabe que en esta zona existen entradas de masas de aire procedentes del Sáhara, características de un modelo desértico, más desfavorable que el marítimo. No obstante, en la mayoría de los casos la altitud del observatorio (se ha tomado 2390 metros para las simulaciones [190], altura promedio de Izaña, con Roque de los Muchachos 6 metros por encima) hace que estos aerosoles no afecten a la propagación al concentrarse en su mayoría a menores alturas [191].

El modo de ejecución de MODTRAN aplicable en este caso es "*radiance w/Scattering*", que computa la radiancia recibida debido a la dispersión de la luz del sol, y para el tipo de camino óptico atmosférico se ha seleccionado "*slant path to space or ground*", apropiado para enlaces descendentes [192, p. 102]. No se han usado modelos de lluvia o nubes, que pueden llegar a bloquear los enlaces, porque las estrategias para minimizar su efecto en FSOC se basan en la diversidad espacial entre sitios no correlacionados estadísticamente (por ejemplo, una disponibilidad de enlace superior al 90 % en un enlace desde espacio profundo requiere un número mínimo de 4 estaciones terrestres distribuidas por la superficie de la Tierra [186, p. 3]). La geometría del camino óptico es "*observer height, zenith angle*", con los 2390 metros mencionados como primer parámetro y un barrido de 0° a 90° en el segundo (que más adelante se limitará a 70° para adaptarse a las condiciones del

---

[20] Establecer un ángulo SEP mínimo implica la necesidad de convivir con un bloqueo del enlace durante una fracción del tiempo de misión. El porcentaje de este bloqueo dependerá de la geometría de la misión, determinada por la ubicación del transmisor. Considerando un promedio para los planetas del Sistema Solar, un ángulo SEP de 5° equivale aproximadamente a una disponibilidad del 98 % [186, p. 2]. Hay que destacar que esta consideración es un tanto conservadora, habiéndose propuesto ángulos SEP menores a 5°.

OLSG). Se ha establecido un número de onda de 6451 cm[-1] (correspondiente a una longitud de onda de 1550,15 nm). El ángulo acimutal "*azimuth angle at observer LOS to Sun*" sirve para definir el plano de observación, por lo que un ángulo de 0° establecerá el plano que se define con las direcciones del cénit al observador y de este al Sol, conteniendo a ambas. Este es el caso habitual de simulación en este tipo de problemas al suponer un caso peor, ya que el efecto de un ángulo diferente (aunque estadísticamente más probable en la práctica) sería de disminución de la radiancia de cielo recibida. En las simulaciones se elimina la radiancia solar directa, de varios órdenes de magnitud por encima de la radiancia del cielo, pero con influencia únicamente en el ángulo subtendido por el Sol en el cielo, de 0,5° de ángulo completo. Es decir, influirá solo a 0,25° alrededor del SZA, y por lo tanto quedará eliminado por la restricción del ángulo mínimo SEP de 5°.

Para simular diferentes situaciones se realizará el mencionado barrido del ángulo de observación para un ángulo solar cenital determinado SZA (del inglés *Solar Zenith Angle*), que marca la dirección del Sol en el cielo. Por ejemplo, en la Figura 99 se ha representado la radiancia del cielo entre 800 y 1700 nm para un ángulo solar cenital de 45˚, con un barrido de ángulos cenitales de observación de 0˚ a 90˚ en pasos de 5˚. De nuevo, como ocurre siempre en estas simulaciones, la curva de mayor radiancia se corresponde con la situación en que el ángulo acimutal es igual a 0˚ y el ángulo cenital solar toma el valor más cercano al ángulo cenital de observación, en este caso exactamente igual a 45˚ en el caso de ambos. Nótese cómo la radiancia a 1550 nm queda unas tres veces por debajo de la radiancia a 1064 nm, lo que ilustra una de las ventajas de esta longitud de onda.

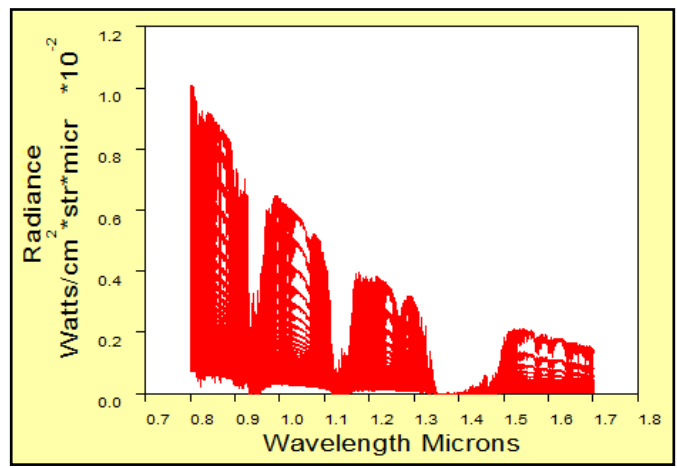

**Figura 99. Radiancia espectral del cielo para un SZA de 45˚, un ángulo acimutal de 0˚ (plano principal) y ángulos cenitales de observación de 0˚ a 90˚ en pasos de 5˚.**

En la Figura 100 se muestra la radiancia espectral del cielo como resultado de simular una masa de aire como la descrita en los párrafos anteriores, particularizando en 1550 nm para un ángulo acimutal de 0°, un ángulo cenital de observación de 0° a 90° y un ángulo solar cenital de 5° a 85°. Se puede comprobar cómo la máxima radiancia se corresponde con la coincidencia del ángulo cenital de observación y el ángulo cenital solar, reduciéndose muy fuertemente cuando la dirección de observación no coincide con el Sol, hasta entre dos y tres órdenes de magnitud cuando hay cerca de 90° entre ambos ángulos. Esto significa que a lo largo de un día de operación las condiciones de ruido de fondo podrán sufrir variaciones drásticas, si bien el caso peor es muy estable. Por esta razón, para realizar las simulaciones del siguiente apartado se seleccionará el valor de caso peor correspondiente al máximo de radiancia.

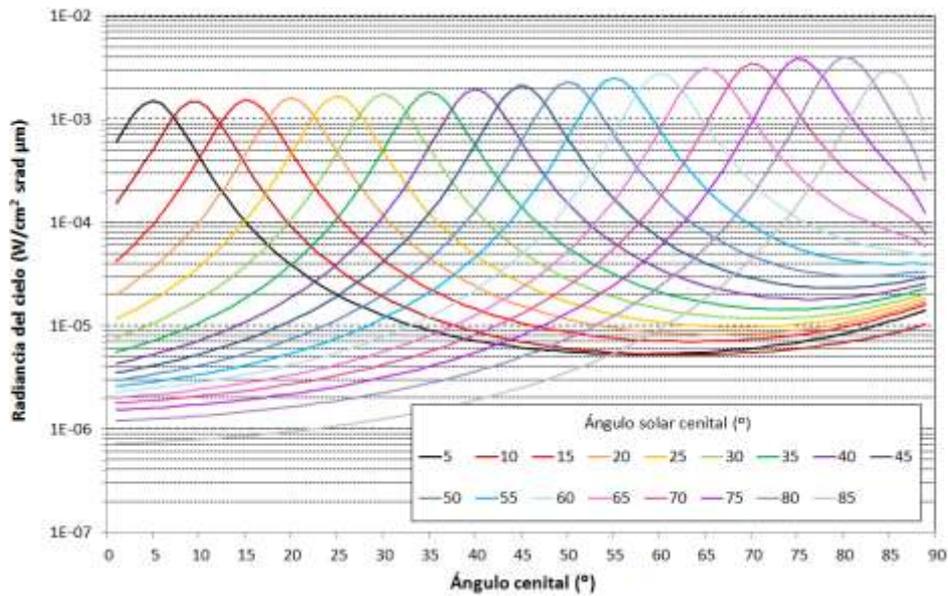

**Figura 100. Radiancia espectral del cielo a 1550 nm para un ángulo acimutal de 0 °, ángulo cenital de observación de 0 ° a 90 ° y ángulo solar cenital de 5 ° a 85 °.**

Si al resultado anterior se le añade, según se indicó al principio de este apartado, las condiciones de un límite máximo de 70° para el ángulo cenital de observación y un ángulo SEP > 5 (es decir, se eliminan los casos en que la distancia entre el ángulo cenital de observación y el SZE quede por debajo de ±5°), se obtiene el resultado de la Figura 101. Si se calcula un promedio de todos los valores máximos de radiancia se obtiene un valor razonable para ser utilizado como caso peor de un enlace diurno. Este cálculo se corresponde con una radiancia de 430 µW/(cm² · srad · µm), que es el valor de radiancia del cielo utilizado en las simulaciones.

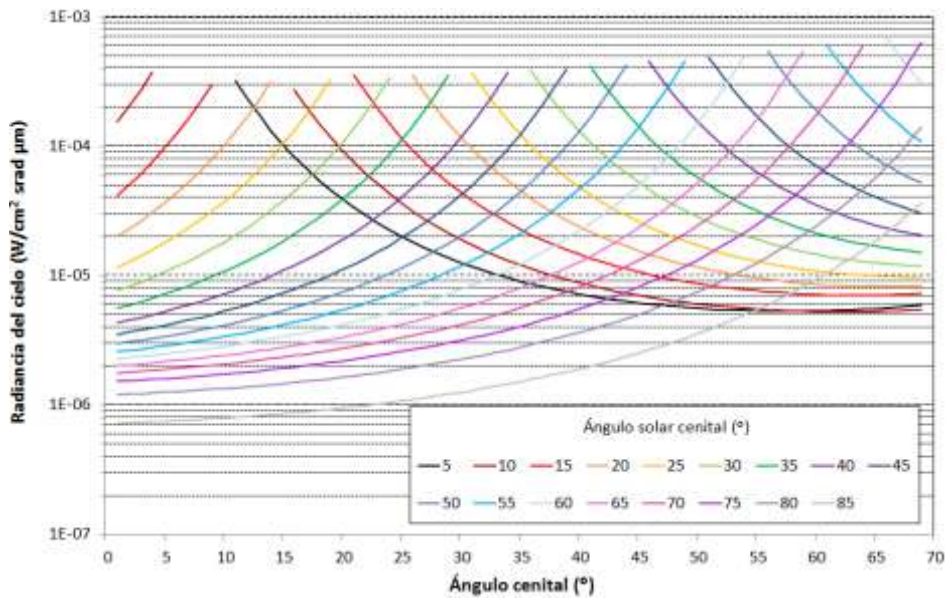

**Figura 101. Radiancia espectral del cielo a 1550 nm para un ángulo acimutal de 0 °, ángulo cenital de observación de 0 ° a 70°, ángulo solar cenital de 5 ° a 85 ° y ángulos SEP > 5 °.**

Para utilizar los resultados de radiancia del cielo es necesario obtener la potencia equivalente de ruido de fondo. Para ello, se utiliza la ecuación (2-51), en la que se comprueba que es necesario establecer, además de las características propias del telescopio, una anchura espectral de detección. Este parámetro viene determinado por el filtrado óptico, llevado a cabo justo antes de la fotodetección. El uso de transmisores láser de espectro extremadamente estrecho permite utilizar anchos de banda muy reducidos, pudiéndose minimizar significativamente el ruido de fondo que se acopla al receptor. Existen diversas tecnologías de filtrado (tales como filtros de resonancia atómica, filtros birrefringentes, filtros Fabry-Perot, filtros de Faraday, etc. [193]) que darán lugar a diferencias en la anchura espectral, la transmisión en la banda de paso, el campo de visión, el tamaño de la apertura y la resolución con la que pueden fabricarse y/o sintonizarse. En aplicaciones de FSOC (incluso utilizando IACT), el campo de visión del filtro no es un problema porque el sistema óptico tiene ya uno muy reducido que siempre será mucho menor al del filtro, y el tamaño de la apertura tampoco es relevante porque la señal óptica justo antes del filtro tiene un tamaño muy pequeño. Los parámetros clave son la transmisión y el ancho de banda. El estado del arte está tendiendo hacia anchuras espectrales próximas a los 0,001 nm con tasas de transmisión cercanas al 90 % [194] [195] [196]. Para los cálculos de potencia de ruido de fondo, como un compromiso conservador, se ha seleccionado un ancho de banda sin pérdidas de 0,01 nm.

En la Tabla 11 se muestran, para cada uno de los telescopios de CTA, los parámetros necesarios para obtener el ruido de fondo. Para ello, se ha utilizado la ecuación (2-15) para calcular el campo de visión $\theta_{FOV}$ (a partir de la longitud focal f y el diámetro del detector d) y la ecuación (2-51) para calcular el ruido de fondo $N_S$ (a partir de la radiancia solar L, el campo de visión $\theta_{FOV}$, el diámetro de la apertura D y la anchura espectral de filtrado $\Delta\lambda$).

**Tabla 11. Parámetros de todos los telescopios de CTA para obtener el ruido de fondo $N_S$.**

| Parámetros Telescopio | L (λ, θ, φ) μW/(cm² srad·μm) | d (cm) | f (m) | θ$_{FOV}$ (mrad) | D (m) | Δλ (nm) | N$_S$ (nW) |
|---|---|---|---|---|---|---|---|
| LST | 430 | 6,2 | 27,6 | 2,24 | 23 | 0,01 | 70,69 |
| MST-DC | 430 | 3,42 | 15,6 | 2,19 | 12 | 0,01 | 18,35 |
| MST-SC | 430 | - | 5,6 | 0,29 | 4,26 | 0,01 | 0,041 |
| SST-DC | 430 | 3,43 | 5,6 | 6,13 | 4 | 0,01 | 15,95 |
| SST-SC GATE | 430 | - | 2,283 | 0,7 | 2 | 0,01 | 0,052 |
| SST-SC ASTRI | 430 | - | 2,15 | 1,05 | 2,51 | 0,01 | 0,183 |

En el siguiente apartado se estudiarán una serie de escenarios diferentes en los que el enlace de comunicación debería operar. La contribución de origen solar en el ruido del cielo es con diferencia [4, p. 168] la predominante en el ruido de fondo acoplado al receptor. Sin embargo, se estudiarán escenarios donde la operación es íntegramente nocturna, y por lo tanto en ausencia de la influencia del Sol. En estos casos, la principal contribución al ruido de fondo la constituye la luz proveniente de otras fuentes como la Luna, los planetas, las estrellas o la luz zodiacal [21]. Como el caso de interés es el más desfavorable, la luz de las estrellas y la zodiacal se ha despreciado para considerar solo la mayor contribución de cada caso. En un enlace nocturno con otro planeta, la mayor aportación la constituye la luz reflejada por el Sol en la superficie de ese planeta (albedo). Se calculará la potencia de ruido $N_M$ recibida a causa del albedo de Marte, por ser la mayor contribución en los escenarios de operación nocturna del siguiente apartado. Para ello se utiliza la ecuación (2-52), donde $I_M$ es la irradiancia de Marte expresada en $W/(cm^2 \cdot \mu m)$ y $\Delta\lambda$ ($\mu m$) es la anchura espectral del filtro óptico.

$$N_M = I_M \pi \left( \frac{D}{2} \right)^2 \Delta\lambda \qquad (2\text{-}52)$$

Si se compara la ecuación (2-52) con la (2-51), se comprueba que la diferencia estriba en que la (2-52) utiliza la irradiancia en lugar de la radiancia, ya que se asume visión directa con Marte, que tiene una extensión angular muy pequeña (un máximo de 0,005° cuando está en oposición con la Tierra). Para calcular la irradiancia $I_M$ de Marte recibida en la Tierra se utiliza la ecuación (2-53) [197, p. 146], donde $A_M$ es el albedo de Marte (del 25 % [198]), $I_S$ es la irradiancia solar a una unidad astronómica (de 28,7 $mW/(cm^2 \cdot \mu m)$ a una longitud de onda de 1,5 $\mu m$ [199, p. 17]), $d_{M\text{-}S}$ es la distancia Sol-Marte en unidades astronómicas (1,52366 UA), $R_M$ es el radio de Marte (3390 km) y $d_{M\text{-}T}$ es la distancia Marte-Tierra (68,6·10^6 km cuando Marte está en oposición, que es el caso calculado aquí, por ser el escenario de operación nocturna). Según estos valores, la irradiancia de Marte es de 46 $pW/(cm^2 \cdot \mu m)$, por lo que asumiendo el mismo filtrado espectral que para el ruido solar ($\Delta\lambda = 0,01$ nm), la potencia de ruido $N_M$ resulta en 445 pW para LST, 120 pW para MST-DC y 7 pW para SST-DC.

$$I_M = A_M \frac{I_S}{d_{M\text{-}S}^2} \left( \frac{R_M}{d_{M\text{-}T}} \right)^2 \qquad (2\text{-}53)$$

## 2.12.3. Enlace basado en IACT bajo diferentes escenarios

En esta tesis se pretende estudiar la influencia en el enlace de comunicaciones del terminal receptor cuando este es un telescopio *Cherenkov*. Para ello, en el cálculo del balance de potencia y la relación señal-ruido, los IACT se caracterizarán utilizando principalmente el campo de visión y la apertura receptora. El cálculo de los términos de la ecuación de enlace que no tienen relación con el terminal receptor se tomarán, en los casos que procedan, del informe del OLSG (descrito en el apartado 2.12). En este informe se proponían una serie de posibles escenarios de enlaces (desde Marte en oposición y conjunción, los puntos de Lagrange L1 y L2, la Luna y órbitas bajas terrestres o LEO, del inglés *Low Earth Orbit*), que serán utilizados en este trabajo por su relevancia y conveniencia.

Como se ha dicho, en el cálculo del balance de potencia y relación señal a ruido del enlace, los IACT se caracterizarán utilizando el campo de visión y la apertura receptora, extrayendo el resto de los términos del mencionado informe del OLSG. La implicación en el balance de potencia tiene que ver únicamente con el tamaño de la apertura de los telescopios, que supone una mejora en términos de potencia, debido a la mayor superficie para recolectar fotones de señal. Como este parámetro solo determina la ganancia de recepción, calculada con la ecuación (2-42), es fácil aislarlo e integrarlo con el resto de parámetros en el balance de potencia. En la Tabla 12 se muestra una recopilación de los parámetros que se han utilizado para calcular el balance de potencia para cada uno de los escenarios contemplados en el OLSG. Estos datos deben considerarse meros ejemplos de algunas misiones típicas, y no enlaces optimizados. Ningún parámetro es arbitrario, pues los no comentados hasta ahora, como la caracterización atmosférica o la transmisión de transmisor y receptor, se han obtenido directamente de diferentes diseños preexistentes o sistemas ya desarrollados.

---

[21] La luz zodiacal proviene del *scattering* que sufre la luz del Sol en las partículas de polvo dispersas por todo el sistema solar. Al igual que la dispersada por las partículas de la atmósfera, su intensidad es mayor a ángulos más cercanos al Sol. Por ello, esta luz es más intensa en el plano de la eclíptica, suponiendo el 60 % de la luz natural durante una noche sin Luna.

**Tabla 12. Parámetros utilizados en el balance de enlace para cada escenario [72].**

| Escenario \ Parámetro | LEO | Luna | Lagrange L1 | Lagrange L2 | Marte oposición | Marte conjunción |
|---|---|---|---|---|---|---|
| Distancia de propagación (km) | $1,3 \cdot 10^3$ | $384 \cdot 10^3$ | $2 \cdot 10^6$ | $2 \cdot 10^6$ | $68,82 \cdot 10^6$ | $400 \cdot 10^6$ |
| Potencia media tx (W) | 0,5 | 0,5 | 1 | 1 | 4 | 4 |
| Diámetro apertura tx (cm) | 8 | 10,76 | 13,5 | 13,5 | 22 | 22 |
| Transmisión transmisor (%) | 50 | 33 | 35 | 35 | 30,3 | 30,3 |
| Pérdidas tx apuntamiento (dB) | 0,11 | 0,31 | 0,08 | 0,08 | 0,05 | 0,05 |
| Tasa binaria (bit/s) | $10 \cdot 10^9$ | $622 \cdot 10^6$ | $120 \cdot 10^6$ | $700 \cdot 10^6$ | $260 \cdot 10^6$ | $764 \cdot 10^3$ |
| Pérdidas por centelleo (dB) | 2 | 1 | 2 | 2 | 0,2 | 0,2 |
| Transmisión atmosférica (%) | 86,2 | 90,3 | 86,2 | 86,2 | 95 | 95 |
| Modulación PPM (símbolos) | (OOK) | 16 | 64 | 16 | 16 | 128 |
| Transmisión receptor (%) | 50 | 46,3 | 35 | 35 | 32,4 | 32,4 |
| Tipo de operación | Diurna y nocturna | Diurna y nocturna | Diurna | Nocturna | Nocturna | Diurna |

En general se puede entender la lógica de estos parámetros de la siguiente forma: a mayor distancia de propagación, la potencia se distribuye en un spot mucho mayor en la Tierra, por lo que es necesario compensar la pérdida de densidad de potencia, para lo cual se incrementa la potencia media de transmisión. Por el mismo motivo, se emplean aperturas de transmisión más grandes con las que reducir la divergencia del haz, minimizando la reducción de densidad de potencia con la distancia. Pese a estas dos estrategias de compensación, al aumentar la distancia se recibe en la Tierra una potencia menor, por lo que es necesario aprovechar mejor la potencia transmitida. Esta es la razón de emplear una modulación con un mayor número de símbolos: hace que para una potencia media determinada se puedan transmitir pulsos de mayor potencia de pico. Como el máximo número posible de slots temporales (símbolos) vendrá normalmente limitado por la mínima anchura de cada pulso, más que por la máxima potencia de pico, el aumento del orden de la modulación implica una reducción de la velocidad del enlace.

En cuanto al cálculo de la relación señal a ruido, la necesidad de usar el campo de visión y la apertura receptora impide utilizar directamente los datos del OLSG, ya que este proporciona directamente el cálculo final de fotones por pulso debidos al ruido de fondo. En este cálculo están incluidos por una parte la radiancia del cielo, con unas condiciones no especificadas, y por otra un campo de visión determinado, así como una apertura específica en la que se han integrado los fotones de ruido, ambos dependientes del telescopio receptor. Para el cálculo de los fotones por pulso de las ecuaciones (2-45) y (2-46), previo a la estimación de la relación señal-ruido, se ha tomado una eficiencia cuántica $\eta_c$ de 75 % como estimación conservadora (los fotodetectores basados en nanohilos explicados en el apartado 2.4.2 han demostrado eficiencias superiores al 90 % en 1550 nm [200]).

En la Figura 102 se muestra la geometría (no a escala) de los diferentes escenarios simulados en este apartado. Observando este diagrama se puede comprobar que existen dos grandes diferencias entre los escenarios contemplados. La primera es la distancia, habiendo hasta cinco órdenes de magnitud entre el más cercano (LEO) y el más lejano (Marte en conjunción). La segunda es la exposición al ruido del cielo, dependiendo del mínimo ángulo SEP que impone cada geometría. En dos de los seis escenarios (Lagrange L2 y Marte en oposición) este ángulo es cercano a los 180° de forma permanente, lo que significa que las comunicaciones se realizan durante la noche, en ausencia de ruido de fondo de origen solar. En otros dos casos (Lagrange L1 y Marte en conjunción) el caso es el opuesto, ya que el Sol siempre está muy cerca de la línea de visión, con los altos niveles de radiancia del cielo que esto implica. Los dos casos restantes (LEO y Luna) suponen situaciones intermedias con alternancia entre operación diurna y nocturna. Por ello, los casos más favorables serán Lagrange L2 (ya que aunque su distancia es mayor que la de LEO o la Luna, nunca se perturba el enlace con el ruido de fondo de origen solar, por lo que el número de fotones de ruido por pulso es muy reducido) y LEO (ya que aunque se alterna operación nocturna y diurna, la distancia será la mínima y el número de fotones recibidos de señal por pulso será muy elevado, especialmente con las grandes aperturas de los telescopios de CTA). Marte en conjunción sería con diferencia el caso más desfavorable, ya que la distancia es máxima, con las elevadas pérdidas de señal que ello implica, y también es máximo el ruido del Sol por su cercanía a la línea de visión.

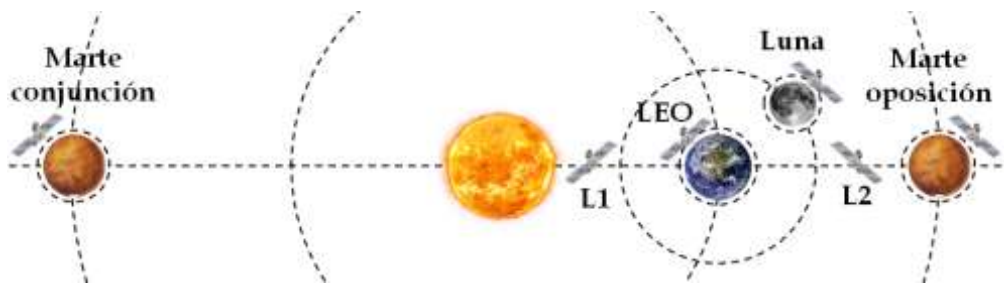

**Figura 102. Diagrama de la geometría de los diferentes escenarios simulados.**

Con el objetivo de ilustrar cómo se comporta cada uno de los telescopios de CTA en un enlace de comunicaciones, es interesante analizarlos en igualdad de condiciones. Es decir, simular su comportamiento en un mismo escenario, utilizando los mismos parámetros en el balance de enlace y el cálculo de fotones de señal y de ruido, a excepción de los característicos de cada telescopio. En la Figura 103 se muestra la relación señal a ruido que permite cada telescopio de CTA para el mismo enlace basado en el escenario Lagrange L1. Se ha elegido este escenario como ejemplo de caso peor en cuanto a ruido solar. Los telescopios se han ordenado de forma ascendente según la SNR obtenida por cada uno. También se han incluido, superpuestos y en el eje secundario, el campo de visión y el área de apertura de cada telescopio (ambos parámetros independientes del escenario) para comparar su influencia en la SNR. Atendiendo al campo de visión se comprueba que los telescopios también aparecen ordenados, aunque de forma inversa. De esto se concluye que en presencia de altos niveles de ruido de fondo es el campo de visión y no el área el que determina la relación señal a ruido del enlace. Este análisis es interesante porque permite determinar qué telescopio se adapta mejor a un entorno ruidoso de comunicaciones. Se observa que en este sentido el mejor de los telescopios de CTA es MST-SC, gracias a su reducido campo de visión, seguido de las dos versiones de SST-SC, que pese a ser dos de los telescopios con menor apertura se comportan incluso mejor que LST debido también a sus reducidos campos de visión. Por último, SST-DC es el que peor SNR obtiene al tener el mayor campo de visión de todos, siendo el único que no es capaz de conseguir los 3 dB mínimos exigidos para realizar la comunicación en este escenario.

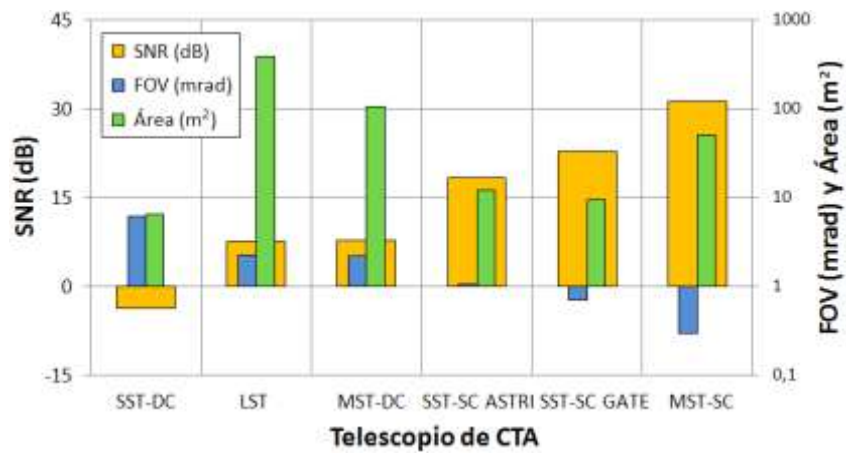

**Figura 103. Campo de visión y área de apertura en comparación con la relación señal a ruido para cada telescopio de CTA en el escenario L1.**

En la Tabla 13 se muestra un resumen detallado del balance de enlace para cada escenario, así como la evaluación de la calidad de cada enlace con su relación señal a ruido. En cada escenario se han elegido preferiblemente los telescopios más pequeños y los telescopios tipo *Davies-Coton* frente a *Schwarzschild-Couder*, en ambos casos para priorizar el menor coste de fabricación, siempre con el objetivo de conseguir una SNR mínima de 3 dB. Para los escenarios con operación diurna se han utilizado los valores de radiancia de cielo de caso peor calculados en el apartado 2.12.2. Respecto a los escenarios de operación nocturna (Lagrange L2 y Marte en oposición), en ambos se han empleado los valores de ruido de fondo correspondientes al albedo de Marte, calculados al final del mismo apartado. Si bien este ruido será la condición normal en el escenario de Marte en oposición, también se ha utilizado en el escenario de Lagrange L2 ya que constituye la mayor fuente de ruido: los demás planetas se encuentran mucho más distantes y Venus nunca es visible en dirección a L2. En la práctica, en este escenario Marte solo cruzará por el campo de visión del receptor de forma esporádica, por lo que supone un caso peor infrecuente. No se ha considerado el efecto del albedo de la Luna porque esta orbita alrededor de la Tierra en un plano que tiene una inclinación de unos 5° respecto al plano de la eclíptica, donde está contenido el punto L2. Esto hace que la Luna solo entre en el campo de visión del receptor en el infrecuente caso en que los puntos donde la Luna cruza la eclíptica (conocidos como nodo ascendente y descendente, con una frecuencia de dos veces al mes) coinciden con la dirección desde la Tierra hacia el punto L2.

**Tabla 13. Componentes del balance de enlace y la relación señal-ruido para cada escenario.**

| Escenario / Parámetro | LEO | Luna | Lagrange L1 | Lagrange L2 | Marte oposición | Marte conjunción |
|---|---|---|---|---|---|---|
| $P_t$ (dBm) | 26,99 | 26,99 | 30 | 30 | 36,02 | 36,02 |
| $L_t$ (dB) | -3,01 | -4,81 | -4,56 | -4,56 | -5,19 | -5,19 |
| $G_t$ (dB) | 104,2 | 106,77 | 108,74 | 108,74 | 112,98 | 112,98 |
| $L_{el}$ (dB) | -260,46 | -309,86 | -324,2 | -324,2 | -354,91 | -370,22 |
| $L_{atm}$ (dB) | -2,64 | -1,44 | -2,64 | -2,64 | -0,42 | -0,42 |
| $G_r$ (dB) | 135,29 (1×SST-DC) | 135,29 (1×SST-DC) | 135,29 (1×SST-DC) | 135,29 (1×SST-DC) | 147,36 (1×MST-DC) | 147,36 (3×MST-SC) |
| $L_r$ (dB) | -3,01 | -3,34 | -4,56 | -4,56 | -4,89 | -4,89 |
| $L_{rap}$ (dB) | -0,11 | -0,31 | -0,08 | -0,08 | -0,05 | -0,05 |
| $P_r$ (dBm) | -2,75 | -50,72 | -62 | -62 | -69,1 | -78,7 |
| SNR (dB) | 45,22 | 3,27 | 7,7 | 25,3 | 6,15 | 3,74 |

En estos cálculos influye de forma importante la ganancia de recepción, impuesta por el tamaño de la apertura del telescopio. Por ello se pueden ver unas elevadas SNR en los escenarios más favorables, en los que se utilizarían telescopios pequeños (40 cm para LEO y 1 metro para L2), en comparación con los 4 metros del SST-DC empleado para esos escenarios. Destaca especialmente la SNR de LEO en la que la diferencia entre ganancias entre el telescopio contemplado por el OLSG y el SST-DC es de 17 dB, y la de L2, donde a la diferencia de ganancias se suma la ausencia de ruido solar (y por lo tanto la indiferencia del campo de visión, el mayor hándicap de un IACT). En el caso de LEO, además influye el alto régimen binario utilizado, ya que debido a la menor distancia no es preciso utilizar una modulación PPM para transmitir pulsos con una mayor potencia de pico. Al utilizar OOK en lugar de PPM se puede dedicar este margen exclusivamente a reducir la duración del pulso. Por ello, la misma duración de pulso da lugar a un mayor régimen binario en OOK que en PPM (con la misma ventaja en relación al menor ruido de fondo, debido a la menor duración de pulso que conlleva aumentar el orden de la modulación PPM). Por ejemplo, en los escenarios de LEO y Luna la duración del pulso es exactamente la misma (y por ello también los fotones por pulso de ruido de fondo) para conseguir un régimen binario 16 veces superior en LEO que en la Luna. Esto se debe a que en la Luna se utiliza una modulación 16-PPM porque la distancia es casi 300 veces mayor (50 dB adicionales de pérdidas), que es necesario superar con una mayor potencia transmitida gracias exclusivamente a la modulación (ya que la potencia media, de 0,5 W, es la misma en ambos casos).

El escenario de la Luna se basa en la misión LLCD (comentada en el apartado 2.4.1). En este sentido, conviene señalar que el terminal remoto puede transmitir a un régimen binario variable entre 40 y 622 Mbit/s, según las condiciones atmosféricas y de ruido de fondo (reduciendo la velocidad cuando las condiciones empeoran). Sin embargo, en los cálculos mostrados en la Tabla 12 se asumió simultáneamente el caso peor de ruido de fondo y el máximo régimen binario de 622 Mbit/s, comprobándose que es posible operar a la vez con ambas condiciones, utilizando un solo telescopio SST-DC. Para un régimen binario de 40 Mbit/s (ajustando el número de símbolos hasta 256 para obtener una misma duración de slot que con 16) se comprueba que se obtiene una ganancia de 12 dB, el margen diseñado para hacer frente a las condiciones variables del canal, y que no es necesario con un telescopio como el SST-DC.

El escenario L2 está basado en la misión *Euclid* de la ESA programada para 2020. En esta misión se planea dedicar 3 horas cada día a descargar los datos científicos acumulados en las restantes 21 horas. Por ello es fundamental una alta velocidad de transmisión, que el OLSG estima en 10 veces la velocidad prevista para radiofrecuencia, es decir, 700 Mbit/s para el enlace óptico. Para conseguir este régimen binario, todos los datos del enlace se han escalado a partir de un transmisor espacial diseñado para un escenario L2 y parcialmente desarrollado por la empresa *RUAG Space*. Como se comentó, este escenario es el más favorable en cuanto a las condiciones de ruido de fondo, y aunque se considerarán como caso peor las de Marte en oposición, la mayor parte del tiempo las condiciones serán más favorables que ese caso. Por esta razón, utilizando solamente un SST-DC se podría diseñar un enlace muy superior al considerado por el OLSG, que asume un telescopio de 1 metro. Por ejemplo, aún manteniendo las condiciones de caso peor y utilizando un solo SST-DC, se podría implementar un enlace como el de LEO utilizando OOK para un régimen binario de 3,7 Gbit/s con una SNR de 19,28 dB, aún con margen de sobra. Se han seleccionado 3,7 Gbit/s (en lugar de los 10 Gbit/s de LEO) porque por encima de este régimen binario el número de fotones por pulso es inferior a uno. Tanto en este escenario como en el de Marte en oposición, con operación nocturna, es donde se obtiene una mayor mejora al utilizar un IACT, ya que al no existir ruido solar, el campo de visión es indiferente, y en este caso no existe diferencia entre un telescopio convencional y un IACT.

Al igual que L2, el escenario L1 también se basa en el mismo transmisor remoto, pero el enlace se ha adaptado a unas condiciones mucho más desfavorables de ruido solar aumentando el orden de la modulación y reduciendo el régimen binario. Las condiciones del enlace están basadas en la misión SOHO (del inglés *SOlar and Heliospheric Observatory*) de la NASA y la ESA, aún en activo desde 1996. La sonda orbita alrededor del punto de Lagrange L1, permitiendo las comunicaciones durante la mayor parte del tiempo, aunque con ángulos SEP muy pequeños. En estas condiciones de ruido solar sería necesario un *array* de cinco telescopios SST-DC para lograr una SNR de 3,24 dB o bien un único MST-DC (como se propone en la Tabla 13) para conseguir cerrar el enlace con un margen adicional de 4,7 dB. Conviene señalar que en general, el comportamiento óptico de un único telescopio es comparable al de un *array* de telescopios de menor apertura pero con un área equivalente similar. Por ello, el efecto de añadir telescopios supone sumar 3 dB por cada nuevo elemento del array en el cálculo del balance de potencia. No obstante, existen algunas condiciones a cumplir, tales como la realización de una correcta sincronización y la superación de la sensibilidad mínima en cada elemento del *array*. Si bien la primera se asume sin más por no suponer un desafío tecnológico (especialmente en CTA que tiene de forma nativa una topología de *array*), la relativa a la sensibilidad de cada telescopio sí se ha tenido en cuenta a la hora de contemplar la utilización de un *array*: cada telescopio debe superar el límite de un fotón recibido por pulso (explicado en el apartado 2.12.1) para poder ser integrado en un *array*.

Las condiciones del escenario de Marte en oposición son similares a L1 pero con una distancia mayor (más de 30 dB adicionales de pérdidas por espacio libre). De nuevo, como el campo de visión no influye en la detección del ruido de fondo, al acoplarse el albedo de Marte de forma íntegra, el enlace no se ve perjudicado por el hándicap de los IACT. Por esta razón, aumentar el tamaño de la apertura no redunda en una mejora de la SNR, por lo que se elige el telescopio más pequeño posible que permita recibir al menos un fotón de señal por pulso. Es por esta razón que no es posible utilizar un SST-DC, ya que aunque la SNR es superior a 3 dB, se detectan 0,2 fotones por pulso. Empleando un MST-DC los fotones por pulso ascienden a 3, lo que hace viable el enlace con un margen adicional de 3,15 dB. En el caso de Marte en conjunción, si bien pese a la gran distancia un solo SST puede detectar 2 fotones de señal por pulso, el número de fotones por pulso de ruido es más de tres órdenes de magnitud superior, lo que proporciona una SNR de menos de -30 dB, haciendo inviable el enlace. El ruido es tan alto en comparación con la señal que para superar tal desproporción serían necesario miles de SST-DC, cientos de MST-DC o decenas de SST-SC. Ninguno de los telescopios de CTA puede conseguir una SNR de 3 dB de forma individual, siendo necesaria la utilización del mejor de los telescopios de CTA (MST-SC) en una configuración de *array* de 3 unidades.

**Tabla 14. Relación señal-ruido (dB) de cada telescopio en cada escenario, incluyendo el número de elementos que serían necesarios en un posible *array* para superar los 3 dB.**

| Escenario / Telescopio | LEO | Luna | Lagrange L1 | Lagrange L2 | Marte oposición | Marte conjunción |
|---|---|---|---|---|---|---|
| SST-DC | 45,22 ✔ | 3,27 ✔ | -3,75 ✖ (5×) 3,24 | 25,30 ✔ | 6,15 ✔ | -35,88 ✖ (7750×) 3,01 |
| LST | 56,52 ✔ | 14,58 ✔ | 7,55 ✔ | 25,30 ✔ | 6,15 ✔ | -33,61 ✖ (4600×) 3,02 |
| MST-DC | 56,67 ✔ | 14,73 ✔ | 7,70 ✔ | 25,30 ✔ | 6,15 ✔ | -24,43 ✖ (560×) 3,06 |
| SST-SC ASTRI | 67,35 ✔ | 25,37 ✔ | 18,35 ✔ | 25,30 ✔ | 6,15 ✔ | -13,78 ✖ (48×) 3,03 |
| SST-SC GATE | 71,75 ✔ | 29,80 ✔ | 22,77 ✔ | 25,30 ✔ | 6,15 ✔ | -9,35 ✖ (18×) 3,20 |
| MST-SC | 80,07 ✔ | 50,03 ✔ | 31,09 ✔ | 25,30 ✔ | 6,15 ✔ | -1,04 ✖ (3×) 3,74 |

Por último, en la Tabla 14 se muestra la relación señal a ruido que proporcionaría cada uno de los telescopios de CTA (ordenados según el resultado de la Figura 103) en cada uno de los escenarios analizados. Se muestra la relación señal a ruido correspondiente a utilizar un único telescopio, y en caso de no superarse los 3 dB mínimos requeridos para cerrar el enlace, se muestra entre paréntesis el número de elementos que requeriría un posible *array* para proporcionar los 3 dB requeridos. Se puede comprobar que si se obvia el escenario de caso peor de Marte en conjunción (para lo cual sería necesario utilizar un array de tres telescopios tipo MST-SC, como solución óptima), cualquiera de los telescopios de CTA sería suficiente para dar soporte a todos de los escenarios contemplados (con la excepción de SST-DC, que no podría soportar el escenario Lagrange L1).

# 2.13. PROPUESTAS PARA MEJORAR LA PSF

En el apartado anterior se ha demostrado que es posible diseñar una estación receptora terrestre para enlaces de comunicaciones desde espacio profundo basada en los telescopios de CTA. Para ello, se han utilizado tal cual estos IACT sin ninguna modificación en su concepción original, salvo las imprescindibles para adaptar su utilización en comunicaciones (básicamente la sustitución de la cámara por un único fotodetector, y el reposicionamiento del plano imagen para enfocar al infinito en lugar de a 10 km). El objetivo de reducir el tamaño de las cámaras en CTA [22], llevó a proponer el diseño *Schwarzschild-Couder*, cuya resolución óptica se adapta suficientemente bien a su uso en FSOC: con un solo SST-SC tipo GATE es posible cubrir las necesidades de todos los escenarios analizados en el apartado 2.12.3 (a excepción de Marte en Conjunción que precisaría de un *array* de 3 MST-SC). Además, el SST-SC no solo es superior en cuanto a resolución óptica en comparación con el MST-DC y LST, sino que es mucho más barato: como se puede observar en la Tabla 15, este telescopio es con diferencia la alternativa de menor coste. Incluso los telescopios MST-DC y LST son opciones muy económicas si se comparan con telescopios astronómicos como GTC y *Keck* o incluso de sus alternativas de bajo coste como *Hobby-Eberly* y SALT. Conviene destacar que los costes de los telescopios de CTA son también mucho menores a los previstos por el OLSG, si bien estos contemplan la cúpula del telescopio, a diferencia los IACT.

El diseño *Schwarzschild-Couder* es muy adecuado para su uso en FSOC, aunque en su tamaño máximo ofrece un diámetro equivalente de menos de cinco metros (MST-SC), lo que hace necesario el uso de varias unidades si se desea utilizar en los escenarios más adversos. La alternativa es utilizar una única unidad de mayor apertura, para lo cual es necesario recurrir a MST-DC o LST. Sin embargo, pese a sus grandes aperturas, la resolución óptica es pobre, por lo que para justificar el mayor coste en relación a SST-SC o MST-SC, deberían adaptarse para su uso en FSOC con una mejor PSF, que permita reducir el campo de visión.

---

[22] Las cámaras dominan el coste de LST y MST-DC [61, p. 2]. Por ejemplo, si cada canal tiene un coste de unos 400 € [39, p. 49], una cámara de 2500 píxeles como la del telescopio LST [201] tendría un coste superior al millón de euros.

Tabla 15. Costes aproximados de instalaciones terrestres
de rayos gamma, astronómicas y de comunicaciones.

| Telescopio | Tipo | Coste |
|---|---|---|
| CTA SST-SC | IACT | <0,5 M€ [61, p. 2] |
| CTA MST-DC | IACT | 1,6 M€ [201, p. 17] |
| CTA LST | IACT | 7,4 M€ [201, p. 17] |
| GTC / Keck | Astronómico | 100 M€ [202] |
| Hobby-Eberly / SALT | Astronómico | 50 M€ [203] |
| OLSG LEO | FSOC | 3,4 M€ [184] |
| OLSG Luna | FSOC | 15,3 M€ [184] |
| OLSG L1 | FSOC | 12,5 M€ [184] |
| OLSG L2 | FSOC | 10,9 M€ [184] |
| OLSG Marte | FSOC | 102,8 M€ [184] |

Conviene no olvidar que el diseño original de estos telescopios se fundamenta en requisitos muy diferentes a los de un receptor de FSOC. En este sentido, es necesario estudiar las formas más eficientes en que ciertas modificaciones pudieran mejorar sus prestaciones en comunicaciones, sin que a consecuencia de ello se vean elevados los costes de forma relevante (ya que la reducción de costes es uno de los objetivos de la utilización de IACT). Este camino debería recorrerse basándose en el análisis del apartado 2.10, comenzando por las aberraciones geométricas, que es el primer límite que debe superar un IACT para mejorar su resolución óptica. El parámetro que determina la resolución óptica en un telescopio de comunicaciones es el campo de visión. Según lo analizado en el apartado 2.11, el campo de visión de los telescopios de CTA, ordenado de menor a mayor, quedaría como sigue: MST-SC (0,29 mrad), SST-SC GATE (0,7 mrad), SST-SC ASTRI (1,05 mrad), MST-DC (2,19 mrad), LST (2,24 mrad) y SST-DC (6,13 mrad). En comparación, el campo de visión que se asume a lo largo del informe del OLSG es 0,02 mrad [72] (independientemente de las aperturas de los telescopios, que van desde los 40 cm hasta los 12 metros de diámetro), por lo que existe margen de mejora, especialmente en los telescopios de foco primario. En estos telescopios se podría conseguir una mejora de la PSF de al menos un orden de magnitud, posiblemente más [204].

A continuación se analizan sucintamente varias alternativas orientadas a conseguir una menor PSF en los telescopios de CTA con el objetivo de reducir su campo de visión, que como se vio en el apartado 2.12.3, es el parámetro que determina la SNR de un enlace en presencia de altos niveles de ruido de fondo. Las propuestas de este apartado pretenden ser una guía de los posibles caminos futuros que se podrían recorrer para alcanzar este objetivo, más que un análisis detallado de dichas propuestas, que excede el propósito de este trabajo.

## 2.13.1. Nuevos espejos para prevenir la aberración

Las aberraciones geométricas tienen su origen en el diseño de la óptica que las genera, y su efecto es impedir que se produzca una imagen puntual a partir de un objeto puntual (obviando el efecto inevitable de la difracción). En general se puede decir que no es posible eliminar completamente todas las aberraciones en un sistema óptico, por lo que el objetivo de un buen diseño es conseguir reducir las aberraciones más relevantes balanceando el resultado final, o bien eliminar o minimizar las que interesen para una aplicación determinada. Por ejemplo, en comunicaciones desde espacio profundo se puede decir que la aberración de coma jugará un papel menos relevante que en astronomía *Cherenkov*. La razón es que los rayos fuera de eje que inciden en el telescopio y que interesa detectar difieren mucho entre ambas aplicaciones: el campo de visión de un IACT puede llegar a los 10° y un telescopio de FSOC como los considerados por el OLSG se limita a 0,001°: una diferencia de cuatro órdenes de magnitud. De las aberraciones primarias, también podrían considerarse menos relevantes para FSOC la distorsión o la curvatura de campo, en comparación con la esférica o el astigmatismo. Por todo lo anterior, el objetivo de un buen diseño óptico para FSOC consistirá más en identificar los requisitos de la aplicación para minimizar las aberraciones relevantes, que en tratar de equilibrar todas las aberraciones. Especialmente en el caso que ocupa, donde el diseño de los telescopios se llevó a cabo con unos requisitos muy diferentes a los necesarios en comunicaciones.

La reducción de costes de fabricación, en efecto, hace que los tres telescopios principales de CTA (LST, MST-DC y SST-DC) estén basados en espejos esféricos, por lo que es esperable la presencia de niveles considerables de aberración. La estrategia más evidente para mejorar la PSF de un IACT es la utilización de una óptica no esférica, más parecida a la que emplean los telescopios astronómicos convencionales. A diferencia de una superficie esférica, cuyo perfil se define completamente con el radio de curvatura, una superficie asférica tiene distintas curvaturas a lo largo de su perfil. Por ejemplo, una lente convexa diseñada para eliminar la corrección esférica tendría una menor curvatura en los puntos más periféricos con el objetivo de que los rayos marginales se curven menos. Las superficies asféricas con simetría rotacional normalmente se definen mediante la ecuación (2-54) que proporciona la altura sagital, donde $c$ es la curvatura en el vértice (inversa al radio de curvatura), $k$ es la constante cónica o de *Schwarzschild* (también definida como $k = -e^2$, siendo $e$ la excentricidad de la sección cónica), $r$ es la distancia radial medida perpendicularmente desde el eje óptico y $a_i r^{2i}$ son los términos asféricos de orden superior, utilizados cuando se desea desviar la superficie de una cónica [205, p.116].

$$z = \frac{r^2 c}{1 + \sqrt{1 - (1+k)c^2 r^2}} + \sum a_i r^{2i} \qquad (2\text{-}54)$$

Por ejemplo, una esfera se define con $k = a_i = 0$, una parábola con $k = -1$ y $a_i = 0$, o la córnea del ojo humano con $k = -0,26$ y $a_i = 0$. Utilizando la ecuación (2-54), cualquier punto de la superficie asférica queda definido por las coordenadas $(z, r)$. Todas las superficies cónicas están libres de aberración esférica para un par de puntos conjugados. Por ejemplo, una esfera forma una imagen sin aberración solo si el objeto está en el centro de la curvatura de la superficie esférica, y en el otro extremo, una parábola forma una imagen sin aberración solo si el objeto está en el infinito. Por ello, para telescopios de foco primario como LST y MST-DC la única alternativa para focalizar un frente de ondas planas sin aberraciones utilizando un solo espejo es el perfil parabólico (asumiendo que no se realiza ninguna corrección de campo, como las explicadas en el apartado 2.13.2).

El caso de LST es el más evidente: su perfil es parabólico y se utilizaron espejos esféricos para definir su superficie únicamente por una cuestión de costes. El diseño parabólico es necesario en un telescopio del tamaño de LST por ser isócrono: los fotones llegan al plano focal de forma simultánea independientemente de en qué zona del reflector hayan impactado. Por el contrario, el diseño esférico de *Davies-Cotton* produce una fuerte dispersión temporal y por ello no se utiliza en grandes aperturas si se precisa una buena resolución temporal, como es el caso tanto en astronomía Cherenkov como en comunicaciones. Por esta razón se utiliza el perfil parabólico en LST, sin embargo sus espejos son esféricos para abaratar costes dado que se consigue una resolución óptica suficiente para eventos *Cherenkov*. Desde el punto de vista de FSOC, el LST proporciona una infraestructura ideal para construir un telescopio parabólico, pero utilizando espejos formados por secciones parabólicas en lugar de esferas. El diseño parabólico es especialmente adecuado para comunicaciones desde espacio profundo, porque carece de aberraciones en el plano focal cuando los rayos llegan alineados con el eje óptico. Su mayor inconveniente es la aberración de coma producida por los rayos fuera de eje. Para FSOC este problema no es tan relevante ya que el campo de visión es muy reducido, y en todo caso, existen soluciones para corregir la aberración de coma cerca del foco (ver apartado 2.13.2).

Desde el punto de vista óptico, un LST con espejos asféricos para conformar una superficie parabólica sería la opción ideal para afrontar cualquier escenario de comunicaciones desde espacio profundo. Ofrecería unas prestaciones sin igual, muy por encima de las consideradas en ningún diseño propuesto en la literatura científica. Con la capacidad de soportar enlaces incluso más allá de Marte, sería un telescopio con una vida útil muy larga, con un papel similar al que jugaron las antenas de 70 metros en la red de espacio profundo (DSN) de la NASA instaladas en Goldstone (EEUU) y Camberra (Australia). No obstante, también existen inconvenientes en esta opción: por una parte, el gran tamaño de estos telescopios implica que la superficie reflectora se vea sometida a intensas deformaciones por efecto de la gravedad, y la PSF puede verse perjudicada (por ejemplo, en HESS la PSF llega a aumentar hasta en un factor 2 entre la máxima y la mínima elevación [169, p. 17]). Por otra parte, la adaptación del telescopio para permitirle llevar a cabo la operación diurna es mucho más compleja y costosa en el caso de un telescopio de un tamaño tan grande. Por último, los LST son los telescopios más costosos (incluso en relación al área reflectora), por lo que requeriría una inversión muy importante. Una primera estimación podría situar esta cifra en torno al coste de un LST convencional, ya que la cámara, innecesaria en FSOC, supone una parte muy importante del coste de este telescopio y esta partida se podría ver compensada por el mayor coste de fabricación de los espejos. Sin embargo, el coste adicional de la adaptación para operación diurna sería muy elevado: en el informe del OLSG se estima en 1.000.000 € para un telescopio de 1 metro [72, p. 73], por lo que el coste total de un LST adaptado a FSOC posiblemente no bajaría de los 10.000.000 €.

En el caso del diseño *Davies-Cotton*, las opciones de adaptación no son tan evidentes como en el caso parabólico de LST. En este caso, el diseño original considera espejos esféricos. Una alternativa a estudiar podría ser la opuesta al LST: si en este se utilizaba un telescopio parabólico con espejos esféricos, en el MST-DC se podría utilizar el telescopio esférico con espejos parabólicos. Según este diseño, cada espejo conformaría la sección de una parábola con una longitud focal ligeramente distinta. Cada sector de la parábola quedaría definido de forma que al situarlo en la posición que le corresponde según el perfil *Davies-Cotton*, formaría la imagen en el mismo plano focal, compartido por todas las parábolas. En la Figura 104 se muestra una ilustración de esta propuesta simplificando para una fila central compuesta solo por tres secciones parabólicas. Se puede observar en rojo el perfil esférico definido por la estructura del telescopio *Davies-Cotton* y en verde, naranja y azul las distintas parábolas de las que se derivaría cada espejo. El esquema se muestra exagerado para apreciar las diferencias de longitud focal y excentricidad de cada una de las parábolas. Se comprueba que cada espejo debe tener una longitud focal menor cuanto más periférica sea su posición con el objetivo de formar la imagen en el mismo plano focal.

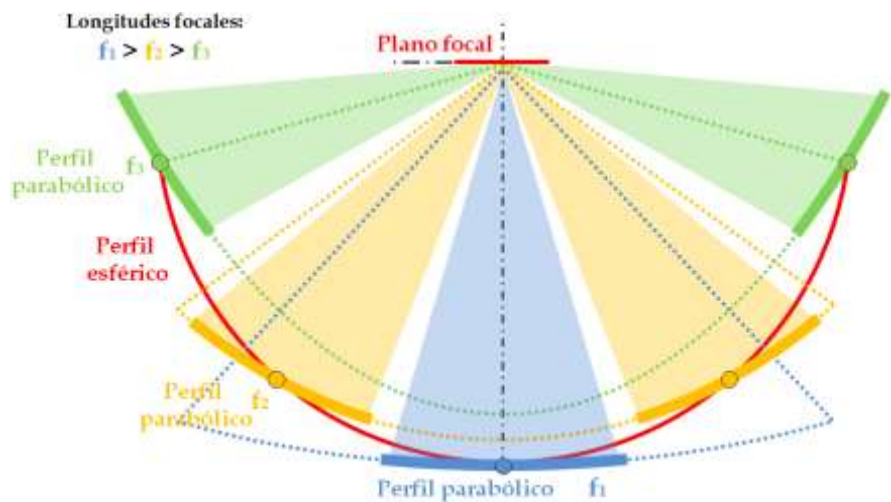

**Figura 104. Ilustración de la propuesta para sustituir los espejos esféricos de un telescopio *Davies-Cotton* por espejos parabólicos.**

De entre las técnicas conocidas de fabricación de espejos para IACT, la técnica de replicado mediante deposición de vidrio (ver apartado 2.6.1) es la preferida en CTA porque abarata la producción de espejos (incluso para superficies asféricas, aunque con ciertos límites) y disminuye sus tiempos de fabricación. Además, la técnica es adecuada para radios de curvatura reducidos. En caso contrario, el vidrio se ve sometido a mucho estrés para adoptar la forma del molde, lo que hace que al liberar el espejo, este tienda a recuperar la forma original deformando su curvatura [206, p. 5]. Por ello, la técnica es adecuada para grandes telescopios como MST-DC o LST debido a sus mayores longitudes focales. Por otra parte, esta técnica es muy eficiente cuando se precisa fabricar un gran número de espejos similares porque aunque un molde es muy caro de fabricar, el coste de cada espejo es mucho menor. Para un telescopio FSOC, ese factor de abaratamiento puede convertirse en un coste adicional si los espejos del telescopio son distintos unos de otros. Este sería el caso si se precisa fabricar una superficie parabólica, para la cual sería necesaria una cantidad prohibitiva de moldes diferentes.

Por lo anterior, la técnica de pulido con diamante (ver apartado 2.6.1) sería la mejor candidata para proporcionar espejos con una buena calidad óptica [207]. Esta técnica, empleada en MAGIC I, se basa en un pulido fino mediante la rotación sobre su propio eje de un cabezal de diamante, que además pivota sobre un eje perpendicular al espejo (Figura 105). Este cabezal rota a una velocidad de 20000 vueltas/hora consiguiendo una gran precisión en la curvatura del espejo: las rugosidades de los espejos originales de MAGIC I están en el orden de varios nanómetros [118, p. 3]. Esta es una precisión cercana a la de los telescopios astronómicos convencionales [33, p. 140], por lo que conseguir la calidad óptica de los espejos de estos es factible utilizando esta técnica.

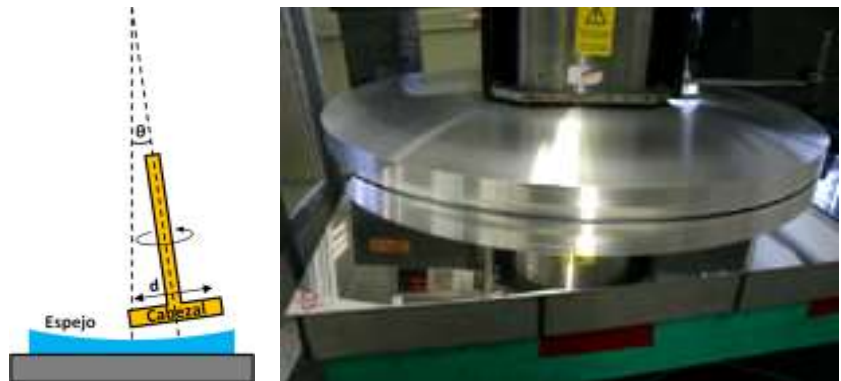

**Figura 105. Técnica de pulido con diamante: ilustración gráfica (izquierda) y aplicación a un espejo de MAGIC I (derecha) [40, p. 73].**

Si d es el diámetro del cabezal de diamante y θ el ángulo sobre el que pivota, el radio de curvatura R se obtiene con la ecuación (2-55). La empresa alemana *LT-Ultra*, encargada de fabricar los espejos para MAGIC I, ha desarrollado una técnica basada en el pulido con diamante que permite girar los espejos de forma que se obtienen radios de curvatura variables. En principio es posible fabricar espejos con curvatura definida por polinomios hasta de grado cuatro, por lo que se podrían producir superficies cónicas como la parábola [208].

$$R = \frac{d/2}{sen(\theta)} \qquad (2\text{-}55)$$

Para los espejos de MAGIC II, el coste de esta técnica fue de 2450 €/m² tardándose un día completo en fabricar cada segmento de 1 m² (este es el mayor inconveniente de esta técnica en CTA). Esto supone un coste cerca de un 50 % mayor que el de la técnica de espejos de vidrio con recubrimiento de aluminio. En CTA, el coste unitario es un aspecto fundamental debido al enorme volumen de fabricación involucrado, y por ello se establece como requisito mantenerse por debajo de los 2000 €/m² [63, p. 4]. Sin embargo, para construir un solo telescopio, el incremento en el coste de la fabricación de espejos no supone un aspecto tan crítico. Por ejemplo, según el coste de fabricación mencionado para MAGIC II, toda la superficie reflectora del mayor de los telescopios de CTA (LST) quedaría por debajo del millón de euros, que supone una pequeña fracción del coste total del telescopio, según se vio al principio del apartado 2.13.

## 2.13.2. Corrector de campo basado en lentes/espejos

Las estrategias para corregir las aberraciones suelen consistir en llevar a cabo un correcto diseño desde el principio. Sin embargo, esta opción no es posible en este caso, al partir de un telescopio ya diseñado, por lo que el margen de corrección será mucho menor. En este sentido, queda descartado un rediseño del perfil del telescopio o de otros parámetros fundamentales, como la longitud focal. Una estrategia eficiente puede ser la utilización de un corrector de campo basado en la minimización de las aberraciones más perjudiciales. Un corrector de campo o corrector de subapertura es un sistema óptico diseñado para mejorar la calidad de imagen corrigiendo las aberraciones de un telescopio cerca del plano focal. Se diseñan cerca del plano focal, con el objetivo de reducir el tamaño de la óptica de corrección y de ahí la denominación de subapertura. Por esta razón, no forman parte de este tipo correctores como el *Schmidt*, estudiado en el apartado 3.10.5, y completamente inviables para las grandes aperturas de los IACT. Para realizar la corrección, una estrategia habitual es la utilización de agrupaciones de lentes esféricas con diferentes perfiles [209]. Esta corrección se basa en reducir la cantidad de desviación que aporta cada superficie de forma que el resultado final es más próximo a la óptica paraxial [23]. Algunos principios en los que se basa esta corrección son una adecuada selección de los perfiles (Figura 106, izquierda) y la división de cada curvatura en más de una (Figura 106, derecha) de forma que los rayos se curven muy poco en cada cambio de índice de refracción. Con estas técnicas es posible reducir el OPD P-V de un frente de onda en varios órdenes de magnitud [205, p. 66].

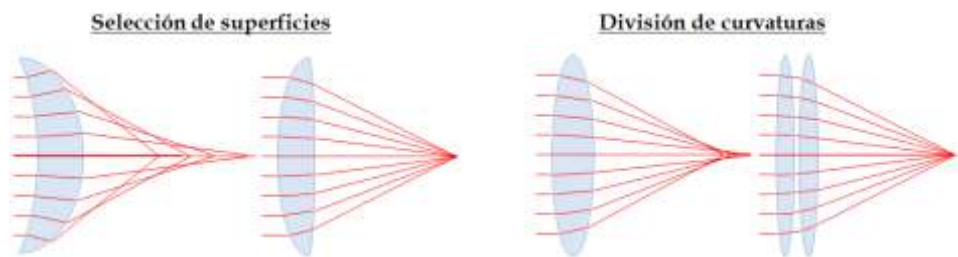

**Figura 106. Corrección de la aberración esférica a una misma longitud focal mediante la selección de perfiles (izquierda) y la división de curvaturas (derecha).**

Existe una variedad de correctores de campo para solucionar problemas específicos [121, p. 325], como el corrector de Ross [210] para minimizar la aberración de coma en telescopios parabólicos, el corrector de Wynne [211] para minimizar el astigmatismo en telescopios tipo Ritchey–Chrètien o el corrector Jones-Bird para minimizar la aberración de telescopios esféricos. Para telescopios con diseños no convencionales, como son el LST o el MST-DC, es necesario realizar un estudio ad-hoc para analizar las soluciones más eficientes de reducir la aberración con correctores de campo.

Aunque las agrupaciones de lentes esféricas son una solución eficiente y de bajo coste para un gran número de aplicaciones, los correctores de campo más utilizados para corregir aberraciones de telescopios se basan en la óptica asférica. Debido al coste de fabricación de este tipo de óptica por su mayor complejidad, el tamaño es determinante. Por ello, un corrector cercano al plano focal es la opción ideal, además de la única en telescopios de foco primario como los de CTA. Por ejemplo, en [212] se demuestra cómo es posible eliminar la aberración esférica de un telescopio esférico mediante un corrector basado en el *Schmidt* pero situado cerca del plano focal, suponiendo un 1 % del coste.

Como ejemplo del tipo de corrector propuesto para los telescopios de CTA, merece la pena mencionar el caso del telescopio *Hobby-Eberly*. Este telescopio fue diseñado con el objetivo de recolectar una gran cantidad de luz (con un diámetro de 9,2 metros de apertura efectiva) con un coste de construcción muy reducido (un 80 % inferior a otros telescopios de tamaño similar). Si bien a la reducción del coste contribuyó el diseño mecánico con un eje de elevación fijo, la óptica también jugó un papel fundamental: su reflector primario tiene una topología segmentada formada por 91 espejos hexagonales y describe un perfil esférico que produce una PSF de varios centímetros (en la línea de los telescopios de CTA simulados en el apartado 2.11). Para la aplicación astronómica a la que se destinaba este telescopio, era necesario corregir la aberración esférica, para lo cual se diseñó un corrector de campo consistente en 4 espejos asféricos con una apertura de 50 cm para el mayor de ellos y una distancia total entre los espejos de los extremos de 1,7 metros, consiguiendo reducir la PSF en unos tres órdenes de magnitud para los rayos en eje [213]. Años más tarde, se sustituyó este corrector por una nueva versión (Figura 107), basada también en 4 espejos asféricos, para permitir corregir la aberración esférica en un campo de visión más amplio (desde los 0,07° originales a 0,37°) [214].

---

[23] Recuérdese que la óptica paraxial se basa en las leyes de Snell utilizando la aproximación de ángulos pequeños, donde el seno de un ángulo es igual al mismo ángulo en radianes. Consecuentemente, al aplicar esta aproximación, desaparecen las aberraciones geométricas, de forma que un objeto puntual se transforma en una imagen puntual (obviando el efecto de la difracción).

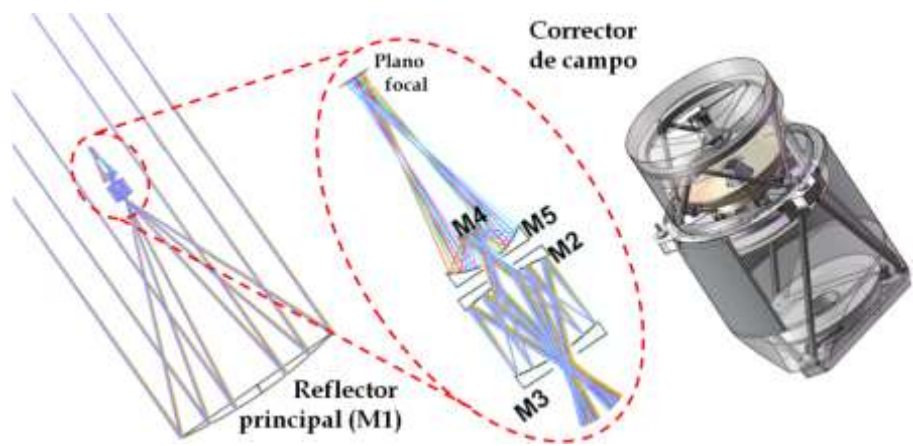

**Figura 107. Corrector de aberración esférica del telescopio *Hobby-Eberly* [214].**

Un ejemplo similar al anterior pero aún más extremo lo constituye el diseño para corregir las aberraciones del cancelado OLT (del inglés *Overwhelmingly Large Telescope*). Este enorme telescopio iba a contar con un reflector primario esférico de 100 metros de diámetro y la corrección también iba a basarse en un corrector de campo con cuatro espejos similar al de *Hobby-Eberly* [215]. Aunque de mucho menor tamaño, es interesante recordar el caso más célebre de corrección de la aberración esférica: el telescopio espacial *Hubble*. Dado que la aberración fue un efecto indeseado debido a un fallo durante la caracterización del espejo primario, se trata de un caso parecido al de la corrección de los IACT, ya que fue necesario diseñar un corrector ad-hoc para un telescopio ya diseñado. El corrector, basado en 2 espejos asféricos, denominado COSTAR (del inglés *Corrective Optics Space Telescope Axial Replacement*), se situó delante de los instrumentos instalados en el *Hubble*, y más tarde fue retirado por pasar a incorporarse de forma nativa en los nuevos instrumentos [216].

Por último, en vista a la similitud del problema, conviene destacar dos ejemplos especialmente interesantes. El primero de ellos es el de un diseño propuesto por el JPL para construir un gran telescopio de foco primario para FSOC desde espacio profundo. Este telescopio tendría una apertura de 10 metros de diámetro, con una relación focal f/1, basada en un reflector segmentado esférico y con un campo de visión deseado de 0,1 mrad. Para conseguir este campo de visión se diseñó un corrector de campo (Figura 108) basado en dos espejos asféricos de 1,47 y 1,1 metros de diámetro y un espaciado de 92,24 cm [161, p. 5].

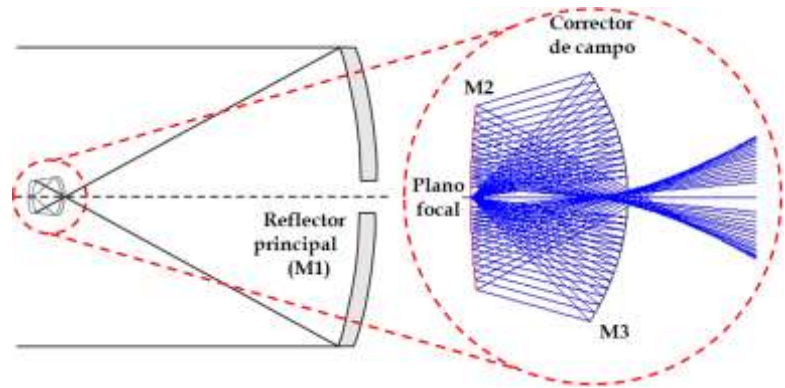

**Figura 108. Corrector de campo propuesto por el JPL para telescopio de FSOC desde espacio profundo basado en reflector esférico de 10 metros [161].**

El segundo ejemplo es otra propuesta del JPL parecida a la anterior, esta vez para corregir las aberraciones de un reflector segmentado de perfil esférico con 8,3 metros de área equivalente, incorporado en la zona central de un radio-telescopio de 34 metros de la red de espacio profundo de la NASA [217]. Este diseño tiene como objetivo lograr un campo de visión de 0,05 mrad y para ello se diseñó un corrector de campo compacto, compuesto por cuatro espejos asféricos de una apertura máxima de 88 cm (M3) y una longitud máxima (distancia entre M3 y M4) de menos de 70 cm (Figura 109). Un corrector de este tipo se podría instalar en el alojamiento previsto para la cámara en los IACT, que en el caso de LST tiene unas medidas de 3,14 × 3,14 m [123, p. 3] y de 3 × 2,8 × 1,8 m en el caso de MST [218, p. 4], aunque también podría ser sustituido por un receptáculo a medida, posiblemente más pequeño y en principio sin estrictas restricciones de peso, ya que las cámaras originales de MST-DC y LST pesan varias toneladas.

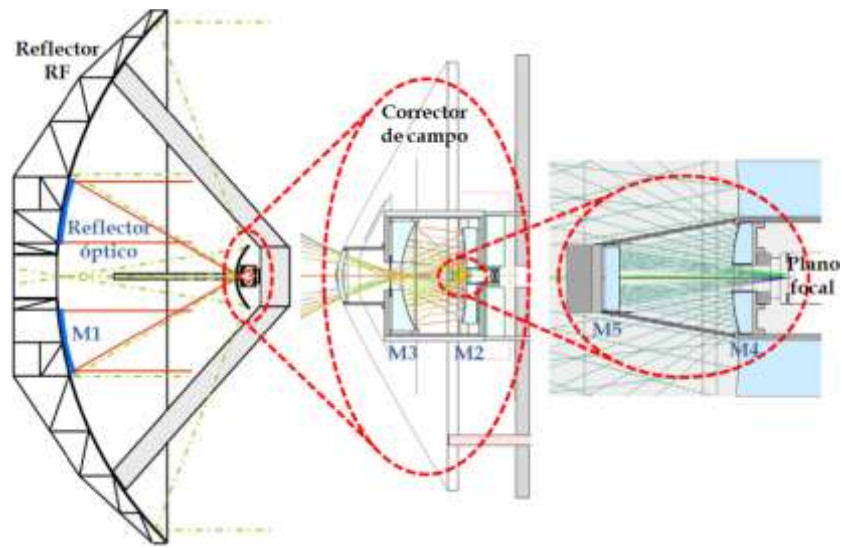

**Figura 109. Corrector de campo propuesto por el JPL para telescopio de 8,3 metros integrado en radio-telescopio de 34 metros [217].**

Por último, dentro de los correctores de campo se podría incluir otra alternativa: la óptica adaptativa (OA). Esta técnica se desarrolló para permitir mejorar la calidad de un sistema óptico, al corregir en tiempo real las perturbaciones introducidas por la atmósfera sobre el frente de onda. Estos sistemas miden muy rápidamente la deformación del frente de onda y lo corrigen usando un espejo deformable controlado por actuadores o una agrupación de espejos orientables. La OA se ha convertido en una herramienta fundamental para la astronomía desde Tierra, al permitir realizar observaciones antes solo accesibles desde el espacio. Un frente de onda, deformado por una óptica como la de los IACT, en principio podría ser igualmente corregido utilizando un sistema de OA.

Para evaluar la aplicabilidad de esta alternativa sería necesario estudiar si es posible realizar la corrección que precisan los telescopios de CTA. Cuanto más aberrado esté el frente de onda a corregir, mayores requisitos se le deberán exigir a los actuadores en cuanto a número de elementos y rango de movimientos. Un punto a favor es que el habitual compromiso entre velocidad y rango dinámico de los actuadores no sería aplicable en este caso, ya que la corrección es estática, o a lo sumo muy lenta. Por otra parte, la ubicación de la corrección es un factor importante en los sistemas de OA. Idealmente deberían ser integrados en el diseño original del telescopio (por ejemplo, en telescopios *Cassegrain* o *Gregorian*, se suele llevar a cabo combinando el sistema corrector en el espejo secundario). Sin embargo, debido a que es una tecnología relativamente reciente, en la gran mayoría de los casos han debido ser añadidos en telescopios ya existentes, como también sería el caso en un IACT.

En [219] el JPL propuso la utilización de un sistema de OA para corregir la aberración de grandes telescopios de bajo coste. Estimaron que un sistema comercial consistente en 400 actuadores podría obtener un coeficiente de *Strehl* de 89 % partiendo de un telescopio de 1,5 metros de diámetro con un coeficiente original de 0,08 %. Pese a que estos sistemas se basan en la monitorización en tiempo real del frente de onda, si se utilizaran para corregir las aberraciones estáticas producidas por la óptica, no sería necesario dedicar una parte de los escasos fotones a esta tarea, pudiéndose basar la corrección en tablas de datos de la caracterización del frente de onda. Esta caracterización podría incluir información relativa a la variación en la deformación del frente de onda con la orientación del telescopio, especialmente relevante en grandes telescopios como LST. Si la corrección del sistema de OA llegara al límite impuesto por la turbulencia (ver apartado 2.10.3), también podría estudiarse la viabilidad de corregir esta, que a 1550 nm es más simple de compensar que en los sistemas visibles habituales [220, p. 19]. Esto sería un valor añadido, pues se lograría superar un límite adicional en la resolución óptica. En [221] se predice una ganancia superior a 8 dB al corregir la turbulencia atmosférica en un enlace desde espacio profundo con 64-PPM en presencia de ruido de fondo a 3° de ángulo SEP.

## 2.14. CONCLUSIONES

- Se ha realizado una propuesta original consistente en la reutilización de la tecnología desarrollada para el observatorio CTA con el objetivo de aprovecharla en la construcción de una o más bases terrenas receptoras para enlaces de comunicaciones ópticas en espacio libre. La relevancia de este resultado queda patente por enmarcarse dentro de una de las líneas de trabajo identificadas como prioritarias por el OLSG en su propósito para desarrollar una red internacional de comunicaciones ópticas por satélite.

- Se ha elaborado una relación de motivaciones que justifican el interés de llevar a cabo la propuesta ofrecida en el primer bloque de esta tesis. Estas motivaciones en general contemplan tanto aspectos técnicos como económicos, cuya convergencia es tan significativa que sugieren que la puesta en práctica de esta idea podría suponer una interesante alternativa al desarrollo ad-hoc de estaciones terrenas receptoras para enlaces de comunicaciones ópticas.

- Se ha llevado a cabo una campaña de medidas experimentales de la reflectividad de espejos de IACT. Los espejos caracterizados incluyen muestras tanto de un telescopio real (MAGIC I) como de los diferentes prototipos contemplados para su utilización en CTA. Esta variedad de espejos permitió caracterizar ejemplos de las diferentes técnicas y materiales que se barajan en CTA. Se concluyó que son perfectamente compatibles con su utilización en comunicaciones en el infrarrojo cercano, la banda espectral de interés en FSOC.

- Se ha llevado a cabo una simulación del telescopio MAGIC II utilizando el software *OpticsLab* basado en óptica geométrica. El resultado de esta simulación proporcionó una primera aproximación al estudio de las limitaciones ópticas que imponen los telescopios *Cherenkov*. Como continuación de esta simulación, se estudió la relación entre la resolución óptica de estos telescopios y su influencia en las prestaciones de un enlace de comunicaciones. Se concluyó que la operación diurna exige reducir el campo de visión del telescopio para minimizar el nivel de ruido de fondo de origen solar acoplado al receptor.

- Se ha elaborado un estudio acerca de las diferentes limitaciones que influyen en el campo de visión. Este estudio incluye todos los efectos que influyen de una u otra manera en la resolución óptica del telescopio, que determina el mínimo campo de visión que es posible establecer. Su relevancia radica en la importancia de identificar el orden de aparición de cada limitación para optimizar los esfuerzos en la modificación de un IACT con el objetivo de reducir su campo de visión. Se concluyó que las aberraciones ópticas constituyen la primera y principal limitación en telescopios *Cherenkov*.

- Se han desarrollado una serie de modelos de los principales telescopios de CTA utilizando el entorno de simulación óptica OSLO. Este software incluye no solo el análisis basado en óptica geométrica, sino también en óptica física y permite llevar a cabo un modelado completo de los telescopios y una caracterización mucho más detallada que *OpticsLab*. Con estos modelos se obtuvieron los diagramas de spot necesarios para calcular el mínimo campo de visión impuesto por la óptica de los telescopios de CTA.

- Se ha llevado a cabo un estudio detallado sobre el comportamiento de los diferentes telescopios de CTA al ser directamente reutilizados como estaciones terrenas receptoras en enlaces de comunicaciones ópticas basados en una serie de distintos escenarios que incluyen una gran variedad de condiciones de operación. Se analizó el comportamiento de cada telescopio en cada escenario y se concluyó que es posible cubrir todas las necesidades de comunicación con diferentes combinaciones de telescopios dependiendo del escenario.

- Se han presentado una serie de propuestas cuyo objetivo es mejorar la resolución óptica en IACT reduciendo el impacto de las aberraciones geométricas, identificadas como la principal limitación de estes telescopios. Si bien los telescopios de CTA pueden ser directamente reutilizados en diferentes escenarios de comunicaciones ópticas en espacio libre, ofrecen un enorme potencial si se realizan algunas adaptaciones con el objetivo de reducir el campo de visión. Para ello, se sugieren y analizan brevemente varios tipos de modificaciones basadas bien en mejorar la óptica de los telescopios o bien en corregir el frente de onda tras haber sido deteriorado por la óptica original.

# 3. Óptica Activa en Distribución Cuántica de Claves

## 3.1. INTRODUCCIÓN

En esta era de la información y en un mundo cada vez más globalizado, las técnicas para intercambiar información de forma segura se han convertido en un asunto de la mayor importancia. La seguridad en las telecomunicaciones, hasta el momento, se ha basado en algoritmos de cifrado que asumen que no existe una forma factible de descifrar la información. Esta suposición se basa en la supuesta ausencia actual de algoritmos o técnicas que comprometan los protocolos existentes, algo que podría cambiar en cualquier momento, sea con la invención de nuevas técnicas criptoanalíticas, el desarrollo de ordenadores mucho más potentes, o la fabricación de computadores cuánticos. La criptografía cuántica pretende resolver esta amenaza con una técnica que permite detectar la intromisión de espías en las comunicaciones entre dos partes. En este caso, a diferencia de la criptografía clásica, la seguridad estaría basada en las leyes inviolables de la mecánica cuántica: que un estado cuántico desconocido no se puede copiar (teorema de no clonación) ni tampoco ser medido sin perturbarlo (principio de incertidumbre de Heisenberg).

Los sistemas de criptografía cuántica, también conocidos como de distribución cuántica de claves, han experimentado un gran desarrollo, especialmente en su alternativa guiada utilizando fibras ópticas como medio de transmisión. Sin embargo, los sistemas no guiados que emplean la atmósfera como medio, presentan una serie de ventajas en relación a la propagación a través de fibra óptica, destacando las menores pérdidas, la ausencia de birrefringencia y la mayor facilidad para desplegar enlaces punto a punto. Estas y otras ventajas los hacen preferibles en determinados escenarios, si bien debido a su posterior surgimiento aún no han alcanzado el grado de desarrollo de la opción guiada, que ya se ha empezado a explotar comercialmente.

Para convertirse en una alternativa real integrable en las redes de comunicación existentes, los sistemas no guiados de criptografía cuántica deben cumplir una serie de requisitos compartidos con otros sistemas más convencionales. Uno de ellos es la operación diurna a plena luz del Sol y en un entorno urbano en presencia de turbulencia atmosférica. La adaptación de estos sistemas de distribución cuántica de claves en espacio libre a estos entornos de trabajo supone un requisito imprescindible al que aún no se ha dado una solución plenamente satisfactoria. Esta tesis explora la problemática relacionada con este tipo de operación y pretende ofrecer una solución integral basándose para ello en un sistema experimental de distribución cuántica de claves actualmente en funcionamiento.

## 3.2. CRIPTOGRAFÍA CLÁSICA

### 3.2.1. Principios de criptografía clásica

La criptografía ha existido desde hace varios miles de años y el propósito básico siempre ha sido alterar un mensaje mediante codificación o cifrado para hacerlo ininteligible ante los intrusos y conservar la confidencialidad del mensaje original con el objetivo de que solo las partes autorizadas puedan acceder a él. Las técnicas criptográficas han ido evolucionando a la par que los ataques a su seguridad (criptoanálisis), aumentando de forma creciente la complejidad, pero en general se puede hablar de un texto plano que es necesario cifrar con algún algoritmo y una clave para transformarlo en un criptograma que más tarde pueda ser descifrado para recuperar el texto plano original mediante otro algoritmo y una clave. Si bien la confidencialidad ha sido el principal objetivo de esta disciplina a lo largo de la historia, con el tiempo han ido surgiendo nuevas propiedades, como la autenticación, integridad y el no repudio. Con el aumento de la capacidad de computación, las técnicas criptográficas han ido aumentando en complejidad, principalmente debido a la creciente facilidad para vulnerar esta seguridad, precisamente propiciada por esta mayor capacidad computacional.

Las técnicas criptográficas modernas se pueden clasificar en dos categorías: cifrados simétricos o de clave secreta y cifrados asimétricos o de clave pública. En los cifrados simétricos, cada par transmisor-receptor normalmente usaría una clave idéntica para cifrar y descifrar. El cifrado del texto plano puede realizarse por bloques o en flujo, siendo la versión en bloques la forma más habitual con protocolos como DES (*Data Encryption Standard*) y sus sustitutos AES (*Advanced Encryption Standard*) y triple-DES. Estos protocolos son muy populares y se usan para una variedad de aplicaciones, desde tarjetas bancarias hasta correo electrónico.

Dado que en los protocolos simétricos cada parte de la comunicación debe poseer la misma clave secreta, la distribución de las claves a través de canales inseguros supone un punto crítico. Este problema lo alivian los protocolos asimétricos, en los que cada extremo de la comunicación tiene asociada una clave que es secreta para todos los demás y otra clave pública de acceso abierto, ambas diferentes pero relacionadas matemáticamente (de forma que no es computacionalmente viable obtener la clave privada a partir de la pública). La clave pública se usaría para cifrar y la privada para descifrar. El esquema de cifrado asimétrico fue propuesto originalmente (si bien años antes había sido inventado de forma independiente y secreta) por Whitfield Diffie y Martin Hellman en 1976 [222], aunque no fue hasta 1978 que Ronald Rivest, Adi Shamir y Len Adleman propusieron su implementación a través del algoritmo RSA [223]. Actualmente existen otros protocolos asimétricos, algunos muy populares como los basados en curvas elípticas.

La mayoría de protocolos asimétricos involucran el uso de operaciones como multiplicaciones y exponenciaciones que son computacionalmente más costosas que el procesado de bloques de los protocolos simétricos. Por ello, los sistemas de uso más extendido son híbridos, por ejemplo cifrando la información de forma rápida con un protocolo simétrico y transmitiendo las claves con un protocolo asimétrico.

La criptografía clásica habitualmente ha empleado los nombres propios de Alice y Bob como los extremos transmisor y receptor del mensaje cifrado, y Eve como el intruso de la comunicación, cuyo objetivo es acceder al texto plano que transmite Alice hacia Bob, asumiendo un canal de comunicación inseguro, al que tiene acceso. Por tradición, la criptografía cuántica ha heredado dicha nomenclatura, por lo que será utilizada en este trabajo de aquí en adelante.

## 3.2.2. El problema de la criptografía clásica

El criptoanálisis es la disciplina que se dedica a buscar debilidades en los sistemas criptográficos para romper su seguridad y acceder a la información secreta. El criptoanálisis ha evolucionado de forma paralela a la criptografía en una carrera para vulnerar la seguridad de cada nuevo sistema criptográfico. Las técnicas han ido cambiando a lo largo de la historia y actualmente ya se basan en sofisticadas técnicas matemáticas, siempre llevadas a cabo por ordenadores.

Para vulnerar la seguridad de los sistemas de criptografía simétrica se puede recurrir al ataque por fuerza bruta (o derivados más sofisticados del mismo), es decir, probar con todas las claves posibles hasta encontrar la correcta. Aunque computacionalmente es un proceso muy costoso, utilizando potentes ordenadores puede ser posible llevarlo a cabo con éxito. Por ejemplo, en 1998 la fundación EFF (*Electronic Frontier Foundation*) fabricó un pequeño sistema que rompía la seguridad de DES en 56 horas de procesado, reduciéndose a 22 horas algunos meses más tarde. Hasta ahora, estos ataques se han evitado aumentando la longitud de las claves y asumiendo que el tiempo necesario para completar con éxito el ataque por fuerza bruta es inviable.

Por otra parte, la seguridad de los protocolos asimétricos se basa en la dificultad computacional de determinadas operaciones matemáticas. En el caso del algoritmo RSA, la factorización de números primos, y en el caso de las curvas elípticas, la resolución de logaritmos discretos. De hecho, si últimamente se ha hecho más popular la criptografía de curvas elípticas es debido a que los métodos numéricos que se han desarrollado para resolver los logaritmos discretos son computacionalmente más eficientes que los desarrollados para factorizar números primos, lo que obliga a usar claves más largas en RSA que en sus equivalentes protocolos de curvas elípticas.

La seguridad de la criptografía está basada en una serie de suposiciones no demostradas: que no existe ningún sistema real que pueda romper la seguridad de los algoritmos en un tiempo razonable. Y por ello, con el aumento de la capacidad de computación, los algoritmos deben ir revisándose, porque lo que en un momento dado es seguro, unos años más tarde puede dejar de serlo. Esto es especialmente cierto con el caso de la computación cuántica. Si bien aún es una disciplina inmadura, su potencial uso en criptoanálisis ya supone una amenaza real a la seguridad de la criptografía actual.

Ya se han creado algoritmos que, en caso de que se desarrollara un auténtico ordenador cuántico, podrían ser aplicados para romper protocolos ampliamente utilizados en la actualidad. Por ejemplo, el algoritmo de Grover [224] permite llevar a cabo ataques por fuerza bruta a una velocidad cuadráticamente mayor que en computación clásica en relación a la longitud de la clave, y el algoritmo de Shor [225] permite factorizar grandes números y encontrar logaritmos discretos en un tiempo que crece cuadráticamente con la longitud de la clave en lugar de exponencialmente en el equivalente clásico. Existe la discusión sobre si el recurso de alargar las claves podría seguir usándose o llegaría un punto en que las claves tendrían una longitud excesivamente larga.

### 3.2.3. Cifrado de Vernam

Pese a la vigencia del panorama descrito, según el cual la seguridad actual de las comunicaciones está en entredicho, tampoco es cierto que todo protocolo criptográfico pueda ser vulnerado. En 1882, Frank Miller propuso por primera vez [226] la libreta de un solo uso (conocido como *one-time pad* en inglés), si bien el protocolo pasó desapercibido y fue reinventado de forma independiente en 1917 por Gilbert Vernam [227], del que recibe el nombre por el que es comúnmente conocido: el cifrado de Vernam. El cifrado propuesto por Vernam no era exactamente el cifrado conocido hoy por cifrado de Vernam, ya que usaba claves que se repetían en bucle, trabajo que fue completado por Joseph Mauborgne para terminar de dar forma al actual cifrado de Vernam. Claude Shannon demostró en 1949 de forma teórica que este cifrado es irrompible [228].

Este cifrado requiere que Alice y Bob compartan inicialmente una clave que solo sea conocida por ellos. Además se deben cumplir tres condiciones para asegurar la seguridad: las claves secretas deben ser aleatorias, no ser reutilizadas nunca y tener la misma longitud que el mensaje a transmitir. Si todas las condiciones se cumplen, este cifrado es el único que hasta la fecha se ha demostrado teóricamente irrompible, incluso ante adversarios con capacidad computacional infinita, y por ello sigue siendo empleado en determinadas situaciones con altas exigencias de seguridad, principalmente en ámbito militar. La operación de cifrado consiste únicamente en una suma módulo 2, implementada mediante una puerta lógica XOR, bit a bit entre la clave y la información a cifrar. El descifrado se realiza aplicando la misma operación bit a bit entre la clave y el mensaje cifrado.

Si bien el cifrado Vernam es el más seguro que existe, no está exento de inconvenientes. Por un lado, el requisito de que la longitud de la clave deba ser igual a la longitud del mensaje impone grandes limitaciones en la cantidad de información que se puede transmitir. Otra dificultad la impone la exigencia de largas claves perfectamente aleatorias, cuya generación es un problema tecnológico no trivial. Por otra parte, las claves deben ser físicamente transportadas entre Alice y Bob de forma totalmente segura, lo que supone un inconveniente mayor cuanto mayor sea la distancia entre ambos. Además, no existe ningún mecanismo para detectar la presencia de un espía durante la comunicación si la clave hubiera sido copiada. La criptografía cuántica ofrece una solución a estos dos últimos problemas, permitiendo hacer uso del cifrado de Vernam al proporcionar una forma segura para transmitir las claves a distancia.

## 3.3. CRIPTOGRAFÍA CUÁNTICA

### 3.3.1. Distribución cuántica de claves

Como se ha visto, la seguridad en criptografía clásica se basa principalmente en tratar de que la complejidad computacional involucrada en los ataques conocidos a los protocolos actuales sea excesiva para llevarlos a cabo con éxito en un tiempo razonable. Por ello, un avance cualitativo en la capacidad de procesamiento de los ordenadores, como el que supondría el desarrollo de un computador cuántico, comprometería gravemente la seguridad de los protocolos actuales. La criptografía cuántica, o más específicamente, la distribución cuántica de claves, supone una forma de proteger las comunicaciones ante esta amenaza, sea directamente gracias a la utilización del cifrado de Vernam, cuyas claves permitiría transmitir en tiempo real, o sea como solución de compromiso con la criptografía clásica usando la criptografía cuántica para sustituir a la distribución de las tradicionales claves de la criptografía simétrica, llevada a cabo actualmente mediante criptografía asimétrica.

A diferencia de la criptografía clásica, donde un bit de información puede estar representado por un 0 o por un 1, la criptografía cuántica se basa, al igual que la computación cuántica, en el concepto de bit cuántico o qubit, que además de tomar el valor 0 ó 1, puede estar en un estado de superposición de ambos. La primera vez que se propone emplear las propiedades de la mecánica cuántica como base para la seguridad es a finales de los '60, cuando Stephen Wiesner sugiere el que sería el concepto clave de la criptografía cuántica: gracias al principio de incertidumbre de Heisenberg, la codificación de información mediante un qubit utilizando una de varias bases conjugadas hace imposible recuperar la información original si no se conoce la base que fue utilizada, ya que un error en la elección de la misma haría que se perdiera la información [229]. De esta propiedad surgió la principal aportación de la distribución cuántica de claves: impedir que la información sea interceptada por un adversario, o más específicamente, detectar cuándo lo está siendo para abortar la comunicación. La primera forma de implementarla en forma de protocolo de criptografía cuántica fue propuesta en 1984 por C.H. Bennett y G. Brassard con el protocolo BB84 [230]. Este protocolo describía una manera de compartir una clave secreta entre Alice y Bob basando la seguridad en que un posible atacante tendría que elegir aleatoriamente la base con la que realizar la medición de cada qubit, lo que destruiría la información cuando la base no coincidiera con la original.

La implementación práctica de este protocolo precisa la utilización de fotones individuales para codificar los qubits, y gracias al teorema de no clonación [231] es imposible realizar dos medidas distintas sobre el mismo estado cuántico. Esto puede llevarse a cabo mediante fuentes de fotones individuales [232] o, más comúnmente, con fuentes de pulsos atenuados [233]. Sin embargo, la utilización de pulsos atenuados es vulnerable a ataques PNSA (del inglés *Photon Number Splitting Attack*), debido a que la técnica de generación de pulsos provoca que a veces se genere más de un fotón por pulso, que podría ser aprovechado para realizar mediciones no detectables. Una solución ante este ataque es la técnica de los estados señuelo (*decoy states* en inglés) [234], consistente en que desde Alice se transmiten aleatoriamente pulsos multifotón y comparando las pérdidas de estos pulsos con las de los pulsos de fotones individuales, es posible detectar una intromisión, en cuyo caso se cancela la comunicación.

Existe una variedad de protocolos de criptografía cuántica, pero todos se pueden agrupar en tres tipos distintos [235, p. 11]: de variable discreta, de variable continua y de fase distribuida. Los protocolos de variable discreta fueron los primeros en desarrollarse y siguen siendo los más frecuentemente implementados. Dentro de esta categoría se enmarca BB84 y B92. Este último protocolo es el implementado en el sistema de distribución cuántica de clave de este trabajo, sin embargo para comprender su funcionamiento es recomendable entender primero el protocolo original en el que se basa, el BB84, explicado en el apartado 3.3.2.

### 3.3.2. Protocolo BB84

El protocolo BB84 utiliza un canal cuántico, en el que se transmiten fotones individuales y al que Eve tiene acceso, y un canal clásico, donde Alice y Bob se comunican de forma autenticada pero asumiendo que Eve puede interceptar toda la comunicación sin ser detectada, es decir, se asume que el canal es público (Figura 110).

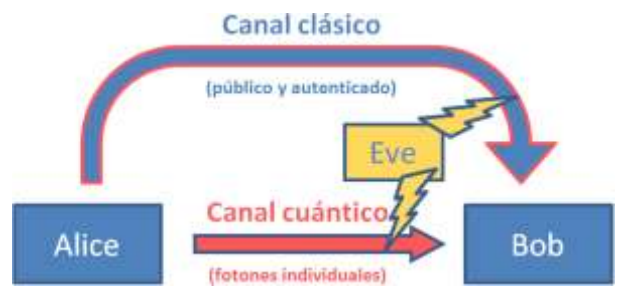

**Figura 110. Esquema básico de un sistema de criptografía cuántica.**

En BB84 se usan dos bases de polarización para codificar la información en los fotones individuales. Estas bases han de ser no ortogonales, de forma que si un fotón se ha codificado en una base, utilizar la base conjugada para medirlo resultaría en una polarización aleatoria, destruyendo la información. Por el contrario, utilizar la misma base en la medición resulta en una medida siempre correcta. Las dos bases conjugadas no ortogonales pueden ser, por ejemplo, rectilínea (polarización lineal a 0° y 90°) y diagonal (polarización lineal a 45° y a -45°) y cada bit, 0 ó 1, se codificará en una polarización distinta. Por ejemplo, En la base rectilínea, el 0 se codificará como 0° y el 1 como 90° y en la diagonal el 0 será 45° y el 1 será -45°.

En el protocolo (Figura 111) se distinguen dos fases: la de comunicación cuántica y la de discusión pública, según el canal de comunicación empleado entre Alice y Bob. Durante la primera, Alice va eligiendo la base de polarización de forma aleatoria para cada bit de información de clave que desea transmitir, codificándolo según corresponda como 0°, 90°, 45° o -45° y guardando la información de qué base se utilizó en cada bit de clave y en qué instante de tiempo se transmitió. En el otro extremo, Bob va midiendo cada fotón recibido, utilizando para ello una base elegida también al azar, y va registrando tanto el resultado de la medida (lo que se denomina la clave en crudo, del inglés *raw key*) como la base empleada y el instante de tiempo en que se recibió. Así termina la fase de comunicación cuántica y comienza la fase de discusión pública. En esta fase, Alice y Bob se comunican a través del canal público, una vez autenticados con alguna técnica clásica, e intercambian las bases utilizadas para codificar cada bit. Dado que ambos las han elegido aleatoriamente, en media coincidirán en un 50 % de los casos. Para los casos restantes en que las bases sean distintas, los bits se descartan, dando lugar a lo que se conoce como clave depurada (del inglés *siftad key*).

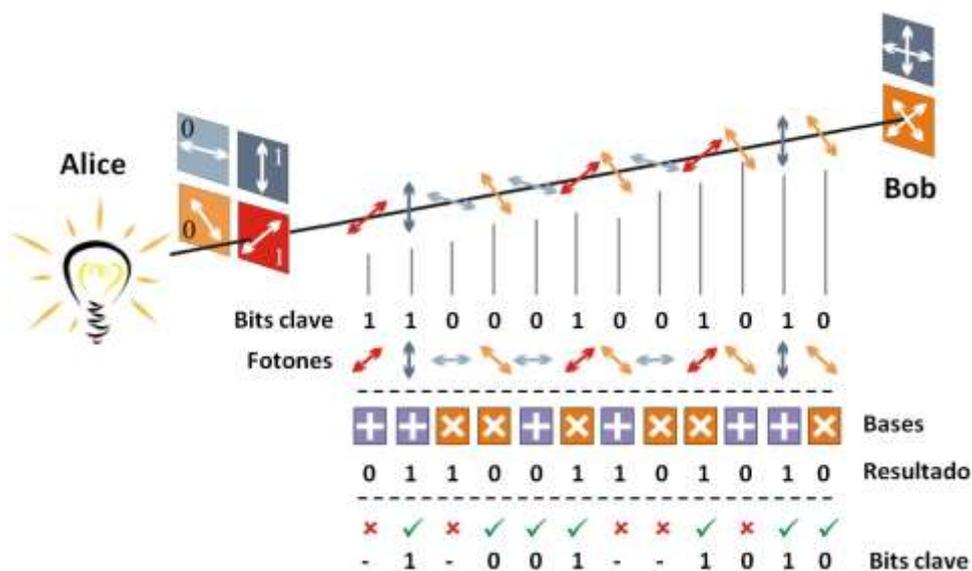

**Figura 111. Esquema de funcionamiento del protocolo BB84.**

Una vez Alice y Bob comparten una clave, antes de poder usarla, es necesario verificar que Eve no la interceptó durante la fase de comunicación cuántica. Para ello, Alice y Bob comparan una pequeña fracción de la clave formada por bits elegidos al azar, para lo cual deben compartir los instantes en que se transmitió/recibió. Comparando la proporción de bits erróneos en relación al total de bits recibidos (relación denominada como tasa de error de bit cuántico, o QBER del inglés *Quantum Bit Error Rate*) es posible determinar si la clave fue interceptada.

Si Eve quisiera copiar la clave, podría medir cada fotón y retransmitirlo tal como fue medido, de forma que, en principio, físicamente no se notaría su presencia. El problema para Eve es que, al igual que Bob, desconoce qué base se usó en la codificación de cada fotón. Por lo tanto, de media acertará en un 50 % de los casos, retransmitiendo correctamente el bit interceptado. Del 50 % restante, tendrá que elegir entre transmitir un 0 o un 1 de forma aleatoria, ya que desconoce el valor del bit, y de media acertará en otro 50 % de los casos. Por ello, en ausencia de otra fuente de errores distinta a esta intromisión, el 75 % de los bits retransmitidos por Eve serán correctos en media. La fortaleza del protocolo BB84 radica en que Alice y Bob pueden asegurar que Eve interceptó la transmisión si, al comparar una fracción significativa de la clave destilada, encuentran que hay un porcentaje de errores igual o mayor al 25 %, es decir QBER ≥ 25 %, en cuyo caso se descarta la clave. En realidad, este límite es aún menor: se ha demostrado teóricamente que es un 11 %, asumiendo condiciones realistas [236].

### 3.3.3. Protocolo B92

En 1987, I. D. Ivanovic demostró que dos estados cuánticos no ortogonales pueden ser distinguidos sin ambigüedad a cambio de añadir pérdidas al sistema [237]. Basado en este principio, C. H. Bennett propuso en 1992 una variante del protocolo BB84: el protocolo B92 [238]. La principal diferencia con BB84 es que se utiliza solamente un estado no ortogonal en cada base complementaria en lugar de dos, con lo que en total resultan dos polarizaciones en lugar de cuatro. Además no es necesaria la reconciliación de las bases, ya que aunque Bob sigue eligiendo la base al azar, ahora puede distinguir sin ambigüedad en los casos en que eligió la base correctamente.

Para comprender este protocolo es conveniente estudiar la implementación de la detección de estados de polarización en Bob (Figura 112). Como en BB84, Alice va codificando los bits de la clave en sendos estados de polarización lineales y formando un ángulo relativo de 45˚ entre sí. En Bob, como en BB84, se va eligiendo aleatoriamente la base de medición. El divisor de haz al 50 % modela en la práctica esta elección aleatoria, ya que un fotón único irá por el camino de transmisión o el de reflexión con una probabilidad del 50 % cada vez, sin tener en cuenta la polarización incidente. Por esta razón, el protocolo B92 será la mitad de rápido que el equivalente BB84, ya que irremediablemente se perderán el 50 % de los fotones en dicha división. Por otra parte, antes de los detectores se disponen polarizadores lineales con una diferencia de orientación de 45˚ entre ambos y a 90˚ de los estados transmitidos. De esta forma cuando uno de los contadores de fotones se dispara, se puede asegurar que el estado recibido es el polarizado a 45˚ de la orientación del analizador previo a ese detector, ya que el polarizador a 90˚ de dicho analizador es bloqueado por el mismo.

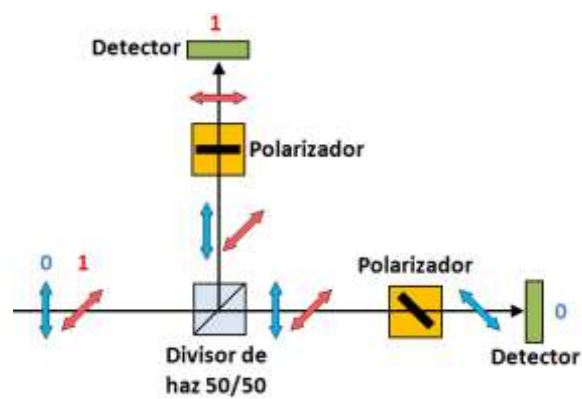

**Figura 112. Implementación del receptor
en Bob utilizando el protocolo B92.**

Cada vez que se recibe un fotón utilizando B92, se puede asegurar que se ha realizado una medida en la base correcta; de otro modo no se detecta nada porque el fotón es bloqueado en uno de los polarizadores. Por ello, en B92 no existe reconciliación de bases y la clave intercambiada será directamente la clave depurada, si bien sí es necesaria una fase de discusión pública tras la fase de comunicación cuántica. En ella, Bob informa a Alice de los instantes en los que midió un fotón que formó parte de la clave. Esta información no compromete la seguridad ya que no se informa del valor de cada bit sino del instante de su recepción. De esta manera, Alice y Bob pueden ponerse de acuerdo sobre qué bits usarán en la clave final para que ambas claves sean idénticas. Previamente, de forma similar a BB84, se ha intercambiado por el canal público una fracción de la clave verificando para cada instante que el bit codificado es idéntico y admitiendo también un cierto valor límite para el QBER a la hora de decidir sobre la presencia de un espía. La fracción de bits que se pueden distinguir sin ambigüedad en Bob es igual a $1 - \cos(\alpha)$, siendo $\alpha$ el ángulo entre ambos estados, que para el caso de 45° es del 29 %.

El protocolo B92 es susceptible de sufrir un ataque a la seguridad conocido como *unambiguous state discrimination* [239], consistente en que si las pérdidas del canal son superiores a 71 %, la presencia de Eve podría pasar totalmente desapercibida si compensara las pérdidas que introduce estableciendo un canal sin ruido entre Eve y Bob. Por ello, el empleo de este protocolo exige un cuidadoso análisis de pérdidas, con el objetivo de eliminar de la clave la información extraída por Eve mediante este ataque. Pese a esta vulnerabilidad y a la disminución de velocidad de B92 respecto a BB84, B92 fue el protocolo seleccionado para la implementación del sistema de distribución cuántica de claves utilizado en este trabajo debido a su mayor simplicidad de desarrollo, si bien siempre se consideró como un paso intermedio orientado a implementar el protocolo BB84 en el sistema final.

### 3.3.4. Reconciliación de claves y amplificación de privacidad

Tras la fase de comunicación cuántica, Alice y Bob han conseguido intercambiar con éxito una clave depurada. Sin embargo, esta clave puede contener errores debido al ruido en el canal de comunicación y a diversos efectos físicos de los dispositivos empleados. Dado que estas claves se van a utilizar para cifrar mensajes en Alice y descifrarlos en Bob, cualquier diferencia entre ambas daría lugar a errores en la comunicación, por lo que es necesario realizar una corrección de errores para reconciliar ambas claves y hacer que sean lo más idénticas posible (por ello, a este proceso también se le conoce como reconciliación de claves), lo que se realiza utilizando el canal público.

Existe una multiplicidad de códigos de corrección de errores que se pueden emplear en esta tarea [240], siendo el esquema de cascada el más ampliamente extendido. Este esquema se basa en un proceso iterativo donde Alice y Bob van dividiendo la clave depurada en bloques, tras lo cual se comprueba la paridad de cada uno. En cada división, los bloques con paridad par se reconocen como libres de errores (bloques idénticos) y no se procesan más, y los de paridad impar se vuelven a dividir en un proceso iterativo hasta la localización y corrección del error. Tras un número de iteraciones, este método logra proporcionar claves idénticas en ambos extremos del enlace.

Eve puede obtener información acerca de la clave si intercepta datos intercambiados durante el proceso de reconciliación, al ser realizado a través de un canal público. Esta información puede eliminarse, mediante el proceso conocido como amplificación de privacidad, con el coste de reducir la longitud de la clave secreta, de forma proporcional a la cantidad de información que potencialmente Eve pueda conocer. El proceso de amplificación de privacidad consiste básicamente en aplicar una operación lógica XOR a pares de bits aleatorios de la clave en Alice y en Bob. El resultado no es anunciado de forma pública, sino que solo se anuncian las posiciones de esos bits, reemplazándolos por el resultado de la operación lógica. Eve no puede obtener ninguna información de este proceso, por lo que su conocimiento sobre la clave se puede disminuir de forma arbitraria.

# 3.4. CRIPTOGRAFÍA CUÁNTICA EN ESPACIO LIBRE

## 3.4.1. Medio de transmisión: guiado vs no guiado

Hasta ahora se ha descrito la distribución cuántica de claves de forma genérica, sin especificar el medio de transmisión. A este medio se le suele denominar canal cuántico, pese a que no es cuántico el canal en sí, sino la información. De igual forma que en las comunicaciones ópticas convencionales, en criptografía cuántica existen dos alternativas: transmisión guiada (mediante fibra óptica) o no guiada (a través del espacio libre). Pese a que los primeros experimentos de QKD fueron no guiados [241], la alternativa más extendida pasó a ser la guiada desde que en 1994 se describiera el primer sistema en fibra utilizando una longitud de onda de 810 nm [242], si bien algo más tarde [243] se empezaron a desarrollar también los sistemas en espacio libre. La opción guiada pudo experimentar un rápido crecimiento debido fundamentalmente a la madurez de las telecomunicaciones convencionales por fibra óptica.

La criptografía cuántica por fibra óptica abre dos caminos diferentes en cuanto a la longitud de onda de trabajo y la tecnología de detección: los detectores de fotones individuales más eficientes comercialmente disponibles están basados en el Silicio, estando su punto de trabajo óptimo, en términos de eficiencia cuántica, situado alrededor de 800 nm. Sin embargo, la tecnología de fibra óptica se ha optimizado alrededor de 1550 nm, donde las pérdidas son las menores posibles. Por ello, los sistemas de QKD en fibra óptica deben elegir el mal menor: la relativamente baja eficiencia de los detectores si se opta por 1550 nm o las relativamente altas pérdidas de la fibra óptica si se opta por 800 nm. Últimamente, parece que el primer camino se está abriendo paso frente al segundo, motivado por los importantes aumentos que se están consiguiendo en la eficiencia de los detectores de InGaAs [14, p. 155]. En cambio, en espacio libre la elección es más directa, ya que la región de máxima eficiencia en detección (800 nm) coincide con un medio con bajas pérdidas, debido a la reducida absorción atmosférica.

La opción guiada tiene una serie de ventajas, además de la mencionada capacidad de aprovechar toda la tecnología desarrollada para comunicaciones ópticas. En relación al espacio libre, la fibra óptica no se ve afectada por factores externos como los debidos a la atmósfera y su gran variabilidad. También está exenta de ruido de fondo de origen solar, que se puede acoplar a los sistemas no guiados. Además no precisa del alineamiento necesario en enlaces en espacio libre, en los que además es necesario mantener en todo momento la línea de visión directa entre los terminales. Sin embargo, las fibras ópticas también presentan inconvenientes, siendo el principal la atenuación, cuando se compara con su alternativa en espacio libre. Las pérdidas debidas a la atenuación de la señal imponen un límite máximo en la distancia de los enlaces, que actualmente se sitúa alrededor de los 300 km [244], aunque con una fuerte dependencia de la tasa de bit, que se ve muy reducida con la distancia. El problema de la atenuación se soluciona de forma trivial en comunicaciones ópticas regenerando y amplificando las señales con un repetidor. Sin embargo, en criptografía cuántica esto continúa siendo inviable por la dificultad para preservar los estados cuánticos sin introducir errores, si bien se están dedicando grandes esfuerzos a esta línea de investigación. Por último, otro problema que presenta la fibra óptica es su birrefringencia. Esta anisotropía óptica modifica la polarización de los estados cuánticos, lo que potencialmente puede degradar los esquemas de QKD basados en la polarización, si bien existen soluciones para evitar el problema.

Las comunicaciones cuánticas en espacio libre carecen en principio de los principales problemas de su alternativa guiada: las pérdidas y la birrefringencia. La atmósfera es un medio esencialmente no birrefringente, por lo que los esquemas basados en codificación en polarización se adaptan bien a él. Por otra parte, una adecuada selección de la longitud de onda asegura unas pérdidas reducidas sobre grandes distancias (se han demostrado factibles enlaces horizontales de distancias superiores a los 300 km [245]). Sin embargo, el medio también impone limitaciones, como pueden ser el ruido ambiental debido sobre todo a la radiación solar que se acopla al sistema y los efectos de la turbulencia atmosférica en el enlace de comunicaciones. Por otra parte, los enlaces no guiados suponen la única alternativa en comunicaciones por satélite (sean entre satélites o entre satélite y Tierra) [246], donde la criptografía cuántica ofrece una solución apropiada para implementar una red segura a nivel global, aliviando la congestión de las redes metropolitanas, así como dotando de seguridad a determinados enlaces estratégicos. Por otro lado, los enlaces horizontales también ofrecen una interesante alternativa, principalmente integrados como parte de una red metropolitana de distribución cuántica de claves, al dotar a estas redes de una mayor flexibilidad y adaptabilidad. Esto se debe a la posibilidad de desplegar enlaces con más facilidad que la instalación de un sistema de fibra óptica. A su vez tienen el inconveniente de sufrir más intensamente los efectos de la atmósfera, al estar situados a muy baja altura y en un entorno de temperaturas muy cambiantes.

## 3.4.2. Descripción del sistema experimental de QKD

A continuación se describe el sistema original de QKD sobre el que se ha realizado este trabajo. El montaje experimental implementa el protocolo B92, explicado en 3.3.3 y un diagrama de dicho sistema puede observarse en la Figura 113, compuesto por los siguientes elementos:

- $V_0$, $V_1$ y $V_{SYNC}$ son láseres tipo VCSEL,
- $A_0$ y $A_1$ son atenuadores de fibra,
- $P_0$ y $P_1$ son polarizadores de alto coeficiente de extinción,
- BS son cubos divisores de haz de 50/50,
- PL es una película divisora de haz de banda ancha,
- DM es un espejo dicroico,

- IF es un filtro interferencial,
- $D_0$ y $D_1$ son detectores de fotones individuales,
- $D_{SYNC}$ es un detector óptico.

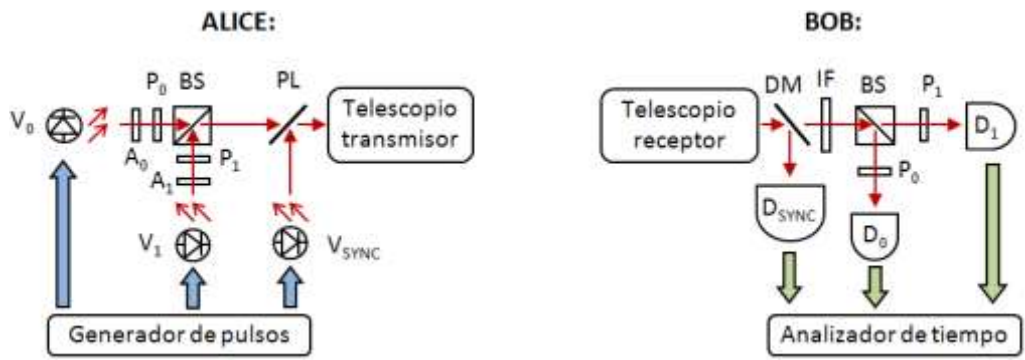

**Figura 113. Diagrama del sistema de QKD incluyendo transmisor (Alice) y receptor (Bob).**

El emisor Alice utiliza un generador de pulsos a frecuencias de GHz para alimentar con una secuencia preprogramada de bits a los drivers que controlan los tres láseres. La emisión de 850 nm a Gbit/s de cada láser $V_0$ y $V_1$ se polariza linealmente con un ángulo relativo entre ambos igual a 45° para codificar los datos binarios de la clave secreta. Ambos láseres se combinan mediante un divisor de haz, dando lugar al denominado canal cuántico, cuya señal es fuertemente atenuada hasta alcanzar el régimen de fotones individuales. Dado que no es posible amplificar la señal cuántica y cada fotón debe dedicarse al protocolo criptográfico, no es posible dividir la señal recibida para extraer información de sincronismo. Para ello se utiliza un tercer láser $V_{SYNC}$ emitiendo a 1550 nm. Esta última señal no se atenúa ya que no se modula ninguna información secreta en ella. Finalmente, la señal de sincronismo se combina con la señal cuántica a 850 nm mediante una película divisora de haz y la señal óptica resultante se transmite colimada mediante un telescopio, empleado para expandir el haz (con el objetivo de disminuir su divergencia), hasta un diámetro de unos 40 mm.

Los fotones que llegan al receptor Bob son recogidos y focalizados mediante un telescopio *Schmidt-Cassegrain* y son separados en longitud de onda mediante un espejo dicroico, que transmite el haz de 850 nm y refleja el de 1550 nm. Tras esta separación, y después de realizar un filtrado espectral mediante el IF, el canal cuántico se vuelve a separar en dos caminos mediante un BS, que modela la elección aleatoria de bases de B92 (al estar el canal cuántico compuesto por fotones individuales, se puede asegurar que cada fotón tendrá un 50 % de probabilidad de ir por cada camino). Finalmente, en cada canal se establece la base de medición mediante polarizadores lineales con un ángulo relativo de 45° entre ambos y, tras detectarlos con sendos contadores de fotones en el caso del canal cuántico y mediante un fotodetector en el caso del canal de sincronismo, los tiempos de llegada de cada bit se registran mediante un analizador de tiempo.

### 3.4.3. Descripción del transmisor Alice

En la Figura 114 se muestra un diagrama detallado del transmisor Alice. En este diagrama, $V_0$, $V_1$ y $V_{SINC}$ representan los diodos láser, $A_0$ y $A_1$ son atenuadores acoplados a fibra, $C_0$, $C_1$ y $C_{SINC}$ son colimadores de fibra a espacio libre, $P_0$ y $P_1$ son polarizadores lineales, $\lambda/4$ es una lámina de cuarto de onda, BS es un cubo divisor de haz, PL es una película divisora de haz, $L_1$ y $L_2$ son lentes tipo doblete y $M_1$ y $M_2$ son espejos metálicos de alta reflectividad.

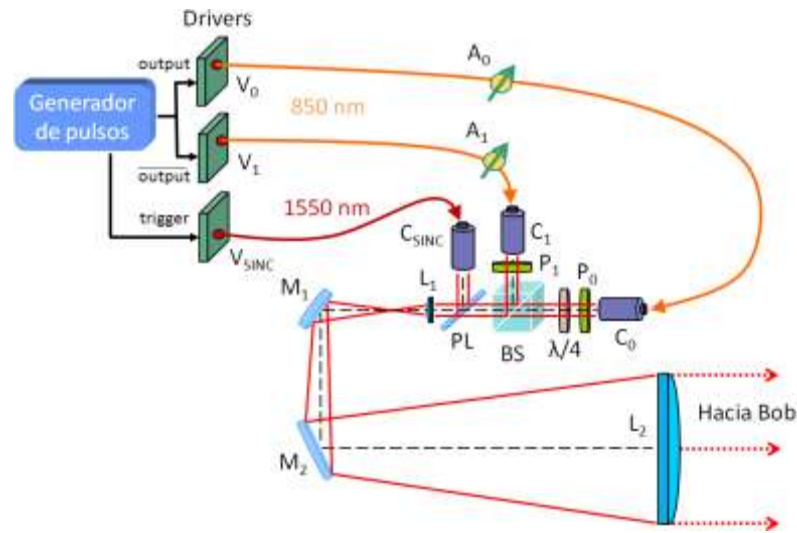

**Figura 114. Diagrama del transmisor Alice.**

La generación de los pulsos que conformarán la clave se lleva a cabo mediante el generador *Agilent 81133A PPG*, que proporciona una frecuencia máxima de 3,33 GHz. La frecuencia de transmisión utilizada es de 1,5 GHz, al ser la máxima frecuencia utilizable sin degradación del QBER debido a las interferencias entre bits adyacentes causadas por el *jitter* del sistema (principalmente debido a los detectores). Las dos salidas de este generador, invertida y no invertida, se utilizan para alimentar a los drivers que controlan los diodos láser de forma que cuando un láser está encendido el otro está apagado. La salida *trigger* alimenta al driver del láser de sincronismo a una frecuencia de 10 MHz (impuesta por la entrada de reloj del analizador de tiempo de Bob). Los diodos láser utilizados, de tipo VCSEL, son el modelo *Finisar HFE4192-582* para 850 nm y *RayCan 1550* para 1550 nm, ambos capaces de superar los 4 Gbps. Estos diodos láser están montados en *drivers* modelo *Maxim MAX3795*, encargados de suministrar a los VCSEL las corrientes de polarización o *bias* y de modulación, ajustadas para minimizar el QBER (ya que estas corrientes determinan la forma en la que se atenúan los pulsos). Cada uno de los VCSEL de 850 nm se acopla mediante fibras monomodo a atenuadores de potencia con el objetivo de alcanzar el régimen de emisión de fotones individuales. Esto se consigue estableciendo una atenuación tal que el número medio de fotones por segundo sea $\mu \sim 0{,}1$ (dividiendo la tasa binaria real en un factor 10), lo que garantiza que solo el 0,5 % de los pulsos emitidos contendrán más de un fotón. Por último, las tres fibras monomodo se conectan directamente a tres colimadores.

Tras la atenuación y la colimación, se establece el estado de polarización de cada base mediante polarizadores de alta extinción. La polarización es necesario establecerla ya en espacio libre, debido a la birrefringencia de las fibras ópticas. Es fundamental conseguir una alta linealidad para minimizar la contribución al QBER de errores debido a la discriminación de la polarización. Sin embargo, en los divisores de haz y en los espejos, al producirse cierta absorción, el índice de refracción se vuelve complejo, lo que hace que los estados lineales se conviertan en elípticos para ángulos de incidencia distintos a 0° ó 90°. Dado que los dos estados de polarización de B92 no son ortogonales, en su implementación uno de ellos conservará bien su polarización y en el otro sufrirá cierta degradación. Por ello, el estado vertical se estableció en el ángulo óptimo que permanecía inalterado, y el diagonal se corrigió mediante una lámina de cuarto de onda, dispuesta tras el polarizador lineal a 45° del vertical. Con dicha corrección, se consiguió una extinción de 1:7000 para el estado vertical y 1:4000 para el estado diagonal tras la corrección.

Finalmente, tras combinar el haz de 1550 nm con el haz de 850 nm, se colima este último a un tamaño mayor (unos 40 mm de diámetro) con el objetivo de disminuir su divergencia. Por último, es necesario alinear los tres haces para conseguir que lleguen al receptor exactamente en la misma posición. Para ello, uno de los colimadores se instaló en una montura fija, utilizando monturas variables en altitud/acimut para los otros dos. Para conseguir la coincidencia de los haces, se les hizo reflejar en un espejo curvo y se monitorizó en una cámara CCD la posición de cada uno de los tres focos, ajustando las monturas hasta lograr que todos fueran coincidentes.

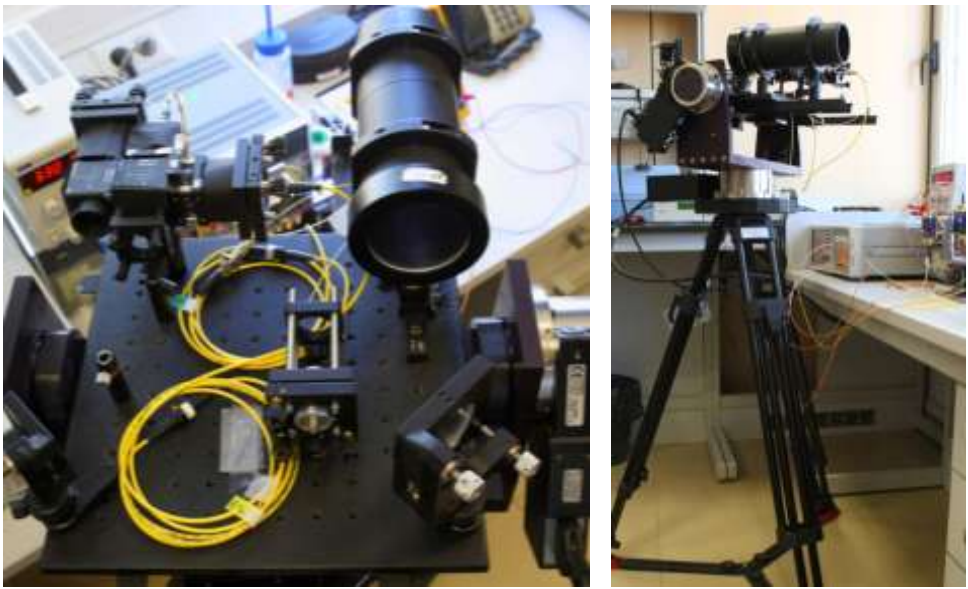

**Figura 115. Imagen del transmisor Alice: montaje óptico (izquierda) y montura y trípode (derecha).**

Todo el montaje de Alice está dispuesto en una pequeña mesa óptica instalada en una montura altacimutal con trípode. Cada eje (Figura 115) está controlado mediante un motor de alta precisión angular (0,00025° para el eje vertical y 0,001° para el horizontal) y programados mediante el controlador *Newport ESP301*. Para aislar a todo el transmisor de la luz de fondo que pudiera acoplarse al sistema, especialmente a través de los espejos, se enclaustró el conjunto dentro de una caja construida con material anti reflectante, y se acopló a la salida un tubo de extensión para disminuir el campo de visión, tras lo cual se pudo medir una reducción del ruido de fondo de unos 6 dB.

### 3.4.4. Descripción del receptor Bob

En la Figura 116 se muestra un diagrama detallado del receptor Bob. En este diagrama, $\lambda/2$ es una lámina de media onda, DM es un espejo dicroico, L es una lente, xyz es un posicionador de tres ejes para acoplar la señal óptica a la fibra, APD (del inglés *Avalanche PhotoDetector*) es un fotodetector de avalancha, IF es un filtro interferencial, BS es un divisor de haz, λ/4 es una lámina de cuarto de onda, P₀ y P₁ son polarizadores lineales y SPAD (del inglés *Single Photon Avalanche Detector*) es un contador de fotones individuales.

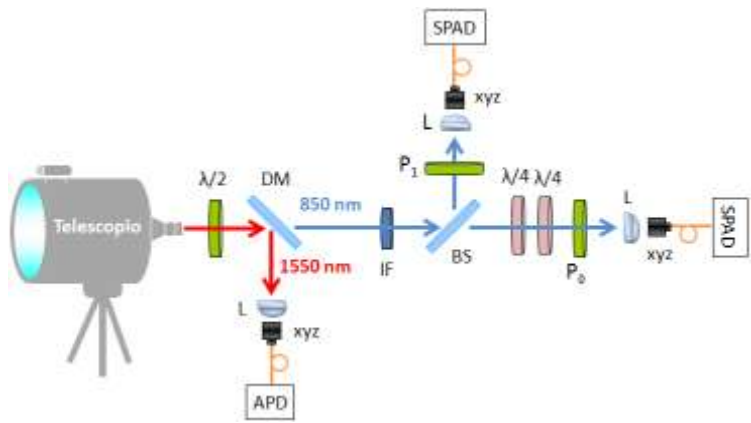

**Figura 116. Diagrama del receptor Bob.**

La fracción de fotones que llegan a Bob desde Alice son recogidos mediante el telescopio *Meade LX200*, que los focaliza junto con lentes de 30 mm de focal para poder acoplarlos a las fibras ópticas de 62,5 μm de diámetro acopladas a los fotodetectores. El espejo dicroico refleja los fotones de 1550 nm, que son enviados hacia una fibra de 200 μm de diámetro conectada a un APD, cuya señal de salida es amplificada con una etapa de transimpedancia de alta velocidad. Tras ser amplificada, se envía a la entrada de 10 MHz de reloj externo del analizador de tiempo para mantener la misma base de tiempos entre Bob y Alice. El analizador de tiempo empleado es el GT658PCI de *GuideTech,* que ofrece un ancho de banda de 400 MHz y una resolución temporal de 75 ps. Su misión es registrar temporalmente los eventos relativos a la llegada de los fotones de 850 nm. Mediante un software desarrollado en *LabView* se puede obtener la lista de tiempos de llegada para cada uno de los dos canales.

El camino de 850 nm se filtra espectralmente para eliminar el ruido de fondo mediante un filtro interferencial de 1 nm de banda de paso centrada en 848 nm y tras la focalización se realiza la detección empleando los contadores de fotones individuales. Existen diversas tecnologías para discriminar fotones, siendo la más común en los sistemas de QKD los detectores de avalancha en modo *Geiger*. Estos detectores son APD diseñados para trabajar a una tensión de polarización por encima de la ruptura, de forma que un solo fotón es capaz de desencadenar el proceso de avalancha y generar una corriente eléctrica. Se trata de una tecnología madura, con alta eficiencia de detección, pocas cuentas oscuras [24], alta resolución temporal y disponible comercialmente. Además, en el caso de los detectores de Silicio (para detectar 850 nm), no es necesaria una refrigeración criogénica, como es el caso del Germanio o el InGaAs, para minimizar las cuentas oscuras. Los SPAD utilizados son el modelo SPCM-AQR-12 de *PerkinElmer*, que puede detectar 15 millones de cuentas por segundo con una eficiencia de detección en torno al 45 % a 830 nm y unas 500 cuentas oscuras por segundo.

En cuanto a la detección óptica de los estados de polarización, los fotones de 850 nm se desvían aleatoriamente hacia cada camino que define una base de medición mediante el divisor de haz. En el camino reflejado el polarizador previo al detector está orientado horizontalmente para bloquear el estado vertical y en el camino transmitido el polarizador está orientado a 90° del estado diagonal para bloquear estos fotones. Este esquema asegura la discriminación determinista de los estados cuánticos a costa de perder el 75 % de los fotones: el 50 % del total se pierden en el divisor de haz, que modela la elección aleatoria de bases, haciendo que la mitad vaya por el camino incorrecto y sean bloqueados por el polarizador correspondiente; y del 50 % de fotones que recorren el camino correcto, un 50 % adicional se pierden en el polarizador al estar a 45° del estado de polarización que se pretende detectar en ese camino.

---

[24] Las cuentas oscuras son las detecciones de fotones que se producen en ausencia de un fotón que las dispare. En consecuencia, se tratan de detecciones erróneas o falsos positivos.

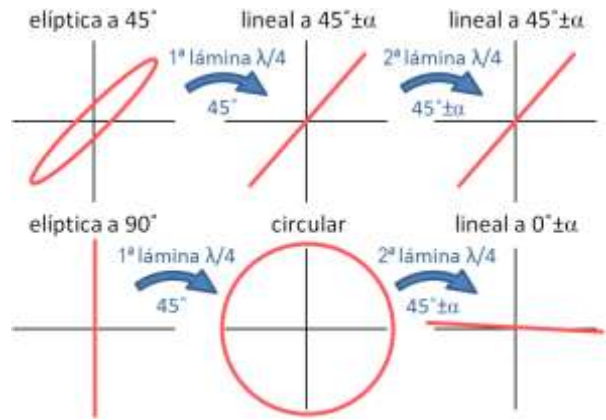

**Figura 117. Efecto de dos láminas de cuarto de onda sobre dos estados polarizados elíptica y linealmente a un ángulo relativo de 45°.**

El primer elemento del montaje óptico es una lámina de media onda. El propósito de esta lámina es alinear las polarizaciones lineales que conforman el haz incidente con el receptor (específicamente con el espejo dicroico, que es el primer elemento transmisivo que se encuentra). Para ello, se ajusta el ángulo de dicha lámina hasta maximizar la extinción del estado vertical, que continuará por el camino reflejado del divisor de haz. Tras este alineamiento la extinción del estado vertical, que era de 1:6700 en Alice, pasa de 1:88 (debido a que el dicroico no está perfectamente alineado con la polarización del haz) a 1:1190. Con este estado optimizado, el problema de la degradación del estado diagonal que se daba en Alice se repite en Bob: tras el espejo dicroico y el divisor de haz, la polarización del estado diagonal se ha degradado, pasando de lineal a elíptica, y es necesario corregirla. Esta corrección se puede realizar utilizando una lámina de cuarto de onda, ajustando su eje respecto a la polarización incidente a un determinado ángulo, hasta convertir la polarización elíptica en lineal. Sin embargo, tras esta lámina el estado vertical se ha transformado en circular. Para hacerlo lineal se utiliza una segunda lámina de cuarto de onda, que no afecta al estado diagonal, ya que está alineado con el eje de la lámina. Tras esta corrección, la elipticidad del estado diagonal se ha eliminado y el estado vertical se ha transformado en prácticamente horizontal, lo que no supone un problema porque sigue existiendo una diferencia de 45° entre ambos (Figura 117).

Las pérdidas totales de Bob fueron medidas como 8,95 dB, siendo la principal aportación debida al recubrimiento antirreflectante del telescopio, que está optimizado para el visible. Considerando estas pérdidas, además de las debidas al protocolo B92 y las debidas a la eficiencia de los detectores, la transmisividad total es de 0,01. Con esta transmisividad, y conociendo la tasa de bit de Alice, es posible estimar la tasa de bit esperada en la clave detectada de Bob. Dado que en Alice se genera una señal de 1,5 Gbit/s y se atenua hasta conseguir 0,1 fotones por pulso, la tasa de emisión de Alice será de 150 Mbps. Tras todas las pérdidas en Bob, que proporcionan una transmisividad de 0,01, la tasa de clave resultante se puede estimar en unos 1,5 Mbps, sin tener en cuenta las pérdidas del canal de comunicación.

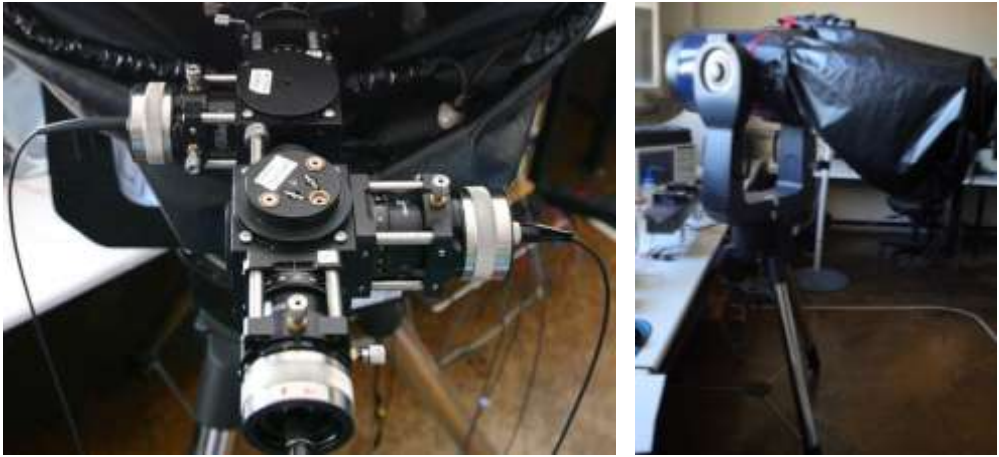

**Figura 118. Imagen del receptor Bob: montaje óptico (izquierda) y montaje en telescopio (derecha).**

# 3.5. PROPAGACIÓN DE HACES DE COMUNICACIONES

## 3.5.1. Propagación en espacio libre de haces gaussianos

El campo eléctrico E de una onda electromagnética que se propaga en el espacio libre debe ser una solución de la ecuación de onda (3-1).

$$\nabla^2 E = \frac{1}{c^2} \cdot \frac{\partial^2 E}{\partial t^2} \qquad (3\text{-}1)$$

En un enlace de comunicaciones ópticas, especialmente al usar un láser como fuente y debido a su alta directividad, el interés se concentra en el estudio del campo cercano al eje óptico, para lo cual se puede emplear la aproximación paraxial. Siguiendo el método habitual en electromagnetismo para simplificar las derivadas parciales de la ecuación de onda con la separación de variables, se obtiene la ecuación de Helmholtz, que se puede resolver utilizando la aproximación paraxial llegando a la ecuación (3-2), que describe el campo eléctrico para el modo menor de la onda electromagnética o TEM$_{00}$, correspondiente al haz gaussiano [247, p. 153].

$$E(r,z) = E_0 \frac{\omega_0}{\omega(z)} \cdot e^{\frac{-r^2}{\omega^2(z)} - ikz - ik\frac{r^2}{2R(z)} + i\xi(z)} \qquad (3\text{-}2)$$

donde r es la distancia radial desde el eje óptico, z es la distancia axial o longitudinal desde el punto donde el haz es más estrecho, llamado "cintura", k es el número de onda, $E_0$ es igual a $|E(0,0)|$, $\omega(z)$ es el radio en el que la amplitud del campo cae $1/e$ de su valor en el eje, $\omega_0$ es el radio de la cintura del haz $\omega_0 = \omega(0)$, R(z) es el radio de curvatura del haz y $\xi$ es un desplazamiento de fase, denominado fase de Gouy, definido según la ecuación (3-3).

$$\xi(z) = \arctan\left(\frac{z}{z_R}\right) \qquad (3\text{-}3)$$

La intensidad promedio I(r, z) del haz viene dada por la ecuación (3-4), donde η representa la impedancia característica del medio, que para espacio libre sería $\eta = \sqrt{\mu_0/\varepsilon_0} \approx 376{,}7\ \Omega$, y $P_0$ es la potencia total transmitida por el haz, cuya propagación se expresa mediante la ecuación (3-5) que define la potencia P (r,z) que atraviesa el círculo de radio r del plano transversal en la posición z.

$$I(r,z) = \frac{|E(r,z)|^2}{2\eta} = I_0 \cdot e^{\frac{-2r^2}{\omega^2(z)}} = \frac{2P_0}{\pi\omega^2(z)} \cdot e^{\frac{-2r^2}{\omega^2(z)}} \qquad (3\text{-}4)$$

$$P(r,z) = P_0 \left(1 - e^{\frac{-2r^2}{\omega^2(z)}}\right) \qquad (3\text{-}5)$$

## 3.5.2. Parámetros de un haz gaussiano

Una vez vista la expresión de la distribución de campo eléctrico e intensidad de un haz gaussiano, se pueden definir una serie de parámetros que determinan su comportamiento geométrico (Figura 119).

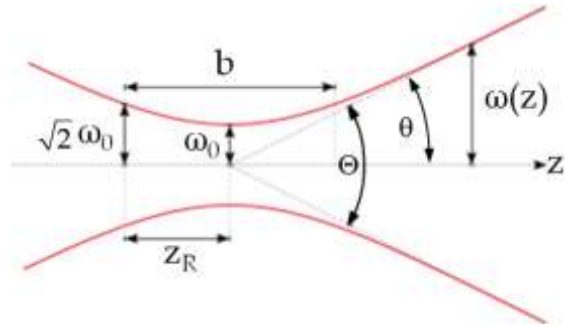

**Figura 119. Principales parámetros de la propagación de un haz gaussiano en espacio libre.**

La anchura del haz o tamaño de spot, definido como su radio ω (z), tomará un valor mínimo $\omega_0$ en la cintura del haz. La variación del tamaño del haz viene definida por la ecuación (3-6).

$$\omega(z) = \omega_0 \sqrt{1 + \left(\frac{z}{z_R}\right)^2} \qquad (3\text{-}6)$$

El parámetro $z_R$ representa la distancia Rayleigh, definida como la distancia desde la cintura del haz hasta donde su área se dobla, es decir, ω $(z = \pm z_R) = \sqrt{2}\ \omega_0$, y depende del tamaño de la cintura y la longitud de onda según la ecuación (3-7).

$$Z_R = \frac{\pi\omega_0^2}{\lambda} \qquad (3\text{-}7)$$

El radio de curvatura R(z) del frente de onda varía en función de la distancia z según la ecuación (3-8).

$$R(z) = z \left[ 1 + \left( \frac{z_R}{z} \right)^2 \right] \qquad (3-8)$$

El radio de curvatura tiene sentido en distancias cercanas a la cintura del haz. Para distancias mucho mayores a la distancia de Rayleigh ($z \gg z_R$), la propagación del haz se define con un ángulo de divergencia, cuyo semiángulo se traza entre el eje óptico y una línea recta que tiene como origen la cintura del haz $\omega_0$. Esta divergencia $\Theta = 2\theta$, viene descrita por su semiángulo $\theta$ en radianes según la ecuación (3-9). Nótese que esta divergencia es menor a la de la ecuación (2-17), ya que aquella incluía el efecto de la difracción de la apertura.

$$\tan \theta = \lim_{z \to \infty} \frac{\omega(z)}{|z|} = \frac{\omega_0}{z_R} = \frac{\lambda}{\pi \omega_0} \qquad (3-9)$$

### 3.5.3. Tamaño del spot focal de un haz gaussiano

Un parámetro fundamental en el diseño de sistemas ópticos es el tamaño de spot en el plano focal. Históricamente, en el campo de la óptica previo a la aparición del láser siempre se había utilizado el diámetro del disco de Airy para calcular el tamaño de spot de un haz enfocado genérico [248]. Para ello, se utilizaba la clásica descripción de un frente de onda plano que ilumina una apertura de forma homogénea. Tras ella, el diámetro del primer anillo (donde la intensidad se hace cero) se podía medir con facilidad y por ello tradicionalmente se utilizó como referencia. Este diámetro d viene dado por la ecuación (3-10), donde $\lambda$ es la longitud de onda, f es la longitud focal del sistema y D la apertura del mismo.

$$d = 2,44 \frac{\lambda f}{D} \qquad (3-10)$$

Sin embargo, la mayoría de los haces que provienen de un láser tienen un perfil de intensidad gaussiano y frecuentemente no iluminan completa y homogéneamente los sistemas ópticos con el objetivo de no perder una gran parte de la intensidad de la fuente. En consecuencia, en estos casos el haz no será truncado por la apertura, no formará anillos de difracción y no podrá utilizarse el modelo del disco de Airy para calcular el tamaño del spot en el foco. En un caso general, en un sistema no limitado por aberraciones, el spot focal tendrá el tamaño d dado por la ecuación (3-11) [249, p. 2.7], donde $\lambda$ es la longitud de onda, K es una constante que depende del factor de truncamiento y de la iluminación de la apertura del sistema, f es su longitud focal y D su apertura.

$$d = K \frac{\lambda f}{D} \qquad (3-11)$$

La constante K, cuando se utiliza una anchura de haz correspondiente a $1/e^2$ de la potencia, se puede aproximar mediante la ecuación empírica (3-12) [250, p. 419], que es válida para T>0,4.

$$K = 1{,}6449 + \frac{0{,}646}{\left(T - 0{,}2816\right)^{1{,}821}} - \frac{0{,}532}{\left(T - 0{,}2816\right)^{1{,}891}} \qquad (3\text{-}12)$$

En la ecuación (3-12), T es el factor de truncamiento definido por la ecuación (3-13), donde $d_0$ es el tamaño del haz a la entrada según el criterio $1/e^2$ y D es el tamaño de la apertura.

$$T = \frac{d_0}{D} \qquad (3\text{-}13)$$

Para T>2 (equivalente a asumir que se ilumina la apertura de forma homogénea, resultado de efectuar un elevado truncamiento sobre el haz gaussiano), el valor de d sería similar al calculado con el modelo de Airy, si bien la constante 2,44 de la ecuación (3-10) se refiere al primer nulo, no al criterio de $1/e^2$ al que se refiere la ecuación (3-12). Para T=0,53, entonces K=2,4 y K·T=1,27=4/$\pi$, por lo que se puede asumir que no se produce truncamiento y se puede usar generalizando para T<0,5 la ecuación (3-14) de un haz gaussiano sin truncamiento [205, p. 204].

$$d = K\frac{\lambda f}{D} = KT\frac{\lambda f}{d_0} = \frac{4}{\pi} \cdot \frac{\lambda f}{d_0} \qquad (3\text{-}14)$$

Para evaluar la importancia de este truncamiento se puede utilizar la ecuación (3-15), que proporciona la pérdida de potencia $P_L$ en función del factor de truncamiento T. Así, manteniendo un mismo tamaño de haz $d_0$, un factor de truncamiento de T=2 permitiría utilizar una apertura de un diámetro 4 veces menor, en comparación con utilizar un factor de truncamiento T=0,5, a costa de perder un 60,65 % de la potencia original en lugar de un 0,03 %.

$$P_L = \exp\left(-2\big/T^2\right) \qquad (3\text{-}15)$$

La conclusión de este estudio del spot focal es que en un enlace de QKD como el de esta tesis, en el que un factor fundamental de diseño es no perder fotones, las aperturas de la óptica del receptor se diseñarán de tal forma que el truncamiento produzca una pérdida de potencia mínima. Para ello, en general se han utilizado aperturas con truncamiento de 0,5 o menores. Una consecuencia importante de este criterio de diseño estriba en que será necesario tener cuidado al obtener el tamaño de spot en el foco del sistema al utilizar programas de diseño óptico. Estos programas habitualmente proporcionan el disco de Airy como el tamaño mínimo del spot en un sistema limitado por difracción donde cualquier otra aberración tiene una influencia menor. Sin embargo, es importante tener en cuenta que esta medida la proporcionan asumiendo que $D = d_0$ (equivalente a reducir el tamaño de la apertura D hasta el tamaño del haz $d_0$ y, por lo tanto, imponer un truncamiento, independientemente de la relación de tamaños de la apertura y el haz que se haya pretendido simular) y utilizando la ecuación (3-10), como en el caso del software *OpticsLab* empleado para todas las simulaciones. Por ello, si se emplea un factor de truncamiento menor a 0,5, deberá utilizarse la ecuación (3-14) considerando el tamaño del haz $d_0$ en lugar del tamaño de la apertura D. Por simplicidad, con el objetivo de utilizar directamente el resultado de la medida del disco de Airy proporcionado por el software de óptica, para calcular los tamaños de spot focal d, en este trabajo se ha corregido la medida del disco de Airy con un factor 1,27/2,44 = 0,52, lo que es equivalente a aplicar la ecuación (3-14).

# 3.6. TURBULENCIA EN ENLACES INTERURBANOS

### 3.6.1. Parámetros diferenciales en enlaces interurbanos

Como se introdujo en el apartado 2.10.3, los efectos de la turbulencia sobre un haz láser que se propaga a través de la atmósfera son variados en función de una serie de factores. Aún hoy, tras décadas de estudio, los efectos de la turbulencia atmosférica sobre un haz láser no están bien modelados. Ya en los años 70, el premio Nobel Richard Feynman señaló a la turbulencia como el "último gran problema sin resolver de la física clásica" [251]. Si bien existen diferentes modelos que pueden ser más realistas en determinados escenarios [252], en este trabajo se ha empleado el modelo clásico que asume una turbulencia homogénea, isótropa, estacionaria basada en el espectro Kolmogorov, que considera escalas espaciales por encima de la denominada escala interior $l_0$ y por debajo de la denominada escala exterior $L_0$. Teniendo en cuenta este modelo de turbulencia, los parámetros fundamentales para el caso estudiado en este trabajo son el tamaño del haz en comparación con las células turbulentas y el perfil atmosférico: horizontal o vertical.

Un enlace horizontal puede verse como un enlace vertical con un elevado ángulo cenital. Así puede comprenderse que generalmente supone un caso extremo del efecto de la turbulencia en la propagación de la luz, si bien tiene sus propias características. La principal es que en un enlace horizontal, el parámetro de estructura del índice de refracción $C_n^2$ se suele considerar un parámetro constante [253, p. 756], a diferencia de lo que ocurre en los enlaces verticales, donde varía con la altura (como se vio en el apartado 2.10.3). Esta variación provoca que según sea el sentido de la propagación, el efecto de la turbulencia sea distinto: en el enlace descendente la turbulencia solo afecta en el último tramo del camino de propagación, a diferencia del enlace ascendente, donde el efecto se produce al inicio y se ve amplificado a lo largo del resto del camino (Figura 120). Un enlace horizontal supone un caso distinto, ya que el efecto es distribuido y acumulativo, y al no disminuir con distancia, esta acumulación puede dar lugar a un efecto más intenso que un enlace vertical. Por ello, enlaces horizontales como el de 150 km entre La Palma y Tenerife se han utilizado repetidamente para simular posibles escenarios de caso peor en comunicaciones por satélite [254, p. 79].

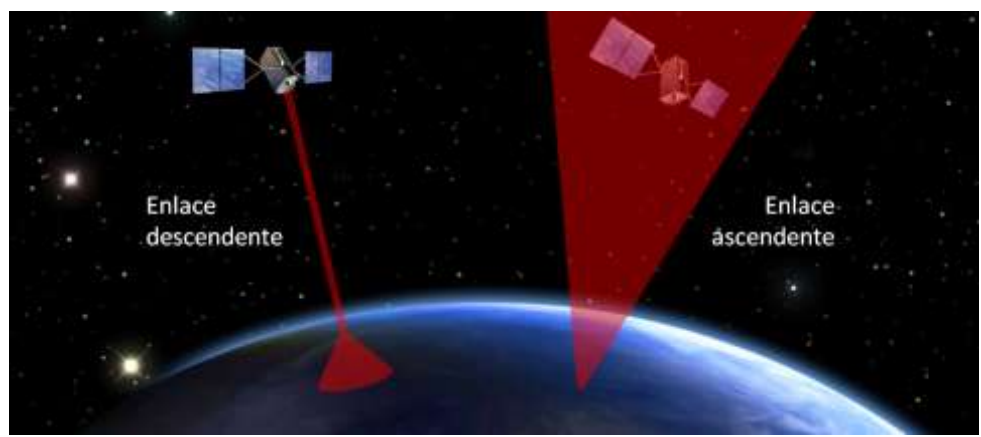

**Figura 120. Refracción del haz en enlaces verticales (ascendente y descendente) debida a la turbulencia atmosférica.**

Otro parámetro fundamental en el efecto de la turbulencia sobre un láser es la relación entre el tamaño del haz y el de las células turbulentas o remolinos. Cuando el tamaño promedio de los remolinos es menor al tamaño del haz láser, el efecto, de alta frecuencia espacial, tiende hacia un ensanchamiento del haz, más allá del debido exclusivamente a la difracción y proporcional a $\lambda/D$, y a una redistribución aleatoria de intensidades en el frente de onda percibida en el receptor como un efecto de centelleo. Cuando el tamaño de los remolinos es mayor que el haz [144, p. 29], el efecto, de baja frecuencia espacial, se traduce en un movimiento aleatorio del haz al final del camino de propagación, conocido como *beam wander* (explicado en el siguiente apartado) y debido a las inhomogeneidades atmosféricas de mayor escala espacial. Este tamaño de los remolinos puede caracterizarse mediante la longitud de coherencia o parámetro de Fried $r_0$, explicado en el apartado 2.10.3. La ecuación (2-25) del parámetro de Fried se puede simplificar en la ecuación (3-16) para el caso de enlaces horizontales [255, p. 4554], siendo k el número de onda (k = $2\pi/\lambda$), L la longitud del enlace y $C_n^2$ el parámetro de estructura del índice de refracción, según se explicó en el apartado 2.10.3.

$$r_0 = 3,02(C_n^2 L k^2)^{-3/5} \qquad (3\text{-}16)$$

En la Figura 121 se muestra la cuantificación del parámetro de Fried para enlaces horizontales de hasta 5 km y distintos tipos de regímenes turbulentos para $\lambda = 1$ μm (como longitud de onda representativa de la zona espectral donde se practica la criptografía cuántica). Se comprueba que la longitud de coherencia $r_0$, en el peor de los casos, será comparable al tamaño de un haz láser en el rango de varios cm, como es habitual en enlaces horizontales interurbanos, y en casi todos los casos mucho mayor, por lo que el *beam wander* será el efecto predominante. Las perturbaciones atmosféricas en este tipo de enlaces se deben principalmente a las turbulencias de gran escala (en comparación con el tamaño del haz), y por ello se suele obviar el efecto de la difracción, pudiéndose utilizar la óptica geométrica en su estudio, lo que permite modelar el *beam wander* como si surgiera de pequeñas variaciones aleatorias en la orientación del transmisor respecto al receptor.

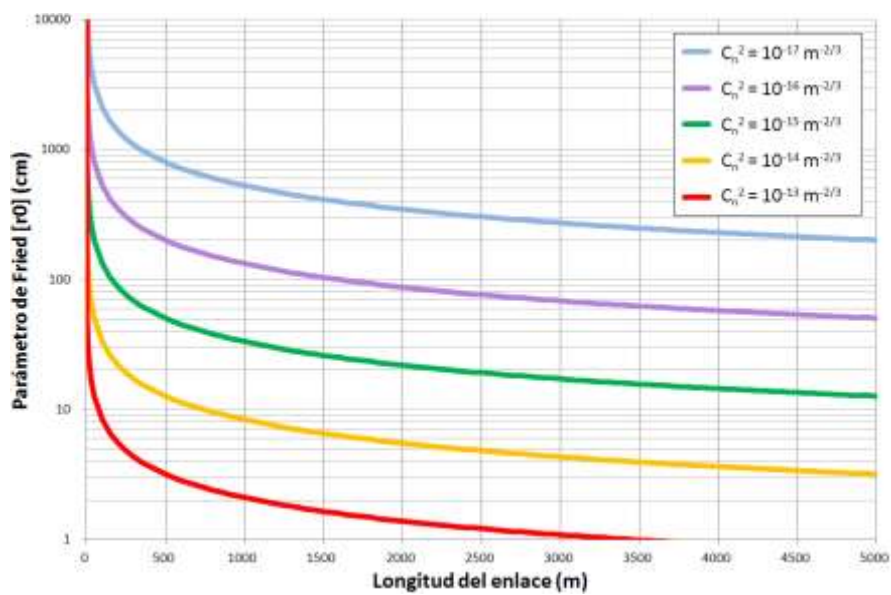

**Figura 121. Parámetro de Fried en enlaces horizontales para λ = 1 μm.**

### 3.6.2. *Beam wander* y spot de corto y largo plazo

Como se introdujo en el apartado anterior, el efecto del *beam wander* consiste en que el haz láser al propagarse por la atmósfera es desviado a causa de las variaciones aleatorias en el índice de refracción del medio. A consecuencia de esto, al final del camino de propagación el ángulo de llegada del haz se ve modificado aleatoriamente y su alineación queda fuera de eje, traduciéndose en un desplazamiento del spot final alrededor del foco del sistema óptico (Figura 122). El centro instantáneo del haz, o punto de máxima intensidad, sufre así desplazamientos en el plano focal en escalas temporales que van desde algunos milisegundos hasta varios cientos de milisegundos, en lo que se conoce como *beam wander*.

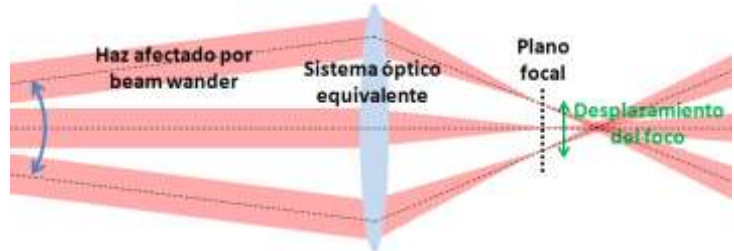

**Figura 122. Efecto de desplazamiento del foco causado por el beam wander.**

El origen de este efecto tiende a localizarse cerca del transmisor más que del receptor, ya que aunque la turbulencia atmosférica fuera exactamente la misma a lo largo de todo el camino (como suele asumirse por simplicidad en enlaces horizontales), cualquier perturbación en los tramos iniciales se ve amplificada durante el resto del trayecto. Esta amplificación será menor cuanto más lejos del transmisor se produzca el efecto por la menor distancia de amplificación y dado que el *beam wander* es un efecto acumulativo, su impacto muy cerca del receptor es mínimo.

Al estudiar la turbulencia de forma estadística, como es la manera habitual, es común el uso del concepto de tamaño de spot de largo plazo $W_{LT}$ como una medida de todos los efectos de la turbulencia sobre un haz determinado en el final del camino de propagación. El spot de largo plazo contempla todos los desplazamientos del spot durante un periodo suficientemente largo como para representar una muestra estadísticamente significativa, lo que se asocia a un tiempo mucho mayor al tiempo de variación medio de la turbulencia. Este tiempo se ha comprobado experimentalmente que es aproximadamente $\Delta T = 2\omega_0 / v$, siendo $\omega_0$ el radio del haz y $v$ la componente transversal de la velocidad de la turbulencia [256, p. 6].

El *beam wander* se puede caracterizar estadísticamente utilizando la varianza $\langle r_c^2 \rangle$ del desplazamiento del centro instantáneo del spot de corto plazo $W_{ST}$ sobre el eje óptico, que acabaría trazando el spot de largo plazo de radio $W_{LT}$ (Figura 123).

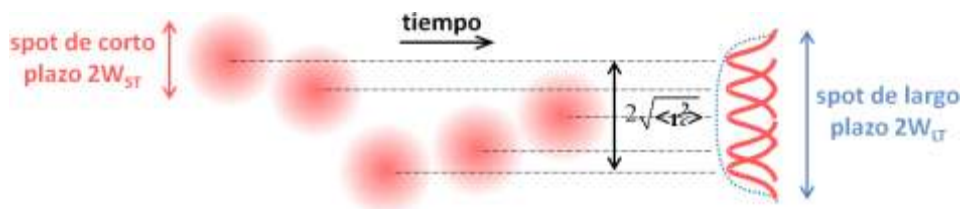

**Figura 123. Visualización del beam wander como la varianza $2\sqrt{\langle r_c^2 \rangle}$ del desplazamiento del centro del spot de corto plazo $2W_{ST}$ inscrito en el spot de largo plazo $2W_{LT}$.**

El spot de largo plazo surge de los efectos de células turbulentas o remolinos de todos los tamaños. Como se ha dicho anteriormente, los remolinos mayores que el tamaño del haz generan los efectos refractivos relacionados con el *beam wander*, por lo que si se eliminan estos efectos del spot de largo plazo, se obtiene una estimación del spot de corto plazo, que se debe a la contribución de los remolinos pequeños y resulta en una influencia puramente difractiva, provocando un ensanchamiento del haz. El spot de corto plazo siempre será mayor al spot original a la salida del transmisor y, si bien esto se debe exclusivamente a la divergencia con la distancia, conviene distinguir dos contribuciones diferentes: por una parte, la relativa a la difracción en la apertura del transmisor, proporcional a $\lambda/D$ (estudiada en el apartado 2.10.1), y por otra, la relativa a las múltiples refracciones en la atmósfera debidas a las células turbulentas más pequeñas, que acumuladas en la distancia, tienen un efecto neto de ensanchamiento del haz.

En la práctica, al haz le afectará la turbulencia en una variedad de formas, ya que los remolinos no tendrán un único tamaño y el efecto será la acumulación de la influencia que tiene cada tamaño de célula turbulenta sobre el haz. Sin embargo, para el tamaño de haz con el que se trabaja en esta tesis, de pocos centímetros de diámetro, se puede suponer razonablemente que el efecto dominante será el del *beam wander* porque normalmente los remolinos tendrán un tamaño similar o superior al del haz. Adicionalmente, cabe señalar que recientemente se ha demostrado [257] que los fenómenos no clásicos involucrados en las comunicaciones cuánticas, como el entrelazamiento y otros, se preservan mucho mejor en canales afectados por *beam wander* que en canales equivalentes con pérdidas constantes.

### 3.6.3. Cálculo de la varianza del beam wander

La relación entre el tamaño de spot de largo plazo $W_{LT}$, el tamaño de spot de corto plazo $W_{ST}$ y la varianza de beam wander $\langle r_c^2 \rangle$ para un haz gaussiano, según lo descrito en el apartado anterior, viene dada por la ecuación (3-17) [258, p. 205] y se puede visualizar en la Figura 124.

$$W_{LT}^2 = W_{ST}^2 + \langle r_c^2 \rangle \qquad (3\text{-}17)$$

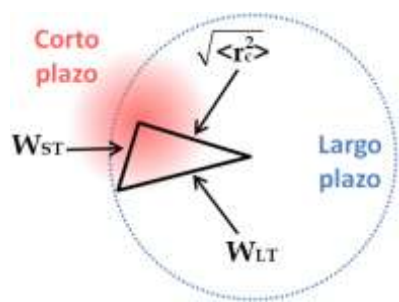

**Figura 124. Relación cuadrática entre spot de largo plazo $W_{LT}$, de corto plazo $W_{ST}$ y su desplazamiento $r_c$.**

La varianza $\langle r_c^2 \rangle$ de un haz colimado depende de la distancia $L$ del enlace, la longitud de onda $\lambda$, el radio del haz en el transmisor $\omega_0$ y el parámetro de Fried $r_0$ (que para un enlace horizontal viene definido por la ecuación (3-16)), según la ecuación (3-18) [258, p. 272]. En la literatura científica esta relación se ha encontrado derivada según diferentes modelos y con diferentes suposiciones, aunque en general son funcionalmente parecidas pero con distintas constantes de escala.

$$\sqrt{\langle r_c^2 \rangle} = 0,69L\left(\frac{\lambda}{2\omega_0}\right)\left(\frac{2\omega_0}{r_0}\right)^{5/6} \qquad (3\text{-}18)$$

Si se sustituye el parámetro de Fried de (3-16) en la ecuación (3-18), se llega a la ecuación (3-19).

$$\langle r_c^2 \rangle = 2,42C_n^2 L^3 \omega_0^{-1/3} \qquad (3\text{-}19)$$

En la Figura 125 se muestra la dependencia de la raíz de la varianza del *beam wander* con la distancia de propagación para distintos regímenes de turbulencia considerando un haz como el del sistema de QKD, igual a 4 cm de diámetro.

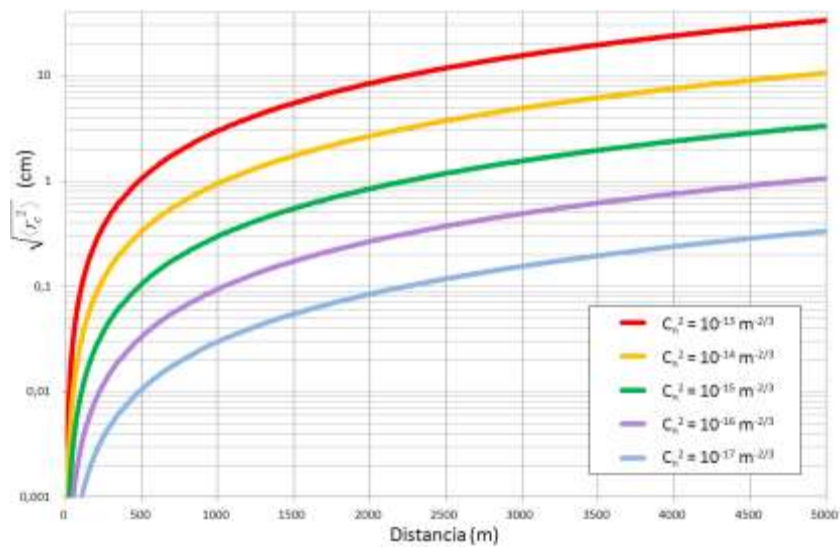

**Figura 125. Raíz de la varianza del *beam wander* con la distancia para distintos regímenes de turbulencia y un haz de 4 cm de diámetro.**

### 3.6.4. Cálculo de spot de largo plazo de un haz gaussiano

El radio del spot de largo plazo $W_{LT}$ se puede aproximar por medio de la ecuación (3-20), que es válida para todo tipo de regímenes turbulentos y para todo tipo de frentes de onda [258, p. 238].

$$W_{LT} \approx \omega\sqrt{1 + 1,63\sigma_R^{12/5}\Lambda} \qquad (3\text{-}20)$$

La ecuación (3-20) tiene su dependencia con el régimen turbulento mediante la varianza de Rytov, dada por la ecuación (3-21).

$$\sigma_R^2 = 1,23C_n^2 k^{7/6} L^{11/6} \qquad (3\text{-}21)$$

La aproximación de Rytov es un método ampliamente utilizado en la descripción del comportamiento de un frente de ondas en un medio turbulento. Fue la primera teoría que, a finales de la década de 1950, e inicialmente orientada a acústica, logró realizar predicciones acordes con los datos experimentales de caminos de propagación de hasta algunos cientos de metros. Previamente los estudios de la propagación de señales electromagnéticas en medios aleatorios como la atmósfera se basaban en la óptica geométrica y solo tenían validez para longitudes muy cortas, en el orden de $kl_0^2$ [259, p. 1670], típicamente de varios metros. Posteriormente se comprobó que la aproximación de Rytov era válida para distancias o turbulencias tales que la varianza de Rytov $\sigma_R^2$ fuese menor a 1 [258, p. 140], sin embargo el modelo se corrigió con otras aportaciones y actualmente sigue siendo empleada para el estudio de la turbulencia.

Para particularizar el spot de largo plazo en un haz gaussiano, es necesario definir los parámetros del haz $\omega$ y $\Lambda$ de la ecuación (3-20). Para este estudio es necesaria la consideración del comportamiento gaussiano del haz más allá del comportamiento de onda plana ilimitada, ya que esta aproximación no considera determinados efectos reales en haces colimados divergentes, como los empleados en este trabajo. Por ejemplo, en la realidad el tamaño del spot en el foco de un sistema nunca será igual a cero (como predice la aproximación de onda plana), incluso en ausencia de aberraciones.

Una onda óptica genérica se puede definir según el campo electromagnético $U(r, L)$, donde $L$ es la distancia de propagación a lo largo del eje longitudinal y $r$ es el vector del plano receptor transversal a la propagación. En el receptor, este campo tomará la forma de la ecuación (3-22) [258, p. 182], con el denominado parámetro de propagación $p(L)$ dado por la ecuación (3-23).

$$U(r, L) = \frac{1}{p(L)} \exp\left( ikL - \frac{r^2}{\omega_0^2} - i\frac{kr^2}{2F} \right) \qquad (3\text{-}22)$$

$$p(L) = 1 - \frac{L}{F_0} + i\frac{2L}{k\omega_0^2} \qquad (3\text{-}23)$$

En estas expresiones, $\omega$ y $F$ representan, respectivamente, el radio del haz en ausencia de turbulencias (donde la amplitud cae $1/e$ respecto a la del eje óptico, (Figura 126, izquierda)) y el radio de curvatura del frente de onda (donde $F = \infty$, $F > 0$ y $F < 0$ corresponden con un haz colimado, convergente y divergente respectivamente (Figura 126, derecha)), siendo $\omega_0$ y $F_0$ los relativos al inicio del camino de propagación, en el transmisor, y $k$ representa al número de onda ($k = 2\pi/\lambda$). La relación entre $\omega$ y $\omega_0$ viene dada por la ecuación (3-24).

$$\omega = \omega_0 \sqrt{\Theta_0^2 + \Lambda_0^2} \qquad (3\text{-}24)$$

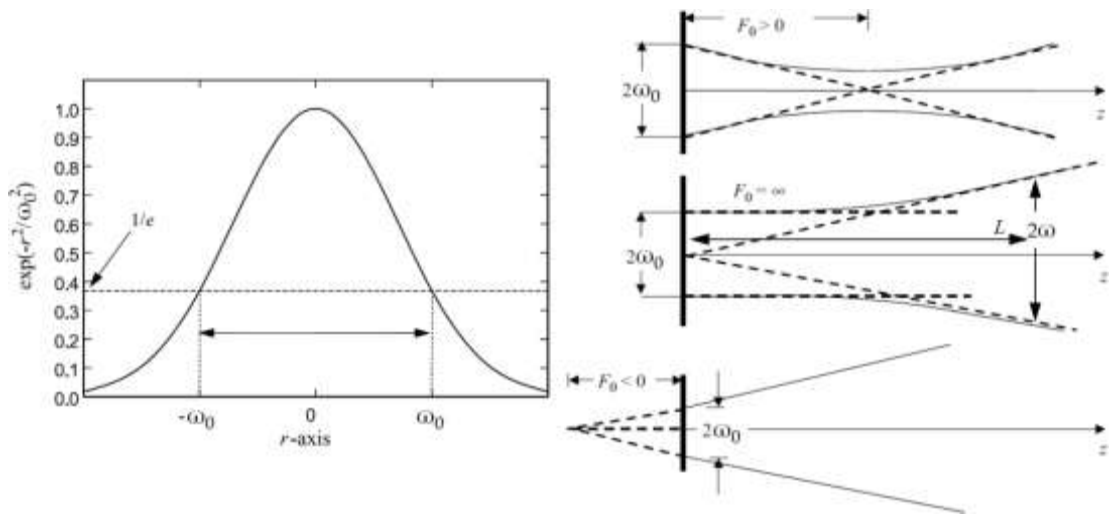

**Figura 126. Parámetros de un haz gaussiano: radio del haz
$\omega_0$ (izquierda) y radio de curvatura $F_0$ (derecha).**

El parámetro de propagación p (L) es un número complejo, y por tanto se puede descomponer en su parte real e imaginaria según la ecuación (3-25), por lo que de la ecuación (3-23) se deduce el valor de los parámetros $\Theta_0$ y $\Lambda_0$ (ecuación (3-26)).

$$p(L) = \Theta_0 + i\Lambda_0 \qquad (3\text{-}25)$$

$$\Theta_0 = 1 - \frac{L}{F_0} \qquad \Lambda_0 = \frac{2L}{k\omega_0^2} \qquad (3\text{-}26)$$

Por conveniencia matemática, se utiliza preferiblemente la parte real e imaginaria de la inversa del parámetro de propagación p (L) según la ecuación (3-27), dando lugar a los parámetros $\Theta$ y $\Lambda$ deducidos en la ecuación (3-28), con lo que ya se dispone de todos los parámetros necesarios para calcular el spot de largo plazo de un haz gaussiano dado en la ecuación (3-20).

$$\frac{1}{p(L)} = \frac{1}{\Theta_0 + i\Lambda_0} = \Theta - i\Lambda \qquad (3\text{-}27)$$

$$\Theta = \frac{\Theta_0}{\Theta_0^2 + \Lambda_0^2} \qquad \Lambda = \frac{\Lambda_0}{\Theta_0^2 + \Lambda_0^2} \qquad (3\text{-}28)$$

Dado que en este trabajo se utilizan siempre haces colimados, para obtener la mínima divergencia con la distancia, el radio de curvatura será $F_0 = \infty$, por lo que $\Theta_0 = 1$. Por lo tanto, las ecuaciones de (3-28) y (3-24) quedan como (3-29).

$$\Theta = \frac{1}{1 + \Lambda_0^2} \qquad \Lambda = \frac{\Lambda_0}{1 + \Lambda_0^2} \qquad \omega = \omega_0\sqrt{1 + \Lambda_0^2} \qquad (3\text{-}29)$$

Con los anteriores desarrollos, ya se dispone de todas las expresiones para calcular el radio del spot de largo plazo $W_{LT}$ según la ecuación (3-30), en función de los parámetros conocidos $\omega_0$, $\sigma_R^2$ y $\Lambda_0$.

$$W_{LT} \approx \omega_0\sqrt{\left[1 + \Lambda_0^2\right]\left[1 + 1{,}63\sigma_R^{12/5}\frac{\Lambda_0}{1 + \Lambda_0^2}\right]} \qquad (3\text{-}30)$$

# 3.7. ESTRATEGIAS PARA CORRECCIÓN DE TURBULENCIA

## 3.7.1. Influencia del ruido de fondo

El sistema descrito en el apartado 3.4 [260] fue probado en las instalaciones del CSIC en dos enlaces distintos: uno de 30 metros de distancia y otro de 300 metros (Figura 127). El principal objetivo de dichos experimentos fue evaluar la influencia de la radiación solar sobre el protocolo cuántico en un ambiente realista de enlace interurbano. Esta influencia fue caracterizada mediante la medida de la tasa de error cuántico, la tasa de clave secreta y la tasa de fotones de ruido de fondo. Las medidas se realizaron durante el día, con condiciones de alta luminosidad, entre primavera y verano.

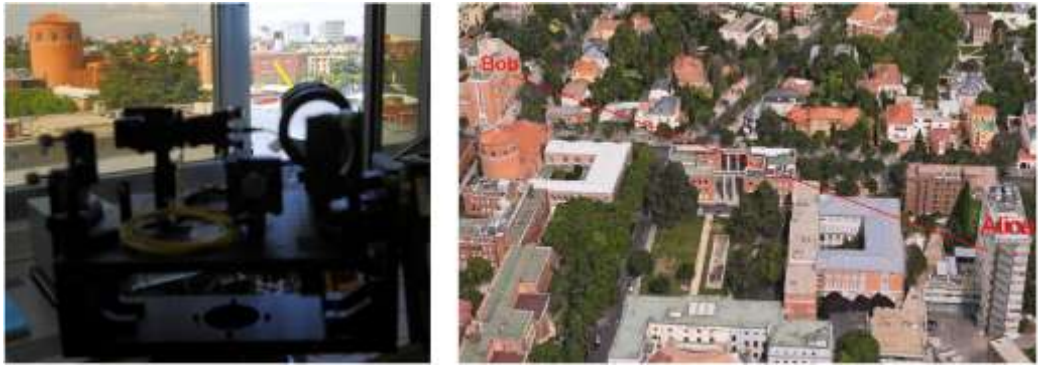

**Figura 127. Enlace de 300 metros ente Alice y Bob en el campus del CSIC en Madrid.**

En la Figura 128 y en la Figura 129 se muestran los resultados obtenidos durante 5 horas para el enlace de 30 metros y para el de 300 metros respectivamente. Se puede observar cómo existe una relación opuesta entre la tasa de error y la tasa de clave secreta, debido a que esta última se calcula a partir de la primera. La tasa de clave depurada se calcula contabilizando temporalmente los eventos recibidos en cada detector como un 1 y como un 0. Estos eventos se registran de forma separada, y mediante la etiqueta temporal que identifica a cada uno se conforma la clave depurada (dado que en B92 no existe proceso para la reconciliación de bases y no existe clave en crudo). Al inicio de la transmisión de esta clave se transmite una secuencia común conocida por Alice y Bob para calcular la tasa de error. Contabilizando en esta muestra de bits el número de errores (en los que Alice y Bob no coinciden) en relación al total de bits recibidos se calcula la tasa de error. Si la tasa de error es mayor que un cierto umbral, que depende del protocolo y la implementación (8 % en el sistema desarrollado), la transmisión no es considerada segura. Por último, la tasa de transmisión de clave secreta se calcula a partir de la tasa de clave depurada y de la tasa de error. En un enlace real, este paso implicaría un proceso de corrección de errores y amplificación de privacidad. Para considerar este efecto en estos experimentos se estimó la pérdida de velocidad considerando un escenario de caso peor en el que se realizan los dos ataques conocidos a los que es susceptible el protocolo B92.

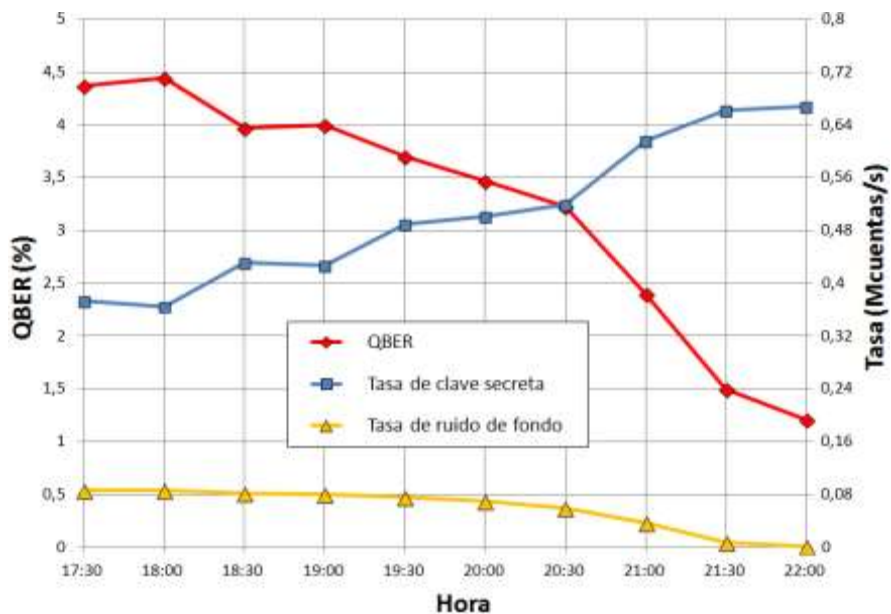

**Figura 128. Tasa de error cuántico, tasa de clave secreta y tasa de ruido de fondo en el enlace de 30 metros con una frecuencia de reloj de 1 GHz.**

En ambas gráficas se puede comprobar la influencia del ruido de fondo sobre la tasa de clave secreta y la tasa de error, observándose un aumento de la primera y disminución de la segunda al reducirse el ruido con la puesta del Sol. Cabe señalar que la diferencia entre las tasas de transmisión de la Figura 128 y la Figura 129 se deben a que en el enlace de 30 metros se utilizó una frecuencia de reloj de 1 GHz y en el de 300 metros una frecuencia de 1,5 GHz. Esto es debido a que la frecuencia óptima de 1,5 GHz se obtuvo posteriormente al experimento del enlace de 30 metros, lo que explica la mayor tasa de transmisión del segundo experimento.

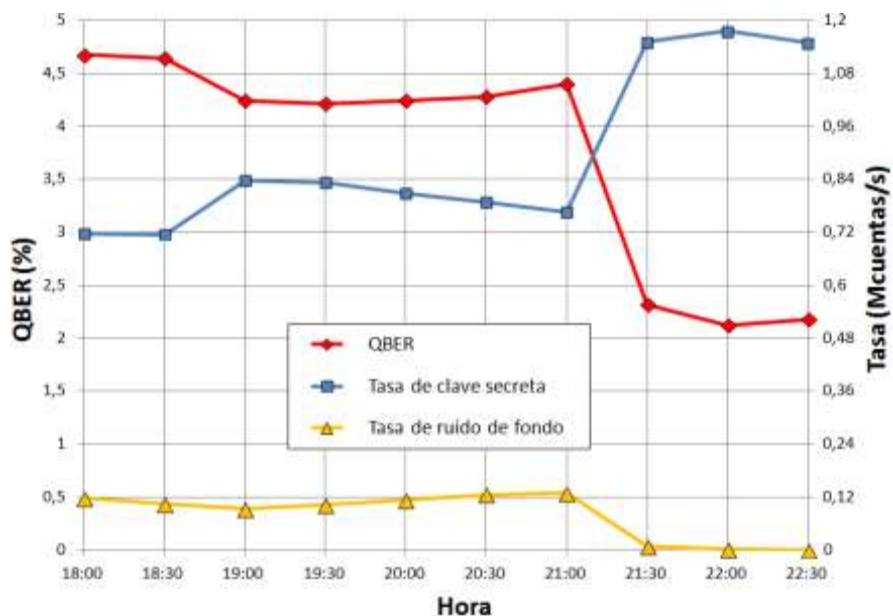

**Figura 129. Tasa de error cuántico, tasa de clave secreta y tasa de ruido de fondo en el enlace de 300 metros con una frecuencia de reloj de 1,5 GHz.**

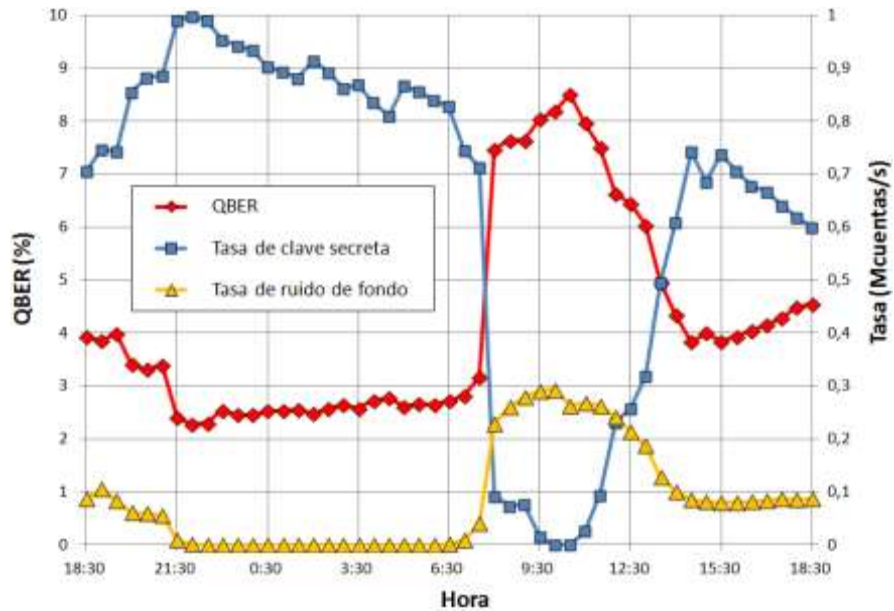

**Figura 130. Tasa de error cuántico, tasa de clave secreta y tasa de ruido de fondo en el enlace de 30 metros durante un día completo.**

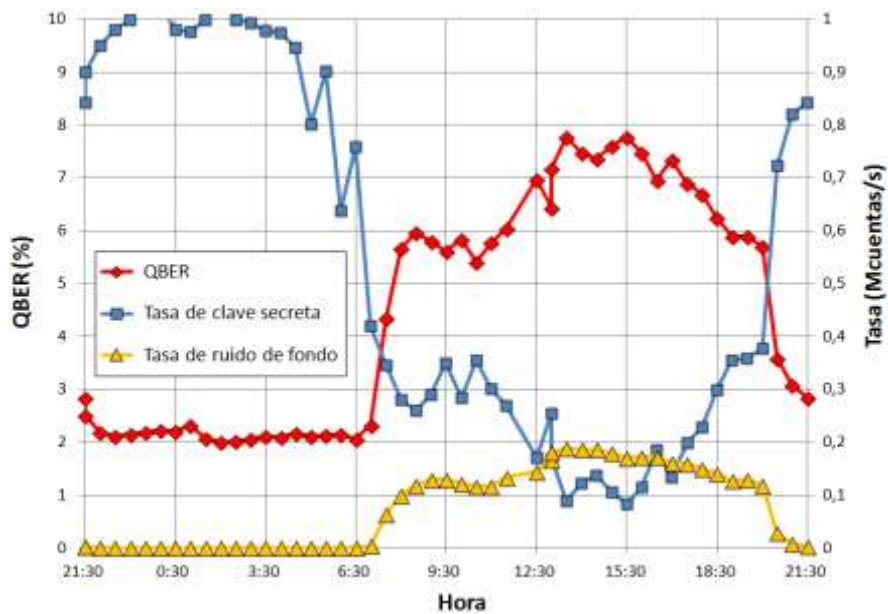

**Figura 131. Tasa de error cuántico, tasa de clave secreta y tasa de ruido de fondo en el enlace de 300 metros durante un día completo.**

En la Figura 130 y en la Figura 131 puede observarse el resultado de los experimentos de 30 y 300 metros durante un día completo. Se puede apreciar no solo la disminución del ruido de fondo con la puesta de Sol sino también una gran subida del mismo tras el amanecer. Este aumento se corresponde con la cercanía del Sol al campo de visión de Bob ocasionando una incidencia más directa de la radiación solar sobre el receptor. Al igual que en las anteriores gráficas, el aumento del ruido de fondo se corresponde directamente con una disminución de la tasa de clave secreta al aumentar la tasa de error, que en determinados momentos llega a superar el umbral de transmisión segura del 8 %. La relación entre el ruido de fondo acoplado al sistema y la tasa de clave secreta se debe a que la extrema sensibilidad de los detectores contabiliza fotones de ruido como fotones válidos, haciendo aumentar la tasa de error.

### 3.7.2. Influencia del *beam wander*

Un haz de luz puede verse desde la perspectiva de un frente de ondas. Antes de entrar al sistema óptico, un frente de ondas proveniente de una fuente en el infinito es plano y perpendicular a la dirección de propagación. Al salir del sistema óptico, el frente de ondas puede verse como esférico de forma que el camino óptico de todos los rayos tiene la misma longitud. Toda desviación de este comportamiento conllevará un error en el frente de onda, que será la diferencia entre el frente de onda real y el de una onda plana o esférica (Figura 132), según qué criterio se utilice. Este error se cuantifica como la media cuadrática o RMS (del inglés *Root Mean Square*) de la desviación sobre toda la superficie del frente de onda y se expresa en nanómetros o en fracciones de la longitud de onda.

Los polinomios de Zernike son un conjunto de polinomios definidos en un círculo de radio unidad que sirven como una técnica analítica útil para trabajar matemáticamente con los errores en el frente de onda. Con un desarrollo de combinaciones lineales de polinomios de Zernike es posible describir cualquier frente de onda $W(r, \varphi)$ y se suelen usan para identificar con datos experimentales la contribución de cada aberración (Figura 133), así como otras fuentes de error como microrrugosidades de la óptica o el efecto de la turbulencia atmosférica, aunque no son prácticos cuando la frecuencia espacial es muy alta.

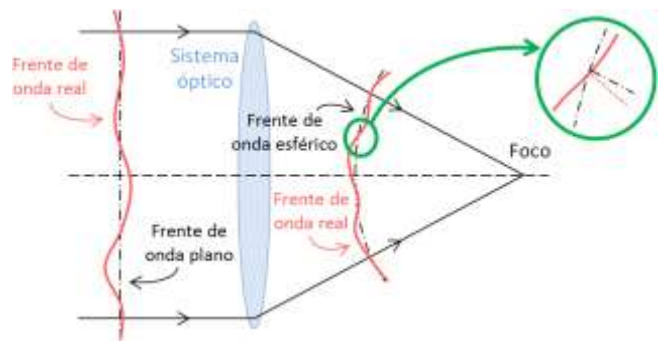

**Figura 132. Errores en el frente de onda.**

$$W(r,\varphi)=\sum_n a_n Z_n(r,\varphi) \qquad (3\text{-}31)$$

$Z_n(r,\varphi)$ es el polinomio n-ésimo de Zernike y $a_n$ es el coeficiente de ese término. Se expresan en coordenadas polares donde $r$ es el radio normalizado y $\varphi$ es el ángulo creciente en sentido anti horario desde la horizontal. Cuanto mayor es el orden del polinomio, mayor es la frecuencia espacial y normalmente menor la amplitud del error.

| Aberración | Polinomio |
|---|---|
| Pistón | $z_1 = 1$ |
| Tilt x | $z_2 = 2r \cdot \cos\varphi$ |
| Tilt y | $z_3 = 2r \cdot \mathrm{sen}\varphi$ |
| Desenfoque | $z_4 = \sqrt{3} \cdot \left(2r^2 - 1\right)$ |

| | | |
|---|---|---|
| Astigmatismo x | $z_5 = \sqrt{6} \cdot r^2 \cdot \cos 2\varphi$ | |
| Astigmatismo y | $z_6 = \sqrt{6} \cdot r^2 \cdot \text{sen} 2\varphi$ | |
| Coma x | $z_7 = \sqrt{8} \cdot \left(3r^2 - 2r\right) \cdot \cos \varphi$ | |
| Coma y | $z_8 = \sqrt{8} \cdot \left(3r^2 - 2r\right) \cdot \text{sen} \varphi$ | |
| Esférica de $3^{er}$ orden | $z_{11} = \sqrt{5} \cdot \left(6r^4 - 6r^2 + 1\right)$ |  |

**Figura 133. Primeros polinomios de Zernike y su identificación con las principales aberraciones.**

El error que induce la turbulencia atmosférica sobre el frente de ondas de un haz láser que se propaga en la atmósfera se puede describir en términos de una serie de polinomios de Zernike. Por otra parte, es posible evaluar el impacto que tendría sobre dicho error global la corrección de cada uno de los polinomios con el objetivo de optimizar un sistema de compensación de la turbulencia atmosférica. En [261] se realiza dicho análisis utilizando un espectro Kolmogorov y considerando los coeficientes $a_n$ como variables aleatorias con distribución gaussiana y media igual a cero. Si se corrigen los N primeros polinomios, el frente de ondas corregido $W_C$ sería el de la ecuación (3-32).

$$W_C = \sum_{n=1}^{N} a_n Z_n \tag{3-32}$$

El error cuadrático medio residual se definiría con la ecuación (3-33).

$$\Delta = \int \left\langle \left(W - W_C\right)^2 \right\rangle \tag{3-33}$$

Sustituyendo la ecuación (3-32) en la (3-33), y teniendo en cuenta que $\langle a_n \rangle = 0$, se puede obtener la ecuación (3-34), que expresa el error cuadrático medio residual del frente de onda cuando se han corregido los primeros N términos.

$$\Delta_N = \left\langle W^2 \right\rangle - \sum_{n=1}^{N} \left\langle \left| a_n \right|^2 \right\rangle \tag{3-34}$$

En la Figura 134 se han representado los primeros 20 términos $\Delta_N$ calculados para la ecuación (3-34) según [261]. Se observa que el impacto de los distintos polinomios es muy poco homogéneo. El mayor error se presenta en los primeros órdenes y correcciones en órdenes superiores no aportan un gran impacto relativo al error global del frente de onda. De hecho, las mayores disminuciones en el error del frente de onda se dan precisamente al corregir los coeficientes relativos al *tilt*, que es el efecto del *beam wander* [262, p. 1431]. En [253] se analiza el efecto de la turbulencia mediante óptica adaptativa evaluando la influencia de cada polinomio de Zernike corregido en términos de tasa de error de un enlace de comunicaciones ópticas en espacio libre. Se concluye que la eliminación del error de los primeros términos produce en todos los casos la mayor mejora, con disminuciones exponencialmente mayores en los menores órdenes. Además, tras estudiar la influencia del tamaño de la apertura receptora sobre la corrección del *beam wander*, se comprueba que la mejora aumenta con aperturas receptoras más pequeñas. La razón de esta dependencia se explicó en 3.6.1 y supone otro motivo para centrarse en la corrección del *beam wander* en telescopios relativamente pequeños como el usado en este trabajo (25,4 cm de diámetro). Por todo ello, se ha optado por limitar la corrección de la turbulencia al efecto producido por el *beam wander*. Los órdenes superiores proporcionarían una mejora menos relevante y precisarían otro tipo de corrección distinta y cualitativamente más compleja que la necesaria para el *tilt*.

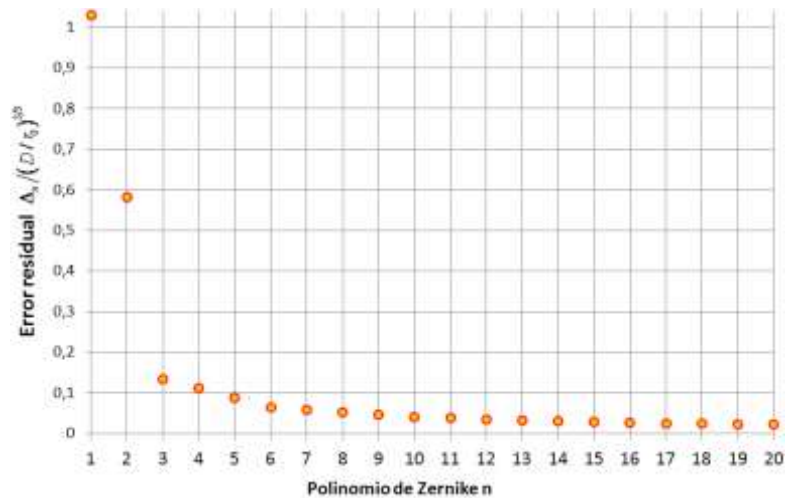

**Figura 134. Error cuadrático medio residual en el frente de onda al corregir los primeros N términos de Zernike.**

En 1985, H. T. Yura y M. T. Tavis [263] describieron una degradación, a la que llamaron anisoplanatismo del centroide, aplicable a los sistemas de corrección de turbulencia basados en medida del centroide de un haz. Esta degradación se debe a la diferencia entre el *tilt* medido por un sensor de centroide, llamado G-tilt (del inglés *Gradient-tilt*) y el *tilt* que introduce un espejo con el objetivo de corregir el G-tilt, llamado Z-tilt (del inglés *Zernike-tilt*). Sin embargo, este efecto solo es observable si otros efectos de la turbulencia se eliminaran, es decir, quedará enmascarado en un sistema que únicamente corrija el *beam wander*. Según [263, p. 765], el efecto es pequeño cuando $D/r_0 < 10$, que es exactamente el caso en este trabajo, si asumimos una apertura $D$ de varios cm y un $r_0$ típico nunca menor a varios cm.

### 3.7.3. Relación entre el ruido de fondo y el *beam wander*

En la tasa de error influyen una variedad de factores, como el ruido de los contadores de fotones y la polarización de cada estado cuántico, pero también la turbulencia atmosférica y el ruido solar de fondo cuando el enlace opera durante el día. Como se verá a continuación, estos dos últimos factores muestran efectos con estrategias de mitigación contrapuestas. Normalmente se utiliza una combinación de filtrado espacial y espectral para reducir el ruido de fondo de origen solar debido al *scattering* en la atmósfera. En el caso del sistema desarrollado, el filtro espectral de 1 nm de ancho, combinado con el filtrado espacial de la fibra óptica de 62,5 µm de diámetro, se consideró suficiente en el primer diseño para reducir el ruido de fondo a niveles que permitieran la recepción de la clave a alta velocidad cuando los niveles de radiación solar no eran demasiado elevados (por ejemplo, al amanecer o atardecer). Sin embargo, cuando la radiación solar alcanzaba intensidades altas, alrededor del mediodía o cuando el sol incidía directamente sobre el receptor, la tasa de error aumentaba considerablemente a niveles cercanos al umbral causando una reducción de la tasa de transmisión de clave secreta. Por ello, un filtrado espacial más exigente se identificó como la principal estrategia para disminuir el ruido de fondo acoplado al sistema durante el día.

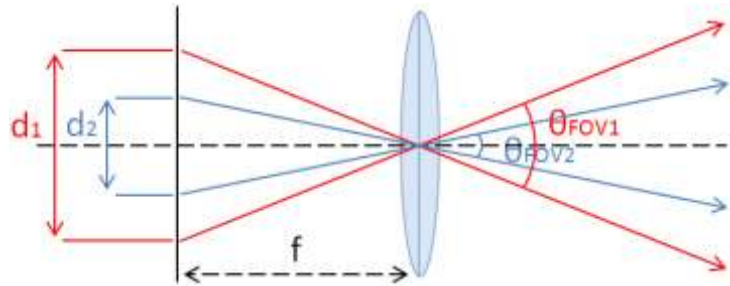

**Figura 135. Reducción del campo de visión con el plano focal.**

La optimización de la relación señal a ruido limitando el ruido de fondo acoplado al receptor es una estrategia clave en cualquier sistema de comunicaciones ópticas en espacio libre diseñado para operar durante el día. El filtrado espacial es una estrategia para reducir este ruido de fondo. Sin embargo, el filtrado espacial implica realizar un truncamiento como el descrito en el apartado 3.5.3, con la consecuente pérdida de potencia. En enlaces convencionales de comunicaciones ópticas es posible aprovechar el filtrado espacial para realizar una optimización de la relación señal a ruido incluso a costa de perder una fracción de la potencia recibida. Sin embargo, en un enlace de QKD es fundamental evitar toda pérdida de fotones porque se traduce directamente en una disminución de la tasa de bit. Por ello, debe evitarse cualquier truncamiento de la señal óptica portadora de los bits de clave. Una solución para realizar un filtrado espacial minimizando la pérdida de potencia es llevarlo a cabo directamente en la fibra óptica acoplada a los detectores de fotones. Esto es posible determinando su tamaño: a menor diámetro del núcleo de la fibra, mayor filtrado espacial. Este filtrado impone una disminución del campo de visión del receptor ya que este es proporcional al tamaño del plano focal (Figura 135), determinado por el tamaño de la fibra óptica, cuyo extremo debe situarse en el foco del sistema. Por ello, el mínimo campo de visión proporcionará el mínimo ruido de fondo acoplado al sistema.

Por otra parte, como se ha visto en el apartado 3.6, el *beam wander* hace que la señal óptica fluctúe en su posición en el plano focal con el tiempo. Si se observa en un plazo de tiempo continuado, el efecto se podría equiparar al clásico círculo borroso debido a las aberraciones o a la difracción. En la práctica, esto obliga a abrir el campo de visión para no perder fotones (Figura 136, izquierda). En consecuencia, el efecto de combatir la pérdida de fotones debida a la turbulencia es opuesto al efecto provocado por limitar el ruido de fondo acoplado al sistema. Por una parte, un campo de visión muy estrecho es muy sensible a las fluctuaciones de posición de los fotones recibidos originadas por el *beam wander*, con lo que gran parte de ellos no se acoplarían a la fibra óptica. Y por otra parte, un campo de visión muy ancho provocaría que un gran número de fotones de ruido se acoplaran a los detectores, introduciendo errores en la transmisión y reduciendo la tasa de clave.

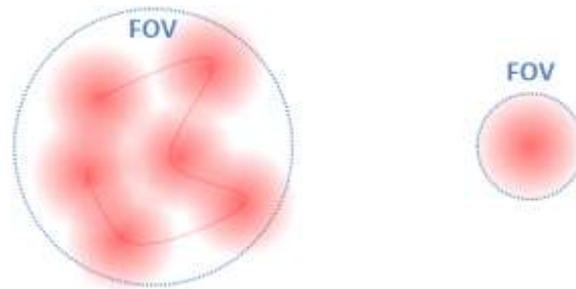

**Figura 136. Campo de visión afectado por *beam wander* (izquierda) y campo de visión con corrección (derecha).**

Por todo lo anterior, si se requiere maximizar la relación señal a ruido, la solución óptima para reducir el ruido de fondo es eliminar el desplazamiento debido al *beam wander* con un sistema de corrección. El objetivo de este sistema sería compensar el movimiento del haz mediante desplazamientos del mismo correspondientes a la medida que se haría en paralelo de su posición instantánea. Un sistema así proporcionaría un spot más estático en el foco, lo que permitiría reducir el campo de visión, con la consiguiente disminución del ruido de fondo acoplado (Figura 136, derecha).

El impacto del *beam wander* es tanto mayor cuanto más largo es el enlace ya que se trata de un efecto acumulativo. Por ello, diferentes técnicas de corrección mediante óptica activa se han venido implementando tradicionalmente en los enlaces convencionales de comunicaciones ópticas en espacio libre [264] [265] [266] [267] [268]. Sin embargo, sigue sin ser una técnica bien explorada en QKD debido a las particularidades que imponen este tipo de enlaces en el sistema de corrección. En estos sistemas, la óptica activa se suele limitar a tareas de alineamiento entre los terminales [269], introduciendo la turbulencia en los estudios teóricos como una fracción de fotones que no llega a acoplarse al receptor [270], o en la práctica, considerando el efecto de la turbulencia como una pérdida en el balance de enlace que se traduce en una disminución de la tasa de transmisión de clave [271], o bien sobredimensionando la apertura del receptor para acomodar las fluctuaciones de la turbulencia atmosférica [272]. También se han realizado varios estudios [273] [274] [275] preliminares de proyectos orientados a desarrollar compensación de turbulencia en QKD, cuyos resultados finales mostrando un sistema de corrección integrado no se han encontrado publicados.

### 3.7.4. Configuraciones para corrección de *beam wander*

Existen varias configuraciones posibles para llevar a cabo la corrección del *beam wander*. Todas ellas están basadas en sistemas clásicos de alineamiento de haces láser y en términos generales consisten en espejos modulables tipo FSM (del inglés *Fast Steering Mirror*) que se controlan con datos proporcionados por detectores sensibles a la posición PSD (del inglés *Position Sensitive Detector*), cerrando el bucle mediante controles de realimentación tipo PID (de Proporcional Integral Derivativo).

El *beam wander* se puede modelar como si fuera originado por una variación angular del haz laser en el transmisor o como una fluctuación en el ángulo de llegada al receptor. Atendiendo a estas dos aproximaciones y considerando el requisito impuesto por un enlace QKD de no perder fotones, se pueden diseñar dos tipos diferentes de técnicas de mitigación: si el diámetro de haz de largo plazo a la entrada del receptor es mayor que su apertura, la corrección debería llevarse a cabo en Alice, porque de otro modo el enlace sufriría de grandes pérdidas ya que el haz no estaría siempre acoplado al receptor; en el caso contrario, en el que incluyendo el efecto del *beam wander*, el haz siempre quedara dentro de la apertura receptora, la corrección podría realizarse en Bob. Atendiendo a este criterio, la turbulencia no implicaría una pérdida significativa de fotones en ningún caso, maximizando así la tasa de transmisión del protocolo cuántico.

Una posible implementación del primero de los esquemas propuestos, correspondiente a la precompensación de la turbulencia en Alice, se muestra en la Figura 137. El objetivo de esta configuración es precompensar en el transmisor el efecto del *beam wander* que afecta al canal de datos. Para ello, se utiliza un canal de seguimiento transmitido desde Bob hacia Alice, en sentido opuesto al de la comunicación cuántica. El esquema del canal de datos es el clásico del protocolo B92 formado en Alice por dos láser a 850 nm correspondientes a las dos polarizaciones lineales que se reciben en dos detectores de fotones (SPAD) en Bob. El canal de seguimiento y el de sincronismo comparten longitud de onda (1550 nm en el sistema desarrollado), distinta a la del canal de datos para evitar interferencias con este, y tienen sentidos de propagación opuestos. Las polarizaciones de los canales de sincronismo y seguimiento son lineales y ortogonales entre sí con el objetivo de aprovechar toda la potencia óptica, al combinarlos y separarlos por completo mediante sendos divisores de haz polarizados (DHP). Estos canales se combinan y separan con el de datos mediante sendos filtros dicroicos en transmisor y receptor, de nuevo optimizando el aprovechamiento de la potencia y especialmente asegurando que no se pierden fotones en el canal de datos. Para realizar la precompensación en Alice se realiza la medida de posición del canal de seguimiento transmitido desde Bob y usando estos datos se compensa el haz de datos transmitido desde Alice modulando para ello el ángulo de salida.

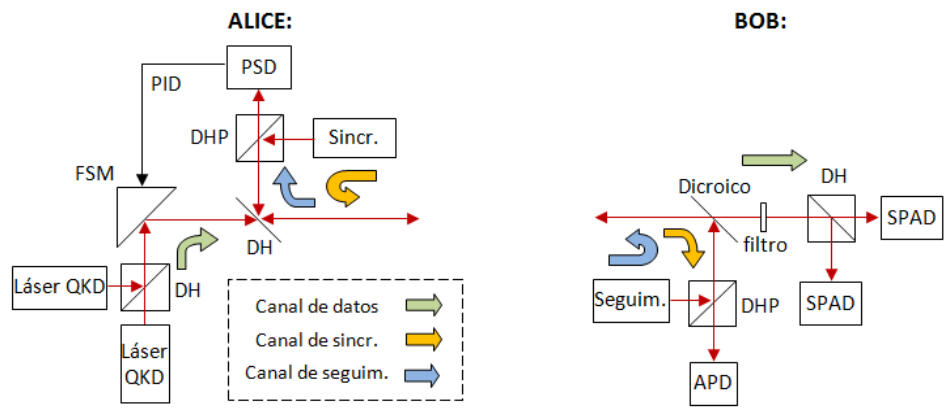

**Figura 137. Esquema de precompensación en Alice para el protocolo B92.**

Esta configuración puede dar lugar a un montaje más complejo por involucrar un láser adicional en sentido contrario a la comunicación y a una longitud de onda distinta al canal de datos para evitar posibles acoplamientos a este canal mediante reflexiones indeseadas en Bob. Además, incluso a una longitud de onda distinta, el empleo de un láser emitiendo una potencia continua podría ser una fuente de errores si sus fotones llegaran a acoplarse a los detectores de Bob. Otro inconveniente de esta configuración es que la precompensación solo se realiza sobre el canal de datos, y el canal de sincronismo sería vulnerable al *beam wander*. Esto normalmente no es un problema, ya que se puede aumentar arbitrariamente la potencia de este canal para recibir en Bob una señal apropiada incluso en presencia de turbulencia, como se pudo comprobar en el experimento del enlace a 300 metros. Sin embargo, este aumento de potencia en la dirección de Bob también podría dar lugar a un aumento de errores por acoplamiento de esta señal en los detectores de fotones individuales. Un inconveniente adicional de este principio lo supone el hecho de que para extraer información del estado instantáneo de la atmósfera es imprescindible que ambos haces sean idénticos y recorran exactamente el mismo camino (aunque en sentidos opuestos). Lograr esto con la precisión requerida podría ser complejo a la hora de alinear y calibrar el sistema. Por último, hay que señalar que esta estrategia es válida dentro de un rango de distancias de propagación. Cuando la distancia es muy larga, los cambios en la atmósfera se producirán más rápido que el tiempo involucrado en llevar a cabo la corrección, es decir, la correlación temporal entre canales en distintos sentidos de propagación se perderá. Este límite dependerá de los parámetros del sistema, pero dado que la escala temporal de la turbulencia atmosférica está en el orden de los milisegundos, la distancia máxima estaría en el orden de los cientos de kilómetros. Pese a los inconvenientes mencionados, esta configuración tiene la ventaja de ser válida para cualquier relación entre el diámetro de largo plazo y la apertura del receptor.

En escenarios donde el diámetro de largo plazo es menor a la apertura del receptor, es posible realizar una compensación más simple que la anterior. Una implementación del segundo de los esquemas propuestos, correspondiente a la compensación en Bob, se muestra en la Figura 138. En esta configuración, los canales de sincronización y de seguimiento se realizan con un solo láser que se transmite desde Alice combinado con el canal de datos mediante un espejo dicroico. En Bob son discriminados espectralmente y el canal de sincronización y seguimiento es dividido de nuevo mediante un divisor de haz hacia un detector para extraer el sincronismo y hacia un detector sensible a la posición. Los datos extraídos de este último detector cierran el bucle de control PID para modular los movimientos del espejo FSM, a la entrada de Bob.

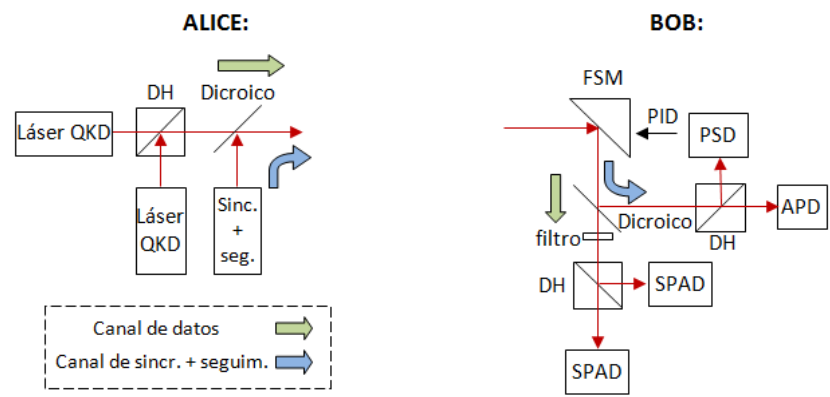

**Figura 138. Esquema de compensación en Bob para el protocolo B92.**

Este esquema, al no ser bidireccional, resulta en una implementación más simple que el primero, por integrar dos canales en un solo láser. Además, evita el riesgo de interferencia que supone el uso de un láser junto a los detectores de fotones. Por otra parte, la corrección del *beam wander* también se realiza sobre el canal de sincronismo, por lo que puede emplearse una potencia de transmisión menor a esta longitud de onda, con menor riesgo de interferencia con el canal de datos, ya que además ambas longitudes de onda están alejadas entre sí. Al igual que la primera configuración, esta también tiene un rango de distancia de propagación. En este caso la limitación proviene del hecho de que el efecto del *beam wander* va aumentando con la distancia y llegará un punto en que el diámetro de largo plazo sea mayor que la apertura receptora, momento a partir del cual se empezarán a perder fotones. Como se verá en el siguiente apartado, el límite de distancia de la configuración de corrección en Bob será considerablemente menor al límite de la configuración de precompensación en Alice, por lo que será necesario evaluar los requisitos de distancia de cada enlace para elegir una estrategia u otra.

### 3.7.5. Selección de la estrategia de corrección

Dependiendo de las características del enlace, la estrategia de compensación en el receptor podría ser suficiente para mitigar el *beam wander*. De lo contrario, debería usarse la estrategia de precompensación en el transmisor. El tipo de corrección a usar dependerá del diseño óptico de Alice y Bob (especialmente los tamaños del haz de salida de Alice y de la apertura colectora de Bob), del régimen de turbulencia (caracterizado por la constante de estructura $C_n^2$) y de la distancia de propagación. En general, el principio de precompensación será aplicable siempre, aunque si es posible debería elegirse la corrección en el receptor con el objetivo de simplificar el sistema, por las razones explicadas en el apartado anterior.

Para un régimen de turbulencia dado, la estrategia de corrección en el receptor se podría extender a mayores distancias si se usara una apertura mayor en el emisor (que permita transmitir un haz de mayor diámetro) y/o una apertura mayor en el receptor. Esta elección de aperturas debe realizarse cuidadosamente ya que existe un óptimo para ambos parámetros. El valor óptimo corresponde a la mínima divergencia del haz causada por difracción y por ensanchamiento del haz debido a la turbulencia y será diferente para cada distancia de propagación. Solo si se elige el par de valores óptimo, el rango de distancia de aplicación para la estrategia de corrección en el receptor será el máximo.

Como se introdujo en el apartado 3.6.4, mediante la ecuación (3-29) se puede calcular el diámetro $2\omega$ de un haz gaussiano de longitud de onda $\lambda$ a una cierta distancia L cuando se transmitió con un diámetro $2\omega_0$. En la Figura 139 se muestra cómo evoluciona esta dependencia con la distancia para $\lambda = 850$ nm, utilizando para ello la mencionada ecuación, así como la (3-26) para introducir la dependencia de $\omega$ con $\Lambda_0$, que a su vez depende también de L, $\lambda$ y $\omega_0$.

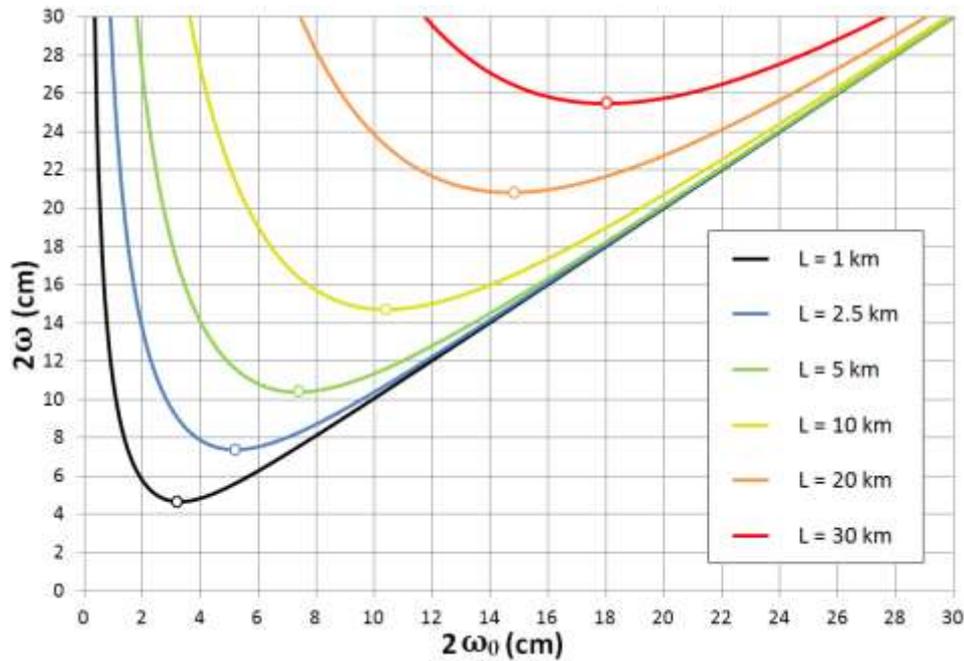

**Figura 139. Diámetro 2ω de un haz gaussiano de λ=850 nm tras propagarse una distancia L al transmitir con un diámetro 2ω₀.**

En la Figura 139 se puede observar que para cada distancia de propagación existe un par $\{2\omega, 2\omega_0\}$ óptimo que proporciona la mínima divergencia. En la Figura 140 se ha representado este par de valores en función de la distancia de propagación.

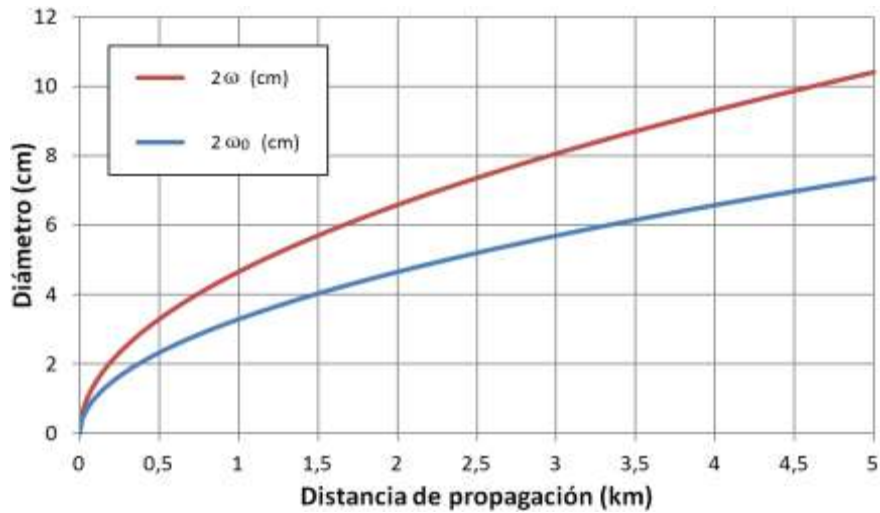

**Figura 140. Par $\{2\omega, 2\omega_0\}$ óptimo en función de la distancia para λ=850 nm.**

De lo anterior se puede concluir que una vez determinada la distancia de propagación, ya es posible determinar qué tamaño deberá tener el haz de salida para minimizar la divergencia. En función de esto debería diseñarse el tamaño de la apertura del transmisor atendiendo al criterio del truncamiento de un haz gaussiano presentado en el apartado 3.5.3. Por ejemplo, para conservar el perfil gaussiano, esta debería ser del doble del haz de salida, es decir, un diámetro de $4\omega_0$.

Para diseñar la apertura receptora óptima no es suficiente con tomar el valor de ω óptimo a la distancia de propagación deseada en la Figura 140, porque este ω no tiene en cuenta el efecto de la turbulencia atmosférica. Para introducir este efecto, se debe emplear el diámetro de largo plazo $W_{LT}$, que es posible calcularlo mediante la ecuación (3-30). Al sustituir $\omega$ por $W_{LT}$ en las dependencias anteriores, el efecto sobre las curvas de la Figura 139 será el de subirlas en el eje vertical permaneciendo el $\omega_0$ en el mismo lugar. Así, para cada distancia, ahora es posible calcular el $\omega$ correspondiente a $\omega_0$ óptimo. En la Figura 141 se muestran los diámetros $\omega_0$ y $\omega$ a $\lambda = 850$ nm en función de la distancia y para distintos regímenes de turbulencia. Usando estos cálculos, es posible diseñar las aperturas óptimas para aprovechar en el receptor la mayor parte de la potencia emitida por el transmisor. Además, sirve como frontera para la utilización de las técnicas de corrección de turbulencia presentadas anteriormente. Por ejemplo, la estrategia de corrección en el receptor podría usarse hasta distancias de 5 km, seleccionando las aperturas óptimas de transmisión y recepción $2\omega_0 = 7{,}4$ cm y $2\omega = 13{,}1$ cm respectivamente y asumiendo un régimen de turbulencia medio ($C_n^2 = 10^{-15}$ m$^{-2/3}$). Obsérvese que si solo se considera el haz transmitido y el recibido en un régimen de turbulencia débil ($C_n^2 = 10^{-16}$ m$^{-2/3}$ y $C_n^2 = 10^{-17}$ m$^{-2/3}$) la Figura 141 prácticamente coincide con la Figura 140.

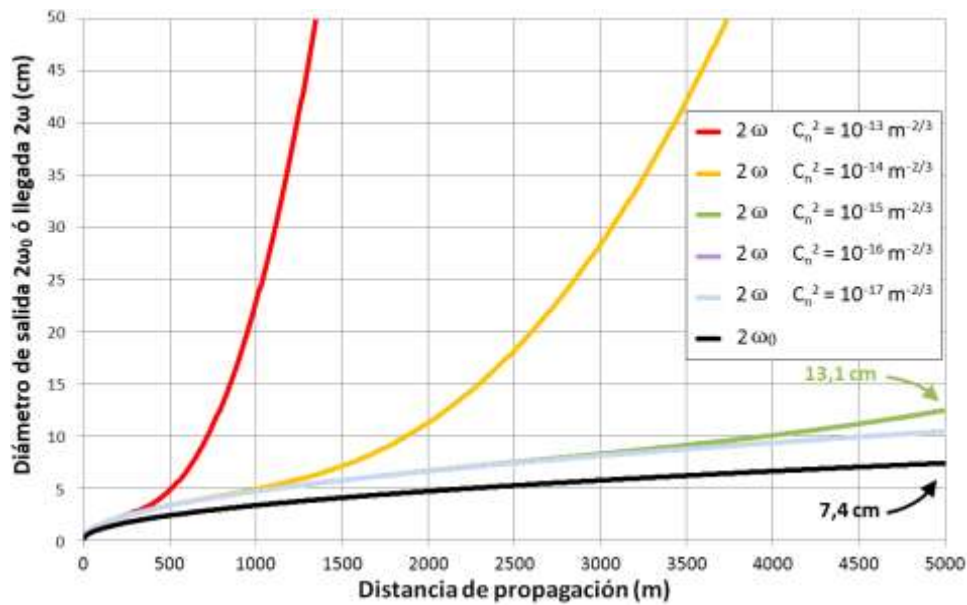

**Figura 141. Pares {2$\omega$, 2$\omega_0$} óptimos en función de la distancia
a $\lambda = 850$ nm y para distintos regímenes de turbulencia.**

Una forma visual para seleccionar la estrategia de corrección de *beam wander* es estudiar la relación entre la apertura receptora y el diámetro de largo plazo. Si esta relación es mayor de 1, entonces el haz siempre se acoplará íntegramente al receptor y por ello la corrección debería realizarse en él. En otro caso, la compensación debería llevarse a cabo en el transmisor para evitar importantes pérdidas. Para cada régimen de turbulencia, se puede calcular una distancia como frontera para la utilización de estas estrategias de corrección. En la Figura 142 se muestra la relación descrita, utilizando los parámetros del sistema de QKD desarrollado: un haz de salida de 4 cm de diámetro y una apertura receptora de 81,5 mm [25]. Se puede comprobar que en un régimen turbulento medio ($C_n^2 = 10^{-15}$ m$^{-2/3}$), es posible realizar la corrección en el receptor hasta distancias menores a 2,45 km (parte superior de la Figura 142). Si se desea operar por encima de esta distancia, se debería usar la precompensación en el emisor (parte inferior de la Figura 142). En el caso peor de turbulencia muy fuerte, la compensación en Bob tendría su límite de distancia a 800 metros, aunque dado que este régimen es poco frecuente, se podría extender la distancia asumiendo un factor de pérdida en los momentos en que tenga lugar semejante régimen de turbulencia.

---

[25] Nótese que es necesario particularizar este cálculo con los valores de estos parámetros, aunque intuitivamente pudiera parecer que no es necesario al estudiar la relación entre ambos, porque el impacto de la turbulencia depende del tamaño del haz.

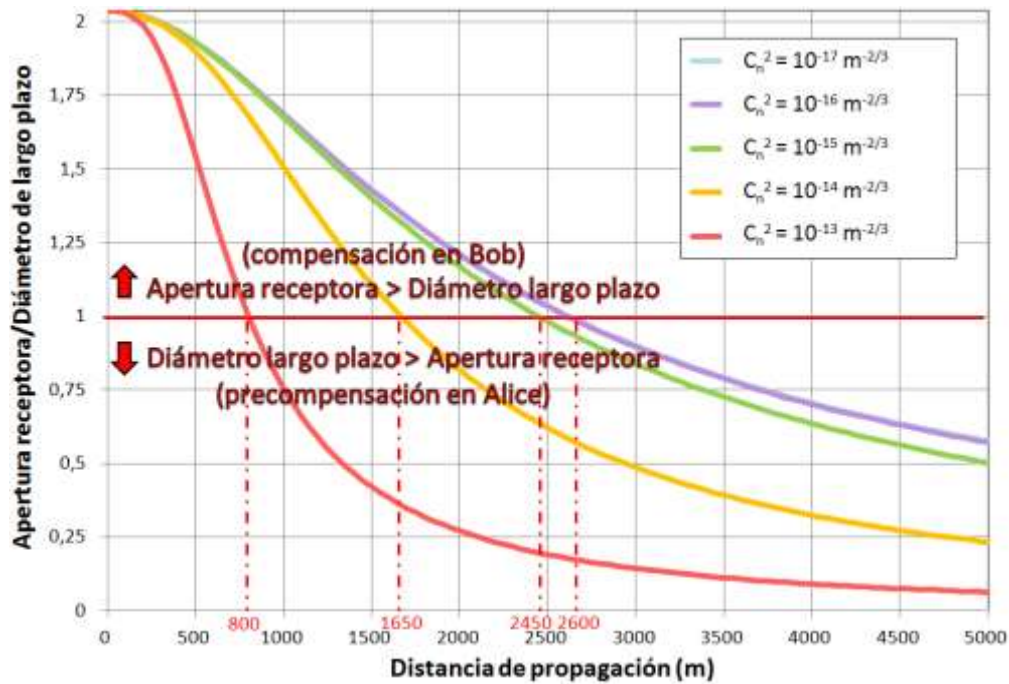

**Figura 142. Relación entre apertura receptora y diámetro de largo plazo en función de la distancia a λ=850 nm para distintos regímenes de turbulencia con un haz transmitido de 4 cm y una apertura receptora de 8 cm.**

# 3.8. COMPONENTES DEL SISTEMA DE CORRECCIÓN

### 3.8.1. Selección del espejo modulable

Como se ha explicado en el apartado anterior, el espejo modulable será un componente fundamental en el sistema de corrección. Se entiende por espejo modulable un espejo cuyos movimientos pueden ser controlados mediante una señal eléctrica. Existen varias tecnologías diferentes que implementan este tipo de espejos, las más importantes de las cuales se describen brevemente a continuación.

- Los espejos piezoeléctricos (Figura 143, a) quizás son los más versátiles, utilizando el efecto piezoeléctrico para conseguir el movimiento del espejo al aplicarle un voltaje de control. Son capaces de mover cargas relativamente pesadas en pequeños rangos de forma muy precisa y a una alta velocidad, limitada por su frecuencia de resonancia.

- Por su simplicidad y bajo coste, otra tecnología muy popular consiste en la utilización de motores (Figura 143, b) paso a paso o de continua para hacer girar los tornillos de precisión que controlan el movimiento de los ejes de una montura óptica de un espejo.

- Una combinación (Figura 143, c) de las dos técnicas anteriores consiste en utilizar el efecto piezoeléctrico no para mover directamente el espejo sino los tornillos que controlan la montura óptica sobre la que va dispuesto el espejo. Las dos últimas técnicas ofrecen una velocidad muy limitada en el movimiento del espejo, aunque más en el caso de los motores que en el de los piezos, ofreciendo los primeros una mayor repetibilidad en el caso de los motores paso a paso. Además, por el sistema de traslación del movimiento, ambas tecnologías introducen una cantidad de ruido elevada para un sistema de control activo, por lo que se utilizan especialmente en montajes que solo requieren de ajustes ocasionales.

- Otra técnica (Figura 143, d) muy empleada utiliza un principio de funcionamiento parecido al de un altavoz de audio. Consiste en bobinas fijadas a una montura que al hacer circular una corriente eléctrica, controlan el movimiento de los imanes que rodean y que van unidos al espejo. Esta configuración asegura un tiempo de vida prácticamente ilimitado y proporciona una alta precisión y velocidad de respuesta.

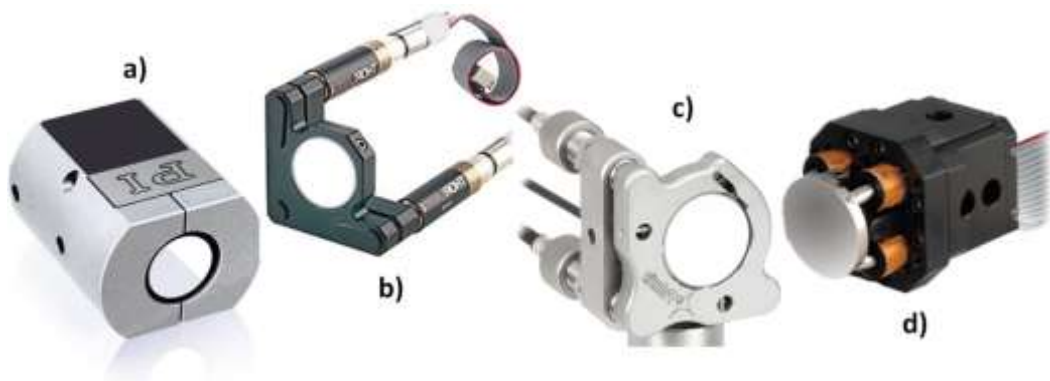

**Figura 143. Tecnologías de implementación de espejos modulables: a) espejos piezoeléctricos, b) monturas motorizadas, c) monturas piezoeléctricas y d) espejos de bobina.**

Los espejos piezoeléctricos y los de bobina son las tecnologías más apropiadas para integrar en un sistema de corrección de *beam wander*. A este tipo de espejos modulables se le conoce comúnmente como FSM (del inglés *Fast/Fine Steering Mirror*). Atendiendo únicamente a las especificaciones, el más apropiado para un sistema como el desarrollado sería el tipo piezoeléctrico debido a que proporciona un ancho de banda mayor, pese a que ofrecer un menor rango angular (esta limitación no debería suponer un problema ya que como se verá los movimientos a corregir serán muy pequeños). Sin embargo, para el sistema de corrección se seleccionó un espejo de bobina por ofrecer unos valores de resolución angular y ancho de banda nominalmente suficientes a un menor coste que el piezoeléctrico.

El FSM seleccionado para el sistema de corrección de turbulencia fue el modelo OIM102 de la empresa *Optics in Motion LLC*. Se trata de un espejo con recubrimiento en oro de 2″ de diámetro (1,8″ de apertura efectiva), con un rango angular total de ± 1,5°, un ancho de banda nominal a 3 dB de 400 Hz y una resolución limitada por la electrónica de control, al tratarse de un dispositivo analógico. El FSM incluye un sensor de posición interno para monitorizar el estado instantáneo del espejo. La información de este sensor proporciona la posición real del FSM, que junto a la posición deseada, se puede utilizar para generar una señal de error de posición que es minimizada mediante un control tipo

PID (introducido en el apartado 2.10.4). En la Figura 144 se muestra el diagrama de bloques del sistema de control del FSM, donde se pueden ver todas las señales de entrada y salida que proporciona. Cuando con el conmutador se selecciona la señal de posición interna del FSM, el sistema actúa del modo descrito anteriormente llevando al FSM a la posición deseada. Sin embargo, existe la posibilidad de utilizar una señal de posición externa como una de las dos entradas del PID. Si a la entrada de posición deseada se la alimenta con una señal DC y como señal de posición externa se utiliza un detector sensible a la posición que monitoriza el haz recibido en Bob, el sistema PID minimizará la señal de error, llevando al haz a la posición determinada por la señal DC. Esta será la estrategia seguida para integrar el FSM en el sistema de corrección de turbulencia de Bob.

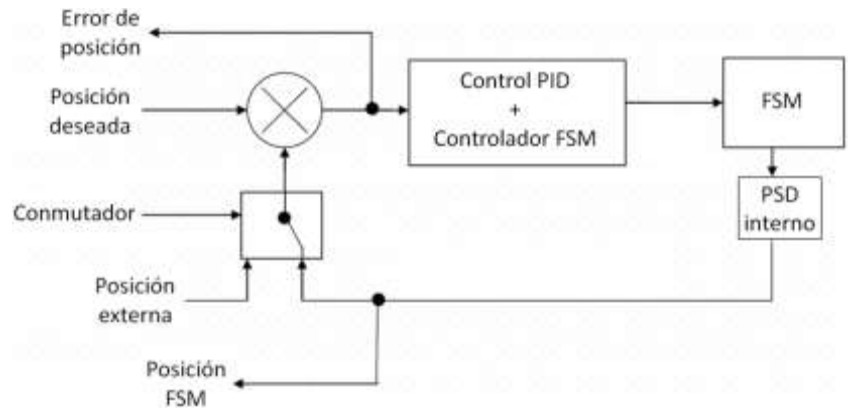

**Figura 144. Diagrama de bloques del sistema de control del FSM.**

## 3.8.2. Caracterización del FSM

Las entradas del controlador del FSM deben ser señales de voltaje DC analógicas de un rango de ±10 V. Para generar estas señales, se seleccionó el sistema de adquisición y generación de señales USB6343 de *National Instruments*, que proporciona cuatro salidas analógicas de ±10 V con 16 bits de resolución (además de 32 entradas analógicas de las mismas características empleadas para adquirir el resto de las señales). La resolución del generador de señales es la que limitará la resolución angular del FSM. En este caso, los 16 bits suponen una resolución de $10\ V/2^{16} = 0,15$ mV. Si el rango angular del espejo es de ±1,5° y se corresponde con un rango eléctrico de ±10 V, entonces la relación entre ambos es de $0,15°/V$ y por lo tanto una resolución eléctrica de $0,15$ mV supone una resolución angular de $0,0000225° = 0,39$ µrad. Esta resolución angular equivale a un desplazamiento del orden de nanómetros a una distancia del orden de centímetros desde el espejo, que es aproximadamente donde se aplicará la corrección. Si idealmente el objetivo del sistema de corrección es mantener el haz reflejado en el FSM dentro de un diámetro igual a una fibra monomodo, del orden de las 10 µm, la resolución angular quedaría entre unos cuatro órdenes de magnitud por debajo de este valor, lo que se considera más que suficiente para llevar a cabo la corrección sin problemas de resolución.

Si bien la función de transferencia del FSM en términos de voltaje aplicado y deflexión angular producida se puede obtener teóricamente de las especificaciones, tal como se ha descrito en el párrafo anterior, como $0,15°/V$, al caracterizarlo experimentalmente se midió una constante de proporcionalidad de $0,3°/V$ (ver apartado 3.8.5). Esta diferencia se explica por el hecho de que el rango angular de ± 1,5° se corresponde con el movimiento mecánico del espejo. Por lo tanto, al traducirlo al haz reflejado, el ángulo de este será del doble (puede visualizarse con un espejo rotado un ángulo de 45° respecto a la normal al eje óptico para conseguir reflejar un haz a 90° del eje óptico). Esto significa que la resolución angular calculada en el párrafo anterior se refiere a la resolución mecánica del FSM y para obtener la resolución de deflexión habría que multiplicar por dos. Esta corrección se obtuvo tras caracterizar el FSM, habiendo seleccionado ya el sistema de adquisición y generación de señales, si bien el amplio margen de cuatro órdenes de magnitud hace despreciable a este factor dos de corrección.

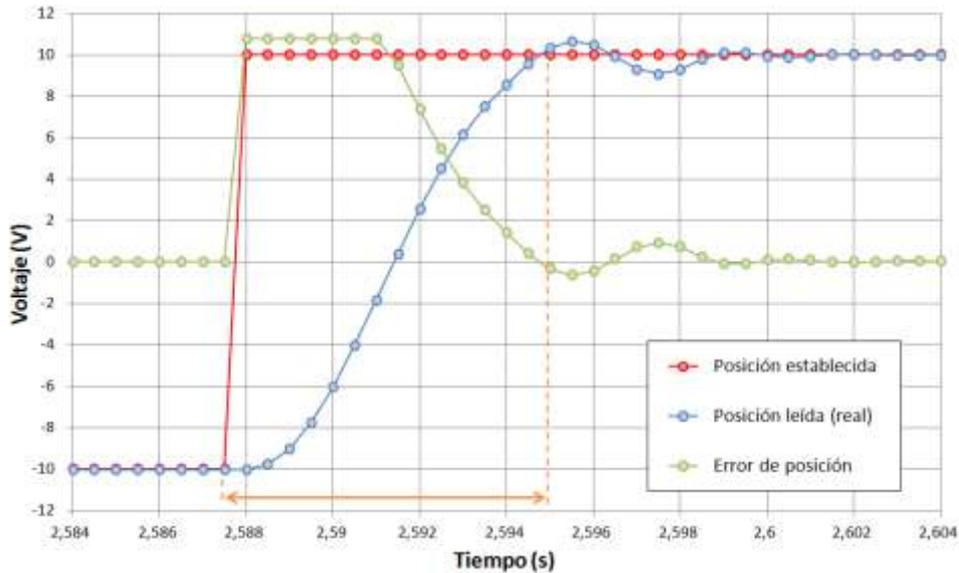

**Figura 145. Medida del tiempo de subida (ts) del FSM ante el estímulo de un escalón.**

Como se ha comentado, el ancho de banda nominal del FSM adquirido es de 400 Hz. Al tratarse de un dato importante para la corrección de la turbulencia, se realizó la caracterización experimental de la velocidad de respuesta con el objetivo de comprobar si era capaz de seguir las variaciones del *beam wander*. Para ello, se desarrolló un programa en *LabView* que proporcionaba como señal de posición una salida de tipo escalón con una amplitud de salto variable y que registraba sincronizadamente la señal de posición real del espejo para los mismos instantes. Para cada amplitud de salto del escalón, se midió el tiempo de subida $t_s$ como el intervalo necesario para pasar del 0 % al 100 % del valor deseado (Figura 145) y se calculó el ancho de banda como BW = 0,35/$t_s$ [276, pp. 3-26]). En la Figura 146 se muestra el resultado, observándose que el valor nominal proporcionado por el fabricante se corresponde con saltos muy pequeños, que por otra parte serán los necesarios en el sistema de corrección al realizarse en el plano focal donde todas las distancias se ven minimizadas. Considerando que el ancho de banda efectivo siempre estará por encima de los 250 Hz (correspondiente a saltos de 1 Vpp, muy superiores a los necesarios en el plano focal), se consideró suficiente para compensar la turbulencia atmosférica, que fue caracterizada espectralmente en otro experimento (ver apartado 3.9.3), donde se comprobó que la mayor parte de las componentes quedaban por debajo de 100 Hz y desaparecían por encima de 300 Hz.

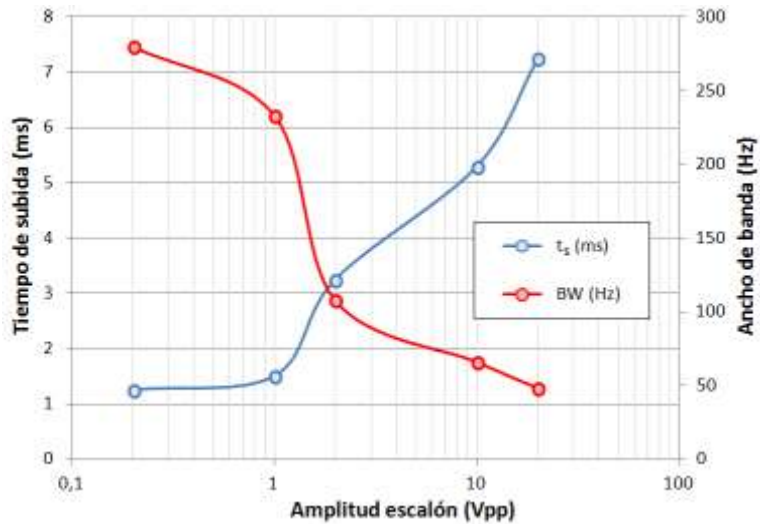

**Figura 146. Tiempo de subida y ancho de banda del FSM como respuesta a un escalón de diferentes amplitudes.**

### 3.8.3. Detectores sensibles a la posición

Otro de los componentes fundamentales del sistema de corrección es el detector sensible a la posición, cuya misión es registrar el efecto del *beam wander* en Bob. El objetivo de cualquier detector sensible a la posición es transformar los movimientos del centroide del haz en el plano transversal a la propagación, ortogonal al eje óptico, en variaciones continuas de una magnitud física medible a través de cuatro electrodos, de cuyas diferencias es posible deducir la posición. Básicamente, existen dos tecnologías: detectores de cuadrantes y detectores de efecto lateral (Figura 147). Sin embargo, sea el tipo que sea, la posición se calcula de la misma forma, comparando las señales de los cuatro electrodos mediante las ecuaciones (3-35) y (3-36). El principio consiste en restar las señales opuestas de cada eje y normalizar con la intensidad total detectada. Dado que la resolución no depende de la relación señal a ruido, este tipo de detectores permite registrar el movimiento con muy baja intensidad de luz.

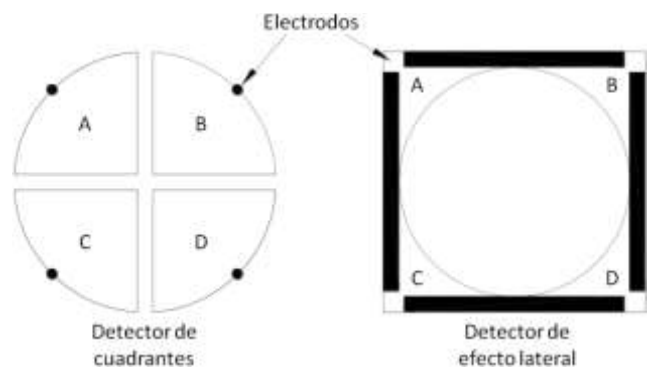

**Figura 147. Tipos de detectores sensibles a la posición.**

$$x = \frac{(B+D)-(A+C)}{A+B+C+D} \qquad (3\text{-}35)$$

$$y = \frac{(A+B)-(C+D)}{A+B+C+D} \qquad (3\text{-}36)$$

Los detectores de cuadrantes (QD, del inglés *Quadrant Detector*) consisten en varias áreas activas (típicamente dos o cuatro según se precise medir un eje o dos), separadas por un hueco o gap intermedio. Los detectores de efecto lateral (PSD, del inglés *Position Sensitive Detectors* [26]) consisten en una sola área activa, en cuyos cuatro extremos están conectados los electrodos. Así, entre cada ánodo y cátodo existirá una resistencia interna que dependerá de la intensidad de la luz incidente y que proporcionará variaciones en la corriente eléctrica. Los QD en general ofrecen una mayor resolución espacial (por debajo de la micra) y los PSD un mayor rango (permiten utilizar toda el área activa con spot pequeños).

Un aspecto fundamental a la hora de elegir una tecnología u otra es cómo será el haz que se desea medir. En general, los PSD permiten medir haces enfocados, ya que no tienen ninguna zona ciega. Sin embargo, el gap de los QD limita el mínimo tamaño del spot, por lo que estos detectores se usan con haces desenfocados. Normalmente, la relación entre el tamaño de spot y el tamaño del detector se determina estableciendo un compromiso entre la resolución y el rango: a menor spot, mayor resolución y menor rango. En la Figura 148 se muestra un esquema del funcionamiento básico de un QD ideal: se puede observar la salida que proporcionaría el detector en el eje horizontal ante las distintas posiciones del centroide de dos haces de intensidad uniforme y tamaños distintos (uno igual al tamaño del detector y el otro igual al tamaño de uno de los segmentos). Se comprueba que existen tres zonas diferentes: el detector proporciona una salida igual a cero cuando el haz está fuera de su área activa, una salida constante cuando el haz está dentro de un solo segmento y una salida lineal cuando está contenido en dos segmentos.

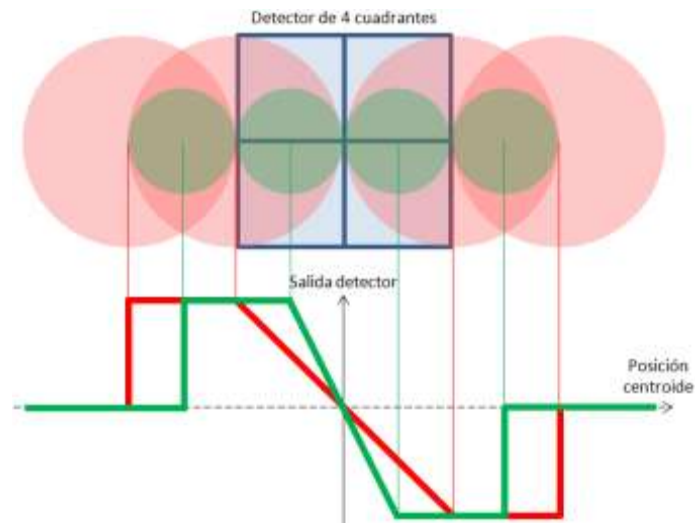

**Figura 148. Funcionamiento básico de un QD ideal.**

Por último, aunque debiera ser el primer aspecto a considerar, se trate de un PSD o un QD, la selección de un dispositivo en particular la determinará el material en el que está fabricada el área activa. Cada material muestra una sensibilidad diferente ante distintos rangos de longitudes de onda, por determinar esta su eficiencia cuántica (Figura 149). En general, para detectar luz visible y de infrarrojo cercano, hasta unos 900 nm, se utiliza el

---

[26] El acrónimo PSD se puede encontrar referido indistintamente a los detectores sensibles a la posición en general (en cuyo caso se emplea LEP para referirse a los de efecto lateral, del inglés *Lateral Effect Photodiode*) o a los detectores de efecto lateral en particular. En esta tesis se ha decidido utilizar la segunda convención, por su mayor frecuencia de aparición en la literatura.

Silicio (Si) y para el rango 1000-1600 nm se utiliza el Arseniuro de Galio e Indio (InGaAs). Por ello, los detectores del canal de datos (850 nm) deberán ser de Si y los del canal de sincronismo y seguimiento (1550 nm) de InGaAs.

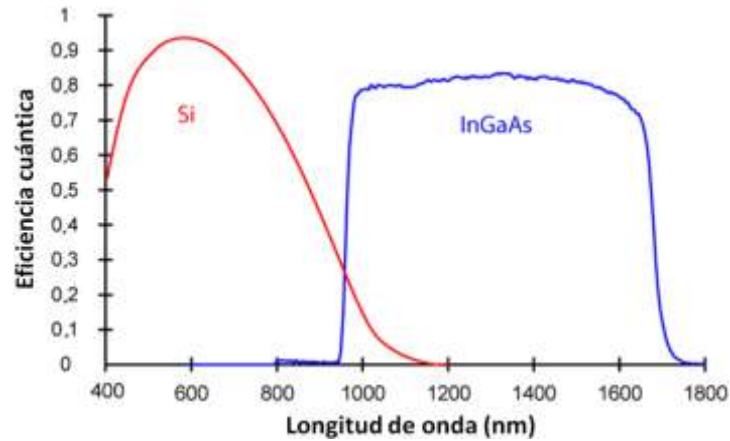

**Figura 149. Eficiencia cuántica del Silicio y el Arseniuro de Galio e Indio en función de la longitud de onda.**

### 3.8.4. Comportamiento real de un detector de cuadrantes

El comportamiento del QD descrito en el apartado 3.8.3 es ideal, si bien es el que siempre se refiere en la literatura relacionada. En la práctica, el comportamiento real dista bastante del ideal debido a diversos factores. El principal es que en general el haz tendrá un perfil gaussiano y no con una distribución de intensidad uniforme circular, como se suele asumir. Esto hace que ni la señal del QD pase de cero a saturada cuando el haz empieza a entrar en el detector ni que el QD sea mucho mayor que la combinación del tamaño del spot focal y su rango de movimientos). La existencia de las faldas del perfil gaussiano hace que el haz sea mucho más visible de lo que se podría intuir. Por ejemplo, si el haz es más grande y cubre totalmente al detector, teóricamente no debería observarse ningún movimiento; y lo mismo se puede decir de si el haz está a simple vista fuera del detector, aunque cerca. En ambos casos, sí es posible detectar cambios de posición ya que el método de cálculo de la posición no consiste en valores absolutos de intensidad, sino relativos (diferencias entre segmentos), e incluso las pequeñas intensidades de las faldas de la gaussiana proporcionan una diferencia suficiente para extraer la información necesaria de la posición del haz.

El perfil gaussiano también tiene influencia en el comportamiento de un QD enfocado. Idealmente, cuando el spot sumado a su movimiento es menor al tamaño del gap, la señal del QD debería ser igual a cero. Sin embargo, pese a que la mayor parte de la energía puede pasar sin ser detectada, las faldas introduciéndose en el QD hacen que las diferencias entre un segmento y otro se puedan detectar pese a su reducida intensidad, obteniéndose una señal de posición distinta a cero. En esta situación, se observa un comportamiento como el de la Figura 150 (derecha), donde se muestra la señal de salida en uno de los ejes del detector IGA-030-QD situado en el plano focal de una lente de 30 mm de focal detectando las variaciones moduladas con el FSM en la posición de un haz (Figura 150, izquierda). Se puede observar la existencia de dos zonas claramente diferenciadas en la respuesta del QD: una correspondiente a valores pequeños de señal donde la respuesta varía lentamente (esta situación se da cuando el máximo de la campana de Gauss del perfil del haz está incluído dentro del gap) y otra de mayor señal con picos de elevada pendiente (correspondiente al momento en que la campana de Gauss sale del gap y cada segmento empieza a detectar una cantidad mucho mayor de potencia óptica).

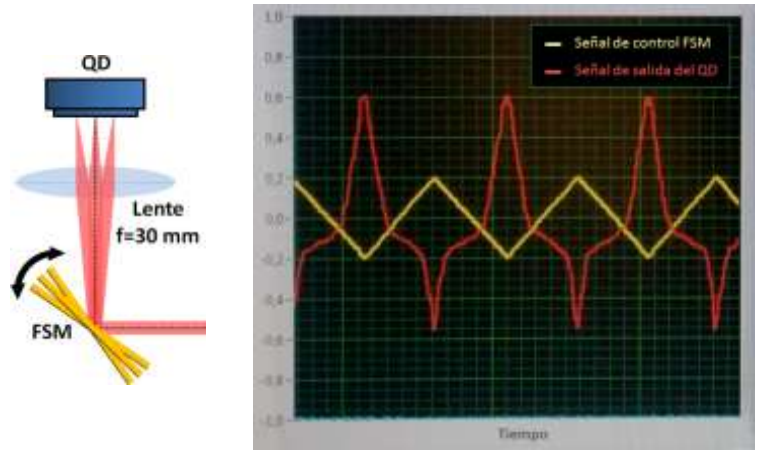

**Figura 150. Montaje del experimento (izquierda) y salida del QD ante variaciones de posición de un haz controlado por el FSM (derecha).**

Por otra parte, la geometría de los haces, generalmente con simetría circular, hace que la zona de variación lenta de la Figura 148 no sea exactamente lineal. Esto se explica porque la relación entre la cantidad de energía (o de área) del haz en cada cuadrante no es exactamente proporcional a la distancia del centroide geométrico al centro del QD, caso que solo se daría si el haz y el QD fueran perfectamente cuadrados. En la Figura 151 se muestra una simulación utilizando trazado de rayos en *OpticsLab* de la salida normalizada del QD seleccionado para el sistema de corrección, modelo IGA-030-QD (de 3 mm de diámetro y 45 µm de gap) ante diferentes diámetros de spot en función de la posición real del centroide respecto al centro del detector en (0, 0). Se puede comprobar cómo la zona idealmente lineal no lo es tal cuando se utilizan haces circulares como los de *OpticsLab*. Para un spot de diámetro $D_{SPOT}$ inferior al tamaño del gap $D_{GAP}$ (40 µm y 45 µm respectivamente), se aprecia cómo la salida del QD es igual a cero, para cualquier posición del centroide respecto al detector. Al considerar *OpticsLab* un perfil de haz perfectamente circular en lugar de gaussiano el spot siempre estará contenido dentro del gap y por tanto no será detectado por el QD. Para un $D_{SPOT} = 60$ µm se aprecia una zona en la que la salida del QD también se anula. Esto se explica por la existencia de una zona central en la que un spot mayor que el gap no sería detectado debido a la mayor superficie que presenta la intersección del gap horizontal y el vertical. El diámetro $D_{central}$ de dicha zona se puede calcular fácilmente como $D_{central} = D_{GAP} \cdot \sqrt{2} = 63{,}64$ µm. Esto quiere decir que la señal del QD ante un spot de un diámetro entre 45 y 63,64 µm pasará directamente del estado de saturación de uno de los dos cuadrantes a cero y luego a la saturación del cuadrante opuesto (siendo estas distancias iguales $\pm (D_{SPOT}/2 - D_{GAP}/2)$, medidas desde el centro geométrico del QD). Solo los spots de diámetro mayor a 63,64 µm describirán una curva entre las zonas de saturación de cuadrantes opuestos.

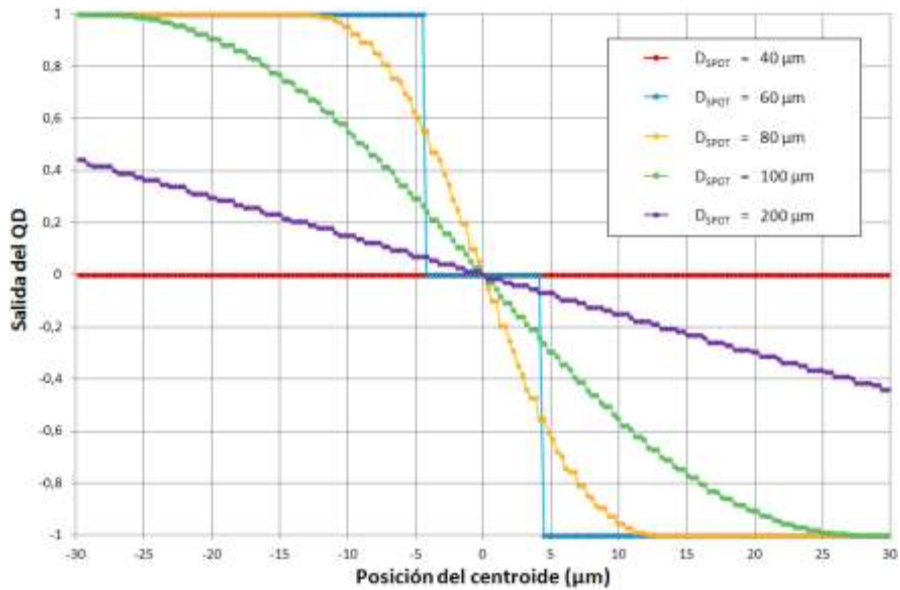

**Figura 151. Simulación de la salida normalizada de IGA-030-QD ante diferentes diámetros de spot D$_{SPOT}$ en función de la posición real del centroide.**

En la Figura 152 se muestra una simulación similar pero con diámetros de spot mayores, y por lo tanto con un centro del spot (eje horizontal) extendido hasta las dimensiones máximas del QD. Se puede comprobar cómo la posición del centroide donde la señal del QD abandona la saturación para proporcionar información sobre la posición coincide con una distancia igual a ± (D$_{SPOT}$/2-D$_{GAP}$/2), como la calculada para la zona ciega del centro del QD. Se puede comprobar que solo cuando el spot tiene un diámetro igual al del QD desaparece la zona de saturación.

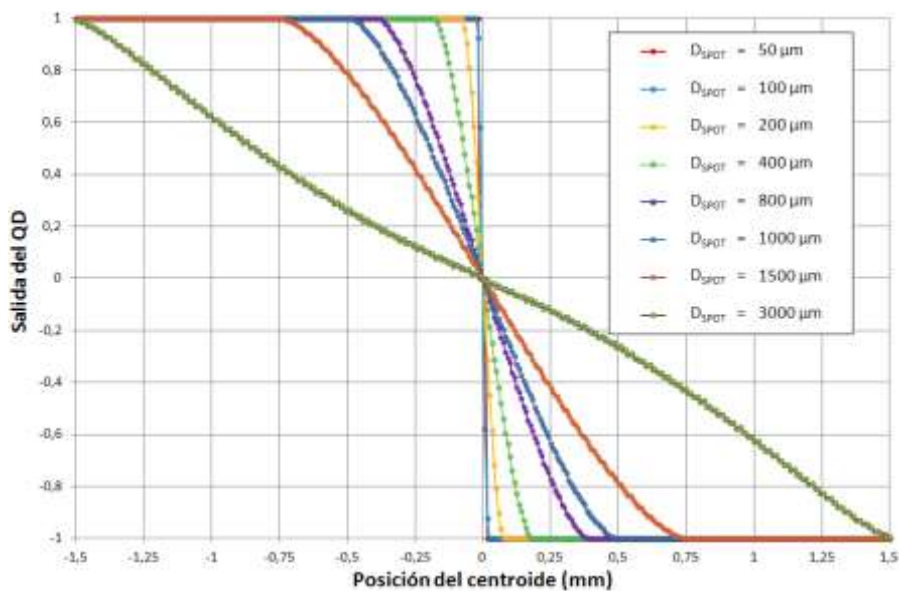

**Figura 152. Simulación de la salida normalizada de IGA-030-QD ante diferentes diámetros de spot D$_{SPOT}$ en función de la posición real del centroide.**

### 3.8.5. Cálculo del tamaño de spot en un QD

Como se ha visto en el apartado anterior, un parámetro fundamental al usar un QD, y por extensión un parámetro fundamental en el sistema de compensación de turbulencia, es el tamaño de spot. Sin embargo, el QD únicamente proporciona información sobre la ubicación del centroide del spot y ninguna información de su tamaño. Utilizando el conocimiento obtenido sobre el comportamiento real de un QD, se ha diseñado un método para deducir el tamaño del spot en un QD únicamente a partir de medidas de la posición del centroide. Este método está basado en la medición de la pendiente de la señal del QD al realizar un barrido de un haz sobre su superficie.

Para un montaje óptico determinado, existirá una relación unívoca entre el tamaño de spot y la pendiente de la señal de posición del QD (ver Figura 152), que puede obtenerse simulando el montaje con distintos tamaños de spot (lo que se consigue estableciendo distintos grados de desenfoque en el QD, siendo el método más simple modificar su posición longitudinalmente en un haz no colimado). Una vez obtenida esta relación para una de estas posiciones, se puede deducir el tamaño de spot a partir de una pendiente medida experimentalmente, utilizando el FSM para mover el haz. Este método no ofrece una medida exacta del tamaño de spot pero sí una estimación suficientemente válida sobre las dimensiones del spot con que se está trabajando, que de otro modo no sería posible obtener. Esto es especialmente cierto cuando se utiliza un haz no colimado, que es el caso más habitual, y se realizan medidas relativas del tamaño de spot de una posición del QD respecto a otra.

A continuación, se describen brevemente los pasos seguidos para realizar dicha estimación:

1. El primer paso es la caracterización del FSM, obteniendo su función de transferencia en términos de voltaje aplicado y deflexión angular producida. Este proceso se realiza midiendo los ángulos de deflexión producidos por una serie de voltajes aplicados al FSM (obteniendo cada vez el ángulo $\alpha$, a partir de la medida del desplazamiento transversal x del haz y de la distancia y al espejo, como $\alpha = \text{atan}(x/y)$). Tras esta serie de medidas, se ajustan los valores a una recta (obteniendo con el ajuste un coeficiente de determinación $R^2$ igual a 0,997, de lo que se concluye que el FSM muestra un comportamiento angular altamente lineal con el voltaje aplicado). Así se obtiene experimentalmente una constante de proporcionalidad igual a 0,293°/V, que es aproximadamente igual al doble de la obtenida a partir de las especificaciones porque este es el ángulo del haz reflejado y no el ángulo mecánico del FSM, que es el dato que proporciona el fabricante.

2. Mediante una simulación con trazado de rayos, es necesario simular el sistema óptico en el que está integrado el QD para obtener la curva de su señal de respuesta en relación al ángulo de deflexión del FSM. Así se obtendrían curvas similares a las de la Figura 152 pero con un eje de abcisas referido al ángulo del FSM en lugar de la posición del centroide. Para cada curva, tras medir el tamaño del spot en la misma posición donde está situada el área activa del QD, es necesario hacer un barrido del FSM en un mismo rango para todas las curvas, registrando su ángulo de deflexión y la salida normalizada del QD. En esta simulación es importante replicar fielmente el montaje óptico real en términos de longitudes focales de los elementos ópticos y sus posiciones relativas, del tamaño de entrada del haz y su divergencia, así como las características reales del QD en cuanto a diámetro del área activa y tamaño del gap.

3. Una vez simuladas las curvas de respuesta del QD, se obtiene la pendiente m de cada una para cada tamaño de spot, realizando un ajuste lineal como y = mx sobre las series de valores obtenidos. El hecho de realizar un ajuste lineal sobre curvas no tendrá mayor relevancia ya que las pendientes equivalentes se medirán en el QD de forma experimental, donde también se obtendrán curvas similares.

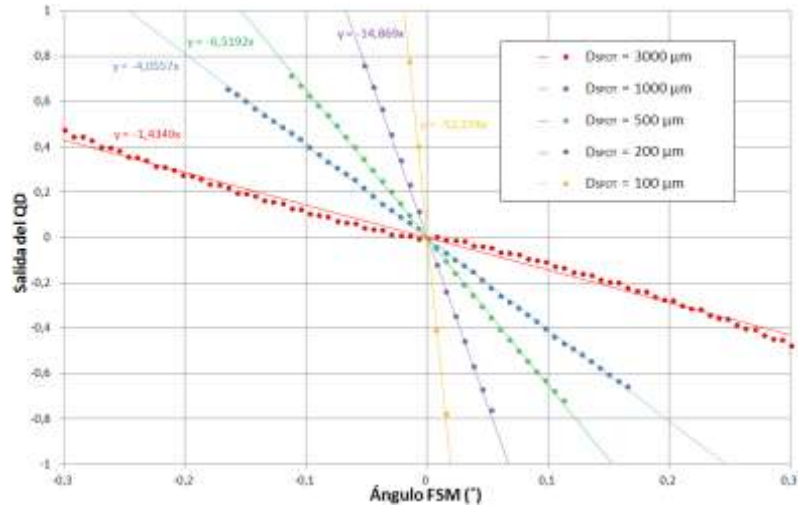

**Figura 153. Simulación de la salida normalizada del QD ante diferentes diámetros de spot en función del ángulo de deflexión del FSM.**

4. Si se ajusta la serie de pares de valores {tamaño de spot, pendiente m} se puede obtener la relación entre el tamaño de spot $D_{SPOT}$ y la pendiente m, que en este caso es de $D_{SPOT} = 3651 \cdot m^{-0.97}$ (Figura 154).

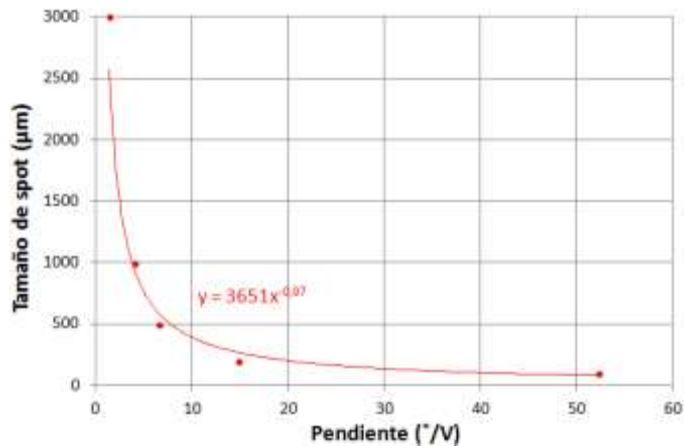

**Figura 154. Tamaño de spot en el QD en función de la pendiente del FSM de deflexión angular frente a voltaje aplicado.**

5. Por último, para obtener la estimación del tamaño de spot con el QD situado en una posición arbitraria sobre el montaje experimental, sería necesario realizar con el FSM un barrido lineal (señal triangular) de igual amplitud al que se llevó a cabo en la simulación (para lo cual se utilizaría la constante de proporcionalidad calculada en el paso 1) y obtener la pendiente siguiendo el mismo procedimiento que en los pasos 2 y 3. En este caso, habría que reescalar ambos ejes: el de ordenadas es necesario normalizarlo para que la salida del QD quede entre -1 y 1 (para lo cual es necesario dividir entre 10 la señal de salida del QD, que tiene un rango de ± 10 V), y para reescalar el de abcisas se utilizaría la constante de proporcionalidad calculada en el paso 1 (convirtiendo los voltios aplicados al FSM en ángulo de deflexión, para obtener un eje como el de la Figura 153). Una vez calculada la pendiente m, resultado de realizar el ajuste lineal sobre los valores de la curva, se puede obtener el tamaño de spot utilizando la ecuación obtenida en el ajuste del paso 4.

### 3.8.6. Caracterización funcional del QD y el PSD

El QD seleccionado para detectar los movimientos de la longitud de onda de 1550 nm fue el IGA-030-QD, del fabricante *EOS Systems*. Este detector está fabricado en InGaAs (sensible a longitudes de onda de 1 a 1,7 μm), tiene 3 mm de diámetro y un tiempo de subida menor a 25 ns (lo que equivale a un ancho de banda de BW = 0,35/25 ns = 14 MHz). Este detector se vende como un componente individual sin electrónica de acondicionamiento, por lo que será necesario convertir las señales de corriente en voltaje, amplificarlas y transformarlas en señales de posición. Según las especificaciones, el espaciado entre elementos es típicamente de 20 μm y de 45 μm como máximo. Para caracterizar este gap, se midió en un microscopio 100×, obteniendo un resultado de 45,66 μm (Figura 155), ligeramente por encima del máximo especificado.

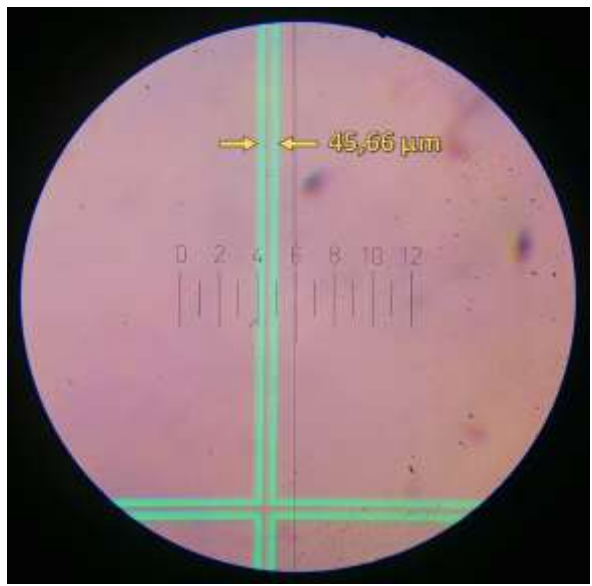

**Figura 155. Captura de microscopio de la separación entre cuadrantes del detector IGA-030-QD.**

Por otra parte, se usó un PSD de Silicio con el objetivo de monitorizar en el camino de 850 nm las correcciones realizadas en el camino de 1550 nm cuando se usa un esquema como el de la Figura 138. El PSD seleccionado fue el módulo C10443-02 de *Hamamatsu*, que incluye la electrónica de preamplificación y conversión de corriente a voltaje. Este módulo incorpora un detector de posición de efecto lineal construido en Silicio (sensible a longitudes de onda de 320 a 1100 nm) de geometría cuadrada con dimensiones de 9×9 mm con un ancho de banda de 3 dB de 16 kHz. El módulo C10443-02 se vende junto al módulo C10460, encargado de llevar a cabo las operaciones descritas en las ecuaciones (3-35) y (3-36), proporcionando una salida analógica de ±10 V de amplitud con una equivalencia de 0,5 mm/V en términos de distancias del centroide medidas en el área activa del PSD.

Para caracterizar el comportamiento del QD y del PSD ante un mismo haz en movimiento, se realizó el montaje de la Figura 156, consistente en crear un camino para cada detector tras su separación mediante un divisor de haz, ambos respondiendo a los mismos movimientos del FSM encargado de desviar controladamente un mismo haz de 650 nm. Se utilizó una longitud de onda visible con el objetivo de simplificar el alineamiento de los distintos componentes ópticos. Si bien el QD teóricamente no está construído para detectar esta longitud de onda, en la práctica debido a su alta sensibilidad, la eficiencia cuántica a 650 nm es reducida pero no nula, por lo que sí es posible detectar esta longitud de onda. Dado que se utiliza un único láser, para evitar la saturación en el PSD se utilizó un polarizador lineal para atenuar la intensidad del láser. Mediante sendas lentes de 30 mm de focal, se hace converger ambos haces y se detectan en los correspondientes detectores que son trasladados longitudinalmente a lo largo de dos carriles para estudiar su comportamiento ante distintos tamaños de spot. El efecto de trasladar los detectores por los carriles será el enfoque/desenfoque de los mismos.

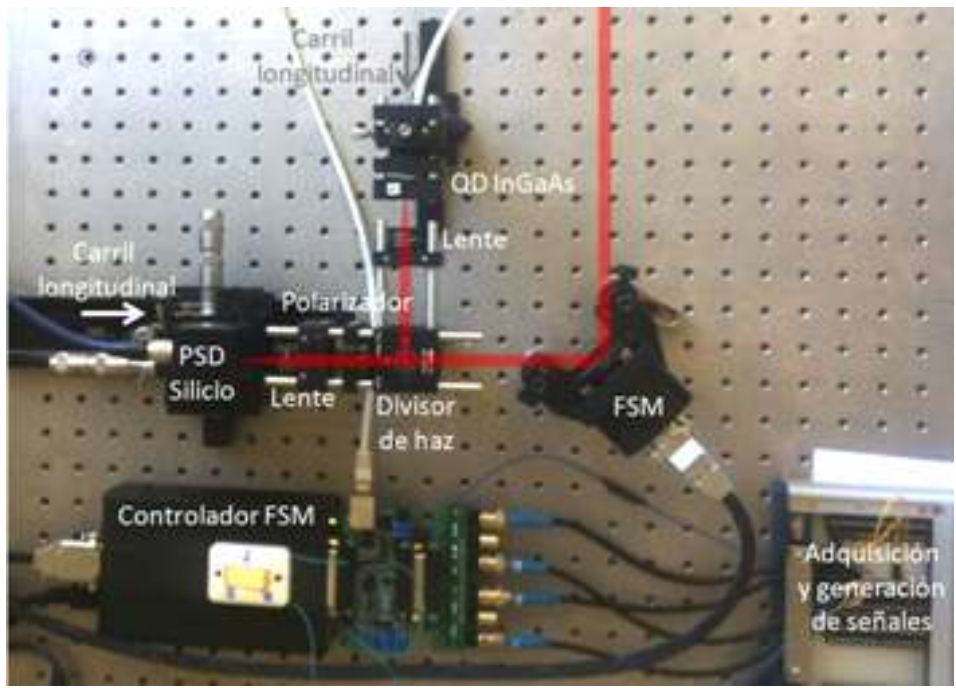

**Figura 156. Montaje óptico para la caracterización funcional del QD y el PSD.**

En la Figura 157 y en la Figura 158 se muestra el resultado de esta caracterización para el PSD y el QD respectivamente. En ambas se muestra la señal de salida de cada detector al trasladarlo longitudinalmente como respuesta ante la rotación del FSM. Las gráficas situadas a la derecha en ambas figuras se obtuvieron experimentalmente utilizando el montaje descrito en el párrafo anterior y las de la izquierda se obtuvieron mediante una simulación de trazado de rayos en *OpticsLab* modelando todos los aspectos del mismo montaje experimental, por lo que se observa la similitud de resultados entre la simulación y el experimento. De uno a otro, solo cambia la frecuencia de oscilación del FSM (que no aporta ninguna información sobre el comportamiento de los detectores) y que la señal de salida simulada y la distancia longitudinal tienen una escala arbitraria (por lo que no se pueden comparar los resultados cuantitativamente, solo cualitativamente, que es el objetivo de esta caracterización funcional).

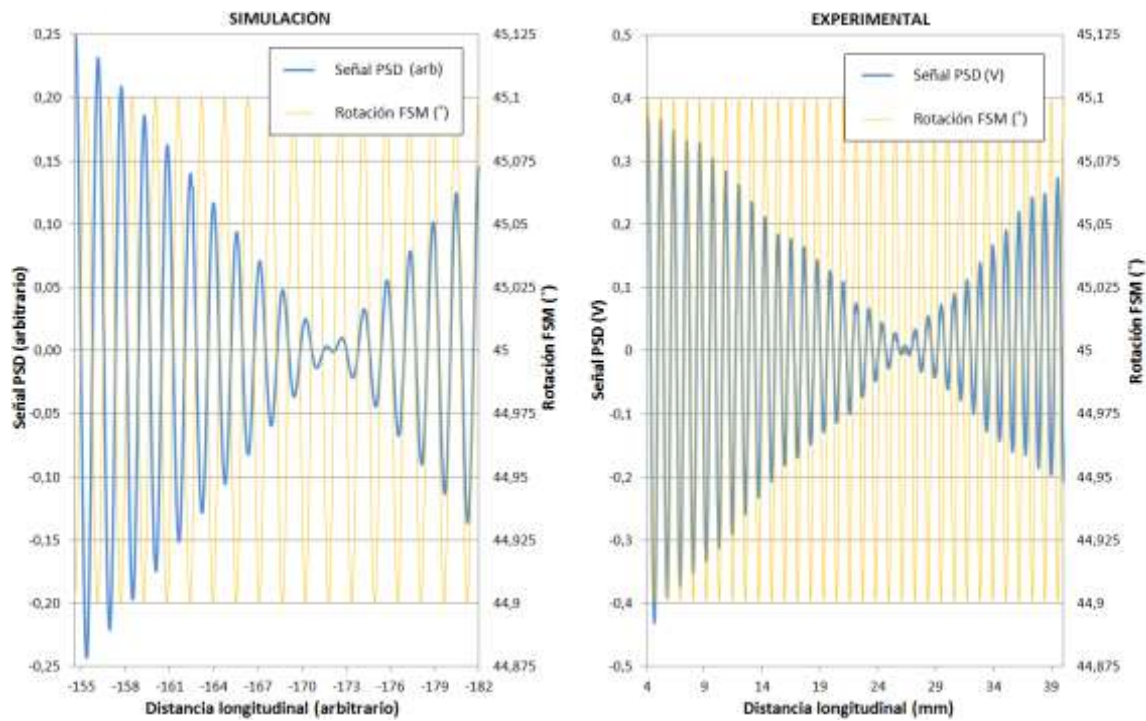

**Figura 157. Simulación y medida experimental de la respuesta cualitativa del PSD al desplazarlo longitudinalmente ante un haz con oscilación sinusoidal.**

En la Figura 157 se observa que al trasladar longitudinalmente el PSD, la amplitud de los movimientos registrados se va atenuando hasta anularse para luego volver a aumentar. El punto donde se anula la señal del PSD no se corresponde con el plano focal, sino con el plano imagen, correspondiente al plano objeto en la superficie del FSM. Hay que tener en cuenta que el comportamiento de un PSD es independiente del tamaño del spot, por lo que la variación observada se corresponde con el movimiento relativo del área activa del detector respecto al plano imagen. Este comportamiento es diferente del mostrado por el QD, que como se vio es muy sensible al tamaño de spot.

En la Figura 158 se puede observar el resultado equivalente para el QD, donde se aprecian dos tipos de movimientos diferentes. Por una parte, se observa el efecto de la variación del tamaño de spot: la evolución de la señal en la parte izquierda de la gráfica, según la cual la amplitud va aumentando hasta llegar a un punto en el que el spot tiene el tamaño mínimo, cuando el QD está en el plano focal. Por otra parte, superpuesto al efecto anterior, se observa un comportamiento similar al del PSD relativo al plano imagen: la evolución de la señal en la parte derecha, según la cual la amplitud va disminuyendo hasta llegar a un punto donde no se aprecia ningún movimiento, cuando el QD está en el plano imagen. Así, se comprueba que las señales del PSD y el QD cambian de fase en relación a la señal de rotación del FSM tras cruzar el plano imagen, lo que permite identificar este punto fácilmente. En el caso del QD, es fácil identificar la posición del plano focal observando la respuesta antes del plano imagen: la señal del QD relativa al movimiento del centroide va aumentando su intensidad hasta llegar al foco y luego empieza a disminuir hasta llegar al plano focal.

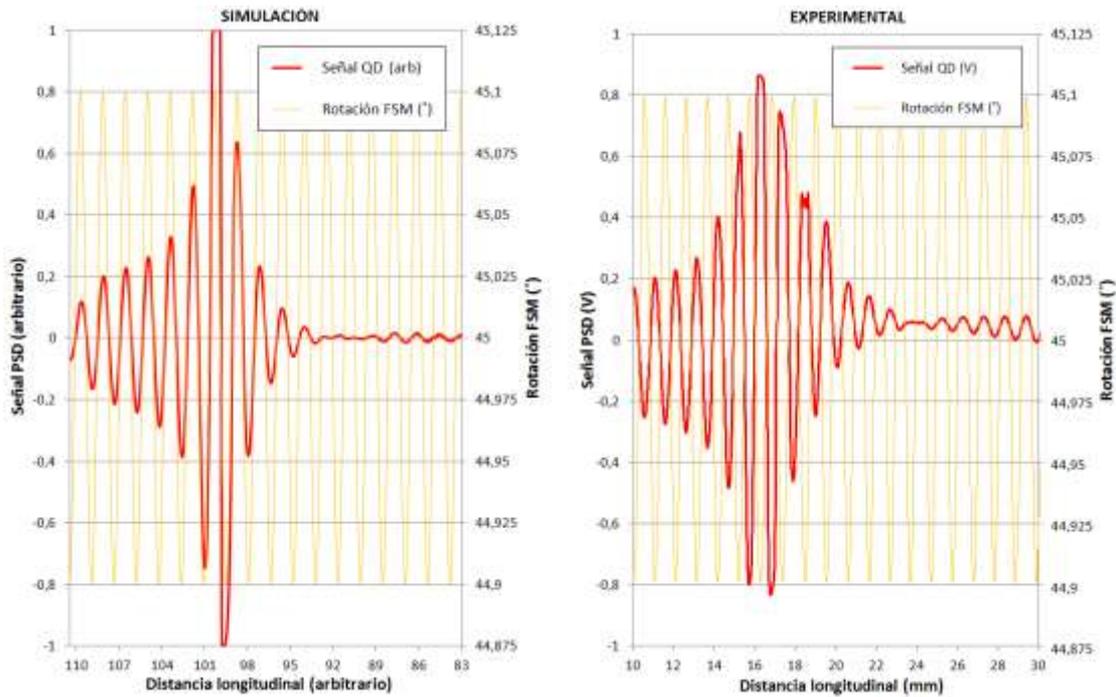

**Figura 158. Simulación y medida experimental de la respuesta cualitativa del QD al desplazarlo longitudinalmente ante un haz con oscilación sinusoidal.**

Como se ha visto, es importante distinguir el plano focal y el plano imagen al utilizar un detector sensible a la posición, y conocer cómo reacciona cada tipo de detector será de ayuda a la hora de diseñar e implementar el sistema de corrección. En el plano focal del sistema óptico, el spot tendrá el mínimo tamaño pero existirá un determinado desplazamiento del centroide. En el plano imagen, el desplazamiento del centroide será el mínimo (nulo ante movimientos puramente angulares) pero el spot no tendrá el mínimo tamaño. Este comportamiento se puede observar en la Figura 159.

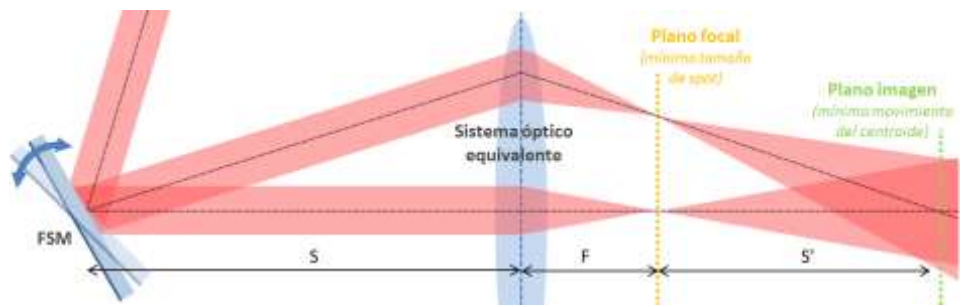

**Figura 159. Plano imagen y plano focal de un sistema óptico ante una variación angular.**

En la Figura 160 se muestra el resultado de una simulación con *OpticsLab* del mismo montaje donde se puede observar por una parte el desplazamiento absoluto en mm del centroide desde el centro del QD (en azul) y por otra el tamaño del spot en mm (en rojo) para cada posición relativa al plano focal de la lente. Se comprueba que el detector llega primero al foco al desplazarlo longitudinalmente y en esa posición el spot se hace mínimo (85 µm, en este caso limitado por la aberración esférica de la lente simulada), y 6,36 mm más lejos se encuentra la otra posición de nulo desplazamiento del centroide pero con un spot 23 veces más grande (1,97 mm). En la Figura 158 se puede comprobar que la distancia entre plano focal y plano imagen obtenida experimentalmente coincide con la obtenida con la simulacion. De forma teórica, la separación entre el plano focal y el plano imagen se puede obtener sin más que utilizar la ecuación (2-4), o fórmula de Gauss para lentes delgadas 1/F=1/S+1/S' [205, p. 17]. En esta caracterización, la distancia focal F de la lente es de 30 mm y la distancia S desde la fuente de la variación angular es 18 cm, por lo que la distancia S' será igual a 36 mm, es decir, una separación de 6 mm, que coincide aproximadamente con el resultado experimental y simulado (la pequeña diferencia entre el valor teórico y el simulado se debe a que la longitud focal equivalente de la lente simulada no era exactamente de 30 mm, debido a la aberración esférica). Esta diferencia entre plano focal e imagen será muy reducida cuando la perturbación angular provenga de la turbulencia atmosférica (al tomar la variable S un valor muy elevado), sin embargo es muy importante tenerla en cuenta a la hora de calibrar el sistema utilizando el FSM ya que es fácil confundir la posición de mínimo desplazamiento con el plano focal.

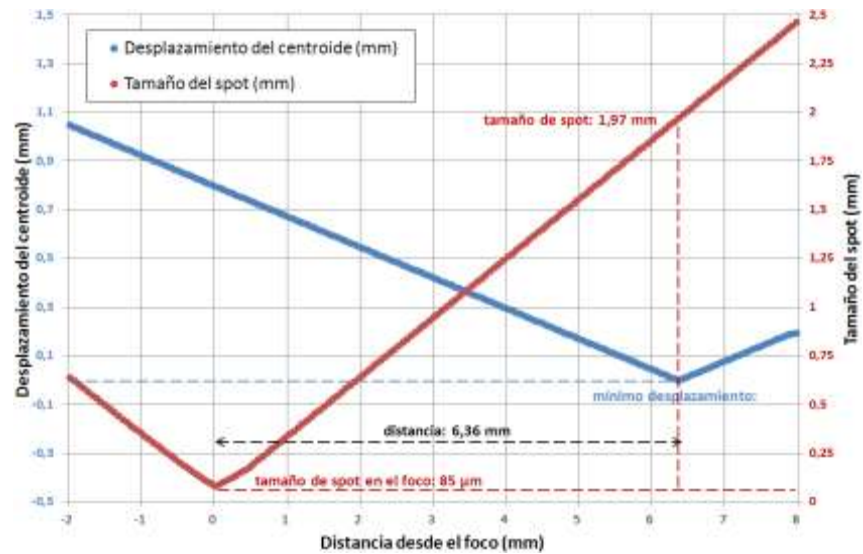

**Figura 160. Simulación del desplazamiento del centroide y del tamaño de spot en función de la posición relativa del QD desde el plano focal.**

### 3.8.7. Electrónica de acondicionamiento del QD

Como se comentó en el apartado anterior, el PSD de Silicio seleccionado para monitorizar las correcciones en el camino de 850 nm incorpora toda la electrónica de acondicionamiento necesaria en un módulo cerrado que proporciona directamente las salidas de voltaje para la posición {x, y}. Para la electrónica del QD, se adquirió una tarjeta ad hoc del mismo fabricante del FSM *Optics In Motion* que incluye básicamente la preamplificacion y las operaciones de las ecuaciones (3-35) y (3-36) para calcular la posición {x, y} a partir de las cuatro señales correspondientes a las salidas del QD. Estas señales se introducen en la tarjeta mediante un cable tipo RJ45 donde van soldadas las cinco salidas del QD: los cuatro ánodos de cada segmento y el cátodo común. En la Figura 161 se muestra el diagrama electrónico básico de esta tarjeta, consistente en una primera etapa de conversión corriente-voltaje y preamplificación y una segunda etapa de cálculo de la posición posición {x, y}, también siguiendo las ecuaciones (3-35) y (3-36) [27]. Utilizando los esquemáticos proporcionados por el fabricante de la tarjeta, se modeló esta parte del circuito utilizando el entorno TINA de *Texas Instruments* de tipo SPICE, con el objetivo de realizar varias simulaciones necesarias para la modificación de una serie de funciones que se irán explicando más adelante.

---

[27] En el esquemático de la Figura 161 no está incluída la normalización de las señales de posición respecto a la suma de las cuatro entradas del detector. Si bien para las simulaciones realizadas esta normalización no fue necesaria, sí está incluída en la electrónica final utilizada en la corrección, para lo cual se emplea el multiplicador analógico AD633AR, usado en configuración de divisor.

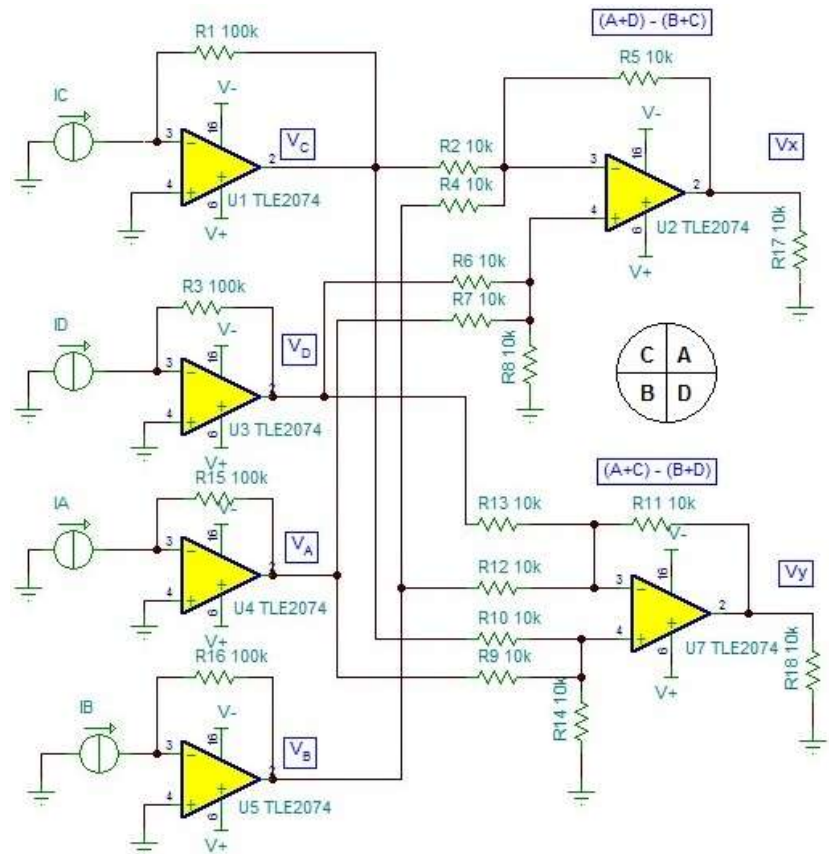

**Figura 161. Electrónica de acondicionamiento y cálculo de la posición del QD.**

Además de las funciones básicas descritas, la tarjeta de *Optics In Motion* tiene algunas características adicionales que serán de utilidad para implementar el sistema de corrección. La más importante es una etapa de amplificación/atenuación, llevada a cabo tras las operaciones para la obtención de la posición {x, y}, que proporciona una ganancia variable a la señales de posición del QD. Esta etapa será importante en la calibración del sistema de corrección para adaptar el rango de la señal de salida del QD al rango de la señal de control del FSM. Ambos rangos deben ser similares en amplitud pero con respuestas inversas (la polaridad de la señal del QD se controla mediante un *jumper*) para realizar correctamente la corrección. En último término, la corrección la realizará el control PID, por lo que no es necesario que ambos rangos sean idénticos, pero la amplificación/ganancia permitirá conseguir la adaptación necesaria de ambos rangos que permita al control PID reducir la diferencia entre la posición del centroide y el centro del QD.

En la Figura 162 se muestra la simulación en TINA del módulo de la respuesta en frecuencia correspondiente a la etapa de preamplificación de las señales de corriente del QD por ser la etapa que limita el ancho de banda del sistema. Como se vio en el apartado anterior, el QD proporciona un ancho de banda a 3 dB de 14 MHz. Sin embargo, se puede comprobar cómo la electrónica de acondicionamiento reduce el ancho de banda efectivo hasta los 15,83 kHz, si bien se comprobará más adelante que sigue quedando muy por encima de las frecuencias de variación de la turbulencia atmosférica que se desea monitorizar.

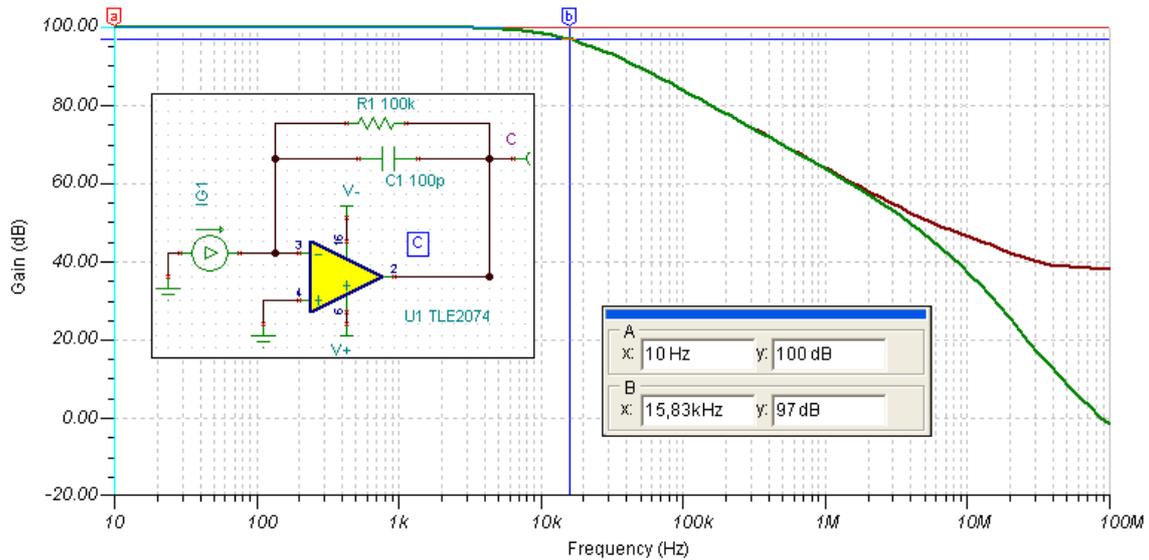

**Figura 162. Ancho de banda a 3 dB de la etapa de acondicionamiento de las señales del QD.**

# 3.9. EFECTOS CROMÁTICOS EN LA CORRECCIÓN

### 3.9.1. Dependencia del *beam wander* con la longitud de onda

En el apartado 3.7.4 se propusieron distintos esquemas para llevar a cabo la corrección del *beam wander*. Ambos esquemas compartían la misma estrategia consistente en analizar las variaciones en la posición del centroide de un haz a una longitud de onda (1550 nm) y utilizar esta información para realizar la corrección de otro haz a una longitud de onda diferente (850 nm). Se puede deducir que cualquier variación en el *beam wander* sufrido por cada haz, es decir, cualquier dependencia con la longitud de onda, se traducirá en un error en la corrección. Por ello, es importante estudiar si tal dependencia existe y cuantificarla en caso positivo.

Por una parte, es común encontrar una dependencia de la longitud de onda en las ecuaciones que caracterizan la turbulencia atmosférica. Sin embargo, cuando se analiza el caso del *beam wander* se comprueba que la dependencia desaparece en el resultado neto. Por ejemplo, si se introduce la expresión del parámetro de Fried, vista en la ecuación (3-16), en la ecuación (3-18), relativa a la varianza de la posición instantánea del spot de corto plazo y se estudia únicamente la influencia de la longitud de onda, agrupando el resto de factores en una constante cte, se obtiene que el *beam wander* tiene un comportamiento estrictamente acromático al desaparecer dicha dependencia, como se puede observar en la ecuación (3-37).

$$\sqrt{\langle r_c^2 \rangle} = 0,69L\left(\frac{\lambda}{2\omega_0}\right)\left(\frac{2\omega_0}{\left(0,16C_n^2k^2L\right)^{-3/5}}\right)^{5/6} \propto \mathrm{cte} \cdot \lambda \left(\frac{1}{\left(\frac{1}{\lambda^2}\right)^{-3/5}}\right)^{5/6} \propto \mathrm{cte} \quad (3\text{-}37)$$

En [273] se realizó un estudio sobre la relación entre un haz de 850 nm y otro de 1550 nm bajo turbulencia atmosférica, mostrando una buena correlación (cercana a 1) entre ambas señales, si bien se obtenían resultados contradictorios de menor correlación a menores distancias de propagación, lo que podría explicarse por desalineamientos entre los dos haces que pudieran hacer diferir ligeramente las trayectorias atmosféricas recorridas. Sin embargo, como suele ser habitual en este tipo de estudios, solo se analiza la intensidad total recibida como resultado de todos los efectos de la turbulencia, y no específicamente el *beam wander*.

Por otra parte, se sabe [96, pp. 22-17] que el coeficiente de atenuación debido al *scattering* atmosférico depende de la longitud de onda, al ser sensible a la relación entre el tamaño de las partículas y la longitud de onda. Cabría esperar que pese a depender de la longitud de onda, una normalización en la medida de la intensidad de cada haz ocultara la dependencia. Sin embargo, como se explica en [277, p. 8], donde se hizo un estudio sobre la atenuación por *scattering* a 850 nm y a 1550 nm, se constató que experimentalmente puede comprobarse que las condiciones climáticas pueden modular esta dependencia de la longitud de onda sobre la atenuación provocada por *scattering* e incluso llegar a eliminarla, por ejemplo con niebla. Por ello, los estudios experimentales habituales que analizan la intensidad de las señales a distintas longitudes de onda no son completamente válidos para extraer conclusiones sobre el *beam wander*.

## 3.9.2. Montaje para medir el *beam wander* a 850 nm y 1550 nm

Por las razones expuestas en el apartado anterior, se decidió llevar a cabo un experimento de laboratorio para medir si el efecto de *beam wander* de la turbulencia dependía de la longitud de onda. Para ello, se utilizó el montaje mostrado en la Figura 163, compuesto por dos fuentes láser, una a 1550 nm y otra a 850 nm, colimadas en haces de idéntico diámetro, igual a 7 mm, y combinados y alineados en el mismo eje óptico mediante un divisor de haz 50/50 (la pérdida de potencia es irrelevante en este experimento). En otra mesa óptica situada a una distancia de 5 metros se situó la parte del receptor compuesta por un espejo dicroico para separar las dos longitudes de onda hacia dos caminos diferentes, en cada uno de los cuales se utiliza una lente similar (una misma longitud focal adaptada a la longitud de onda de cada camino) para focalizar el haz en un detector de posición. En el caso del camino de 1550 nm se utilizó un detector de InGaAs de cuatro cuadrantes de 3 mm de diámetro (el IGA-030-QD, descrito en el apartado 3.8.6) y en el caso del camino de 850 nm se utilizó un detector de Silicio de 10 mm de tipo efecto lateral (el C10443-02, descrito en el apartado 3.8.6).

El detector de InGaAs fue seleccionado para medir la posición del canal de seguimiento del sistema final, por lo que se eligió con una alta sensibilidad. Debido a esto, parte de la potencia de 850 nm se acoplaba a este detector incluso después de separarla con el espejo dicroico. Por ello, se situó un polarizador lineal antes del detector de InGaAs y otro alineado a 90° justo después del colimador del haz de 850 nm con el objetivo de anular esta componente. Antes del espejo dicroico, se hace reflejar a los haces en un FSM, empleado en la calibración previa. Por último, para simular la turbulencia, se utilizó inmediatamente después de la combinación de los haces, un calentador de aire orientado hacia el camino óptico. Esta es una práctica habitual y aceptada [278, p. 5] para simular el efecto que tendría la turbulencia atmosférica en un camino óptico mucho más largo.

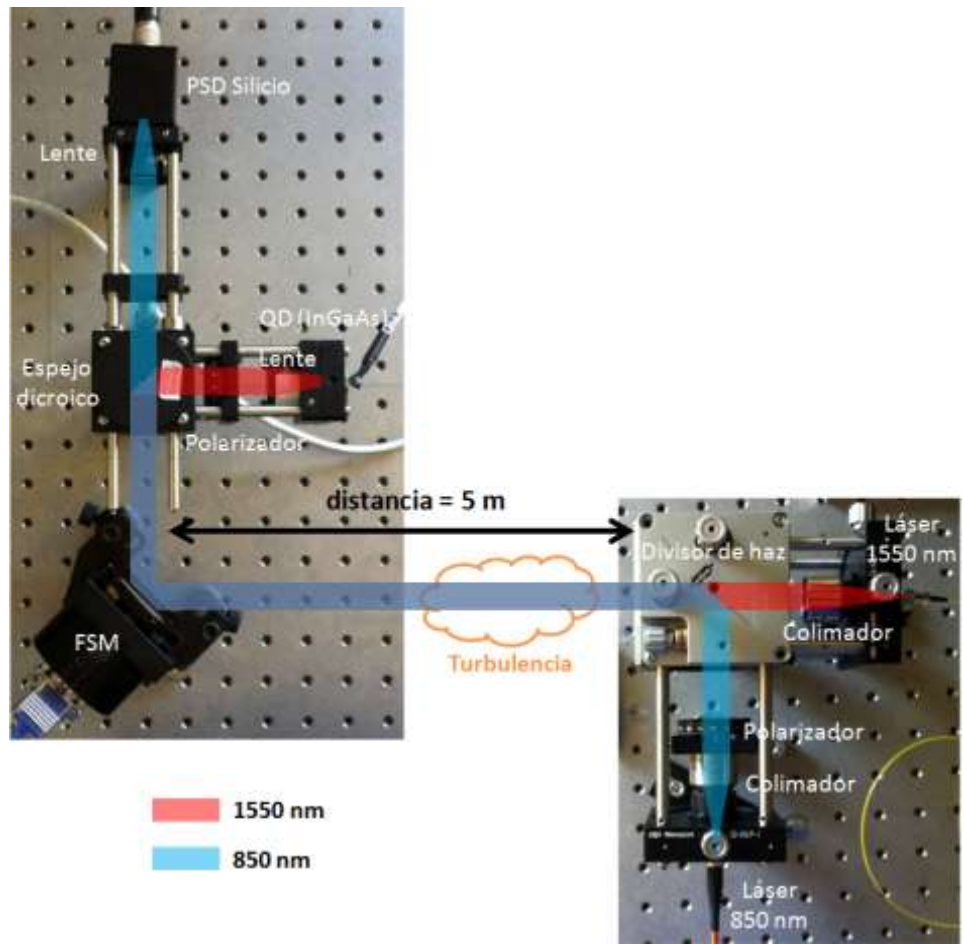

Figura 163. Montaje de laboratorio para medir el *beam wander* a 850 nm y a 1550 nm.

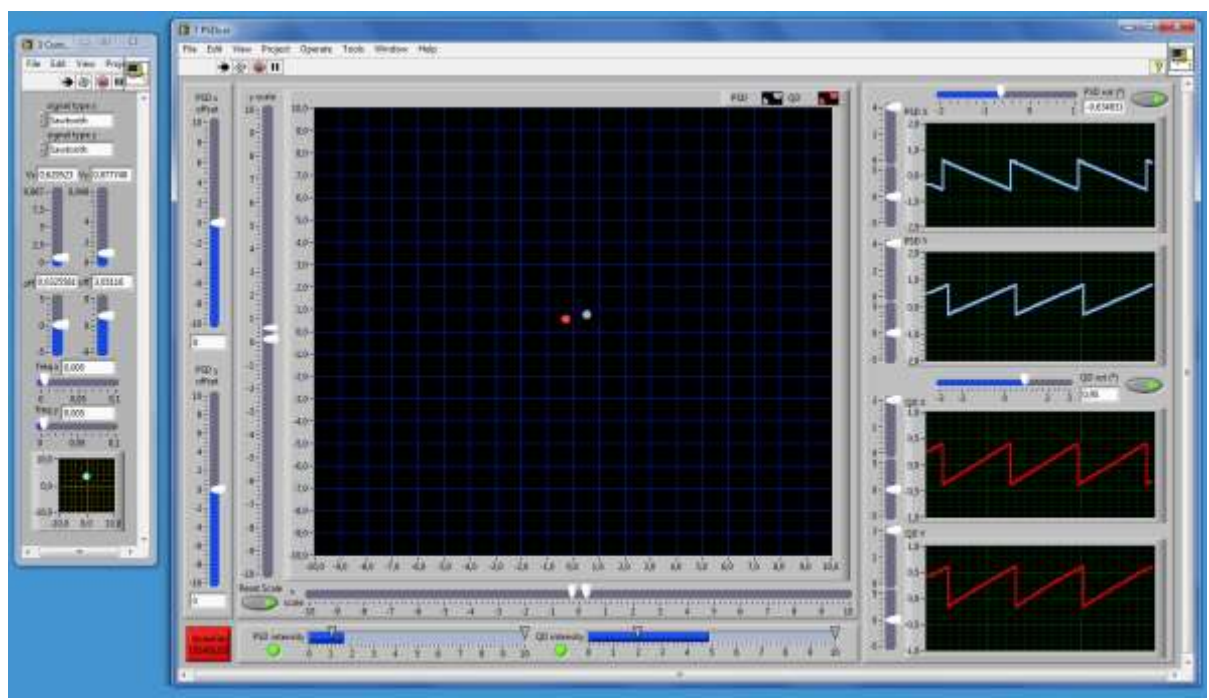

Figura 164. Aplicación programada en *LabView* para la calibración del montaje.

La calibración de este sistema consiste en conseguir detectar el mismo movimiento en cada detector, usando para ello el espejo modulable como patrón de calibración. Para ello, se programó una aplicación en *LabView* consistente en la generación de movimientos programables (forma de onda, amplitud, offset y frecuencia) sobre el espejo (Figura 164, izquierda) y la representación de las señales (para cada detector: intensidad, amplitud del eje x y amplitud del eje y) de los detectores (Figura 164, derecha). En la captura de pantalla, se comprueba cómo en la ventana de la izquierda se está generando una señal de diente de sierra con la que mover el espejo a idéntica frecuencia en ambos ejes y cómo en la ventana de la derecha se está registrando el mismo movimiento en ambos detectores (a excepción de que el eje x del camino de 1550 nm está invertido respecto al de 850 nm, debido a la reflexión en el espejo dicroico). Una vez alcanzado este estado, se puede considerar que el sistema está calibrado y es posible empezar a registrar los movimientos de los haces debidos a la turbulencia.

Dos aspectos fundamentales para medir el *beam wander* de forma correcta utilizando dos longitudes de onda son la colimación y el alineamiento de los haces. Si ambos haces no están perfectamente alineados, no estarán recorriendo el mismo camino óptico y por lo tanto se verán afectados por fenómenos turbulentos diferentes. Por otra parte, si la colimación de ambos haces no es idéntica, cada haz se verá afectado de forma diferente y se registrarán distintos comportamientos al final de la propagación. Ambos ajustes se comprobaron críticos en la calibración del sistema, y cualquier desajuste de estos provoca comportamientos totalmente distintos en ambos haces, que pueden fácilmente dar lugar a equívocos porque no tienen ninguna relación con comportamientos cromáticos.

La misma importancia que tienen en este experimento los dos factores mencionados en el párrafo anterior, la tiene en la implementación del sistema de corrección, por lo que es necesario atender a estos parámetros de forma muy cuidadosa en la calibración del sistema final. Para realizar un correcto alineamiento, se utilizaron dos láseres de 650 nm en ambos caminos (de 850 nm y 1550 nm), de forma que inicialmente fuera posible observar los haces, ya que en los experimentos con turbulencia ninguno de los haces es visible a simple vista. Para la colimación, se utilizaron sendos colimadores de idéntico haz de salida, diseñados para cada longitud de onda, con el objetivo de que no hubiera ninguna diferencia entre ambos.

### 3.9.3. Resultados de *beam wander* a 850 nm y 1550 nm

Para interpretar los resultados obtenidos en el experimento descrito, es importante considerar el ancho de banda de cada sistema de detección. La diferencia en ancho de banda de los detectores empleados es muy elevada, de unos tres órdenes de magnitud (16 kHz en el caso del detector de 850 nm y 13,6 MHz en el caso del detector de 1550 nm, ya que el tiempo de subida $t_s$ es menor a 25 ns y el ancho de banda se calcula como $BW = 0,35/t_s$ [276, pp. 3-26]). Sin embargo, el ancho de banda final viene determinado también por la electrónica posterior de adquisición y amplificación. Según se calculó en el apartado 3.8.7, el ancho de banda de la electrónica del detector de 1550 nm es de 15,83 kHz, muy por debajo del ancho de banda del detector pero por encima de cualquier fenómeno turbulento previsto. En el caso del detector de 850 nm no es posible calcular el ancho de banda de la electrónica porque no se dispone de un modelo del circuito. Por ello, se decidió medir directamente la respuesta en frecuencia de ambos circuitos.

En la Figura 165 se muestra el módulo de la transformada rápida de Fourier (FFT), calculado a partir de las señales obtenidas por los detectores de 850 nm y de 1550 nm, al observar la misma turbulencia. Sabiendo que el sistema detector de 1550 nm está captando todas las componentes espectrales de la turbulencia (ya que todas las componentes están muy por debajo de su ancho de banda nominal), se puede comprobar que el de 850 nm está recortando una parte importante si se le compara con el de 1550 nm. Además, para descartar que la diferencia en la respuesta en frecuencia sea debida a algún comportamiento cromático de la turbulencia, se realizó el mismo experimento utilizando una sola longitud de onda registrada en ambos detectores. Para ello, se eliminaron los polarizadores, con lo que el detector de 1550 nm pudo detectar también la señal de 850 nm (por la razón explicada anteriormente). Tras calcular el espectro de ambas señales, se obtuvo exactamente el mismo comportamiento en frecuencia en ambos sistemas detectores que el observado con dos longitudes de onda, por lo que se puede concluir que la diferencia se debe a la electrónica de acondicionamiento del detector de 850 nm. Conviene señalar que este detector se pretende utilizar únicamente para monitorizar, y que es el de 1550 nm el destinado a la corrección, por lo que esta limitación de ancho de banda no se dará en el sistema de compensación de turbulencia.

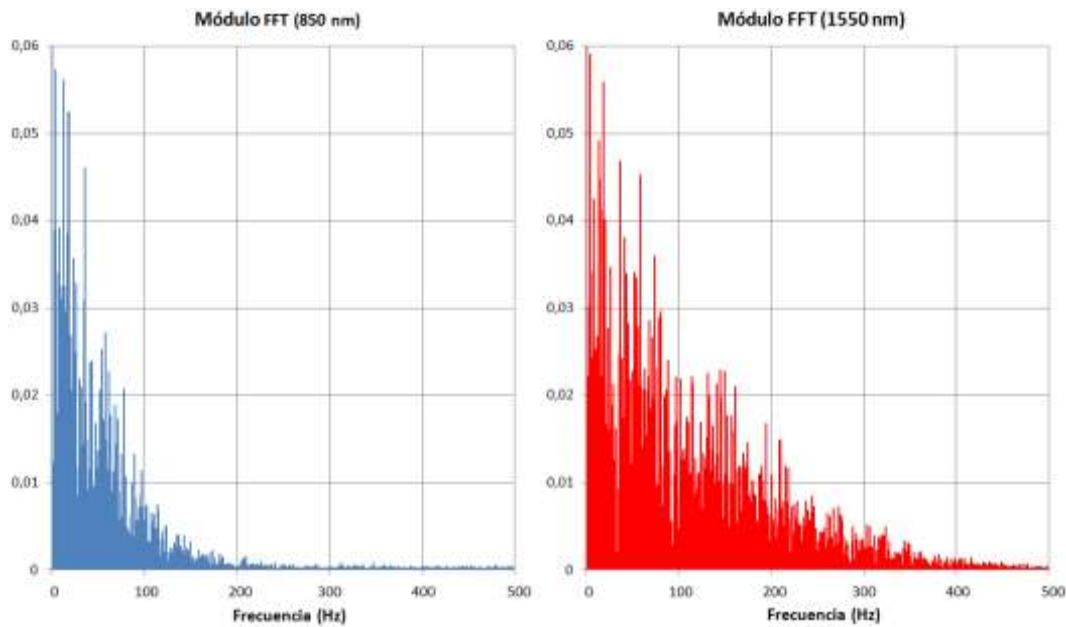

**Figura 165. Módulo de la transformada rápida de Fourier (FFT) de la señal del detector de 850 nm (izquierda) y del detector de 1550 nm (derecha) observando la misma turbulencia.**

En la Figura 166 se muestra la respuesta del detector de 850 nm y de 1550 nm de la posición instantánea del centroide (en uno de los dos ejes) de un solo haz sometido a la turbulencia descrita anteriormente, durante un intervalo de 50 ms. En la Figura 167 se muestra durante el mismo tiempo la respuesta de cada detector a cada uno de los dos haces de sus correspondientes longitudes de onda tras separarlos mediante el espejo dicroico. En el último caso ambos haces están alineados, igualmente colimados y sometidos a la misma turbulencia. Se puede comprobar que se obtiene una respuesta muy similar en ambos casos: el movimiento del centroide es el mismo en ambos detectores salvo en los cambios rápidos (altas frecuencias). Al observar la baja frecuencia (que es la única que es posible registrar de igual forma con ambos detectores) se comprueba que el movimiento del centroide es el mismo en ambos detectores, ya se use una misma longitud de onda o dos distintas provenientes de dos haces diferentes, y es solo cuando se producen cambios muy rápidos que el detector de 850 nm no es capaz de seguirlos.

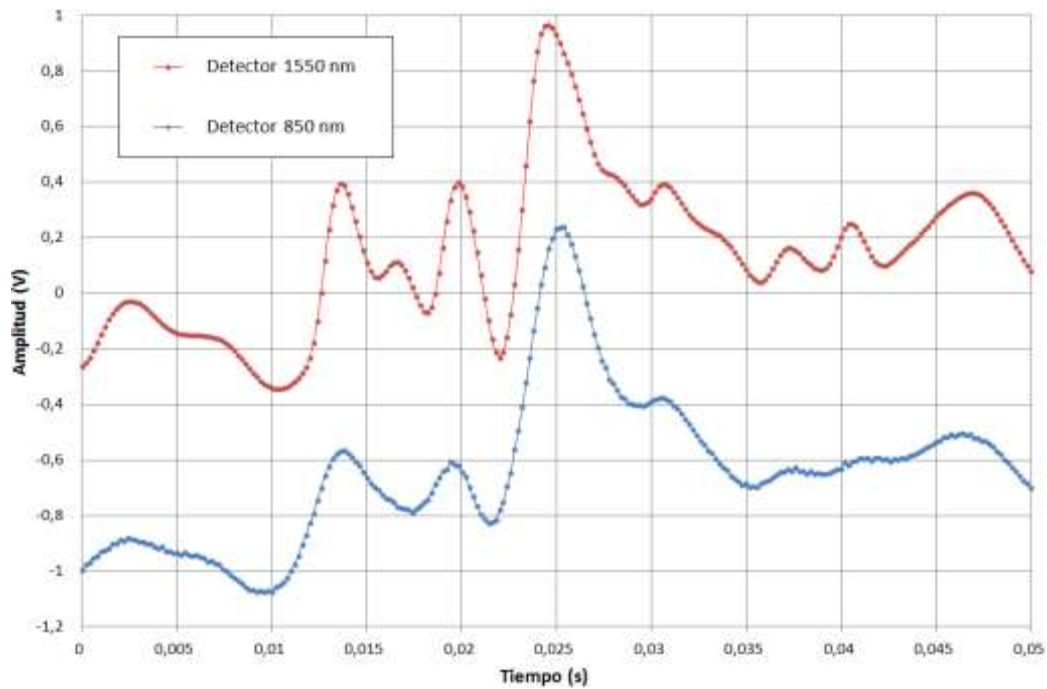

**Figura 166. Respuesta del detector de 850 nm (azul) y de 1550 nm (rojo) de posición instantánea del centroide de un solo haz de 850 nm sometido a la turbulencia.**

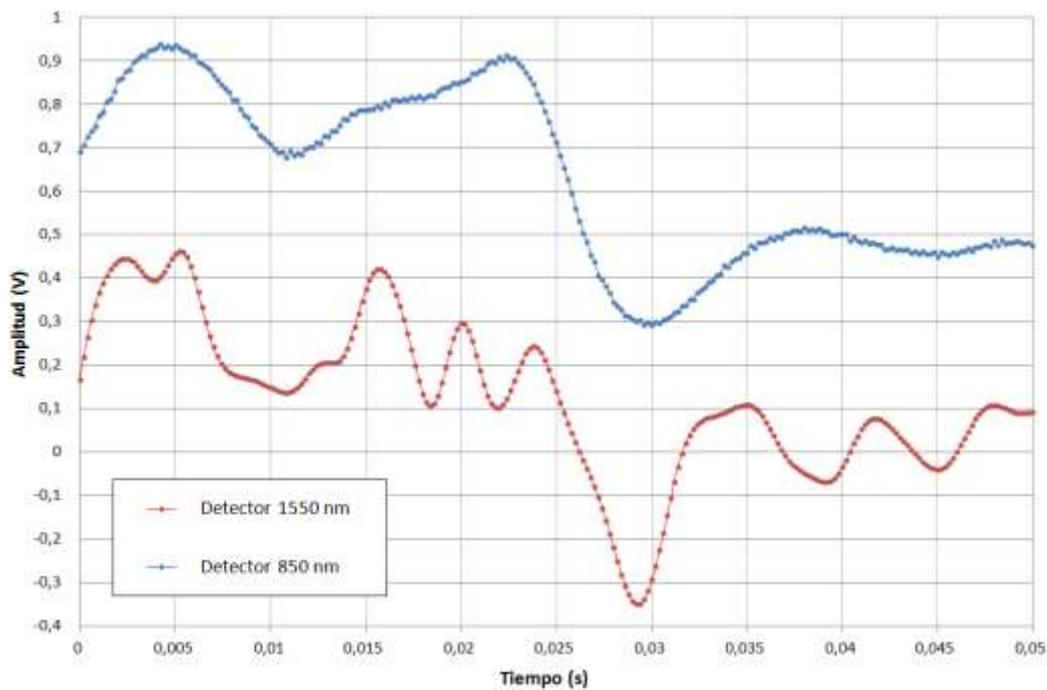

**Figura 167. Respuesta del detector de 850 nm (azul) y de 1550 nm (rojo) de posición instantánea del centroide de dos haces alineados de 850 nm y de 1550 nm sometidos a la misma turbulencia.**

Para cuantificar el parecido entre las señales de ambos detectores con el objetivo de estudiar la correlación entre las dos longitudes de onda frente a la turbulencia, se ha calculado el coeficiente de correlación de *Pearson* entre ambas señales. El coeficiente de *Pearson* [279, p. 292] ofrece la relación lineal entre dos variables aleatorias, proporcionando un resultado similar al cálculo de la covarianza pero independiente de la escala de medida de las variables, lo que simplifica el tratamiento de los datos, que al venir de diferentes detectores tienen distinta escala.

Para intentar ponderar más las bajas frecuencias que las altas al calcular la correlación, se ha aplicado una media móvil de 20 términos sobre las series de datos de ambos detectores, lo que equivale a realizar un filtrado paso bajo sobre las señales. El resultado de la aplicación de este filtrado se muestra en la Figura 168 para el caso de un mismo haz de 850 nm registrado por ambos detectores y en la Figura 169 para el caso de un haz de 850 nm y otro de 1550 nm, cada uno detectado en su correspondiente detector.

El coeficiente de correlación en el caso de utilizar la misma longitud de onda es del 98 % y en el caso de dos longitudes de onda distintas del 86 %. Esta diferencia se relacionó con la dificultad de calibrar perfectamente el sistema cuando se utilizan dos longitudes de onda en lugar de la misma ya que en el resultado de la correlación influían pequeños ajustes sobre cada camino óptico. Por ello, una correlación tan alta como 86 % entre dos longitudes de onda tan diferentes como 850 nm y 1550 nm permite concluir que el efecto real del *beam wander* será similar sobre ambas.

Como se dijo, una estrategia común para simular la turbulencia atmosférica en el laboratorio es utilizar calentadores de aire orientados hacia el haz de salida del experimento. Si bien cualitativamente el efecto de esta turbulencia es el mismo que el que se esperaría en un experimento de propagación atmosférica, esto no es cierto en términos cuantitativos. En cuanto a la amplitud, este tipo de modelado de la turbulencia ofrece el equivalente a un camino mucho más largo que el del experimento, permitiendo registrar el efecto del *beam wander* en un camino óptico tan corto como el de los experimentos de laboratorio descritos. Esto pudo comprobarse midiendo la amplitud media de la turbulencia artificial en laboratorio comparada con la medida por el mismo sistema en una propagación natural atmosférica de unos 30 metros. En el último caso, la amplitud registrada por los detectores de posición resultó ser de más del doble, comprobándose el efecto amplificador que tiene la distancia sobre el *beam wander*.

Si bien la amplitud de la turbulencia natural es una variable que depende de la distancia, no ocurre lo mismo con su espectro. En cuanto al comportamiento en frecuencia, realizando el experimento anterior pudo comprobarse una gran discrepancia: en la Figura 170 se muestra una comparación entre el módulo de la FFT de la turbulencia artificial y la natural. En ambos casos se utilizó el detector de 1550 nm con el objetivo de registrar todas las componentes espectrales de la turbulencia sin ningún recorte (el ancho de banda de este sistema es de unos 16 kHz). Se puede comprobar cómo el espectro de la turbulencia artificial es más extenso que el de la turbulencia natural, cuyas componentes desaparecen casi totalmente por encima de 200 Hz, quedando la gran mayoría por debajo de 100 Hz. Este resultado coincide con medidas de turbulencia en caminos mucho más largos [268, p. 365], lo que sugiere que el FSM seleccionado en el apartado 3.8.1 (con un ancho de banda nominal a 3 dB de 400 Hz) es suficiente para realizar la corrección del *beam wander* en un entorno de turbulencia atmosférica real.

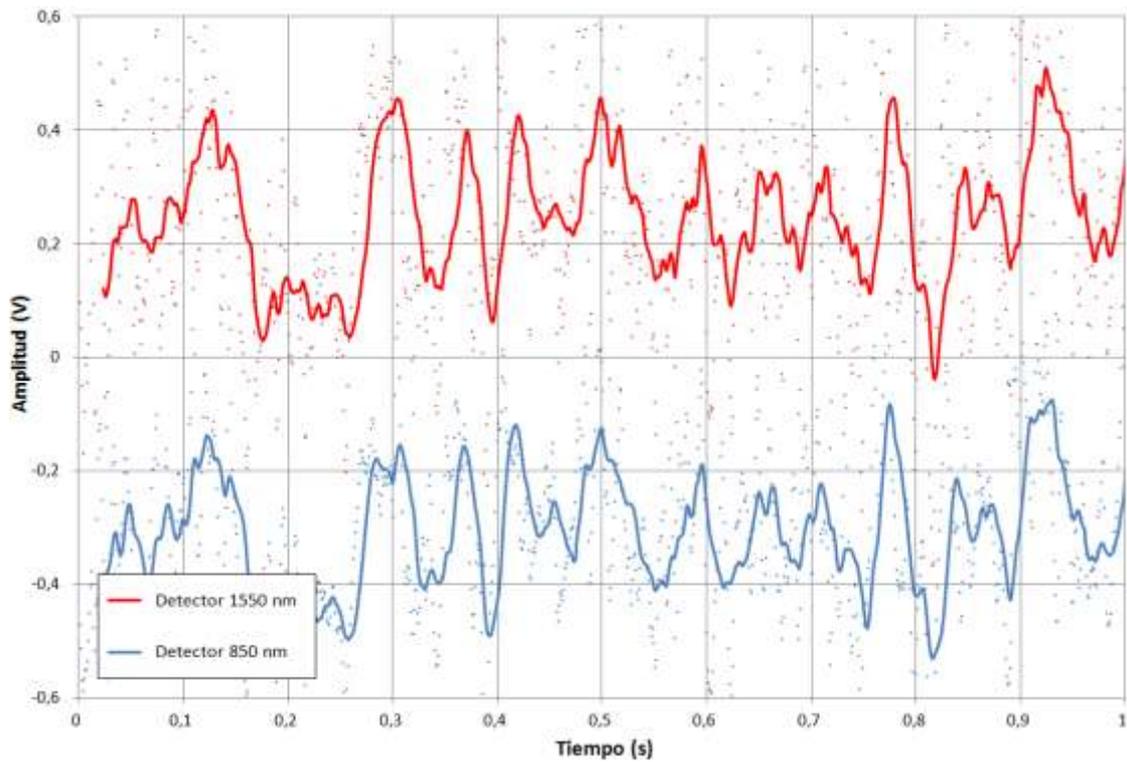

**Figura 168. Respuesta del detector de 850 nm (azul) y de 1550 nm (rojo) de una media móvil de 20 pasos a la posición instantánea del centroide de un solo haz de 850 nm sometido a una turbulencia artificial.**

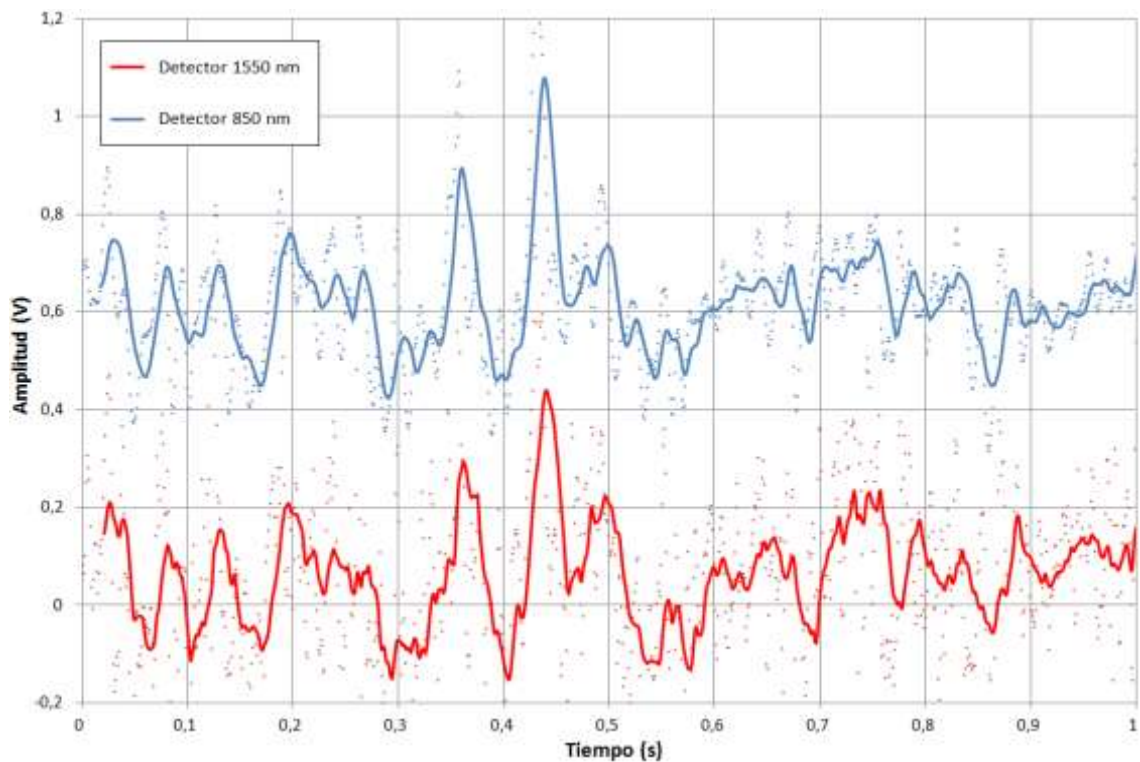

**Figura 169. Respuesta del detector de 850 nm (azul) y de 1550 nm (rojo) de una media móvil de 20 pasos a la posición instantánea del centroide de dos haces alineados de 850 nm y de 1550 nm sometidos a la misma turbulencia.**

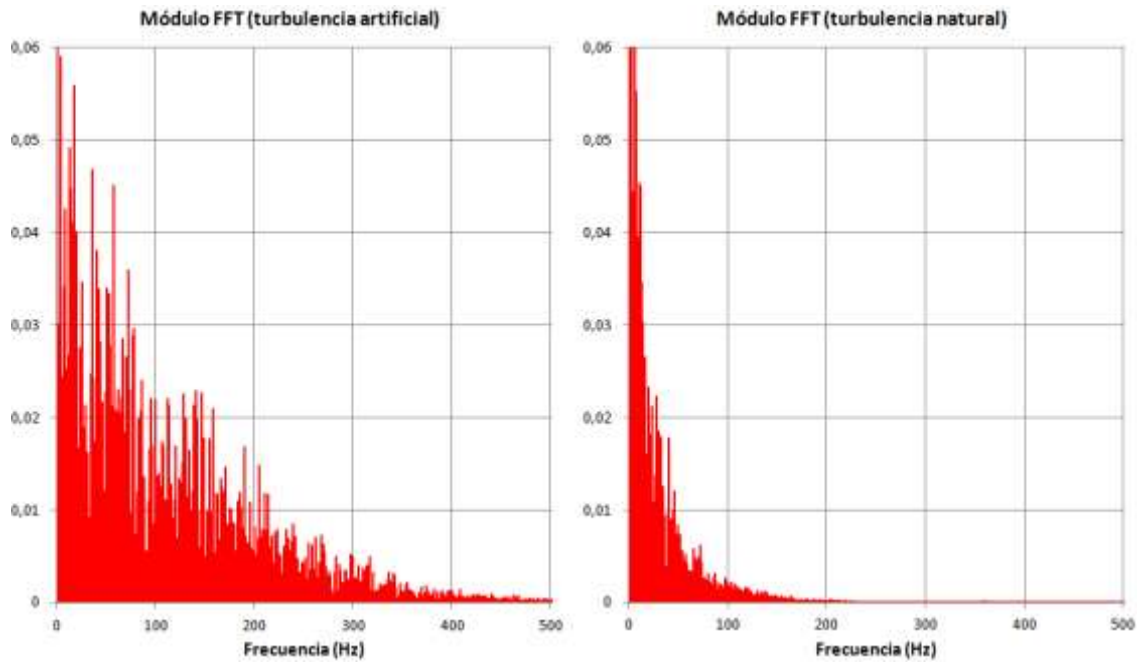

**Figura 170. Módulo de la transformada rápida de Fourier (FFT) de la señal del detector de 1550 nm observando una turbulencia artificial (izquierda) producida por un calentador de aire y una natural (derecha) producida en un camino de propagación de unos 30 metros.**

# 3.10. TELESCOPIO RECEPTOR EN BOB

### 3.10.1. Descripción del telescopio utilizado

El telescopio receptor usado en Bob para recoger la señal transmitida desde Alice es el Meade LX200-ACF (Figura 171) de 10 pulgadas (25,4 cm). Se trata de un telescopio tipo *Schmidt–Cassegrain*, consistente en un espejo primario esférico de 263,5 mm (de mayor diámetro al nominal para proporcionar las 10 pulgadas de apertura en todo el campo de visión), un espejo secundario hiperbólico y una placa correctora asférica tipo *Schmidt* frente al primario [280].

El diseño *Schmidt–Cassegrain* es un compromiso entre el *Cassegrain* simple y el *Ritchey–Chrétien*. Si bien el primero consiste en un espejo primario parabólico y un secundario hiperbólico, consiguiendo eliminar la aberración esférica aunque presentando aberración de coma, el segundo está formado por dos espejos hiperbólicos y es capaz de obtener imágenes libres de aberración sobre un amplio campo de visión. El inconveniente de ambos es la complejidad de fabricación de grandes aperturas parabólicas o hiperbólicas. Para solucionarlo, el Meade LX200-ACF utiliza un espejo primario esférico, de más simple construcción, y una placa asférica correctora de la aberración esférica de dicho primario. Este conjunto logra una calidad óptica similar a la de un primario hiperbólico desde el punto de vista del secundario, que sí es realmente hiperbólico.

En un sistema óptico como el diseñado en este trabajo, en el que se busca reducir el campo de visión al mínimo, las buenas prestaciones de este telescopio en cuanto a aberración de coma son poco importantes, ya que se espera una extensión angular muy reducida, si bien ofrecerá una gran inmunidad frente a pequeños desalineamientos con el terminal transmisor. No obstante, el uso de un telescopio de gran calidad óptica está justificado por el hecho de que al buscar la reducción del campo de visión, se pretende obtener el mínimo spot posible, para lo cual es fundamental la inmunidad ante aberraciones, especialmente esférica, que pudieran aumentar su extensión.

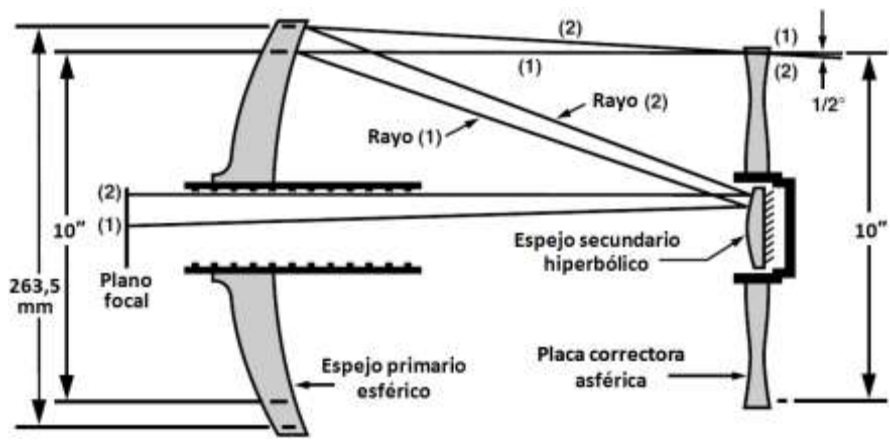

**Figura 171. Diagrama del telescopio Meade LX200-ACF de 10".**

## 3.10.2. Capacidad de enfoque del telescopio

Un inconveniente a la hora de reutilizar telescopios astronómicos en sistemas de comunicaciones puede ser la necesidad de recolimar el haz recibido para llevarlo al montaje óptico, para lo cual se ha de emplear óptica adicional que puede introducir pérdidas y aberraciones. En el caso del Meade LX200-ACF, es posible establecer la posición del foco a una distancia arbitraria, dentro de un rango amplio. Esto es posible mediante la modificación de la distancia entre el espejo secundario y el primario, al permitir trasladar este último. Así, al acercar el primario al secundario (que es fijo), el foco se alejaría del telescopio, obteniendo un aumento efectivo de la distancia focal total.

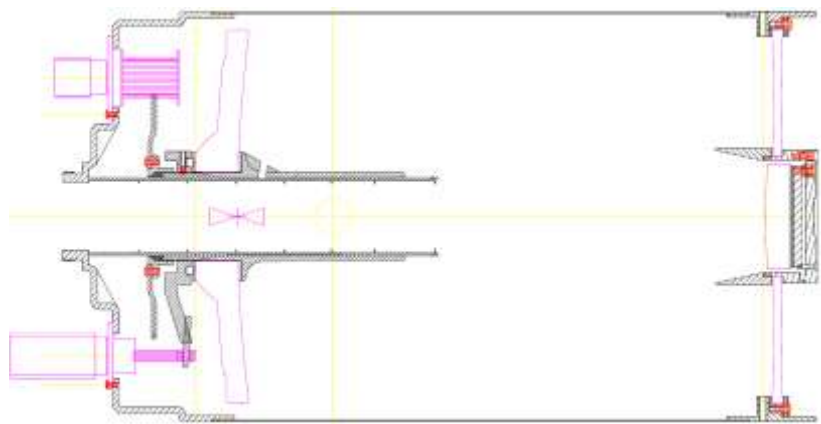

**Figura 172. Vista horizontal del telescopio Meade LX200-ACF de 10" [281].**

El enfoque variable se proporciona con la finalidad de enfocar objetos cercanos (tanto como 7,62 metros según especificaciones, lo que correspondería a una focal de 1600 mm [280]), además de al infinito (que se correspondería con la focal nominal del telescopio de 2500 mm). Sin embargo, como se explicará a continuación, se pudo comprobar que este telescopio permite extender la focal aún más, lo que será una ventaja para el diseño del receptor. En la Figura 172 se puede observar el funcionamiento de dicha propiedad de enfoque: al ser girado el mando inferior, es posible desplazar el espejo primario, actuando el mando superior a modo de bloqueo de dicho desplazamiento.

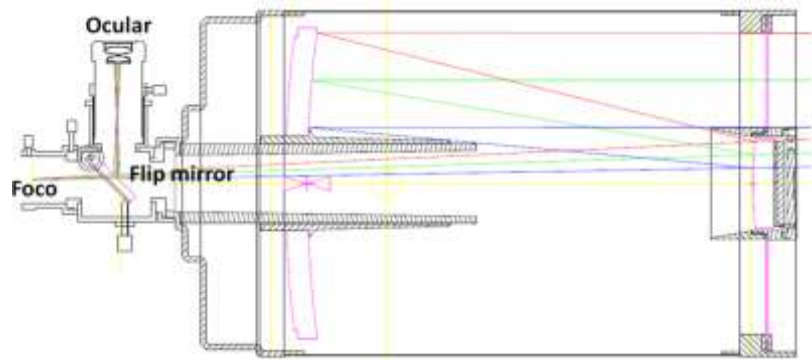

**Figura 173. Caminos ópticos seleccionables mediante el *flip mirror*.**

Para caracterizar la capacidad de enfoque del telescopio, se utilizó la Luna a modo de objeto en el infinito para determinar la posición del plano focal en el que la distancia focal es 2500 mm. Para ello, modificando la distancia del espejo primario, se enfocó visualmente la Luna en el ocular del telescopio. Esto significa que el foco del telescopio, es decir, la imagen de la Luna, estará a una distancia del ocular igual a su focal en dirección hacia el telescopio. Después se desvió la imagen mediante un *flip mirror* (un espejo de dos posiciones separadas por 45°) hacia el eje óptico del telescopio (Figura 173). Al enfocar el infinito utilizando el ocular se puede asegurar que el telescopio está ajustado para que su focal sea igual a 2500 mm. Por lo tanto, para localizar el plano focal (coincidente con el plano imagen por la gran distancia hacia el objeto), un método efectivo consiste en utilizar una pantalla blanca a lo largo del eje óptico y proyectar la imagen de la Luna sobre la misma. Para cada distancia primario-secundario, al modificar la longitud focal, el foco estará en el lugar donde la Luna aparezca nítidamente sobre dicha pantalla y así será muy fácil determinar su posición exacta. Mediante este procedimiento, se pudo establecer la distancia entre el plano focal del telescopio y el espejo secundario, que será utilizada en los siguientes apartados.

## 3.10.3. Simulación del telescopio receptor

Para llevar a cabo el diseño óptico de Bob es importante disponer de un modelo realista del telescopio receptor utilizado en la práctica. Con una simulación fiel de dicho telescopio es posible simplificar en gran medida la tarea de diseño. Además, se verá que disponer de una simulación completa del sistema óptico, permite incluir el efecto del *beam wander*, así como caracterizar el sistema de corrección directamente en funcionamiento sobre dicho modelo. Uno de los objetivos del diseño óptico de Bob será optimizar el tamaño y posición del spot en el foco del sistema completo, para reducir al mínimo el campo de visión que verán las fibras ópticas acopladas a los contadores de fotones del canal cuántico. Para ello, es necesario realizar una correcta correspondencia entre el telescopio real y el simulado.

Meade proporciona muy escasa información respecto a las características técnicas de la óptica del telescopio. Más allá de la descrita en el apartado 3.10.1, sabiendo que se trata de un diseño *Schmidt–Cassegrain* con un primario esférico de 263,5 mm de diámetro y un secundario hiperbólico, para poder modelarlo adecuadamente es necesario conocer los radios de curvatura de ambos espejos, así como la asfericidad de la placa correctora y las distancias relativas. En general, las longitudes focales de cada uno de los espejos de un telescopio de dos elementos formado por un reflector primario cóncavo y un secundario convexo, responden a la ecuación (3-38), siendo c la curvatura de cada espejo, r su radio de curvatura y f su longitud focal. Adoptando la convención de que ambos radios de curvatura sean negativos, entonces $f_1$ será positivo y $f_2$ negativo.

$$f_1 = -\frac{r_1}{2} = -\frac{1}{2c_1} \quad ; \quad f_2 = \frac{r_2}{2} = \frac{1}{2c_2} \qquad (3\text{-}38)$$

Las distancias focales $f_1$ y $f_2$ están relacionadas entre sí mediante la longitud focal equivalente F del telescopio y la distancia L entre los vértices de ambos espejos según la ecuación (3-39) [282].

$$F = \frac{f_1 \cdot f_2}{f_1 + f_2 - L} \qquad (3\text{-}39)$$

En un telescopio de varios elementos separados entre sí, la longitud focal equivalente o EFL (del inglés *Effective Focal Length*) es la referida a una lente delgada que tuviera la misma apertura que su primario y que produzca un haz refractado convergente con el mismo ángulo que el producido por el sistema de varios elementos. El foco equivalente de dicha longitud focal resultante se suele denominar foco *Cassegrain*. En la Figura 174 se puede observar que la focal equivalente de un *Schmidt-Cassegrain* es mayor que las focales individuales [28] y que su suma, lo que en relación al tamaño del telescopio proporciona una gran distancia focal (típicamente entre unas tres y cuatro veces más). A continuación se describen los pasos que se siguieron para conseguir la simulación del *Schmidt-Cassegrain*.

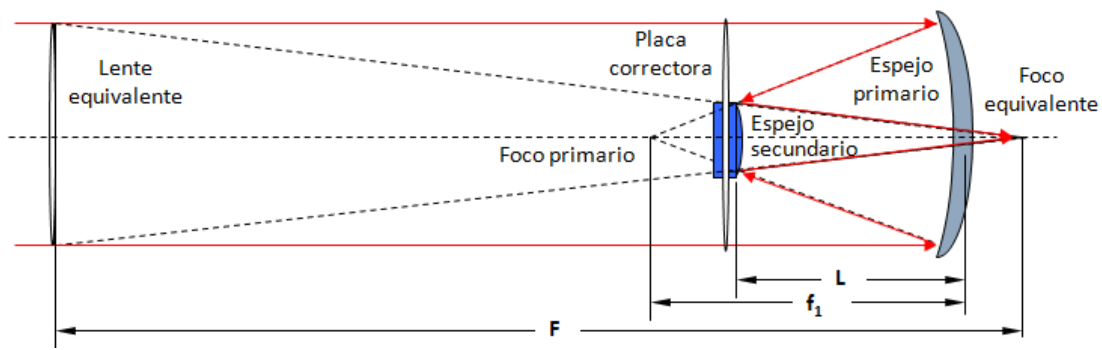

**Figura 174. Longitud focal equivalente F en un telescopio *Schmidt-Cassegrain* formado por una placa correctora, un espejo primario de focal $f_1$ y un secundario de focal $f_2$, separados una distancia L.**

## 3.10.4. Diseño de un telescopio *Cassegrain*

Al carecer de las especificaciones ópticas del Meade LX200 con las que poder modelarlo directamente en la simulación, se llevó a cabo un proceso de diseño para obtenerlas a partir de los datos de que se disponía. Se decidió modelar inicialmente un telescopio *Cassegrain*, debido a que su diseño es más directo y está bien caracterizado, para posteriormente realizar un ajuste sobre los datos que se desconocen y obtener el telescopio *Schmidt-*

---

[28] Nótese que la focal $f_2$ no es igual a la distancia entre el vértice del espejo secundario y el foco equivalente. Será una distancia mayor a esta, ya que al secundario le llega un haz enfocado, no uno colimado. Esta focal no se ha dibujado en la figura para no perjudicar la legibilidad de la misma.

*Cassegrain*, más aproximado al telescopio real. Este proceso se describirá en los siguientes párrafos.

El primer paso para diseñar un telescopio *Cassegmin* es definir las características ópticas del espejo primario. Un valor imprescindible para definirlo es el radio de curvatura. Pese a que Meade no proporciona este dato de forma oficial, fue posible obtener la relación focal del reflector primario por comunicación personal de dos de sus distribuidores de forma separada, coincidiendo dicho valor en $f_1/1,95$. Dado que la relación focal expresa la distancia focal respecto a la apertura y se conoce que el diámetro es 263,5 mm, se puede concluir que la longitud focal del primario es $f_1 = 513,825$ mm, y mediante la ecuación (3-38), se obtiene el radio de curvatura $r_1 = 1027,65$ mm.

Dado que el espejo primario de un *Cassegrain* tiene un perfil parabólico, quedará completamente definido por su radio de curvatura y su apertura. Una vez modelado dicho reflector, es necesario diseñar el secundario. El diseño de este espejo ya depende de la distancia relativa al primario, que viene determinada por L. Otras distancias importantes (Figura 175) que se usarán en el diseño son la distancia S, definida desde el foco *Cassegmin* hasta el vértice del reflector primario, y la distancia B, definida desde el foco *Cassegmin* hasta el vértice del secundario. Según estas definiciones, se puede deducir la ecuación (3-40).

$$L = B - S \qquad (3\text{-}40)$$

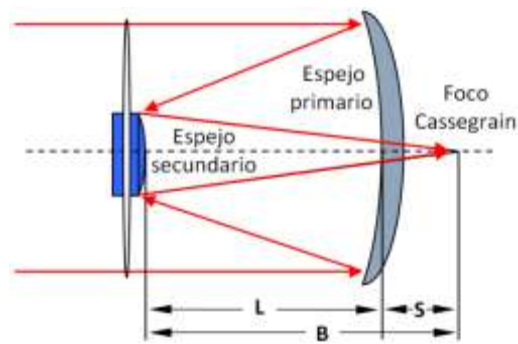

**Figura 175. Distancias importantes en el diseño de un telescopio *Cassegraín*.**

El diseño del telescopio es muy sensible a la distancia entre los espejos, por lo que deberá ser ajustada cuidadosamente. Como condición de partida, se midió físicamente la distancia B, que es la más accesible al telescopio real. Habiendo determinado la posición exacta del foco *Cassegrain* en el experimento de la Luna, la determinación de la posición del vértice del secundario es mucho más fiable que la del primario gracias a los planos disponibles del telescopio (Figura 172) y a que su posición es fija, a diferencia del primario. Una vez determinada la distancia B (que es igual a 623,5 mm), es posible calcular la separación L entre primario y secundario mediante la ecuación (3-41) [133, p. 393] y asumiendo una F = 2500 mm, que es la focal equivalente que se corresponde con el enfoque al infinito (que es donde se ha medido físicamente el foco *Cassegmin*, obteniendo la distancia B utilizada en dicha ecuación).

$$L = \frac{f_1 \cdot (F - B)}{F} \qquad (3\text{-}41)$$

Para diseñar el espejo secundario es necesario definir dos parámetros: el radio de curvatura $r_2$ y la constante cónica $k_2$. El radio de curvatura $r_2$ se determina mediante la ecuación (3-42) [283, p. 8], dependiendo de las longitudes L y B. Por su parte, la constante cónica $k_2$ viene dada por la ecuación (3-43) [133, p. 395] y depende de la excentricidad e, que a su vez depende de la focal $f_1$, de la distancia S y de la distancia L. Al tratarse de un espejo hiperbólico el valor de esta constante será menor que -1.

$$r_2 = -\frac{2 \cdot L \cdot B}{F - L - B} \qquad (3\text{-}42)$$

$$k_2 = -e^2 = -\left(\frac{f_1 + S}{2L - f_1 + S}\right)^2 \qquad (3\text{-}43)$$

### 3.10.5. Diseño del corrector asférico *Schmidt*

Un telescopio *Schmidt-Cassegrain* únicamente se diferencia del clásico *Cassegrain* en que en este último el espejo primario tiene forma parabólica y en el primero tiene forma esférica, cuya aberración esférica se corrige añadiendo una lente asférica correctora a la entrada del telescopio (Figura 176).

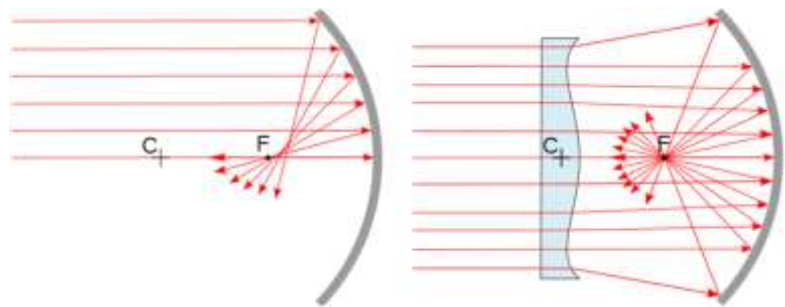

**Figura 176. Foco afectado por aberración esférica (izquierda) y foco corregido mediante una lente *Schmidt* (derecha).**

El perfil descrito por una lente *Schmidt* se corresponde con el de la ecuación (3-44), donde x es la coordenada del eje óptico y y es la distancia al eje óptico en cualquiera de los ejes ortogonales, ya que tiene simetría cilíndrica. Los coeficientes $A_1$, $A_2$ y $A_3$ dependen del índice de refracción de la lente, de la longitud focal del espejo esférico, de su apertura y de la posición de la zona neutra (la zona más delgada de la lente). Al tratarse de una lente, el corrector *Schmidt* introduce aberración esférica (así como aberraciones residuales como astigmatismo y coma). Esto se consigue minimizar construyendo la lente lo más delgada posible y eligiendo una posición óptima para la zona neutra, que en toda placa *Schmidt* se sitúa a un 86,6 % del eje óptico.

$$x = A_1 \cdot y^2 + A_2 \cdot y^4 + A_3 \cdot y^6 \qquad (3\text{-}44)$$

Para obtener los coeficientes $A_n$ es necesario realizar un proceso de minimización de la aberración esférica en *OpticsLab*, tras el cual se obtuvieron los coeficientes $A_1 = -1,14640231645997 \cdot 10^{-5}$, $A_2 = 4,280784651203688 \cdot 10^{-10}$ y $A_3 = 6,269973124890084 \cdot 10^{-16}$. Este diseño se realizó para $\lambda = 546,07$ nm, por ser una longitud de onda media del espectro visible, comúnmente utilizada en diseño de sistemas ópticos en el visible, como es el caso de este telescopio.

Dado que el primario del *Schmidt-Cassegrain* está formado por el conjunto espejo primario + lente correctora, su focal equivalente $f_{1\ primario+corrector}$ no es exactamente igual a la calculada a partir del radio de curvatura del espejo primario (que era $r_1 = 1027{,}65$ mm). Por ello, para que el espejo secundario vea la focal $f_1$ deseada (idealmente $f_1 = r_1/2 = 513{,}825$ mm), es necesario calcular el radio de curvatura efectivo $r_1''$ que da lugar a una focal lo más parecida posible a esta. En la Figura 177 se muestra la caracterización mediante simulación con *OpticsLab* de dicha longitud focal $f_1$ en función del radio de curvatura $r_1$ nominal del primario y el valor que consigue la focal deseada del conjunto espejo primario + lente correctora, que resultó ser de $r_1'' = 1031{,}63$ mm (una diferencia de 3,98 mm respecto al teórico), consiguiendo en el modelo final simulado del primario una focal $f_{1\ primario+corrector} = 513{,}81$ mm (una diferencia de 15 μm respecto a la ideal). Nótese que para cada par de valores $\{r_1'', f_{1\ primario+corrector}\}$, es necesario rediseñar la lente *Schmidt* al depender esta de $r_1''$. Los valores de los coeficientes proporcionados en el párrafo anterior se corresponden con el diseño final, una vez obtenida la $f_{1\ primario+corrector} = 513{,}81$ mm.

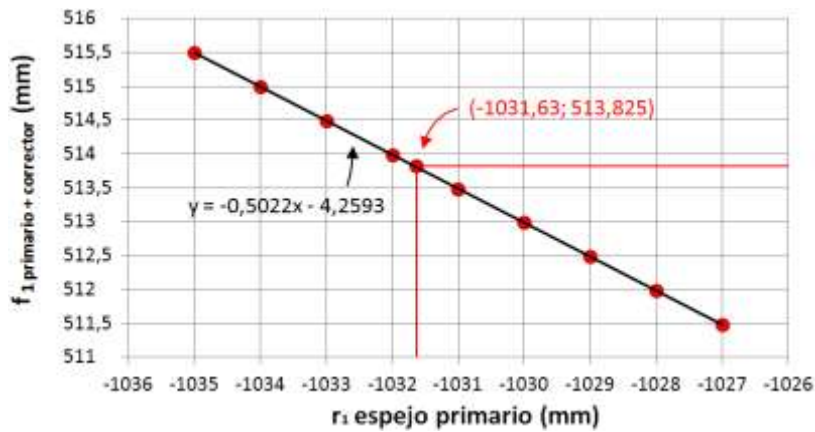

**Figura 177. Longitud focal del conjunto primario+corrector para cada radio de curvatura del espejo primario.**

## 3.10.6. Ajuste para modelar el *Schmidt-Cassegrain*

Una vez obtenida $f_{1\ primario+corrector}$, mediante el procedimiento de ajuste de $r_1''$ y diseño de la lente *Schmidt* seguido en el apartado anterior, ya es posible calcular el resto de parámetros del telescopio. Mediante la ecuación (3-41), conocidas $f_{1\ primario+corrector} = 513{,}81$ mm, $B = 623{,}5$ mm y $F = 2500$ mm, se puede obtener $L = 385{,}66$ mm. Utilizando la ecuación (3-42), se obtiene $r_2 = -322{,}59$ mm, y mediante (3-43), se obtiene $k_2 = -2{,}302448742$. Con todos estos parámetros ya es posible simular completamente una primera aproximación al telescopio *Schmidt-Cassegrain*. Sin embargo, dado que se han utilizado las ecuaciones de diseño del *Cassegrain*, la posición del foco equivalente no es exactamente la medida en el telescopio real (determinada por la distancia B), existiendo una diferencia de 26,41 mm. Para eliminar dicho error se ajustó el radio de curvatura del espejo secundario $r_2$ hasta lograr situar el foco *Cassegrain* en su posición correcta. El valor real de $r_2$ es desconocido, así como $L$ y $k_2$, que dependen de $r_2$. Mediante las correspondientes simulaciones para varias series de dichos valores se obtuvo la dependencia entre el error en la posición del foco *Cassegrain* respecto al medido en el telescopio real en función del radio de curvatura del secundario (Figura 178). Se concluyó que para eliminar dicho error, el radio del secundario debía ser $r_2$ = -336,62 mm (lo que proporciona una $f_2$ = -168,31 mm), con L = 380,09 mm y $k_2$ = -2,390215259 mm.

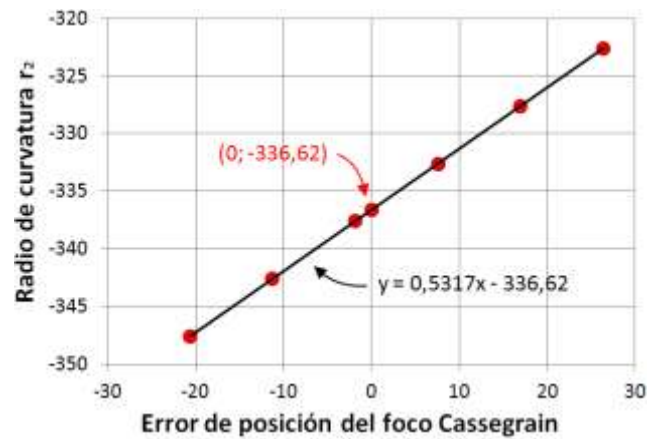

**Figura 178. Error en la posición del foco *Cassegrain*
en función del radio de curvatura del secundario.**

En la Figura 179 se muestra el modelo final del Meade LX200 simulado en *OpticsLab* con los parámetros calculados en los apartados anteriores, obteniendo un tamaño de spot en el foco de 41,9 μm (con un haz de entrada uniforme y de tamaño igual a la apertura), en la línea de lo esperado en un telescopio de estas características, si se compara con otros similares disponibles como ejemplos en *OpticsLab*. Como comprobación adicional, el único de los parámetros del diseño que es posible estimar en el telescopio real es la separación entre espejos L, que no fue impuesta en el proceso de diseño, habiéndose obtenido un valor muy acorde con la distancia real que se puede estimar en el telescopio.

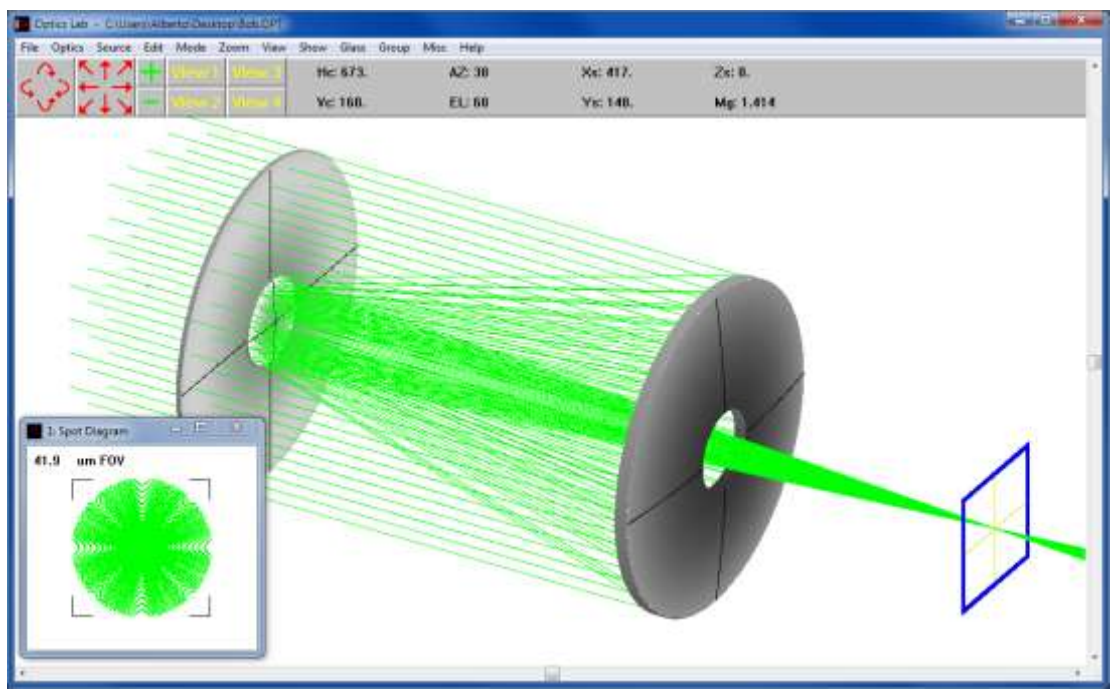

**Figura 179. Simulación con *OpticsLab* del telescopio Meade LX200-ACF de 10''.**

Por último, en la Tabla 16 se resumen las características ópticas del telescopio Meade LX200-ACF 10'' deducidas a lo largo de este apartado.

**Tabla 16. Resumen de las características ópticas del telescopio Meade LX200-ACF de 10''.**

| Magnitud | Valor |
|---|---|
| Apertura (pulgadas) | 10 |
| Longitud focal equivalente F (mm) | 2500 |
| Diámetro espejo primario (mm) | 263,5 |
| Diámetro corrector *Schmidt* (pulgadas) | 10 |
| Coeficiente A1 corrector *Schmidt* | $-1,14640231645997 \cdot 10^{-5}$ |
| Coeficiente A2 corrector *Schmidt* | $4,280784651203688 \cdot 10^{-10}$ |
| Coeficiente A3 corrector *Schmidt* | $6,269973124890084 \cdot 10^{-16}$ |
| Radio de curvatura primario $r_1$ (mm) | 1031,63 |
| Radio de curvatura secundario $r_2$ (mm) | -336,62 |
| Constante cónica secundario $k_2$ | -2,390215259 |
| Distancia primario-secundario L (mm) | 380,09 |

### 3.10.7. Rango de la focal del telescopio

Como se explicó en el apartado 3.10.2, el Meade LX200 permite modificar su distancia focal equivalente mediante un mando de control que traslada el espejo primario. Así, es posible reducir su focal aumentando la distancia entre el primario y el secundario, de forma que un objeto en el infinito quedaría enfocado en el interior del telescopio. Si bien la máxima focal debería ser 2500 mm (que permite enfocar en el ocular objetos en el infinito), en la práctica es posible seguir aumentándola de forma que al enfocar al infinito el foco se traslade a lo largo del eje óptico en dirección saliente desde el telescopio. Para caracterizar dicho rango de focales, se realizó el siguiente experimento. En primer lugar, se buscó la focal máxima, que corresponde a la mínima distancia entre espejos, establecida manualmente, y se determinó la posición donde la proyección de la Luna aparecía enfocada en la pantalla. Esta posición determina la B máxima (623,5 mm, valor determinado en el apartado 3.10.4, más una distancia de 805 mm que se desplazó la imagen de la Luna, es decir, el foco). Desde esa posición, se fue aumentando la distancia primario-secundario mediante el mando de control y registrando el valor de B para cada distancia. A su vez, se midieron las vueltas necesarias para cada posición del foco hasta llegar a la original de focal igual a 2500 mm (correspondiente a B = 623,5 mm). Entre ambas posiciones, se registraron 15 vueltas y desde este punto hasta la mínima focal (máxima distancia entre espejos) otras 22 vueltas trasladando el foco hacia dentro del telescopio.

Utilizando las ecuaciones de un telescopio *Cassegrain* vistas en el apartado 3.10.4, es posible estimar el rango de movimiento del primario en relación al secundario (el rango de L), lo que permite conocer el rango de longitudes focales que proporciona el telescopio. Para ello, se partió del telescopio enfocado en el infinito (F = 2500 mm) y se utilizó la ecuación (3-39) para calcular el rango de focales F en función de la separación entre espejos L. Conocidas las focales de ambos espejos, para obtener el rango total de F es necesario establecer los límites de L. El límite inferior de L se obtuvo directamente de las medidas experimentales descritas en el párrafo anterior, donde se pudo medir la distancia B (entre el secundario y el foco *Cassegrain*), de la que es posible identificar la L necesaria para obtener esa misma B resolviendo el sistema de ecuaciones de dos incógnitas definido por las ecuaciones (3-39) y (3-41). El límite superior no es posible estimarlo a partir de la medida del foco, debido a que este se encuentra físicamente en el interior del telescopio. Como el movimiento del primario lo controla el giro de un tornillo, es posible asumir que hay una relación lineal entre las vueltas de dicho tornillo y la distancia recorrida por el espejo. De esta manera, es posible determinar la máxima separación L, obteniendo un rango total de 40,75 mm, que proporciona la capacidad de establecer focales desde 1452 hasta 5069 mm (Figura 180).

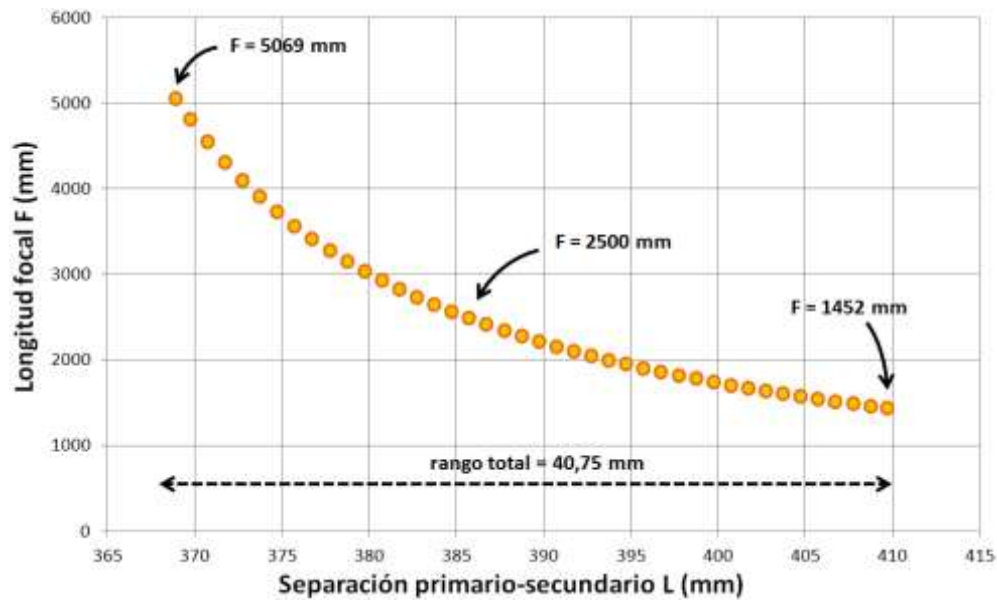

**Figura 180. Rango de longitud focal equivalente del telescopio
en función de la separación entre espejos.**

En la ecuación (3-45) se muestra la dependencia entre la separación entre espejos L en función de la longitud focal equivalente F (ambos en mm) según un ajuste polinómico de cuarto orden con un coeficiente de determinación $R^2 = 0,9999$.

$$L = 510,51 - 0,115 \cdot F + 4 \cdot 10^{-5} F^2 - 7 \cdot 10^{-9} F^3 + 4 \cdot 10^{-13} F^4 \qquad (3-45)$$

### 3.10.8. Optimización de la focal del telescopio

Una vez modelado completamente el telescopio y establecido el rango total de separación entre espejos L, es posible modificarlo en la simulación para realizar el diseño más conveniente. En este caso, como ya se explicó, el objetivo es minimizar la extensión del campo de visión, por lo que se buscará la posición donde el foco tenga el menor tamaño. Sin embargo, es necesario hacer varias consideraciones: por una parte, hay que tener en cuenta dónde irá situado el montaje óptico. En el primer diseño de Bob los componentes orientados a implementar el protocolo de QKD se dispusieron en línea con el eje óptico "colgando" del telescopio por la parte trasera (Figura 181). No obstante, al ir añadiendo más componentes al montaje, esta disposición demostró no ser la óptima, ya que el peso provocaba desalineamientos y carecía de estabilidad.

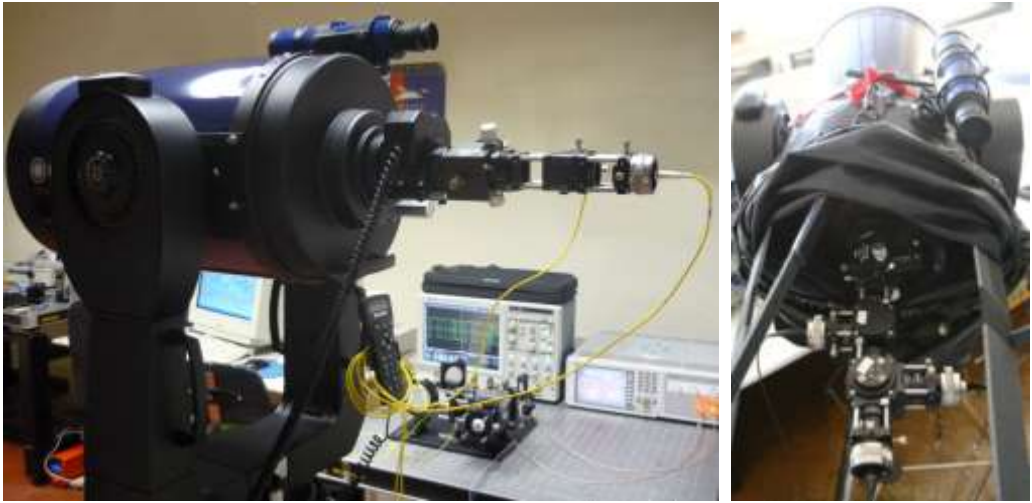

**Figura 181. Primer diseño de Bob con el montaje óptico "colgando" del telescopio.**

Para solventar los problemas del primer diseño de Bob se optó por desplazar el montaje a la parte superior del telescopio. Dicho traslado es posible llevarlo a cabo haciendo uso de la capacidad de enfoque del telescopio, que permite situar el foco en un punto más lejano. Para ello, se ancló una pequeña mesa óptica a dicha zona, trasladando la señal óptica a la misma mediante el propio *flip mirror* del telescopio y mediante el FSM.

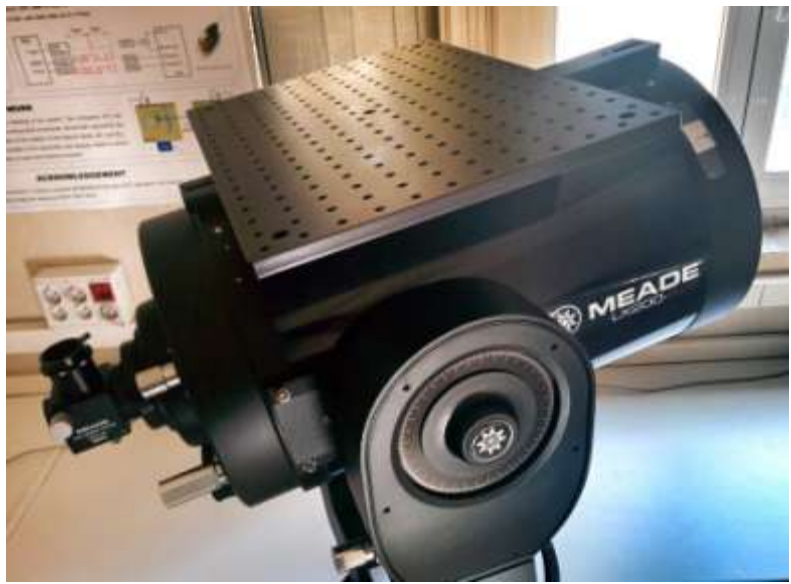

**Figura 182. Mesa óptica para nuevo montaje de Bob sobre el telescopio.**

En los apartados anteriores se había considerado una longitud de onda media del espectro visible de 546,07 nm. A partir de ahora, para realizar el diseño del sistema de QKD, se utilizará la longitud de onda de trabajo de $\lambda = 850$ nm. En la Figura 183 se muestra el diagrama de spot en el plano focal del telescopio cuando $F = 2500$ mm para esta longitud de onda cuando se ilumina su apertura de forma homogénea por un haz de rayos paralelos al eje óptico. El diagrama de spot resulta ser de 28 μm de diámetro, ligeramente limitado por aberración: se puede observar el disco de Airy circunscrito dentro del diagrama de spot. Cuando se aumenta la distancia focal del telescopio, moviendo el primario, el tamaño de spot tiende a aumentar. Mediante el modelo simulado, se pudo comprobar que dicho tamaño llega hasta casi 1 mm en la zona del montaje óptico. Para reducirlo, será necesario utilizar una óptica de focalización antes de la fibra óptica, para lo cual se utilizarán lentes tipo doblete asférico que serán capaces de minimizar las aberraciones en el plano focal. Para un mismo telescopio, además de la lontigud focal del mismo, esta óptica final será la mayor responsable del tamaño del spot focal, por lo que deberá realizarse un diseño cuidadoso. El objetivo siempre será obtener el menor spot focal posible, para lo cual se deberá encontrar el punto óptimo donde se cruce la aportación de la aberración y la difracción. Esto se debe a que ambas tienen causas opuestas: en general, la aberración dominará sobre la difracción al disminuir la longitud focal y aumentar el tamaño del spot de entrada, y viceversa.

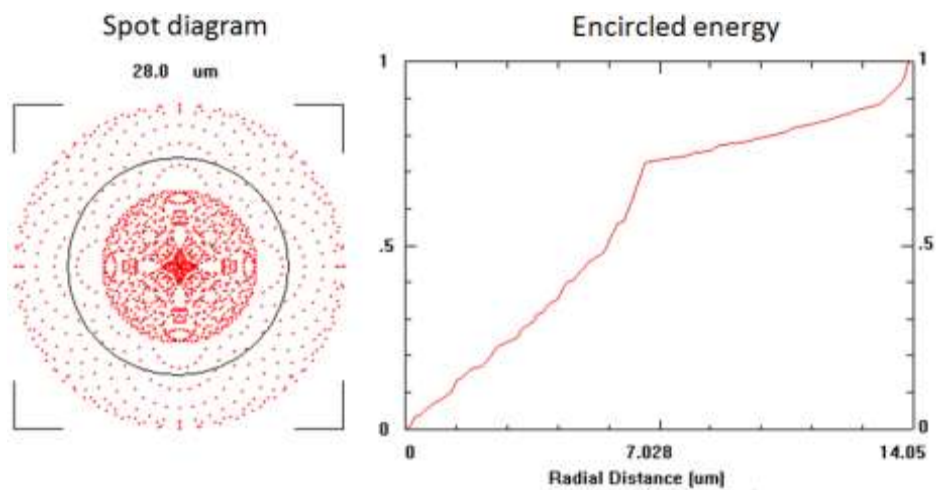

**Figura 183. Simulación del diagrama de spot del telescopio Meade LX200 para λ = 850 nm y F = 2500 mm.**

# 3.11. SISTEMA DE CORRECCIÓN EN BUCLE ABIERTO

### 3.11.1. Discusión de la estrategia

La estrategia más simple e intuitiva que se puede implementar para llevar a cabo la corrección de la turbulencia consiste en realizar un bucle abierto entre la señal del centroide de cualquier detector sensible a la posición y la señal invertida de control de un FSM. Se trata de un bucle abierto ya que el proceso aplica la corrección de forma ciega, únicamente monitorizando la perturbación sin observar el resultado y basándose en una calibración previa. En la Figura 184 se muestra un diagrama de bloques de esta estrategia, que es una variante del diagrama de la corrección en Bob de la Figura 138. La mayor diferencia es que en esta configuración los movimientos del FSM sobre el haz recibido solo tienen efecto en el canal de datos, usándose el otro canal únicamente para monitorizar el beam wander (además de para recibir la información de sincronismo). Por lo tanto, en este esquema solo se corrige el canal de datos, quedando el de sincronismo y seguimiento sin corregir. Para acoplar al detector de la señal de sincronismo el haz recibido y la mayor parte de su variación transversal debida al *beam wander*, sería necesario emplear una fibra de mayor diámetro. Si bien esto es un inconveniente de este esquema, no se trata de algo crítico ya que en el canal de sincronismo y seguimiento se emplea un láser de mayor potencia y esta pérdida no debería suponer ninguna limitación sobre el enlace.

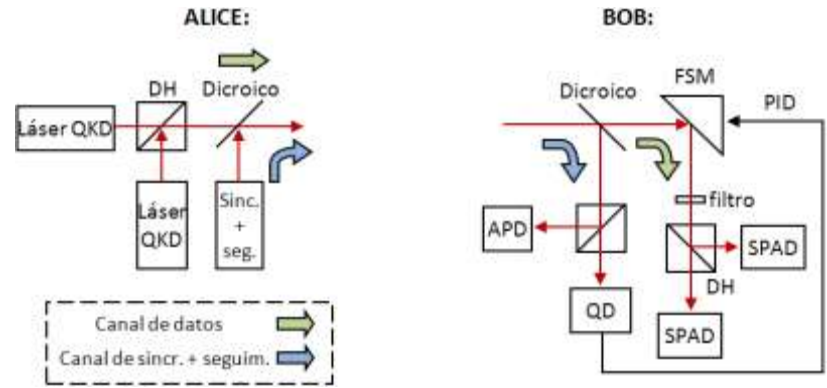

**Figura 184. Diagrama de bloques del sistema de corrección en bucle abierto.**

El funcionamiento básico de este montaje consiste en alimentar directamente al FSM con la señal de posición del QD. Como de lo que se trata es de compensar los movimientos detectados en el QD, el FSM deberá moverse de forma opuesta. Para ello, este esquema precisa una calibración previa basada en adaptar la amplitud de la señal del QD necesaria para compensar los movimientos con el FSM. El método para realizarlo consiste en utilizar un control de ganancia en la electrónica de acondicionamiento del QD para amplificar/atenuar su señal hasta conseguir que la señal en el canal de datos permanezca inmóvil. En la Figura 185 se muestra una simulación con trazado de rayos en *OpticsLab*, donde se observa el resultado del sistema ya calibrado. En la simulación se aplica una perturbación angular con movimiento sinusoidal al haz de llegada y para uno de los ejes se registra el movimiento mediante los detectores QD y PSD, así como la rotación angular instantánea del FSM. Se puede comprobar que en ausencia de corrección, el FSM permanece inmóvil y una vez se activa la corrección, éste pasa a realizar movimientos de sentido opuesto a la señal del QD, logrando anular la señal del PSD.

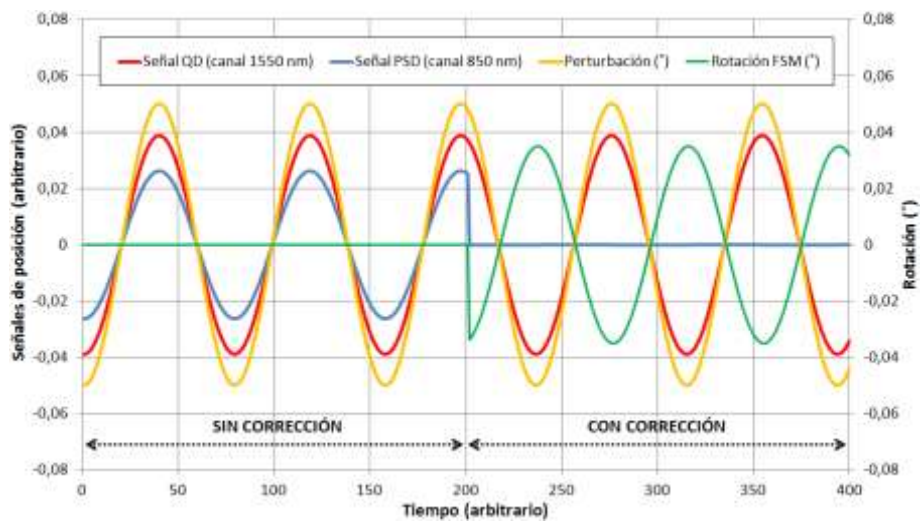

**Figura 185. Simulación en *OpticsLab* del sistema corrección en bucle abierto aplicando una perturbación angular al haz recibido.**

## 3.11.2. Ventajas e inconvenientes

Además de la simplicidad de su implementación, la gran ventaja de este esquema es que permite monitorizar la turbulencia en el canal de seguimiento mediante un QD desenfocado mientras se está aplicando la corrección en el plano focal del canal de datos.

Como se verá en el apartado 3.12, esto supone una ventaja muy importante que simplifica en gran medida la implementación y calibración del sistema de corrección. No obstante, el hecho de no poder cerrar el bucle, monitorizando la corrección, presenta una serie de inconvenientes que harán que este esquema no sea el más apropiado en la mayoría de las ocasiones. A continuación se describen los problemas que presenta esta estrategia y por los cuales se descartó como sistema de corrección:

1. Pese a que el efecto del *beam wander* se suele modelar como una perturbación localizada en el transmisor, en realidad el origen no es un único punto, sino que está distribuido a lo largo de todo el camino óptico. Si la calibración se realiza utilizando una variación angular en el transmisor, el sistema de corrección podrá corregir correctamente el *beam wander* originado a esa distancia, pero no el generado a cualquier otra. Esto es así porque el sistema óptico verá el origen de la variación angular como el plano objeto y tanto este como el plano imagen correspondiente dependen de la distancia. El objetivo de la calibración es ajustar la señal del QD para que, esté en el plano que esté, el FSM corrija el *beam wander* en el plano focal, que es fijo. En la Figura 186 se aprecia este fenómeno, donde se puede ver cómo una misma variación angular producida a distintas distancias forma una imagen en un plano distinto cada vez. En la práctica, el *beam wander* producirá un plano imagen equivalente, resultado de la aportación del efecto de la variación angular originada a cada distancia, y por lo tanto distinto al plano focal. Sin embargo, cabe señalar que este problema tiene un efecto limitado, debido a que la máxima variación angular observada en Bob será la originada a la mayor distancia, debido al efecto acumulativo, y los planos imagen relativos a planos objeto muy alejados están más próximos entre sí que los más cercanos, determinando en mayor medida la posición del plano equivalente. Por ello, al calibrar el sistema según este plano imagen equivalente, se observará una mayor corrección, aunque nunca perfecta debido a la aportación de perturbaciones localizadas en distancias más cercanas al receptor.

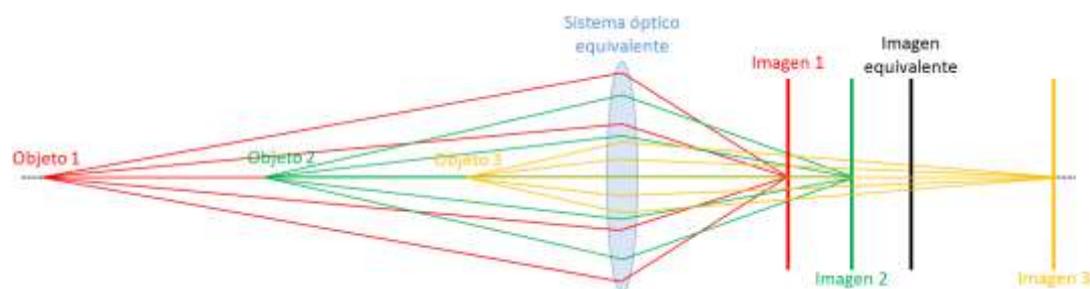

**Figura 186. Relación entre plano imagen y origen del *beam wander* en función de la distancia.**

2. El efecto del *beam wander* se suele modelar como una variación angular con la fuente en el transmisor. En la realidad, ni tiene un origen único como se ha explicado, ni tampoco es una variación única y estrictamente angular. Si bien en general el origen del *beam wander* sí es angular, el efecto resultante tiene también una componente traslacional, transversal al eje óptico. Esta componente quedaría totalmente eliminada en el plano focal de un sistema sin corrección: se puede modelar como un haz colimado en el que cada rayo representa una componente traslacional distinta y cuyo spot alcanza el mínimo tamaño en el foco. Como el objetivo de esta estrategia de corrección es eliminar la componente angular, y para ello se realiza la calibración, como efecto colateral negativo se tendrá una amplificación de la componente traslacional. Si bien un sistema sin corrección con el detector en el foco idealmente no tiene componente traslacional, pero sí componente angular, un sistema con corrección que sigue esta estrategia elimina la componente angular a costa de aumentar la presencia de la traslacional. La corrección de ambas componentes simultáneamente solo se podría llevar a cabo si los detectores de los dos caminos estuvieran en el plano focal (o en cualquier otro plano pero siempre el mismo), eliminando así la principal ventaja de este esquema (la posibilidad de corregir utilizando un QD desenfocado). Al igual que en el punto 1, este problema tiene un efecto limitado ya que estadísticamente la principal componente del *beam wander* será siempre angular, y por ello se podrá llevar a cabo la corrección pero nunca de forma óptima.

3. Por último, al realizar la corrección se introducirá una componente adicional de error siempre que los ejes del FSM no estén perfectamente alineados con los del QD. Si existe un desalineamiento entre ambos, la corrección de cada eje introduciría un error en el otro, que sería máximo cuando la diferencia entre ambos fuera de 45°. Si bien los dos primeros problemas que presenta esta estrategia se deben a la naturaleza de la turbulencia atmosférica y por lo tanto son inevitables, el tercero se puede minimizar con un buen alineamiento entre FSM y QD que elimine las interferencias entre sus ejes.

### 3.11.3. Implementación y calibración

En la Figura 187 se muestra el montaje óptico de este esquema con sus principales componentes. Por simplicidad en el alineamiento, y habiendo demostrado en el apartado 3.9 la buena correlación entre las longitudes de onda de 850 nm y 1550 nm, en este montaje y en los posteriores se utilizó solamente una longitud de onda visible de 650 nm. Como se explicó, la alta sensibilidad del QD permite detectar la señal de 650 nm, aún estando fabricado en InGaAs, y para solucionar la diferencia de intensidad detectada en cada detector se utilizó un polarizador en el camino de 850 nm, utilizado a modo de atenuador. Al utilizar la misma longitud de onda, las lentes empleadas para focalizar el haz en cada camino (de 850 nm y de 1550 nm) fueron la misma: un doblete acromático AC254-030-B-ML de *Thorlabs* de 30 mm de longitud focal y 1 pulgada de apertura, optimizado para el rango de 650 a 1050 nm.

El montaje de la Figura 187 se implementó en la mesa óptica acoplada al telescopio de Bob mostrada en la Figura 182. La parte de Alice consistió en un láser acoplado a fibra de 650 nm y colimado a 9 mm de diámetro. En la Figura 188 se muestra una imagen de satélite del enlace en el que se hicieron las medidas experimentales de este esquema de corrección. Se trata de un enlace de 30 metros de distancia entre dos laboratorios de la cuarta planta del Instituto Leonardo Torres Quevedo del CSIC, en Madrid. Esta distancia en un entorno urbano como el del centro de Madrid fue suficiente para registrar una *beam wander* considerable en Bob de forma natural (sin necesidad de simularlo de forma artificial), al tiempo que el alineamiento se puede realizar sin la complejidad que implica hacerlo en enlaces de mayor distancia.

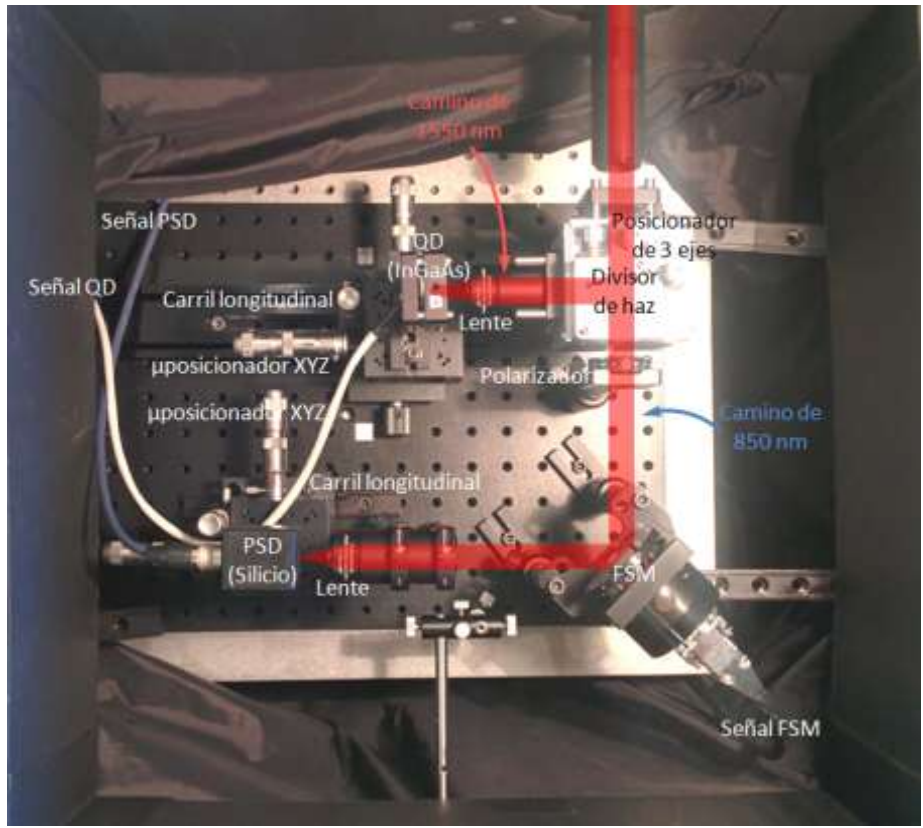

**Figura 187. Montaje óptico del sistema corrección en bucle abierto.**

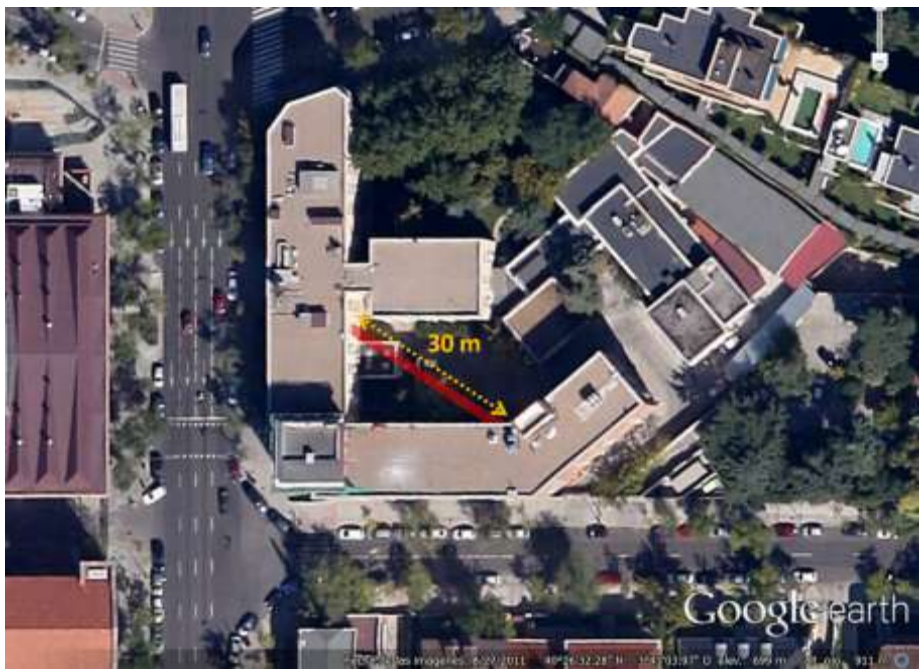

**Figura 188. Enlace de 30 metros entre laboratorios del Instituto Leonardo Torres Quevedo del CSIC en Madrid.**

En el sistema final los fotones que se propagan por el camino de 850 nm (el canal de datos) únicamente lo registrarán los contadores de fotones, es decir, no existe ningún PSD monitorizando la corrección a 850 nm, sino que la señal se focaliza directamente en las fibras ópticas que van acopladas a los SPAD. En las pruebas del sistema de corrección se sustituyeron estos SPAD por un PSD con el que monitorizar la señal que se estaba corrigiendo. Una dificultad que impone el uso del PSD para esta tarea es la de situarlo en el plano focal. Si bien encontrar el foco es una tarea más directa utilizando las fibras ópticas del sistema final (se trata de microposicionar el conector de la fibra hasta que se maximice la señal recibida utilizando un medidor de potencia conectado a la fibra en lugar del SPAD), el PSD no proporciona ninguna información que permita conocer la posición del plano focal. El PSD proporciona la intensidad total recibida como la suma de sus cuatro salidas, sin embargo de esta información no se puede deducir la posición del foco ya que el tamaño del área activa es mucho mayor que el spot focal y por lo tanto la potencia no varía al desplazar longitudinalmente el PSD. La técnica utilizada para enfocar el PSD consistió en desplazar transversalmente el haz en Alice mientras se buscaba en Bob la posición del PSD donde sus señales {x, y} tenían la mínima amplitud, lo que únicamente sucede en el plano focal: se puede ver como si cada posición transversal del haz fueran distintos rayos paralelos que van todos a coincidir en el foco en un haz colimado.

La calibración del sistema de corrección consiste en ajustar la ganancia de la señal del QD que alimenta al FSM para controlar sus movimientos. Tras invertirla mediante un *jumper* para que realice los movimientos opuestos a los registrados, se va modificando la ganancia hasta conseguir minimizar la señal del PSD, lo que significa que se está compensando la turbulencia. Por las razones explicadas en el apartado anterior, no es posible realizar la calibración utilizando un movimiento controlado desde Alice, ya que así no se consigue modelar correctamente el efecto real del *beam wander*. Este tipo de calibración fue la primera que se intentó, comprobando que no se conseguía ninguna corrección apreciable sobre la turbulencia real, lo que llevó a descubrir los problemas de esta estrategia explicados en el apartado anterior, demostrando que el modelo de *beam wander* que se suele utilizar en los estudios teóricos (de origen puramente angular y localizado en la fuente) no es completamente correcto y no debe ser aplicado para corregir turbulencia atmosférica real, al usar determinadas estrategias, como la de bucle abierto.

### 3.11.4. Resultados experimentales

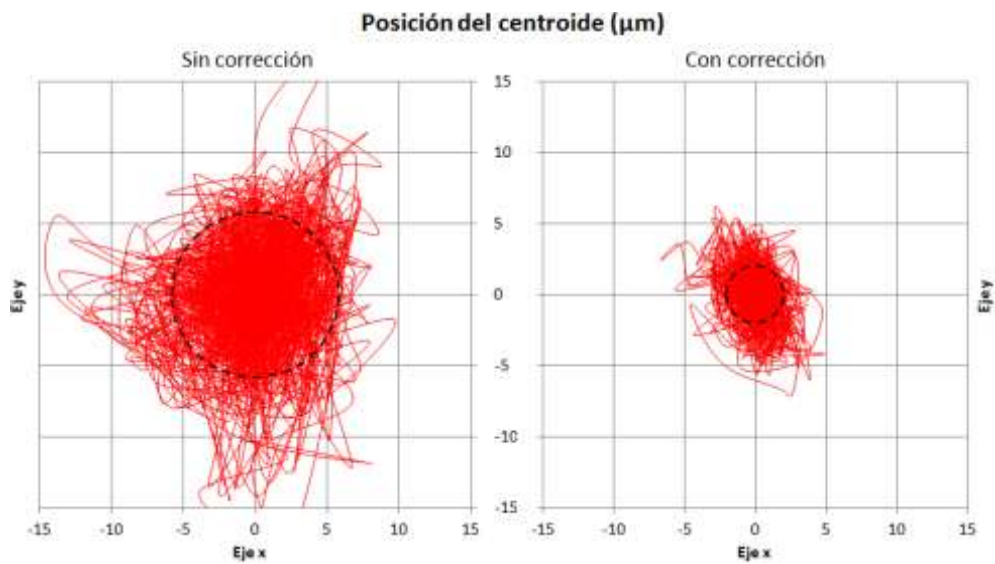

**Figura 189. En rojo, la posición instantánea del centroide en el PSD (canal de datos) sin y con corrección durante un intervalo de 30 segundos en un enlace de 30 metros con turbulencia natural. En negro, el diámetro que contiene el 90 % de los eventos.**

En la Figura 189 se muestra una captura durante un intervalo de 30 segundos de la posición del centroide observado por el PSD en el plano focal, sin y con corrección, en el enlace de 30 metros con turbulencia natural. El estado sin corrección (a la izquierda) se corresponde con la desactivación del bucle entre el QD y el FSM, dejando a éste último fijo en una posición.

La forma más eficaz para evaluar cuánto se dispersa la señal con el beam wander es realizar un histograma para cuantificar el número de muestras registradas de la señal del PSD en relación a la distancia al centro del detector. Este análisis proporciona una medida de cuánto se mueve el haz o cuánto tiempo pasa a cada distancia del centro. En la Figura 190 se puede observar el histograma realizado a partir de los datos de la señal mostrada en la Figura 189. Considerando el diámetro que contiene el 90 % de los eventos, el sistema de corrección consigue pasar de 11,4 µm a 4,6 µm: un factor de mejora de 2,5 veces el diámetro, que se traduce en una disminución del área en más de 6 veces.

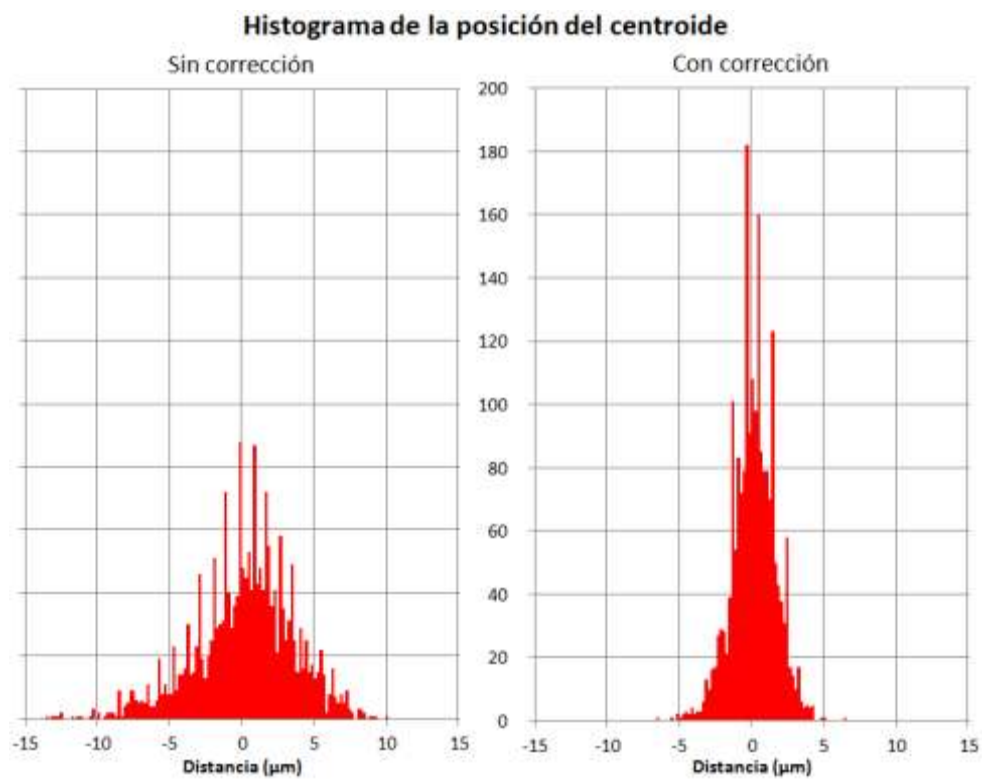

**Figura 190. Histograma de la posición del centroide en el PSD (canal de datos) sin y con corrección durante un intervalo de 30 segundos en un enlace de 30 metros con turbulencia natural.**

En la Figura 191 muestra el resultado del mismo experimento pero añadiendo turbulencia artificial a la turbulencia natural. Para ello se utilizó el calentador de aire empleado en el apartado 3.9.2, dispuesto tras el transmisor con el objetivo de simular una turbulencia mayor. Se puede observar que ahora la distancia al centro sin corrección es mayor que en el caso anterior (más de un 60 % mayor), si bien tras la corrección es similar. Como resultado, se produce una diminución de casi 9 veces en el área sin corrección respecto al área con corrección, en comparación con la mejora de 6 veces de la turbulencia natural. Este mejor comportamiento se puede explicar porque la turbulencia artificial añadida al enlace tiene un comportamiento más parecido al ideal consistente en una perturbación angular localizada cerca del transmisor en el inicio de la propagación.

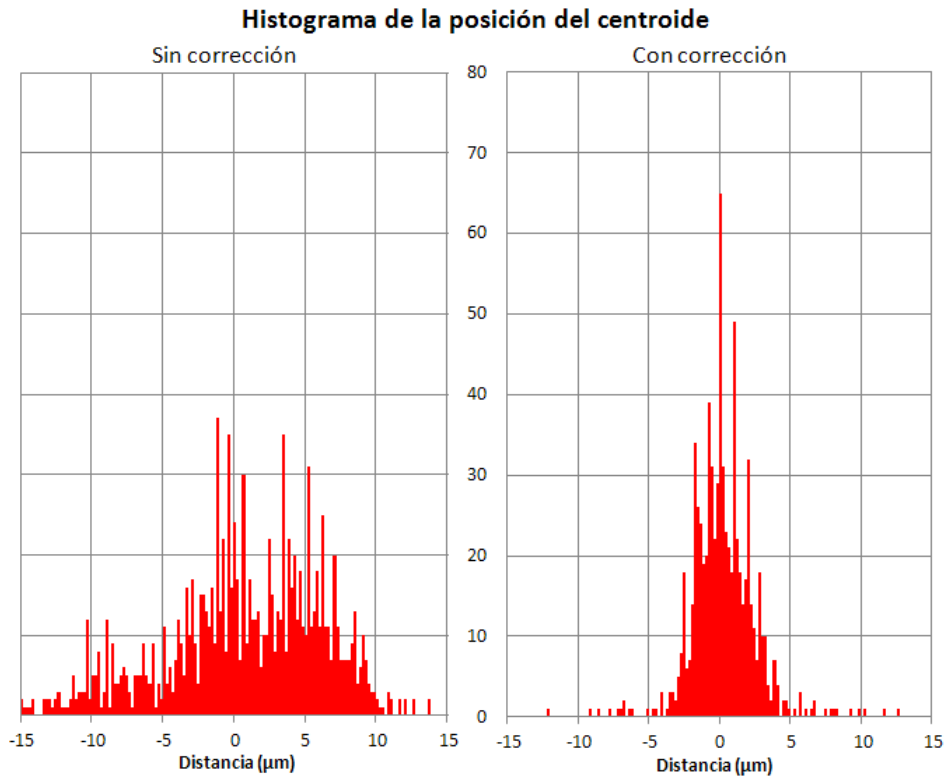

**Figura 191. Histograma de la posición del centroide en el PSD (canal de datos) sin y con corrección durante un intervalo de 20 segundos en un enlace de 30 metros con turbulencia natural y artificial.**

# 3.12. SISTEMA DE CORRECCIÓN EN BUCLE CERRADO

### 3.12.1. Discusión de la estrategia

La estrategia de corrección en bucle abierto, vista en el apartado 3.11, presenta una serie de problemas importantes cuyo origen es la naturaleza de la propia turbulencia. Utilizando una corrección en bucle cerrado es posible solventar estos problemas. Al monitorizar el resultado de la compensación de la turbulencia en tiempo real, por una parte se evita la necesidad de una calibración previa eliminando la sensibilidad del sistema de corrección ante una eventual descalibracion. Y más importante aún, se independiza el sistema de corrección de los problemas relacionados con la naturaleza de la turbulencia, ya que el objetivo será trasladar exactamente la misma corrección que se está monitorizando en el canal de seguimiento al canal de datos. Para ello, surje la necesidad de que los detectores de ambos canales (el extremo de la fibra óptica en el caso del canal de datos) estén exactamente en el mismo plano. Dado que el objetivo último es acoplar el haz a la fibra óptica, la apertura de esta estará situada en el plano focal del canal de datos, y por lo tanto el detector de posición en el canal de seguimiento también deberá situarse en el plano focal, lo que a su vez será el origen de una serie de problemas que se discutirán a continuación.

En la Figura 192 se muestra el diagrama de bloques del sistema de corrección en bucle cerrado. Se puede comprobar que la principal diferencia con el esquema de corrección en bucle abierto (Figura 184) es que el FSM es el primer elemento del montaje, por lo que sus movimientos afectan a los dos canales. El bucle de realimentación entre el QD y el FSM se cierra mediante un control PID, que asegura que se produzca la corrección en el canal de seguimiento y en el de sincronismo. Para obtener la corrección también en el canal de datos será necesario situar las fibras ópticas en el mismo plano equivalente, aunque en la práctica la calibración será la opuesta: se colocarán las fibras ópticas en el plano focal del receptor y posteriormente se situará el QD a la distancia necesaria para conseguir la misma corrección en los canales de seguimiento y de datos.

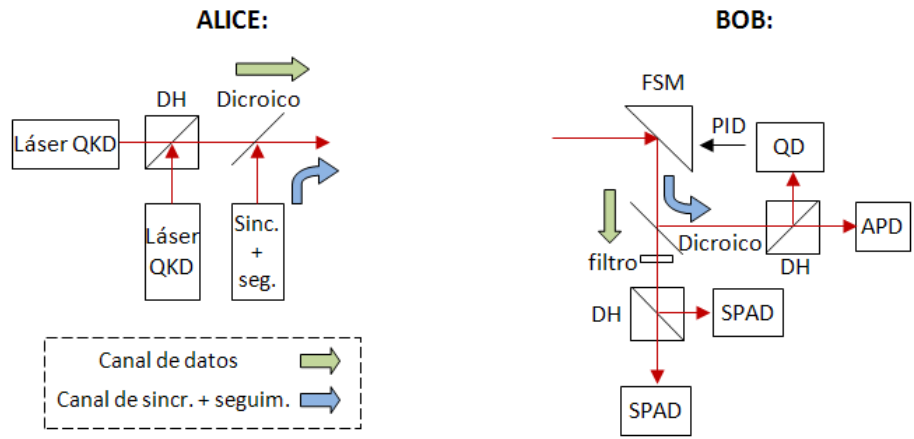

**Figura 192. Diagrama de bloques del sistema de corrección en bucle cerrado.**

## 3.12.2. Ventajas e inconvenientes

Como se adelantó en el apartado anterior, la primera ventaja de este esquema de corrección es que lo hace independiente de un proceso de calibración, necesario en el esquema de corrección en bucle abierto, eliminando además los problemas de aquella estrategia. Al trasladar la misma corrección de un canal a otro, deja de tener importancia el tipo o el origen de las variaciones del haz recibido: el FSM estará compensando en tiempo real cualquier tipo de desplazamiento respecto al centro del plano focal. Por otra parte, con la corrección en bucle cerrado, el canal de sincronismo también se beneficiará de la compensación de la turbulencia y será posible utilizar menos potencia en el láser de sincronismo, o bien emplear un detector menos sensible en este canal.

El principal problema que presenta la corrección en bucle cerrado surge del hecho de que al trasladarse la corrección del canal de seguimiento al de datos, se hace imprescindible situar los detectores de ambos canales en el mismo plano. El canal de seguimiento siempre estará corregido ya que es donde se monitorizan las perturbaciones del haz mediante el QD que cierra el bucle con el FSM, pero el canal de datos únicamente reflejará esta corrección si el detector (la fibra óptica en este caso) se sitúa exactamente en el mismo plano de corrección que el plano en el que está el QD. De otro modo, no se realizará adecuadamente la corrección en la fibra óptica y se registrarán movimientos del centroide, porque esta estrategia proporciona la corrección en un único plano en cada canal (es la contrapartida a la ventaja de corregir cualquier tipo de movimiento independientemente de su tipo y origen). Este comportamiento se puede visualizar en la Figura 193, donde se muestran tres rayos recibidos con distintos colores y distintos ángulos y posiciones para lo cual el FSM debe rotar su posición con el objetivo de dirigirlos hacia el mismo punto del plano de corrección en el canal de seguimiento. Puede comprobarse que debido a que la distancia entre el divisor de haz y el plano de corrección es la misma en ambos canales ($L_s = L_d$), ambos detectores están en la misma posición relativa. Si $L_d \neq L_s$, entonces en el canal de datos no se estarán corrigiendo las perturbaciones del haz recibido.

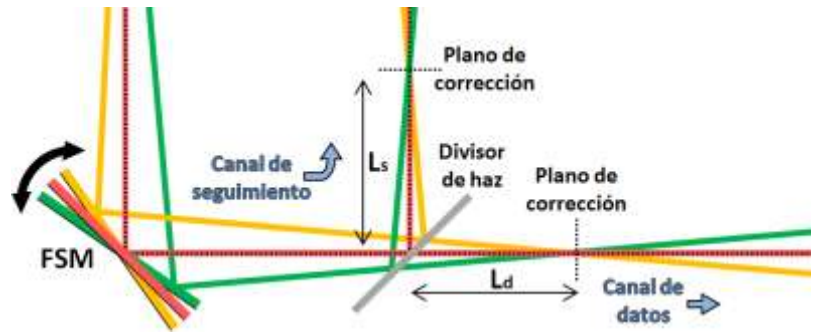

**Figura 193. Plano de corrección en cada canal según
la estrategia de corrección en bucle cerrado.**

A continuación se muestra el resultado de una simulación de óptica de rayos realizada con *OpticsLab* (Figura 194, derecha), donde se aprecia la dependencia del desplazamiento del centroide en cada canal (de seguimiento y de datos) con las diferencias entre $L_s$ y $L_d$ al corregir perturbaciones angulares empleando el esquema de corrección en bucle cerrado (Figura 194, izquierda). En esta simulación se utilizó un QD ideal con un gap igual a cero, lo que permite realizar la corrección en el plano focal en ambos canales, si bien el resultado sería el mismo usando un QD con gap distinto de cero y eligiendo cualquier otro plano distinto al focal. Se puede comprobar cómo el desplazamiento del centroide es igual a cero en el canal de seguimiento suponiendo una corrección ideal (ignorando el transitorio del PID) y cómo para el canal de datos existe una única posición (cuando $L_d/L_s = 1$) donde el desplazamiento del centroide es igual a cero.

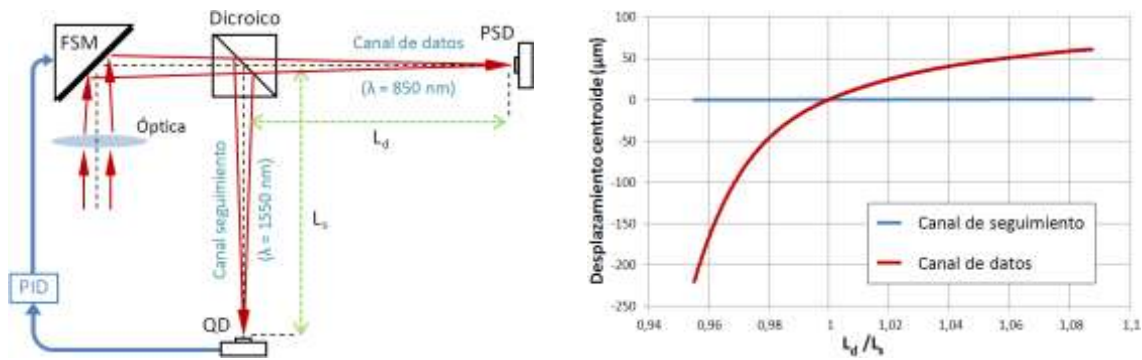

**Figura 194. Desplazamiento del centroide en función de la posición relativa del detector de cada canal
(derecha) al corregir perturbaciones angulares con un esquema en bucle cerrado (izquierda).**

Por otra parte, dado que en el canal de datos las fibras ópticas estarán obligatoriamente en el plano focal por ser donde el spot alcanza su menor tamaño (condición necesaria para utilizar el mínimo campo de visión posible), en el canal de seguimiento deberá situarse el QD también en el plano focal para conseguir que $L_d = L_s$. En la Figura 193 se presenta solo el movimiento del haz por simplicidad, sin embargo en la realidad los haces estarán focalizados por el sistema óptico receptor. El objetivo será entonces hacer coincidir el plano focal con el plano de corrección en el canal de seguimiento, situando también las fibras ópticas del canal de datos en el plano focal. El problema que presenta la coincidencia de estos planos es que, como se vio al estudiar el comportamiento real de un QD en el apartado 3.8.4, en general no se cumplirá uno de los requisitos básicos para utilizar correctamente este tipo de detectores: que el spot incidente sea mayor que el gap entre cuadrantes. Al situar el QD en el plano focal del sistema óptico, se estará tratando de utilizar el menor tamaño de spot para monitorizar la turbulencia, que es lo opuesto al uso recomendado de este tipo de detector. Surge así una incompatibilidad a causa de la necesidad de hacer coincidir los dos planos en los que se sitúan los detectores en cada canal, buscando un spot muy pequeño en el canal de datos y uno relativamente grande en el de seguimiento.

El último elemento del sistema óptico será una lente de focalización y en general tendrá una longitud focal corta con el objetivo de alcanzar un óptimo entre la aberración y la difracción introducida al haz recibido. En el sistema implementado se eligió una longitud focal de 30 mm como óptima para reducir el tamaño del spot focal. Con esta corta longitud focal, se hace necesario utilizar una lente diferente en cada canal en lugar de una única lente para los dos canales situada antes del divisor de haz. Nótese que el divisor de haz de la Figura 193 estará implementado en la práctica con un espejo dicroico que separará las longitudes de onda de cada canal. Por ello, normalmente las longitudes focales serán ligeramente diferentes en ambos canales aún utilizando lentes con la misma focal nominal en cada rango espectral. Tras simular con OSLO las lentes específicas que se utilizaron a cada longitud de onda, se obtuvieron diferencias muy pequeñas, en el orden de varias micras. No obstante, esta diferencia de la longitud focal para alcanzar el mismo plano equivalente en cada canal es despreciable, ya que lo que importa es la equivalencia de los planos, no la distancia física, por lo que incluso podrían utilizarse lentes diferentes en cada canal, con tal de que los planos fueran equivalentes.

Si bien los problemas que presentaba el sistema de corrección en bucle abierto no le impedían llegar a obtener una corrección de la turbulencia, el problema del sistema de corrección en bucle cerrado hace imposible cualquier corrección si no se toma alguna medida para solucionar el problema descrito en los párrafos anteriores. A continuación se presentan tres posibles soluciones para poder utilizar la estrategia de corrección de bucle cerrado manteniendo los detectores en el plano focal.

### 3.12.3. Solución 1: utilización de un PSD

Una solución intuitiva puede ser la utilización de un PSD en lugar de un QD en el canal de seguimiento. Tal como se describió en el apartado 3.8.3, los PSD están compuestos por un único elemento fotosensible y por ello no presentan ningún gap que restrinja el tamaño del spot que se puede usar en la detección. Estos dispositivos son menos comunes y más caros que los QD, especialmente los fabricados en InGaAs (el material necesario para detectar los fotones del canal de seguimiento en $\lambda = 1550$ nm). Otro inconveniente de estos detectores es que carecen de una alta linealidad, si bien las nuevas técnicas de fabricación, basadas en distorsionar la forma del área activa, consiguen solventar este problema en gran medida. Además, al usar el PSD en el plano focal con un spot muy pequeño, se usa solo una reducida fracción del área activa, lo que proporciona una linealidad mucho más alta que si se utilizara toda la superficie.

El PSD que se seleccionó para realizar las pruebas experimentales fue el EOS IGA-050-PSD del fabricante *EOS Systems*, que presenta un área activa de 5 mm, una resistencia entre electrodos de $R = 300\,\Omega$ y una capacitancia de $C = 2500$ pF, lo que proporciona un ancho de banda limitado por la frecuencia de corte a 3 dB dada por $f_c = 1/(2\pi RC) \approx 200$ kHz, más que suficiente para detectar todas las componentes espectrales de la turbulencia atmosférica. Una vez adquirido este dispositivo, se detectó un problema en su funcionamiento consistente en que las salidas de corriente de los cuatro electrodos proporcionaban intensidades no equilibradas entre sí. Esto se traducía en la existencia de un desplazamiento en la posición, calculada empleando las señales de salida del PSD, en relación con la posición real del haz incidente en su superficie. Dicho desplazamiento era lo suficientemente significativo como para poder apreciar a simple vista que un haz incidiendo en el centro del dispositivo producía unas señales de salida que no se correspondían con dicha posición. El propósito del PID es situar el haz en {0, 0}, por lo que al perseguir este objetivo utilizando un PSD con un desplazamiento, lo que se consigue es llevar al haz a una posición física muy alejada del centro del detector. Si bien no hubo éxito al intentar obtener una explicación satisfactoria del fabricante, otro profesional con experiencia en este tipo de detectores [284] confirmó que es un defecto que se suele dar debido a pequeñas diferencias en las resistencias de cada electrodo.

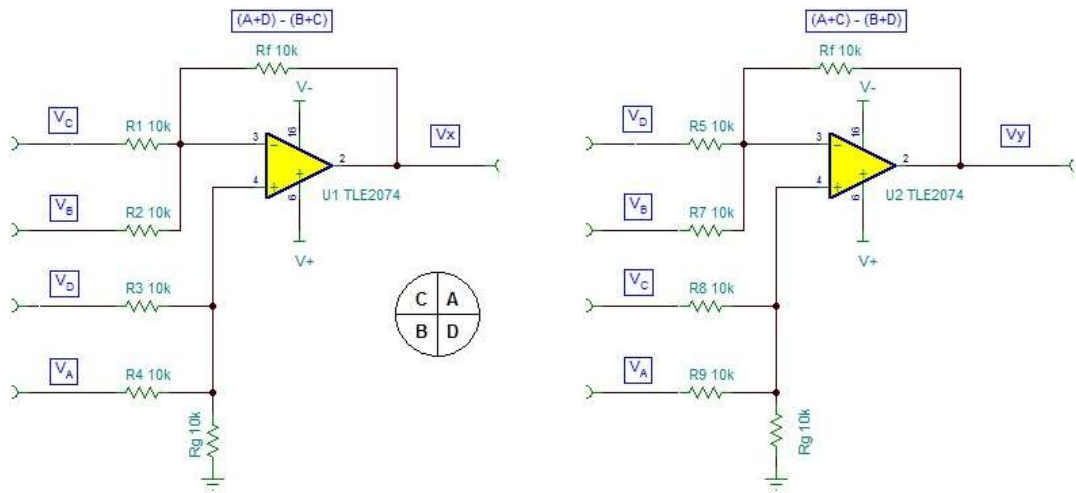

**Figura 195. Diagrama electrónico del cálculo de la posición { $V_x$, $V_x$}
a partir de las salidas del PSD convertidas a voltaje {$V_A$, $V_B$, $V_C$, $V_D$}.**

En la Figura 195 se muestra un diagrama electrónico del cálculo de la posición para cada coordenada {x, y} (a excepción de la etapa inicial de conversión corriente-voltaje de cada electrodo y de la etapa final del cociente entre el resultado de {$V_x$, $V_y$} y la suma total), a partir de las cuatro salidas del PSD convertidas a voltaje $V_A$, $V_B$, $V_C$ y $V_D$. Las coordenadas $V_x$ y $V_y$ se pueden obtener a partir de las ecuaciones (3-46) y (3-47).

$$V_x = -V_C \frac{R_f}{R_1} - V_B \frac{R_f}{R_2} + \left(1 + \frac{R_f}{R_1 \mid\mid R_2}\right) \cdot \left[V_D \frac{R_g \mid\mid R_4}{R_g \mid\mid R_4 + R_3} + V_A \frac{R_g \mid\mid R_3}{R_g \mid\mid R_3 + R_4}\right] \quad (3\text{-}46)$$

$$V_y = -V_D \frac{R_f}{R_1} - V_B \frac{R_f}{R_2} + \left(1 + \frac{R_f}{R_1 \mid\mid R_2}\right) \cdot \left[V_C \frac{R_g \mid\mid R_4}{R_g \mid\mid R_4 + R_3} + V_A \frac{R_g \mid\mid R_3}{R_g \mid\mid R_3 + R_4}\right] \quad (3\text{-}47)$$

Si todas las resistencias son iguales, entonces $V_x$ y $V_y$ se simplificaría en las ecuaciones (3-48) y (3-49) y al situar un haz en el centro del detector se obtendría $V_x = V_y = 0$.

$$V_x = (V_A + V_D) - (V_B + V_C) \quad (3\text{-}48)$$

$$V_y = (V_A + V_C) - (V_B + V_D) \quad (3\text{-}49)$$

En la práctica, se comprueba que $V_x \neq V_y \neq 0$, por lo que las señales $V_A$, $V_B$, $V_C$ y $V_D$ no están balanceadas. Aunque se trata de un problema fácil de resolver mediante *software*, la integración con el control PID para realizar la corrección en tiempo real exige una solución *hardware*. En la Figura 196 se muestran los diagramas electrónicos correspondientes a dos posibles soluciones *hardware*, ambas consistentes en la suma mediante potenciómetros de una tensión de *offset* para conseguir calibrar el PSD.

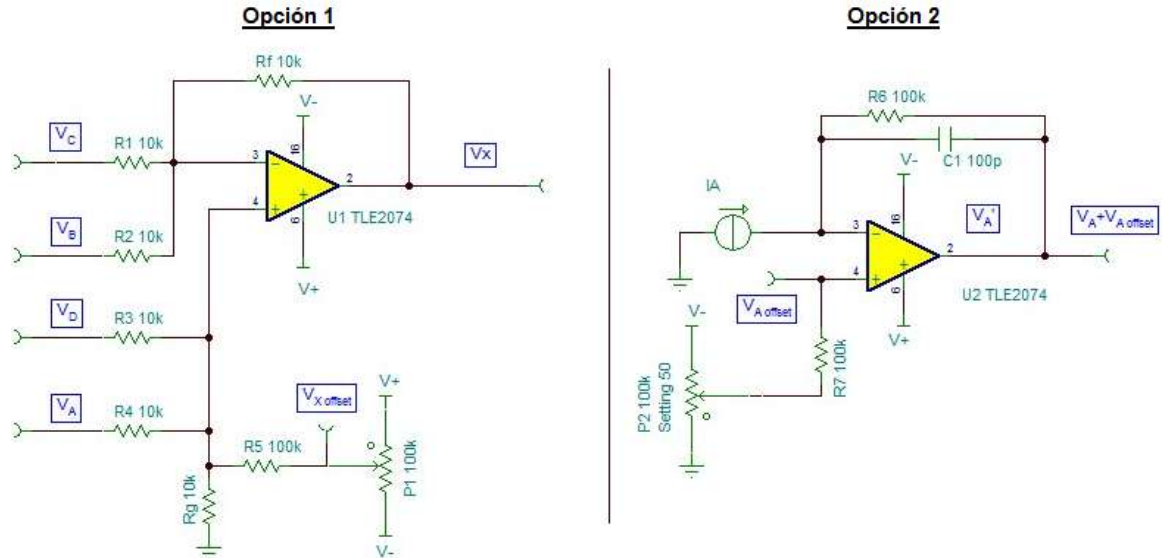

**Figura 196. Diagrama electrónico de las dos opciones para solucionar el desplazamiento en la posición del PSD.**

La opción 1 (Figura 196, izquierda) consiste en la suma de una tensión de *offset* directamente sobre las señales $V_x$ y $V_y$ para conseguir anularlas cuando el haz incide en el centro del PSD. La función de transferencia de este circuito (solo para $V_x$, ya que $V_y$ tendría una expresión equivalente) viene dada por la ecuacion (3-50).

$$V_x = -C\frac{R_f}{R_1} - B\frac{R_f}{R_2} + \left(1 + \frac{R_f}{R_1 \mid\mid R_2}\right) \cdot \left[ D \frac{R_g \mid\mid R_4 \mid\mid R_5}{R_g \mid\mid R_4 \mid\mid R_5 + R_3} + \right.$$

$$\left. + A\frac{R_g \mid\mid R_3 \mid\mid R_5}{R_g \mid\mid R_3 \mid\mid R_5 + R_4} + V_{X\,offset}\frac{R_g \mid\mid R_3 \mid\mid R_4}{R_g \mid\mid R_3 \mid\mid R_4 + R_5} \right] \tag{3-50}$$

Simplificando y sustituyendo los valores de las resistencias (todas iguales a 10 kΩ, excepto $R_5$) se llega a las ecuaciones (3-51) y (3-52), con k y n definidos por las ecuaciones (3-53) y (3-54) y todos los valores de las resistencias en kΩ.

$$V_x = k\left(V_A + V_D\right) - \left(V_B + V_C\right) + n \cdot V_{X\,offset} \tag{3-51}$$

$$V_y = k\left(V_A + V_C\right) - \left(V_B + V_D\right) + n \cdot V_{Y\,offset} \tag{3-52}$$

$$k = 3\frac{5 \mid\mid R_5}{5 \mid\mid R_5 + 10} \tag{3-53}$$

$$n = \frac{10}{3,33 + R_5} \tag{3-54}$$

Esta solución presenta el inconveniente de que la corrección se realiza directamente sobre el resultado de cada coordenada $V_x$ y $V_y$, y por lo tanto al realizar el cociente con $V_A + V_B + V_C + V_D$ (ver ecuaciones (3-35) y (3-36)), el efecto del *offset* no quedaría incluído en la normalización. Esto hace que la calibración del circuito dependa de la potencia recibida, por lo que se puede decir que en general la solución no es válida.

La opción 2 (Figura 196, derecha) consiste en la suma de un *offset* sobre cada una de las entradas de voltaje $V_A$, $V_B$, $V_C$ y $V_D$ para obtener unas nuevas entradas $V_A'$, $V_B'$, $V_C'$ y $V_D'$, calibradas previamente a realizar la normalización. Así, las señales $V_x$ y $V_y$ vienen dadas por las ecuaciones (3-55) y (3-56), con $V_A'$, $V_B'$, $V_C'$ y $V_D'$ definidas en las ecuaciones (3-57), (3-58), (3-59) y (3-60).

$$V_x = (V_A' + V_D') - (V_B' + V_C') \tag{3-55}$$

$$V_y = (V_A' + V_C') - (V_B' + V_D') \tag{3-56}$$

$$V_A' = V_A + V_{A\,offset} \tag{3-57}$$

$$V_B' = V_B + V_{B\,offset} \tag{3-58}$$

$$V_C' = V_C + V_{C\,offset} \tag{3-59}$$

$$V_D' = V_D + V_{D\,offset} \tag{3-60}$$

Pese a ser más compleja de implementar esta segunda opción, fue la elegida por conseguir una señal de posición independiente de la potencia recibida, al normalizar la señal de posición con las señales que contienen el *offset*. Por ello, este circuito precisa una única calibración, consistente en actuar sobre los cuatro potenciómetros $P_1$, $P_2$, $P_3$ y $P_4$ hasta conseguir que un haz situado físicamente en el centro del PSD proporcione una señal de salida $V_x = V_y = 0$.

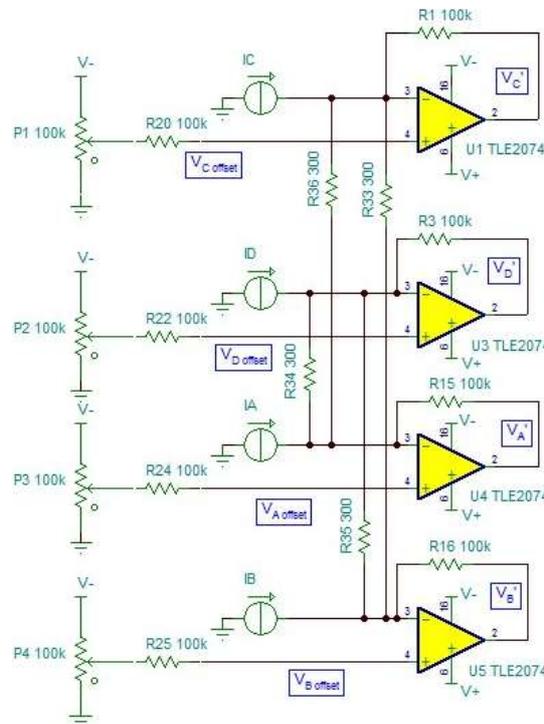

**Figura 197. Diagrama electrónico del circuito final, incluyendo la corrección de desplazamiento de posición del PSD.**

En la Figura 197 se muestra el diagrama electrónico que incluye la corrección del desplazamiento de la posición del PSD. En este diagrama se incluye el circuito equivalente para considerar las resistencias entre los cuatro electrodos del PSD (R33, R34, R35 y R36, igual a 300 Ω, según especificaciones del dispositivo) ya que se descubrió que se trata de un efecto importante a la hora de comparar los resultados experimentales con los simulados. Debido a este efecto, cada una de las tensiones de salida $V_A'$, $V_B'$, $V_C'$ y $V_D'$, que deberían depender únicamente de su propio *offset*, dependen también de las tensiones de *offset* de otros eletrodos ajenos según las ecuaciones (3-61), (3-62), (3-63) y (3-64).

$$V_A' = -I_A \cdot R15 + V_{Aoffset}\left(1 + \frac{R15}{R34\|R36}\right) - V_{Coffset}\frac{R15}{R15\|R36} - V_{Doffset}\frac{R15}{R34\|R15} \quad (3\text{-}61)$$

$$V_B' = -I_B \cdot R16 + V_{Boffset}\left(1 + \frac{R16}{R33\|R35}\right) - V_{Coffset}\frac{R16}{R33\|R16} - V_{Doffset}\frac{R16}{R16\|R35} \quad (3\text{-}62)$$

$$V_C' = -I_C \cdot R1 + V_{Coffset}\left(1 + \frac{R1}{R33\|R36}\right) - V_{Aoffset}\frac{R1}{R1\|R36} - V_{Boffset}\frac{R1}{R33\|R1} \quad (3\text{-}63)$$

$$V_D' = -I_D \cdot R3 + V_{Doffset}\left(1 + \frac{R3}{R34\|R36}\right) - V_{Aoffset}\frac{R3}{R34\|R3} - V_{Boffset}\frac{R3}{R3\|R36} \quad (3\text{-}64)$$

Dado que los circuitos de acondicionamiento y cálculo estaban ya integrados en una placa de circuito impreso, para incluir la corrección del desplazamiento tuvo que diseñarse una placa separada con los componentes necesarios y las conexiones a la placa original. En la Figura 198 se muestra el diagrama de estas conexiones (izquierda) y una imagen de su implementación e integración con el resto del circuito (derecha). La parte de la Figura 197 incluída en esta placa es la de la generación de los *offset* que se introducen por las entradas no inversoras de los operacionales, es decir, la parte que incluye las resistencias R20, R22, R24 y R25 (que en la Figura 198 aparece como R3, R4, R1 y R2, respectivamente) y los potenciómetros P1, P2, P3 y P4.

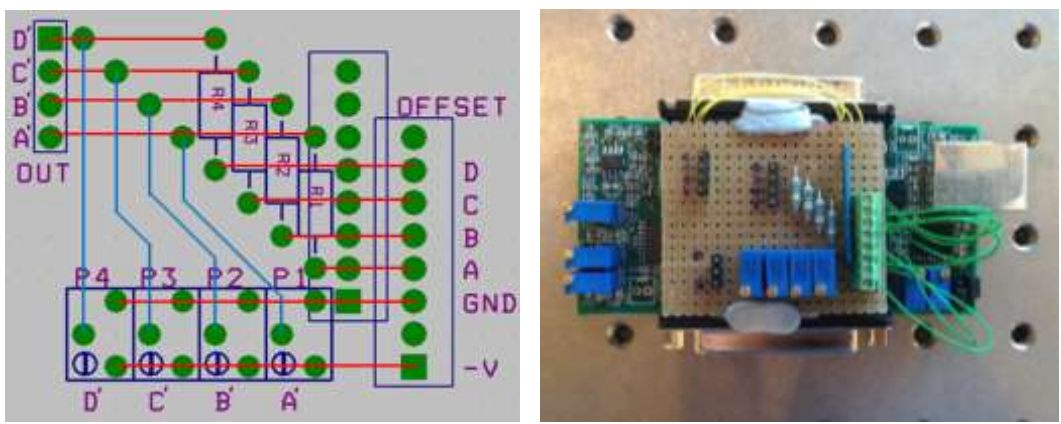

**Figura 198. Diagrama de la placa diseñada para la corrección de la posición del PSD (izquierda) e imagen de la placa implementada e integrada con el resto del circuito (derecha).**

Tras la implementación de la placa descrita, fue posible llevar a cabo la corrección del desplazamiento de la señal del PSD mediante una calibración inicial y el PSD fue integrado en el esquema de corrección de turbulencia de la Figura 192 en lugar del QD. Tal como se explicó en el apartado 3.12.2, el PSD debe situarse en el plano focal del sistema óptico en el canal de seguimiento, de la misma forma que la fibra óptica en el canal de datos, con el objetivo de trasladar la misma corrección de un canal a otro. Esto tiene un efecto sobre el funcionamiento del PSD distinto al que tiene sobre el QD. Por una parte, la primera diferencia entre ambos es la que motivó su uso en esta configuración de la corrección: que al reducir el tamaño del spot por debajo del gap, el QD deja de ser utilizable. Sin embargo, esta disminución tiene otro efecto distinto sobre estos dispositivos: la disminución del movimiento real del centroide sobre el área activa. Si se considera que el origen de la turbulencia está a una distancia muy lejana (lo que en general es una suposición razonable, especialmente si se considera que las perturbaciones más distantes tienen mayor influencia que las más cercanas), entonces el plano imagen (donde el movimiento del centroide es el mínimo) y el plano focal (donde el tamaño del spot es el mínimo) coinciden aproximadamente en la misma posición. Por ello, el PSD se habrá de situar en un lugar donde el spot además de ser muy pequeño se moverá muy poco. El efecto de esto sobre el PSD es distinto al efecto sobre el QD: si en el QD se traduce en un aumento de la pendiente de la respuesta de salida en función de la posición real (una mayor sensibilidad ante los desplazamientos del haz, pero el mismo rango de amplitud que para un spot mayor con mayores movimientos), en el PSD se traduce únicamente en una disminución del rango de la señal de salida (un menor movimiento significará una menor amplitud), ya que el PSD es insensible al tamaño del spot. Dado que estos movimientos son muy pequeños (en el orden de micras a decenas de micras), la señal proporcionada por el PSD es muy reducida y precisa de una gran amplificación para obtener una señal apropiada para controlar el FSM.

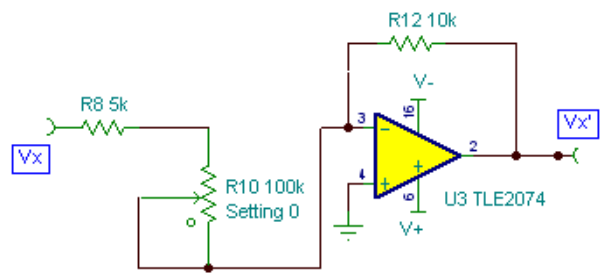

**Figura 199. Ganancia de las señales de posición.**

$$V_x' = V_x \cdot G_x = V_x \frac{R12}{R8 + R10} \qquad (3\text{-}65)$$

En la Figura 199 se muestra el circuito de control de ganancia para el eje x, cuya función de transferencia viene dada por la ecuación (3-65). Usando los valores de las resistencias mostrados, la ganancia $G_x$ tendrá un rango de $\{0,1 - 2\}$. Una ganancia máxima igual al doble de la señal $V_x$ del PSD resulta insuficiente para controlar el FSM. Dado que el circuito de amplificación está integrado en la placa original, para aumentar la ganancia con la menor interferencia se recurrió a disminuir el denominador de la ecuación (3-65), conectando en paralelo con R8 una resistencia de 1 kΩ soldada directamente sobre ella y obteniendo así una resistencia equivalente de 833 Ω, en lugar de 5 kΩ. En la Figura 200 se puede apreciar a la izquierda la resistencia original R8 (el código 4991 significa $499 \cdot 10^1$ Ω ≈ 5 kΩ) y a la derecha la nueva resistencia soldada sobre R8 (el código 102, leído al revés en la imagen, significa $10 \cdot 10^2$ Ω = 1 kΩ) y la resistencia R9 correspondiente al eje y. De esta forma, se consigue aumentar el rango de ganancia desde $\{0,1 - 2\}$ hasta $\{0,1 - 12\}$.

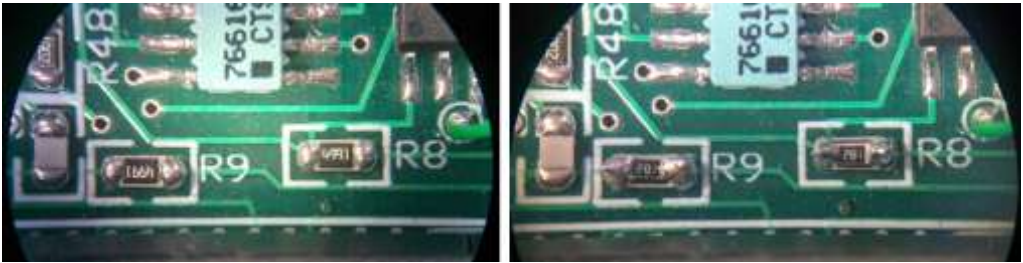

**Figura 200. Aumento de la ganancia mediante la conexión en paralelo de resistencias.**

Una vez calibrado el PSD para eliminar el desplazamiento y ajustada la ganancia al nivel adecuado para controlar el FSM, fue posible empezar a corregir el *beam wander* utilizando la estrategia de bucle cerrado. En la Figura 201 se muestra el resultado de la corrección de la turbulencia artificial generada en un experimento de laboratorio. Este resultado se corresponde con el histograma de la posición del centroide detectado por el PSD en el plano focal del canal de seguimiento durante un periodo de 6 segundos sin y con corrección. Dentro del histograma se puede observar una captura de la posición real del centroide en el plano focal durante el mismo intervalo. Considerando el diámetro que contiene el 90 % de los eventos, el sistema de corrección consigue una mejora (una reducción) de un factor 1,5, que se traduce en una disminución del área en 2,4 veces.

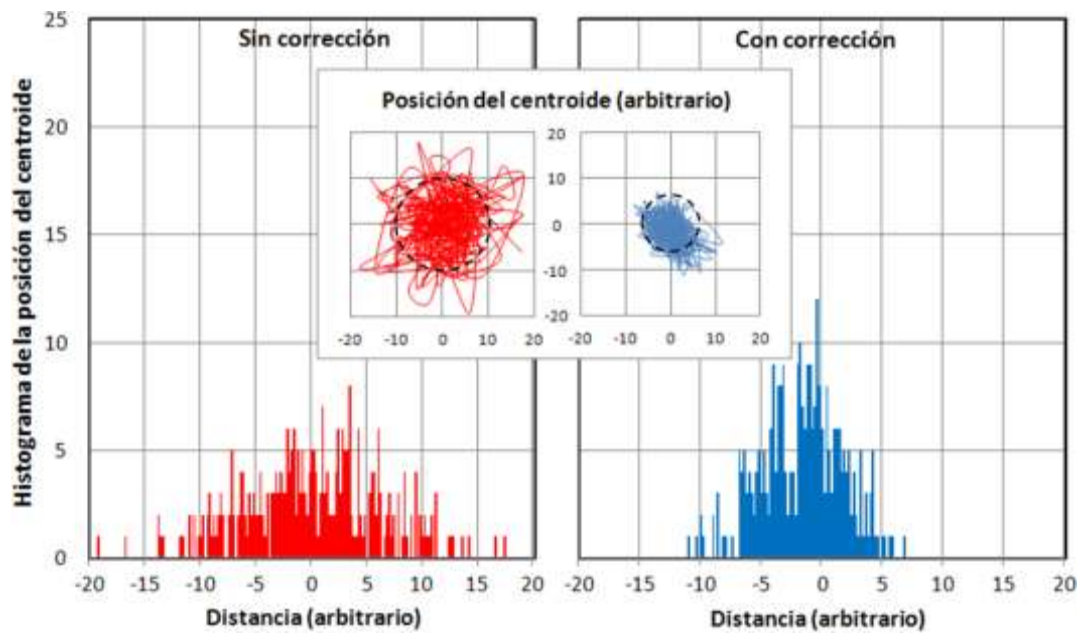

**Figura 201. Histograma de la posición del centroide en el plano focal durante un periodo de 6 segundos sin y con corrección. Dentro, en rojo y azul la posición real del centroide en el plano focal del canal de seguimiento para el mismo intervalo y en negro el diámetro que contiene el 90 % de los eventos.**

Esta corrección es peor que la del sistema de corrección en bucle abierto debido a la respuesta del PSD al situarlo en el plano focal. Los movimientos reales del centroide se reducen tanto que la señal de posición generada por el PSD es muy pequeña y la gran amplificación necesaria para obtener una señal de control del FSM adecuada, hace que la amplitud del ruido electrónico sea comparable con la amplitud de la señal de posición. Para solucionar este problema y hacer que esta opción sea viable para la corrección de la turbulencia sería necesario realizar un nuevo diseño electrónico del circuito de acondicionamiento de las señales que elimine el ruido antes de la amplificación. Dado que esto implicaba el desarrollo de una nueva placa, se optó por seguir utilizando la actual y recurrir a las estrategias descritas en los siguientes apartados.

### 3.12.4. Solución 2: QD con aumento de la longitud focal

Una vez comprobados los problemas de utilizar un PSD para corregir el *beam wander* en el plano focal, se decidió regresar al QD. La principal diferencia de utilizar un QD en el plano focal respecto a utilizar un PSD consiste en que el rango de amplitudes de salida proporcionado por el QD no depende únicamente del rango de movimientos sobre el área activa, sino también del tamaño del spot incidente, siendo el PSD insensible a este. En la Figura 202 se muestra a la izquierda el comportamiento de la señal de salida del QD ante movimientos solo limitados por el área activa de haces de distintos tamaños. Se comprueba que al disminuir el tamaño de spot, el rango de amplitudes de la señal de salida no cambia y solo cambia el rango de movimientos que es posible detectar: menor rango cuanto más pequeño es el spot, por alcanzarse antes la saturación. No obstante, esto puede no presentar un problema al utilizar un QD ya que el efecto de acercar el QD al plano focal no solo es el de disminuir el spot, sino también el de reducir el rango de movimientos. Esto es así porque el plano focal y el plano imagen coinciden en la misma posición cuando el origen de la turbulencia es muy distante. Por ello, enfocar un QD que monitoriza un movimiento producido a gran distancia en general no hace que su rango de amplitudes de salida disminuya, sino que de hecho puede aumentar porque a igualdad de movimientos, un spot más pequeño produce un rango de salida mayor. En la Figura 202 a la derecha se comprueba este efecto: cuando el rango de movimientos es constante, disminuir el tamaño del spot incidente significa aumentar el rango de la señal de salida del QD.

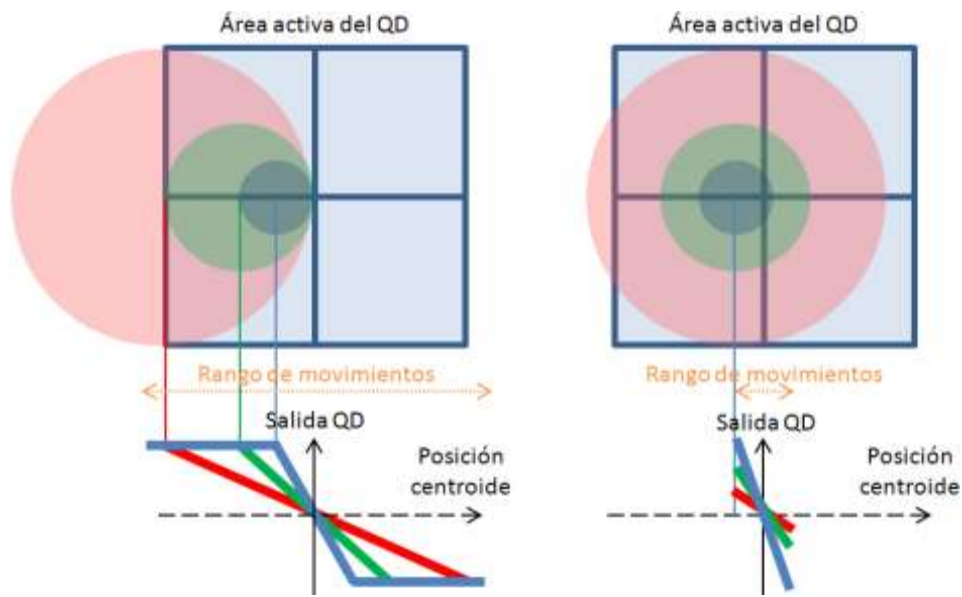

**Figura 202. Comportamiento de un QD ante variaciones en el tamaño del spot con movimiento ilimitado (izquierda) y con movimiento limitado (derecha).**

Por otra parte, el inconveniente del QD es que presenta una gran sensibilidad a la relación entre el tamaño del spot incidente y el tamaño del gap. Hasta el punto que si el tamaño de spot llega a ser mucho menor que el tamaño de gap, podrá no obtenerse ninguna señal. Este es el caso cuando se sitúa el detector en el plano focal, donde además de reducirse al mínimo el spot, también se reduce al mínimo el movimiento. Por ejemplo, según se vio en el apartado 3.11.4, el diámetro que contiene el 90 % de los fotones en presencia de turbulencia natural medido en el plano focal en el experimento de 30 metros es de unas 10 µm. El diámetro del spot focal del mismo sistema utilizado en el experimento está sobre las 30 µm según una simulación con óptica de rayos. Sumando ambos diámetros se puede obtener una primera estimación del tamaño que tendría el spot de largo plazo trasladado al foco, de unas 40 µm. Como se calculó en el apartado 3.8.4, el QD utilizado presenta una zona de gap con un diámetro de 63,64 µm dentro de la cual no se detecta nada. Según estos cálculos, el QD sería incapaz de detectar ningún movimiento si se situara en el plano focal: si el haz estuviera alineado con el centro del QD, la salida sería igual a cero y en otro caso la salida sería la de saturación constante del correspondiente cuadrante.

El problema descrito tiene una posible solución: aumentar el tamaño del spot en el QD. Si bien la forma más evidente de conseguir este aumento es alejando el QD del plano focal, existen otras formas de aumentar el spot sin desenfocar el QD. La que se propone en este apartado consiste en aumentar la longitud focal equivalente del sistema óptico. Si en el canal de datos interesa que el sistema óptico produzca un spot lo más pequeño posible para así utilizar el menor diámetro de fibra óptica limitando el ruido que se acopla al sistema, en el canal de seguimiento no existe ninguna razón por la que no pueda trabajarse con un spot de mayor tamaño, siempre y cuando ambos estén en el mismo plano (el plano focal). Como se vio en el apartado 3.5.3, el spot focal es proporcional a la longitud focal, por lo que para aumentarlo basta con aumentar la focal equivalente del sistema óptico. El problema de esta solución es que para conseguir un tamaño de spot mayor al gap es necesario utilizar una focal muy larga ya que la dependencia es lineal. Por ejemplo, en el experimento de los 30 metros, para obtener un spot focal de 100 µm (aproximadamente el doble del tamaño de gap), utilizando la ecuación (3-14) con una longitud de onda de 1550 nm y un haz incidente de 9 mm, sería necesario utilizar una focal de 450 mm (en comparación con la focal de 30 mm que se usa en el experimento). Para simplificar el montaje se decidió implementar uno consistente en un objetivo de microscopio 10× (modelo M-10× de *Newport*), que cumple la misma función de aumentar la focal equivalente en un montaje más compacto (Figura 203).

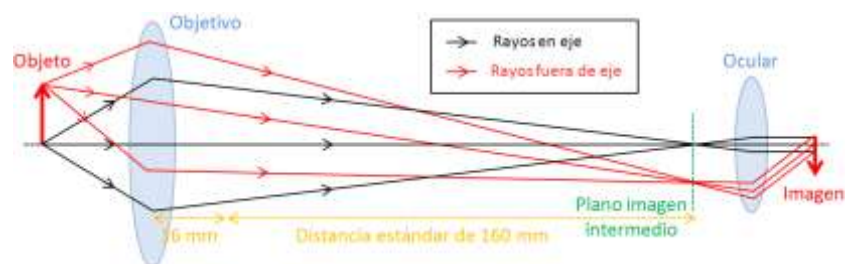

**Figura 203. Comportamiento óptico de un microscopio genérico de 10×.**

En el montaje, el objetivo se sitúa tras la lente de 30 mm, que se encarga de focalizar el haz colimado hasta un tamaño suficiente para introducirlo por la apertura del objetivo, que es de 5,5 mm y debe ser mayor que la suma del spot, de su máximo desplazamiento y de un margen de seguridad para evitar truncamientos. Utilizando la ecuación (3-39), es posible calcular la focal equivalente F del sistema compuesto por la lente de focal igual a 30 mm y el objetivo de microscopio de focal igual a 16 mm. Para ello es necesario calcular la separación L entre ambas, que será igual a la suma de la distancia focal de la lente (30 mm) y la distancia del objetivo de microscopio a su plano objeto para formar una imagen 176 mm hacia el otro extremo de la lente (esta distancia se obtiene sumando los 16 mm de la focal del objetivo más los 160 mm de la longitud estandarizada de tubo para todos los objetivos de microscopio). Para calcular la distancia al plano objeto se utiliza la ecuación (2-4), conocida la focal del objetivo (16 mm) y la distancia del plano imagen (176 mm), resultando en 17,6 mm. Por lo tanto, la distancia L entre la lente y el objetivo de microscopio será 30 mm + 17,6 mm = 47,6 mm. Con estos datos, se obtiene que el conjunto lente+objetivo presenta una focal equivalente igual a 300 mm, con lo que se ha conseguido el propósito inicial de aumentar la distancia focal en un montaje más compacto.

El montaje se dispone de forma que la distancia entre el objetivo de microscopio y el QD se mantiene fija a 176 mm. Así, situando el foco de la lente de 30 mm en el plano objeto del objetivo de microscopio, a 17,6 mm del mismo, se obtiene en la superficie del QD la imagen del foco de la lente de 30 mm. Como una vez establecida la distancia de 176 mm entre objetivo y QD, el conjunto se deja fijo, para ajustar la distancia entre el objetivo y la lente, se desplazará esta longitudinalmente hasta obtener el foco original en el QD de unas diez veces el tamaño del foco original. El spot focal del sistema original no limitado por aberración era de unas 10 μm (las 30 μm referidas antes se refieren al tamaño real del spot focal, que está limitado por aberración), por lo que al multiplicarlo por diez, se convertiría en unas 100 μm, aproximadamente el doble que el tamaño del gap. También es posible aumentar la distancia entre el objetivo de microscopio y el QD para obtener un spot focal aún mayor, aunque más alejado del montaje por lo que podría no ser conveniente. En lugar de esto, si se desea aumentar el tamaño del foco es preferible recurrir a un objetivo de microscopio de mayor aumento. En la Figura 204 se muestra el montaje óptico del sistema de corrección de bucle cerrado con el montaje del objetivo de microscopio para aumentar la longitud focal equivalente.

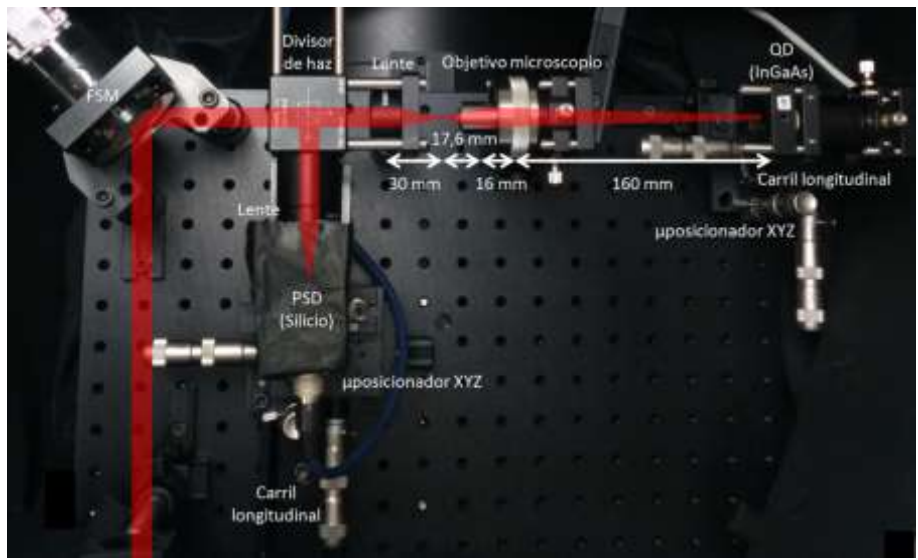

**Figura 204. Montaje óptico del sistema corrección en bucle cerrado con el QD en el plano focal con aumento de la longitud focal.**

La estrategia descrita permite utilizar el QD en el plano focal, aunque también presenta algunos inconvenientes. El primero es que la alineación del nuevo montaje se hace más compleja. La larga longitud focal obtenida hace que cualquier pequeño desalineamiento desplace al spot del eje óptico al trasladar longitudinalmente la lente. Además, esta traslación es necesario hacerla muchas veces para situar al QD en su posición focal. Este proceso supone otra dificultad de este montaje, ya que con un spot mucho mayor que el gap ya no se observan los comportamientos característicos de un QD en el plano focal que permiten localizar el foco fácilmente. Como la longitud focal es muy larga, el tamaño del spot varía de forma muy lenta, por lo que es más difícil e imprecisa la búsqueda del plano focal [29]. Por otra parte, el poder multiplicador del objetivo también actúa sobre los movimientos del spot, haciéndolos mucho mayores. Esto suponía deshacer la modificación del circuito de amplificación explicada en el apartado anterior, ya que en este caso sería necesaria una mayor atenuación, al ser insuficiente la ganancia mínima para obtener un nivel de señal del QD adecuado para controlar el FSM. Debido a estos inconvenientes, se decidió seguir utilizando la misma electrónica sin modificación en el montaje explicado en el siguiente apartado, que también solventa el problema de utilizar un QD en el foco pero de una forma más simple.

Merece la pena señalar que las soluciones explicadas en este apartado y en el siguiente son compatibles con la más obvia: disminuir el tamaño del gap. Las 45 μm del IGA-030-QD utilizado en esta tesis es un tamaño bastante común en este tipo de detectores, si bien existen comercialmente dispositivos con un menor tamaño de gap. En general, a menor tamaño de área activa en el detector, más fácil es encontrar un menor tamaño de gap. Por ejemplo, el IGA-010-QD de *EOS Systems* es una versión similar al empleado en esta tesis pero de menor área activa (1 mm en lugar de 3 mm) y un tamaño de gap algo menor (<35 μm). Este fue el menor gap encontrado en InGaAs de forma comercial, aunque en Silicio sí existen alternativas con menor gap al ser un material más extendido en este tipo de detectores. Por ejemplo, el 84-614 de *Edmund Optics*, con 0,5 mm de área activa, proporciona 13 μm de tamaño de gap, lo que demuestra que la tecnología para conseguir tamaños muy reducidos de gap existe. En [285] explican que es factible construir un QD con ~10 μm de gap partiendo de un fotodetector de área activa única, desarrollo que llevaron a cabo motivados por la no disponibilidad comercial de QD en InGaAs basados en detectores APD (de mayor sensibilidad) en lugar de PIN. En principio no debería haber problema con fabricar un QD con un gap incluso menor a partir de un fotodetector simple sobre el que se litografiara mediante láser un gap muy estrecho. La fotolitografía láser es una tecnología muy extendida en la actualidad. Se han conseguido resoluciones en el orden de varios nanómetros [286] y de forma comercial existen equipos con capacidad suficiente para obtener un gap tan pequeño como 1 μm sin problemas de interferencias entre pistas [287].

## 3.12.5. Solución 3: QD con aumento de las aberraciones

Para aumentar el spot focal existe una alternativa más simple que el aumento de la longitud focal: el aumento de las aberraciones. En general, el límite de difracción es el límite al que todo sistema óptico aspira a llegar y normalmente el más difícil de alcanzar. Para que un sistema esté limitado por difracción antes es necesario eliminar todas las aberraciones que afectan a la luz que lo atraviesa. Es la primera limitación que hay que eliminar en cualquier sistema que se pretenda optimizar y es el factor que más ha limitado la resolución desde los inicios de la óptica. En general, en casi todos los sistemas ópticos la aberración será el primer elemento presente antes de su optimización. Por ello, aumentar la aberración es el recurso más directo para obtener un spot focal de mayor diámetro.

---

[29] Para ello se precisa realizar movimientos de traslación del láser paralelos al eje óptico e ir moviendo longitudinalmente la lente de 30 mm hasta encontrar la posición de la lente que consigue el menor movimiento en la superficie del QD, lo que significa que está en el plano focal.

Existen tres tipos básicos de aberraciones: cromáticas, en eje y fuera de eje (ver apartado 2.10.5). Las cromáticas no podrán utilizarse para aumentar el spot focal, ya que el objetivo es detectar una única longitud de onda y por lo tanto no existirá ninguna aberración de este tipo. Las aberraciones fuera de eje únicamente afectan a los rayos incidentes con ángulos distintos al eje óptico. Presentan el inconveniente de que generan un spot con un grado de deformación dependiente del ángulo, por lo que en general la forma y tamaño del mismo será muy variable, con mayor extensión del spot cuanto mayor sea el ángulo con que incida el haz. Así, podría darse el caso de que para ángulos pequeños (lo que se puede equiparar con una baja intensidad de *beam wander*) el spot tenga un tamaño insuficiente en comparación con el gap, aunque sí lo tuviera con mayores ángulos. Por ello, es preferible utilizar una aberración en eje y de ellas la más importante es la aberración esférica primaria. Esta aberración no puede eliminarse en una lente única (por lo que se utilizan agrupaciones de lentes), y en caso de usarse se emplean formas especiales para reducir su efecto. Por ello, lo más adecuado para aumentar la aberración esférica será una lente única de forma perfectamente circular.

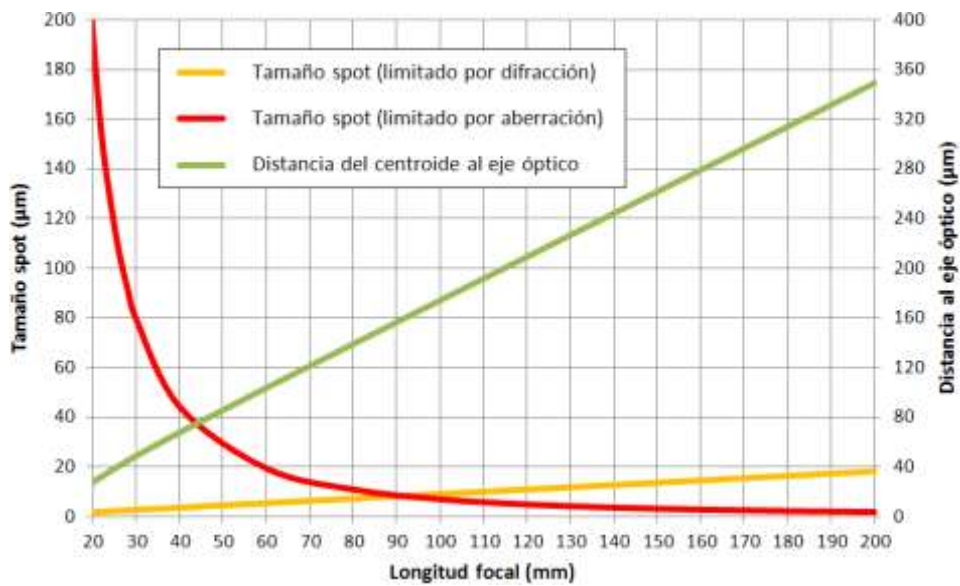

**Figura 205. Dependencia con la longitud focal del tamaño de spot focal (limitado por difracción y por aberración) y de la distancia del centroide al eje óptico.**

El efecto de la aberración esférica primaria es proporcional a la cuarta potencia del diámetro de la lente e inversamente proporcional a la tercera potencia de la longitud focal. Esto supone una ventaja adicional sobre la opción de aumentar la longitud focal, ya que el spot focal aumentará más rápidamente al reducir la longitud focal que al aumentarla (si bien en el siguiente párrafo se explicará que esta reducción tiene límites). En la Figura 205 se muestra el resultado de una simulación con óptica de rayos en *OpticsLab* para ilustrar la dependencia con la longitud focal del tamaño del spot en el foco de una lente y de la distancia del centroide al eje óptico (al aplicar un desplazamiento angular al haz). En esta simulación se utilizó un haz de 9 mm como el del experimento de los 30 metros y una lente plano-convexa de 15 mm de diámetro para acomodar un *beam wander* sin truncamiento del haz. Se muestra separadamente el límite del spot focal según el límite de difracción (ecuación (3-14)) y según la aberración esférica, por lo que el tamaño real del spot sería siempre el mayor de estos valores. Se puede comprobar cómo una reducción de la longitud focal consigue un mayor aumento del spot, al predominar la aberración, que aumenta más rápidamente. Por otra parte, se puede observar que al disminuir la distancia focal, también disminuye la distancia del centroide al eje óptico, es decir, los movimientos del haz se reducen con la focal, si bien de forma lineal. Además, en esta simulación se ha mantenido un diámetro de la lente constante, por lo que también es posible mantener la focal y aumentar el diámetro para obtener más aberración esférica.

Según se ha visto, lo recomendable será utilizar una lente con un diámetro lo mayor posible y una focal lo más corta posible, es decir lentes llamadas "lentas", con una reducida relación focal f/D. Este tipo de lentes no son comunes, precisamente por su mala calidad de imagen, y en general no se suelen encontrar lentes de relación focal menor a 2. Es decir, interesa utilizar una lente con longitud focal corta, pero entonces el tamaño también será necesariamente muy pequeño. Por ejemplo, en el experimento de los 30 metros, una lente simple plano-convexa de 10 mm de focal proporciona un spot mínimo de casi 1 mm, pero a costa de reducir su diámetro por debajo del propio haz incidente, lo que la hace inutilizable. Para utilizar una lente de un diámetro del doble del haz de llegada, la menor focal pasa a ser cercana a los 30 mm utilizados en la lente original, aunque si se elige una plano-convexa en lugar de un doblete, la aberración esférica será mayor. No obstante, incluso así, el spot focal no es mucho mayor que el tamaño del gap, por lo que no es la mejor solución. Para poder llevar a cabo una prueba de concepto de esta estrategia evitando la fabricación de una lente ad hoc, se recurrió a una solución más simple: utilizar el mismo doblete original pero invertido (girado 180°), lo que hace que el spot focal pase de las menos de 30 μm a casi 170 μm (según una simulación con óptica de rayos), lo que ya es mucho mayor que el tamaño del gap y permitirá posicionar el QD en el plano focal.

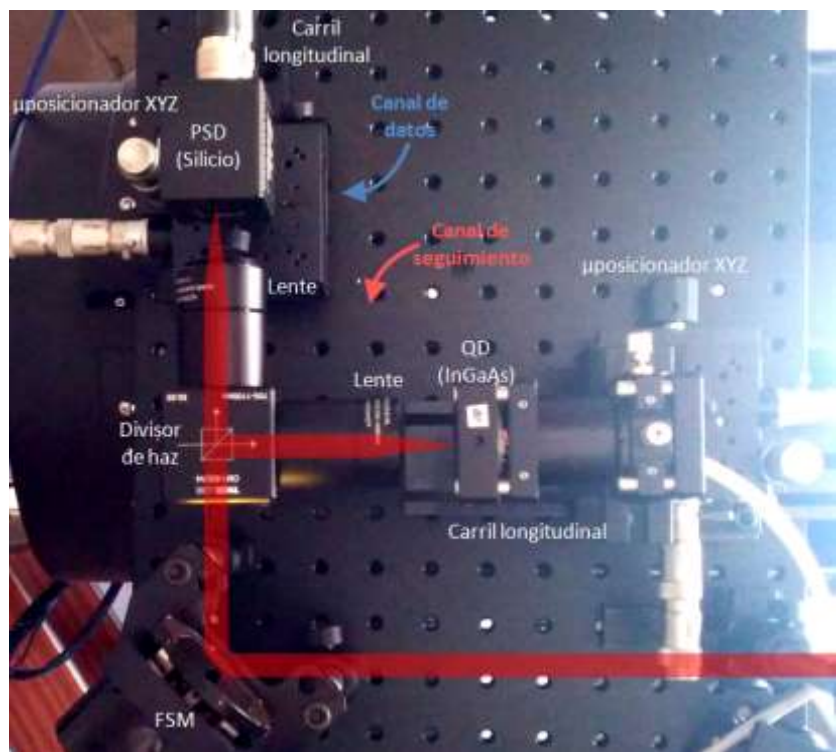

**Figura 206. Montaje óptico del sistema corrección en bucle cerrado con QD en el plano focal con aumento de la aberración.**

En la Figura 206 se muestra el montaje óptico del sistema de corrección en bucle cerrado utilizando un QD en el plano focal del canal de seguimiento, monitorizando un spot con la aberración aumentada. En los experimentos realizados con este montaje se verificó el comportamiento esperado, ya que se pudo situar el QD en el plano focal. El spot mínimo tras el doblete invertido se observa a simple vista que ha aumentado de tamaño. Además, al enfocar el QD realizando oscilaciones del haz con el FSM, se obtiene una señal de respuesta uniforme, sin las variaciones que provoca la presencia del gap (véase la Figura 150). Por otra parte, la amplitud de la señal de posición proporcionada por el QD es suficiente como para que la etapa de ganancia pueda proporcionar una señal adecuada para controlar el FSM, disponiendo de un rango adicional de ganancia disponible. Esto indica que es posible seguir aumentando el tamaño del spot focal sin necesidad de aumentar la ganancia máxima y por ello manteniendo un nivel de ruido muy por debajo del introducido en la opción del PSD.

Utilizando este montaje, fue posible realizar una buena corrección del *beam wander* natural en el experimento de los 30 metros en el canal de seguimiento. No obstante, esta estrategia presenta un inconveniente adicional que es necesario solventar. Si bien en el canal de datos se debería observar la misma corrección que en el canal de seguimiento, esto no ocurre de forma perfecta. Esto se debe a la no linealidad de la respuesta del QD ante un spot que no es mucho mayor que el gap. Si se observa la Figura 151, se puede comprobar que la zona en la que la respuesta es lineal aumenta con el tamaño del spot debido a que el efecto del gap disminuye. Al aumentar la relación de tamaños entre el spot y el gap, la energía perdida en el gap es cada vez menor y así aumenta la linealidad de la respuesta del QD. Por ello, lo ideal sería seguir aumentando el tamaño del spot hasta un tamaño aproximado a la mitad de un cuadrante para acomodar el desplazamiento debido al *beam wander* y conseguir disminuir el efecto del gap. En la prueba de concepto, según la simulación de óptica de rayos, el spot focal es unas 4 veces mayor que el gap, y según el criterio propuesto se podría conseguir un factor 17,5, por lo que hay mucho margen de mejora utilizando una lente más adecuada.

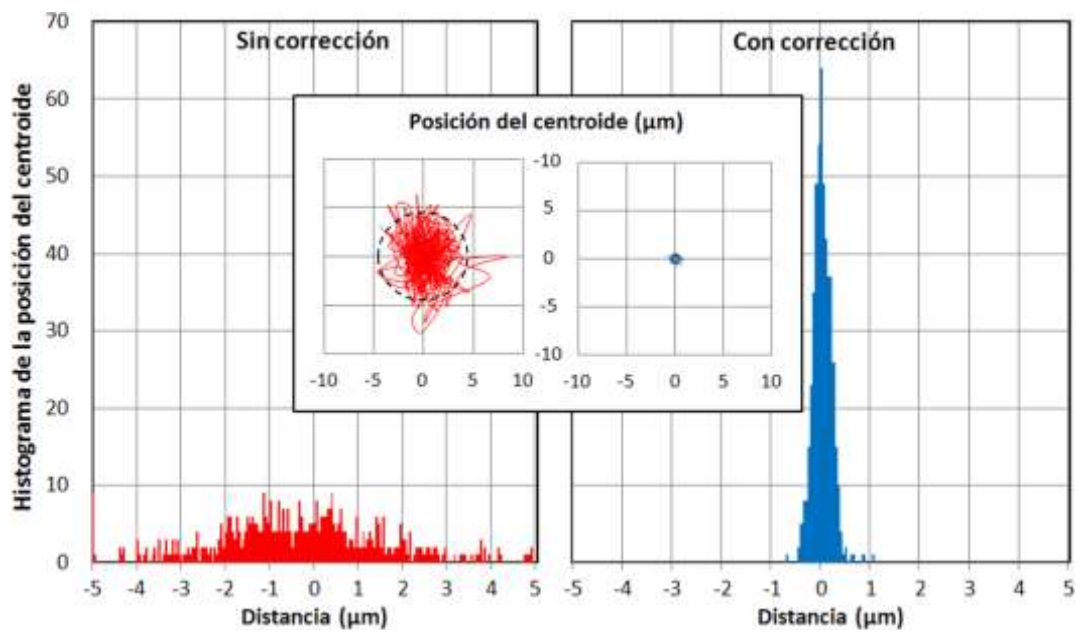

**Figura 207. Histograma de la posición del centroide en el plano focal del canal de seguimiento durante un periodo de 15 segundos sin y con corrección. Dentro, en rojo y azul la posición real del centroide en el plano focal para el mismo intervalo y en negro el diámetro que contiene el 90 % de los eventos.**

En la Figura 207 se presenta el resultado de la corrección ante la turbulencia natural del experimento de los 30 metros proporcionado por el QD en el canal de seguimiento en forma de histograma de la posición del centroide en el plano focal. Dentro, se puede observar la posición real del centroide durante el mismo intervalo de 15 segundos. Para convertir las señales de voltaje del QD en distancias se realizó una correspondencia entre la amplitud media del *beam wander* medido con el PSD C10443-02 de *Hamamatsu* (que proporciona una señal de salida calibrada de 0,5 mm/V) en el mismo plano focal donde se hicieron las medidas con el QD. En este experimento la corrección de la turbulencia proporcionó una mejora del diámetro que contiene el 90 % de los fotones de un factor 8,6 (se pasa de un diámetro de 9,5 μm sin corrección a uno de 1,1 μm con corrección), que se traduce en una disminución del área de 74 veces, el mejor de los resultados obtenidos durante la aplicación de las diferentes estrategias de corrección.

### 3.12.6. Evaluación de la potencial mejora en el sistema de QKD

Utilizando los resultados de la prueba de concepto del apartado anterior, es posible evaluar la potencial mejora que la corrección de la turbulencia aportaría sobre parámetros del protocolo cuántico como la tasa cuántica de error de bit (QBER) y la tasa de transmisión de clave secreta (SKR). Hay que señalar que el experimento de los 30 metros se realizó utilizando un montaje óptico simplificado con el objetivo de evaluar únicamente el sistema de corrección, por ello la evaluación de la mejora será solo una estimación de la que se conseguiría con el sistema completo de QKD. Por ejemplo, en el experimento no se utilizaron telescopios en Alice ni en Bob debido a la corta distancia del enlace. En Alice se utilizó un colimador acoplado a fibra que proporciona un haz de 9 mm de diámetro de salida, en lugar de los 40 mm del sistema original, lo que incrementa el efecto del *beam wander*, ya que este es inversamente proporcional al tamaño del haz (ver el apartado 3.6.3). En Bob fue suficiente con una lente de 1 pulgada de diámetro para recolectar toda la potencia del láser sin truncamiento e incluyendo todos los movimientos transversales debidos a la turbulencia. Se podría argumentar que en un enlace real la distancia de propagación sería mayor y por lo tanto también sería mayor el movimiento del centroide, lo que unido a que el haz es más grande exige usar telescopios. Sin embargo, este es un efecto opuesto al que determina el mayor diámetro del haz en un enlace real (si se observa la ecuación (3-19), se puede comprobar la relación inversa entre el tamaño de haz y el *beam wander*). Además, un mayor *beam wander* en el plano focal, donde su amplitud ya está muy atenuada, se puede afrontar aumentando el área activa del QD y ajustando las constantes del PID para optimizar la respuesta del FSM ante una mayor amplitud. Por estas razones, en lugar del factor de corrección de ~9 obtenido en la prueba de concepto experimental, se decidió utilizar un rango más realista en la estimación de la potencial mejora en el protocolo cuántico: como caso peor se consideró un factor de corrección igual a 5 y uno igual a 10 como caso mejor.

Dado que el rango de corrección {5-10} se refiere a la distancia del centroide al eje óptico, este factor se debería aplicar sobre el parámetro $r_c$ (introducido en el apartado 3.6.3). Sin embargo, este parámetro está referido al final del trayecto de propagación, justo a la entrada del terminal receptor, y el factor de corrección está calculado en el plano focal de Bob. Por ello, será necesario trasladar $r_c$ al plano focal, parámetro al que se denominará $r_c'$. Además, para realizar la estimación de mejora sobre el protocolo cuántico, el parámetro que interesa conocer es el radio de largo plazo $W_{LT}$, también trasladado al foco como $W_{LT}'$, que es el que permite evaluar cómo disminuye el área focal con la corrección del *beam wander*. Así, $W_{LT}'$ podrá calcularse a partir de $r_c'$ y de $W_{ST}'$ según la ecuación (3-17), para lo cual será necesario también calcular el radio de corto plazo en el foco $W_{ST}'$. Todo esto se hará considerando regímenes de turbulencia de intensidad media ($C_n^2 = 10^{-15}$ m$^{-2/3}$) a alta ($C_n^2 = 10^{-13}$ m$^{-2/3}$), por corresponderse con los casos donde más interesa realizar la corrección de la turbulencia. Además, utilizando la ecuación (3-19), conociendo la distancia de propagación L y el tamaño del haz transmitido $W_0$, a partir de las medidas del *beam wander* tomadas con el PSD en el experimento de los 30 metros, se pudo estimar un régimen de turbulencia fuerte.

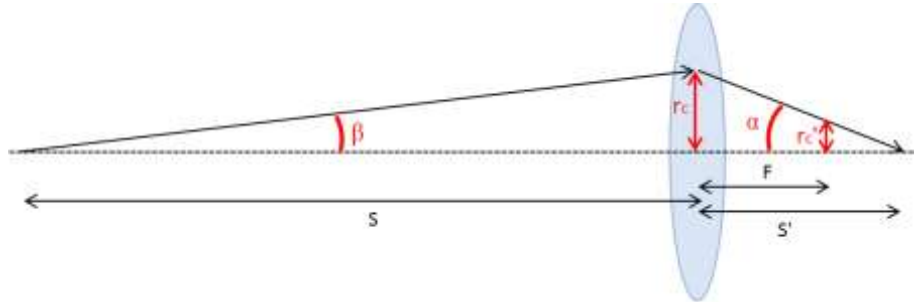

**Figura 208. Traslación de $r_c$ al plano focal para obtener $r_c'$.**

Para llevar a cabo el procedimiento anterior es necesario trasladar los parámetros $r_c$ y $W_{ST}$ de la entrada de Bob a $r_c'$ y $W_{ST}'$ en el plano focal. En la Figura 208 se ilustra la proyección de $r_c$ hacia $r_c'$. Utilizando las ecuaciones (3-66) y (3-67), deducidas a partir de la Figura 208, a través de la ecuación (3-68), es posible llegar a la relación entre $r_c$ y $r_c'$ de la ecuación (3-69). Esta ecuación permite trasladar desde la entrada del sistema óptico una distancia o movimiento $r_c$ al plano focal, obteniendo $r_c'$, y partiendo de la distancia de propagación S y de la longitud focal equivalente F. Para introducir F en esta ecuación, se utilizó la distancia focal del sistema óptico simulado en el apartado 3.10, incluyendo el telescopio y la lente de focalización, lo que medido en *OpticsLab* proporciona una focal equivalente de F = 1,68 m. Por equivalencia, la misma expresión es válida para calcular $W_{ST}'$ sin más que sustituir $r_c'$ por $W_{ST}'$ y $r_c$ por $W_{ST}$.

$$\tan(\alpha) = \frac{r_c'}{S'-F} = \frac{r_c}{S'} \tag{3-66}$$

$$\tan(\beta) = \frac{r_c}{S} \tag{3-67}$$

$$\frac{1}{F} = \frac{1}{S} + \frac{1}{S'} \tag{3-68}$$

$$r_c' = r_c \frac{(S-F)^2}{F^3 S} \tag{3-69}$$

Utilizando la ecuación (3-19), es posible calcular el *beam wander* $r_c$ a la entrada de Bob en función del régimen de turbulencia, la distancia de propagación L y el tamaño del haz transmitido $\omega_0$ (esta vez de 20 mm de radio, el haz empleado en el sistema final). Mediante la ecuación (3-69), se traslada al plano focal para obtener $r_c'$, utilizando la distancia focal equivalente F de Bob, y ya es posible aplicar el rango de reducción {5—10} sobre $r_c'$ para introducir la corrección. Para obtener $W_{LT}'$, falta calcular el parámetro $W_{ST}$ relacionado con el ensanchamiento del haz debido solo a efectos difractivos de la turbulencia y por la propagación en espacio libre. Esta contribución se calcula mediante la ecuación (3-17), restando de $W_{LT}$ la contribución de $r_c$. Una vez obtenido $W_{ST}'$, trasladado al plano focal el parámetro $W_{ST}$, ya es posible calcular $W_{LT}'$ sin y con corrección (para lo cual en el segundo caso hay que reducir el $r_c'$ según el rango {5—10} antes de añadirlo al $W_{ST}'$). En la Figura 209 se muestra el diámetro focal de largo plazo $2W_{LT}'$ en función de la distancia de propagación sin corrección (línea delgada) y con corrección (línea gruesa, representando el rango de mejora) para los tres regímenes de turbulencia mencionados anteriormente. Cada régimen de turbulencia está representado hasta la distancia máxima de aplicación que se calculó en el apartado 3.7.5, siendo de 800 metros, 1650 metros y 2450 metros para $C_n^2 = 10^{-13}$ m$^{-2/3}$, $C_n^2 = 10^{-14}$ m$^{2/3}$ y $C_n^2 = 10^{-15}$ m$^{2/3}$ respectivamente.

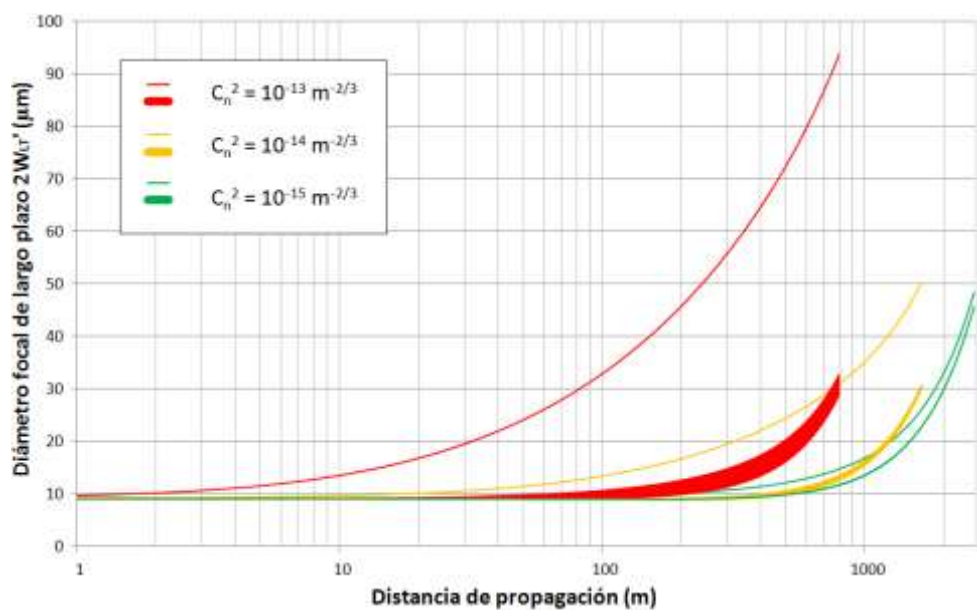

**Figura 209. Diámetro focal de largo plazo en función de la distancia de propagación sin corrección (línea delgada) y con corrección (línea gruesa) para tres regímenes de turbulencia distintos.**

En la Figura 210 se muestra la mejora del área en el plano focal obtenida por el sistema de corrección en función de la distancia de propagación. El área es el parámetro más relevante para evaluar la mejora que proporciona la corrección porque es proporcional a la potencia recibida, tanto de señal como de ruido. Recuérdese que el objetivo del sistema de corrección es disminuir la potencia de ruido acoplada para disminuir la tasa de errores del protocolo cuántico. Por una parte, se puede observar cómo la corrección aumenta con la intensidad de la turbulencia sufrida por el enlace. Esto se explica debido a que la potencial mejora es menos apreciable cuanta menos turbulencia afecta al enlace: en el caso extremo, un sistema de corrección de turbulencia proporcionaría una escasa mejora a un enlace sin apenas turbulencia. Por otra parte, se aprecia cómo en general al aumentar la distancia, la mejora del sistema de corrección es mayor debido a que la intensidad del *beam wander* aumenta con la distancia. Llegado a un punto, ya no sigue aumentando porque el tamaño del haz recibido, que también aumenta con la distancia, hace que la corrección de su posición sea menos relevante al disponer de menor movimiento disponible.

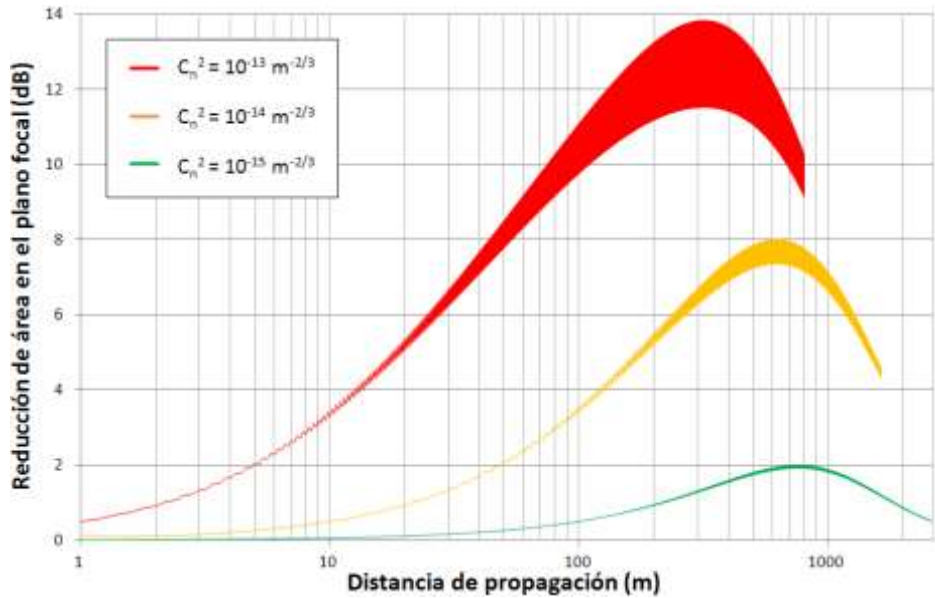

**Figura 210. Mejora en el área focal en función de la distancia de propagación para tres regímenes de turbulencia distintos con sus distancias máximas correspondientes para un factor de corrección de la turbulencia de {5–10}.**

Para estimar los parámetros QBER y SKR del protocolo cuántico se utilizaron datos experimentales recogidos por el sistema de QKD sin corrección de turbulencia en el experimento de los 300 metros descrito en el apartado 3.7.1. El QBER se puede calcular como un término fijo más un término variable según la ecuación (3-70) [288, p. 361], donde $\kappa$ es la contribucion fija al error, debida a los componentes pasivos en el sistema (por ejemplo, la contribución debida a los desalineamientos de la polarización en las reflexiones por espejos y transmisiones por componentes como espejos dicroicos, láminas retardadoras, etc.). La componente variable está compuesta por D, que es la tasa de error de detección en fotones por segundo en ausencia de transmisión de datos (debida principalmente al ruido ambiental y a las cuentas oscuras de los detectores) y B es la tasa de fotones por segundo recibidos en el receptor durante la transmisión de datos.

$$QBER = \kappa + \frac{D}{2B} \tag{3-70}$$

El término fijo $\kappa$ se calculó como un promedio de las medidas de QBER en ausencia total de luz ambiental, durante la noche, y resultó ser 2,14 %. Para el término variable del QBER se simuló D y B a partir de datos reales medidos con el sistema a 300 metros: en concreto, los medidos a la 13:30 h, con máxima tasa de error D de 190000 cuentas por segundo y 1,7 millones de cuentas por segundo de tasa de datos recibidos B. Para estimar la tasa B recibida a distintas distancias se utilizó el código de transmisión atmosférica MODTRAN (para una atmósfera con aerosol urbano y visibilidad de 5 km, que proporciona una absorción de 2,048 dB/km). De esta forma, es posible simular la tasa B a distintas distancias partiendo del dato real a 300 metros. Por otro lado, se asumió que el diámetro del núcleo de la fibra óptica acoplada al SPAD en Bob es igual a $2W_{LT}'$ corregido (calculado previamente). Por lo tanto, partiendo del dato experimental medido con una fibra de diámetro 62,5 μm a 300 metros, es fácil estimar la ganancia o pérdida que se obtiene utilizando una fibra de un diámetro distinto $2W_{LT}'$. Esta ganancia o pérdida se aplica sobre las tasas D y B y se puede calcular el QBER en función de la distancia sin corrección (considerando solo la atenuación atmosférica) y con corrección (considerando la atenuación atmosférica y el efecto del *beam wander*). La SKR se estimó mediante el mismo procedimiento que en [260], pero utilizando las D y B simuladas anteriormente.

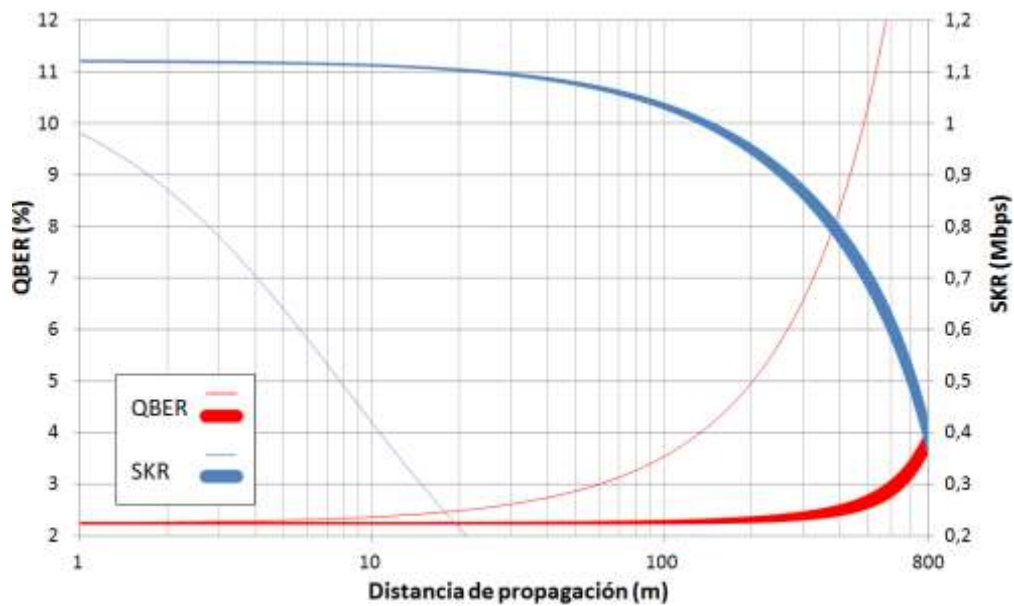

**Figura 211. Tasa cuántica de error de bit (QBER) y tasa de clave secreta (SKR) sin corrección (línea delgada) y con corrección (línea gruesa) para un régimen de turbulencia fuerte ($C_n^2 = 10^{-13}$).**

En la Figura 211 se muestra el resultado de los cálculos descritos para obtener QBER y SKR sin corrección (línea delgada) y con corrección (línea gruesa) bajo un régimen de turbulencia fuerte ($C_n^2 = 10^{-13}$ m$^{-2/3}$), que es el más parecido al régimen en el que se realizaron los experimentos. Se puede comprobar cómo el QBER mejora significativamente con la corrección del *beam wander* al aumentar la distancia de propagación, de igual modo que el SKR. En este régimen de turbulencia, el QBER se ve reducido hasta en un 80 % y el SKR se consigue mantener hasta en tres órdenes de magnitud por encima del equivalente sin corrección, que se degrada rápidamente al aumentar la distancia de propagación. En otros regímenes menos turbulentos, la potencial mejora es menor: el QBER podría reducirse hasta en un 50 % y un 10 % respectivamente para régimen de turbulencia media y débil, y el SKR se mantiene hasta dos órdenes de magnitud por encima del caso sin corrección para turbulencia media y un orden de magnitud para turbulencia débil.

# 3.13. SISTEMA DE CORRECCIÓN DOBLE

## 3.13.1. Discusión de la estrategia

El mayor inconveniente de las estrategias explicadas en los apartados anteriores es que todas pretenden eliminar el *beam wander* sin estabilizar el haz por completo. Lo que se persigue es estabilizar un único plano de los infinitos que constituyen el eje óptico, y esta es la causa de la larga serie de dificultades que se han ido detallando. Sin embargo, existe una estrategia que permite eludir estos inconvenientes, precisamente por tener como objetivo la estabilización del haz a lo largo de todo un trayecto y no en un solo plano. Esta estrategia está inspirada en las técnicas de alineamiento de láser empleadas en los montajes ópticos de laboratorio y consiste en la utilización de dos espejos, en lugar de solo uno. La utilización de dos espejos permite que cualquier haz pueda alinearse simultáneamente en dos planos distintos, lo que consigue la estabilización completa del haz.

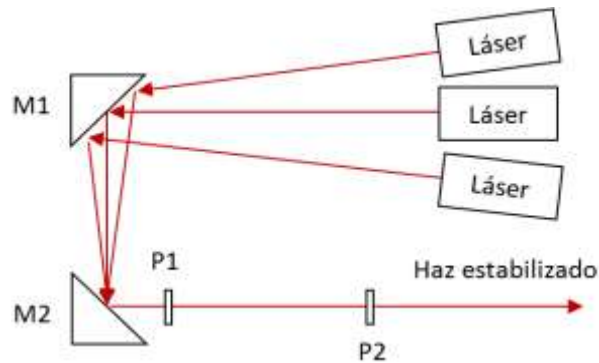

**Figura 212. Técnica de alineamiento *walking the beam*.**

En la Figura 212 se muestra un esquema básico de la técnica conocida en inglés como *walking the beam*. El montaje consiste en dos espejos M1 y M2 y en dos *pinholes* P1 y P2 (aunque estos pueden ser sustituidos por dos pantallas para observar la posición del haz). Esta técnica permite que cualquier haz que incida sobre el espejo M1 sea alineado sobre un mismo eje óptico a la salida de M2, a través de una combinación de movimientos angulares en los dos ejes (altitud y acimut) de ambos espejos. El espejo M2 controlaría únicamente el ángulo de salida del haz que pivotaría sobre un punto fijo sobre su superficie. Los ajustes sobre el espejo M1 tendrían dos efectos distintos. El más importante es la traslación del haz que resulta en un desplazamiento del punto sobre el que pivota el haz sobre la superficie del espejo M2. Sin embargo, la traslación no es pura porque el espejo M1 también modifica el ángulo de incidencia del haz sobre M1, cambiando el ángulo que tiene a la salida de M2.

El nombre de la técnica, que podría traducirse como "acompañar al haz", se debe al hecho de que el alineamiento se consigue tras un proceso iterativo. La necesidad de este proceso iterativo se debe al doble efecto que introduce el espejo M1. Esta técnica permite alinear cualquier haz con un montaje determinado tras solo unas pocas iteraciones (dependiendo de la resolución precisada), que tras algo de práctica se realizan de forma simultánea con una mano en cada espejo. El procedimiento consiste en centrar el haz sobre el *pinhole* P1 actuando sobre el espejo M1 y luego corregir en ángulo con el espejo M2 centrando el haz sobre P2. Como al actuar sobre M2, se habrá modificado el ángulo de salida, el haz ya no estará centrado en P1, por lo que es necesario repetir el proceso varias veces.

La separación entre los espejos M1 y M2 influye en el alineamiento del haz: una mayor distancia hace que predomine en mayor medida el efecto de la traslación sobre el haz que provoca M1 sobre el efecto angular. En la Figura 213 se muestra el resultado de una simulación con óptica de rayos del promedio de las posiciones del centroide del haz en los dos planos P1 y P2, que se obtiene para una distancia genérica entre los espejos M1 y M2 igual a 1, para el doble y para diez veces la misma. Se puede observar que en solo seis iteraciones esta técnica reduce la distancia del centroide al eje óptico en hasta tres órdenes de magnitud (dependiendo de la separación entre espejos). Además, se comprueba que esta reducción aumenta con la distancia entre espejos, o de forma equivalente, hacen falta menos iteraciones para alcanzar una misma alineación.

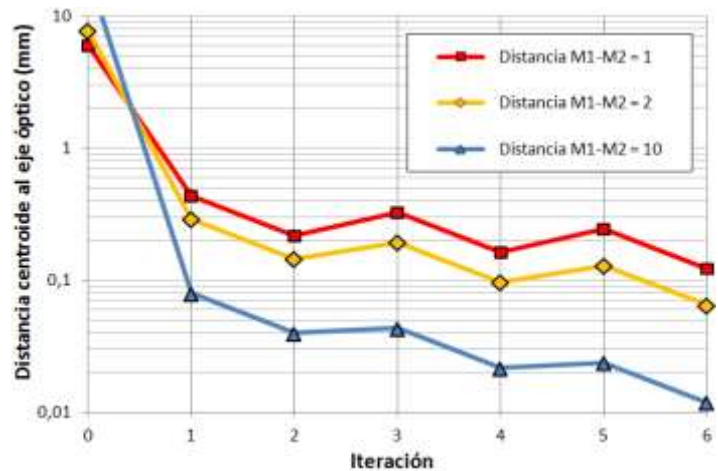

**Figura 213. Influencia de la distancia entre espejos en la estabilización del haz.**

La estrategia de doble corrección consigue alinear un haz a lo largo de todo el camino óptico que se inicia a la salida del espejo M2. Por una parte, esto elimina todos los problemas de la estrategia de corrección en bucle abierto, vista en el apartado 3.11. Recuérdese que esta estrategia solo permitía corregir un tipo de perturbación en el haz (las debidas a movimientos o bien angulares o bien de traslación) y solo la originada en un punto del espacio determinado. La corrección con dos espejos consigue asegurar que se eliminen ambos tipos de perturbaciones e independientemente de dónde se originen. Por otra parte, esta estrategia también elimina los inconvenientes de la corrección en bucle cerrado, vista en el apartado 3.12. Esta estrategia eliminaba las perturbaciones de cualquier tipo (angulares y de traslación) y provenientes de cualquier origen, pero únicamente en un plano del eje óptico. Esto presentaba muchos problemas de implementación, debido especialmente a la dificultad de utilizar un PSD o un QD en el plano focal, condición necesaria ya que las fibras ópticas conectadas a cada SPAD en Bob irán situadas en el plano focal del canal de datos. Con la estrategia de doble corrección, se verá que la corrección del canal de datos se puede conseguir en cualquier punto. Esto permitirá situar la óptica de focalización hacia las fibras ópticas acopladas a los SPAD en cualquier punto, evitando la compleja calibración necesaria en la corrección en bucle cerrado.

### 3.13.2. Implementaciones y prueba de concepto

En la Figura 214 se muestran dos posibles montajes para implementar la estrategia de la estabilización del haz mediante doble corrección. Tomando como referencia el montaje de la Figura 212, los espejos M1 y M2 se implementarían mediante los espejos FSM 1 y FSM 2 y los pinhole P1 y P2 mediante los detectores de posición QD 1 y QD 2, utilizando sendos controles PID 1 y PID 2 para conseguir alinear el haz en cada QD. El resultado final es idéntico utilizando ambas implementaciones ya que las dos consiguen estabilizar a la salida del FSM 2 cualquier haz de entrada incidente sobre el FSM 1. Los dos montajes se basan en corregir mediante el FSM 1 el movimiento de traslación del haz sobre la superficie del FSM 2 y mediante el FSM 2 el ángulo de salida. Para ello, cada QD controla los movimientos del FSM correspondiente mediante un bucle de realimentación PID.

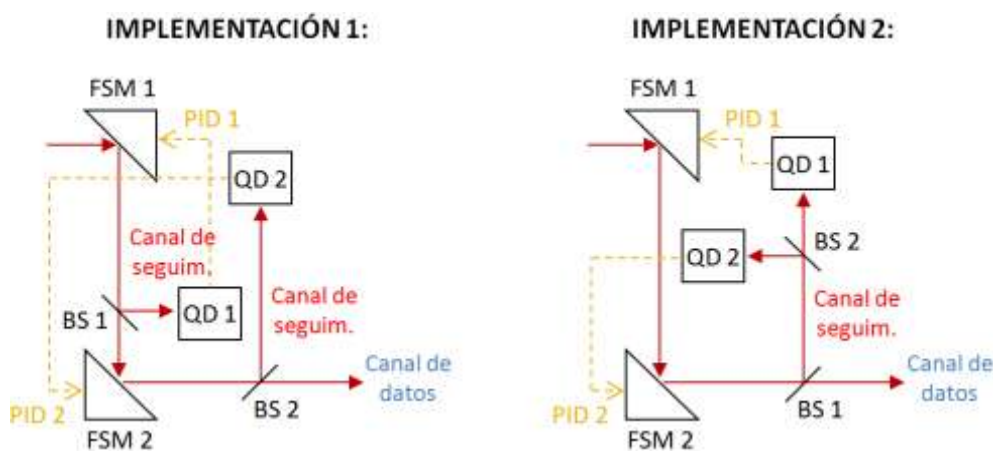

Figura 214. Diferentes implementaciones de la estrategia de doble corrección.

En la técnica explicada en el apartado 3.13.1 se requería seguir de forma manual un protocolo secuencial e iterativo hasta conseguir el alineamiento. Para adaptar la técnica al sistema de corrección de *beam wander* con el objetivo de automatizar su operación, se diseñó una estrategia consistente en independizar la corrección que lleva a cabo cada control PID. Esta estrategia precisa una calibración muy simple, que consiste en determinar la posición del QD 1 que lo hace insensible a los movimientos del FSM 2. Para conseguirlo, es necesario formar en el QD 1 la imagen del haz incidente en la superficie del FSM 2 y así los movimientos del FSM 1 no interfieren en el bucle PID 2 que se dedica a controlar únicamente el ángulo del haz de salida. Al crear una imagen de FSM 2 en QD 1 se consigue que el PID 1 imponga en la superficie del FSM2 un primer plano estabilizado. El segundo plano estabilizado lo proporcionará el bucle PID 2 y con dos planos donde el haz está inmóvil se consigue estabilizarlo por completo a la salida del sistema de corrección.

Aunque ambas implementaciones consiguen su propósito, la implementación 1 no es compatible con el sistema de QKD diseñado. Se puede comprobar que en la implementación 1 el canal de datos atraviesa dos divisores de haz y en la implementación 2 solo uno, si bien el canal de seguimiento atraviesa los dos divisores en ambas implementaciones. Conviene recordar que el canal de seguimiento funciona a una longitud de onda diferente al canal de datos, por lo que para no introducir pérdidas en el canal de datos, se hace necesaria la utilización de un espejo dicroico para su división. Si se separan ambos canales en BS1 (implementación 1) utilizando un espejo dicroico, y utilizando otro similar en BS 2, entonces no quedará potencia para detectar el canal de seguimiento en el QD 2. Si en BS1 se utiliza un divisor de haz convencional (en cuyo caso en BS 2 se utilizaría un espejo dicroico), se estarían introduciendo pérdidas sobre el canal de datos que degradarían el rendimiento del protocolo cuántico. Sin embargo, la implementación 2 permite utilizar un espejo dicroico en el lugar del divisor de haz BS 1 y así no introducir ninguna pérdida adicional sobre el canal de datos. Por ello, esta implementación es la elegida para integrarla en el sistema de QKD [30].

---

[30] Cabe mencionar una posible solución para utilizar la implementación 1 en un sistema de QKD: utilizar en BS 1 un espejo dicroico capaz de transmitir al 100 % la longitud de onda del canal de datos y al 50 % la del canal de seguimiento. La inexistencia de tal componente óptico de forma comercial exigiría su fabricación ad-hoc, lo que unido a que la implementación 1 no aporta en principio ninguna ventaja sobre la implementación 2, hace que esta última sea la opción preferida.

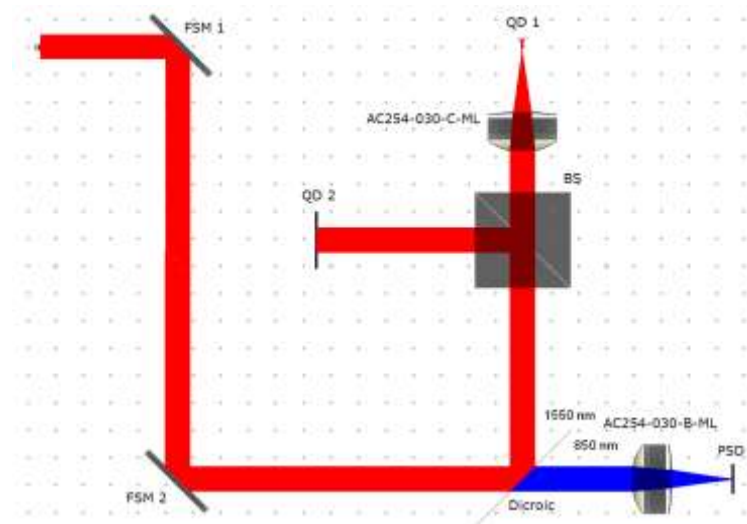

**Figura 215. Montaje en *OpticsLab* para la prueba de concepto.**

En la Figura 215 se muestra una simulación del montaje óptico de la implementación 2, llevado a cabo en *OpticsLab* para simular una prueba de concepto utilizando óptica de rayos. El haz de entrada, con un diámetro de 9 mm, está compuesto por las dos longitudes de onda de los canales de datos (850 nm, en azul) y de seguimiento y sincronismo (1550 nm, en rojo). Para la formación en QD 1 de la imagen del haz en la superficie de FSM 2 se modeló la lente AC254-030-C-ML de *Thorlabs* empleada en los experimentos, que se trata de un doblete acromático de 30 mm de focal optimizado para el rango 1050-1620 nm. El QD 2 es un detector de 2 cm de diámetro sin óptica de focalización, ya que no es necesario en términos del alineamiento del haz. No obstante, sí podría ser necesario utilizar una lente en caso de que el área del QD 2 fuera menor a los movimientos del spot. Para focalizar el haz del canal de datos, tras separar la longitud de onda de 850 nm mediante el espejo dicroico, se modeló una lente AC254-030-B-ML, de características equivalentes al otro doblete pero optimizado [31] para el rango 650-1050 nm. En el plano focal de esta lente, en el lugar donde iría el extremo de la fibra óptica acoplada al SPAD, se colocó un PSD para monitorizar los movimientos del haz, que tiene unas 20 μm de diámetro. La única calibración que precisa este sistema es la descrita al comienzo de este apartado: la determinación de la posición del QD 1 que hace que no detecte los movimientos del FSM 2. En esta posición del QD 1, el haz tiene un diámetro de 1360 μm (mucho mayor que el correspondiente al plano focal), que es un tamaño ideal para monitorizar sus movimientos con el QD de 3 mm, tal como se explicó al final del apartado 3.12.5.

Sobre el sistema de la Figura 215 se programó en *OpticsLab* un bucle de realimentación entre la posición del centroide del QD 1 y el ángulo de rotación del FSM 1, y equivalentemente entre QD 2 y FSM 2, mediante un control puramente integral (suficiente para realizar la corrección). Una vez ajustadas sendas constantes I para cada bucle de control, este sistema es capaz de llevar a cabo la estabilización de cualquier haz de entrada mediante ambos bucles de forma simultánea, proporcionando un haz estabilizado a la salida del FSM 2. Para realizar la prueba de concepto se programó una perturbación periódica sobre la posición y el ángulo del haz de entrada. La forma de la perturbación es una señal cuadrada para observar claramente el comportamiento del sistema de corrección ante cambios repentinos en la posición y ángulo del haz de entrada. En la parte izquierda de la Figura 216 se muestra la respuesta del sistema ante la perturbación angular y en la parte derecha la respuesta ante la perturbación en posición. Se puede comprobar que tras el correspondiente transitorio, el sistema proporciona una señal de posición estabilizada tanto en cada uno de los QD como en el haz de salida, monitorizado por el PSD. La diferencia de amplitudes entre las señales de ambos QD se debe a la ausencia de focalización en el QD 2, lo que hace que los movimientos detectados sean mayores. Sobre la implementación de este sistema, esta diferencia se manifestaría en que entre el bucle QD 1 y FSM 1 se precisaría una ganancia distinta a la del bucle QD 2 y FSM 2.

---

[31] La optimización de un doblete en un rango espectral determinado se refiere tanto al recubrimiento anti-reflectante como a los parámetros del diseño óptico que proporcionan la longitud focal nominal.

Nótese que la lente AC254-030-B-ML está colocada en una posición arbitraria tras el espejo dicroico. En cualquier posición del eje óptico la señal del PSD estaría igualmente estabilizada ya que todo el haz lo está, lo que constituye la gran ventaja de esta estrategia de corrección. Con la estrategia de corrección en bucle abierto (apartado 3.11), una corrección como esta es imposible de conseguir, ya que el sistema solo puede calibrarse para perturbaciones de posición o de ángulo, pero no de ambos tipos a la vez; y la estrategia de corrección en bucle cerrado (apartado 3.12) únicamente podría conseguir esta corrección de forma ideal cuando el QD y el PSD se situaran en el mismo plano focal, con todos los problemas que conlleva situar un QD en el plano focal, y además el sistema sería muy sensible a su calibración.

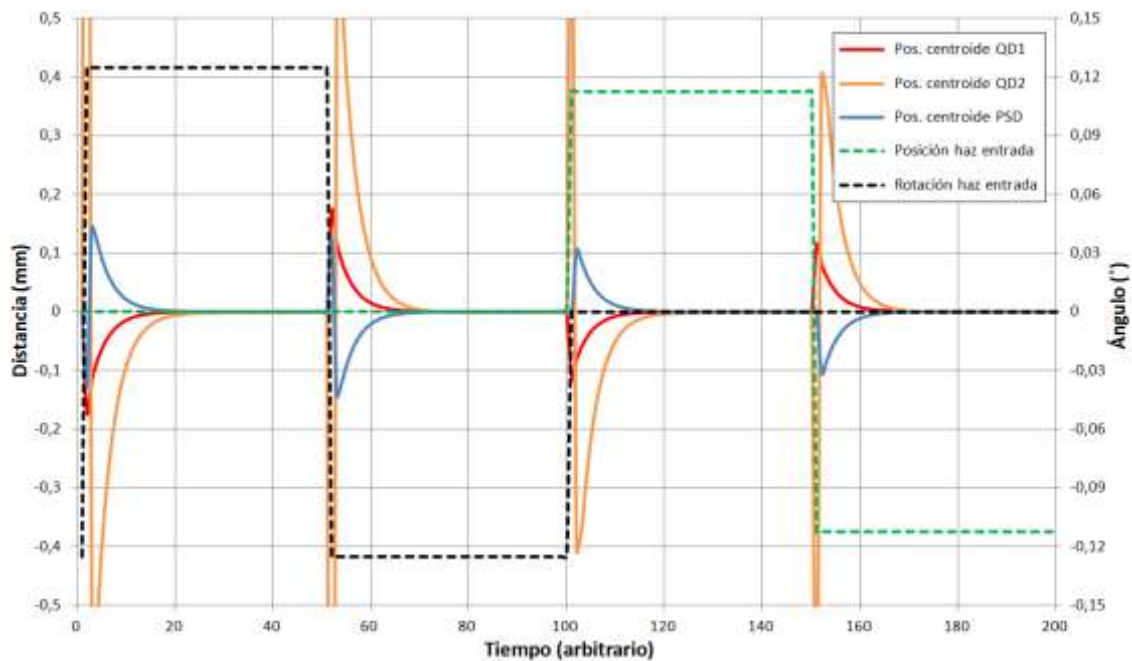

**Figura 216. Simulación de la posición del centroide en los QD del sistema de corrección y del centroide del PSD para el haz estabilizado ante perturbaciones del haz en posición y ángulo.**

## 3.13.3. Integración con el sistema de QKD

Una vez realizada la demostración de concepto, y al no haber sido posible montar un experimento de laboratorio por falta de material, se llevó a cabo una simulación del sistema de corrección de *beam wander* integrado en el sistema de QKD, para realizar un diseño lo más óptimo posible. El diagrama de bloques básico del sistema completo, incluyendo la corrección de turbulencia, se muestra en la Figura 217. Como el canal de seguimiento y el de sincronismo utilizan la misma longitud de onda, el canal de sincronismo se puede separar utilizando un divisor de haz convencional (BS 2). Además, como esta separación se realiza tras el FSM 2, el canal de sincronismo también estará estabilizado. Esto permite aprovechar mejor la potencia óptica, lo que es especialmente importante en el montaje de doble corrección ya que se añade una división adicional respecto a las estrategias anteriores, al precisarse dos caminos distintos para monitorizar los movimientos del haz en el canal de seguimiento:

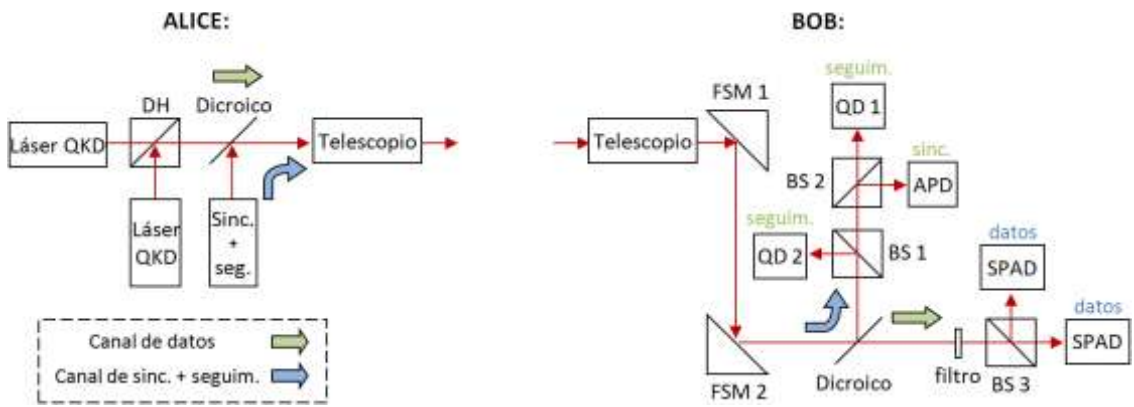

**Figura 217. Diagrama de bloques del sistema de QKD con el montaje para corrección doble.**

Como se explicó en el apartado 3.10.8, el montaje de Bob está situado en la parte superior del telesopio (ver Figura 182). Para llevar el haz recibido y focalizado por el telescopio de Bob hasta este montaje, en primer lugar es necesario colimarlo. Para ello, se utilizó una reducción de haz tipo Kepler (Figura 218), consistente en dos lentes convergentes con focos coplanares. Con el telescopio actuando como primera lente ($L_1$), se utilizó un doblete invertido como segunda lente ($L_2$).

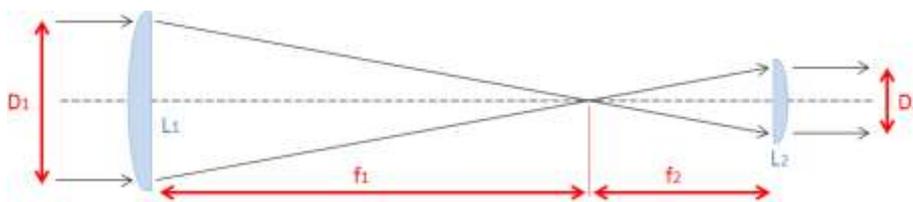

**Figura 218. Reductor de haz tipo Kepler.**

Para obtener un haz de unos 4 mm de diámetro ($D_2$), compatible con las dimensiones de la óptica empleada en Bob, la reducción debe tener un factor ~10 si se asume un haz de entrada al telescopio de unos 4 cm de diámetro ($D_1$) [32]. Utilizando la ecuación (3-71) y asumiendo una focal del telescopio de 2500 mm, la longitud focal de la segunda lente debería ser de 250 mm. El valor más próximo a 250 mm para un doblete comercial es de 300 mm y por ello se eligió la lente AC254-300-B-ML de *Thorlabs* como L2. Este doblete está situado en la zona del montaje encima del telescopio, por lo que para hacer coincidir ambos focos, es necesario modificar la separación entre el espejo primario y secundario del telescopio para aumentar la longitud focal equivalente del mismo. Con una reducción de 10 y una focal para la segunda lente de 300 mm, se precisará en el telescopio una focal equivalente de 3000 mm, para lo que es necesario separar los espejos unos 37 cm (ver apartado 3.10.7). Tras introducir esta separación en la simulación y ajustar la posición del doblete, se obtuvo un haz de salida de 4,01 mm.

---

[32] Este spot de llegada al telescopio desprecia el efecto de la divergencia del haz debida a la propagación por espacio libre, que es muy reducida hasta algunos cientos de metros (por ejemplo, a 300 m el spot ha aumentado solo 2 mm en relación a sus 4 cm de diámetro). Un haz más ancho en Bob (sea debido a una mayor distancia y/o a un menor tamaño del haz de salida de Alice) precisaría una reducción mayor, que es posible conseguir gracias a la capacidad de enfoque del telescopio (ver apartado 3.10.7) y al consiguiente desplazamiento de la segunda lente para hacer coincidir ambos focos.

$$\text{factor de reducción} = \frac{D_1}{D_2} = \frac{f_1}{f_2} \tag{3-71}$$

El doblete AC254-300-B-ML está optimizado en el rango de 650 a 1050 nm. Si bien el haz que atraviesa esta lente está compuesto por dos longitudes de onda, el hecho de estar optimizado en el rango 650-1050 nm, únicamente influirá en que el canal de 1550 nm sufrirá algunas pequeñas pérdidas adicionales debidas a reflexiones en la lente. Sin embargo, en el canal de 1550 nm se dispone de potencia suficiente, a otra longitud de onda precisamente para no interferir con el canal de datos, y el objetivo de este diseño óptico es no introducir pérdidas en el canal de 850 nm.

En la Figura 219 se muestra el modelo de *OpticsLab* del sistema de QKD con el sistema de corrección integrado. La simulación está hecha a escala, con todas las distancias reales. Toda la parte del montaje óptico, que comienza en la lente L2, iría rotada 90° en torno a su eje óptico en la implementación real. En la simulación se ha mantenido el mismo plano por claridad y simplicidad, ya que no afecta a las distancias entre elementos.

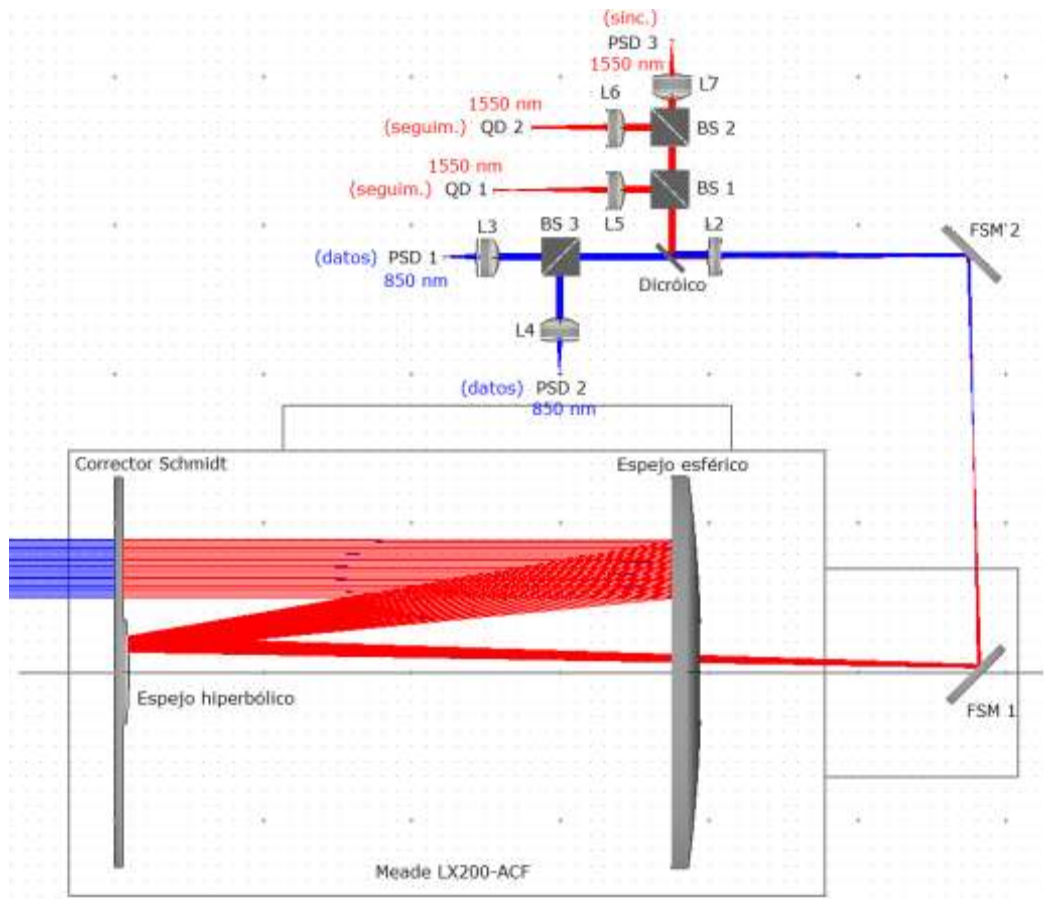

**Figura 219. Simulación en *OpticsLab* del montaje óptico del sistema de QKD con el sistema de corrección de *beam wander* integrado.**

El haz colimado a la salida de L2 (AC254-300-B-ML) tiene un diámetro de 4 mm, que con los dobletes L3 y L4 (AC254-030-B-ML), se focaliza hacia las fibras ópticas hasta un tamaño de 7,23 µm. Este spot está limitado por difracción y no por aberraciones (ya que el spot focal debido solo a aberraciones es igual a 3,71 µm). Este tamaño tan reducido permitirá disminuir aún más el campo de visión y así el ruido de fondo acoplado al sistema.

Por otra parte, el utilizar un haz tan estrecho con el objetivo de minimizar las aberraciones en el canal de datos hace al spot focal muy pequeño también en el canal de seguimiento. Además, como el QD 1 está situado en el plano imagen del FSM 2, y el plano focal y el plano imagen están muy próximos al ser la distancia entre el FSM 2 y el QD 1 relativamente grande, el spot que llega al QD 1 es demasiado pequeño en relación al gap. El objetivo es aumentarlo para minimizar los efectos del gap en la detección de los movimientos del haz. Para conseguirlo se utilizó el doblete L5 (AC254-075-C-ML), que proporciona una longitud focal de 75 mm, la mayor focal que permiten las dimensiones del montaje óptico. La razón de elegir una longitud focal lo más larga posible es conseguir un mayor spot en el plano imagen de la lente L5. Esto se puede comprobar con la ecuación (3-72), donde se observa que al aumentar la focal F, aumenta el tamaño de la imagen y' (el spot en el QD 1) en relación al objeto y (el haz en la superficie del FSM 2), siempre que S (distancia de la lente al objeto) sea mayor que F, como es el caso (es decir, debe ser una imagen real, no virtual). La ecuación (3-72) se puede deducir fácilmente a partir de la ecuación (3-68) y de la ecuación (3-73), que define la magnificación de una lente.

$$y' = y\left(\frac{F}{S - F}\right) \qquad\qquad (3\text{-}72)$$

$$M = \frac{y'}{y} = \frac{S'}{S} \qquad\qquad (3\text{-}73)$$

La ecuación (3-72) demuestra que al aumentar la focal, el spot en el plano imagen aumentará de tamaño. Sin embargo, esto no tiene en cuenta el efecto de las aberraciones, que es el opuesto: disminuyen, reduciendo el tamaño del spot, al aumentar la focal. En este caso, con un haz de entrada tan estrecho y una focal tan larga, la lente está limitada por difracción, por lo que prevalece la primera relación y todo aumento de la focal provocará un aumento del spot.

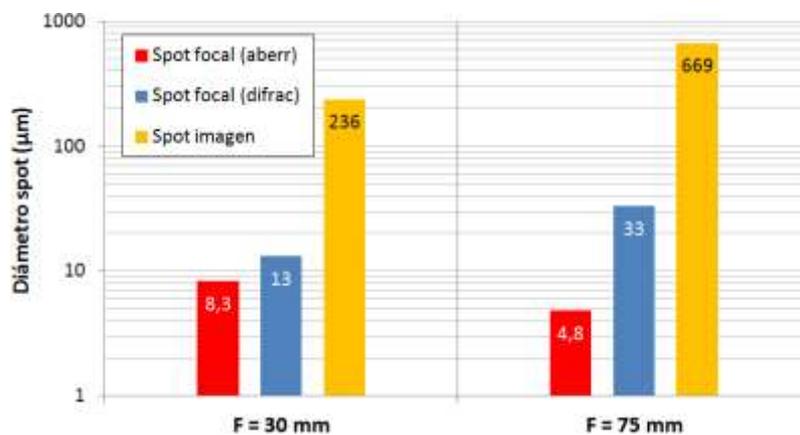

**Figura 220. Diámetro del spot focal e imagen del FSM 2 en el QD 1 utilizando un doblete de 30 mm de focal y de 75 mm.**

En la Figura 220 se muestra el tamaño del spot focal e imagen para el doblete de 30 mm (izquierda) y el de 75 mm (derecha), obtenido mediante una simulación utilizando el montaje de la Figura 219. Se comprueba que incluso la lente de 30 mm está limitada por difracción (ver el spot focal), por lo que al aumentar la focal hasta 75 mm, el spot focal e imagen aumentan de tamaño. La distancia entre el foco y la imagen es de 1,74 mm en el caso de la lente de 30 mm y de 12,53 mm en el caso de la de 75 mm, con un aumento del tamaño del spot en un factor ~3. Con este diseño es posible obtener un spot imagen de 669 μm, unas 15 veces mayor que el tamaño del gap.

A diferencia de la prueba de concepto, para focalizar el haz de 4 mm sobre el QD 2, se ha utilizado una lente para hacer los dos bucles similares. También por similitud, se ha utilizado el mismo doblete que para la focalización del haz en QD 1 (si bien cualquier otra longitud focal sería igualmente válida) y obteniéndose un tamaño similar al que se obtuvo en QD 2. Es importante tener en cuenta que esta técnica de estabilización se basa en conseguir la estabilización simultáneamente en dos planos diferentes. Por lo tanto, no se puede situar al QD 2 en el plano imagen de FSM 2 ya que en ese caso ambos bucles intentarían estabilizar el mismo plano.

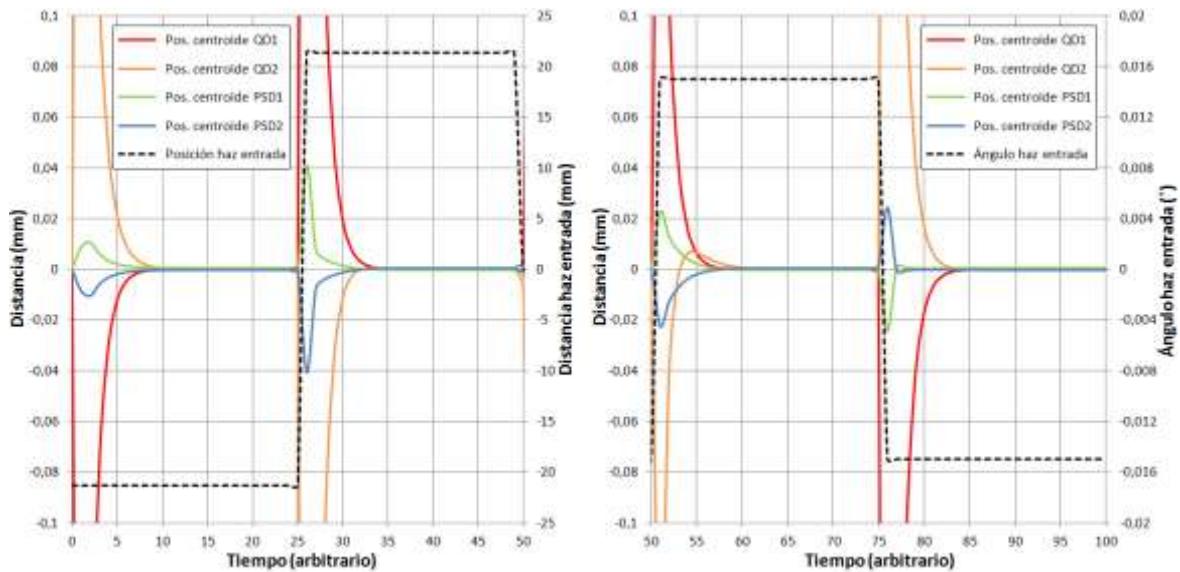

**Figura 221. Simulación de la posición del centroide en los QD del canal de seguimiento y en los PSD del canal de datos ante perturbaciones del haz en posición (izquierda) y ángulo (derecha).**

En la Figura 221 se muestra el resultado de la simulación del sistema de corrección con los dos controles PID funcionando simultáneamente. Con el objetivo de modelar el *beam wander* de forma realista, sobre el haz recibido se ha impuesto una perturbación correspondiente al caso peor de turbulencia fuerte estudiado en el apartado 3.7.5. Además, para considerar los dos tipos de origen del *beam wander*, se ha simulado por separado el caso en el que toda la variación tiene origen angular y el caso en el que toda la variación tiene origen traslacional. Si se utiliza la ecuación (3-19), considerando un régimen de turbulencia fuerte ($C_n^2 = 10^{-13}$ m$^{-2/3}$), una distancia de propagación L de 800 metros y un diámetro del haz $2\omega_0$ de 40 mm (el mismo que en el sistema real, ver apartado 3.7.5), el máximo desplazamiento del centroide del haz debido al *beam wander* tendrá un radio $\sqrt{\langle r_c^2 \rangle}$ a la entrada del telescopio de 21,37 mm. La parte de la izquierda de la Figura 221 representa el caso en el que se considera que la turbulencia tiene un origen puramente traslacional (equivalente a un desplazamiento transversal del haz con un recorrido máximo de ±21,37 mm) y en la parte de la derecha el caso en el que el origen es puramente angular (equivalente a una variación angular en Alice (ver Figura 208) igual a $\beta = \arctan(r_c/L) = \arctan(0{,}2137/800) = \pm 0{,}015°$, que se traduce en un desplazamiento en Bob igual a ± 21,37 mm a 800 metros de distancia de Alice).

Para monitorizar la corrección obtenida, en los planos focales donde irían las fibras ópticas acopladas (tanto a los dos caminos del canal de datos como al canal de sincronismo), se han utilizado detectores PSD (PSD 1, PSD 2 y PSD 3, respectivamente). Se puede comprobar que, sea ante perturbaciones de posición o de ángulo (y también ante ambas simultáneamente, aunque no se muestra en la Figura 221), el sistema de doble corrección consigue eliminar el consiguiente desplazamiento de los spots, llevándolos siempre al eje óptico. En la simulación se aprecia una mayor amplitud en las señales de posición de los QD que en los PSD, lo que se explica por el hecho de que los últimos están situados en el plano focal (a diferencia de los QD), donde los movimientos son menores. Por último, en la simulación se pudo verificar que la corrección conseguida en el canal de datos (y también en el canal de sincronismo) es independiente de la posición de las lentes L3 y L4 (y L7 en el caso del canal de sincronismo), ya que el sistema consigue estabilizar el haz por completo en cualquier posición tras el espejo FSM 2.

# 3.14. CONCLUSIONES

- Se ha analizado la influencia del ruido de fondo y del *beam wander* en un sistema de comunicaciones cuánticas con operación diurna en un entorno urbano en presencia de turbulencia atmosférica. Tras establecer la relación entre ambos fenómenos, se determinó que tienen efectos opuestos en las prestaciones del enlace de comunicación, ya que la técnica para reducir las pérdidas debidas al ruido de fondo (una disminución en el campo de visión), aumenta las pérdidas debidas al *beam wander*, y viceversa. Este comportamiento es especialmente perjudicial en sistemas de QKD, en los que no se puede compensar el ruido de fondo con aumentos de potencia óptica debido a que se transmiten fotones individuales.

- Con el objetivo de reducir el efecto del ruido de fondo y del *beam wander* de forma simultánea, se ha propuesto un sistema de estabilización del haz de llegada. Este sistema permite mitigar el efecto del *beam wander*, al compensar los movimientos del haz recibido con el objetivo de estabilizarlo en el plano focal en un área menor que en presencia de *beam wander*. Esto permite reducir el campo de visión del sistema receptor con el objetivo de minimizar el ruido de fondo acoplado a los detectores del sistema de QKD.

- Para implementar la corrección del *beam wander*, se han propuesto dos estrategias diferenciadas por dónde se realiza la compensación: en el transmisor (Alice) o en el receptor (Bob). Ambas estrategias se basan en la utilización de un canal de seguimiento, adicional a los canales de datos y sincronismo inherentes al sistema de QKD. Se ha discutido cómo debería seleccionarse una u otra estrategia en función de las características del sistema de QKD, de la turbulencia atmosférica y de la distancia del enlace.

- Teniendo en cuenta el sistema de QKD desarrollado y las distancias involucradas, se ha elegido la estrategia de corrección en el receptor debido a su mayor simplicidad y adecuación a un sistema de QKD. Para llevar a cabo su implementación, se han seleccionado y caracterizado los diferentes componentes necesarios, dando una especial importancia al espejo modulable (FSM) y el detector de posición (del cual se han analizado los fundamentos de las dos diferentes tecnologías existentes: detector de efecto lateral PSD y detector de cuadrantes QD).

- Se ha llevado a cabo una serie de experimentos orientados a verificar la independencia del *beam wander* con la longitud de onda. Este es un aspecto crítico en el sistema de corrección propuesto, ya que el canal de datos y el de seguimiento utilizan diferentes longitudes de onda. El canal de seguimiento se emplea para corregir el canal de datos, debido a la imposibilidad de monitorizar este último, por transmitir fotones individuales, dedicados en exclusiva al protocolo cuántico. Tras estos experimentos, se determinó que existe una alta correlación entre ambos canales, lo que demuestra que la corrección propuesta es factible.

- Se ha analizado una posible implementación del sistema de corrección basada en un bucle abierto entre la señal de posición del QD y la señal del FSM. Esta estrategia, pese a ser muy simple, presenta una serie de problemas que imposibilitan obtener una buena corrección del *beam wander*. Se ha analizado la naturaleza de dichos problemas y se ha demostrado la técnica en un experimento con turbulencia natural y artificial.

- Se ha descrito una implementación basada en un bucle cerrado que soluciona los problemas de la estrategia de bucle abierto. Esta estrategia sí permite realizar la corrección del *beam wander*, pero presenta una serie de dificultades técnicas, que se han analizado en detalle. Estas dificultades derivan del hecho de que la estabilización del haz se lleva a cabo en un único plano. Para solucionarlas, se han propuesto diferentes alternativas de implementación, realizando distintos experimentos para demostrar su funcionamiento. Basado en los resultados experimentales, se ha llevado a cabo una evaluación de la potencial mejora de esta estrategia en el sistema de QKD en términos de tasa de error (QBER) y tasa de transmisión (SKR).

- Se ha propuesto una estrategia de corrección basada en estabilizar un haz recibido en dos planos distintos (en contraste con el único plano de estabilización proporcionado por la estrategia en bucle cerrado). Esto consigue estabilizar el haz por completo a lo largo de todo el camino óptico, lo que simplifica en gran medida la calibración del sistema de corrección y su integración en el sistema de QKD. Se ha analizado el fundamento de esta estrategia, realizado una simulación del concepto y se ha propuesto una implementación detallada para integrarla en el sistema de QKD. Por falta de material para llevar a cabo la implementación, se ha llevado a cabo una simulación que demuestra su correcto funcionamiento. Para ello, se han utilizado todos los componentes del sistema de QKD, y se ha diseñado toda la óptica necesaria para implementar la estrategia.

# **Bibliografía**